\documentclass[a4paper,11pt,twoside,openright]{scrbook}
\usepackage{lmodern}
\usepackage[T1]{fontenc}
\usepackage[utf8x]{inputenc}
\usepackage{amsmath}
\usepackage{amssymb}
\usepackage{color}
\usepackage{graphicx}
\usepackage{caption}
\usepackage{subcaption}
\usepackage{float}
\usepackage{youngtab}
\usepackage{cite}
\usepackage{mathtools}
\usepackage{pdflscape}
\usepackage{changepage}
\usepackage{rotating}
\usepackage[english,german]{babel}

\usepackage[toc,page]{appendix}

\usepackage{chngcntr}
\counterwithout{footnote}{chapter}

\usepackage[thinlines]{easybmat}
\usepackage{etex}
\usepackage{enumerate}
\usepackage{amsmath,amssymb}
\usepackage{bm}
\usepackage{epic}
\usepackage{pdfpages}

\numberwithin{equation}{section}
\numberwithin{figure}{section}

\usepackage{tikz}
\usetikzlibrary{decorations.markings}
\usetikzlibrary{arrows}

\usepackage[dvips]{pstricks}
\psset{xunit=.5pt,yunit=.5pt,runit=.5pt}

\makeatletter
\renewcommand{\@chapapp}{}

\makeatother

\usepackage[bookmarksnumbered=true,pdfa]{hyperref}
\newcommand{\mon}{M}

\newcommand{\Nsym}[1]{$\mathcal{N}={#1}$~super Yang-Mills}

\newcommand{\sln}{\ensuremath{\mathfrak{sl}({n})}}
\newcommand{\gln}{\ensuremath{\mathfrak{gl}({n})}}
\newcommand{\gl}[1]{\ensuremath{\mathfrak{gl}({#1})}}
\newcommand{\rap}{\ensuremath{\theta}}
\newcommand{\inh}{\ensuremath{v}}
\newcommand{\rfund}{\ensuremath{\mathbf{R}}}
\newcommand{\perm}{\ensuremath{\mathbf{P}}}
\newcommand{\brt}{\ensuremath{z}}
\newcommand{\osca}{\ensuremath{\mathbf{a}}}
\newcommand{\oscb}{\ensuremath{\mathbf{b}}}
\newcommand{\bvac}{\ensuremath{\vert\Omega\rangle}}
\newcommand{\twist}{\mathcal{D}}

\newcommand{\tm}{{\bf T}}
\newcommand{\Am}{A}

\newcommand{\tmf}{{\bf T}^f}

\def\Tiny{ \font\Tinyfont = cmr10 at 4pt \relax  \Tinyfont}

\newcommand{\ffbox}{\Tiny\yng(1)}

\DeclareMathOperator{\qdet}{qdet}

\DeclareMathOperator{\sgn}{sgn}
\DeclareMathOperator{\tr}{tr}

\newcommand{\magm}{m}

\newcommand{\s}{\mathbf{s}}
\newcommand{\bs}{\bar{\mathbf{s}}}

\newcommand{\spec}{z}

\newcommand{\magn}{x}
\newcommand{\inhdiff}{u}
\newcommand{\inht}{v}

\newcommand{\epb}{j}
\newcommand{\epe}{i}

\newcommand{\graph}{\mathbf{G}}
\newcommand{\inhtset}{\mathbf{v}}
\newcommand{\brtset}{\mathbf{z}}
\newcommand{\magnset}{\mathbf{x}}
\newcommand{\repset}{\boldsymbol{\Lambda}}
\newcommand{\rapset}{\boldsymbol{\theta}}
\newcommand{\stateset}{\boldsymbol{\alpha}}

\newcommand{\sites}{L}
\newcommand{\lines}{N}
\newcommand{\brts}{m}
\newcommand{\dsites}{K}

\newcommand{\dint}{d}

\newcommand{\sfrac}[2]{{\textstyle\frac{#1}{#2}}}
\newcommand{\half}{\sfrac{1}{2}}
\newcommand{\osch}{\mathbf{h}}

\newcommand{\specbaz}{z}

\newcommand{\Qop}{{\bar{\bf{Q}}}}

\newcommand{\Lop}{{\mathcal L}}

\newcommand{\Lbf}{{L}}

\newcommand{\Dbb}{{\mathbb D}}
\newcommand{\Dbf}{{\mathcal D}}
\newcommand{\Rbf}{{\mathbf R}}
\newcommand{\Pbf}{{\mathbf P}}
\newcommand{\Xbf}{\bar{\mathbf X}}
\newcommand{\beq}{\begin{equation}}
\newcommand{\eeq}{\end{equation}}

\newcommand{\oscalg}{{\mathcal H}}

\newcommand{\Gbb}{{\mathbb G}}

\newcommand{\E}{{J}}
\newcommand{\oE}{\hat{J}}
\newcommand{\oJ}{\hat{\mathbf J}}

\newcommand{\oscbp}{\bar{\mathbf b}}
\newcommand{\oscbm}{{\mathbf b}}%

\newcommand{\oscbpo}{\bar{\mathbf b}^{[1]}}
\newcommand{\oscbmo}{{\mathbf b}^{[1]} }
\newcommand{\oscbpt}{\bar{\mathbf b}^{[2]}}
\newcommand{\oscbmt}{{\mathbf b}^{[2]} }
\newcommand{\p}{{p}}

\renewcommand{\a}{{a}}
\renewcommand{\b}{{b}}
\newcommand{\g}{{c}}

\newcommand{\e}{e}

\newcommand{\ac}{{\textsc a}}
\newcommand{\bc}{{\textsc b}}
\newcommand{\gc}{{\textsc c}}

\newcommand{\rep}{{\Lambda}}
\newcommand{\m}{{\lambda}}

\newcommand{\ds}{{\displaystyle}}
\newcommand{\bea}{\begin{eqnarray}}
\newcommand{\eea}{\end{eqnarray}}

\newcommand{\ntr}{{\widehat{\tr}}} 

\newcommand{\Rfi}{{L}_I}

\newcommand{\Rli}{\mathcal{R}_{ I}}

\begin{document}
\selectlanguage{german}
\thispagestyle{empty}
\pagenumbering{roman}

{\large
\vfill
\begin{center}\parindent0pt


\vfill
\vspace*{1cm}
%
 {\LARGE \bf Q-operators, Yangian invariance and the\\ quantum inverse scattering method}
\\
\vspace{1cm}

{\large D i s s e r t a t i o n}\\[5 pt]
{\small (\"Uberarbeitete Fassung)}\\[25pt]
Eingereicht an der\\[10pt]
Mathematisch-Naturwissenschaftlichen Fakult\"at\\[10pt]
der Humboldt-Universit\"at zu Berlin\\[10pt]
von\\[10pt]
{\large \bf Rouven Frassek}\\[1 pt]

    \texttt{
      \href{mailto:rouven.frassek@durham.ac.uk}{rouven.frassek@durham.ac.uk}}



  \vspace{2\baselineskip}

    \textit{
      Department of Mathematical Sciences, Durham University,\\
      South Road, Durham DH1 3LE, United Kingdom\\
      \vspace{0.5\baselineskip}
      Institut für Mathematik und Institut für Physik, Humboldt-Universität zu Berlin,\\
      IRIS-Adlershof, Zum Großen Windkanal 6, 12489 Berlin, Germany\\
      \vspace{0.5\baselineskip}
      Max-Planck-Institut für Gravitationsphysik, Albert-Einstein-Institut,\\
      Am Mühlenberg 1, 14476 Potsdam, Germany\\
      \vspace{0.5\baselineskip}
    }

    \vspace{2\baselineskip}

\end{center}

\thispagestyle{empty}

\newpage
\phantom{blank}
\newpage
\topskip0pt
\vspace*{\fill}
\begin{center}
\section*{\LARGE Zusammenfassung}
\end{center}

\noindent
Inspiriert von den integrablen Strukturen der schwach gekoppelten planaren $\mathcal{N}=4$ Super-Yang-Mills-Theorie studieren wir Q-Operatoren und Yangsche Invarianten. Wir geben eine \"Ubersicht der Quanten-Inverse-Streumethode zusammen mit der Yang-Baxter Gleichung welche zentral f\"ur diesen systematischen Zugang zu integrablen Modellen ist. Den Fokus richten wir auf rationale integrable Spinketten und Vertexmodelle. Wir besprechen einige ihrer bekannten Gemeinsamkeiten und wie sie durch Bethe-Ansatz-Methoden mit Hilfe sogenannter Q-Funktionen gel\"ost werden k\"onnen. Die mathematische Struktur, die diesen integrablen Modellen zu Grunde liegt ist als Yangian bekannt. Um dem Leser das Auftreten des Yangians in diesem Kontext in Erinnerung zu rufen, geben wir einen \"Uberblick \"uber dessen {\small RTT}-Realisierung. Der Hauptteil basiert auf den urspr\"unglichen Publikationen des Autors. Zuerst konstruieren wir Q-Operatoren, deren Eigenwerte zu den Q-Funktionen rationaler homogener Spinketten f\"uhren, welche insbesondere in der Erforschung des spektralen Problems der planaren  $\mathcal{N}=4$ Super-Yang-Mills-Theorie eine wichtige Rolle spielen. Die Q-Operatoren werden als Spuren gewisser Monodromien von R-Operatoren, welche  die Yang-Baxter Gleichung l\"osen, eingef\"uhrt. Unsere Konstruktion erlaubt es uns die Hierarchie der kommutierenden Q-Operatoren und ihre funktionalen Beziehungen herzuleiten. 
Mit dem Ziel die herk\"ommliche Konstruktion der Transfermatrix umgehen zu k\"onnen, studieren wir wie der n\"achste-Nachbarn Hamiltonoperator, sowie h\"ohere lokale Ladungen direkt aus den Q-Operatoren extrahiert werden k\"onnen.
Danach widmen wir uns der Formulierung der Yangschen Invarianzbedingung, wie sie auch im Zusammenhang mit Baumgraphen die bei der Berechnung von Streuamplituden in der  $\mathcal{N}=4$ Super-Yang-Mills-Theorie auftreten, innerhalb der {\small RTT}-Realisierung. Dies erlaubt es uns den algebraischen Bethe-Ansatz anzuwenden und die dazugeh\"origen Bethe Gleichungen herzuleiten, welche f\"ur die Konstruktion der Eigenzust\"ande die Yangsche Invarianz aufweisen, relevant sind. Die Komponenten dieser Eigenzust\"ande der von uns betrachteten Spinketten k\"onnen au\ss erdem als Zustandssummen gewisser zweidimensionaler Vertexmodelle angesehen werden. Zudem analysieren wir die Verbindung zwischen den Eigenzust\"anden der von uns betrachteten Spinketten und den oben genannten Baumgraphen der vierdimensionalen Eichtheorie. Schlussendlich diskutieren wir die von uns vorgelegten Ergebnisse und deren Folgen im Hinblick auf die Erforschung der planaren $\mathcal{N}=4$ Super-Yang-Mills-Theorie.
\\
\\
\\
\noindent
{\bf Schlagw\"orter:}\\
Quanten-Inverse-Streumethode, Bethe ansatz, Q-Operatoren, Yangsche Invarianz, $\mathcal{N}=4$  Super-Yang-Mills-Theorie.

\newpage
\phantom{blank}
\newpage
\topskip0pt
\vspace*{\fill}
\begin{center}
\section*{\LARGE Abstract}
\end{center}
\selectlanguage{english}
\noindent
Inspired by the integrable structures appearing in weakly coupled planar $\mathcal{N}=4$ super Yang-Mills theory, we study Q-operators and Yangian invariants of rational integrable spin chains.
We review the quantum inverse scattering method ({\small QISM}) along with the Yang-Baxter equation which is the key relation in this systematic approach to study integrable models. Our main interest concerns rational integrable spin chains and lattice models. We recall the relation among them and how they can be solved using Bethe ansatz methods incorporating so-called Q-functions. The basic mathematical structure underlying these integrable models is known as the Yangian. In order to remind the reader how the Yangian emerges in this context, an overview of its so-called {\small RTT}-realization is provided.
The main part is based on the author's original publications. 
Firstly, we construct Q-operators whose eigenvalues yield the Q-functions for rational homogeneous spin chains that, in particular, play an important role in the study of the spectral problem of planar $\mathcal{N}=4$ super Yang-Mills theory.
The Q-operators are introduced as traces over certain monodromies of R-operators obtained from the Yang-Baxter equation. Our construction allows us to derive the hierarchy of commuting Q-operators and the functional relations among them. 
With the aim of being able to avoid the ordinary transfer matrix construction, we study how the nearest-neighbor Hamiltonian and in principle also higher local charges can be extracted from the Q-operators directly. 
Secondly, we formulate the Yangian invariance condition, also studied in relation to tree-level scattering amplitudes of $\mathcal{N}=4$ super Yang-Mills theory, in the {\small RTT}-realization. We find that Yangian invariants can be interpreted as special eigenvectors of certain inhomogeneous spin chains. This allows us to apply the algebraic Bethe ansatz and derive the corresponding Bethe equations that are relevant to construct the invariants. Furthermore, we examine the connection between the Yangian invariant spin chain eigenstates whose components can be understood as partition functions of certain two-dimensional lattice models and tree-level scattering amplitudes of the four-dimensional gauge theory. 
Finally, we conclude and discuss some future directions and implications of our studies for planar $\mathcal{N}=4$ super Yang-Mills theory.
\\
\\
\\
\noindent
{\bf Keywords:}\\
Quantum inverse scattering method, Bethe ansatz, Q-operators, Yangian invariance, $\mathcal{N}=4$ super Yang-Mills theory.
\newpage
\phantom{blank}
\newpage
\selectlanguage{english}
\newpage
\vfill
\section*{List of own publications}
\vspace{0.5cm}
{Bethe Ansatz for Yangian Invariants: Towards Super Yang-Mills Scattering Amplitudes}{\newline  R. Frassek, N. Kanning, Y. Ko and M. Staudacher}\newline{Published in Nucl.Phys.B 883 (2013) 373-424}{\newline}{\href{http://arxiv.org/abs/1312.1693}{\color{black}arXiv:1312.1693}}{}
\vspace{1cm}
\newline{From Baxter Q-Operators to Local Charges}{\newline R. Frassek, C. Meneghelli}\newline{Published in J.Stat.Mech. (2013) P02019}{\newline}{\href{http://arxiv.org/abs/arXiv:1207.4513}{\color{black}arXiv:1207.4513}}
\vspace{1cm}
\newline{Baxter Operators and Hamiltonians for "nearly all" Integrable Closed gl(n) Spin Chains}{\newline R. Frassek, T. \L{}ukowski, C. Meneghelli and M. Staudacher}\newline{Published in Nucl.Phys.B 874 (2013) 620-646}{\newline}{\href{http://arxiv.org/abs/1112.3600}{\color{black}arXiv:1112.3600}}
\vspace{1cm}
\newline{Oscillator Construction of su(n|m) Q-Operators}{\newline R. Frassek, T. \L{}ukowski, C. Meneghelli and M. Staudacher}\newline{Published in Nucl.Phys.B 850 (2011) 175-198}{\newline}{\href{http://arxiv.org/abs/1012.6021}{\color{black}arXiv:1012.6021}}
\vspace{1cm}
\newline{Baxter Q-Operators and Representations of Yangians}
{\newline V. V. Bazhanov, R. Frassek, T. \L{}ukowski, C. Meneghelli and M. Staudacher}
\newline{Published in Nucl.Phys.B 850 (2011) 148-174}
{\newline}{\href{http://arxiv.org/abs/1010.3699}{\color{black}arXiv:1010.3699}}
\vfill
\setcounter{tocdepth}{2}
\tableofcontents
\chapter{Introduction}\label{ch:intro}
\pagenumbering{arabic}
 \paragraph{Particle Physics and the Standard Model\smallskip{}
}

\noindent

\noindent The development of quantum field theory in the second half
of the \textit{$20$th} century marks a milestone in the history of
theoretical physics. The electromagnetic, weak and strong nuclear
forces are uniformly described in this framework by the Standard Model
of particle physics \cite{aitchinson2007}. The predictive power of
the model is remarkable. 
The basic theoretical concept used to calculate probabilities of particle scatterings from the Standard Model Lagrangian are Feynman diagrams.
They describe how the colliding particles interact on a microscopic
scale. This method proves to be very powerful in certain energy regimes
and is in perfect agreement with experiment. However, its perturbative
nature makes strongly coupled regimes hard to access analytically.
A key example for this phenomenon is Quantum Chromodynamics ({\small QCD})
which describes the strong interaction between quarks and gluons.
In contrast to Quantum Electrodynamics, {\small QCD} is strongly coupled
at low energies. 

\paragraph{Integrability in gauge theories\smallskip{}
}

\noindent

\noindent There is growing amount of evidence that {\small QCD} and in general Yang-Mills theories possess additional hidden symmetries at the quantum level that are not manifest in the Lagrangian formulation. 
Remarkably, in certain regimes this extra symmetry allows to describe Yang-Mills theories as an integrable system\,\footnote{The first example of the appearance of an integrable model in effective {\small QCD} goes back to 1993 and can be found in \cite{Lipatov1994571-574a,Faddeev1995}. For $\mathcal{N}=4$ super Yang-Mills see also \cite{Minahan:2002ve,BeisertNucl.Phys.B670:439-4632003}}.
In such situations the integrable structure may allow to overcome
the difficulties of accessing the strongly coupled regimes of gauge
theories analytically.
The study of integrable models is a research line of theoretical
physics on its own. Their simplicity and beauty attracted many physicists from different research fields. Interestingly, also Richard Feynman himself got fascinated by models that can be solved by Bethe ansatz methods\,\footnote{\textit{``I got really fascinated by these $1+1$-dimensional models
that are solved by the Bethe ansatz and how mysteriously they jump
out at you and work and you don\textquoteright{}t know why. I am trying
to understand all this better.''} \cite{Batchelor2007}%
}. In particular, these models can often be understood on a non-perturbative
or exact level. As such they appear in statistical physics but also
allow the study of certain $1+1$-dimensional quantum field theories
in all regimes of the coupling using the thermodynamic Bethe ansatz
({\small TBA}) \cite{Yang1969,Gaudin1971,Takahashi1971,Zamolodchikov1990,Destri1992,Klumper1991,Klumper1992,Dorey1996}.
To understand how integrability appears in $3+1$-dimensional gauge theories it is convenient to study $\mathcal{N}=4$ super Yang-Mills theory, the maximally supersymmetric gauge theory in four dimensions, which provides an ideal testing ground. In particular, there is strong evidence that this theory is integrable in the 't Hooft limit! 
Furthermore, the {\small AdS/CFT} duality allows to actually compare results
obtained using integrability methods to the strongly and weakly coupled
regime of the {\small AdS/CFT} system; further details are contained in the next paragraph. Of course,
the results obtained for $\mathcal{N}=4$ super Yang-Mills theory cannot always be carried over directly to other gauge theories
like {\small QCD}. As mentioned above, the maximal amount of supersymmetry makes the theory in many ways simpler.
However, a complete understanding of this theory is conceptually extremely important. To describe physical quantities of  gauge theories with less supersymmetry as perturbations around their exact counterparts in $\mathcal{N}=4$ super Yang-Mills is certainly one of the most exciting prospects of this subject.

\paragraph{{\small AdS/CFT}, integrability and the thermodynamic Bethe ansatz\smallskip{}
}

\noindent

\noindent The {\small AdS/CFT} correspondence \cite{MaldacenaAdv.Theor.Math.Phys.2:231-2521998,WittenAdv.Theor.Math.Phys.2:253-2911998,GubserPhys.Lett.B428:105-1141998}
has been one of the most important advances in string theory of the
last decades. It states the equivalence of certain superstring theories
on anti de Sitter space-time ({\small AdS}) and supersymmetric conformal
quantum field theories ({\small CFT}). The {\small AdS/CFT} duality is
of strong/weak type, which means that it provides a way to describe
the strongly coupled regime of gauge theories using weakly coupled
string theory and vice versa. By now there exist several
realizations of the correspondence. As mentioned above, we focus on
the duality between the maximally supersymmetric gauge theory $\mathcal{N}=4$
super Yang-Mills in four dimensions and type {\small IIB} superstring
theory compactified on $AdS_{5}\times S^{5}$. The discovery of integrable
structures in the gauge theory shed new light on this duality. An overview on this subject can be found in the recent review collection \cite{Beisert2010a}. It was noticed that in the 't Hooft limit, where non-planar Feynman diagrams are suppressed, the spectrum of local operators of $\mathcal{N}=4$ super Yang-Mills
can be calculated using integrability techniques. More precisely,
the dilatation operator can be identified with the nearest-neighbor
Hamiltonian of a $\mathfrak{psu}(2,2\vert4)$ integrable
super spin chain at one-loop order of the 't Hooft coupling constant \cite{Minahan:2002ve,BeisertNucl.Phys.B670:439-4632003}.
Therefore, powerful Bethe ansatz methods can be applied to obtain
the one-loop anomalous dimensions. This integrable structure originates
from a hidden infinite-dimensional Yangian symmetry algebra which
typically appears in integrable models. The strongly coupled regime
of the gauge theory is described by a non-linear string sigma model
which is known to be classically integrable. According to the {\small AdS/CFT}
dictionary, the string energy is equal to the anomalous dimension.
Thus, the question whether integrability holds in the intermediate regime
immediately arises. Assuming integrability, asymptotic Bethe ansatz
equations and the dispersion relation were written down for the all-loop
{\small AdS/CFT} system. However, the so called wrapping effects are
not captured by these equations. They have to be taken into account
whenever the order of perturbation theory exceeds the length of the
operator under investigation \cite{AmbjornNucl.Phys.B736:288-3012006,Bajnok2009}. The way out is the thermodynamic Bethe
ansatz ({\small TBA}) \cite{Bombardelli2009,Arutyunov2009,Gromov2010}. The violation of the asymptotic Bethe equations can be described
in terms of L\"{u}scher corrections and yields the exact spectrum predicted
by the thermodynamic Bethe ansatz.

\paragraph{Baxter Q-operators\label{qop-1}}

\noindent

\noindent To solve the {\small TBA} equations of the spectral problem
additional input regarding the analytic structure of so called Y-functions
is needed. This system of equations can be reduced to a finite set
of non-linear integral equations ({\small FiNLIE}) in terms of Wronskians
of Q-functions, see \cite{Gromov2014a} and references therein. Therefore, the analytic structure of the Y-functions
is inherited from the Q-functions which are the eigenvalues of certain
unknown Q-operators. A direct construction of the operators would
already contain the information of the analytic structure! Originally,
the concept of the Q-operator was introduced by R.J. Baxter in 1971 in order
to calculate exactly the partition function of the eight-vertex model
\cite{Baxter:1972hz}. The method of functional relations and commuting
transfer matrices originated in this work. In a series of papers~\cite{Bazhanov2010a,Bazhanov2010,Frassek2010,Frassek2011},
my collaborators and I constructed the Q-operators for homogeneous
spin chains from fundamental principles. The construction follows
the quantum inverse scattering method employing
degenerate solutions of the Yang-Baxter equation as generating objects
along the lines of~\cite{BazhanovCommun.Math.Phys.177:381-3981996,Bazhanov1997,Bazhanov1999}. The hierarchy
of Q-operators was derived in this way. The functional relations among
them are best understood in the notion of the Hasse diagram~\cite{Tsuboi2009}.
The Bethe equations follow without making an ansatz for the wave function.
The Q-operators for the $\mathfrak{psu}(2,2\vert4)$ super spin chain
describing the complete one-loop level of the planar {\small AdS/CFT}
system can be deduced from the work presented here. Furthermore, we show how to extract the nearest-neighbor
Hamiltonian and in principle also higher local charges from Q-operators
directly in the non-supersymmetric case \cite{Frassek2013}. This circumvents the paradigm
of constructing the transfer matrix with equal representations in
quantum and auxiliary space and underlines the strength of the Q-operator
construction. Furthermore, the energy formula for the spin chain in
terms of Bethe roots is proven for a large class of representations.

\paragraph{An integrability framework for scattering amplitudes\label{amps}}

\noindent

\noindent In the last years, there has been a lot of progress in the
understanding of scattering amplitudes of four-dimensional quantum
field theories and especially in $\mathcal{N}=4$ super Yang-Mills
\cite{Roiban2011}. In particular, the duality between scattering
amplitudes and Wilson loops led to deep insights concerning the symmetries
of the theory. More specifically, it was shown that the superconformal
symmetry and the dual superconformal symmetry combine into a Yangian
symmetry~\cite{Drummond2009}. It is the same Yangian which was found
in the spectral problem. This holds for planar tree-level scattering
amplitudes but also manifests itself in the loop-integrands, see e.g. \cite{trnkaphd}.
The Yangian, as defined by Drinfeld, arises as a consequence of the Yang-Baxter equation underlying
one-dimensional quantum integrable models in the so-called rational
case. Thus the Yangian structure appearing in the
four-dimensional scattering amplitudes naturally suggests a
possible reformulation of the problem as an integrable model.
The Gra\ss mannian integral
formulation provides a convenient way to construct scattering amplitudes.
It was argued that all amplitudes can be obtained from on-shell building
blocks using the {\small BCFW} recursion relation~\cite{Arkani-Hamed2012}.
Furthermore, it was shown how the Hamiltonian of the $\mathfrak{psu}(2,2\vert4)$
spin chain is closely related to tree-level amplitudes in~\cite{Zwiebel2011}.
The Yang-Baxter equation already appeared
in the work of~\cite{Arkani-Hamed2012} and was further exploited
in~\cite{Ferro2012,Ferro:2013dga}. Here the basic building blocks of the amplitudes
were deformed by spectral parameters and directly derived from the
bootstrap equation\,%
\footnote{The bootstrap equation is well known in the context of $1+1$-dimensional
integrable quantum field theories. It describes the formation of bound
states, see e.g.~\cite{zoltanbook}. %
}. It was speculated that at one-loop level the spectral
parameters can be reinterpreted as {\small IR}-regulators of the amplitudes.
However, while
integrability has been essential for the (conjectured) solution of the
spectral problem, in the scattering problem it has not yet directly
led to any practical advantages in computations, with the notable
exception of the recent approach of
\cite{Basso:2013vsa,Basso:2013aha,Basso2014b,Basso2014,Papathanasiou:2013uoa}. The reason
is that the associated large integrability toolbox, the quantum
inverse scattering method, is so far available only for the
calculation of anomalous dimensions. Its application usually leads to
powerful Bethe ansatz methods. In contradistinction, apparently no
such methods exist to-date for directly exploiting Yangian invariance.
In this thesis we study how Yangian invariants can be constructed using a Bethe ansatz, see \cite{Frassek2013b}. We derive a special set of Bethe equations 
that allow to construct eigenstates of a spin chain that are Yangian invariant from the algebraic Bethe
ansatz. These spin chain eigenstates are the finite-dimensional analogues of the
scattering amplitudes! Furthermore, they can be interpreted as the
partition function on a certain lattice similar to the one studied by Baxter in the context of the perimeter Bethe ansatz~\cite{Baxter:1987}. 
\newpage
\section*{Overview}

\noindent

\noindent 
This thesis is organised in two parts.
\paragraph{Introductory part}
Chapter~\ref{ch:ybe} serves as an introduction to the algebraic Bethe ansatz and also contains a short introduction to the coordinate Bethe ansatz. In Chapter~\ref{ch:2dlattice} we present some aspects of two-dimensional lattice models and their relation to the algebraic Bethe ansatz. In particular, we discuss the notion of crossing, unitarity and the formation of bound states which is commonly studied in relation to integrable quantum field theories in the context of lattice models. The algebraic Bethe ansatz is endowed with a rich mathematical structure. We introduce the Yangian and discuss some of its properties that are of importance for us in Chapter~\ref{ch:yangian}.
\paragraph{Research part}
The second part of the thesis is based on our original contributions. 
In Chapter~\ref{ch:qop} we construct Q-operators for rational homogeneous spin chains in the framework of the quantum inverse scattering method and discuss how the Hamiltonian and higher local charges can be obtained from them. In Chapter~\ref{ch:BAforYI} we discuss how the amplitude problem as mentioned above can be implemented in this framework. Furthermore, we use the algebraic Bethe ansatz to construct finite-dimensional counterparts of the actual tree-level scattering amplitudes. Apart from that, we provide several appendices. In particular, Appendix~\ref{susylax} contains the R-operators for Q-operators of rational homogeneous supersymmetric $\gl{n\vert m}$ spin chains. Appendix~\ref{app:comp}   contains a curious relation between the special points of an $\gln$-invariant R-matrix and the {\small BCFW} recursion relation. Last but not least, in Appendix~\ref{app:hop}, we give a concise discussion of the ``hopping'' representation of a spin chain Hamiltonian as it also appears in the spectral problem of $\mathcal{N}=4$ super Yang-Mills theory and present a neat way to rewrite it. 

\newpage
\vspace{0.5cm}
\noindent 
The second part of this thesis is based on the articles~\cite{Bazhanov2010,Frassek2010,Frassek2011,Frassek2013,Frassek2013b}.   
 
 \begin{itemize}
\item Chapter~\ref{ch:qop} and Appendix~\ref{app:hfq} are based on \cite{Bazhanov2010, Frassek2011,Frassek2013}. We derive the R-operators for Q-operators from the Yang-Baxter equation and derive the functional relations among the Q-operators. Furthermore, we present a method to extract the Hamiltonian and in principle also higher local charges directly from the Q-operators.
\item Chapter~\ref{ch:BAforYI} is based on~\cite{Frassek2013b}. We formulate the condition for  Yangian invariance in the RTT-realization. This enables us to study Yangian invariants in the framework of the algebraic Bethe ansatz. Furthermore, we present a relation between tree-level scattering amplitudes in \Nsym{4} and spin chain eigenstates. 
\item In Appendix~\ref{susylax} we give the expression for the Lax operators used to construct the Q-operators for $\gl{n|m}$ obtained in \cite{Frassek2010}. Furthermore, it contains the unpublished Lax operators for Q-operators with more general representations of  $\gl{n|m}$ in the quantum space.
\item Appendix~\ref{app:hop} contains an expression for the Hamiltonian as given by N.~Beisert in \cite{Beisert:2004ry} using certain ``hopping`` operators that can also be used to express the corresponding R-matrix, see also Section~\ref{sec:osc-examples}. Furthermore, we present a way to write the Hamiltonian building on the form of its action given by B.~Zwiebel in \cite{Zwiebel2011}.
\item Appendix~\ref{app:comp} gives a receipe to construct Yangian invariants from the building blocks introduced in  Section~\ref{sec:osc-examples}. Furthermore, we discuss special points of the R-matrix in relation to scattering amplitudes. At these points the R-matrix can be decomposed into invariants of fewer legs. 
\item In Appendix~\ref{app:su3} we present the coordinate wave function for a $\gl3$ invariant spin chain with fundamental and anti-fundamental representations at each site of the quantum space.

\end{itemize}

 \begin{center}
  - - - - - - - - -
\end{center}
\noindent
Text elements of the author's publications \cite{Bazhanov2010,Frassek2010,Frassek2011,Frassek2013,Frassek2013b} have been used in this thesis.  

\chapter{Bethe ansatz techniques}\label{ch:ybe}
In this chapter we introduce the basic framework of the quantum inverse scattering method (QISM). This systematic approach to study integrable models was proposed by the Leningrad School led by Ludvig Faddeev in the late 1970s \cite{Faddeev1980}. The Yang-Baxter equation plays a key role in this discussion which has been the fundamental equation underlying integrable systems in two-dimensional statistical physics and quantum integrable field theories since the 1960s. For a short but instructive summary of the developments until the late 80s see e.g. \cite{Jimbo1989}. 

Here we focus on rational solutions to the Yang-Baxter equation.  We will outline how these solutions can be obtained and then employ them to construct families of commuting operators. Furthermore, we discuss how the spectrum of the latter can be found from the algebraic Bethe ansatz. The discussion closely follows the survey of Ludvig Faddeev \cite{Faddeev2007} which provides an excellent introduction to integrable spin chains. The topics that are relevant for the following chapters of this thesis will be treated more explicitly than others. In particular, we turn our attention to rational inhomogeneous spin chains and Q-functions which will reappear at several points in this thesis. In particular, we will see in Chapter~\ref{ch:qop} how the Q-operators whose eigenvalues yield the Q-functions can be constructed directly from the Yang-Baxter equation.  In addition we discuss how the eigenvalues of the commuting family of operators and their eigenvectors can be obtained from the solution of the Bethe equations. Also we will introduce a graphical notation which is helpful to understand the structures of the equations and furthermore makes the relation to two-dimensional vertex models that are discussed in Chapter~\ref{ch:2dlattice} apparent.

At the end of this chapter we review the coordinate Bethe ansatz invented by Hans Bethe in 1931 \cite{Bethe1931}. It yields a different representation of the eigenfunctions of the family of commuting operators and is more intuitive than the algebraic Bethe ansatz. This representation reappears in Chapter~\ref{ch:BAforYI} where Baxter's perimeter Bethe ansatz is discussed.
\section{The Yang-Baxter equation}\label{sec:ybe}
 An R-matrix acts on the tensor product of two vector spaces 
\begin{equation}\label{defr}
 R(u):V\otimes V'\rightarrow V\otimes V'\,,
\end{equation} 
where $V$ and $V'$ are not necessarily finite-dimensional. It depends on the complex spectral parameter $u$. We consider the tensor product of three vector spaces $V_1\otimes V_2 \otimes V_3$ and define the action of $R_{ij}$ with $i,j=1,2,3$ and $i\neq j$ on $V_1\otimes V_2 \otimes V_3$ using \eqref{defr} with $V=V_i$ and $V'=V_j$ imposing that $R_{ij}$ acts trivially as the identity on the space $V_k$ with $k\neq i,j$. To be precise,  $R_{12}$ acts on the space $V_1\otimes V_2\otimes V_3$ as
\begin{equation}
 R_{12}(u)=R(u)\otimes \mathbb{I}\,,
\end{equation} 
where $V=V_1$, $V'=V_2$ and $\mathbb{I}$ denotes the identity on operator acting on $V_3$. The analogue holds true for $R_{13}$ and $R_{23}$ which act as an identity in space $2$ and $1$ respectively. Using the notation introduced above the Yang-Baxter equation can conveniently be written as
\begin{equation}\label{ybe1}
 R_{12}(u_1-u_2)R_{13}(u_1-u_3)R_{23}(u_2-u_3)= R_{23}(u_2-u_3)R_{13}(u_1-u_3)R_{12}(u_1-u_2)\,.
\end{equation} 
In the following we will often encounter vector spaces formed out of the multiple tensor product $\bigotimes_{i} V_i$. The operator acting non-trivially as the R-matrix $R$ on two spaces $V_i$ and $V_j$ will then also be denoted by $R_{ij}$ and no reference to the spaces where it acts trivially will be made. Our glossary will not distinguish between $R$ and $R_{ij}$ and we call $R_{ij}$ an R-matrix.

It is convenient to think about an R-matrix as a vertex of two intersecting straight lines that indicate the two different spaces on which $R_{ij}$ acts non-trivially
\begin{equation}\label{Rvert}
 R_{ij}(u_i-u_j)=\input{content/pictures/rmatrix.tex}\,.
\end{equation} 
The spaces on which $R_{ij}$ acts trivially are not indicated in the graphical notation \eqref{Rvert}.
We associated a spectral parameter to spaces $i$ and $j$ and assume that the R-matrix depends on the difference of those. Also we have introduced an arrow to each line. In this context the arrow directions distinguish between multiplication from the left and from the right. Elements multiplied to the R-matrix \eqref{Rvert} from the right are attached to endpoints with outpointing arrows and vice versa. As an example we consider the product of two R-matrices $R_{12}$ and $R_{13}$ sharing the common space $1$:
\begin{equation}\label{tworex}
  R_{12}(u_1-u_2)R_{13}(u_1-u_3)=\input{content/pictures/2rmatrix.tex}\,.
\end{equation} 
Using the diagrammatic notation introduced the Yang-Baxter equation can be depicted as 
\begin{align}\label{pic:ybe}
 \begin{aligned}
  \includegraphics[scale=0.70]{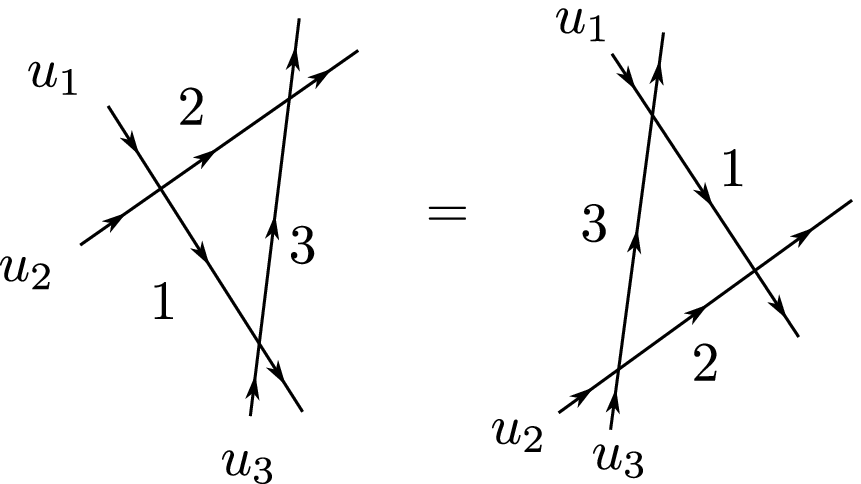}
 \end{aligned}
 \,,
\end{align}
compare also \eqref{tworex}. Thus, we can think about the Yang-Baxter equation in a graphical way, namely as the condition that the line $3$ can be moved through the vertex formed by the intersecting lines $1$ and $2$. This has some crucial implications on the partition functions of two-dimensional vertex models that will be discussed in Chapter \ref{ch:2dlattice}. It is known as Z-invariance.

Naively, the Yang-Baxter equation \eqref{ybe1} contains three spectral parameters each of which is associated to a line. However, we imposed that each R-matrix depends on the difference of two spectral parameters. The difference property reduces the number of free parameters which becomes obvious if we define the variables $u$ and $v$ satisfying
\begin{equation}\label{rapid}
 u=u_1-u_3,\quad v=u_2-u_3,\quad u-v=u_1-u_2\,.
\end{equation} 
This choice of spectral parameters is often more convenient and will be used in the following. However, in terms of the diagrammatic description and in the context of vertex models it is more natural to think about one spectral parameter per space or per line in the graphical notation. Under the choice of variables \eqref{rapid} the Yang-Baxter equation \eqref{ybe1} takes the form
\begin{equation}\label{ybeshort}
R_{12}(u-v)R_{13}(u)R_{23}(v)= R_{23}(v)R_{13}(u)R_{12}(u-v)\,.
\end{equation} 

Finding explicit solutions to the Yang-Baxter equation is a rather difficult problem and a general solution is unknown\,\footnote{There are systematic ways to obtain R-matrices from representations of quantum groups such as Yangians and quantum affine algebras. These methods are based on the existence of a universal R-matrix \cite{Drinfeld1985,Drinfeld1988a,Jimbo1986,Jimbo1986a}. Explicit formulas were obtained in \cite{Khoroshkin1991} and evaluated in e.g. \cite{Boos2010a}}. 
In general, \eqref{ybeshort} yields an overdetermined set of equations. Let us restrict to the case where $\dim V_i=d$ for $i=1,2,3$. The R-matrices can then be written as
\begin{equation}
 R(u)=\sum_{a_{1,2},b_{1,2}=1}^d\, \langle a_1,a_2\vert R(u)\vert b_1,b_2\rangle\,e_{a_1b_1}\otimes e_{a_2b_2}\,,
\end{equation} 
where $\langle a_1,a_2\vert R(u)\vert b_1,b_2\rangle$ denotes the components of the R-matrix and $e_{ab}$ are $d\times d$ matrices with the entry $1$ in the $a^{\text{th}}$ row and $b^{\text{th}}$ column
\begin{equation}
 \left(e_{ab}\right)_{ij}=\delta_{ai}\delta_{bj}\,.
\end{equation} 
Furthermore, we restrict to the case where $R_{23}$, $R_{13}$ and $R_{12}$ are the same operators acting on different spaces. In this example, we end up with $d^6$ equations for $d^4$ unknowns. The $d^6$ equations correspond to the elements of the $d^3\times d^3$ matrices that arise from the product of the three R-matrices on each side of the Yang-Baxter equation \eqref{ybeshort}. The unknowns are given by the $d^4$ components $\langle a_1,a_2\vert R(u)\vert b_1,b_2\rangle$.
\section{A first solution to Yang-Baxter}
We will now encounter the first solution to the Yang-Baxter equation \eqref{ybeshort}. We define the R-matrix
\begin{equation}\label{fundr}
 \rfund(u)=u+\perm\,.
\end{equation} 
It is an ${n^2}\times n^2$ matrix acting on the tensor product of the two copies of ${\mathbb C}^n$
\begin{equation}\label{Ract}
 \rfund(u):{\mathbb C}^n\otimes {\mathbb C}^n\rightarrow {\mathbb C}^n\otimes {\mathbb C}^n\,.
\end{equation} 
Here $\perm$ is the permutation operator. Its action on $w_1\otimes w_2\in {\mathbb C}^n\otimes {\mathbb C}^n$ is given by
\begin{equation}\label{permop}
 \perm(w_1\otimes w_2)=w_2\otimes w_1\,.
\end{equation} 
The permutation operator can conveniently be written in terms of the elementary matrices $e_{ab}$ with $a,b=1,\ldots,n$. It takes the form
\begin{equation}
 \perm=\sum_{a,b=1}^n e_{ab}\otimes e_{ba}\,.
\end{equation} 
The R-matrix in \eqref{fundr} can then be decomposed as
\begin{equation}\label{rfunddec}
 \rfund(u)=u\,\mathbb{I}\otimes\mathbb{I}+\sum_{a,b=1}^n e_{ab}\otimes e_{ba}\,,
\end{equation} 
where $\mathbb{I}$ is the $n\times n$ identity matrix which we did not write out explicitly in \eqref{fundr}. This decomposition makes the action of the R-matrix on each of the two spaces in the tensor product \eqref{Ract} manifest. For $\rfund_{ij}$ defined through \eqref{rfunddec} as discussed in Section~\ref{sec:ybe} the Yang-Baxter equation \eqref{ybe1} with $V_i={\mathbb C}^n$ is satisfied. Namely, using the spectral parameters $u$ and $v$ as defined in \eqref{rapid} it holds that
\begin{equation}\label{ybeshortfund}
 \rfund_{12}(u-v)\rfund_{13}(u)\rfund_{23}(v)= \rfund_{23}(v)\rfund_{13}(u)\rfund_{12}(u-v)\,.
\end{equation} 
To check that the R-matrix \eqref{fundr} is indeed a solution to the Yang-Baxter equation it is convenient to expand both sides of the equation in terms of the spectral parameters. As $u$ and $v$ are arbitrary all coefficients of this expansion have to agree at any order. Here we sketch how the Yang-Baxter equation can be shown diagrammatically. As the R-matrix \eqref{fundr} only contains the identity and the permutation it is convenient to introduce the diagrammatic notation
\begin{equation}\label{frmatrix}
 \rfund(u)=\input{content/pictures/frmatrix.tex}\,,
\end{equation} 
where the first term denotes the identity times the spectral parameter $u$ and the second the permutation.
Substituting this expression into the Yang-Baxter equation \eqref{ybeshort} we can compare the permutations that arise from the R-matrix to all powers of the spectral parameter graphically, cf. \eqref{pic:ybe}. Here we start with the lowest order where all R-matrices are given by the permutation operator:
\begin{align}\label{pic:ybeo1}
 \begin{aligned}
  \includegraphics[scale=0.70]{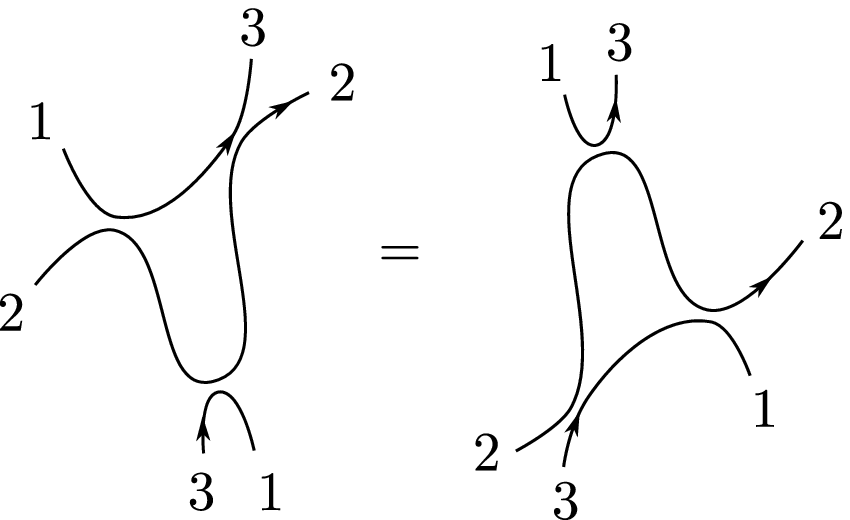}
 \end{aligned}
 \,.
\end{align} 
As we can read off from this diagram the left and the right hand side of \eqref{pic:ybe} reduce to the permutations 
\begin{equation}
 1\rightarrow 3,\quad  2\rightarrow 2,\quad 3\rightarrow 1\,.
\end{equation} 
To complete the proof one has to check the Yang-Baxter equation \eqref{ybeshort} to higher orders of the spectral parameters. It is rather easy to see that it is satisfied for the orders $u^2v,uv^2,u^2,v^2$ as at most one permutation is involved. It remains to do the analysis for the terms proportional to $u$, $v$ and $uv$ which involve sums of different diagrams. Here, we check the order $u$ and leave the two remaining for the reader:
\begin{align}\label{pic:ybeo2}
 \begin{aligned}
  \includegraphics[scale=0.70]{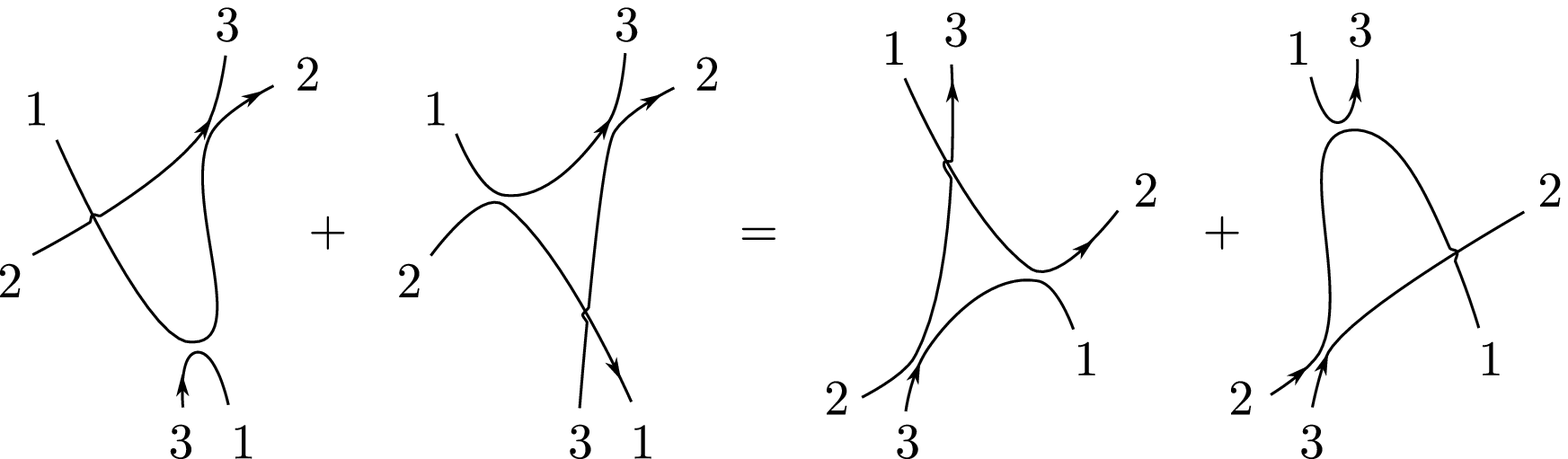}
 \end{aligned}
 \,.
\end{align} 
Thus also at this order the Yang-Baxter equation \eqref{ybeshort} is satisfied. Both sides reduce to the permutations
\begin{equation}
  1\rightarrow 2,\quad  2\rightarrow 3;\quad 3\rightarrow 1\quad\text{and}\quad 1\rightarrow 3,\quad  2\rightarrow 1,\quad 3\rightarrow 2\,.
\end{equation} 
\section{Lax matrices}\label{sec:lax}
As mentioned previously, solutions to the Yang-Baxter relation are usually hard to find.
However, we can use known solutions to 
determine further solutions in a systematic way \cite{Faddeev2007}. Here we will use the R-matrix $\rfund$ defined in \eqref{fundr} to demonstrate this mechanism. Let us consider the Yang-Baxter equation \eqref{ybeshort} with $V_1=\mathbb{C}^n$ and $V_2=\mathbb{C}^n$, leaving $V_3$ unspecified\,\footnote{This equation is sometimes referred to as the RLL-relation. As we will see in Chapter~\ref{ch:yangian} it provides a defining relation of the Yangian algebra $\mathcal{Y}(\gln)$.}
\begin{equation}\label{rll}
 \rfund_{12}(u-v){\mathbf L}_{13}(u){\mathbf L}_{23}(v)= {\mathbf L}_{23}(v){\mathbf L}_{13}(u)\rfund_{12}(u-v)\,.
\end{equation} 
Here ${\mathbf L}_{13}$ (${\mathbf L}_{23}$) is an $n\times n$ matrix acting on the first (second) space with arbitrary operatorial entries acting on the third space $V_3$. Furthermore, inspired by the form of the R-matrix $\rfund$ we require that the leading term in the spectral parameter of ${\mathbf L}$ is proportional to the identity. As we will see later, relaxing this condition yields solutions relevant for the construction of Q-operators. However, here we make the ansatz
\begin{equation}\label{laxan}
 {\mathbf L}(u)=u+\sum_{a,b=1}^ne_{ab}\otimes J_{ba}\,,
\end{equation} 
where for convenience we suppressed the identity operators in the two different spaces in the term proportional to the spectral parameter. In the following we refer to the R-matrices ${\mathbf L}$ satisfying \eqref{rll} as Lax matrices (or Lax operators).
We will now use the explicit expression for R-matrix introduced in \eqref{fundr} to derive certain algebraic constraints on the operatorial entries of the Lax matrices. After substituting \eqref{fundr} into Yang-Baxter equation \eqref{rll} we obtain the relation
\begin{equation}\label{preyangcom}
\begin{split}
 (u-v)[{\mathbf L}_{13}(u),{\mathbf L}_{23}(v)]= \perm_{12}\left({\mathbf L}_{13}(v){\mathbf L}_{23}(u)-{\mathbf L}_{13}(u){\mathbf L}_{23}(v)\right)\,,
\end{split}
\end{equation} 
where $\perm_{12}$ acts as a permutation on the spaces $1$ and $2$ and trivially on $3$, cf. \eqref{permop}. After substituting the ansatz for the Lax matrix \eqref{laxan} into the relation above we find that its operatorial entries $J_{ba}$ have to satisfy the algebraic relations
\begin{equation}\label{glnalg}
 [J_{ab},J_{cd}]=\delta_{cb}J_{ad}-\delta_{ad}J_{cb}\,.
\end{equation} 
These commutation relations define a $\mathfrak{gl}_n$-algebra with the $\gln$-generators which will be labeled as $J_{ab}$ in the following.
Thus, for any representation of $\gln$-generators $J_{ab}$ labeled by the Dynkin weights $\Lambda=(\lambda^1,\ldots,\lambda^n)$, see Appendix~\ref{app:gln}, the Lax operator
\begin{equation}\label{lax}
  {\mathbf L}_\Lambda(u)\equiv{\mathbf L}(u)=u+\sum_{a,b=1}^ne_{ab}\otimes J_{ba}\,,
\end{equation} 
is as a solution to the Yang-Baxter equation \eqref{rll}.
Furthermore, for the fundamental representation $\Lambda=(1,0,\ldots,0)$ denoted by the Young diagram~$\square$ we recover the R-matrix introduced in \eqref{fundr}
\begin{equation}\label{laxundrr}
{\mathbf L}_\square (u)= \rfund(u)\,.
\end{equation} 
\section{R-matrices}
\label{sec:rmatrices}
We have derived the R-matrix (Lax matrix) with the fundamental representation in first space and representation $\Lambda$ in the second. Naturally, there arises the question for the R-matrix with two arbitrary representations $\Lambda_1$ and $\Lambda_2$. This object can implicitly be defined through the Yang-Baxter equation intertwining two Lax matrices of different representations as given in \eqref{lax}. We denote the space associated to it using the representation labels as $V_\square\otimes V_{\Lambda_1}\otimes V_{\Lambda_2}$.  The Yang-Baxter equation reads
\begin{equation}\label{rinter}
 {\mathbf L}_{\Lambda_2}(u-v){\mathbf L}_{\Lambda_3}(u)R_{\Lambda_2,\Lambda_3}(v)= R_{\Lambda_2,\Lambda_3}(v){\mathbf L}_{\Lambda_3}(u){\mathbf L}_{\Lambda_2}(u-v)\,,
\end{equation}  
where the index carried by the representation labels $\Lambda$ implies the space in which the R-matrix acts non-trivially. The space in the fundamental representation is suppressed in this notation. Furthermore, when restricting to the fundamental representation in the first space we have
\begin{equation}\label{laxr}
 R_{\square,\Lambda}(u)=\mathbf{L}_{\Lambda}(u)\quad\text{and}\quad  R_{\square,\square}(u)=\rfund(u)\,.
\end{equation} 
There are several methods to obtain solutions to this equation, see e.g. \cite{Kulish1981,MacKay1991,Zabrodin1997,Frassek2013b}. A convenient way to determine such R-matrices is to think about them as a Yangian invariants, see Chapter~\ref{ch:BAforYI} where a neat expression for the R-matrix with arbitrary symmetric representations is derived. Here we do not want to go into the details of this method and only discuss the $\gl2$-invariant R-matrix which was obtained in \cite{Kulish1981}. Therefore, we first note that $R_{\Lambda_1,\Lambda_2}$ is $\gln$-invariant
\begin{equation}\label{invr}
 [J_{ab}\otimes\mathbb{I}+\mathbb{I}\otimes J_{ab},R_{\Lambda_1,\Lambda_2}(v)]=0\,.
\end{equation} 
Here the generators and identity operators in the first (second) space of the tensor product are understood to be in the representation $\Lambda_1$ ($\Lambda_2$).
This follows from the expansion of the Yang-Baxter equation \eqref{rinter} in powers of $u$.
It tells us that the R-matrix evaluated on any state in the multiplet of an irreducible representation in the tensor product of the two representations $\Lambda_1$ and $\Lambda_2$ yields the same result. Thus once evaluated on all highest weight states it is completely determined. 

A comprehensive derivation of the $\gl2$-invariant R-matrix following the logic discussed above can be found in \cite{Faddeev2007}. For finite-dimensional representations $\Lambda=(\lambda^1,\lambda^2)$ with $\lambda^1>\lambda^2$ it can conveniently be written as
\begin{equation}\label{rss}
R_{\Lambda,\Lambda}(u;{\mathbb S})=\kappa(u)(-1)^{\mathbb S}\,\frac{\Gamma(1+u+\mathbb S)}{\Gamma(1-u+\mathbb S)}\,,
\end{equation} 
 where $\kappa$ is a normalization that cannot be determined from the Yang-Baxter equation. The operator $\mathbb S$ takes the eigenvalues $0,1,2,\ldots,(\lambda^1-\lambda^2)$ measuring the total spin $(\lambda_\text{tot}^1-\lambda_\text{tot}^2)/2$ of the irreducible representations $\Lambda_\text{tot}$ in the tensor product of the two spaces of the R-matrix. As an example we consider the R-matrix in the fundamental representation \eqref{fundr}. In this case the tensor product decomposition of the two fundamental representations is given by
\begin{equation}\label{tensdec}
 2\otimes2 =3\oplus 1\,,
\end{equation} 
where for the triplet we have ${\mathbb S}=1$ and for the singlet ${\mathbb S}=0$. After substituting these values into \eqref{rss} we obtain
\begin{equation}
R_{\square,\square}(u;1)=\kappa(u)\frac{\Gamma(1+u)}{\Gamma(1-u)}\frac{u+1}{u-1},\quad  R_{\square,\square}(u;0)=\kappa(u)\frac{\Gamma(1+u)}{\Gamma(1-u)}\,.
\end{equation}  We can check this result by expressing the R-matrix \eqref{fundr} in the proper eigenbasis according to \eqref{tensdec} with the help of the Clebsch-Gordan coefficients\,\footnote{In this case the normalization in \eqref{rss} is fixed to be $\kappa(u)=\frac{\Gamma(1-u)}{\Gamma(1+u)}(u-1)$.
}.
\section{A commuting family of operators}\label{sec:comfam}
Solutions to the Yang-Baxter equation can be used to generate commuting families of operators. They are physically important as the local charges and in particular the nearest-neighbor spin chain Hamiltonian belong to them. The general principle to obtain commuting families of operators is to construct certain monodromy matrices out of R-matrices. In the following we discuss representatives of the family that are constructed from the Lax operators defined in \eqref{lax}. These are of distinguished importance as the algebraic Bethe ansatz is most conveniently formulated within this framework. The monodromy is built out of Lax operators ${\mathbf L}_{\Lambda_i}$ with $i=1,\ldots,L$ multiplied in the common space which carries the fundamental representation of $\gln$:
\begin{equation}\label{moninh}
 \mathcal{M}(z;{\bf v})={\mathbf L}_{\Lambda_1}(z-v_1){\mathbf L}_{\Lambda_2}(z-v_2){\mathbf L}_{\Lambda_3}(z-v_3)\cdots {\mathbf L}_{\Lambda_L}(z-v_L)\,.
\end{equation} 
Here we used the variable $z$ to denote the spectral parameter carried by the common so-called auxiliary space. The parameters $v_i$ are assigned to the quantum space and are collectively denoted by the set ${\bf v}$. 
The monodromy above acts on the auxiliary space, which is in the fundamental representation, and the tensor product of $L$ spaces in the representations $\Lambda_1,\ldots,\Lambda_L$ of $\gln$ which we will refer to as quantum space
\begin{equation}
 \mathcal{M}(z;{\bf v}):V_\square\otimes \bigotimes_{i=1}^L V_{\Lambda_i}\rightarrow V_\square\otimes \bigotimes_{i=1}^L V_{\Lambda_i}\,.
\end{equation} 
Hence, we can think about \eqref{moninh} as an $n\times n$ matrix in the auxiliary space containing operatorial entries that act in the quantum space. Diagrammatically we can write the monodromy as 
\begin{equation}\label{pic:smon}
  \mathcal{M}(z;{\bf v})=\,\,\input{content/pictures/smon.tex}\,,
\end{equation} 
where again we used the representation label $\Lambda_i$ to indicate space $i$. The label ``$\text{aux}$'' denotes the auxiliary space. In the context of spin chains it is common to call $z$ the spectral parameter and $v_1,\ldots,v_L$ inhomogeneities. A transfer matrix is built as the trace of the monodromy in the auxiliary space
 \begin{equation}\label{transf}
   \tm (z;{\bf v})=\tr_{\text{aux}}  \mathcal{M}(z;{\bf v})\,.
 \end{equation}
Thus $\tm$ acts purely on the quantum space. Its diagrammatic expression is given by 
\begin{equation}\label{pic:transm}
  \tm(z;{\bf v})=\,\,\input{content/pictures/transm.tex}\,,
\end{equation} 
where the trace closes the auxiliary space, cf. \eqref{pic:smon}. In the following we suppress the ${\bf v}$-dependence. 

An important property of the transfer matrix is that it commutes with itself for any, a priori different, values of the spectral parameters $x$ and $y$
\begin{equation}\label{tcom}
 [\tm(x),\tm(y)]=0\,.
\end{equation} 
The proof of \eqref{tcom} is based on the Yang-Baxter equation \eqref{rll}. First we note that the monodromy satisfies the so-called RTT-relation\,\footnote{Unfortunately the choice of our notation does not reflect the origin of this name.}
\begin{equation}\label{rMM}
\rfund_{12}(x-y) \mathcal{M}_1(x)\mathcal{M}_2(y)=\mathcal{M}_2(y)\mathcal{M}_1(x)\rfund_{12}(x-y)\,,
\end{equation} 
where we suppressed the dependence of the inhomogeneities in the monodromy. The labels $1$ and $2$ denote the auxiliary spaces of the two monodromies and the multiplication in the quantum space is according to the RLL-relation \eqref{rll}. The RTT-relation \eqref{rMM} can conveniently be shown diagrammatically using the so-called train track argument \cite{Faddeev1995b} as explained in the following. Rewriting \eqref{rMM} graphically yields
\begin{align}\label{pic:train}
 \begin{aligned}
  \includegraphics[scale=0.80]{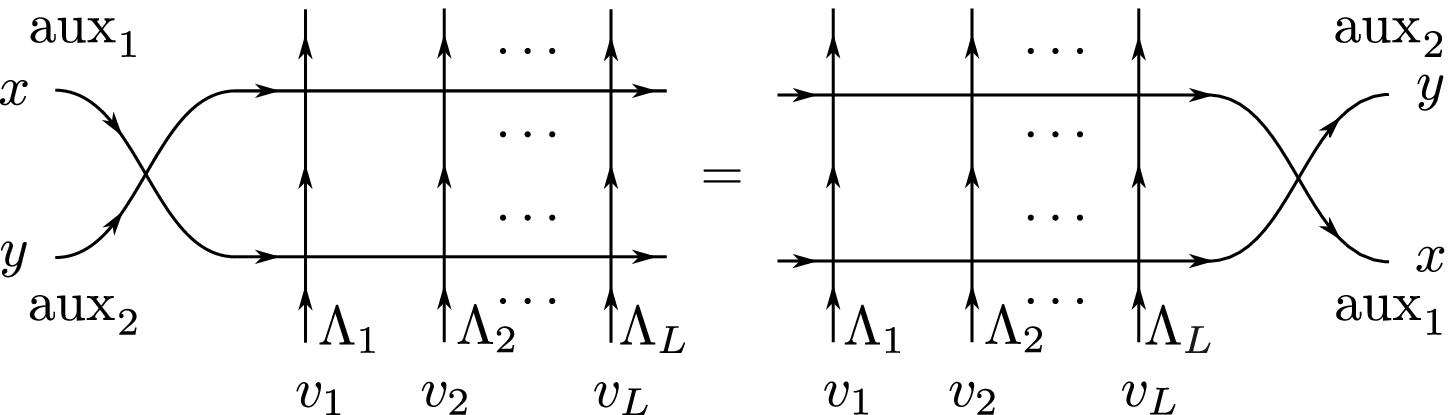}
 \end{aligned}
 \,.
\end{align}
For the case discussed here the first and the second auxiliary space are in the fundamental representation of $\gln$.
Now, using the appropriate Yang-Baxter equation \eqref{rll}, cf. also the diagrammatic representation in \eqref{pic:ybe}, we can shift all lines through the vertex formed by the intersecting lines $1$ and $2$ and find that by construction \eqref{rMM} holds true. In the next step we use that $\rfund$ is invertible for generic values of the spectral parameter $z$. Its inverse is given by
\begin{equation}\label{invrfund}
 \rfund^{-1}(z)=\frac{z-P}{z^2-1}\,.
\end{equation} 
Then we can rewrite \eqref{rMM} as
\begin{equation}\label{rtt2}
\mathcal{M}_1(x)\mathcal{M}_2(y)=\rfund_{12}^{-1}(x-y)\,\mathcal{M}_2(y) \mathcal{M}_1(x)\,\rfund_{12}(x-y)\,.
\end{equation} 
The final step is to take the trace of the the equation above in the first and second auxiliary space. Again the diagrammatics is convenient to convince oneself that the R-matrix $\rfund$ cancels with its inverse under the trace and one finds
\begin{equation}
 \tm(x) \tm(y)= \tm(y) \tm(x)\,,
\end{equation} 
which completes the proof of \eqref{tcom}.

Furthermore, we would like to note that the transfer matrix $\tm$ is $\gln$-invariant
\begin{equation}\label{tinv}
  [\tm(z),J^{\,\text{tot}}_{ab}]=0\,.
\end{equation} 
Here we use the notation 
\begin{equation}\label{geni}
J^{\,\text{tot}}_{ab}=\sum_{i=1}^L J_{ab}^{(i)}\quad\text{with}\quad J_{ab}^{(i)}=\underbrace{\mathbb{I}\otimes\ldots\otimes\mathbb{I}}_{i-1}\otimes J_{ab}\otimes \underbrace{\mathbb{I}\otimes\ldots\otimes\mathbb{I}}_{L-i}
\end{equation}
to denote the $\gln$ generator acting non-trivially only on the  $i^\text{th}$ component in the tensor product of the quantum space. To show that the transfer matrix is $\gln$-invariant we consider of the RTT-relation \eqref{rMM} in the spectral parameter $y$. The leading terms of the monodromy are given by
\begin{equation}
 \mathcal{M}(y)=y^L+y^{L-1}\sum_{i=1}^L (e_{ab}\otimes J_{ba}^{(i)} -v_i)+\ldots\,.
\end{equation} 
Thus, one obtains the invariance condition 
\begin{equation}\label{moninv}
 [\mathcal{M}(x),e_{ab}\otimes \mathbb{I}+\mathbb{I}\otimes J^\text{tot}_{ab}]=0\,.
\end{equation} 
After taking the trace in the auxiliary space of \eqref{moninv} one finds that $\tm$ is $\gln$-invariant, cf. \eqref{tinv}. 

\section{Further members of the family}\label{sec:further}
The derivation of the commutativity relation for the transfer matrix in \eqref{tcom} relies on the existence of the RLL-relation \eqref{rll} and the existence of an inverse of the R-matrix intertwining the two monodromies. The family of commuting operators is extended by transfer matrices that are built from monodromies with other representations in the auxiliary space. This becomes clear from the train argument presented in \eqref{pic:train}. Previously, both auxiliary spaces were in the fundamental representation. However, from the type of Yang-Baxter equation presented in \eqref{rinter} one can show that \eqref{pic:train} holds for different representations in the auxiliary space, see also \cite{Zabrodin1997}. Following the same logic as before one finds that additional transfer matrices can be defined as 
 \begin{equation}
   \tm_{\text{aux}} (z)=\tr \mathcal{M}_{\text{aux}} (z)\,.
 \end{equation} 
These extend the family of commuting operators
 \begin{equation}
[ \tm_{\text{aux}_1}(x), \tm_{\text{aux}_2}(y)]=0\,,
\end{equation} 
and in particular commute with the transfer matrix \eqref{transf}.
The transfer matrices constructed in this way are not independent but satisfy certain functional relations. In particular for finite dimensional representations in the auxiliary space any transfer matrix can be expressed only in terms of the ones constructed from a totally symmetric or totally antisymmetric representation in the auxiliary space. This formula is known as quantum Jacobi-Trudi formula \cite{Bazhanov1990}.
In particular using the quantum Jacobi-Trudi formula it can be shown that the transfer matrices $\tm_{a,s}$ for representations corresponding to rectangular Young diagrams with Dynkin labels
\begin{equation}
 \Lambda=(\underbrace{s,\cdots,s}_a,0,\cdots,0)\,,
\end{equation} 
in the auxiliary space satisfy the Hirota equation \cite{Krichever1997}
\begin{equation}\label{hiroota}
\tm_{a,s}(z+1) \tm_{a,s}(z-1)= \tm_{a+1,s}(z) \tm_{a-1,s}(z)+ \tm_{a,s+1}(z) \tm_{a,s-1}(z)\,.
\end{equation} 
This bilinear difference equation depends on the three variables $a$, $s$ and $z$. It is well known from the theory of classical solitons \cite{Hirota1981,Miwa1981} and plays an important role in the current study of the all loop anomalous dimensions of single trace operators in \Nsym4 \cite{Gromov2010a}.

\section{Hamiltonian and shift operator}\label{spinchainham}
A spin chains is usually characterized by a Hamiltonian that describes its energy spectrum. One of the most famous examples of an integrable Hamiltonian is the one written down by Heisenberg \cite{Heisenberg1928} for the closed XXX$_\frac{1}{2}$ spin chain
\begin{equation}\label{su2_ham}
 \mathbf{H}_{\text{XXX}}=4\sum_{i=1}^L\left(\frac{1}{4}-{\vec S}_i\,{\vec S}_{i+1}\right)\quad\text{with} \quad {\vec S}_{L+1}={\vec S}_1\,,
\end{equation}  
where ${\vec S}_i=\frac{1}{2}\,{\vec \sigma}_i$, with the usual Pauli matrices ${\vec \sigma}$ at site $i$, compare \eqref{geni} where we used upper indices to indicate the sites. The space of states is given by the $L$-fold tensor product
\begin{equation}
 \underbrace{{\mathbb C}^2\otimes\cdots\otimes{\mathbb C}^2}_{L}\,.
\end{equation} 
From \eqref{su2_ham} we see that the Heisenberg XXX$_\frac{1}{2}$ spin chain Hamiltonian describes nearest-neighbor interactions as the term that is summed over acts non-trivially only in the spaces $i$ and $i+1$. 
To see how this physical quantity is related to the commuting family of operators introduced previously we concentrate on the monodromy 
\begin{equation}\label{fundmon}
 \mathcal{M}(z)=\rfund_{\,\text{a},1}(z)\rfund_{\,\text{a},2}(z)\rfund_{\,\text{a},3}(z)\cdots\rfund_{\,\text{a},L}(z)\,.
\end{equation} 
Note that we have set all inhomogeneities equal to zero and recall that the quantum space of the monodromy above is given by 
\begin{equation}
  \underbrace{{\mathbb C}^n\otimes\cdots\otimes{\mathbb C}^n}_{L}\,,
\end{equation} 
as we focus on the fundamental representation of $\gln$ with the R-matrix given in \eqref{fundr}. This example might be misleading as for the fundamental representation the operator ${\mathbf L}$ coincides with $\rfund$, cf. \eqref{laxundrr}. However, we would like to stress that the monodromy that yields the Hamiltonian following the approach outlined here\,\footnote{In Chapter~\ref{ch:qop} we will discuss how the Hamiltonian can be extracted from another member of the commuting family of operators; namely the Q-operator.} is characterized by the property that it is built from the R-matrices that carry the same representations at each site of the quantum space and in the auxiliary space. To recover the Hamiltonian \eqref{su2_ham} we will set $n=2$ in the end.

Let us first define the transfer matrix with equal representation for the auxiliary as for each site of the quantum space using \eqref{fundmon}
\begin{equation}\label{transT}
\tmf(z)=\tr \mathcal{M}(z)\,.
\end{equation} 
The crucial property of this transfer matrix is that it turns into a string of permutations at a special point of the spectral parameter, compare \eqref{fundr} or \eqref{frmatrix}. The operator at this special point of the transfer matrix is called shift operator $\mathbf U$ for reasons that will become obvious soon. It is given by
\begin{equation}\label{shiftop}
 {\textbf U}= \tmf(0)={\textbf P}_{L,L-1}{\textbf P}_{L-1,L-2}\cdots {\textbf P}_{2,1}\,.
\end{equation} 
Its action on a state $w\in{\mathbb C}^n$ at site $i$ is given by
\begin{equation}
 {\textbf U}\,w_i=w_{i-1}\,.
\end{equation} 
For completeness and later purposes we also give its diagrammatic expression
\begin{equation}\label{shiftpic}
 {\mathbf U}=\,\input{content/pictures/shiftop.tex}\,.
\end{equation} 
The Hamiltonian can now be obtained from the expansion of the transfer matrix around the shift point
\begin{equation}\label{texpansion}
 \tmf(z)=\mathbf{U}\,(\mathbb{I}+z\,\mathbf{H}+\ldots)\,.
\end{equation} 
It is given by the logarithmic derivative of the transfer matrix at this point
\begin{equation}\label{logder}
 {\textbf H}=\left. \frac{\partial}{\partial z}\ln\tmf(z)\right|_{z=0}\,.
\end{equation} 
By construction all operators in the expansion \eqref{texpansion} belong to the commuting family of operators. The terms proportional to the spectral parameter are also referred to as local charges. As we have seen the Hamiltonian is of nearest neighbor type. The range of interaction in \eqref{logder} grows with the order of the spectral parameter.
Here we would like to stress again that at the special points the R-matrices in the transfer matrix reduce to the permutation. This is a priori only possible if each site of the quantum space carries the same representation as the auxiliary space. An instructive calculation then shows that for our case the Hamiltonian can be written as
\begin{equation}\label{hamiltonian}
  {\textbf H}=\sum_{i=1}^L \mathcal{H}_{i,i+1}\,,
\end{equation} 
with the Hamiltonian density
\begin{equation}\label{hamden}
\mathcal{H}_{i,i+1}=\mathbf{P}_{i,i+1}\,.
\end{equation} 
Here we refer the reader to \cite{Faddeev2007} where further details are presented. We obtain the Hamiltonian of the XXX$_\frac{1}{2}$ spin chain when restricting to $n=2$ using $\mathbf{P}_{i,i+1}=(\mathbb{I}+4\vec S_i\vec S_{i+1})/2$. Furthermore, we need to add a term proportional to the identity
\begin{equation}\label{hamrel}
\mathbf{H}_\text{XXX}=2(L- \mathbf{H})\,.
\end{equation} 
However, this term does not spoil the commutativity within the family of operators and only shifts the eigenvalues by a constant. 

Without any proof we also would like to mention that the Hamiltonian density can directly be obtained from the logarithmic derivative of the R-matrix at the permutation point \cite{Faddeev2007}
\begin{equation}\label{logderR}
 \mathcal{H}=\left.\frac{\partial}{\partial z}\ln{\textbf R}(z)\right|_{z=0}\,.
\end{equation} 
This can conveniently be used to obtain the Hamiltonian of spin chains with other representations in the quantum space. Using the R-matrix introduced in \eqref{rss} we obtain the Hamiltonian density
\begin{equation}\label{hamJ}
  \mathcal{H}_{i,i+1}=2(-1)^{2\,\mathbb{S}}\,\psi({\mathbb S}+1)\,,
\end{equation}
where $\psi$ denotes the digamma function. It reduces to the harmonic numbers for integer values $n$
\begin{equation}
 \psi(n)=h_{n-1}-\gamma_e\quad\quad\text{with}\quad\quad h_{n}=\sum_{k=1}^n\frac{1}{k}\,,
\end{equation} 
and the Euler-Mascheroni constant $\gamma_e$. We also would like to note that the action of the Hamiltonian in terms of generators defined at each site as in \eqref{su2_ham} can be obtained from \eqref{hamJ} through a Lagrange interpolation, see \cite{Faddeev2007}. The same is true for the R-matrix in \eqref{rss}.

\section{Algebraic Bethe ansatz}
To diagonalize the spin chain Hamiltonian and the corresponding family of commuting operators we discuss the algebraic Bethe ansatz. It heavily relies on the fact that two commuting operators have the same eigenspace. The representative of the commuting family of operators that is commonly diagonalized by the Bethe ansatz is the transfer matrix with the fundamental representation in the auxiliary space. Its monodromy was introduced in \eqref{moninh}. In the following we present the algebraic Bethe ansatz in detail for $\gl2$-invariant spin chains and give the central equations for the $\gln$ case. Finally, we discuss how it is used to obtain the eigenvalues of the shift operator and Hamiltonian.
\subsection{The ABCD of the algebraic Bethe ansatz}\label{sec:gl2}
Now, we review the algebraic Bethe ansatz for $\gl2$-invariant inhomogeneous spin chains. The corresponding monodromy was introduced in \eqref{moninh} for $\gln$. We have already argued that in the case of $\gl2$ the monodromy can be written as a $2\times2$ matrix in the auxiliary space 
\begin{equation}\label{gl2mon}
  \mathcal{M}(z)=\left(\begin{array}{cc}
                 \Am(z)&B(z)\\
                 C(z)&D(z)\\
                \end{array}\right)\,,
\end{equation} 
where the operatorial entries $\Am,B,C$ and $D$ act purely in the quantum space. For convenience we suppressed the inhomogeneities. Using this notation the transfer matrix can be written as
\begin{equation}\label{ApD}
 \tm(z)=\Am(z)+D(z)\,.
\end{equation} 
The algebraic Bethe ansatz relies on the existence of a reference state such that
\begin{equation}\label{actmon}
 \Am(z)\bvac=\alpha(z)\bvac,\quad  D(z)\bvac=\delta(z)\bvac,\quad  C(z)\bvac=0\,.
\end{equation}
For $\bvac$ being the $L$-fold tensor product of highest weight states $\vert\sigma_i\rangle$ at each site of the spin chain 
\begin{equation}\label{bethevac}
 \bvac=\bigotimes_{i=1}^L\vert\sigma_i\rangle
\end{equation} 
with 
\begin{equation}
 J_{11}^{(i)}\bvac=s_i\bvac\,,\quad\quad J_{22}^{(i)}\bvac=0\,,
\end{equation} 
i.e. $\Lambda_i=(s_i,0)$, compare \eqref{geni}, one finds that the eigenvalues of the operators $\Am$ and $D$ are given by
\begin{equation}\label{evs}
 \alpha(z)=\prod_{i=1}^L(z-v_i+s_i),\quad\quad\delta(z)=\prod_{i=1}^L(z-v_i)\,.
\end{equation} 
Furthermore, one can check that the condition for $C$ in \eqref{actmon} is satisfied. Now the ansatz for the eigenvectors of the commuting family of operators is given by
\begin{equation}\label{su2BV}
\vert\Phi\rangle=B(z_1)B(z_2)\cdots B(z_\magm)\bvac  \,.             
\end{equation} 
To motivate this ansatz we note that the commutation relations of the operator $B$ with the Cartan elements of $\gl2$ realized on the quantum space of the spin chain take the form 
\begin{equation}
 [J^{\,\text{tot}}_{11},B(z)]=-B(z)\,,\quad\quad [J^{\,\text{tot}}_{22},B(z)]=+B(z)\,.
\end{equation} 
Thus, they act as lowering/raising operators. The relations above are a direct consequence of \eqref{moninv}. The Cartan elements evaluated on the ansatz \eqref{su2BV} are given by
\begin{equation}
 J^{\,\text{tot}}_{11}\vert\Phi\rangle=(\sum_{i=1}^Ls_i-\magm)\vert\Phi\rangle,\quad\quad  J^{\,\text{tot}}_{22}\vert\Phi\rangle=\magm\vert\Phi\rangle\,.
\end{equation} 
The magnon number $\magm$ counts the excitations on the highest weight state $\bvac$, or equivalently the number of B operators acting on the vacuum $\bvac$. Now, the idea is to act with the transfer matrix on the ansatz for the eigenfunctions \eqref{su2BV}. From the RLL-relation \eqref{rll} we derive the following so-called fundamental commutation relations
\begin{equation}\label{Bcom}
 [B(x),B(y)]=0\,,
\end{equation}
\begin{align} 
 \Am(x)B(y)=\frac{1+y-x}{y-x}\,B(y)\Am(x)-\frac{1}{y-x}\,B(x)\Am(y)\,,\label{abcom}\\
 \nonumber\\
 D(x)B(y)=\frac{1+x-y}{x-y}\,B(y)D(x)-\frac{1}{x-y}\,B(x)D(y)\,.\label{dbcom}
\end{align}
By commuting $\Am$ and $D$ through all B operators using the relations above one can show that
\begin{equation}\label{acta}
 \Am(z)\vert\Phi\rangle=\alpha(z)\frac{Q(z-1)}{Q(z)}\vert\Phi\rangle+\sum_{j=1}^\magm\prod_{k\neq j}^m\frac{\alpha(z_j)}{z_j-z}\frac{Q(z_j-1)}{z_j-z_k}\vert \Phi_j\rangle\,,
\end{equation} 
\begin{equation}\label{actd}
 D(z)\vert\Phi\rangle=\delta(z)\frac{Q(z+1)}{Q(z)}\vert\Phi\rangle+\sum_{j=1}^\magm\prod_{k\neq j}^m\frac{\delta(z_j)}{z_j-z}\frac{Q(z_j+1)}{z_j-z_k}\vert \Phi_j\rangle\,,
\end{equation} 
with
\begin{equation}
 \vert\Phi_j\rangle=B(z_1)\cdots B(z_{j-1})B(z)B(z_{j+1})\cdots B(z_\magm)\vert\Omega\rangle\,.
\end{equation} 
Here we introduced the so-called Q-function
\begin{equation}\label{qfct}
 Q(z)=\prod_{i=1}^\magm(z-z_i)\,.
\end{equation} 
It is a polynomial of degree $\magm$ in the spectral parameter $z$ with zeros located at the so-called Bethe roots $z_i$. Their operatorial form will be constructed in Chapter~\ref{ch:qop} from the Yang-Baxter equation. From \eqref{acta} and \eqref{actd} we find that the transfer matrix \eqref{ApD} is diagonal on $\vert \Phi\rangle$ if the Bethe roots satisfy the condition
\begin{equation}\label{prebae}
 \alpha(z_k)Q(z_k-1)+\delta(z_k)Q(z_k+1)=0\,.
\end{equation} 
Then the non-diagonal terms in \eqref{acta} and \eqref{actd} cancel. These equations are the famous Bethe equations. Using the explicit expressions for $\alpha$, $\delta$ \eqref{qfct} and the Q-function \eqref{evs} we write them in the more familiar form
\begin{equation}\label{bae}
 \prod_{i=1}^L\frac{z_k-v_i+s_i}{z_k-v_i}=-\prod_{j=1}^\magm\frac{z_k-z_j+1}{z_k-z_j-1}
\end{equation} 
where $z_k\neq v_i$ and $z_k\neq z_j+1$. The eigenvalues $\tau$ of the transfer matrix $\tm$ are given by the so-called Baxter equation
\begin{equation}\label{baxeq}
 \tau(z)=\alpha(z)\frac{Q(z-1)}{Q(z)}+\delta(z)\frac{Q(z+1)}{Q(z)}\,.
\end{equation} 
Each solution to the Bethe equations \eqref{bae} yields an eigenvalue of the transfer matrix using the Baxter equation \eqref{baxeq} and the corresponding eigenvector using \eqref{su2BV}. In the following we will sometimes use the term off-shell Bethe vector if we want to stress that we are talking about \eqref{su2BV} for general values $z_i$ and on-shell Bethe vector if the variables $z_i$ satisfy the Bethe equations \eqref{bae}. We note that the Bethe equations can be obtained from the Baxter equation \eqref{baxeq}: By construction the transfer matrix and therefore also its eigenvalues are polynomials in the spectral parameter. Therefore, taking the residue at $z=z_k$ with $k=1,\ldots,\magm$ in \eqref{baxeq} we obtain \eqref{prebae}. In the functional Bethe ansatz which is not discussed here the Baxter equations can be seen as the fundamental relation \cite{Sklyanin2008a,Sklyanin,Niccoli2013}. As discussed in Chapter~\ref{ch:qop} the Baxter equation \eqref{baxeq} also holds on the level of operators.
Last but not least, we would like to note that the operators $B$ map between the highest weight states of the irreducible representations appearing in the tensor product decomposition of the spin chain if the Bethe equations \eqref{bae} are satisfied, see \cite{Faddeev2007}. Other states in the multiplets can be obtained by descending from the highest weight states with the appropriate lowering operators. It is discussed in Section~\ref{twist} how descendants of the highest weight states can be obtained directly from the Bethe ansatz by breaking the $\gl2$ symmetry of the transfer matrix. 
\subsection{Elements of  the nested ABA}\label{sec:naba}
The nested algebraic Bethe ansatz (ABA) was introduced in \cite{Kulish1983a}. To diagonalize the transfer matrix \eqref{transf} of $\gln$-invariant spin chains one has to go through $n-1$ nesting levels. In Section~\ref{sec:gl2} we outlined the algebraic Bethe ansatz for $\gl2$. Here, the main results were the Baxter equation \eqref{baxeq}, the off-shell Bethe vector \eqref{su2BV} and of course the Bethe equations \eqref{bae}. The construction of the Bethe vectors follows the same reasoning as before. Starting from a vacuum $\bvac$ the eigenstates can be constructed from the upper triangular entries of the monodromy matrix. At every nesting step $i$ the length of the spin chain is increased by $\magm_{i-1}$ which denotes the corresponding magnon number at the level $i-1$. The main difficulty that arises here compared to the contained $\gl2$ case is to keep track of the index structure carried by the creation operators. As we are mostly interested in the functional relations, we do not give the explicit form of the eigenvectors and refer the reader to \cite{Kulish1983a}. Nevertheless, we provide a diagrammatic form of the $\gl3$ Bethe vectors in Appendix~\ref{app:bvaspf}. 

Following \cite{Kulish1983a}, see also Chapter~\ref{ch:qop}, we find that the eigenvalues of the transfer matrix \eqref{transf} are given by 
\begin{equation}\label{baxtergln}
 \tau(z)=\sum_{i=1}^n \mu_i(z)\frac{Q_{i-1}(z+1)}{Q_{i-1}(z)}\frac{Q_i(z-1)}{Q_i(z)}\,.
\end{equation} 
This formula reduces to the Baxter equation introduced in \eqref{baxeq} for $n=2$.
The Q-functions introduced above are given by
\begin{equation}\label{qfuncgln}
 Q_i(z)=\prod_{k=1}^{\magm_i}(z-z_k^{(i)})\,,
\end{equation} 
where $z_k^{(i)}$ denotes the $k^\text{th}$ Bethe root of the $i^\text{th}$ nesting level. Furthermore, we have $Q_0=Q_n=1$. The eigenvalues of the diagonal elements of the monodromy matrix $\mathcal{M}_{ii}$ on the vacuum $\bvac$ which is constructed in analogy to \eqref{actmon} are given by
\begin{equation}\label{tvac}
 \mu_i(z)=\prod_{k=1}^L(z-v_k+\lambda_k^{i})\,.
\end{equation} 
For the case of $n=2$ we identify $\mu_1(z)=\alpha(z)$ and $\mu_2(z)=\delta(z)$. 
Instead of calculating all the unwanted terms as in the $\gl2$ case we directly extract the Bethe equations from \eqref{baxtergln} following the same logic as discussed at the end of the previous section. For general values of the Bethe roots $z_k^{(i)}$ we obtain $n-1$ nested Bethe equations which are coupled among themselves
\begin{equation}\label{nbaes}
 \frac{\mu_i(z_k^{(i)})}{\mu_{i+1}(z_k^{(i)})}=-\frac{Q_{i-1}(z_k^{(i)})}{Q_{i-1}(z_k^{(i)}+1)}\frac{Q_{i}(z_k^{(i)}+1)}{Q_{i}(z_k^{(i)}-1)}\frac{Q_{i+1}(z_k^{(i)}-1)}{Q_{i+1}(z_k^{(i)})}
\end{equation} 
for $i=1,\ldots,n-1$. They agree with the Bethe equations presented in \eqref{bae} for $n=2$.
\subsection{Observables}\label{sec:obs}
As already mentioned, the shift operator and the Hamiltonian belong to the commuting family of operators. Therefore, they can be diagonalized using the eigenvectors introduced in Section~\ref{sec:gl2}. The goal of this section is to write their eigenvalues in terms of the Bethe roots in the spirit of the Baxter equation \eqref{baxeq}, see also \eqref{baxtergln}. In particular, throughout this section all inhomogeneities are set to zero. Furthermore, we focus on the case of fundamental representations in the quantum space which is particularly simple as the monodromy \eqref{moninh} reduces to the monodromy \eqref{fundmon} that can be used to extract the local charges. Namely for $\Lambda_i=(1,0,\ldots,0)$ we have
\begin{equation}
\tmf(z)= \tm(z)\,.
\end{equation} 
In this case we obtain the desired formulas from the Baxter equation \eqref{baxtergln}. We have seen in \eqref{shiftop} that the transfer matrix reduces to shift operator at $z=0$. Therefore, from \eqref{baxtergln} we determine the eigenvalues of the shift operator to be 
\begin{equation}\label{abashift}
 U=\tau(0)=\frac{Q_1(-1)}{Q_1(0)}=\prod_{k=1}^{\magm_1}\frac{z_k^{(1)}+1}{z_k^{(1)}}\,,
\end{equation} 
where we used the explicit form of the Q-function \eqref{qfuncgln} and the entries on the diagonal of the transfer matrix on the vacuum \eqref{tvac}.
Following the same logic we can also derive the energy eigenvalues of the Hamiltonian \eqref{logder}. Taking the logarithmic derivative of the Baxter equation \eqref{baxtergln} at the shift point one finds
\begin{equation}\label{abaenergy}
 H=E=L+\frac{Q_1'(-1)}{Q_1(-1)}-\frac{Q_1'(0)}{Q_1(0)}=L+\sum_{k=1}^{\magm_1}\left(\frac{1}{z_k^{(1)}}-\frac{1}{1+z_k^{(1)}}\right)\,.
\end{equation} 
As we can see from \eqref{abashift} and \eqref{abaenergy}, the physical observables only depend on the first level Bethe roots $z_k^{(1)}$. This is a consequence of the representation that we are looking at. In terms of Dynkin diagrams we only excited one node. We say that the roots corresponding to this node in the Dynkin diagram are momentum carrying.
\section{Twist}\label{twist}
At the end of Section~\ref{sec:gl2} we briefly discussed that using the algebraic Bethe ansatz as presented above one only obtains highest weight states by acting with the creation operators \eqref{su2BV} and subsequently solving the Bethe equations \eqref{bae}. This is an artifact of the $\gln$-invariance of the spin chain.  In this section we will break this invariance employing a so-called twist in the auxiliary space while preserving the integrability condition \eqref{tcom}, see \cite{DeVega1984}. In this way we are able to obtain descendants of the highest weight states mentioned above\,\footnote{The issue of completeness of the Bethe ansatz, i.e. if and how all eigenvalues and eigenvectors can be obtained from the Bethe ansatz, is still an active field of research, see e.g. \cite{Nepomechie2014,Baxter:2001sx} and references therein.}.

We define the monodromy matrix 
\begin{equation}
 \mathcal{M}_\twist(z)=\twist\,\mathbf{L}_{\Lambda_1}(z-v_1)\cdots\mathbf{L}_{\Lambda_L}(z-v_L)\,,
\end{equation} 
and the corresponding transfer matrix
\begin{equation}
 \tm_\twist(z)=\tr  \mathcal{M}_\twist(z)\,.
\end{equation} 
Here the twist $\twist$ is an $n\times n$ matrix that purely acts in the auxiliary space. To preserve the integrability condition
\begin{equation}\label{tcomd}
 [\tm_\twist(z), \tm_\twist(z')]=0\,,
\end{equation} 
we require that the R-matrix that appeared in the RTT-relation \eqref{rMM} commutes with the tensor product of two twist matrices
\begin{equation}
 [\rfund(z),\twist\otimes\twist]=0\,.
\end{equation} 
In general, this property is true for any non-degenerate matrix $\twist$.
The proof of \eqref{tcomd} is in analogy to the proof of \eqref{tcom}, see also \cite{Yung1995}. As before we can study the commutation relations of the transfer matrix with the $\gln$ generators. Following the same logic as presented at the end of Section~\ref{sec:comfam} one finds that the $\gln$-invariance is broken
\begin{equation}
 [\tm_\twist(z), \sum_{i=1}^n J_{ab}^{(i)}]=[\mathcal{M}_\twist(z),\twist]_{ba}\,.
\end{equation} 
However, for a diagonal twist $\twist_{ab}=d_a\delta_{ab}$ one finds that 
\begin{equation}
 [\tm_\twist(z), \sum_{i=1}^{n} J_{aa}^{(i)}]=0\,.
\end{equation} 
From this follows that the magnon number $\magm$ is still a ``good'' quantum number to label the eigenstates. The twist manifests itself in the Hamiltonian and in the Bethe equations. In the diagonal case we find that the Bethe equations for $\gl2$ get modified as
\begin{equation}\label{baetwist}
 \frac{d_1}{d_2}\,\prod_{i=1}^L\frac{z_k-v_i+s_i}{z_k-v_i}=\prod_{j=1}^\magm\frac{z_k-z_j+1}{z_k-z_j-1}\,,
\end{equation} 
cf. \eqref{bae}, while the $\gln$-case is discussed in \cite{Kulish1983a}. The Hamiltonian can be obtained from the logarithmic derivative of the appropriate transfer matrix, cf. \eqref{logder}. Using 
 \begin{equation}
  \tmf_\twist(0)={\bf U}=\perm_{L,L-1}\perm_{L-1,L-2}\cdots \perm_{2,1}\mathcal{D}_1\,,
 \end{equation} 
 and
  \begin{equation}
  \begin{split}
   \left. \frac{\partial}{\partial z} \tmf_\twist(z)\right|_{z=0} =&\sum_{i=2}^{L-1}\perm_{L,L-1}\cdots \perm_{i+2,i+1} \perm_{i+1,i-1}\perm_{i-1,i-2}\cdots \perm_{2,1}\,\mathcal{D}_1\\
     &+\perm_{L-1,L-2}\cdots \perm_{2,1}\,\mathcal{D}_1
     +\mathcal{D}_L\,\perm_{L,L-1}\cdots \perm_{3,2}\,,
  \end{split}
 \end{equation} 
 where ${\mathcal D}_i$ denotes the twist acting non-trivially on the $i^\text{th}$ space, we find the twist dependent Hamiltonian 
 \begin{equation}\label{sln-ham}
  {\bf H}=\sum_{i=1}^L\perm_{i,i+1}\quad\text{with}\quad   \perm_{L,L+1}=\twist_L\,\perm_{L,1}\,\twist_L^{-1}\,,
 \end{equation} 
 compare also \eqref{hamiltonian} and \eqref{hamden}. The energy formula \eqref{abaenergy} remains unaffected and the twist dependence of the energy enters via the Bethe roots, see also Chapter~\ref{ch:qop}.
 
\section{Coordinate Bethe ansatz}\label{sec:cba}
In this section we discuss the coordinate Bethe ansatz which provides a different and more explicit representation of the Bethe vectors \eqref{su2BV}. It reveals certain physical structure that allows us to interpret the eigenfunctions as a superposition of single particle wave-functions. In contrast to the algebraic Bethe ansatz the coordinate Bethe ansatz is much more pragmatic. Here, our starting point is the nearest neighbor Hamiltonian that we want to diagonalize. Here we concentrate on the Hamiltonian of the closed $\gl2$-invariant spin chain for the fundamental representation given in \eqref{su2_ham}.
It is convenient to rewrite the Hamiltonian in terms of the permutation operator $\perm$ such that \eqref{su2_ham} takes the form
\begin{equation}\label{su2_ham_perm}
 \mathbf{H}_\text{XXX}=2\,\sum_{i=1}^L\left(1-\perm_{i,i+1}\right)\,.
\end{equation}  
Bethe made an educated guess for the eigenvectors of the Hamiltonian in \eqref{su2_ham_perm}. To motivate his ansatz\,\footnote{In the following we use the notation $\vert \sigma_i\rangle\leftrightarrow\vert\downarrow\rangle$ and $J_{21}\vert \sigma_i\rangle\leftrightarrow\vert\uparrow\rangle$, cf. \eqref{bethevac}.
} we first note that 
\begin{equation}\label{singwave_su2}
 \vert \psi \rangle=\sum_{x=1}^L e^{ipx}\vert \downarrow\cdots\stackrel{x}{\uparrow}\cdots\downarrow\rangle\,,
\end{equation}
is an eigenvector of \eqref{su2_ham_perm} when neglecting boundary terms, i.e. in the infinite volume. This can be seen by acting with \eqref{su2_ham_perm} on \eqref{singwave_su2} using the action of the permutation on the basis vectors 
\begin{equation}
 \perm_{x,x+1}\vert \downarrow\cdots\stackrel{x}{\uparrow}\cdots\downarrow\rangle=\vert \downarrow\cdots\stackrel{x+1}{\uparrow}\cdots\downarrow\rangle\,.
\end{equation} 
The energy eigenvalue corresponding to the Hamiltonian in \eqref{su2_ham_perm} is given by
\begin{equation}\label{energy_su2_one}
 E_{\text{XXX}}=4(1-\cos p)\,.
\end{equation} 
Now the ansatz is to take a superposition of the free waves in \eqref{singwave_su2} and add a term to describe the scattering
\begin{equation}{\label{wave_su2}}
 \vert \psi \rangle=\sum_{1\leq x_1<\ldots<x_\magm\leq L}\psi(x_1,\ldots,x_\magm)\vert \downarrow\cdots\stackrel{x_1}{\uparrow}\cdots\stackrel{x_\magm}{\uparrow}\cdots\downarrow\rangle\,,
\end{equation}
with
\begin{equation}\label{wave_comp_su2}
 \psi(x_1,\ldots,x_\magm)=\sum_P {\rm A}(P) e^{ip_{P(1)}x_1+\ldots+ip_{P(\magm)}x_\magm}\,,
\end{equation} 
where the sum goes over all permutations $P$ of $(1,\ldots,\magm)$. The positions of the magnons along the spin chain are labeled by $x_i$ that take values from the set $\{1,\ldots,L\}$. The amplitude term ${\rm A}(P)$ does in principle depend on the momenta $p_i$. To determine it one again has to study the action of the Hamiltonian on (\ref{wave_su2}). Ignoring boundary terms at site $1$ and $L$ one finds that (\ref{wave_su2}) is an eigenfunction for
\begin{equation}\label{cbaampli}
 {\rm A}(P)={\rm A}(P(1),\ldots,P(\magm))=\prod_{1\leq j<k\leq \magm}\frac{z_{P(j)}-z_{P(k)}+1}{z_{P(j)}-z_{P(k)}}\,,
\end{equation} 
see e.g. \cite{Karbach1997,Ovchinnikov2010}. Here we introduced the rapidities $z_i$ which will become the Bethe roots. They are related to the momenta via 
\begin{equation}
 p_k=\frac{1}{i}\ln\frac{z_k+1}{z_k}\,.
\end{equation} 
Let us also note that the total momentum is related to the eigenvalues of the shift operator via
\begin{equation}
 U=e^{i \sum_{k=1}^\magm p_k}\,.
\end{equation} 
The corresponding energy eigenvalues are simply the sum of the one particle energies \eqref{energy_su2_one}
\begin{equation}\label{energy_su2}
 E_\text{XXX}=4\sum_{k=1}^\magm(1-\cos p_k)=-2\sum_{k=1}^\magm\left(\frac{1}{z_k}-\frac{1}{z_k+1}\right)\,,
\end{equation} 
compare \eqref{abaenergy} and \eqref{abashift} as well as \eqref{hamrel}.
We have diagonalized the Hamiltonian in the infinite volume and will now impose the periodic boundary conditions. This will lead to a rather familiar restriction on the momenta $p_i$ or equivalently $z_i$.

To impose periodicity let us first consider the simple example of $\magm=3$ magnons on a spin chain of length $L+1$. In this example the magnons are located at the positions $x_1=2,x_2=3,x_3=L+1$. As in the algebraic Bethe ansatz we have to identify site $L+1$ with site $1$. Then the cyclicity condition on the components of the wave-function \eqref{wave_comp_su2} reads
\begin{equation}
  \psi(2,3,L+1)\stackrel{!}{=} \psi(1,2,3)\,.
\end{equation} 
This generalizes to an arbitrary $\magm$ magnon configuration as
\begin{equation}
  \psi(x_2,x_3,\ldots,x_\magm,L+x_1)\stackrel{!}{=} \psi(x_1,\ldots,x_\magm)\,.
\end{equation} 
To obtain the constraints on the momenta  we substitute \eqref{wave_comp_su2} into the condition and obtain
\begin{equation}\label{impose_per}
\sum_P {\rm A}(P) e^{ip_{P(1)}x_2+\ldots+ip_{P(\magm)}(x_1+L)}\stackrel{!}{=}\sum_P {\rm A}(P) e^{ip_{P(1)}x_1+\ldots+ip_{P(\magm)}x_\magm}\,.
\end{equation} 
By shifting the permutation according to $P(i)={\tilde P}(i+1)$ and $P(\magm)={\tilde P}(1)$ we can rewrite the left hand side of \eqref{impose_per} and find
\begin{equation}
\sum_{\tilde P} {\rm A}(\tilde P) e^{ip_{\tilde P(1)}x_1+\ldots+ip_{\tilde P(\magm)}x_\magm}e^{ip_{\tilde P(1)}L}\stackrel{!}{=}\sum_P {\rm A}(P) e^{ip_{P(1)}x_1+\ldots+ip_{P(\magm)}x_\magm}\,.
\end{equation} 
By comparing the single terms in the sum we find that this equation is satisfied for all $x_i$ if
\begin{equation}\label{cyctr}
e^{ip_{P(1)}L}=\frac{{\rm A}(P(1),\ldots,P(\magm))}{ {\rm A}(P(2),\ldots,P(\magm),P(1))\,}\,.
\end{equation} 
These are the Bethe equations. To rewrite them in the way as presented in \eqref{bae} we note that the fraction of the amplitude terms is related to the so-called S-matrix via
\begin{equation}\label{smatrix_su2}
 \frac{{\rm A}(1,2,\ldots,j,j+1,\ldots,\magm)}{{\rm A}(1,2,\ldots,j+1,j,\ldots,\magm)}=\frac{z_j-z_{j+1}+1}{z_j-z_{j+1}-1}=S(z_j,z_{j+1})\,,
\end{equation} 
from which follows that
\begin{equation}\label{aind}
 \frac{{\rm A}(P(1),\ldots,P(\magm))}{{\rm A}(P(2),\ldots,P(\magm),P(1))}=\prod_{j=2}^\magm S(z_{P(1)},z_{P(j)})\,.
\end{equation} 
Using \eqref{aind} to rewrite \eqref{cyctr} we find
\begin{equation}\label{baemon}
 e^{ip_kL}=-\prod_{j=1}^\magm S(z_k,z_j)\,.
\end{equation} 
In terms of the Bethe roots $z_k$ this is equation reads
\begin{equation}
 \left(\frac{z_k+1}{z_k}\right)^L=-\prod_{j=1}^\magm \frac{z_k-z_{j}+1}{z_k-z_{j}-1}\quad\quad k=1,\ldots,\magm \,,
\end{equation} 
which agrees with \eqref{bae}.
The left hand side of \eqref{baemon} is sometimes called the driving term. The right hand side is the scattering part. A way to think about \eqref{baemon} is that a particle scattered through the chain only picks up a phase factor.
If the Bethe equations which appear as the cyclicity condition are satisfied \eqref{wave_su2} is an eigenvector of the closed spin chain Hamiltonian \eqref{su2_ham} with the corresponding eigenvalue \eqref{energy_su2}. 
By dividing the wave-function in \eqref{wave_comp_su2} by ${\rm A}(1,2,\ldots,\magm)$ and 
using the relation given in \eqref{smatrix_su2} subsequently one obtains the more intuitive expression of the wave-function expressed via the S-matrix
\begin{equation}
 \tilde\psi(x_1,\ldots,x_\magm)=e^{ip_{1}x_1+ip_{2}x_2+\ldots+ip_{\magm}x_\magm}+S(z_2,z_1)e^{ip_{2}x_1+ip_{1}x_2+\ldots+ip_{\magm}x_\magm}+\ldots\,.
\end{equation} 
This representation makes the picture of particles moving along the chain and picking up scattering terms when they are interchanged manifest. This factorization into two particle scatterings is a typical sign of integrability.
\section{More on CBA}\label{sec:morecba}
It is far from obvious that the coordinate wave-function in \eqref{wave_su2} coincides with the expression obtained through the algebraic Bethe ansatz \eqref{su2BV} for $v_i=0$ and $s_i=1$. Nevertheless, as we have seen, they diagonalize the same Hamiltonian. In the following we state the wave-function for finite-dimensional representations of $\gl2$ with inhomogeneities such that the results in Section \ref{sec:gl2} and \ref{sec:cba} coincide. These results were checked experimentally using Mathematica. For a proof we refer the reader to appendix 3.E in \cite{Essler2005}, where the generalized two site model, see e.g. \cite{Korepin1997}, is used to obtain the wave-function from the algebraic Bethe ansatz. 

We find that the coordinate wave-function is related to the Bethe vector of the homogeneous $s=1$ $\gl2$ spin chain obtained via the algebraic Bethe ansatz by the overall factor
\begin{equation}\label{cbaaba}
 \vert\Phi^\text{hom}_{s=1}\rangle=\prod_{k=1}^\magm\frac{z_k^L}{z_k+1}\,\vert\psi\rangle_{\text{CBA}}\,.
\end{equation} 
Note, that the overall normalization above is symmetric under permutations of the Bethe roots and as such can be written inside the sum, cf. \eqref{wave_comp_su2}. The off-shell Bethe vector $\vert\Phi^\text{hom}_{s=1}\rangle$ was given in terms of the $B$ operators in \eqref{su2BV}. Using \eqref{cbaaba} we rewrite it as
\begin{equation}\label{barecbavec1}
\vert\Phi^\text{hom}_{s=1}\rangle =\sum_{1\leq x_1<\ldots<x_\magm\leq L}\Phi^\text{hom}_{s=1}(x_1,\ldots,x_\magm)\,e_{21}^{(x_1)}\cdots e_{21}^{(x_\magm)}\vert\Omega\rangle\,,
\end{equation} 
with
\begin{equation}
\Phi^\text{hom}_{s=1}(x_1,\ldots,x_\magm)=\sum_P {\rm A}(P) \prod_{k=1}^\magm \left(z_{P(k)}+1\right)^{x_k-1}\left(z_{P(k)}\right)^{L-x_k}\,.
\end{equation} 
Furthermore, substituting
\begin{equation}\label{su2inh}
 \Phi^\text{inh}_{s=1}(x_1,\ldots,x_\magm)=\sum_P {\rm A}(P) \prod_{k=1}^\magm \prod_{j=1}^{x_k-1}\left(z_{P(k)}-v_j+1\right)\prod_{j=x_k+1}^{L}\left(z_{P(k)}-v_j\right)\,.
\end{equation} 
into \eqref{barecbavec1} yields the Bethe vectors $\vert\Phi^\text{inh}_{s=1}\rangle$ for $s_i=1$ and inhomogeneities.

For completeness we state the wave-function for the $\gl2$ spin chain with symmetric representations labeled by $s_j$ at each site with inhomogeneities. It is given by
\begin{equation}
 \vert \Phi^\text{inh}\rangle =\sum_{1\leq x_1\leq\ldots\leq x_\magm\leq L}\Phi^\text{inh}(x_1,\ldots,x_\magm)\,J_{21}^{(x_1)}\cdots J_{21}^{(x_\magm)}\vert\Omega\rangle\,,
\end{equation} 
with
\begin{equation}
 \Phi^\text{inh}(x_1,\ldots,x_\magm)=\omega\sum_P {\rm A}(P) \prod_{k=1}^\magm \prod_{j=1}^{x_k-1}\left(z_{P(k)}-v_j+s_j\right)\prod_{j=x_k+1}^{L}\left(z_{P(k)}-v_j\right)\,,
\end{equation}
and
\begin{equation}
\omega=\prod_{i=1}^L \frac{1}{\omega_{i}!}\,.
\end{equation}
The variables $\omega_{i}$ count the multiplicity of the site  $i$ appearing in the set $\{x_1,\ldots,x_\magm\}$. And as such counts the excitations per site and takes values $0\leq \omega_i\leq s_i$.

The wave-function for the fundamental representation of $\gln$ is known. It can be expressed in a rather intuitive way in terms of nested wave-functions of $n-1$ particle types, see Appendix~\ref{app:su3} for the case of $\gl{3}$. We refer the interested reader to \cite{jorgephd} where the $\gln$- case is reviewed. However, in Appendix~\ref{app:su3} we present the wave function for a $\gl{3}$-invariant spin chain with the representations $\Lambda=(1,0,0)$ and $\Lambda'=(1,1,0)$ in the quantum space.

\chapter{2d solvable lattice models}\label{ch:2dlattice}
In the previous chapter we discussed how the Hamiltonian along with the family of commuting operators of one-dimensional quantum integrable spin chains can be diagonalized using Bethe ansatz methods. In the current chapter we focus on classical statistical systems in two dimensions defined on a lattice. We review how these at first sight unconnected topics are closely related.

In the following, we encounter different kinds of lattices. Most of them are built from line segments, for short lines in the following, that are allowed to intersect as long as not more than two lines meet in one point. We call such an intersection a lattice site or vertex. Additionally, we will also introduce certain trivalent vertices which can be obtained from the ordinary (four-valent) vertices. They naturally appear in the fusion procedure of R-matrices and in the so-called bootstrap equation. In the further sections we will then assign orientations and rapidities to the lines. Furthermore, to each lattice edge we associate a state. In this way various configurations depending on the edge variables can be assigned to a vertex. To each of these configurations we associate a Boltzmann weight that can be thought of as the probability with which such a given vertex configuration can appear. The main observable that we are interested in is the partition function. The partition function is given by the sum over all possible interior vertex configurations where the states at the boundary are held fixed. The result will then in general simply be a complex number. 

Lattice models appeared in various research fields. One of the most famous and early applications was the modelling of ice. The solution of the ice-type models goes back to Lieb who was able to connect it to the Bethe ansatz in 1967 \cite{Lieb1967,Lieb1967a}. Let us consider a rectangular lattice as shown in Figure~\ref{fig:rect}. 
\begin{figure}[ht]
  \centering
    \includegraphics[width=0.4\textwidth]{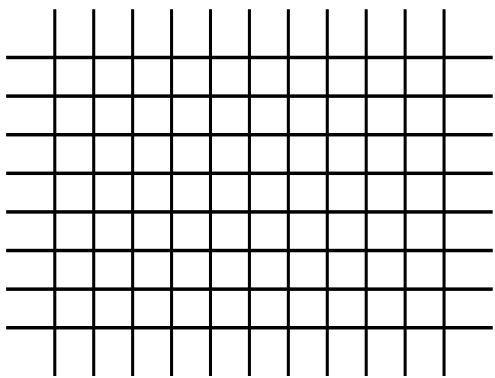}
  \caption{A rectangular lattice.}
  \label{fig:rect}
\end{figure}
At each vertex we can imagine an oxygen atom connecting by hydrogen atoms between the lattice sites to its neighbours. We can impose different boundary conditions on such a lattice. As shown in Figure~\ref{fig:zyl} we can identify the left and right endpoints of the lattice lines to obtain a cylindrical lattice topology. After identifying the end points at the top and bottom we end up with a torus as shown in Figure~\ref{fig:tor}. 
\begin{figure}[ht]
        \centering
        \begin{subfigure}[b]{0.3\textwidth}
                \includegraphics[width=\textwidth]{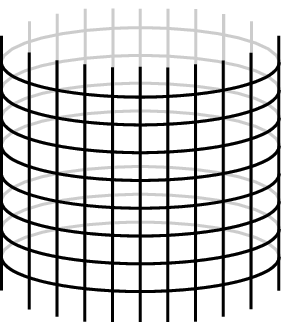}
                \caption{}
                   \label{fig:zyl}
                   \end{subfigure}
                   \quad\quad\quad\quad
         \begin{subfigure}[b]{0.4\textwidth}
                \includegraphics[width=\textwidth]{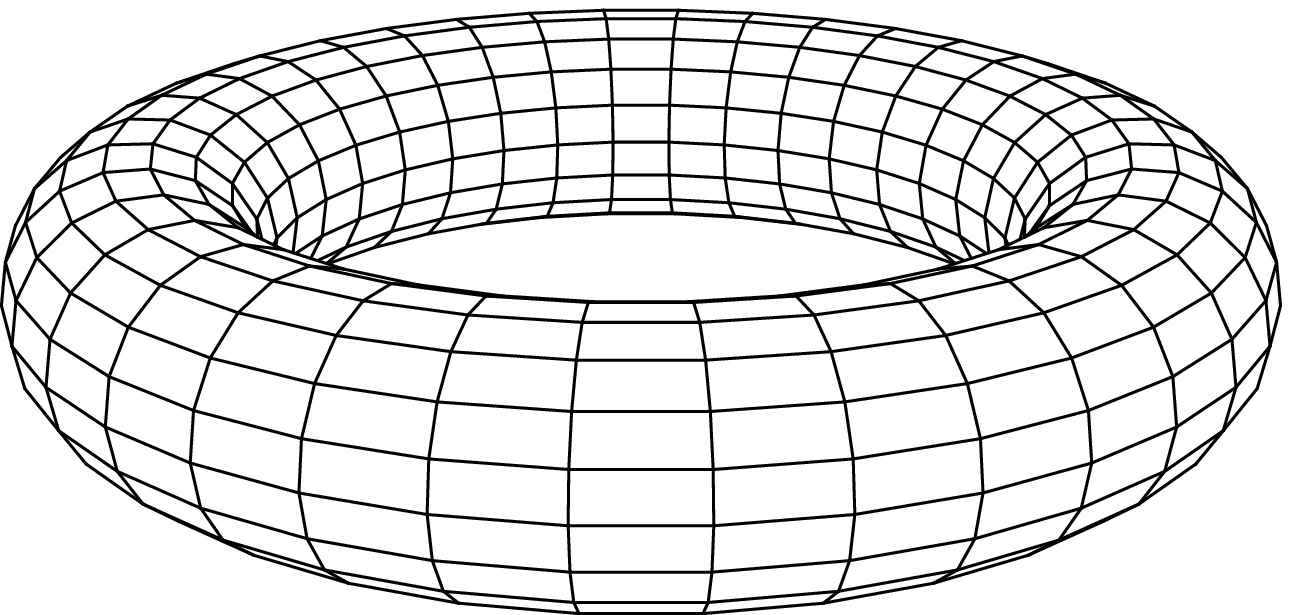}
                \caption{}
                \label{fig:tor}
                \end{subfigure}
                \caption{Rectangular lattice with (a) cylindrical (b) toroidal topology.}
                \label{Figguu}
\end{figure}
Apart from rectangular lattices we can also have other shapes. The lattice in Figure~\ref{fig:3pt} contains three rectangular lattices that are connected by non-intersecting lines. Interestingly, as explained in Appendix~\ref{app:bvaspf}, for certain boundary configurations  each rectangular lattice can be thought of as a Bethe vector of a $\gl2$-invariant spin chain as presented in \eqref{su2BV}. The whole lattice in Figure~\ref{fig:3pt} corresponds to a contraction of three such Bethe vectors. Recently there has been a lot of interest in calculating such objects in the context of {\small AdS/CFT}. The partition function is intimately related to the structure constants of three-point correlation functions of certain operators in \Nsym{4}, see e.g. \cite{Escobedo2010,Foda2011}. 
\begin{figure}[ht]
        \centering
        \begin{subfigure}[b]{0.4\textwidth}
               \includegraphics[width=\textwidth]{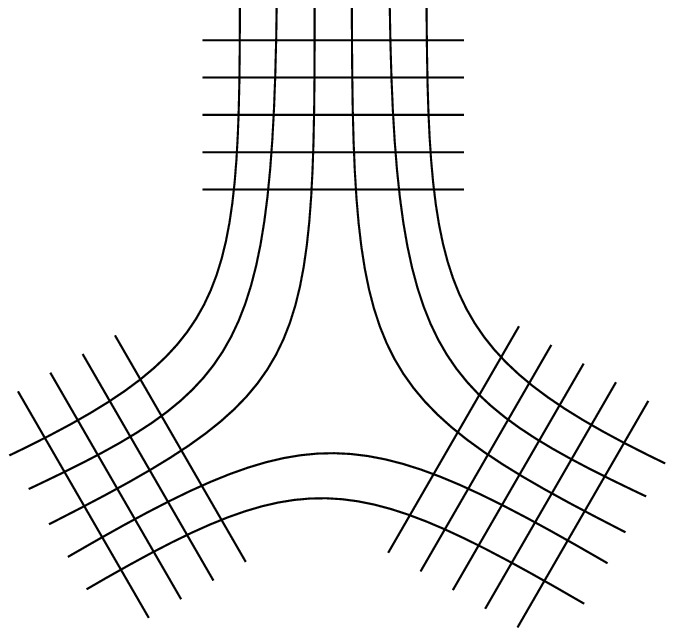}
  \caption{}
  \label{fig:3pt}
                   \end{subfigure}
                   \quad\quad\quad\quad
         \begin{subfigure}[b]{0.4\textwidth}
               \includegraphics[width=\textwidth]{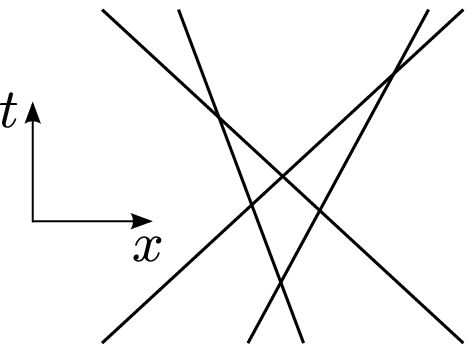}
  \caption{}
  \label{fig:xtlattice}
                \end{subfigure}
                \caption{(a): Non-intersecting contraction of three rectangular lattices. (b): Worldlines in $2d$ spacetime.}
\end{figure}
Another research field on its own are $2d$ integrable quantum field theories that also bear a close relation to lattice models. Here the lattice is given by straight lines that do not meet asymptotically. We think about each line as the world line of a particle parametrized by one space and one time-coordinate $(x,t)$. Initially the particles are well separated at $t\rightarrow-\infty$, then they scatter at the intersections of the lattice to create the end state at $t\rightarrow +\infty$. The particle number is conserved. Such a lattice is depicted in Figure~\ref{fig:xtlattice}. In the context of Baxter's perimeter Bethe ansatz, see Chapter~\ref{ch:BAforYI}, we will study a vertex model that is slightly more general than the one shown in Figure~\ref{fig:xtlattice}. Namely, we loosen the condition that the lines are not allowed to intersect asymptotically, see also Figure~\ref{subfig:uncrazya}. 
\section{The six-vertex model}\label{sec:6vmodel}
As its name indicates, the six-vertex model consist of six different vertices. It is common to denote them with in or outpointing arrows at each edge of a vertex that represent the states as mentioned above. In this case only vertices with as many ingoing arrows as outgoing ones are allowed which constrains the total number of vertices to six. This is often referred to as the ice-rule, see also Section~\ref{sec:relr}. It incorporates the fact that each oxygen atom shares a binding with two hydrogen atoms. To each of these six vertex configurations we associate one of the three different Boltzmann weights $a,b,c$ that can be thought of as the probability of each of these local configurations. In addition there is an extra $\mathbb{Z}_2$ symmetry which leaves the weights invariant under the exchange of all arrow directions such that 
\begin{equation}\label{6v1}
 a[u,v]=
\psset{xunit=.5pt,yunit=.5pt,runit=.5pt}
\begin{pspicture}[shift=-40](76.5,75.67536926)
{
\newrgbcolor{curcolor}{0 0 0}
\pscustom[linewidth=1,linecolor=curcolor]
{
\newpath
\moveto(50.5,50.17536926)
\lineto(50.5,75.17536926)
}
}
{
\newrgbcolor{curcolor}{1 1 1}
\pscustom[linestyle=none,fillstyle=solid,fillcolor=curcolor]
{
\newpath
\moveto(50.5,74.99136926)
\lineto(54.5,68.07136926)
\lineto(46.5,68.07136926)
\lineto(50.5,74.99136926)
\closepath
}
}
{
\newrgbcolor{curcolor}{0 0 0}
\pscustom[linewidth=1,linecolor=curcolor]
{
\newpath
\moveto(50.5,74.99136926)
\lineto(54.5,68.07136926)
\lineto(46.5,68.07136926)
\lineto(50.5,74.99136926)
\closepath
}
}
{
\newrgbcolor{curcolor}{0 0 0}
\pscustom[linewidth=1,linecolor=curcolor]
{
\newpath
\moveto(50.5,24.17536926)
\lineto(50.5,31.65099926)
\lineto(50.5,49.17536926)
}
}
{
\newrgbcolor{curcolor}{1 1 1}
\pscustom[linestyle=none,fillstyle=solid,fillcolor=curcolor]
{
\newpath
\moveto(50.5,31.46699926)
\lineto(54.5,24.54699926)
\lineto(46.5,24.54699926)
\lineto(50.5,31.46699926)
\closepath
}
}
{
\newrgbcolor{curcolor}{0 0 0}
\pscustom[linewidth=1,linecolor=curcolor]
{
\newpath
\moveto(50.5,31.46699926)
\lineto(54.5,24.54699926)
\lineto(46.5,24.54699926)
\lineto(50.5,31.46699926)
\closepath
}
}
{
\newrgbcolor{curcolor}{0 0 0}
\pscustom[linewidth=1,linecolor=curcolor]
{
\newpath
\moveto(25,49.67536926)
\lineto(32.47563,49.67536926)
\lineto(50,49.67536926)
}
}
{
\newrgbcolor{curcolor}{1 1 1}
\pscustom[linestyle=none,fillstyle=solid,fillcolor=curcolor]
{
\newpath
\moveto(32.29163,49.67536926)
\lineto(25.37163,45.67536926)
\lineto(25.37163,53.67536926)
\lineto(32.29163,49.67536926)
\closepath
}
}
{
\newrgbcolor{curcolor}{0 0 0}
\pscustom[linewidth=1,linecolor=curcolor]
{
\newpath
\moveto(32.29163,49.67536926)
\lineto(25.37163,45.67536926)
\lineto(25.37163,53.67536926)
\lineto(32.29163,49.67536926)
\closepath
}
}
{
\newrgbcolor{curcolor}{0 0 0}
\pscustom[linewidth=1,linecolor=curcolor]
{
\newpath
\moveto(51,49.67536926)
\lineto(76,49.67536926)
}
}
{
\newrgbcolor{curcolor}{1 1 1}
\pscustom[linestyle=none,fillstyle=solid,fillcolor=curcolor]
{
\newpath
\moveto(75.816,49.67536926)
\lineto(68.896,45.67536926)
\lineto(68.896,53.67536926)
\lineto(75.816,49.67536926)
\closepath
}
}
{
\newrgbcolor{curcolor}{0 0 0}
\pscustom[linewidth=1,linecolor=curcolor]
{
\newpath
\moveto(75.816,49.67536926)
\lineto(68.896,45.67536926)
\lineto(68.896,53.67536926)
\lineto(75.816,49.67536926)
\closepath
}
}
{
\newrgbcolor{curcolor}{0 0 0}
\pscustom[linestyle=none,fillstyle=solid,fillcolor=curcolor]
{
\newpath
\moveto(51,49.17538188)
\lineto(50,49.17538188)
\lineto(50,50.17538188)
\lineto(51,50.17538188)
\closepath
}
}
{
\newrgbcolor{curcolor}{0 0 0}
\pscustom[linestyle=none,fillstyle=solid,fillcolor=curcolor]
{
\newpath
\moveto(0.5,50.16253926)
\lineto(0.5,49.16253926)
\lineto(15.5,49.16253926)
\lineto(15.5,50.16253926)
\lineto(0.5,50.16253926)
\closepath
}
}
{
\newrgbcolor{curcolor}{0 0 0}
\pscustom[linestyle=none,fillstyle=solid,fillcolor=curcolor]
{
\newpath
\moveto(11.5,49.67536926)
\lineto(9.5,47.67536926)
\lineto(16.5,49.67536926)
\lineto(9.5,51.67536926)
\lineto(11.5,49.67536926)
\closepath
}
}
{
\newrgbcolor{curcolor}{0 0 0}
\pscustom[linewidth=0.5,linecolor=curcolor]
{
\newpath
\moveto(11.5,49.67536926)
\lineto(9.5,47.67536926)
\lineto(16.5,49.67536926)
\lineto(9.5,51.67536926)
\lineto(11.5,49.67536926)
\closepath
}
}
{
\newrgbcolor{curcolor}{0 0 0}
\pscustom[linestyle=none,fillstyle=solid,fillcolor=curcolor]
{
\newpath
\moveto(6.76562367,36.50349248)
\curveto(6.98437367,35.69099248)(7.70312367,35.15974248)(8.54687367,35.15974248)
\curveto(9.23437367,35.15974248)(9.70312367,35.62849248)(10.01562367,36.25349248)
\curveto(10.35937367,36.97224248)(10.60937367,38.19099248)(10.60937367,38.22224248)
\curveto(10.60937367,38.44099248)(10.45312367,38.44099248)(10.39062367,38.44099248)
\curveto(10.17187367,38.44099248)(10.17187367,38.34724248)(10.10937367,38.06599248)
\curveto(9.82812367,36.94099248)(9.45312367,35.59724248)(8.60937367,35.59724248)
\curveto(8.20312367,35.59724248)(7.98437367,35.84724248)(7.98437367,36.50349248)
\curveto(7.98437367,36.94099248)(8.23437367,37.87849248)(8.39062367,38.59724248)
\lineto(8.95312367,40.75349248)
\curveto(9.01562367,41.03474248)(9.20312367,41.78474248)(9.29687367,42.09724248)
\curveto(9.39062367,42.56599248)(9.57812367,43.31599248)(9.57812367,43.44099248)
\curveto(9.57812367,43.78474248)(9.29687367,43.97224248)(9.01562367,43.97224248)
\curveto(8.92187367,43.97224248)(8.39062367,43.94099248)(8.23437367,43.28474248)
\curveto(7.85937367,41.81599248)(6.98437367,38.31599248)(6.73437367,37.28474248)
\curveto(6.70312367,37.19099248)(5.92187367,35.59724248)(4.45312367,35.59724248)
\curveto(3.42187367,35.59724248)(3.23437367,36.50349248)(3.23437367,37.22224248)
\curveto(3.23437367,38.34724248)(3.79687367,39.90974248)(4.29687367,41.28474248)
\curveto(4.54687367,41.90974248)(4.64062367,42.19099248)(4.64062367,42.56599248)
\curveto(4.64062367,43.44099248)(4.01562367,44.19099248)(3.01562367,44.19099248)
\curveto(1.10937367,44.19099248)(0.39062367,41.28474248)(0.39062367,41.12849248)
\curveto(0.39062367,40.90974248)(0.57812367,40.90974248)(0.60937367,40.90974248)
\curveto(0.82812367,40.90974248)(0.82812367,40.97224248)(0.92187367,41.28474248)
\curveto(1.42187367,43.00349248)(2.20312367,43.75349248)(2.95312367,43.75349248)
\curveto(3.14062367,43.75349248)(3.45312367,43.72224248)(3.45312367,43.09724248)
\curveto(3.45312367,42.62849248)(3.23437367,42.03474248)(3.10937367,41.75349248)
\curveto(2.35937367,39.75349248)(1.95312367,38.53474248)(1.95312367,37.56599248)
\curveto(1.95312367,35.65974248)(3.32812367,35.15974248)(4.39062367,35.15974248)
\curveto(5.70312367,35.15974248)(6.42187367,36.06599248)(6.76562367,36.50349248)
\closepath
}
}
{
\newrgbcolor{curcolor}{0 0 0}
\pscustom[linestyle=none,fillstyle=solid,fillcolor=curcolor]
{
\newpath
\moveto(64.8749992,7.78474248)
\curveto(64.8749992,8.87849248)(64.3437492,9.19099248)(63.9999992,9.19099248)
\curveto(63.4999992,9.19099248)(62.9999992,8.65974248)(62.9999992,8.22224248)
\curveto(62.9999992,7.97224248)(63.0937492,7.84724248)(63.3124992,7.62849248)
\curveto(63.7499992,7.22224248)(63.9999992,6.72224248)(63.9999992,6.00349248)
\curveto(63.9999992,5.15974248)(62.7812492,0.59724248)(60.4687492,0.59724248)
\curveto(59.4374992,0.59724248)(58.9687492,1.28474248)(58.9687492,2.34724248)
\curveto(58.9687492,3.44099248)(59.5312492,4.90974248)(60.1249992,6.56599248)
\curveto(60.2812492,6.90974248)(60.3749992,7.19099248)(60.3749992,7.56599248)
\curveto(60.3749992,8.44099248)(59.7499992,9.19099248)(58.7499992,9.19099248)
\curveto(56.8749992,9.19099248)(56.1249992,6.28474248)(56.1249992,6.12849248)
\curveto(56.1249992,5.90974248)(56.3124992,5.90974248)(56.3437492,5.90974248)
\curveto(56.5624992,5.90974248)(56.5624992,5.97224248)(56.6562492,6.28474248)
\curveto(57.2499992,8.28474248)(58.0937492,8.75349248)(58.6874992,8.75349248)
\curveto(58.8437492,8.75349248)(59.1874992,8.75349248)(59.1874992,8.12849248)
\curveto(59.1874992,7.62849248)(58.9687492,7.06599248)(58.8437492,6.72224248)
\curveto(57.9687492,4.40974248)(57.7187492,3.50349248)(57.7187492,2.62849248)
\curveto(57.7187492,0.47224248)(59.4687492,0.15974248)(60.3749992,0.15974248)
\curveto(63.7187492,0.15974248)(64.8749992,6.75349248)(64.8749992,7.78474248)
\closepath
}
}
{
\newrgbcolor{curcolor}{0 0 0}
\pscustom[linestyle=none,fillstyle=solid,fillcolor=curcolor]
{
\newpath
\moveto(50,15.50628926)
\lineto(50,0.50628926)
\lineto(51,0.50628926)
\lineto(51,15.50628926)
\lineto(50,15.50628926)
\closepath
}
}
{
\newrgbcolor{curcolor}{0 0 0}
\pscustom[linestyle=none,fillstyle=solid,fillcolor=curcolor]
{
\newpath
\moveto(50.5,11.49999926)
\lineto(52.5,9.49999926)
\lineto(50.5,16.49999926)
\lineto(48.5,9.49999926)
\lineto(50.5,11.49999926)
\closepath
}
}
{
\newrgbcolor{curcolor}{0 0 0}
\pscustom[linewidth=0.5,linecolor=curcolor]
{
\newpath
\moveto(50.5,11.49999926)
\lineto(52.5,9.49999926)
\lineto(50.5,16.49999926)
\lineto(48.5,9.49999926)
\lineto(50.5,11.49999926)
\closepath
}
}
\end{pspicture}=
\psset{xunit=.5pt,yunit=.5pt,runit=.5pt}
\begin{pspicture}[shift=-40](76.5,75.67536926)
{
\newrgbcolor{curcolor}{0 0 0}
\pscustom[linestyle=none,fillstyle=solid,fillcolor=curcolor]
{
\newpath
\moveto(0.5,50.16253926)
\lineto(0.5,49.16253926)
\lineto(15.5,49.16253926)
\lineto(15.5,50.16253926)
\lineto(0.5,50.16253926)
\closepath
}
}
{
\newrgbcolor{curcolor}{0 0 0}
\pscustom[linestyle=none,fillstyle=solid,fillcolor=curcolor]
{
\newpath
\moveto(11.5,49.67536926)
\lineto(9.5,47.67536926)
\lineto(16.5,49.67536926)
\lineto(9.5,51.67536926)
\lineto(11.5,49.67536926)
\closepath
}
}
{
\newrgbcolor{curcolor}{0 0 0}
\pscustom[linewidth=0.5,linecolor=curcolor]
{
\newpath
\moveto(11.5,49.67536926)
\lineto(9.5,47.67536926)
\lineto(16.5,49.67536926)
\lineto(9.5,51.67536926)
\lineto(11.5,49.67536926)
\closepath
}
}
{
\newrgbcolor{curcolor}{0 0 0}
\pscustom[linestyle=none,fillstyle=solid,fillcolor=curcolor]
{
\newpath
\moveto(6.76562367,36.50349248)
\curveto(6.98437367,35.69099248)(7.70312367,35.15974248)(8.54687367,35.15974248)
\curveto(9.23437367,35.15974248)(9.70312367,35.62849248)(10.01562367,36.25349248)
\curveto(10.35937367,36.97224248)(10.60937367,38.19099248)(10.60937367,38.22224248)
\curveto(10.60937367,38.44099248)(10.45312367,38.44099248)(10.39062367,38.44099248)
\curveto(10.17187367,38.44099248)(10.17187367,38.34724248)(10.10937367,38.06599248)
\curveto(9.82812367,36.94099248)(9.45312367,35.59724248)(8.60937367,35.59724248)
\curveto(8.20312367,35.59724248)(7.98437367,35.84724248)(7.98437367,36.50349248)
\curveto(7.98437367,36.94099248)(8.23437367,37.87849248)(8.39062367,38.59724248)
\lineto(8.95312367,40.75349248)
\curveto(9.01562367,41.03474248)(9.20312367,41.78474248)(9.29687367,42.09724248)
\curveto(9.39062367,42.56599248)(9.57812367,43.31599248)(9.57812367,43.44099248)
\curveto(9.57812367,43.78474248)(9.29687367,43.97224248)(9.01562367,43.97224248)
\curveto(8.92187367,43.97224248)(8.39062367,43.94099248)(8.23437367,43.28474248)
\curveto(7.85937367,41.81599248)(6.98437367,38.31599248)(6.73437367,37.28474248)
\curveto(6.70312367,37.19099248)(5.92187367,35.59724248)(4.45312367,35.59724248)
\curveto(3.42187367,35.59724248)(3.23437367,36.50349248)(3.23437367,37.22224248)
\curveto(3.23437367,38.34724248)(3.79687367,39.90974248)(4.29687367,41.28474248)
\curveto(4.54687367,41.90974248)(4.64062367,42.19099248)(4.64062367,42.56599248)
\curveto(4.64062367,43.44099248)(4.01562367,44.19099248)(3.01562367,44.19099248)
\curveto(1.10937367,44.19099248)(0.39062367,41.28474248)(0.39062367,41.12849248)
\curveto(0.39062367,40.90974248)(0.57812367,40.90974248)(0.60937367,40.90974248)
\curveto(0.82812367,40.90974248)(0.82812367,40.97224248)(0.92187367,41.28474248)
\curveto(1.42187367,43.00349248)(2.20312367,43.75349248)(2.95312367,43.75349248)
\curveto(3.14062367,43.75349248)(3.45312367,43.72224248)(3.45312367,43.09724248)
\curveto(3.45312367,42.62849248)(3.23437367,42.03474248)(3.10937367,41.75349248)
\curveto(2.35937367,39.75349248)(1.95312367,38.53474248)(1.95312367,37.56599248)
\curveto(1.95312367,35.65974248)(3.32812367,35.15974248)(4.39062367,35.15974248)
\curveto(5.70312367,35.15974248)(6.42187367,36.06599248)(6.76562367,36.50349248)
\closepath
}
}
{
\newrgbcolor{curcolor}{0 0 0}
\pscustom[linestyle=none,fillstyle=solid,fillcolor=curcolor]
{
\newpath
\moveto(64.8749992,7.78474248)
\curveto(64.8749992,8.87849248)(64.3437492,9.19099248)(63.9999992,9.19099248)
\curveto(63.4999992,9.19099248)(62.9999992,8.65974248)(62.9999992,8.22224248)
\curveto(62.9999992,7.97224248)(63.0937492,7.84724248)(63.3124992,7.62849248)
\curveto(63.7499992,7.22224248)(63.9999992,6.72224248)(63.9999992,6.00349248)
\curveto(63.9999992,5.15974248)(62.7812492,0.59724248)(60.4687492,0.59724248)
\curveto(59.4374992,0.59724248)(58.9687492,1.28474248)(58.9687492,2.34724248)
\curveto(58.9687492,3.44099248)(59.5312492,4.90974248)(60.1249992,6.56599248)
\curveto(60.2812492,6.90974248)(60.3749992,7.19099248)(60.3749992,7.56599248)
\curveto(60.3749992,8.44099248)(59.7499992,9.19099248)(58.7499992,9.19099248)
\curveto(56.8749992,9.19099248)(56.1249992,6.28474248)(56.1249992,6.12849248)
\curveto(56.1249992,5.90974248)(56.3124992,5.90974248)(56.3437492,5.90974248)
\curveto(56.5624992,5.90974248)(56.5624992,5.97224248)(56.6562492,6.28474248)
\curveto(57.2499992,8.28474248)(58.0937492,8.75349248)(58.6874992,8.75349248)
\curveto(58.8437492,8.75349248)(59.1874992,8.75349248)(59.1874992,8.12849248)
\curveto(59.1874992,7.62849248)(58.9687492,7.06599248)(58.8437492,6.72224248)
\curveto(57.9687492,4.40974248)(57.7187492,3.50349248)(57.7187492,2.62849248)
\curveto(57.7187492,0.47224248)(59.4687492,0.15974248)(60.3749992,0.15974248)
\curveto(63.7187492,0.15974248)(64.8749992,6.75349248)(64.8749992,7.78474248)
\closepath
}
}
{
\newrgbcolor{curcolor}{0 0 0}
\pscustom[linestyle=none,fillstyle=solid,fillcolor=curcolor]
{
\newpath
\moveto(50,15.50628926)
\lineto(50,0.50628926)
\lineto(51,0.50628926)
\lineto(51,15.50628926)
\lineto(50,15.50628926)
\closepath
}
}
{
\newrgbcolor{curcolor}{0 0 0}
\pscustom[linestyle=none,fillstyle=solid,fillcolor=curcolor]
{
\newpath
\moveto(50.5,11.49999926)
\lineto(52.5,9.49999926)
\lineto(50.5,16.49999926)
\lineto(48.5,9.49999926)
\lineto(50.5,11.49999926)
\closepath
}
}
{
\newrgbcolor{curcolor}{0 0 0}
\pscustom[linewidth=0.5,linecolor=curcolor]
{
\newpath
\moveto(50.5,11.49999926)
\lineto(52.5,9.49999926)
\lineto(50.5,16.49999926)
\lineto(48.5,9.49999926)
\lineto(50.5,11.49999926)
\closepath
}
}
{
\newrgbcolor{curcolor}{0 0 0}
\pscustom[linewidth=1,linecolor=curcolor]
{
\newpath
\moveto(50.5,49.1753521)
\lineto(50.5,24.1753521)
}
}
{
\newrgbcolor{curcolor}{1 1 1}
\pscustom[linestyle=none,fillstyle=solid,fillcolor=curcolor]
{
\newpath
\moveto(50.5,24.3593521)
\lineto(46.5,31.2793521)
\lineto(54.5,31.2793521)
\lineto(50.5,24.3593521)
\closepath
}
}
{
\newrgbcolor{curcolor}{0 0 0}
\pscustom[linewidth=1,linecolor=curcolor]
{
\newpath
\moveto(50.5,24.3593521)
\lineto(46.5,31.2793521)
\lineto(54.5,31.2793521)
\lineto(50.5,24.3593521)
\closepath
}
}
{
\newrgbcolor{curcolor}{0 0 0}
\pscustom[linewidth=1,linecolor=curcolor]
{
\newpath
\moveto(50.5,75.1753521)
\lineto(50.5,67.6997221)
\lineto(50.5,50.1753521)
}
}
{
\newrgbcolor{curcolor}{1 1 1}
\pscustom[linestyle=none,fillstyle=solid,fillcolor=curcolor]
{
\newpath
\moveto(50.5,67.8837221)
\lineto(46.5,74.8037221)
\lineto(54.5,74.8037221)
\lineto(50.5,67.8837221)
\closepath
}
}
{
\newrgbcolor{curcolor}{0 0 0}
\pscustom[linewidth=1,linecolor=curcolor]
{
\newpath
\moveto(50.5,67.8837221)
\lineto(46.5,74.8037221)
\lineto(54.5,74.8037221)
\lineto(50.5,67.8837221)
\closepath
}
}
{
\newrgbcolor{curcolor}{0 0 0}
\pscustom[linewidth=1,linecolor=curcolor]
{
\newpath
\moveto(76,49.6753521)
\lineto(68.52437,49.6753521)
\lineto(51,49.6753521)
}
}
{
\newrgbcolor{curcolor}{1 1 1}
\pscustom[linestyle=none,fillstyle=solid,fillcolor=curcolor]
{
\newpath
\moveto(68.70837,49.6753521)
\lineto(75.62837,53.6753521)
\lineto(75.62837,45.6753521)
\lineto(68.70837,49.6753521)
\closepath
}
}
{
\newrgbcolor{curcolor}{0 0 0}
\pscustom[linewidth=1,linecolor=curcolor]
{
\newpath
\moveto(68.70837,49.6753521)
\lineto(75.62837,53.6753521)
\lineto(75.62837,45.6753521)
\lineto(68.70837,49.6753521)
\closepath
}
}
{
\newrgbcolor{curcolor}{0 0 0}
\pscustom[linewidth=1,linecolor=curcolor]
{
\newpath
\moveto(50,49.6753521)
\lineto(25,49.6753521)
}
}
{
\newrgbcolor{curcolor}{1 1 1}
\pscustom[linestyle=none,fillstyle=solid,fillcolor=curcolor]
{
\newpath
\moveto(25.184,49.6753521)
\lineto(32.104,53.6753521)
\lineto(32.104,45.6753521)
\lineto(25.184,49.6753521)
\closepath
}
}
{
\newrgbcolor{curcolor}{0 0 0}
\pscustom[linewidth=1,linecolor=curcolor]
{
\newpath
\moveto(25.184,49.6753521)
\lineto(32.104,53.6753521)
\lineto(32.104,45.6753521)
\lineto(25.184,49.6753521)
\closepath
}
}
{
\newrgbcolor{curcolor}{0 0 0}
\pscustom[linestyle=none,fillstyle=solid,fillcolor=curcolor]
{
\newpath
\moveto(50,50.17537)
\lineto(51,50.17537)
\lineto(51,49.17537)
\lineto(50,49.17537)
\closepath
}
}
\end{pspicture},\quad\quad b[u,v]=
\psset{xunit=.5pt,yunit=.5pt,runit=.5pt}
\begin{pspicture}[shift=-40](76.5,75.67536926)
{
\newrgbcolor{curcolor}{0 0 0}
\pscustom[linestyle=none,fillstyle=solid,fillcolor=curcolor]
{
\newpath
\moveto(0.5,50.16253926)
\lineto(0.5,49.16253926)
\lineto(15.5,49.16253926)
\lineto(15.5,50.16253926)
\lineto(0.5,50.16253926)
\closepath
}
}
{
\newrgbcolor{curcolor}{0 0 0}
\pscustom[linestyle=none,fillstyle=solid,fillcolor=curcolor]
{
\newpath
\moveto(11.5,49.67536926)
\lineto(9.5,47.67536926)
\lineto(16.5,49.67536926)
\lineto(9.5,51.67536926)
\lineto(11.5,49.67536926)
\closepath
}
}
{
\newrgbcolor{curcolor}{0 0 0}
\pscustom[linewidth=0.5,linecolor=curcolor]
{
\newpath
\moveto(11.5,49.67536926)
\lineto(9.5,47.67536926)
\lineto(16.5,49.67536926)
\lineto(9.5,51.67536926)
\lineto(11.5,49.67536926)
\closepath
}
}
{
\newrgbcolor{curcolor}{0 0 0}
\pscustom[linestyle=none,fillstyle=solid,fillcolor=curcolor]
{
\newpath
\moveto(6.76562367,36.50349248)
\curveto(6.98437367,35.69099248)(7.70312367,35.15974248)(8.54687367,35.15974248)
\curveto(9.23437367,35.15974248)(9.70312367,35.62849248)(10.01562367,36.25349248)
\curveto(10.35937367,36.97224248)(10.60937367,38.19099248)(10.60937367,38.22224248)
\curveto(10.60937367,38.44099248)(10.45312367,38.44099248)(10.39062367,38.44099248)
\curveto(10.17187367,38.44099248)(10.17187367,38.34724248)(10.10937367,38.06599248)
\curveto(9.82812367,36.94099248)(9.45312367,35.59724248)(8.60937367,35.59724248)
\curveto(8.20312367,35.59724248)(7.98437367,35.84724248)(7.98437367,36.50349248)
\curveto(7.98437367,36.94099248)(8.23437367,37.87849248)(8.39062367,38.59724248)
\lineto(8.95312367,40.75349248)
\curveto(9.01562367,41.03474248)(9.20312367,41.78474248)(9.29687367,42.09724248)
\curveto(9.39062367,42.56599248)(9.57812367,43.31599248)(9.57812367,43.44099248)
\curveto(9.57812367,43.78474248)(9.29687367,43.97224248)(9.01562367,43.97224248)
\curveto(8.92187367,43.97224248)(8.39062367,43.94099248)(8.23437367,43.28474248)
\curveto(7.85937367,41.81599248)(6.98437367,38.31599248)(6.73437367,37.28474248)
\curveto(6.70312367,37.19099248)(5.92187367,35.59724248)(4.45312367,35.59724248)
\curveto(3.42187367,35.59724248)(3.23437367,36.50349248)(3.23437367,37.22224248)
\curveto(3.23437367,38.34724248)(3.79687367,39.90974248)(4.29687367,41.28474248)
\curveto(4.54687367,41.90974248)(4.64062367,42.19099248)(4.64062367,42.56599248)
\curveto(4.64062367,43.44099248)(4.01562367,44.19099248)(3.01562367,44.19099248)
\curveto(1.10937367,44.19099248)(0.39062367,41.28474248)(0.39062367,41.12849248)
\curveto(0.39062367,40.90974248)(0.57812367,40.90974248)(0.60937367,40.90974248)
\curveto(0.82812367,40.90974248)(0.82812367,40.97224248)(0.92187367,41.28474248)
\curveto(1.42187367,43.00349248)(2.20312367,43.75349248)(2.95312367,43.75349248)
\curveto(3.14062367,43.75349248)(3.45312367,43.72224248)(3.45312367,43.09724248)
\curveto(3.45312367,42.62849248)(3.23437367,42.03474248)(3.10937367,41.75349248)
\curveto(2.35937367,39.75349248)(1.95312367,38.53474248)(1.95312367,37.56599248)
\curveto(1.95312367,35.65974248)(3.32812367,35.15974248)(4.39062367,35.15974248)
\curveto(5.70312367,35.15974248)(6.42187367,36.06599248)(6.76562367,36.50349248)
\closepath
}
}
{
\newrgbcolor{curcolor}{0 0 0}
\pscustom[linestyle=none,fillstyle=solid,fillcolor=curcolor]
{
\newpath
\moveto(64.8749992,7.78474248)
\curveto(64.8749992,8.87849248)(64.3437492,9.19099248)(63.9999992,9.19099248)
\curveto(63.4999992,9.19099248)(62.9999992,8.65974248)(62.9999992,8.22224248)
\curveto(62.9999992,7.97224248)(63.0937492,7.84724248)(63.3124992,7.62849248)
\curveto(63.7499992,7.22224248)(63.9999992,6.72224248)(63.9999992,6.00349248)
\curveto(63.9999992,5.15974248)(62.7812492,0.59724248)(60.4687492,0.59724248)
\curveto(59.4374992,0.59724248)(58.9687492,1.28474248)(58.9687492,2.34724248)
\curveto(58.9687492,3.44099248)(59.5312492,4.90974248)(60.1249992,6.56599248)
\curveto(60.2812492,6.90974248)(60.3749992,7.19099248)(60.3749992,7.56599248)
\curveto(60.3749992,8.44099248)(59.7499992,9.19099248)(58.7499992,9.19099248)
\curveto(56.8749992,9.19099248)(56.1249992,6.28474248)(56.1249992,6.12849248)
\curveto(56.1249992,5.90974248)(56.3124992,5.90974248)(56.3437492,5.90974248)
\curveto(56.5624992,5.90974248)(56.5624992,5.97224248)(56.6562492,6.28474248)
\curveto(57.2499992,8.28474248)(58.0937492,8.75349248)(58.6874992,8.75349248)
\curveto(58.8437492,8.75349248)(59.1874992,8.75349248)(59.1874992,8.12849248)
\curveto(59.1874992,7.62849248)(58.9687492,7.06599248)(58.8437492,6.72224248)
\curveto(57.9687492,4.40974248)(57.7187492,3.50349248)(57.7187492,2.62849248)
\curveto(57.7187492,0.47224248)(59.4687492,0.15974248)(60.3749992,0.15974248)
\curveto(63.7187492,0.15974248)(64.8749992,6.75349248)(64.8749992,7.78474248)
\closepath
}
}
{
\newrgbcolor{curcolor}{0 0 0}
\pscustom[linestyle=none,fillstyle=solid,fillcolor=curcolor]
{
\newpath
\moveto(50,15.50628926)
\lineto(50,0.50628926)
\lineto(51,0.50628926)
\lineto(51,15.50628926)
\lineto(50,15.50628926)
\closepath
}
}
{
\newrgbcolor{curcolor}{0 0 0}
\pscustom[linestyle=none,fillstyle=solid,fillcolor=curcolor]
{
\newpath
\moveto(50.5,11.49999926)
\lineto(52.5,9.49999926)
\lineto(50.5,16.49999926)
\lineto(48.5,9.49999926)
\lineto(50.5,11.49999926)
\closepath
}
}
{
\newrgbcolor{curcolor}{0 0 0}
\pscustom[linewidth=0.5,linecolor=curcolor]
{
\newpath
\moveto(50.5,11.49999926)
\lineto(52.5,9.49999926)
\lineto(50.5,16.49999926)
\lineto(48.5,9.49999926)
\lineto(50.5,11.49999926)
\closepath
}
}
{
\newrgbcolor{curcolor}{0 0 0}
\pscustom[linewidth=1,linecolor=curcolor]
{
\newpath
\moveto(51,49.67536926)
\lineto(76,49.67536926)
}
}
{
\newrgbcolor{curcolor}{1 1 1}
\pscustom[linestyle=none,fillstyle=solid,fillcolor=curcolor]
{
\newpath
\moveto(75.816,49.67536926)
\lineto(68.896,45.67536926)
\lineto(68.896,53.67536926)
\lineto(75.816,49.67536926)
\closepath
}
}
{
\newrgbcolor{curcolor}{0 0 0}
\pscustom[linewidth=1,linecolor=curcolor]
{
\newpath
\moveto(75.816,49.67536926)
\lineto(68.896,45.67536926)
\lineto(68.896,53.67536926)
\lineto(75.816,49.67536926)
\closepath
}
}
{
\newrgbcolor{curcolor}{0 0 0}
\pscustom[linewidth=1,linecolor=curcolor]
{
\newpath
\moveto(25,49.67536926)
\lineto(32.47563,49.67536926)
\lineto(50,49.67536926)
}
}
{
\newrgbcolor{curcolor}{1 1 1}
\pscustom[linestyle=none,fillstyle=solid,fillcolor=curcolor]
{
\newpath
\moveto(32.29163,49.67536926)
\lineto(25.37163,45.67536926)
\lineto(25.37163,53.67536926)
\lineto(32.29163,49.67536926)
\closepath
}
}
{
\newrgbcolor{curcolor}{0 0 0}
\pscustom[linewidth=1,linecolor=curcolor]
{
\newpath
\moveto(32.29163,49.67536926)
\lineto(25.37163,45.67536926)
\lineto(25.37163,53.67536926)
\lineto(32.29163,49.67536926)
\closepath
}
}
{
\newrgbcolor{curcolor}{0 0 0}
\pscustom[linewidth=1,linecolor=curcolor]
{
\newpath
\moveto(50.5,75.17536926)
\lineto(50.5,67.69973926)
\lineto(50.5,50.17536926)
}
}
{
\newrgbcolor{curcolor}{1 1 1}
\pscustom[linestyle=none,fillstyle=solid,fillcolor=curcolor]
{
\newpath
\moveto(50.5,67.88373926)
\lineto(46.5,74.80373926)
\lineto(54.5,74.80373926)
\lineto(50.5,67.88373926)
\closepath
}
}
{
\newrgbcolor{curcolor}{0 0 0}
\pscustom[linewidth=1,linecolor=curcolor]
{
\newpath
\moveto(50.5,67.88373926)
\lineto(46.5,74.80373926)
\lineto(54.5,74.80373926)
\lineto(50.5,67.88373926)
\closepath
}
}
{
\newrgbcolor{curcolor}{0 0 0}
\pscustom[linewidth=1,linecolor=curcolor]
{
\newpath
\moveto(50.5,49.17536926)
\lineto(50.5,24.17536926)
}
}
{
\newrgbcolor{curcolor}{1 1 1}
\pscustom[linestyle=none,fillstyle=solid,fillcolor=curcolor]
{
\newpath
\moveto(50.5,24.35936926)
\lineto(46.5,31.27936926)
\lineto(54.5,31.27936926)
\lineto(50.5,24.35936926)
\closepath
}
}
{
\newrgbcolor{curcolor}{0 0 0}
\pscustom[linewidth=1,linecolor=curcolor]
{
\newpath
\moveto(50.5,24.35936926)
\lineto(46.5,31.27936926)
\lineto(54.5,31.27936926)
\lineto(50.5,24.35936926)
\closepath
}
}
{
\newrgbcolor{curcolor}{0 0 0}
\pscustom[linestyle=none,fillstyle=solid,fillcolor=curcolor]
{
\newpath
\moveto(50.00001262,49.17536926)
\lineto(50.00001262,50.17536926)
\lineto(51.00001262,50.17536926)
\lineto(51.00001262,49.17536926)
\closepath
}
}
\end{pspicture}=
\psset{xunit=.5pt,yunit=.5pt,runit=.5pt}
\begin{pspicture}[shift=-40](76.5,75.67536926)
{
\newrgbcolor{curcolor}{0 0 0}
\pscustom[linestyle=none,fillstyle=solid,fillcolor=curcolor]
{
\newpath
\moveto(0.5,50.16253926)
\lineto(0.5,49.16253926)
\lineto(15.5,49.16253926)
\lineto(15.5,50.16253926)
\lineto(0.5,50.16253926)
\closepath
}
}
{
\newrgbcolor{curcolor}{0 0 0}
\pscustom[linestyle=none,fillstyle=solid,fillcolor=curcolor]
{
\newpath
\moveto(11.5,49.67536926)
\lineto(9.5,47.67536926)
\lineto(16.5,49.67536926)
\lineto(9.5,51.67536926)
\lineto(11.5,49.67536926)
\closepath
}
}
{
\newrgbcolor{curcolor}{0 0 0}
\pscustom[linewidth=0.5,linecolor=curcolor]
{
\newpath
\moveto(11.5,49.67536926)
\lineto(9.5,47.67536926)
\lineto(16.5,49.67536926)
\lineto(9.5,51.67536926)
\lineto(11.5,49.67536926)
\closepath
}
}
{
\newrgbcolor{curcolor}{0 0 0}
\pscustom[linestyle=none,fillstyle=solid,fillcolor=curcolor]
{
\newpath
\moveto(6.76562367,36.50349248)
\curveto(6.98437367,35.69099248)(7.70312367,35.15974248)(8.54687367,35.15974248)
\curveto(9.23437367,35.15974248)(9.70312367,35.62849248)(10.01562367,36.25349248)
\curveto(10.35937367,36.97224248)(10.60937367,38.19099248)(10.60937367,38.22224248)
\curveto(10.60937367,38.44099248)(10.45312367,38.44099248)(10.39062367,38.44099248)
\curveto(10.17187367,38.44099248)(10.17187367,38.34724248)(10.10937367,38.06599248)
\curveto(9.82812367,36.94099248)(9.45312367,35.59724248)(8.60937367,35.59724248)
\curveto(8.20312367,35.59724248)(7.98437367,35.84724248)(7.98437367,36.50349248)
\curveto(7.98437367,36.94099248)(8.23437367,37.87849248)(8.39062367,38.59724248)
\lineto(8.95312367,40.75349248)
\curveto(9.01562367,41.03474248)(9.20312367,41.78474248)(9.29687367,42.09724248)
\curveto(9.39062367,42.56599248)(9.57812367,43.31599248)(9.57812367,43.44099248)
\curveto(9.57812367,43.78474248)(9.29687367,43.97224248)(9.01562367,43.97224248)
\curveto(8.92187367,43.97224248)(8.39062367,43.94099248)(8.23437367,43.28474248)
\curveto(7.85937367,41.81599248)(6.98437367,38.31599248)(6.73437367,37.28474248)
\curveto(6.70312367,37.19099248)(5.92187367,35.59724248)(4.45312367,35.59724248)
\curveto(3.42187367,35.59724248)(3.23437367,36.50349248)(3.23437367,37.22224248)
\curveto(3.23437367,38.34724248)(3.79687367,39.90974248)(4.29687367,41.28474248)
\curveto(4.54687367,41.90974248)(4.64062367,42.19099248)(4.64062367,42.56599248)
\curveto(4.64062367,43.44099248)(4.01562367,44.19099248)(3.01562367,44.19099248)
\curveto(1.10937367,44.19099248)(0.39062367,41.28474248)(0.39062367,41.12849248)
\curveto(0.39062367,40.90974248)(0.57812367,40.90974248)(0.60937367,40.90974248)
\curveto(0.82812367,40.90974248)(0.82812367,40.97224248)(0.92187367,41.28474248)
\curveto(1.42187367,43.00349248)(2.20312367,43.75349248)(2.95312367,43.75349248)
\curveto(3.14062367,43.75349248)(3.45312367,43.72224248)(3.45312367,43.09724248)
\curveto(3.45312367,42.62849248)(3.23437367,42.03474248)(3.10937367,41.75349248)
\curveto(2.35937367,39.75349248)(1.95312367,38.53474248)(1.95312367,37.56599248)
\curveto(1.95312367,35.65974248)(3.32812367,35.15974248)(4.39062367,35.15974248)
\curveto(5.70312367,35.15974248)(6.42187367,36.06599248)(6.76562367,36.50349248)
\closepath
}
}
{
\newrgbcolor{curcolor}{0 0 0}
\pscustom[linestyle=none,fillstyle=solid,fillcolor=curcolor]
{
\newpath
\moveto(64.8749992,7.78474248)
\curveto(64.8749992,8.87849248)(64.3437492,9.19099248)(63.9999992,9.19099248)
\curveto(63.4999992,9.19099248)(62.9999992,8.65974248)(62.9999992,8.22224248)
\curveto(62.9999992,7.97224248)(63.0937492,7.84724248)(63.3124992,7.62849248)
\curveto(63.7499992,7.22224248)(63.9999992,6.72224248)(63.9999992,6.00349248)
\curveto(63.9999992,5.15974248)(62.7812492,0.59724248)(60.4687492,0.59724248)
\curveto(59.4374992,0.59724248)(58.9687492,1.28474248)(58.9687492,2.34724248)
\curveto(58.9687492,3.44099248)(59.5312492,4.90974248)(60.1249992,6.56599248)
\curveto(60.2812492,6.90974248)(60.3749992,7.19099248)(60.3749992,7.56599248)
\curveto(60.3749992,8.44099248)(59.7499992,9.19099248)(58.7499992,9.19099248)
\curveto(56.8749992,9.19099248)(56.1249992,6.28474248)(56.1249992,6.12849248)
\curveto(56.1249992,5.90974248)(56.3124992,5.90974248)(56.3437492,5.90974248)
\curveto(56.5624992,5.90974248)(56.5624992,5.97224248)(56.6562492,6.28474248)
\curveto(57.2499992,8.28474248)(58.0937492,8.75349248)(58.6874992,8.75349248)
\curveto(58.8437492,8.75349248)(59.1874992,8.75349248)(59.1874992,8.12849248)
\curveto(59.1874992,7.62849248)(58.9687492,7.06599248)(58.8437492,6.72224248)
\curveto(57.9687492,4.40974248)(57.7187492,3.50349248)(57.7187492,2.62849248)
\curveto(57.7187492,0.47224248)(59.4687492,0.15974248)(60.3749992,0.15974248)
\curveto(63.7187492,0.15974248)(64.8749992,6.75349248)(64.8749992,7.78474248)
\closepath
}
}
{
\newrgbcolor{curcolor}{0 0 0}
\pscustom[linestyle=none,fillstyle=solid,fillcolor=curcolor]
{
\newpath
\moveto(50,15.50628926)
\lineto(50,0.50628926)
\lineto(51,0.50628926)
\lineto(51,15.50628926)
\lineto(50,15.50628926)
\closepath
}
}
{
\newrgbcolor{curcolor}{0 0 0}
\pscustom[linestyle=none,fillstyle=solid,fillcolor=curcolor]
{
\newpath
\moveto(50.5,11.49999926)
\lineto(52.5,9.49999926)
\lineto(50.5,16.49999926)
\lineto(48.5,9.49999926)
\lineto(50.5,11.49999926)
\closepath
}
}
{
\newrgbcolor{curcolor}{0 0 0}
\pscustom[linewidth=0.5,linecolor=curcolor]
{
\newpath
\moveto(50.5,11.49999926)
\lineto(52.5,9.49999926)
\lineto(50.5,16.49999926)
\lineto(48.5,9.49999926)
\lineto(50.5,11.49999926)
\closepath
}
}
{
\newrgbcolor{curcolor}{0 0 0}
\pscustom[linewidth=1,linecolor=curcolor]
{
\newpath
\moveto(50.00001,49.67536926)
\lineto(25.00001,49.67536926)
}
}
{
\newrgbcolor{curcolor}{1 1 1}
\pscustom[linestyle=none,fillstyle=solid,fillcolor=curcolor]
{
\newpath
\moveto(25.18401,49.67536926)
\lineto(32.10401,53.67536926)
\lineto(32.10401,45.67536926)
\lineto(25.18401,49.67536926)
\closepath
}
}
{
\newrgbcolor{curcolor}{0 0 0}
\pscustom[linewidth=1,linecolor=curcolor]
{
\newpath
\moveto(25.18401,49.67536926)
\lineto(32.10401,53.67536926)
\lineto(32.10401,45.67536926)
\lineto(25.18401,49.67536926)
\closepath
}
}
{
\newrgbcolor{curcolor}{0 0 0}
\pscustom[linewidth=1,linecolor=curcolor]
{
\newpath
\moveto(76.00001,49.67536926)
\lineto(68.52438,49.67536926)
\lineto(51.00001,49.67536926)
}
}
{
\newrgbcolor{curcolor}{1 1 1}
\pscustom[linestyle=none,fillstyle=solid,fillcolor=curcolor]
{
\newpath
\moveto(68.70838,49.67536926)
\lineto(75.62838,53.67536926)
\lineto(75.62838,45.67536926)
\lineto(68.70838,49.67536926)
\closepath
}
}
{
\newrgbcolor{curcolor}{0 0 0}
\pscustom[linewidth=1,linecolor=curcolor]
{
\newpath
\moveto(68.70838,49.67536926)
\lineto(75.62838,53.67536926)
\lineto(75.62838,45.67536926)
\lineto(68.70838,49.67536926)
\closepath
}
}
{
\newrgbcolor{curcolor}{0 0 0}
\pscustom[linewidth=1,linecolor=curcolor]
{
\newpath
\moveto(50.50001,24.17536926)
\lineto(50.50001,31.65099926)
\lineto(50.50001,49.17536926)
}
}
{
\newrgbcolor{curcolor}{1 1 1}
\pscustom[linestyle=none,fillstyle=solid,fillcolor=curcolor]
{
\newpath
\moveto(50.50001,31.46699926)
\lineto(54.50001,24.54699926)
\lineto(46.50001,24.54699926)
\lineto(50.50001,31.46699926)
\closepath
}
}
{
\newrgbcolor{curcolor}{0 0 0}
\pscustom[linewidth=1,linecolor=curcolor]
{
\newpath
\moveto(50.50001,31.46699926)
\lineto(54.50001,24.54699926)
\lineto(46.50001,24.54699926)
\lineto(50.50001,31.46699926)
\closepath
}
}
{
\newrgbcolor{curcolor}{0 0 0}
\pscustom[linewidth=1,linecolor=curcolor]
{
\newpath
\moveto(50.50001,50.17536926)
\lineto(50.50001,75.17536926)
}
}
{
\newrgbcolor{curcolor}{1 1 1}
\pscustom[linestyle=none,fillstyle=solid,fillcolor=curcolor]
{
\newpath
\moveto(50.50001,74.99136926)
\lineto(54.50001,68.07136926)
\lineto(46.50001,68.07136926)
\lineto(50.50001,74.99136926)
\closepath
}
}
{
\newrgbcolor{curcolor}{0 0 0}
\pscustom[linewidth=1,linecolor=curcolor]
{
\newpath
\moveto(50.50001,74.99136926)
\lineto(54.50001,68.07136926)
\lineto(46.50001,68.07136926)
\lineto(50.50001,74.99136926)
\closepath
}
}
{
\newrgbcolor{curcolor}{0 0 0}
\pscustom[linestyle=none,fillstyle=solid,fillcolor=curcolor]
{
\newpath
\moveto(50.99999738,50.17536926)
\lineto(50.99999738,49.17536926)
\lineto(49.99999738,49.17536926)
\lineto(49.99999738,50.17536926)
\closepath
}
}
\end{pspicture},
\end{equation} 
\begin{equation}\label{6v2}
c[u,v]=
\psset{xunit=.5pt,yunit=.5pt,runit=.5pt}
\begin{pspicture}[shift=-40](76.5,75.67536926)
{
\newrgbcolor{curcolor}{0 0 0}
\pscustom[linestyle=none,fillstyle=solid,fillcolor=curcolor]
{
\newpath
\moveto(0.5,50.16253926)
\lineto(0.5,49.16253926)
\lineto(15.5,49.16253926)
\lineto(15.5,50.16253926)
\lineto(0.5,50.16253926)
\closepath
}
}
{
\newrgbcolor{curcolor}{0 0 0}
\pscustom[linestyle=none,fillstyle=solid,fillcolor=curcolor]
{
\newpath
\moveto(11.5,49.67536926)
\lineto(9.5,47.67536926)
\lineto(16.5,49.67536926)
\lineto(9.5,51.67536926)
\lineto(11.5,49.67536926)
\closepath
}
}
{
\newrgbcolor{curcolor}{0 0 0}
\pscustom[linewidth=0.5,linecolor=curcolor]
{
\newpath
\moveto(11.5,49.67536926)
\lineto(9.5,47.67536926)
\lineto(16.5,49.67536926)
\lineto(9.5,51.67536926)
\lineto(11.5,49.67536926)
\closepath
}
}
{
\newrgbcolor{curcolor}{0 0 0}
\pscustom[linestyle=none,fillstyle=solid,fillcolor=curcolor]
{
\newpath
\moveto(6.76562367,36.50349248)
\curveto(6.98437367,35.69099248)(7.70312367,35.15974248)(8.54687367,35.15974248)
\curveto(9.23437367,35.15974248)(9.70312367,35.62849248)(10.01562367,36.25349248)
\curveto(10.35937367,36.97224248)(10.60937367,38.19099248)(10.60937367,38.22224248)
\curveto(10.60937367,38.44099248)(10.45312367,38.44099248)(10.39062367,38.44099248)
\curveto(10.17187367,38.44099248)(10.17187367,38.34724248)(10.10937367,38.06599248)
\curveto(9.82812367,36.94099248)(9.45312367,35.59724248)(8.60937367,35.59724248)
\curveto(8.20312367,35.59724248)(7.98437367,35.84724248)(7.98437367,36.50349248)
\curveto(7.98437367,36.94099248)(8.23437367,37.87849248)(8.39062367,38.59724248)
\lineto(8.95312367,40.75349248)
\curveto(9.01562367,41.03474248)(9.20312367,41.78474248)(9.29687367,42.09724248)
\curveto(9.39062367,42.56599248)(9.57812367,43.31599248)(9.57812367,43.44099248)
\curveto(9.57812367,43.78474248)(9.29687367,43.97224248)(9.01562367,43.97224248)
\curveto(8.92187367,43.97224248)(8.39062367,43.94099248)(8.23437367,43.28474248)
\curveto(7.85937367,41.81599248)(6.98437367,38.31599248)(6.73437367,37.28474248)
\curveto(6.70312367,37.19099248)(5.92187367,35.59724248)(4.45312367,35.59724248)
\curveto(3.42187367,35.59724248)(3.23437367,36.50349248)(3.23437367,37.22224248)
\curveto(3.23437367,38.34724248)(3.79687367,39.90974248)(4.29687367,41.28474248)
\curveto(4.54687367,41.90974248)(4.64062367,42.19099248)(4.64062367,42.56599248)
\curveto(4.64062367,43.44099248)(4.01562367,44.19099248)(3.01562367,44.19099248)
\curveto(1.10937367,44.19099248)(0.39062367,41.28474248)(0.39062367,41.12849248)
\curveto(0.39062367,40.90974248)(0.57812367,40.90974248)(0.60937367,40.90974248)
\curveto(0.82812367,40.90974248)(0.82812367,40.97224248)(0.92187367,41.28474248)
\curveto(1.42187367,43.00349248)(2.20312367,43.75349248)(2.95312367,43.75349248)
\curveto(3.14062367,43.75349248)(3.45312367,43.72224248)(3.45312367,43.09724248)
\curveto(3.45312367,42.62849248)(3.23437367,42.03474248)(3.10937367,41.75349248)
\curveto(2.35937367,39.75349248)(1.95312367,38.53474248)(1.95312367,37.56599248)
\curveto(1.95312367,35.65974248)(3.32812367,35.15974248)(4.39062367,35.15974248)
\curveto(5.70312367,35.15974248)(6.42187367,36.06599248)(6.76562367,36.50349248)
\closepath
}
}
{
\newrgbcolor{curcolor}{0 0 0}
\pscustom[linestyle=none,fillstyle=solid,fillcolor=curcolor]
{
\newpath
\moveto(64.8749992,7.78474248)
\curveto(64.8749992,8.87849248)(64.3437492,9.19099248)(63.9999992,9.19099248)
\curveto(63.4999992,9.19099248)(62.9999992,8.65974248)(62.9999992,8.22224248)
\curveto(62.9999992,7.97224248)(63.0937492,7.84724248)(63.3124992,7.62849248)
\curveto(63.7499992,7.22224248)(63.9999992,6.72224248)(63.9999992,6.00349248)
\curveto(63.9999992,5.15974248)(62.7812492,0.59724248)(60.4687492,0.59724248)
\curveto(59.4374992,0.59724248)(58.9687492,1.28474248)(58.9687492,2.34724248)
\curveto(58.9687492,3.44099248)(59.5312492,4.90974248)(60.1249992,6.56599248)
\curveto(60.2812492,6.90974248)(60.3749992,7.19099248)(60.3749992,7.56599248)
\curveto(60.3749992,8.44099248)(59.7499992,9.19099248)(58.7499992,9.19099248)
\curveto(56.8749992,9.19099248)(56.1249992,6.28474248)(56.1249992,6.12849248)
\curveto(56.1249992,5.90974248)(56.3124992,5.90974248)(56.3437492,5.90974248)
\curveto(56.5624992,5.90974248)(56.5624992,5.97224248)(56.6562492,6.28474248)
\curveto(57.2499992,8.28474248)(58.0937492,8.75349248)(58.6874992,8.75349248)
\curveto(58.8437492,8.75349248)(59.1874992,8.75349248)(59.1874992,8.12849248)
\curveto(59.1874992,7.62849248)(58.9687492,7.06599248)(58.8437492,6.72224248)
\curveto(57.9687492,4.40974248)(57.7187492,3.50349248)(57.7187492,2.62849248)
\curveto(57.7187492,0.47224248)(59.4687492,0.15974248)(60.3749992,0.15974248)
\curveto(63.7187492,0.15974248)(64.8749992,6.75349248)(64.8749992,7.78474248)
\closepath
}
}
{
\newrgbcolor{curcolor}{0 0 0}
\pscustom[linestyle=none,fillstyle=solid,fillcolor=curcolor]
{
\newpath
\moveto(50,15.50628926)
\lineto(50,0.50628926)
\lineto(51,0.50628926)
\lineto(51,15.50628926)
\lineto(50,15.50628926)
\closepath
}
}
{
\newrgbcolor{curcolor}{0 0 0}
\pscustom[linestyle=none,fillstyle=solid,fillcolor=curcolor]
{
\newpath
\moveto(50.5,11.49999926)
\lineto(52.5,9.49999926)
\lineto(50.5,16.49999926)
\lineto(48.5,9.49999926)
\lineto(50.5,11.49999926)
\closepath
}
}
{
\newrgbcolor{curcolor}{0 0 0}
\pscustom[linewidth=0.5,linecolor=curcolor]
{
\newpath
\moveto(50.5,11.49999926)
\lineto(52.5,9.49999926)
\lineto(50.5,16.49999926)
\lineto(48.5,9.49999926)
\lineto(50.5,11.49999926)
\closepath
}
}
{
\newrgbcolor{curcolor}{0 0 0}
\pscustom[linewidth=1,linecolor=curcolor]
{
\newpath
\moveto(50.5,49.17536926)
\lineto(50.5,24.17536926)
}
}
{
\newrgbcolor{curcolor}{1 1 1}
\pscustom[linestyle=none,fillstyle=solid,fillcolor=curcolor]
{
\newpath
\moveto(50.5,24.35936926)
\lineto(46.5,31.27936926)
\lineto(54.5,31.27936926)
\lineto(50.5,24.35936926)
\closepath
}
}
{
\newrgbcolor{curcolor}{0 0 0}
\pscustom[linewidth=1,linecolor=curcolor]
{
\newpath
\moveto(50.5,24.35936926)
\lineto(46.5,31.27936926)
\lineto(54.5,31.27936926)
\lineto(50.5,24.35936926)
\closepath
}
}
{
\newrgbcolor{curcolor}{0 0 0}
\pscustom[linewidth=1,linecolor=curcolor]
{
\newpath
\moveto(25,49.67536926)
\lineto(32.47563,49.67536926)
\lineto(50,49.67536926)
}
}
{
\newrgbcolor{curcolor}{1 1 1}
\pscustom[linestyle=none,fillstyle=solid,fillcolor=curcolor]
{
\newpath
\moveto(32.29163,49.67536926)
\lineto(25.37163,45.67536926)
\lineto(25.37163,53.67536926)
\lineto(32.29163,49.67536926)
\closepath
}
}
{
\newrgbcolor{curcolor}{0 0 0}
\pscustom[linewidth=1,linecolor=curcolor]
{
\newpath
\moveto(32.29163,49.67536926)
\lineto(25.37163,45.67536926)
\lineto(25.37163,53.67536926)
\lineto(32.29163,49.67536926)
\closepath
}
}
{
\newrgbcolor{curcolor}{0 0 0}
\pscustom[linewidth=1,linecolor=curcolor]
{
\newpath
\moveto(76,49.67536926)
\lineto(68.52437,49.67536926)
\lineto(51,49.67536926)
}
}
{
\newrgbcolor{curcolor}{1 1 1}
\pscustom[linestyle=none,fillstyle=solid,fillcolor=curcolor]
{
\newpath
\moveto(68.70837,49.67536926)
\lineto(75.62837,53.67536926)
\lineto(75.62837,45.67536926)
\lineto(68.70837,49.67536926)
\closepath
}
}
{
\newrgbcolor{curcolor}{0 0 0}
\pscustom[linewidth=1,linecolor=curcolor]
{
\newpath
\moveto(68.70837,49.67536926)
\lineto(75.62837,53.67536926)
\lineto(75.62837,45.67536926)
\lineto(68.70837,49.67536926)
\closepath
}
}
{
\newrgbcolor{curcolor}{0 0 0}
\pscustom[linewidth=1,linecolor=curcolor]
{
\newpath
\moveto(50.5,50.17536926)
\lineto(50.5,75.17536926)
}
}
{
\newrgbcolor{curcolor}{1 1 1}
\pscustom[linestyle=none,fillstyle=solid,fillcolor=curcolor]
{
\newpath
\moveto(50.5,74.99136926)
\lineto(54.5,68.07136926)
\lineto(46.5,68.07136926)
\lineto(50.5,74.99136926)
\closepath
}
}
{
\newrgbcolor{curcolor}{0 0 0}
\pscustom[linewidth=1,linecolor=curcolor]
{
\newpath
\moveto(50.5,74.99136926)
\lineto(54.5,68.07136926)
\lineto(46.5,68.07136926)
\lineto(50.5,74.99136926)
\closepath
}
}
{
\newrgbcolor{curcolor}{0 0 0}
\pscustom[linestyle=none,fillstyle=solid,fillcolor=curcolor]
{
\newpath
\moveto(49.99998738,50.175354)
\lineto(50.99998738,50.175354)
\lineto(50.99998738,49.175354)
\lineto(49.99998738,49.175354)
\closepath
}
}
\end{pspicture}=
\psset{xunit=.5pt,yunit=.5pt,runit=.5pt}
\begin{pspicture}[shift=-40](76.5,75.67536926)
{
\newrgbcolor{curcolor}{0 0 0}
\pscustom[linestyle=none,fillstyle=solid,fillcolor=curcolor]
{
\newpath
\moveto(0.5,50.16253926)
\lineto(0.5,49.16253926)
\lineto(15.5,49.16253926)
\lineto(15.5,50.16253926)
\lineto(0.5,50.16253926)
\closepath
}
}
{
\newrgbcolor{curcolor}{0 0 0}
\pscustom[linestyle=none,fillstyle=solid,fillcolor=curcolor]
{
\newpath
\moveto(11.5,49.67536926)
\lineto(9.5,47.67536926)
\lineto(16.5,49.67536926)
\lineto(9.5,51.67536926)
\lineto(11.5,49.67536926)
\closepath
}
}
{
\newrgbcolor{curcolor}{0 0 0}
\pscustom[linewidth=0.5,linecolor=curcolor]
{
\newpath
\moveto(11.5,49.67536926)
\lineto(9.5,47.67536926)
\lineto(16.5,49.67536926)
\lineto(9.5,51.67536926)
\lineto(11.5,49.67536926)
\closepath
}
}
{
\newrgbcolor{curcolor}{0 0 0}
\pscustom[linestyle=none,fillstyle=solid,fillcolor=curcolor]
{
\newpath
\moveto(6.76562367,36.50349248)
\curveto(6.98437367,35.69099248)(7.70312367,35.15974248)(8.54687367,35.15974248)
\curveto(9.23437367,35.15974248)(9.70312367,35.62849248)(10.01562367,36.25349248)
\curveto(10.35937367,36.97224248)(10.60937367,38.19099248)(10.60937367,38.22224248)
\curveto(10.60937367,38.44099248)(10.45312367,38.44099248)(10.39062367,38.44099248)
\curveto(10.17187367,38.44099248)(10.17187367,38.34724248)(10.10937367,38.06599248)
\curveto(9.82812367,36.94099248)(9.45312367,35.59724248)(8.60937367,35.59724248)
\curveto(8.20312367,35.59724248)(7.98437367,35.84724248)(7.98437367,36.50349248)
\curveto(7.98437367,36.94099248)(8.23437367,37.87849248)(8.39062367,38.59724248)
\lineto(8.95312367,40.75349248)
\curveto(9.01562367,41.03474248)(9.20312367,41.78474248)(9.29687367,42.09724248)
\curveto(9.39062367,42.56599248)(9.57812367,43.31599248)(9.57812367,43.44099248)
\curveto(9.57812367,43.78474248)(9.29687367,43.97224248)(9.01562367,43.97224248)
\curveto(8.92187367,43.97224248)(8.39062367,43.94099248)(8.23437367,43.28474248)
\curveto(7.85937367,41.81599248)(6.98437367,38.31599248)(6.73437367,37.28474248)
\curveto(6.70312367,37.19099248)(5.92187367,35.59724248)(4.45312367,35.59724248)
\curveto(3.42187367,35.59724248)(3.23437367,36.50349248)(3.23437367,37.22224248)
\curveto(3.23437367,38.34724248)(3.79687367,39.90974248)(4.29687367,41.28474248)
\curveto(4.54687367,41.90974248)(4.64062367,42.19099248)(4.64062367,42.56599248)
\curveto(4.64062367,43.44099248)(4.01562367,44.19099248)(3.01562367,44.19099248)
\curveto(1.10937367,44.19099248)(0.39062367,41.28474248)(0.39062367,41.12849248)
\curveto(0.39062367,40.90974248)(0.57812367,40.90974248)(0.60937367,40.90974248)
\curveto(0.82812367,40.90974248)(0.82812367,40.97224248)(0.92187367,41.28474248)
\curveto(1.42187367,43.00349248)(2.20312367,43.75349248)(2.95312367,43.75349248)
\curveto(3.14062367,43.75349248)(3.45312367,43.72224248)(3.45312367,43.09724248)
\curveto(3.45312367,42.62849248)(3.23437367,42.03474248)(3.10937367,41.75349248)
\curveto(2.35937367,39.75349248)(1.95312367,38.53474248)(1.95312367,37.56599248)
\curveto(1.95312367,35.65974248)(3.32812367,35.15974248)(4.39062367,35.15974248)
\curveto(5.70312367,35.15974248)(6.42187367,36.06599248)(6.76562367,36.50349248)
\closepath
}
}
{
\newrgbcolor{curcolor}{0 0 0}
\pscustom[linestyle=none,fillstyle=solid,fillcolor=curcolor]
{
\newpath
\moveto(64.8749992,7.78474248)
\curveto(64.8749992,8.87849248)(64.3437492,9.19099248)(63.9999992,9.19099248)
\curveto(63.4999992,9.19099248)(62.9999992,8.65974248)(62.9999992,8.22224248)
\curveto(62.9999992,7.97224248)(63.0937492,7.84724248)(63.3124992,7.62849248)
\curveto(63.7499992,7.22224248)(63.9999992,6.72224248)(63.9999992,6.00349248)
\curveto(63.9999992,5.15974248)(62.7812492,0.59724248)(60.4687492,0.59724248)
\curveto(59.4374992,0.59724248)(58.9687492,1.28474248)(58.9687492,2.34724248)
\curveto(58.9687492,3.44099248)(59.5312492,4.90974248)(60.1249992,6.56599248)
\curveto(60.2812492,6.90974248)(60.3749992,7.19099248)(60.3749992,7.56599248)
\curveto(60.3749992,8.44099248)(59.7499992,9.19099248)(58.7499992,9.19099248)
\curveto(56.8749992,9.19099248)(56.1249992,6.28474248)(56.1249992,6.12849248)
\curveto(56.1249992,5.90974248)(56.3124992,5.90974248)(56.3437492,5.90974248)
\curveto(56.5624992,5.90974248)(56.5624992,5.97224248)(56.6562492,6.28474248)
\curveto(57.2499992,8.28474248)(58.0937492,8.75349248)(58.6874992,8.75349248)
\curveto(58.8437492,8.75349248)(59.1874992,8.75349248)(59.1874992,8.12849248)
\curveto(59.1874992,7.62849248)(58.9687492,7.06599248)(58.8437492,6.72224248)
\curveto(57.9687492,4.40974248)(57.7187492,3.50349248)(57.7187492,2.62849248)
\curveto(57.7187492,0.47224248)(59.4687492,0.15974248)(60.3749992,0.15974248)
\curveto(63.7187492,0.15974248)(64.8749992,6.75349248)(64.8749992,7.78474248)
\closepath
}
}
{
\newrgbcolor{curcolor}{0 0 0}
\pscustom[linestyle=none,fillstyle=solid,fillcolor=curcolor]
{
\newpath
\moveto(50,15.50628926)
\lineto(50,0.50628926)
\lineto(51,0.50628926)
\lineto(51,15.50628926)
\lineto(50,15.50628926)
\closepath
}
}
{
\newrgbcolor{curcolor}{0 0 0}
\pscustom[linestyle=none,fillstyle=solid,fillcolor=curcolor]
{
\newpath
\moveto(50.5,11.49999926)
\lineto(52.5,9.49999926)
\lineto(50.5,16.49999926)
\lineto(48.5,9.49999926)
\lineto(50.5,11.49999926)
\closepath
}
}
{
\newrgbcolor{curcolor}{0 0 0}
\pscustom[linewidth=0.5,linecolor=curcolor]
{
\newpath
\moveto(50.5,11.49999926)
\lineto(52.5,9.49999926)
\lineto(50.5,16.49999926)
\lineto(48.5,9.49999926)
\lineto(50.5,11.49999926)
\closepath
}
}
{
\newrgbcolor{curcolor}{0 0 0}
\pscustom[linewidth=1,linecolor=curcolor]
{
\newpath
\moveto(50.99998474,49.67538262)
\lineto(75.99998474,49.67538262)
}
}
{
\newrgbcolor{curcolor}{1 1 1}
\pscustom[linestyle=none,fillstyle=solid,fillcolor=curcolor]
{
\newpath
\moveto(75.81598474,49.67538262)
\lineto(68.89598474,45.67538262)
\lineto(68.89598474,53.67538262)
\lineto(75.81598474,49.67538262)
\closepath
}
}
{
\newrgbcolor{curcolor}{0 0 0}
\pscustom[linewidth=1,linecolor=curcolor]
{
\newpath
\moveto(75.81598474,49.67538262)
\lineto(68.89598474,45.67538262)
\lineto(68.89598474,53.67538262)
\lineto(75.81598474,49.67538262)
\closepath
}
}
{
\newrgbcolor{curcolor}{0 0 0}
\pscustom[linewidth=1,linecolor=curcolor]
{
\newpath
\moveto(50.49998474,24.17538262)
\lineto(50.49998474,31.65101262)
\lineto(50.49998474,49.17538262)
}
}
{
\newrgbcolor{curcolor}{1 1 1}
\pscustom[linestyle=none,fillstyle=solid,fillcolor=curcolor]
{
\newpath
\moveto(50.49998474,31.46701262)
\lineto(54.49998474,24.54701262)
\lineto(46.49998474,24.54701262)
\lineto(50.49998474,31.46701262)
\closepath
}
}
{
\newrgbcolor{curcolor}{0 0 0}
\pscustom[linewidth=1,linecolor=curcolor]
{
\newpath
\moveto(50.49998474,31.46701262)
\lineto(54.49998474,24.54701262)
\lineto(46.49998474,24.54701262)
\lineto(50.49998474,31.46701262)
\closepath
}
}
{
\newrgbcolor{curcolor}{0 0 0}
\pscustom[linewidth=1,linecolor=curcolor]
{
\newpath
\moveto(50.49998474,75.17538262)
\lineto(50.49998474,67.69975262)
\lineto(50.49998474,50.17538262)
}
}
{
\newrgbcolor{curcolor}{1 1 1}
\pscustom[linestyle=none,fillstyle=solid,fillcolor=curcolor]
{
\newpath
\moveto(50.49998474,67.88375262)
\lineto(46.49998474,74.80375262)
\lineto(54.49998474,74.80375262)
\lineto(50.49998474,67.88375262)
\closepath
}
}
{
\newrgbcolor{curcolor}{0 0 0}
\pscustom[linewidth=1,linecolor=curcolor]
{
\newpath
\moveto(50.49998474,67.88375262)
\lineto(46.49998474,74.80375262)
\lineto(54.49998474,74.80375262)
\lineto(50.49998474,67.88375262)
\closepath
}
}
{
\newrgbcolor{curcolor}{0 0 0}
\pscustom[linewidth=1,linecolor=curcolor]
{
\newpath
\moveto(49.99998474,49.67538262)
\lineto(24.99998474,49.67538262)
}
}
{
\newrgbcolor{curcolor}{1 1 1}
\pscustom[linestyle=none,fillstyle=solid,fillcolor=curcolor]
{
\newpath
\moveto(25.18398474,49.67538262)
\lineto(32.10398474,53.67538262)
\lineto(32.10398474,45.67538262)
\lineto(25.18398474,49.67538262)
\closepath
}
}
{
\newrgbcolor{curcolor}{0 0 0}
\pscustom[linewidth=1,linecolor=curcolor]
{
\newpath
\moveto(25.18398474,49.67538262)
\lineto(32.10398474,53.67538262)
\lineto(32.10398474,45.67538262)
\lineto(25.18398474,49.67538262)
\closepath
}
}
{
\newrgbcolor{curcolor}{0 0 0}
\pscustom[linestyle=none,fillstyle=solid,fillcolor=curcolor]
{
\newpath
\moveto(50,49.17537)
\lineto(50,50.17537)
\lineto(51,50.17537)
\lineto(51,49.17537)
\closepath
}
}
\end{pspicture}\,.
\end{equation} 
Here the complex parameters $u$ and $v$ are the rapidities or spectral parameters assigned to the lines. The white arrows label the different states and the black arrows are used to fix their orientation. 
In the following, we consider the so-called rational case of the six-vertex model where the Boltzmann weights are given up to an overall normalizations by
\begin{equation}\label{6vbw}
 a[u,v]=u-v+1,\quad b[u,v]=u-v,\quad c[u,v]=1\,.
\end{equation} 
Each of them depends on the difference of the two rapidities or spectral parameters $u$ and $v$.
We would like to study the partition function of a rectangular lattice as shown in Figure~\ref{fig:rect}. Therefore, we first have to introduce an orientation and spectral parameters (inhomogeneities) to each of the lines. We assign the rapidities $u_i$ to the horizontal lines and $v_i$ to the vertical ones. As indicated by the external arrows in Figure~\ref{subfig:rap} the orientation of the lines is from left to right and bottom to top. 
\begin{figure}[ht]
        \centering
        \begin{subfigure}[b]{0.3\textwidth}
                \includegraphics[width=\textwidth]{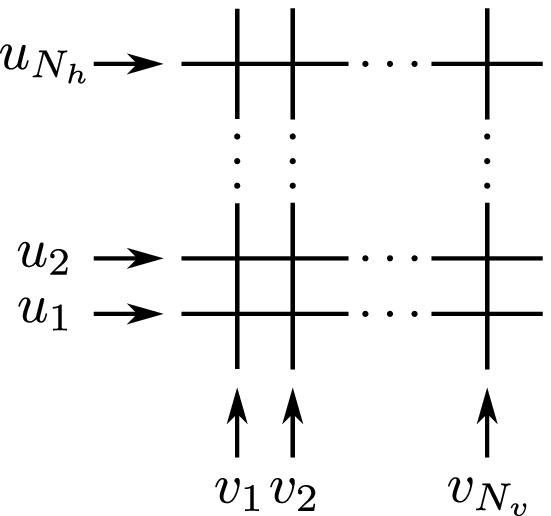}
                \caption{}
                \label{subfig:rap}
                   \end{subfigure}
                   \quad\quad \quad\quad
         \begin{subfigure}[b]{0.3\textwidth}
                \includegraphics[width=\textwidth]{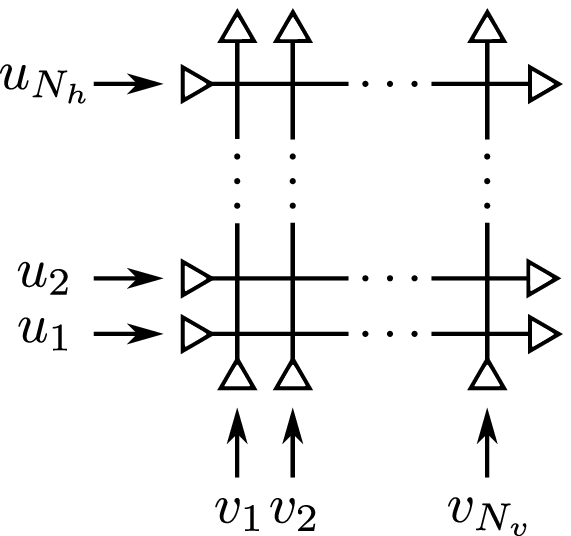}
                \caption{}
                \label{fig:dwbc}
                \end{subfigure}
                \caption{Rectangular lattice (a) with fixed boundary configuration (b).}
\end{figure}
Once we fix all configuration arrows, i.e. white arrows, at the boundary we can calculate the partition function using the Boltzmann weights defined in \eqref{6v1} and \eqref{6v2}. It is defined by
\begin{equation}\label{pfunc}
 Z=\underset{\text{states}}{\sum_{\text{internal}}}\prod_{i=1}^{N_h}\prod_{j=1}^{N_v}w[u_i,v_j]\,.
\end{equation} 
Depending on the internal states, i.e. white arrows at the edges of a vertex, the weight $w$ is given by $a,b$ or $c$ as introduced in \eqref{6vbw}. Summing over all possible state configurations in the interior of the lattice then yields the partition function.
As an example let us consider the case given in Figure~\ref{fig:dwbc}. In this case the partition function is particularly easy to calculate as there is only one configuration of internal arrows allowed at each vertex. This can be seen from \eqref{6v1} and \eqref{6v2}. Starting from the lower left vertex in Figure~\ref{fig:dwbc} we see that the only allowed local configuration is the first vertex with weight $a$ in \eqref{6v1}. As a consequence the same is true for the next vertex on the right.  Now proceeding line by line from bottom to top we find that the sum in \eqref{pfunc} is trivial and obtain
\begin{equation}
 Z=\prod_{i=1}^{N_h}\prod_{j=1}^{N_v} a[u_i,v_i]=\prod_{i=1}^{N_h}\prod_{j=1}^{N_v} (u_i-v_j+1)\,.
\end{equation}
In general for an arbitrary lattice as the ones given in the beginning of this chapter we can associate an orientation and a spectral parameter to each line and calculate the partition function on that lattice after fixing a certain boundary configuration via the formula
\begin{equation}\label{pfunn}
 Z=\underset{\text{states}}{\sum_{\text{internal}}}\prod_{\text{vertices}}w[\text{vertex}]\,.
\end{equation} 
Here the argument of the Boltzmann weight $w$ depends on the orientation and rapidities of the two crossing lines. For convenience we suppressed the dependence on the state variables.
\section{Relation to R-matrix}\label{sec:relr}
The careful reader might have noticed that the Boltzmann weights of the six-vertex model \eqref{6vbw} are related to the entries of the $\gl2$-invariant R-matrix introduced in \eqref{fundr} via
\begin{equation}\label{rweight}
 \mathbf{R}(u_i-u_j)=\left(\begin{array}{cccc}
                 a[u_i,u_j]&0&0&0\\
                 0&b[u_i,u_j]&c[u_i,u_j]&0\\
                0&c[u_i,u_j]&b[u_i,u_j]&0\\
                0&0&0&a[u_i,u_j]
                \end{array}\right)\,.
\end{equation} 
To relate the graphical notation used in the previous chapter to the one used in Section~\ref{sec:6vmodel} we note that in Chapter \ref{ch:ybe} an orientation was introduced at each line of a vertex to distinguish left and right multiplication in each of the spaces on which the R-matrix acts non-trivially, compare \eqref{Rvert}. However, these black arrows on the lines introduced for this purpose should not be confused with the white arrows used to label the different states of the six-vertex model. The black arrows correspond to the black arrows in \eqref{6v1} and \eqref{6v2} and the white arrows translate into the row and column indices of the matrix \eqref{rweight} acting on the tensor product $\mathbb{C}^2\otimes\mathbb{C}^2$. If a white arrow points into the same direction as the arrow that denotes the orientation of the line we replace it by the state label $1$. If it is pointing in the opposite direction we assign the state label $2$. Thus, in this notation the vertex configurations of the six-vertex model in \eqref{6v1} and \eqref{6v2} are depicted as
\begin{align}\label{pic:6vstate1}
 \begin{aligned}
a[u_i,u_j]
 \end{aligned}
 =
 \begin{aligned}
  \includegraphics[scale=0.70]{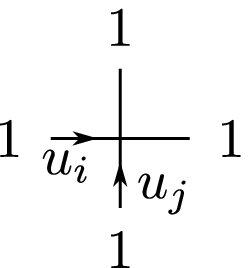}
 \end{aligned}
 =
  \begin{aligned}
  \includegraphics[scale=0.70]{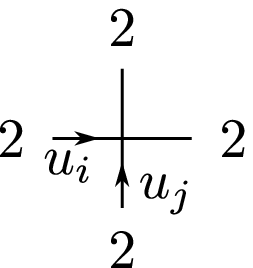}
 \end{aligned}
 \,,\quad
  \begin{aligned}
b[u_i,u_j]
 \end{aligned}
 =
 \begin{aligned}
  \includegraphics[scale=0.70]{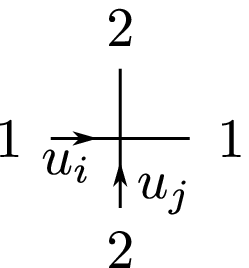}
 \end{aligned}
 =
  \begin{aligned}
  \includegraphics[scale=0.70]{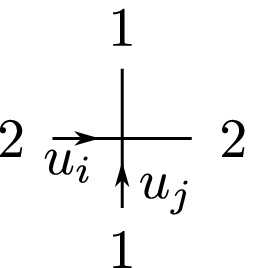}
 \end{aligned}
 \,,
\end{align}
\begin{align}\label{pic:6vstate2}
 \begin{aligned}
c[u_i,u_j]
 \end{aligned}
 =
 \begin{aligned}
  \includegraphics[scale=0.70]{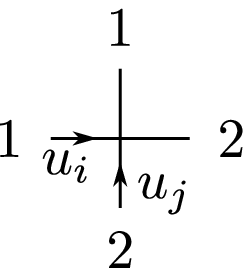}
 \end{aligned}
 =
  \begin{aligned}
  \includegraphics[scale=0.70]{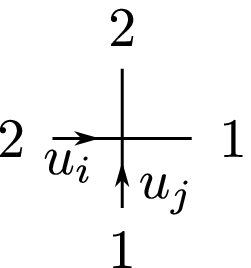}
 \end{aligned}
 \,.
\end{align}
Here, according to \eqref{rweight}, all other vertex configurations are identical to zero. Furthermore, we would like to note at this point that the ice-rule translates into the charge conservation condition $[J^{(1)}_{aa}+J^{(2)}_{aa},\rfund_{12}]=0$ with $a=1,2$ which follows from the $\gl2$-invariance of the R-matrix. As a consequence the states assigned to the bottom and left edge have to reappear at the top and right of each vertex.  

In general, we can think about the entries of an arbitrary $\gln$-invariant R-matrix denoted by $\langle \alpha_1,\alpha_2\vert R_{\Lambda_1\Lambda_2}(u_1-u_2)\vert \beta_1, \beta_2\rangle$ as Boltzmann weights. Here the indices $\alpha_{1,2}$ and $\beta_{1,2}$ take the values $1,\ldots,\dim \Lambda_{1,2}$, where $ \Lambda_{1,2}$ denotes the dimension of the representation $\Lambda_{1,2}$ in space $1,2$ respectively. Graphically this is denoted as
\begin{align}\label{rcomps}
 \langle \alpha_1,\alpha_2\vert R_{\Lambda_1\Lambda_2}(u_1-u_2)\vert \beta_1, \beta_2\rangle=
 \begin{aligned}
  \includegraphics[scale=0.80]{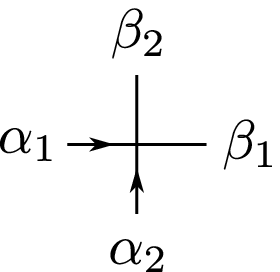}
 \end{aligned}
 \,,
\end{align}
compare \eqref{rweight}. This notation has the advantage that the relation to our previous studies becomes more transparent. Whenever we talk about the components of the R-matrices, i.e. Boltzmann weights, we indicate the corresponding indices at the end of the lines as shown in \eqref{rcomps}.
Here we have neglected the rapidities, they are assigned in the same way as in \eqref{Rvert}. 
\section{Partition functions and QISM}
\label{pfaqism}
We are now in the position to study the partition function of the six-vertex model using the QISM. In particular, we can write the partition function in \eqref{pfunc} for a given boundary configuration as a product of monodromies of length $N_v$ introduced in \eqref{moninh} with inhomogeneities $v_1,\ldots,v_{N_v}$ and fundamental representation of $\gl2$ in quantum and auxiliary space
\begin{equation}
\begin{split}
Z(\{a\},\{b\},\{c\},\{d\})=&\\
\langle a_1,\ldots a_{N_h},b_1,\ldots b_{N_v}\vert&\mathcal{M}(u_1,{\mathbf v})\cdots \mathcal{M}(u_{N_h},{\mathbf v})\vert c_1,\ldots c_{N_h},d_1,\ldots d_{N_v}\rangle\,.
\end{split}
\end{equation} 
Here the set of indices $a,b,c,d=1,2$ label the configuration at the boundary, compare Figure~\ref{subfig:rap}. In our notation $a$ and $b$ are associated to the external edges of the lattice to the left and bottom respectively. The external states at the right and top are labeled by $c$ and $d$. For cylindrical boundary conditions in the horizontal direction we contract the indices $a_i$ and $c_i$ which yields the transfer matrix introduced in \eqref{transf} or pictorially in \eqref{pic:transm}
\begin{equation}
\sum_{a=1}^n\langle a\vert \mathcal{M}(u,{\mathbf v})\vert a \rangle=\tr \mathcal{M}(u,{\mathbf v})=\tmf(u,{\mathbf v})\,.
\end{equation} 
The partition function defined on a cylinder as shown in Figure~\ref{fig:zylrap} for a given boundary configuration takes the form 
\begin{equation}\label{Zcyl}
\begin{split}
Z(\{b\},\{d\})=\langle b_1,\ldots b_{N_v}\vert\tmf(u_1,{\mathbf v})\cdots \tmf(u_{N_h},{\mathbf v})\vert d_1,\ldots d_{N_v}\rangle\,.
\end{split}
\end{equation} 
This is the same transfer matrix as the one we have diagonalized using the Bethe ansatz in Chapter~\ref{ch:ybe}. Let us now consider the partition function on a toroidal topology as shown in Figure~\ref{fig:tor}. It can be obtained from \eqref{Zcyl} by contracting the indices that now correspond to the edges in the vertical direction
\begin{equation}\label{ztor}
Z_{N_v,N_h}({\mathbf v},{\mathbf u})=\sum_{\{b\}}\langle b_1,\ldots b_{N_v}\vert\tmf(u_1,{\mathbf v})\cdots \tmf(u_{N_h},{\mathbf v})\vert b_1,\ldots b_{N_v}\rangle\,.  
\end{equation} 
The partition function is invariant under basis transformations. Hence it can be written in terms of eigenvalues of the transfer matrices
\begin{equation}\label{ztor2}
  Z_{N_v,N_h}({\mathbf v},{\mathbf u})=\sum_{\alpha=1}^{2^{N_v}}\prod_{i=1}^{N_h}\tau_\alpha(u_i,{\mathbf v})\,,
\end{equation} 
see e.g. \cite{Baxter2007,Reshetikhin2010}.
Here $\tau_\alpha(u_i,{\mathbf v})$ denotes the $\alpha^{\text{th}}$ eigenvalue of the transfer matrix $\tmf(u_i,{\mathbf v})$. Thus we obtained an expression for the partition function defined on a torus parametrized by Bethe roots satisfying the corresponding Bethe equations. Again, as discussed previously in Section~\ref{twist}, the issue of completeness of the Bethe equations arises in this context. Furthermore, we would like to note that our derivation of \eqref{ztor2} heavily relied on the the quantum inverse scattering method which was not at hand in 1967 when Lieb presented his solution. His derivation is along the lines of Bethe's original approach from 1931 presented in Section~\ref{sec:cba}.

We end this section with the example shown in Figure~\ref{fig:zylrapcros} where some vertical lines are oriented in the opposite direction in order to prepare the reader for the discussion in Chapter~\ref{ch:BAforYI}. Let $ \tilde{\mathcal M}(u,{\mathbf v})$ be the monododromy describing a horizontal layer with an opposite orientation at site $i$.  It takes the form
\begin{equation}\label{strtr}
 \tilde{\mathcal M}(u,{\mathbf v})=\ldots\rfund_{\,\text{a},i-1}(u-v_{i-1})(\perm_{i,\,\text{a}}\rfund_{\,\text{a},i}(v_i-u)\perm_{i,\,\text{a}})^{t_i}\rfund_{\,\text{a},i+1}(u-v_{i+1})\ldots\,,
\end{equation} 
where $t_i$ denotes the transposition in the $i^\text{th}$ vertical space. The permutation operators $\perm$ ensure that the vertex is oriented properly and the transposition is necessary to write the partition function as a product of transfer matrices, compare \eqref{Zcyl}. Using crossing symmetry this transfer matrix can be rewritten in the familiar form. In particular, this is discussed in the next sections.
\begin{figure}[ht]
        \centering
        \begin{subfigure}[b]{0.3\textwidth}
                \includegraphics[width=\textwidth]{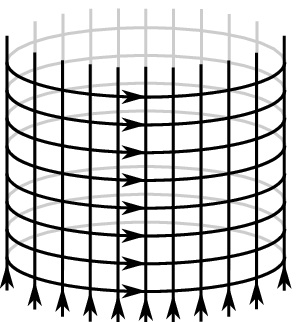}
                \caption{}
                \label{fig:zylrap}
                   \end{subfigure}
                   \quad\quad\quad\quad
         \begin{subfigure}[b]{0.3\textwidth}
                \includegraphics[width=\textwidth]{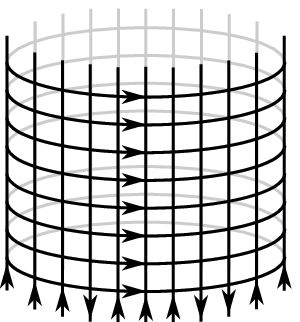}
                \caption{}
                \label{fig:zylrapcros}
                \end{subfigure}
                \caption{Cylindrical topology of a square lattice (a) of mixed orientation (b).}
\end{figure}
\section{Symmetries of the R-matrix}
We would like to study some of the discrete symmetries of the R-matrix in \eqref{rweight}. For this purpose we extend our analysis to the $\gln$-invariant R-matrix in the fundamental representation as introduced in \eqref{fundr}. It contains the same weights as the $\gl2$-invariant R-matrix but in total there are $n(2n-1)$ vertices. In components the only difference lies in the range of the indices. We have
\begin{equation}\label{Rcomp}
 \langle a_1,a_2\vert\rfund(u_1-u_2)\vert b_1,b_2\rangle=(u_1-u_2)\delta_{a_1b_1}\delta_{a_2b_2}+\delta_{a_1b_2}\delta_{a_2b_1}
\end{equation} 
with $a_i,b_i=1,\ldots,n$. 
From \eqref{Rcomp} we see that the Boltzmann weights are left invariant under the transformation $a_{1}\leftrightarrow b_1,a_{2}\leftrightarrow b_2$:
\begin{equation}
\langle b_1,b_2\vert\rfund(u_1-u_2)\vert a_1,a_2\rangle=\langle a_1,a_2\vert\rfund(u_1-u_2)\vert b_1,b_2\rangle\,.
\end{equation} 
This reflects the symmetry under transposition in both spaces $1$ and $2$ of the R-matrix.
Furthermore, we also find that \eqref{Rcomp} is left invariant under substituting $a_{1}\leftrightarrow a_2,b_{1}\leftrightarrow b_2$ and therefore
\begin{equation}
\langle a_2,a_1\vert\rfund(u_1-u_2)\vert b_2,b_1\rangle=\langle a_1,a_2\vert\rfund(u_1-u_2)\vert b_1,b_2\rangle\,.
\end{equation}
Note that here the indices belonging to space $1$ on the left hand side of the equation are identified with the indices of space $2$ on the right hand side and vice versa. This identification of $\rfund_{12}$ with $\rfund_{21}$ is possible as we have the same representation in the spaces $1$ and $2$. The symmetries discussed above can be encoded into the pictorial equivalence relation
\begin{align}
 \begin{aligned}
  \includegraphics[scale=0.80]{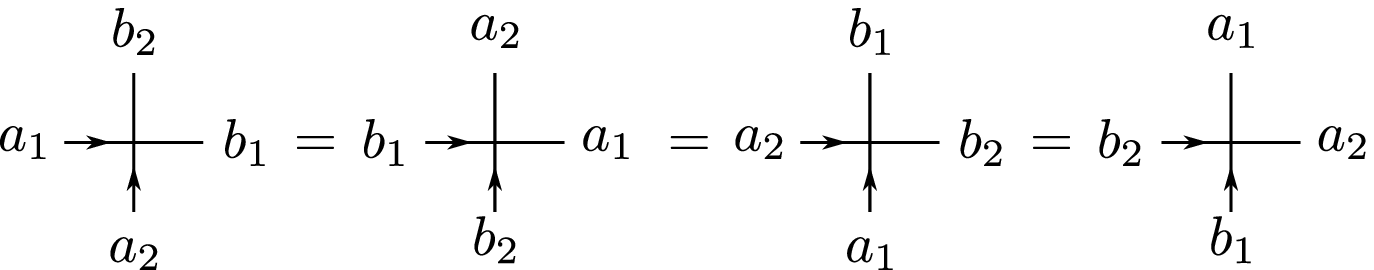}
 \end{aligned}
 \,.
\end{align}
\section{The anti-fundamental representation}
In the following we introduce the R-matrix $R_{\ffbox,\bar\ffbox}$ with the fundamental representation in the first space and the anti-fundamental in the second. The generators of the anti-fundamental representation are of the form
\begin{equation}\label{antifu}
 J_{ab}=-e_{ba}\,.
\end{equation} 
The representation is labeled by $\Lambda=(0,\ldots,0,-1)$ which is denoted by $\bar\ffbox$ in the following.
It is straightforward to show that this choice indeed satisfies the $\gln$ algebra \eqref{glnalg}, see also next section. Let us again start with the $\gl2$ case. In analogy to \eqref{rweight} we can write the R-matrix as 
\begin{equation}\label{rmatbar}
 R_{\ffbox,\bar\ffbox}(u_i-u_j)=\left(\begin{array}{cccc}
                 \bar a[u_i,u_j]&0&0&\bar c[u_i,u_j]\\
                 0&\bar b[u_i,u_j]&0&0\\
                0&0&\bar b[u_i,u_j]&0\\
                \bar c[u_i,u_j]&0&0&\bar a[u_i,u_j]
                \end{array}\right)\,,
\end{equation} 
where we used the definition of the Lax operators in \eqref{lax}. For short we denote this R-matrix as $\bar\rfund$, keeping in mind that
\begin{equation}
 \bar\rfund(u)\equiv  \rfund_{\ffbox,\bar\ffbox}(u)=\mathbf{L}_{\bar\ffbox}(u)\,.
\end{equation} 
The entries of the R-matrix \eqref{rmatbar} are given by
\begin{equation}
 \bar a[u_i,u_j]=u_i-u_j-1,\quad \bar b[u_i,u_j]=u_i-u_j,\quad \bar c[u_i,u_j]=-1\,.
\end{equation} 
As before we proceed to the corresponding $\gln$-invariant R-matrix. Again, the only difference is the range of the indices and we find the general expression
\begin{equation}\label{Rcompb}
\langle a_1,a_2\vert\bar\rfund(u_1-u_2)\vert b_1,b_2\rangle=(u_1-u_2)\delta_{a_1b_1}\delta_{a_2b_2}-\delta_{a_1a_2}\delta_{b_1b_2}
\end{equation} 
for $a_i,b_i=1,\ldots,n$. Diagrammatically we denote this R-matrix as
\begin{align}\label{Rcompbfig}
  \begin{aligned}
\langle a_1,a_2\vert\bar\rfund(u_1-u_2)\vert b_1,b_2\rangle
 \end{aligned}
 =
 \begin{aligned}
  \includegraphics[scale=0.80]{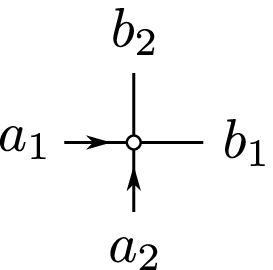}
 \end{aligned}
 \,,
\end{align}
where we introduced the white dot to distinguish it from the R-matrix $\rfund$ introduced previously, cf. \eqref{rcomps}. The vertical line corresponds to the space in the anti-fundamental representation. This R-matrix has the same symmetries under the exchange of indices as the $\rfund$. Namely under the transformation $a_{1}\leftrightarrow b_1,a_{2}\leftrightarrow b_2$ we find
\begin{equation}
\langle b_1,b_2\vert\bar\rfund(u_1-u_2)\vert a_1,a_2\rangle=\langle a_1,a_2\vert\bar\rfund(u_1-u_2)\vert b_1,b_2\rangle
\end{equation} 
and moreover the transformation $a_{1}\leftrightarrow a_2,b_{1}\leftrightarrow b_2$ yields
\begin{equation}\label{antiperm}
\langle a_2,a_1\vert\bar\rfund(u_1-u_2)\vert b_2,b_1\rangle=\langle a_1,a_2\vert\bar\rfund(u_1-u_2)\vert b_1,b_2\rangle\,.
\end{equation} 
Here we note that the dimension of the fundamental and anti-fundamental representation are the equal, thus the indices of space $1$ and $2$ take the same values which makes the identification in \eqref{antiperm} possible. Diagrammatically we write these relations as
\begin{align}
 \begin{aligned}
  \includegraphics[scale=0.80]{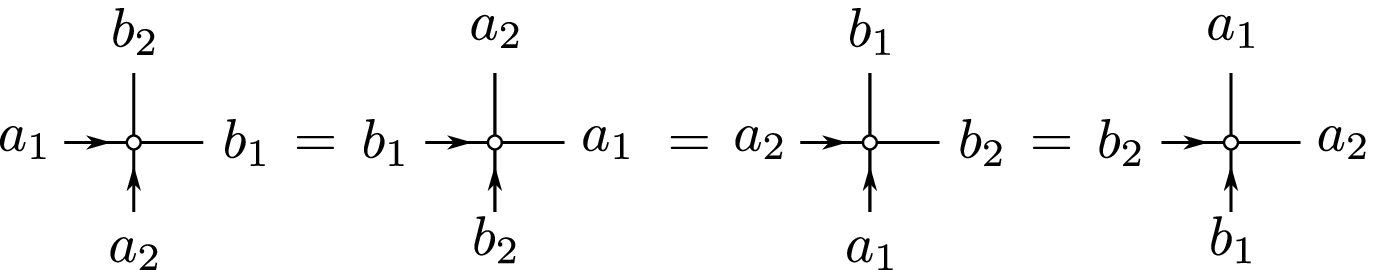}
 \end{aligned}
 \,.
\end{align}
\section{Crossing symmetry}\label{sec:crosfun}
The goal of this section is to relate the R-matrix with fundamental representation of $\gln$ in both spaces with the R-matrix $R_{\ffbox,\bar\ffbox}$ whose second space is in the anti-fundamental representation. The relation can be motivated from the existence of the $\mathbb{Z}_2$ outer automorphism\,\footnote{Note that for the case $n=2$ the two representations are equivalent.} of $\gln$
\begin{equation}\label{outaut}
 J_{ab}\rightarrow\bar J_{ab}=-J_{ba}\,.
\end{equation} 
Under the action of the automorphism the Dynkin labels transform as
\begin{equation}\label{dyntrans}
 \Lambda=(\lambda_1,\ldots,\lambda_n)\rightarrow \bar\Lambda=(-\lambda_n,\ldots,-\lambda_1)\,,
\end{equation} 
compare also to the choice of the generators in \eqref{antifu}.
First, we allow a normalization of the R-matrix that we did not include up to now as it cannot be determined from the Yang-Baxter equation. However, in the following it plays an important role. Thus, we rewrite the R-matrices introduced previously as
\begin{equation}\label{Rn}
\rfund(u)\rightarrow \rfund(u)=f(u)(1+\frac{1}{u}\perm)
\end{equation} 
and
\begin{equation}\label{bRn}
\bar\rfund(u)\rightarrow \bar\rfund(u)=g(u)(1-\frac{1}{u}\mathbf{K})
\end{equation} 
with $\mathbf{K}=\sum_{a,b=1}^ne_{ab}\otimes e_{ab}$ 
. From \eqref{Rcomp} and \eqref{Rcompb}, where we expressed the R-matrices under study in terms of components, we find that the two R-matrices can be related via
\begin{equation}\label{precrosrfund}
\langle b_1,a_2\vert\rfund(u_1-u_2)\vert a_1,b_2\rangle=\langle a_1,a_2\vert\bar\rfund(u_2-u_1)\vert b_1,b_2\rangle\,.
\end{equation}
This condition naturally imposes the relation among the normalizations of the R-matrices in \eqref{Rn} and \eqref{bRn}
\begin{equation}\label{crossnorm}
 f(u_1-u_2)=g(u_2-u_1)\,.
\end{equation} 
Using the symmetries discussed previously we note that \eqref{precrosrfund} can be rewritten as
\begin{equation}\label{crossingrfundeq}
\langle a_1,a_2\vert\bar\rfund(u_2-u_1)\vert b_1,b_2\rangle=\langle b_2,a_1\vert\rfund(u_1-u_2)\vert a_2,b_1\rangle\,.
\end{equation} 
Thus graphically this relation can be expressed as
\begin{align}\label{crossingrfund}
  \begin{aligned}
\langle a_1,a_2\vert\bar\rfund(u_2-u_1)\vert b_1,b_2\rangle
 \end{aligned}
 =
 \begin{aligned}
  \includegraphics[scale=0.80]{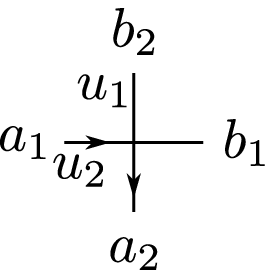}
 \end{aligned}
 \,,
\end{align}
which can be interpreted as a ``rotation'' of the vertex given in \eqref{rcomps} by $90^\circ$. Crossing relations like the one in \eqref{crossingrfund} naturally appear in the study of $2d$ integrable quantum field theories \cite{Zamolodchikov1979,Berg1978}. We use this property to relate the partition function of the vertex model in Figure~\ref{fig:zylrapcros} to the commuting family of operators.
Using the notation $\rfund_{\,\text{a},\bar i}$ for denoting the R-matrix $\bar \rfund$ with the antifundamental representation in space $i$, we find that the monodromy matrix discussed in \eqref{strtr} for $\gl2$ can be rewritten as 
\begin{equation}
 \tilde {\mathcal M}(u,{\mathbf v})=\ldots\rfund_{\,\text{a},i-1}(u-v_{i-1})\rfund_{\,\text{a},\bar i}(u-v_i)\rfund_{\,\text{a},i+1}(u-v_{i+1})\ldots
\end{equation} 
which fits into the monodromies studied in the previous chapter and therefore can be diagonalized using the Bethe ansatz. This holds true generally for the extension to $\gln$ along the lines as discussed in this section. 

Let us also consider the partition function on the torus defined in \eqref{ztor}. As a consequence of \eqref{crossingrfundeq} there exist two equivalent ways to write the partition function in terms of transfer matrices. In Section \ref{sec:relr} we thought about the partition function as the trace of $N_h$ transfer matrices of length $N_v$, however we can also think about it as the trace of $N_v$ transfer matrices of length $N_h$. It yields a representation of the partition function 
\begin{equation}
  Z_{N_v,N_h}({\mathbf v},{\mathbf u})=\tr \bar T(v_{N_v},{\bf u})\bar T(v_{N_v-1},{\bf u})\cdots\bar T(v_{1},{\bf u})
\end{equation} 
with the transfer matrices $\bar T$ formed out of the R-matrices in \eqref{bRn}
\begin{equation}\label{tbar}
 \bar T(v,{\bf u})=\tr \rfund_{\text{a},\bar 1}(v-u_1)\cdots\rfund_{\text{a},\bar {N}_h}(v-u_{N_h})\,.
\end{equation} 
Again, this prescription applies in the general framework of this section but it has an important implication for the case of $\gl2$ where the outer automorphism discussed in \eqref{outaut} becomes an inner automorphism. Therefore, the anti-fundamental representation can be directly related to the fundamental. To be precise, for
\begin{equation}
\frac{g(u+\frac{1}{2})}{f(u-\frac{1}{2}) }=\frac{u+\frac{1}{2}}{u-\frac{1}{2}}\,,
\end{equation} 
we find that
\begin{equation}\label{selfcros}
R_{\ffbox,\bar\ffbox}(u)=S\,R_{\ffbox,\ffbox}(u-1)\,S^{-1}
               \,,
\end{equation} 
with the similarity transformation containing the Levi-Civita symbol
\begin{equation}
S= \left(\begin{array}{cc}
                +1&0\\
                 0&+1\\
                \end{array}\right)\otimes
 \left(\begin{array}{cc}
                0&+1\\
                 -1&0\\
                \end{array}\right)\,.
\end{equation} 
See also \cite{Reshetikhin1983}, where this relation is discussed for the trigonometric case.
For general $\gln$ such a relation does not exist as for $n\neq 2$ the anti-fundamental and fundamental representation are not equivalent.
After substituting \eqref{selfcros} into \eqref{tbar} the similarity transformations in the auxiliary space cancel each other under the trace and we obtain
\begin{equation}
 \bar T(v,{\bf u})=\tr \rfund_{\text{a},1}(v-u_1-1)\cdots\rfund_{\text{a}, {N}_h}(v-u_{N_h}-1)\,.
\end{equation} 
Having in mind the integrability condition \eqref{tcom} we conclude that
\begin{equation}
  Z_{N_v,N_h}({\mathbf v},{\mathbf u})= Z_{N_h,N_v}({\mathbf u}+1,{\mathbf v})\,,
\end{equation} 
where we denoted the collective shift in the variables $u_i$ by ${\mathbf u}+1$.
This so-called modular identity is an important step when studying the thermodynamic limit \cite{Destri1995,Reshetikhin2010}.
\section{Z-invariance}\label{sec:zinv}
Now, we consider a slight generalization of the lattice presented in Figure~\ref{fig:xtlattice}. The new feature that appears is that the lines are allowed to intersect asymptotically. We assign rapidities and orientations to the lines as shown in Figure~\ref{subfig:uncrazya} such that all lines point from the bottom to the top. The major point we want to make here is that the lines can be moved freely as long as their endpoints do not intersect. This can change the form of the lattice but will not change the partition function. This is a consequence of the Yang-Baxter equation 
\begin{align}
 \begin{aligned}
  \includegraphics[scale=0.80]{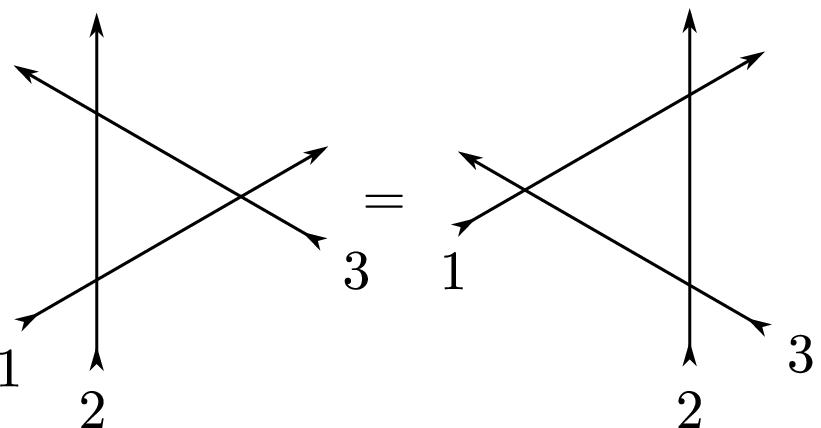}
 \end{aligned}
 \,,
\end{align}
which we studied in detail for certain R-matrices in the previous chapter, compare e.g. Figure~\ref{pic:ybe}. The property that the partition function \eqref{pfunn} is invariant when the lattice is changed accordingly is usually referred to as $Z$-invariance, see \cite{Baxter1978} and in particular \cite{Zamolodchikov1990a} for a detailed discussion\,\footnote{We thank Patrick Dorey for pointing out this reference.}. As a consequence the partition function is completely determined by the boundary data which contains the external state labels and the enumeration of the lines. Thus, we do not have to care about the actual lattice configuration inside as shown in Figure~\ref{subfig:zinv} as long as we don't allow self- nor double-crossings of the lines.
\begin{figure}[ht]
        \centering
        \begin{subfigure}[b]{0.3\textwidth}
                \includegraphics[width=\textwidth]{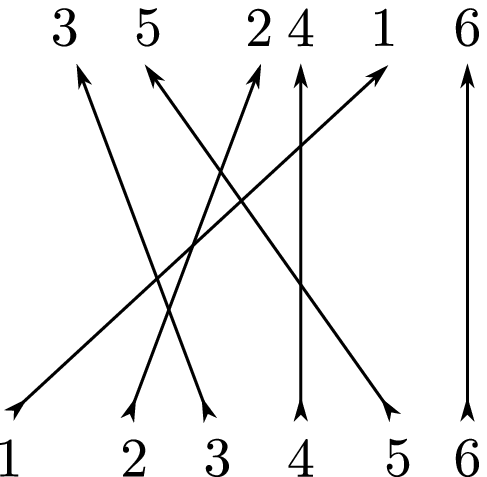}
                \caption{}
                \label{subfig:uncrazya}
                   \end{subfigure}
                   \quad\quad \quad\quad
         \begin{subfigure}[b]{0.3\textwidth}
                \includegraphics[width=\textwidth]{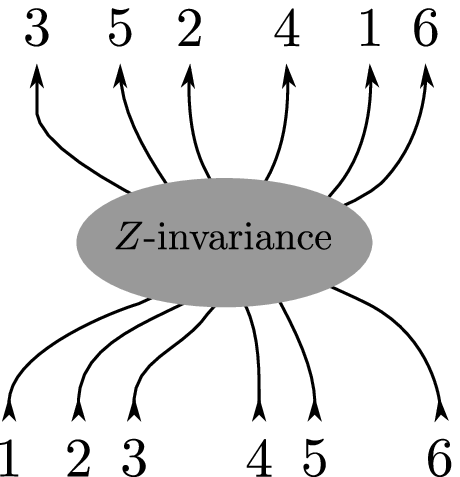}
                \caption{}
                \label{subfig:zinv}
                \end{subfigure}
                \caption{(a) Arbitrary lattice built out of straight lines (b) Z-invariance}
\end{figure}
\section{Unitarity}\label{sec:unitarity}
As a consequence of the Yang-Baxter relation \eqref{ybe1} there naturally arises a symmetry among R-matrices. This symmetry relates the R-matrix $R_{12}$ to the one where spaces $1$ and $2$ are exchanged, namely $R_{21}$. To obtain this relation we  multiply the Yang-Baxter equation \eqref{ybe1} from both sides with the inverse of $R_{12}$ and obtain
\begin{equation}\label{invybe}
\begin{split}
R_{12}(u_1-u_2)^{-1}R_{23}&(u_2-u_3)R_{13}(u_1-u_3)\\
&=R_{13}(u_1-u_3)R_{23}(u_2-u_3) R_{12}(u_1-u_2)^{-1}\,.
\end{split}
\end{equation} 
Exchanging the spaces $1$ and $2$ in \eqref{ybe1} we observe that the inverse of $R_{12}$ satisfies the same Yang-Baxter equation as $R_{21}$. Thus, assuming that the solution is unique, we conclude that they are proportional to each other
\begin{equation}\label{proprel}
 R_{21}(u_2-u_1)\sim R_{12}(u_1-u_2)^{-1}\,.
\end{equation} 

It is convenient to fix the normalization in such a way that \eqref{proprel} becomes an equality. This yields the unitarity relation
\begin{equation}\label{unitarity}
 R_{12}(u_1-u_2) R_{21}(u_2-u_1)=1\,.
\end{equation} 
Graphically this equation is depicted in Figure~\ref{fig:unit}.
\begin{figure}[ht]
  \centering
    \includegraphics[width=0.4\textwidth]{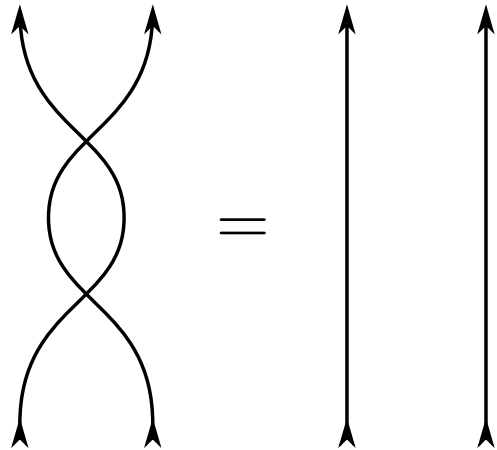}
  \caption{Unitarity relation.}
  \label{fig:unit}
\end{figure}
It has a natural interpretation in terms of $2d$ integrable field theory where we can think of two particles scattering and rescatter again, see eg. \cite{zoltanbook,Dorey1996c}.

\subsection{An example}
In this section we consider a general lattice as introduced in Figure~\ref{subfig:uncrazya} but allow for multiple intersections of the lines as shown in Figure~\ref{fig:crazy}. We concentrate on the $\gln$-invariant R-matrices with fundamental and anti-fundamental representations in either of the two spaces as introduced in \eqref{Rn} and \eqref{bRn}. 
Let us now consider the lattice presented in Figure~\ref{fig:crazy}. As shown in the figure we allow to associate different, i.e. fundamental and anti-fundamental, representations of $\gln$ to each line. In this sense it is more convenient to think about Figure~\ref{fig:crazy} as a product of R-matrices instead of a lattice with Boltzmann weights. Also, as discussed below, the fact that the structure of the lattice can be changed according to the unitarity relation depicted in Figure~\ref{fig:unit} makes this interpretation more natural.
\begin{figure}[ht]
  \centering
    \includegraphics[width=0.5\textwidth]{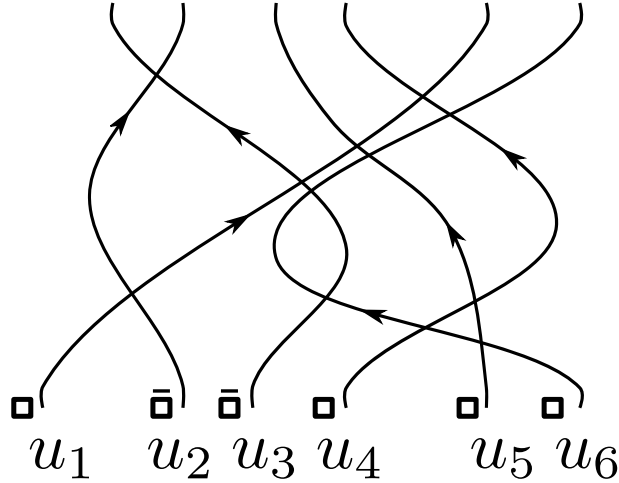}
  \caption{Arbitrary lattice with multiple intersections. As indicated by the Young diagrams, the spaces carrying rapidities $u_2$ and $u_3$ are in the anti-fundamental representation while the others are in the fundamental.}
  \label{fig:crazy}
\end{figure}
We already encountered two of the $\gln$-invariant R-matrices that appear in our example, i.e. 
\begin{equation}
 \rfund(u)=R_{\ffbox,\ffbox}(u),\quad\quad \bar\rfund(u)=R_{\ffbox,\bar\ffbox}(u)\,,
\end{equation} 
as defined in \eqref{Rn} and \eqref{bRn} containing the normalizations $f$ and $g$. In addition we introduce the two R-matrices
\begin{equation}\label{blubbi3}
 R_{\bar\ffbox,\ffbox}(u)\sim R_{\ffbox,\bar\ffbox}(-u)^{-1}=\frac{1}{g(-u)}\left(1-\frac{1}{u+n}{\bf K}\right)\,,
\end{equation} 
which  can be obtained from the relation \eqref{proprel} and 
\begin{equation}\label{blubbi4}
R_{\bar\ffbox,\bar\ffbox}(u)\sim R_{\ffbox,\ffbox}(u)\,.
\end{equation} 
 that can be derived from the Yang-Baxter equation \eqref{ybe1} rather quickly noting that $R_{\bar\ffbox,\bar\ffbox}$ intertwines two R-matrices $R_{\bar\ffbox,\ffbox}(u)$ and using \eqref{blubbi3}. Furthermore, we fix a relative normalization such that
\begin{equation}\label{blubbi}
 R_{\bar\ffbox,\ffbox}(u)=R_{\ffbox,\bar\ffbox}(u+n)\,.
\end{equation} 
This equation without the shift in the spectral parameter is fulfilled trivially for the R-matrix $R_{\ffbox,\ffbox}$ as both spaces are in the same representation. The same holds true for $R_{\bar\ffbox,\bar\ffbox}$.
Pictorially, this relation can be written as in Figure~\ref{fig:symrel} for $\Lambda=\bar\ffbox$.

Let us impose the unitarity condition in \eqref{unitarity} to determine the normalization $f$ and $g$ introduced previously. In principle, there are four ways to assign the representations to the two lines in Figure~\ref{fig:unit}:
$(\ffbox,\ffbox)$, $(\ffbox,\bar\ffbox)$, $(\bar\ffbox,\ffbox)$, $(\bar\ffbox,\bar\ffbox)$. However, the unitarity conditions for $(\ffbox,\bar\ffbox)$ and $(\bar\ffbox,\ffbox)$ are equivalent. Furthermore, it follows from \eqref{blubbi4} that also the cases $(\ffbox,\ffbox)$ and $(\bar\ffbox,\bar\ffbox)$ yield the same relation. First, we consider the case where we have two fundamental representations. We impose
\begin{equation}
R_{\ffbox,\ffbox}(u_1-u_2)R_{\ffbox,\ffbox}(u_2-u_1)=1\,,
\end{equation} 
and obtain the condition 
\begin{equation}\label{normf}
 f(u_1-u_2)f(u_2-u_1)=\frac{(u_1-u_2)^2}{(u_1-u_2)^2-1}\,,
\end{equation} 
for the normalization $f$. Following the same strategy for the case where we have one fundamental representation and one anti-fundamental
\begin{equation}
R_{\ffbox,\bar\ffbox}(u_1-u_2)R_{\bar\ffbox,\ffbox}(u_2-u_1)=1\,,
\end{equation} 
 with the normalization $g$, cf. \eqref{bRn} and \eqref{blubbi}, we obtain
\begin{equation}\label{normg}
 g(u_1-u_2)g(u_2-u_1+n)=1\,.
\end{equation} 
The normalizations defined implicitly by \eqref{normf} and \eqref{normg} allow to disentangle the lines in Figure~\ref{fig:crazy} according to Figure~\ref{fig:unit}. We find that with this choice the lattice given in Figure~\ref{fig:crazy} is equivalent to the one presented in Figure~\ref{subfig:uncrazya}. In this sense we can generalize the notion of $Z$-invariance discussed in Section \ref{sec:zinv}. 

Last but not least, we would like to give an explicit solution that satisfies the unitarity conditions \eqref{normf}, \eqref{normg} and additionally the crossing relation \eqref{crossingrfund} of the $\gln$-invariant R-matrices discussed here. It takes the form
\begin{equation}\label{crosnorm1}
 f(u)=\frac{\Gamma\left(\frac{1-u}{n}\right)\Gamma\left(\frac{n+u}{n}\right)}{\Gamma\left(-\frac{u}{n}\right)\Gamma\left(\frac{1+n+u}{n}\right)}\,.
\end{equation} 
This normalization is not unique as one can multiply it with arbitrary functions $h$ that satisfy
\begin{equation}
 h(u_1-u_2)h(u_2-u_1)=1\,,\quad\quad  h(u_1-u_2)h(u_2-u_1-n)=1\,.
\end{equation} 
Here $h$ is the so-called CDD factor \cite{Castillejo1956} which is of crucial importance in $2d$ integrable field theories. In particular different CDD factors may yield different theories. As this will not be crucial in the following we do not elaborate further on this subject. Further details can be found in e.g. \cite{Karowski1979,zoltanbook,Dorey1996c}.
\section{Crossing and unitarity for Lax operators}\label{sec:croslax}
For later purposes we discuss the crossing and unitarity relation for the Lax operators introduced in \eqref{lax} to construct the spin-chain monodromy \eqref{moninh}. These R-matrices have the fundamental representation of $\gln$ in the auxiliary space and representation $\Lambda$ in the quantum space, cf. \eqref{laxr}.
As in Section~\ref{sec:crosfun} we allow for a normalization $f_\Lambda$ that cannot be obtained from the Yang-Baxter equation
\begin{equation}\label{normlax}
 R_{\ffbox,\Lambda}(u_1-u_2)=f_\Lambda(u_1-u_2)\left(1 +\frac{1}{u_1-u_2}\sum_{a,b=1}^n\,e_{ab}J_{ba}\right)\,.
\end{equation}   
To determine $ R_{\Lambda,\ffbox}$ we follow the method introduced in Section~\ref{sec:unitarity} and determine the inverse of the Lax operator \eqref{normlax}. For representations that satisfy the condition\,\footnote{See also Section~\ref{sec:Projection-property-of} for further details on this type of representations.}
\begin{equation}\label{genrec}
 \sum_{c=1}^n J_{ca}J_{bc}=\gamma J_{ba}+\sigma\delta_{ab}\mathbb{I}\,,
\end{equation} 
with $\gamma,\sigma\in\mathbb{C}$ and the identity $\mathbb{I}$ in the appropriate representation $\Lambda$, one obtains 
\begin{equation}
 R_{\ffbox,\Lambda}(u)^{-1}=\frac{u(u+\gamma)}{(u^2+u \gamma-\sigma)f_\Lambda(u)}\left(1 -\frac{1}{u+\gamma}\sum_{a,b=1}^n\,e_{ab}J_{ba}\right)\,.
\end{equation} 
Using the proportionality relation \eqref{proprel} then yields the R-matrix $ R_{\Lambda,\ffbox}$.
In addition we impose the relation 
\begin{equation}\label{symree}
R_{\ffbox,\Lambda}(u_1-u_2)= R_{\Lambda,\ffbox}(u_1-u_2+\gamma)
\end{equation} 
to fix the normalization, see Figure~\ref{fig:symrel}.
\begin{figure}[ht]
  \centering
    \includegraphics[width=0.5\textwidth]{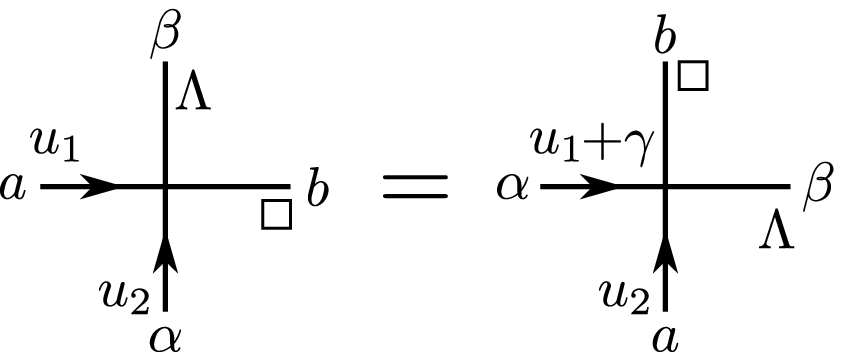}
  \caption{Symmetry relation.}
  \label{fig:symrel}
\end{figure}
Now, imposing the unitarity condition as introduced in \eqref{unitarity}
\begin{equation}
 R_{\ffbox,\Lambda}(u_1-u_2)R_{\Lambda,\ffbox}(u_2-u_1)=1\,,
\end{equation} 
we obtain a difference equation for the normalization $f$
\begin{equation}\label{unit_fund}
 f_\Lambda(-u)f_\Lambda(u-\gamma )=\frac{u(u-\gamma)}{u^2-u\gamma-\sigma}\,,
\end{equation} 
cf. \eqref{normf}.

Next, we introduce $R_{\ffbox,\bar{\Lambda}}$ using the algebra automorphism $ \bar {J}_{ab}=-J_{ba}$ in \eqref{outaut}.
From \eqref{lax} we find that it can be written as
\begin{equation}
 R_{\ffbox,\bar{\Lambda}}(u)=f_{\bar\Lambda}(u)\left(1 -\frac{1}{u}\,e_{ab}J_{ab}\right)\,.
\end{equation}   
Here we would like to stress that we employ the generators $J_{ab}$ of representation $\Lambda$ to realize the R-matrix of the cojugate representation $\bar\Lambda$ as discussed in \eqref{dyntrans}. 
Following the same reasoning as before we determine $R_{\bar{\Lambda},\ffbox}$ through the inverse of $ R_{\ffbox,\bar{\Lambda}}$ and fix the normalization such that 
\begin{equation}
R_{\ffbox,\bar{\Lambda}}(u)= R_{\bar{\Lambda},\ffbox}(u-n+\gamma)\,,
\end{equation} 
where we used the condition on the representation introduced in \eqref{genrec}.
For the R-matrices $R_{\ffbox,\bar{\Lambda}}$ and $R_{\bar{\Lambda},\ffbox}$ the unitarity relation reads
\begin{equation}
 R_{\ffbox,\bar{\Lambda}}(u)R_{\bar{\Lambda},\ffbox}(-u)=1\,.
\end{equation} 
It yields the difference equation
\begin{equation}\label{laxnb}
 f_{\bar\Lambda}(-u)f_{\bar\Lambda}(u-\mu)=\frac{u(u-\mu)}{u^2-u\mu-\nu}\,,
\end{equation} 
for the normalization $ f_{\bar\Lambda}$. Here we introduced the complex numbers $\mu$ and $\nu$ via
\begin{equation}
  \sum_{c=1}^n J_{bc}J_{ca}=\mu J_{ba}+\nu\delta_{ab}\mathbb{I}\,.
\end{equation} 
Their relation to $\gamma$ and $\sigma$ is given by
\begin{equation}\label{genrel}
 \mu=n+\gamma\quad \quad \nu\mathbb{I}=\sigma\mathbb{I}-\sum_{c=1}^nJ_{cc}\,.
\end{equation} 
For the fundamental representation we find the values
\begin{equation}
\sigma=1,\quad \quad \mu=n,\quad \quad \gamma=\nu=0\,.
\end{equation} 
Thus, from \eqref{unit_fund} and \eqref{laxnb} we reproduce as special cases the difference equations derived earlier in \eqref{normf} and \eqref{normg}.

In the end we present the crossing relation for Lax operators. This relation was discussed in Section~\ref{sec:crosfun} for the fundamental/anti-fundamental representation of $\gln$. In general for an arbitrary representation $\Lambda$ with $\langle \alpha\vert J_{ab}\vert\beta\rangle=\langle\beta\vert J_{ba}\vert \alpha\rangle$ we obtain the crossing equation 
\begin{equation}\label{croslaxx}
   \langle a,\beta\vert R_{\ffbox,\Lambda}(u_1-u_2)\vert b,\alpha\rangle=\langle a,\alpha\vert R_{\ffbox,\bar{\Lambda}}(u_2-u_1)\vert b,\beta\rangle\,,
\end{equation} 
and thus the relation between their normalizations
\begin{equation}\label{crossss}
 f_{\bar\Lambda}(u_1-u_2)=f_{\Lambda}(u_2-u_1)\,,
\end{equation} 
compare \eqref{crossnorm}. 

\section{Three-vertices from R-matrices}\label{sec:three}
In this section we introduce certain intertwiners that can be interpreted as three-vertices. We focus on a rather simple example and postpone a more general discussion to the next section and Chapter~\ref{ch:BAforYI}. Let us recall the R-matrix with the fundamental representation of $\gln$ in quantum and auxiliary space 
\begin{equation}
 \rfund(z)=z+\perm\,.
\end{equation} 
Apart from the permutation point which, as we discussed in Chapter~\ref{ch:ybe}, is of crucial importance to obtain local charges in the framework of the quantum inverse scattering method there are two special points of the spectral parameter where the rank of the R-matrix is reduced. This can be seen if one rewrites the R-matrix in terms of projectors, see e.g. \cite{Fateev1994},
\begin{equation}\label{rfedec}
 \rfund(z)=2(z+1)P_{+}+2(z-1)P_{-}\,,
\end{equation} 
with $P_{\pm}=(1\pm\perm)/2$.
One immediately sees that these points are located at $z=\pm1$. We will be interested in the case $z=1$ where the 
R-matrix reduces to the projector $P^2_{+}=P_{+}$ on the symmetric representation in the tensor product decomposition of two fundamental representations
\begin{equation}\label{ytff}
 {\scriptsize\yng(1)}\otimes{\scriptsize\yng(1)}={\scriptsize\yng(2)}\oplus{\scriptsize\yng(1,1)}\,.
\end{equation} 

The projector $P_{+}$ can be factorized into the product of two rectangular matrices
\begin{equation}\label{zerleg}
Y_L\cdot Y_R=P_{+}\,,\quad\quad Y_R=Y^t_L\,,
\end{equation} 
which are related to each other by transposition. Here $Y_L$ is an $n^2\times\frac{n(n+1)}{2}$ matrix that can explicitly be written as
\begin{equation}\label{YL}
 Y_L=\sum_{a,b=1}^n\left(\frac{1}{\sqrt{2}}+\delta_{ab}(1-\frac{1}{\sqrt{2}})\right)\vert {\scriptsize{ \young(a)}}\rangle\otimes\vert{\scriptsize\young(b)}\rangle\langle{\scriptsize\young(ab)}\vert\,.
\end{equation} 
Here we employed the notion of Young tableaux to denote the normalized basis vectors, see e.g.  \cite{Georgi1999,Hamermesh1989}. As an example let us consider the $\gl2$ case. For $n=2$ we find
\begin{equation}
 P_{+}=\left( \begin{array}{cccc}
1 & 0 & 0&0 \\
0 & \frac{1}{2} & \frac{1}{2}&0 \\
0 & \frac{1}{2} &\frac{1}{2}&0 \\
0 & 0 & 0&1 \end{array} \right)=\left( \begin{array}{ccc}
1 & 0 & 0 \\
0 & \frac{1}{\sqrt{2}} &0 \\
0 & \frac{1}{\sqrt{2}} &0 \\
0 & 0 &1 \end{array} \right)\cdot\left( \begin{array}{cccc}
1 & 0 & 0&0 \\
0 & \frac{1}{\sqrt{2}} & \frac{1}{\sqrt{2}}&0 \\
0 & 0 & 0&1 \end{array} \right)=Y_L\cdot Y_R\,.
\end{equation} 
From the tensor structure of $Y_L$ in \eqref{YL} we see that it provides a linear map $f_Y$
\begin{equation}
 f_Y:\mathbb{C}^n\otimes\mathbb{C}^n\rightarrow\mathbb{C}^\frac{n(n+1)}{2}\,,
\end{equation} 
where $\frac{n(n+1)}{2}$ is the dimension of the symmetric representation in \eqref{ytff}.  Thus, in the following we denote $Y_{L,R}$ as three vertices
\begin{align}\label{3fvertex}
 \begin{aligned}
Y_R
 \end{aligned}
 =
 \begin{aligned}
  \includegraphics[scale=0.80]{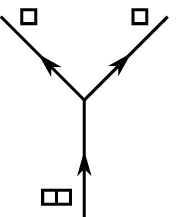}
 \end{aligned}
 \,,\quad\quad\quad
  \begin{aligned}
Y_L
 \end{aligned}
 =
 \begin{aligned}
  \includegraphics[scale=0.80]{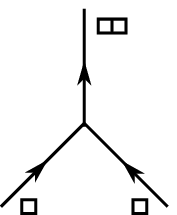}
 \end{aligned}
 \,.
\end{align}
It follows that the R-matrix \eqref{rfedec} at the special point can be written diagrammatically as
\begin{align}
 \begin{aligned}
 \frac{1}{4}\,\rfund(1)
 \end{aligned}
 =
 \begin{aligned}
  \includegraphics[scale=0.80]{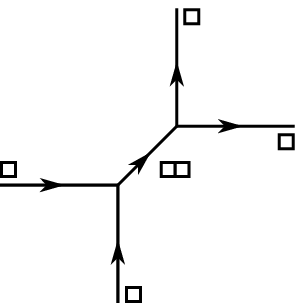}
 \end{aligned}
 \,.
\end{align}
Both vertices $Y_L$ and $Y_R$ are invariant under permutation of the two spaces that are in the fundamental representation
\begin{equation}
 \perm \cdot Y_L=Y_L\,,\quad\quad   Y_R\cdot\perm =Y_R\,,
\end{equation} 
and from the projection property of $P_+$ it follows that
\begin{equation}\label{YYid}
 Y_R\cdot Y_L=\mathbb{I}_{\frac{n(n+1)}{2}\times \frac{n(n+1)}{2}}\,,
\end{equation}
which we denote graphically as
\begin{align}
 \begin{aligned}
  \includegraphics[scale=0.80]{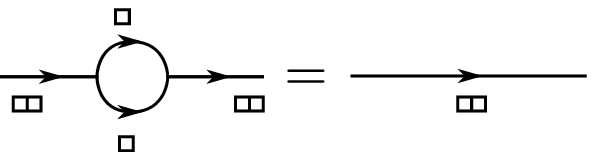}
 \end{aligned}
 \,.
\end{align}
The three-vertices introduced in \eqref{3fvertex} will be studied on more general grounds in the next section and in particular in Chapter~\ref{ch:BAforYI}. At this point, we will also come back to the discussion of the special points of the R-matrix.
 
\section{Bootstrap equation}\label{sec:bstrap}
It is interesting that the three-vertices $Y_{L,R}$ introduced in the previous section satisfy a Yang-Baxter like equation. Namely, it holds that
\begin{equation}\label{bootsym1}
{\bf L}_{\ffbox}(z-u_1)\, {\bf L}_{\ffbox}(z-u_1+1)\,Y_L=Y_L\, {\bf L}_{\Tiny\yng(2)}(z-u_1)\,.
\end{equation}
In this case, the corresponding relation involving $Y_R$ can be obtained by transposing \eqref{bootsym1} in all spaces. Furthermore, here transposition in space 1, which is suppressed in the notation above, relates the equation to one involving crossed Lax matrices, cf. \eqref{boottt} where this operation inverts the orientation of line 1.
For a proof of these relations we refer the reader to Chapter~\ref{ch:BAforYI}. 
At this point we would like to stress that the so-called Bootstrap equation \cite{Zamolodchikov1990b} above is not a consequence of the Yang-Baxter equation. As we will see in Chapter~\ref{ch:BAforYI} the Yang-Baxter equation can be reconstructed from it in certain cases. We introduce it through the generic form
\begin{equation}
 R_{12}(u-v_2)R_{13}(u-v_3)\mathbf{Y}=\mathbf{Y}R_{14}(u-v_4)
\end{equation} 
where ${\bf Y}$ is some projector from $V_2\otimes V_3$ onto a subspace $V_4$, cf. \eqref{YL}. In addition there is a constraint on the spectral parameters $v_i$ that depends on the representations under study. 
 Pictorially, we denote the bootstrap equation as
\begin{align}\label{boottt}
  \begin{aligned}
  \includegraphics[scale=0.80]{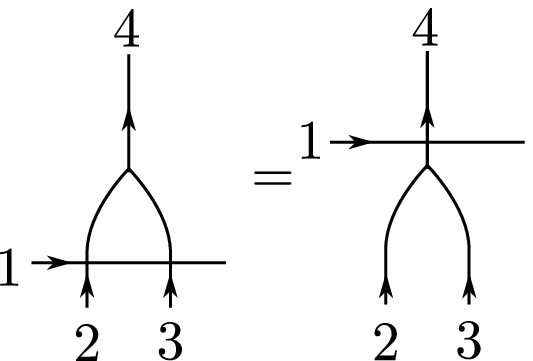}
 \end{aligned}
 \,.
\end{align}
In $2d$ integrable field theories we think about two particles $2$ and $3$ that form a bound state $4$.

The fusion procedure of R-matrices, see e.g. \cite{Zabrodin1997}, can be understood in this framework. Multiplying the bootstrap equation \eqref{bootsym1} with $Y_R$ from the left and subsequently using the relation in \eqref{YYid} we find an expression for Lax matrix with representation $\Lambda=(2,0,\ldots,0)$ in terms of the two Lax matrices ${\bf L}_{\ffbox}$
\begin{equation}\label{fusionsym}
{\bf L}_{\Tiny\yng(2)}(z-u_1)=Y_R {\bf L}_{\ffbox}(z-u_1) {\bf L}_{\ffbox}(z-u_1+1)Y_L\,.
\end{equation} 
Diagrammatically this relation can be depicted as
\begin{align}
 \begin{aligned}
  \includegraphics[scale=0.80]{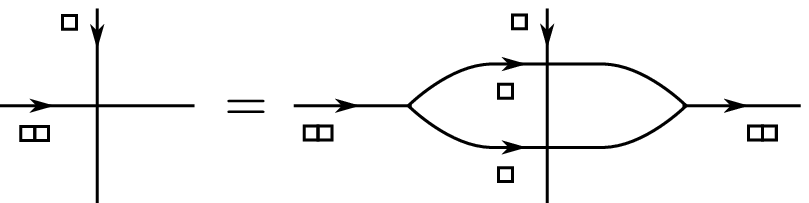}
 \end{aligned}
 \,,
\end{align}
where the vertical space denotes the auxiliary space of the Lax operators.
A nice feature of the bootstrap equation and the fusion procedure is that we can generate the crossing normalizations of the R-matrices. Following \cite{Zabrodin1997} we obtain the crossing normalization $f_\s$ for Lax operators with symmetric representations $\s=(s,0,\ldots,0)$ from $f$ introduced in \eqref{crosnorm1} for the case $\Lambda=(1,0,\ldots,0)$
\begin{equation}\label{symcnorm}
 f_\s(u)=\prod_{k=1}^sf(u+k-1)=\frac{\Gamma\left(\frac{1-u}{n}\right)\Gamma\left(\frac{n+u}{n}\right)}{\Gamma\left(-\frac{s-1+u}{n}\right)\Gamma\left(\frac{s+n+u}{n}\right)}\,.
\end{equation} 
It can be checked that this function indeed solves the constraints on the normalization for $\gamma=s-1$ and $\sigma=s$ as discussed in Section~\ref{sec:croslax}.
For the example above we find that
\begin{equation}
 f_{\Tiny\yng(2)}(u)=f(u)f(u+1)=\frac{\Gamma\left(\frac{1-u}{n}\right)\Gamma\left(\frac{n+u}{n}\right)}{\Gamma\left(-\frac{1+u}{n}\right)\Gamma\left(\frac{2+n+u}{n}\right)}\,,
\end{equation} 
here the terms $f(u)$ and $f(u+1)$ arise from the two Lax matrices in the fundamental representation, cf. \eqref{fusionsym}.

\chapter{Yangian}\label{ch:yangian}
  In this chapter we discuss aspects of the Yangian. This infinite-dimensional Hopf algebra, named after C.N. Yang, was introduced by Drinfeld in 1985 \cite{Drinfeld1985,Drinfeld1988a}. It naturally appears in the context of the Yang-Baxter equation and the quantum inverse scattering method. There are three different realizations of the Yangian, usually refered to as Drinfeld's first realization, Drinfeld's second realization and the RTT-realization. We are mainly interested in the RTT-realization which is closely tied to the RTT-relation \eqref{rMM}. We discuss how the Yangian algebra arises in this framework and study certain automorphisms and anti-automorphisms. Furthermore, we introduce the quantum determinant which generates the center of the Yangian. We introduce the coproduct and discuss its action on the Yangian generators to motivate the relation between the RTT-realization and Drinfeld's first realization that can be found in Appendix~\ref{app:1rel}. Last but not least we discuss the Yangian generators of rational spin chains and exemplify the use of the coproduct and antipode.

Here we will closely follow the presentation in \cite{molevbook} and refer the reader to this reference for proofs and further studies. Other sources are \cite{Essler2005,Gomez1996,Chari1994a,Korepin1997,Kuramoto2009,Bernard1993,MacKay2005}.

\section{Yang-Baxter algebra}
We start our discussion with the Yang-Baxter equation \eqref{ybe1} written in the form
\begin{equation}\label{rtt}
\rfund(z_1-z_2)( \mon(z_1)\otimes \mathbb I)(\mathbb I\otimes   \mon(z_2))=(\mathbb I\otimes   \mon(z_2))(  \mon(z_1)\otimes \mathbb I)  \rfund(z_1-z_2)\,.
\end{equation} 
Here $\rfund$ is the R-matrix in the fundamental representation \eqref{fundr} acting on the spaces $1$ and $2$. Furthermore, we assume that the remaining two R-matrices are $n\times n$ matrices with operatorial entries 
\begin{equation}
M(z)=\sum_{a,b=1}^nM_{ab}(z)\otimes e_{ab}\,.
\end{equation} 
For later purposes we suppressed the indices indicating on which spaces the R-matrices act trivially and used the notation
\begin{equation}
 R_{12}(z)=\rfund\,,\quad R_{13}(z)=\mon(z)\otimes\mathbb I\,,\quad R_{23}(z)=\mathbb I\otimes\mon(z)\,,
\end{equation}
where the identity indicates the trivial action on the space $2$ and $1$, respectively.
The Yang-Baxter equation \eqref{rtt} is a defining relation for the Yangian algebra $\mathcal{Y}(\mathfrak{gl}_n)$ \cite{Drinfeld1985}. Using our previous result \eqref{preyangcom} we eliminate the dependence on the spaces $1$ and $2$ and obtain the commutation relations
\begin{equation}\label{preyang}
 (z_1-z_2)[\mon_{ab}(z_1),\mon_{cd}(z_2)]=\mon_{cb}(z_2)\mon_{ad}(z_1)-\mon_{cb}(z_1)\mon_{ad}(z_2)\,,
\end{equation} 
for the entries of the matrix $M$ introduced above. The formal Laurent expansion 
\begin{equation}\label{monexpansion}
\mon_{ab}(z)=\mon^{[0]}_{ab}+\mon_{ab}^{[1]}z^{-1}+\mon_{ab}^{[2]}z^{-2}+\cdots
\end{equation} 
with $\mon^{[0]}_{ab}=\delta_{ab}$ yields the generators of the Yangian $\mathcal{Y}(\mathfrak{gl}_n)$. As mentioned previously, we will discuss certain degenerate solutions with $\mon^{[0]}_{ab}\neq\delta_{ab}$ that are relevant for the construction of Q-operators in Chapter~\ref{ch:qop}. Using \eqref{preyang} and the geometric series expansion, we obtain a system of relations for the Yangian generators $\mon^{[r]}_{ab}$. They satisfy the Yangian algebra
\begin{equation}\label{yangalg}
 [\mon^{[r]}_{ab},\mon^{[s]}_{cd}]=\sum_{q=1}^{\min(r,s)}\left(\mon_{cb}^{[r+s-q]}\mon_{ad}^{[q-1]}-\mon_{cb}^{[q-1]}\mon_{ad}^{[r+s-q]}\right)\,.
\end{equation}
By construction, these relations are equivalent to the RTT-relation \eqref{rtt}. 

Let us recall the Lax operator \eqref{lax}. This solution to the Yang-Baxter equation  was constructed from the ansatz given in \eqref{laxan}. As discussed any solution to the Yang-Baxter equation can be multiplied by a function. Thus dividing \eqref{laxan} by the spectral parameter we find that our ansatz is equivalent to the case where 
\begin{equation}\label{evaluationrep}
 M^{[r]}_{ab}=0\,,\quad \text{for}\quad r>2\,. 
\end{equation} 
In this case the commutation relations of the Yangian algebra \eqref{yangalg} reduces to the $\gln$ commutation relations as discussed in Section~\ref{sec:lax}. In the mathematical literature this representation is usually referred to as the evaluation homomorphism from the Yangian to the universal enveloping algebra of $\gln$.
\section{Automorphisms}
We will now discuss certain operations on the monodromy
\begin{equation}
\mon(z)\mapsto \tilde\mon(z)\,,
\end{equation} 
such that $\tilde\mon$ still satisfies the Yang-Baxter equation \eqref{rtt} and therefore leave \eqref{yangalg} invariant. As discussed previously any solution of the Yang-Baxter equation can be multiplied by a scalar function and remains a solution. For 
\begin{equation}
 f(z)=1+f_1z^{-1}+f_2z^{-2}+\cdots\,,
\end{equation} 
we find that the Yangian algebra \eqref{yangalg} remains invariant. Thus the operation 
\begin{equation}\label{multaut}
\mon(z)\mapsto f(z)\mon(z)\,.
\end{equation} 
is an algebra automorphism. Furthermore, we have the freedom of shifting the spectral parameter 
\begin{equation}\label{shiftaut}
\mon(z)\mapsto \mon(z+c)\,,
\end{equation} 
by some complex number $c$. This leaves the Yangian algebra \eqref{yangalg} invariant as the R-matrix in \eqref{rtt} enjoys the difference property, namely it only depends on the difference of the spectral parameters $z_1-z_2$. Additionally, for any invertible $N\times N$ matrices $B_{1,2}$ which act trivially in the quantum space $V$ one has the automorphism 
\begin{equation}\label{multa}
 \mon(z)\mapsto B_1\,\mon(z)\,B_2\,.
\end{equation} 
This follows from the property 
\begin{equation}\label{gl2inv}
\rfund(z_1-z_2)( B_{1,2}\otimes B_{1,2})=( B_{1,2}\otimes B_{1,2})\rfund(z_1-z_2)\,.
\end{equation} 
\section{Anti-automorphism and antipode}
The anti-automorphisms map the solutions $M$ of the Yang-Baxter equation \eqref{rtt} to solutions of the equation
\begin{equation}\label{inv2ybe}
\rfund(z_1-z_2)(\mathbb I\otimes   \bar\mon(z_2))(\bar\mon(z_1)\otimes \mathbb I)=(\bar\mon(z_1)\otimes \mathbb I) (\mathbb I\otimes   \bar\mon(z_2)) \rfund(z_1-z_2)\,.
\end{equation} 
The crucial difference in this equation compared to \eqref{rtt} is the order of multiplication in the quantum space. It is rather straightforward to see that the inversion 
\begin{equation}\label{antipodee}
\mon(z)\mapsto \mon^{-1}(z)\,,
\end{equation} 
satisfies \eqref{inv2ybe}. We discussed a similar phenomenon in Chapter~\ref{ch:2dlattice}, see \eqref{invybe}. The mapping in \eqref{antipodee} is usually referred to as the antipode.
Additionally, we have the transposition in the auxiliary space
\begin{equation}
\mon(z)\mapsto \mon^{t_a}(z)\,.
\end{equation} 
We can show that \eqref{inv2ybe} is satisfied by transposing \eqref{rtt} in the two auxiliary spaces and by using the symmetry of the R-matrix $\rfund^{t}=\rfund$. Furthermore, one has the reflection
\begin{equation}
\mon(z)\mapsto \mon(-z)\,,
\end{equation} 
which follows from the relation
\begin{equation}
\rfund(z)\rfund(-z)=1-z^2\,,
\end{equation} 
cf.~\eqref{invrfund} and the automorphism \eqref{multaut}. In particular, the fact that the composition of two anti-automorphisms yields an automorphism can be used to fix a certain normalization for $\mon$. This was discussed in detail for the Lax operators in Section~\ref{sec:croslax}. For completeness we give the analogues of the defining relations of the Yangian algebra \eqref{yangalg}
\begin{equation}\label{yangalgbar}
 [\bar\mon^{[r]}_{ab},\bar\mon^{[s]}_{cd}]=\sum_{q=1}^{\min(r,s)}\left(\bar\mon_{ad}^{[r+s-q]}\bar\mon_{cb}^{[q-1]}-\bar\mon_{ad}^{[q-1]}\bar\mon_{cb}^{[r+s-q]}\right)\,.
\end{equation}

\section{Quantum determinant and $\mathcal{Y}(\mathfrak{sl}(n))$}
As we have seen in Section~\ref{sec:three} the R-matrix $\rfund$ becomes a projector on at the special points $z=\pm1$ on the symmetric and antisymmetric representation in the tensor product decomposition \eqref{ytff}. Previously we were interested in the case $z=1$ where the R-matrix becomes a symmetrizer. The definition of quantum determinant incorporates the other special point $z=-1$ where the R-matrix reduces to the anti-symmetrizer
\begin{equation}\label{asyr}
 \rfund(-1)=-4P_-=-2(1-\perm)\,.
\end{equation} 
This property together with the RTT-relation \eqref{rtt}, which simplifies for the choice of the spectral parameters as in \eqref{asyr} (compare \eqref{abcom} and \eqref{dbcom}), motivates the definition of the quantum determinant. Here we only present the main properties of the quantum determinant and refer the reader to \cite{molevbook,Korepin1997,Essler2005,Kuramoto2009} for their derivations. From the form of the R-matrix $\rfund$ and the RTT-relation \eqref{rtt} it follows that the quantum determinant is given by
\begin{equation}
\qdet \mon(z)=\sum_P \sgn(P) \mon_{P(1)1}(z)\mon_{P(2)2}(z+1)\cdots\mon_{P(n)n}(z+n-1)\,.                 
\end{equation} 
Here the sum goes over all permutations $P$ of $(1,\ldots,n)$. The quantum determinant of $\mon$ belongs to the center of $\mathcal{Y}(\gln)$  
\begin{equation}\label{qdetcom}
 [\qdet \mon(z_1),\mon_{ab}(z_2)]=0\,.
\end{equation} 
Expanding the quantum determinant in inverse powers of the spectral parameter yields
\begin{equation}
\qdet \mon(z)=1+\sum_{k=1}^\infty \frac{d_k}{z^k}\,.
\end{equation} 
As a consequence of \eqref{qdetcom}, the elements $d_i$ commute among themselves and with all Yangian generators. They generate the center of $\mathcal{Y}(\gln)$. In analogy to ordinary Lie groups one defines $\mathcal{Y}(\sln)$ to be isomorphic to the quotient 
\begin{equation}\label{ysl}
 \mathcal{Y}(\sln)\simeq \mathcal{Y}(\gln)/(\qdet\mon(z)=1)\,.
\end{equation} 

\section{Comultiplication}\label{sec:cop}
The product in the auxiliary space of any two solutions of the RTT-relation \eqref{rtt} yields another solution of the RTT-relation. This can be shown directly using the RTT-relation \eqref{rtt} or diagrammatically using the train argument presented in \eqref{pic:train}. We relied on this important property of the Yang-Baxter equation when constructing the commuting family of operators in Chapter~\ref{ch:ybe}. In the mathematical literature this structure is usually referred to as the coproduct. It should not be confused with the multiplication in the quantum space that appears in the RTT-relation. In the RTT-realization of the Yangian the coproduct is simply given by matrix multiplication in the auxiliary space and the tensor product in the quantum space
\begin{equation}\label{coprod}
 \Delta (\mon_{ab}(z))=\sum_{c=1}^n\mon_{ac}(z)\otimes\mon_{cb}(z)\,.
\end{equation} 
As mentioned above, the coproduct satisfies the compatibility condition
\begin{equation}\label{rttcop}
\begin{split}
\rfund(z_1-z_2)( \Delta(\mon(z_1))&\otimes \mathbb I)(\mathbb I\otimes   \Delta(\mon(z_2)))\\
&=(\mathbb I\otimes   \Delta(\mon(z_2)))( \Delta( \mon(z_1))\otimes \mathbb I)  \rfund(z_1-z_2)\,.
\end{split}
\end{equation} 
Furthermore, it can be shown that the quantum determinant introduced in the previous section is comultiplicative 
\begin{equation}
 \Delta(\qdet M(z))=\Delta(\qdet M(z))\otimes\Delta(\qdet M(z))\,.
\end{equation} 
The coproduct, together with the antipode introduced in \eqref{antipodee} and the counit
\begin{equation}
 \epsilon(\mon_{ab}(z))= \delta_{ab}\,,
\end{equation} 
are characteristic structures of a Hopf algebra. We will not further elaborate on the Hopf algebra structure of the Yangian as it is not necessarily needed to follow the subsequent presentation and refer the reader to the literature. For a mathematical approach see e.g. \cite{Chari1994a,Kassel1995} as well as the PhD thesis \cite{lucyphd} and for a physicist's presentation see e.g. \cite{Essler2005,Gomez1996}.

We would like to make a few more comments at this point that we find helpful to understand the relation of the RTT-realization to Drinfeld's first realization of the Yangian that is discussed in Appendix~\ref{app:1rel}.
In order to do so, we present the action of the coproduct on the Yangian generators introduced in \eqref{monexpansion} which can be obtained from the expansion of \eqref{coprod} using \eqref{monexpansion}. In general it can be written as
\begin{equation}\label{copmon}
\begin{split}
  \Delta (\mon_{ab}^{[r]})&=\sum_{c=1}^n\sum_{s=0}^r\mon_{ac}^{[s]}\otimes\mon_{cb}^{[r-s]}\\&=\mon_{ab}^{[r]}\otimes\mathbb{ I}+\mathbb{I}\otimes\mon_{ab}^{[r]}+\sum_{k=1}^n\sum_{s=1}^{r-1}\mon_{ac}^{[s]}\otimes\mon_{cb}^{[r-s]}\,.
\end{split}
\end{equation} 
Hence, we see that for $r=1$ we obtain
\begin{equation}
  \Delta (\mon_{ab}^{[1]})=\mon_{ab}^{[1]}\otimes\mathbb{ I}+\mathbb{I}\otimes\mon_{ab}^{[1]}\,,
\end{equation} 
which is familiar from the representation theory of Lie algebras and quantum mechanics. For higher orders we get a deformation of this term as for example for $r=2$
\begin{equation}
  \Delta (\mon_{ab}^{[2]})=\mon_{ab}^{[2]}\otimes\mathbb{ I}+\mathbb{I}\otimes\mon_{ab}^{[2]}+\sum_{c=1}^n\mon_{ac}^{[1]}\otimes\mon_{cb}^{[1]}\,.
\end{equation} 
The action of the antipode $S$ on the first two Yangian generators is given by 
 \begin{equation}\label{antii1q}
  S(M_{ab}^{[1]})=-M_{ab}^{[1]}\,,
 \end{equation} 
 and
 \begin{equation}\label{antii2q}
 \begin{split}
  S(M_{ab}^{[2]})=-M_{ab}^{[2]}+\sum_{c=1}^nM_{ac}^{[1]}M_{cb}^{[1]}\,.
 \end{split}
 \end{equation} 
Using the ordinary multiplication (multiplication in the quantum space) of the Hopf algebra that we did not introduce explicitly we can verify \eqref{antii1q} and \eqref{antii2q}. 
Knowing the Yangian algebra \eqref{yangalg} and the action of the coproduct and antipode on the Yangian generators we can in principle forget about the auxiliary space and the RTT-relation. However, naively we still have to deal with infinitely many generators. The beauty of Drinfeld's first realization is that the Yangian $\mathcal{Y}(\sln)$ is characterized by the first two levels and an additional algebraic constraint called Serre relations. Solving these equations and imposing the structure introduced previously one can reformulate the Yangian without the need of a Yang-Baxter equation. The explicit connection between the two realizations of the Yangian is presented in great detail for the case of $\gl2$ in \cite{Essler2005}. We are not aware of a reference where this has been worked out for $\gln$. In Chapter~\ref{ch:BAforYI} we will discuss some advantages and disadvantages of Drinfeld's first realization and the RTT-realization.
\section{Yangian and spin chains}
Let us recall the spin chain monodromy introduced in \eqref{moninh} with all inhomogeneities set to zero. As shown graphically in \eqref{pic:train}, this monodromy satisfies the RTT-relation \eqref{rtt} and therefore yields a realization of the Yangian algebra \eqref{yangalg}. 
Defining
\begin{equation}
\check{\mathcal{M}}_{ab}(z)=\frac{1}{z^L} \mathcal{M}_{ab}(z)
\end{equation} 
we can extract the form of the Yangian generators from the expansion in terms of the spectral parameter
\begin{equation}
 \check{\mathcal{M}}_{ab}(z)=\delta_{ab}+z^{-1} \check{\mathcal{M}}^{[1]}_{ab}+z^{-2} \check{\mathcal{M}}^{[2]}_{ab}+\ldots\,,
\end{equation} 
cf. \eqref{monexpansion}.
We have already seen in Section~\ref{sec:comfam} that the term proportional to $z^{L-1}$ is given by
\begin{equation}
 \check{\mathcal{M}}^{[1]}_{ab}=\sum_{i=1}^L J_{ba}^{(i)}\,.
\end{equation} 
The term proportional to $z^{L-2}$ is bi-local, i.e. it acts on non-adjacent sites of the quantum space and takes the form
\begin{equation}\label{lev2}
 \check{\mathcal{M}}^{[2]}_{ab}=\sum_{1\leq k<j \leq L} \sum_{c=1}^n J_{ca}^{(k)}J_{bc}^{(j)}\,.
\end{equation} 

To end this chapter, we present the action of the coproduct and the antipode for spin chains. The coproduct in \eqref{copmon} adds sites to the spin chain. For convenience we introduce the notation 
\begin{equation}
  \check{\mathcal{M}}^{[1],[2]}_{ab}\rightarrow  \check{\mathcal{M}}^{[1],[2]}_{ab}(L)\,.
\end{equation} 
It is rather easy to verify that 
\begin{equation}
 \begin{split}
 \check{\mathcal{M}}^{[1]}_{ab}(L+L')&=\sum_{i=1}^{L+L'} J_{ba}^{(i)}=\check{\mathcal{M}}^{[1]}_{ab}(L)\otimes {\mathbb I}+{\mathbb I}\otimes\check{\mathcal{M}}^{[1]}_{ab}(L')\,.
\end{split}
\end{equation} 
For the second level we find
\begin{equation}
\begin{split}
 \check{\mathcal{M}}^{[2]}_{ab}(L+L')&=\sum_{1\leq k<j \leq L+L'} \sum_{c=1}^n J_{ca}^{(k)}J_{bc}^{(j)}\\
 &=\check{\mathcal{M}}^{[2]}_{ab}(L)\otimes {\mathbb I} +{\mathbb I}\otimes\check{\mathcal{M}}^{[2]}_{ab}(L')+\sum_{c=1}^n \check{\mathcal{M}}^{[1]}_{ac}(L)\otimes \check{\mathcal{M}}^{[1]}_{cb}(L')\,,
\end{split}
 \end{equation} 
which is compatible with \eqref{lev2}.

The action of the antipode on the first and second level Yangian generators can be deduced from the inverse of the monodromy which can be constructed from the inverse Lax operators introduced in Section~\ref{sec:croslax}. Here we would like to remind the reader that we focussed on a certain class of representations satisfying the relation \eqref{genrec}. 
The inverse of the monodromy can then be written as
\begin{equation}
 \mathcal{M}(z)^{-1}=R_{\ffbox,\Lambda_L}(z)^{-1}R_{\ffbox,\Lambda_{L-1}}(z)^{-1}\cdots R_{\ffbox,\Lambda_1}(z)^{-1}\,.
\end{equation} 
Expanding the inverse monodromy above yields the action of the antipode 
 \begin{equation}\label{antii1}
  S(M_{ab}^{[1]})=-\sum_{i=1}^LJ_{ba}^{(i)}=-M_{ab}^{[1]}\,,
 \end{equation} 
 and
 \begin{equation}\label{antii2}
 \begin{split}
  S(M_{ab}^{[2]})&=\sum_{L\leq k<j \leq 1} \sum_{c=1}^n J_{ca}^{(k)}J_{bc}^{(j)}+\gamma\sum_{i=1}^LJ_{ba}^{(i)}+\sigma L\delta_{ab}\\
  &=-M_{ab}^{[2]}+\sum_{c=1}^nM_{ac}^{[1]}M_{cb}^{[1]}\,.
 \end{split}
 \end{equation}
These relations are in agreement with the general formulas given in \eqref{antii1q} and \eqref{antii2q} and naturally generalize to inhomogeneous spin chains as the ones studied in Chapter~\ref{ch:BAforYI}. In particular, for a certain normalization of the monodromy matrix the Yangian generators for the inhomogeneous spin chain can simply be obtained by shifting the $\gln$ generators at each site $J_{ab}^{(i)}\rightarrow J_{ab}^{(i)}-v_i\,\delta_{ab}$.
\chapter{Q-operators}\label{ch:qop}
In this chapter we present a systematic approach to the construction of Baxter Q-operators for rational homogeneous spin chains. This program is motivated
by the desire to gain a deeper understanding of the integrable structure of the free/planar
{\small AdS/CFT} system, where spin chains appear in the weak coupling limit \cite{Beisert2010a}. Although the R-operators provided in the Appendix~\ref{susylax} can be used to construct the Q-operators relevant for the full one loop spin chain of {\small AdS/CFT} there has not been any direct application of the method presented in the following, yet. However, the current techniques used to determine anomalous dimensions are based on the imposed analytic structure of all loop Q-functions, see \cite{Gromov2014a} and references therein. Therefore, one may hope that it is possible to construct Q-operators at any value of the 't Hooft coupling leading to the all loop Q-functions and containing the full information about the eigenspace. 

We start the chapter with a review of the Q-operator construction for $\gl2$ spin chains carried out in \cite{Bazhanov2010} and discuss how the functional relations among the Q-operators and the Bethe equations can be derived.
Generalizing the construction to $\gln$ spin chains, we develop a new approach to Baxter Q-operators by relating
them to the theory of Yangians, which are the simplest examples for
quantum groups.
A historic overview of the subject with references to earlier works can be found in Section~5 of \cite{Bazhanov2010}. Furthermore, a significant part of the PhD thesis \cite{carlophd} is dedicated to the Q-operator construction for $\gln$ invariant spin chains, cf. Section~\ref{sec:Q-opgln}. In particular, it contains a careful analysis of the R-operators used in the Q-operator construction which we do not discuss here in its full extent. The method to obtain local charges from the Q-operators presented in Section~\ref{sec:qtoh} as well as the Lax operators \eqref{superlax} are not contained in \cite{carlophd}.
Following the construction of Bazhanov, Lukyanov and Zamolodchikov \cite{BazhanovCommun.Math.Phys.177:381-3981996,Bazhanov1997,Bazhanov1999}, we study
certain degenerate solutions of the Yang-Baxter equation connected
with harmonic oscillator algebras.
These infinite-state solutions of the Yang-Baxter equation
serve as elementary, ``partonic'' building blocks for
other solutions via the standard fusion procedure.
After a reviewing the Q-operator construction for $\gl2$ compact spin chains with the fundamental representation at each site of the quantum space we consider $\gln$ compact spin chains 
and derive the full hierarchy of operatorial functional equations for
all related commuting transfer matrices and Q-operators\,\footnote{See also \cite{Chicherin2012a,Chicherin2012b,Kazakov2010} for independent approaches.}.
This leads to a systematic and transparent solution of these chains, where
the nested Bethe equations are derived in an entirely algebraic
fashion, without any reference to the traditional Bethe ansatz
techniques.
Furthermore, we discuss the Q-operator construction for spin chains with more general representations of $\gln$ at each site of the quantum space and derive the corresponding Bethe equations.
In the last part of this Chapter, we discuss how the shift operator and the Hamiltonian enter the hierarchy
of Baxter Q-operators in the example of $\gln$ homogeneous
spin-chains. We find that a reduced set of
Q-operators can be used to obtain local charges. The mechanism relies
on projection properties of the corresponding R-operators $\mathcal{R}$
on a highest/lowest weight state of the quantum space. It is intimately
related to the ordering of the oscillators in the auxiliary space.
We introduce a diagrammatic language that makes these
properties manifest. This approach circumvents
the paradigm of constructing the transfer matrix with equal representations
in quantum and auxiliary space and underlines the strength of the
Q-operator construction.

We would like to stress that in Chapter~\ref{ch:ybe} and \ref{ch:BAforYI} we build monodromies from R-matrices where the first space was taken to be the auxiliary space. This is a matter of conventions and one can equally well built them from R-matrices where the second space is the auxiliary space. In Section~\ref{sec:croslax} we introduced the symmetry relation \eqref{symree} that relates the two monodromies above if the auxiliary space is in the fundamental representation and the quantum space satisfies the relation \eqref{genrec}. In this case we find 
\begin{equation}\label{tbart}
{\bf T}(z)=\tr R_{\ffbox,\Lambda_1}(z)\cdots R_{\ffbox,\Lambda_L}(z)=\tr R_{\Lambda_1,\ffbox}(z+\gamma)\cdots R_{\Lambda_L,\ffbox}(z+\gamma)=\bar{\bf T}(z+\gamma)\,,
\end{equation} 
where for convenience all representations at the sites were taken to be equal.
Thus, the two transfer matrices are related by a shift in the spectral parameter and therefore belong to the same family of commuting operators. In the following we focus on $\bar{\bf T}$ type transfer matrices following the conventions used in \cite{Bazhanov2010,Bazhanov2010a,Frassek2010,Frassek2011,Frassek2013}.

\section{Q-operators in a nutshell}\label{sec:qnut}
In the following we focus on the Q-operator construction for $\gl2$-invariant homogeneous spin chains with fundamental representation at each site of the quantum space. Following \cite{Bazhanov2010a,Bazhanov2010}, see also \cite{Beisert2010a}, we employ certain solutions of the Yang-Baxter equation \eqref{rll} to derive the Baxter equation and the Bethe equations without referring to a vacuum nor making an ansatz for the eigenstates, cf. Section~\eqref{sec:gl2}. 

In Section~\ref{sec:lax}, we derived the Lax operators ${\bf L}_\Lambda$ from the Yang-Baxter equation making the ansatz \eqref{laxan}. Later on we saw that this solution yields a realization of the Yangian, cf. Chapter~\ref{ch:yangian}. In the following we concentrate on the partonic Lax operators 
\begin{equation}\label{plax}
L_{\,\ffbox,+}(z)=\left(\begin{array}{cc}
              z-\osch&\bar\osca\\
	      -\osca&1
             \end{array}\right)\quad  \text{and} \quad L_{\,\ffbox,-}(z)=\left(\begin{array}{cc}
              1&-\osca\\
	      \bar\osca&z-\osch
             \end{array}\right)\,,
\end{equation}
with
\begin{equation}\label{oscalk}
[\osca,\bar\osca]=1,\quad \osch=\bar\osca\osca+\half\,.
\end{equation}
As shown in Section~\ref{sec:solution}, they satisfy the Yang-Baxter equation
\begin{equation}\label{rllpart}
 \rfund(u-v)({L}_{\,\ffbox,\pm}(u)\otimes\mathbb{I})(\mathbb{I}\otimes{L}_{\,\ffbox,\pm}(v))=(\mathbb{I}\otimes{L}_{\,\ffbox,\pm}(v)) ({L}_{\,\ffbox,\pm}(u)\otimes\mathbb{I})\rfund(u-v)\,,
\end{equation} 
cf. \eqref{rtt}, where the third space has been realized using the harmonic oscillator algebra \eqref{oscalk}.
In contrast to the Lax operators introduced previously, the term proportional to the spectral parameter $z$ is not the identity matrix.

It is easy to show that the product of the two solutions \eqref{plax} is related to the ordinary Lax operator ${\bf L}_\Lambda$ as introduced in Section~\ref{sec:lax}. For this purpose we multiply the two partonic Lax operators \eqref{plax} in the fundamental space and take the tensor product in the oscillator space. We obtain the factorization formula
\begin{equation}\label{shortfacsl2}
L_{\,\ffbox,+}^{[1]}(z_1)\,L_{\,\ffbox,-}^{[2]}(z_2)=\mathcal{S}\, \mathcal{L}_{\,\ffbox,\Lambda}(z)\,\mathbb{G}\,\mathcal{S}^{-1}\,,
\end{equation} 
where the additional labels carried by $L_{\,\ffbox,\pm}$ indicate the two copies of the oscillators $(\osca^{[1]},\bar\osca^{[1]})$ and $(\osca^{[2]},\bar\osca^{[2]})$ and
\begin{equation}\label{laxg}
 \mathcal{L}_{\,\ffbox,\Lambda}(z)=\left(\begin{array}{cc}
z+{\bf J}_{11}\,\,&{\bf J}_{21}\\
{\bf J}_{12}\,\,&z+{\bf J}_{22}\\
\end{array}\right)\,,\quad\quad \mathbb{G}=\left(\begin{array}{cc}
1\,\,&-\osca^{[2]}\\
0\,\,&1\\
\end{array}\right)\,.
\end{equation} 
The $\gl2$-generators on the left-hand side of \eqref{laxg} are realized by the oscillators $(\osca^{[1]},\bar\osca^{[1]})$ in Holstein-Primakoff form \cite{Holstein1940} as
\begin{align}\label{hp}
{\bf J}_{11}&=\lambda^1-\bar\osca^{[1]}\osca^{[1]}\,,&{\bf J}_{21}&=\bar\osca^{[1]}\left(\bar\osca^{[1]}\osca^{[1]}-(\lambda^1-\lambda^2)\right)\,,\\
{\bf J}_{12}&=-\osca^{[1]}\,,&{\bf J}_{22}&=\lambda^2+\bar\osca^{[1]}\osca^{[1]}\,,\nonumber
\end{align}
satisfying the $\gl2$ commutation relations as given in \eqref{glnalg}.
We denote the Lax operator ${\bf L}_{\ffbox,\Lambda}$ using a calligraphic letter ${\mathcal L}_{\ffbox,\Lambda}$ to stress that the $\gl2$ generators ${\bf J}_{ab}$ are realized using oscillators as above. 
Furthermore, we defined the complex variables
\begin{equation}
 z_1=z+\lambda^1+\half,\quad z_2=z+\lambda^2-\half\,,
\end{equation} 
as well as the similarity transformation acting solely on the oscillators
\begin{equation}\label{gl2sim}
         \mathcal{S}=e^{\bar\osca^{[1]}\bar\osca^{[2]}}      \,.                                  
\end{equation} 
Following the vertex notation introduced in Chapter~\ref{ch:2dlattice}, we can write the factorization formula \eqref{shortfacsl2} using the notation $L_\pm=L_{\,\ffbox,\pm}$ and $\mathcal{L}_\Lambda=\mathcal{L}_{\,\ffbox,\Lambda}$ as shown in Figure~\ref{fig:fact}.
\begin{figure}
\begin{center}
    \includegraphics[scale=0.80]{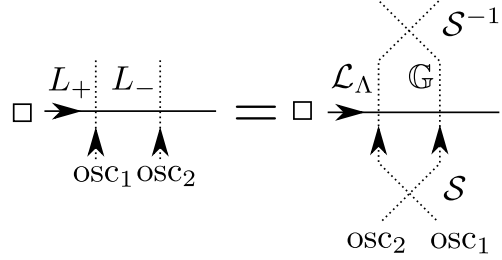}
\end{center}
   \caption{Graphical representation of the factorization formula \eqref{shortfacsl2}. The dashed lines denote the oscillator spaces $\text{osc}_{1,2}$.}
\label{fig:fact}
\end{figure}
Using the Holstein-Primakoff realization of the generators \eqref{hp}, we fix the highest weight state to be the Fock vacuum
\begin{equation}
 \vert \sigma\rangle=\vert 0\rangle\quad\text{with} \quad\osca\vert0\rangle=0\,,
\end{equation} 
where we suppressed the index carried by the oscillators. The action of the $\gl2$ generators \eqref{hp} on the highest weight state is then given by 
\begin{equation}\label{hwsrep}
 {\bf J}_{11}\vert\sigma\rangle=\lambda^1\vert\sigma\rangle\,,\quad{\bf J}_{22}\vert\sigma\rangle=\lambda^2\vert\sigma\rangle\,,\quad{\bf J}_{12}\vert\sigma\rangle=0\,,
\end{equation} 
and we label the representation as $\Lambda=(\lambda^1,\lambda^2)$. Furthermore, we note that for a given positive integer $k$ a representation with $\lambda^1-\lambda^2=k$ is finite-dimensional and the lowest weight state  annihilated by ${\bf J}_{21}$  is of the form $\vert k\rangle \sim \bar\osca^k\vert0\rangle$, cf. Figure~\ref{modulsub}.

The factorization formula \eqref{shortfacsl2} is of prior importance to derive the functional relations among the Q-operators and the transfer matrices with fundamental representations in the quantum space and various representations of $\gl2$ in the auxiliary space.
A similar formula where the order of the partonic Lax operators is reversed can 
be found in \cite{Bazhanov2010a}.
Like the transfer matrices in Chapter~\ref{ch:ybe}, the Q-operators are constructed as traces over monodromies built out of the partonic Lax operators \eqref{plax}. They are defined as
\begin{equation}\label{qopsl2}
\bar{\textbf Q}_\pm(z)=e^{\pm i z\Phi }\,Z_\pm^{-1}\,{\tr}\, e^{\mp 2 i\Phi\bar\osca\osca}\,L_{\,\ffbox_1,\pm}(z)\cdots L_{\,\ffbox_L,\pm}(z)\,,
\end{equation} 
with the trace explicitly given by
\begin{equation}\label{trehce}
 \tr (X)=\sum_{p=0}^\infty\langle p\vert X\vert p\rangle\,,\quad \text{with}\quad \langle p\vert q\rangle=\delta_{p,q}\,,
\end{equation} 
and the normalization
\begin{equation}
 Z_\pm=\tr e^{\mp 2 i\Phi\bar\osca\osca}=\pm\frac{e^{\pm i\Phi}}{e^{+i\Phi}-e^{-i\Phi}}\,.
\end{equation} 
Here the auxiliary space is the Fock space generated by the oscillators $(\osca,\bar\osca)$, while the quantum space is given by the $L$-fold tensor product of $\mathbb{C}^2$. Hence, \eqref{qopsl2} is a $2^L\times 2^L$ matrix. Naively, the trace over the infinite-dimensional Fock space diverges. Thus, we introduce a twist containing the complex parameter $\Phi$ in analogy to Section~\ref{twist}, see also Section~\ref{sec:traces} for further details. 

Using the factorization formula \eqref{shortfacsl2}, one can show that the Q-operators defined in \eqref{qopsl2} are related to the transfer matrix\,\footnote{Note that we choose the auxiliary space to be the second space of the Lax operators. To obtain a monodromy whose auxiliary space is constructed from the first space of the Lax operators via the fusion procedure the degenerate solutions presented in Appendix~\ref{otherlax} have to be employed.}
\begin{equation}\label{tplusss}
 \bar{\textbf T}^+_\Lambda=\tr \,e^{i\Phi({\bf J}_{11}-{\bf J}_{22})}\mathcal{L}_{\,\ffbox_1,\Lambda}(z)\cdots\mathcal{L}_{\,\ffbox_L,\Lambda}(z)\,,
\end{equation} 
via
\begin{equation}\label{TQQ}
2i\sin(\Phi)\,\bar{\textbf T}^+_\Lambda(z)=\bar{\textbf Q}_+(z+\lambda^1+\half)\,\bar{\textbf Q}_-(z+\lambda^2-\half)\,.
\end{equation} 
Here we used the cyclicity of the trace and the relation
\begin{equation}
\tr\, e^{2i\Phi\bar\osca\osca}\,\mathbb{G}\otimes\ldots\otimes \mathbb{G}=Z_-\cdot\mathbb{I}\,.
\end{equation} 
Diagrammatically it is easy to see that the similarity transformations cancel each other, see Figure~\ref{fig:tracefac}.
\begin{figure}
\begin{center}
    \includegraphics[scale=0.80]{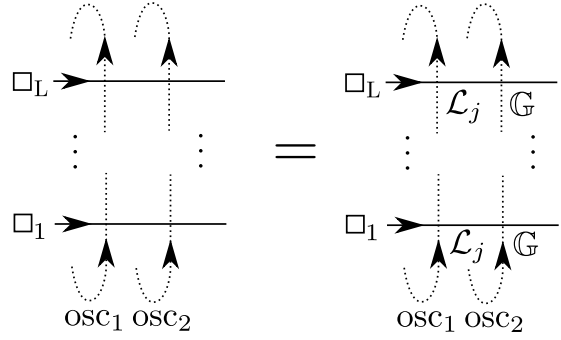}
\end{center}
   \caption{Graphical representation of \eqref{TQQ}. All similarity transformations cancel.}
\label{fig:tracefac}
\end{figure}

 However, when taking the trace over the complete Fock space, cf. \eqref{tplusss}, also states $\vert p\rangle$ with $p>k$ contribute. These have to be subtracted by hand. As indicated in Figure~\ref{modulsub}, the superfluous states are given by the module with highest weight representation $\Lambda=(\lambda^2-1,\lambda^1+1)$, i.e. the module defined through
 \begin{equation}
 {\bf J}_{11}\vert\sigma\rangle=(\lambda^2-1)\vert\sigma\rangle\,,\quad{\bf J}_{22}\vert\sigma\rangle=(\lambda^1+1)\vert\sigma\rangle\,,\quad{\bf J}_{12}\vert\sigma\rangle=0\,,
\end{equation} 
see also \cite{Beisert:2004ry}.
The transfer matrix for finite-dimensional representations $\Lambda=(\lambda^1,\lambda^2)$ in terms of the transfer matrices defined in \eqref{tplusss} then reads
\begin{equation}
\bar{\textbf T}_{\Lambda}(z)=\bar{\textbf T}^+_{(\lambda^1,\lambda^2)}(z)-\bar{\textbf T}^+_{(\lambda^2-1,\lambda^1+1)}(z)\,.
\end{equation} 
As a consequence of the factorization formula \eqref{shortfacsl2} and the resulting relation \eqref{TQQ}, we obtain the transfer matrix in terms of the Q-operators
 \begin{equation}\label{qqrels}
\Delta(\Phi)\bar{\textbf T}_\Lambda(z)=\bar{\textbf Q}_+(z+\lambda^1+\half)\,\bar{\textbf Q}_-(z+\lambda^2-\half)-\bar{\textbf Q}_+(z+\lambda^2-\half)\,\bar{\textbf Q}_-(z+\lambda^1+\half)\,,
\end{equation} 
with $\Delta(\Phi)=2i\sin(\Phi)$.
\begin{figure}
  \includegraphics[scale=0.60]{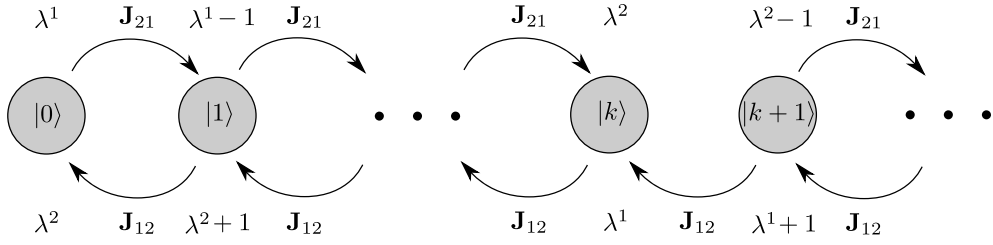}
  \caption{Reducible highest weight module where each state is labeled by the corresponding eigenvalues of the Cartan elements. The operators ${\bf J}_{21}$ and ${\bf J}_{12}$ act as raising and lowering operators, respectively.}
\label{modulsub}
\end{figure}
The relation above is the main result of this section. It can be shown that the operators defined in this section commute with each other
\begin{equation}
 [{\bar{\textbf Q}}_\pm(z),{\bar{\textbf Q}}_\pm(z')]=0\,,\quad [{\bar{\textbf T}}_\Lambda(z),{\bar{\textbf Q}}_\pm(z')]=0\,,\quad [{\bar{\textbf Q}}_\mp(z),{\bar{\textbf Q}}_\pm(z')]=0\,.
\end{equation} 
Following the logic presented in Section~\ref{sec:comfam}, the first relation follows from the Yang-Baxter equation \eqref{rllpart}, the second and third one are more difficult to prove. We refer the reader to the discussion of the appropriate Yang-Baxter relations in \cite{carlophd}. To derive the Baxter equation, cf.~\eqref{baxeq}, we evaluate \eqref{qqrels} for the fundamental representation $\Lambda=(1,0)$ and for the trivial representation $\Lambda=(0,0)$. Combining the two resulting relations we obtain the Baxter equation
 \begin{equation}\label{baxeqbar}
\bar{\textbf Q}_\pm(z)\bar{\textbf T}_{\ffbox}(z-\half)=\bar{\textbf Q}_\pm(z+1)\,\bar{\textbf T}_0(z-\half)+\bar{\textbf Q}_\pm(z-1)\,\bar{\textbf T}_0(z+\half)\,.
\end{equation} 
Here we introduced the notation $\bar{\textbf T}_{(1,0)}=\bar{\textbf T}_{\ffbox}$ and $\bar{\textbf T}_{(0,0)}=\bar{\textbf T}_{0}$.
By construction the eigenvalues of the operators $\bar{\textbf Q}_\pm$ are polynomials of the form
\begin{equation}
 \bar Q_\pm(z)=e^{\pm iz\Phi}\prod_{i=1}^{m_\pm}(z-z_i^\pm)\,,
\end{equation} 
with $ m_++m_-=L$. Note, that we did not prove that $\bar{\bf{Q}}_\pm$ is indeed diagonalizable. In order to do so, one should derive the appropriate Yang-Baxter relation to show that $\bar{\bf{Q}}_\pm$ is a normal operator, i.e. it commutes with its transpose, compare Section~\ref{sec:comfam}.
Following the argumentation in Chapter~\ref{ch:ybe}, we obtain the Bethe equations 
\begin{equation}
\bar Q_\pm(z^\pm_k+1)\,\bar T_0(z^\pm_k-\half)+\bar Q_\pm(z^\pm_k-1)\,\bar T_0(z^\pm_k+\half)=0\,,
\end{equation} 
from the Baxter equation \eqref{baxeqbar}. Noting that by construction $\bar{\bf T}_0(z)$ is diagonal with eigenvalues\,\footnote{In this chapter we use the letter $T$ to denote the eigenvalues of the transfer matrix instead of $\tau$ .} given by $\bar T_0(z)=z^L$ and using the explicit form of the Q-functions we find
\begin{equation}\label{bethebar}
\left(\frac{z^\pm_k+\half}{z^\pm_k-\half}\right)^L=-e^{\pm 2i\Phi}\prod_{i=1}^{m_\pm}\frac{z^\pm_k-z^\pm_i-1}{z^\pm_k-z^\pm_i+1}
\end{equation} 
The Bethe equations can also be derived directly from the functional relations \eqref{qqrels} for $\Lambda=(0,0)$, see Section~\ref{sec:bees}.

To end this section let us connect our results to Chapter~\ref{ch:ybe}. Noting that 
\begin{equation}
 \bar{\bf T}_{\ffbox}(z)={\bf T}_{\ffbox}(z)\,,
\end{equation} 
cf. Section~\ref{sec:unitarity}, we obtain a slightly modified energy formula from the logarithmic derivative of the diagonal form of \eqref{baxeqbar}
\begin{equation}
 E=L+\sum_{i=1}^{m_\pm}\left(\frac{1}{z^\pm_i-\half}-\frac{1}{z^\pm_i+\half}\right)\,,
\end{equation} 
while the form of the Hamiltonian \eqref{logder} coincides.
This more symmetric form of the energy formula and the Bethe equations \eqref{bethebar} with $\Phi\rightarrow 0$ can be recast into the form introduced in \eqref{abaenergy} and \eqref{bae} with vanishing inhomogeneities by shifting the Bethe roots  
\begin{equation}
 z_i^\pm=\tilde z_i^\pm+\half\,.
\end{equation} 
This naturally also yields a redefinition of the Q-functions which are polynomials with zeros located at the Bethe roots
\begin{equation}
 \bar{ Q}_\pm(z)={Q}_\pm(z-\half)\,,
\end{equation} 
such that we recover the Baxter equation \eqref{baxeq} from \eqref{baxeqbar}. In Chapter~\ref{ch:ybe}, we only introduced one Q-function. The reason for this is that we fixed the reference state $\vert\Omega\rangle$ to be the $L$-fold tensor product of highest weight states, cf. \eqref{bethevac}, and used the operators $B$ to create the eigenstates \eqref{su2BV}. However, we could have chosen the reference state to be the $L$-fold tensor product of lowest weight states and create the Bethe vectors with the operator $C$, cf. \eqref{gl2mon}. This symmetry is reflected by the existence of the two Q-operators/functions, see also Section~\ref{sec:func} and Section~\ref{sec:bees}. Furthermore, later on we will discuss how the Hamiltonian can be obtained from the Q-operators. In particular, for the fundamental representation in the quantum space it is obtained in Appendix~\ref{fundhamq}.

 \section{More on Q-operators}\label{sec:Q-opgln}

In the previous section we reviewed the construction of Q-operators for $\gl2$-invariant spin chains with a twist. In this section we generalize this construction for $\gln$-invariant spin chains where the twist again plays an important role. We classify a set of degenerate solutions of the Yang-Baxter equation which constitutes the main ingredients to construct the full hierarchy of Q-operators for $\gln$-invariant spin chains. We start with the construction of Q-operators with the fundamental representation at each site of the quantum space. A generalized factorization formula, cf. \eqref{shortfacsl2} and \eqref{TQQ}, allows us to construct the hierarchy of Q-operators and derive their functional relations. The Bethe equations follow from this construction. In Section~\ref{sec:oscpart} we present the R-operators that are the building blocks of Q-operators with arbitrary representations of $\gln$ at each site of the quantum space and derive the corresponding Bethe equations.


\subsection{Representations  of Yangians}
\label{sec:solution}
The $\gln$ spin chain Hamiltonian \eqref{sln-ham} with twisted
boundary conditions commutes with a large commuting family of operators $\bar{\bf T}$ and
$\bar{\bf Q}$. In the following we explicitly construct these
operators via traces of certain monodromy matrices associated with
infinite-dimensional representations of the harmonic oscillator
algebra. To do this we
need to find appropriate solutions of the Yang-Baxter equation  
\begin{equation}
\label{YB-main}
\rfund(z_1-z_2)( \mon(z_1)\otimes \mathbb I)(\mathbb I\otimes   \mon(z_2))=(\mathbb I\otimes   \mon(z_2))(  \mon(z_1)\otimes \mathbb I)  \rfund(z_1-z_2)\,,
\end{equation}
where $\Rbf(z)$ is an $n^2 \times n^2$ matrix, 
\begin{equation}
{\Rbf}(z):\qquad 
{\mathbb C}^n \otimes {\mathbb C}^n \to {\mathbb C}^n \otimes
{\mathbb C}^n, \qquad 
{\Rbf}(z)
=z + \Pbf\, ,\label{Ryang}
\end{equation}
acting in the direct product of two $n$-dimensional spaces 
${\mathbb C}^n \otimes\,{\mathbb C}^n$, cf. \eqref{rtt}.

We will now show that there exist further first order operators, different 
from the Lax operators \eqref{lax} and present their complete classification. For the case of $\gl2$ these operators were introduced in \eqref{plax}.  To begin, let us recall a symmetry of the Yang-Baxter equation
\eqref{YB-main}. From the Yangian algebra \eqref{yangalg} it follows that the elements 
$\mon_{ab}^{[0]}$ are central, i.e.~they commute among themselves and
with all $\mon_{ab}^{[r]}$ for $r\ge1$. Therefore, we may regard $\mon^{[0]}$ as a numerical $n\times n$ matrix. Applying the automorphism \eqref{multa}, this
matrix can always be brought to a diagonal form
\beq
\mon^{[0]}=\mbox{diag}
\big(\underbrace{1,1,\ldots,1}_{\p},
\underbrace{0,0,\ldots,0}_{(n-\p)}\big),\qquad
\p=1,2,\ldots, n\, , 
\label{L0}
\eeq
where $\p$ is an integer $1\le \p \le n$. The number $\p$ 
coincides with the rank of the matrix $\mon^{[0]}$.
Evidently, if $\p=n$, the leading term in the series expansion
\eqref{monexpansion} is the unit matrix. This case is well studied in the existing
representation theory, see e.g. \cite{molevbook}. In fact, the assumption that the series
\eqref{monexpansion} starts with the unit matrix is usually included into the
definition of the Yangian. Here we will not make this assumption,
and consider the more general case with arbitrary $1\le \p \le n$.

Let us concentrate on the  
simple case when the series \eqref{monexpansion} truncates after the second
term, i.e.~assume that all $\mon_{ab}^{[r]}=0$ for $r\ge2$. It is convenient to write the
only remaining non-trivial coefficient $\mon^{[1]}$ as a block matrix 
\beq 
\mon^{[1]}=\left(
\begin{BMAT}(e){c.c}{c.c}
{\rm A}_{\a\b}\ &{\rm B}_{\a\dot\b}\\
{\rm C}_{\dot\a\b}&{\rm D}_{\dot \a \dot\b} 
\end{BMAT}\, \right)\ ,\label{L1}
\eeq
where ${\rm A},{\rm B},{\rm C}$ and ${\rm D}$ are operator-valued matrices of dimensions 
$\p\times \p$, $\p\times (n-\p)$, $(n-\p)\times \p$ and $(n-\p)\times (n-\p)$,
respectively. We furthermore assume that all undotted indices run over the
values $\{1,2,\ldots,\p\}$, whereas their dotted counterparts take the values $\{\p+1,\ldots,n\}$:
\beq
1\le\a,\b\le \p,\qquad
\p+1\le\dot\a,\dot\b\le n\ .
\eeq

Substituting \eqref{L0} and
\eqref{L1} into \eqref{yangalg}, one realizes  that the elements 
${\rm D}_{\dot\a\dot\b}$ are central, i.e.~they commute among
themselves and with all other elements of $\mon^{[1]}$. 
The other commutation relations read
\beq
\begin{array}{rclrcl}
\ds[{\rm A}_{\a\b},{\rm A}_{\g\e}] 
&\;=\;&\delta_{\a\e}\,{\rm A}_{\g\b}-\delta_{\g\b}\,{\rm A}_{\a\e}\,,\quad&
\ds[{\rm A}_{\a\b},{\rm B}_{\g\dot\g}]&\;=\;&-\delta_{\b\g}\,{\rm B}_{\a\dot\g}\,,\quad \\[.4cm]
\ds[{\rm A}_{\a\b},{\rm C}_{\dot\g\g}]&\;=\;&+\delta_{\a\g}\,{\rm C}_{\dot\g\b}\,,&
\ds[{\rm B}_{\a\dot\b},{\rm C}_{\dot\a\b}]&=&\delta_{\a\b}\,{\rm D}_{\dot\a\dot\b}\,,\\[.4cm]
[{\rm B}_{\a\dot\b},{\rm B}_{\g\dot\e}]&=&0\,,&       
[{\rm C}_{\dot\a\b},{\rm C}_{\dot\g\e}]&=&0\, .
\end{array}\label{abc-comm}
\eeq
Using the remaining freedom of making the
transformations \eqref{multa}, which do not affect the form of $\mon^{[0]}$ in 
\eqref{L0}, one
can bring the matrix ${\rm D}$ to a diagonal form with zeros and ones on
the diagonal, in analogy to \eqref{L0}. Here we are only interested in highest 
weight representations of the algebra \eqref{abc-comm}. These
representations admit a definition of the trace, as required for the
construction of transfer matrices in Section~\ref{sec:traces} below.
For this reason we only consider the non-degenerate case\,\footnote{It appears that for $\det {\rm D}=0$ the algebra \eqref{abc-comm} 
does not admit a definition for a suitable trace as needed for the construction of
transfer matrices commuting with the Hamiltonian \eqref{sln-ham}.}, 
$\det {\rm D}\not=0$, where the diagonal form of ${\rm D}$ coincides with the
$(n-\p) \times (n-\p)$ unit matrix 
\beq
{\rm D}_{\dot\a\dot\b}=\delta_{\dot\a\dot\b}, \qquad
\p+1\le\dot\a,\dot\b\le n\ . 
\eeq
The resulting algebra \eqref{abc-comm} can be realized as a direct
product of the algebra $\gl{\p}$ with $\p(n-\p)$ copies of the harmonic oscillator 
algebra:
\beq
{\mathcal A}_{n,\p}=\gl{\p}\otimes \oscalg^{\otimes
 \p(n-\p)}\,. \label{product} 
\eeq

Therefore, we introduce $p(n-p)$ independent  
oscillator pairs $(\oscbm_{\dot\a\b},\oscbp_{\b\dot\a})$,\ 
where $\dot\a=p+1,\ldots,n$ and $\b=1,\ldots,p$, 
satisfying the relations
\beq
\qquad [\oscbm_{\dot\a\b},\oscbp_{\g\dot\e}]=
\delta_{\dot\a\dot\e}\,\delta_{\b\g}\,.\label{osc} 
\eeq
Furthermore, let $\E_{\a\b}$, \
$\a,\b=1,2,\ldots,p$ denote the generators of the algebra
$\gl{p}$ defined by \eqref{glnalg}, where $n$ is replaced by $\p$. 
The generators $\E_{\a\b}$ commute with all oscillators in
\eqref{osc}.  
The connection of the algebra \eqref{abc-comm} with the product
\eqref{product}  
is established by the following relations
\beq
{\rm A}_{\a\b}=\oE_{\a\b}-\sum_{\dot\g=\p+1}^n \big(\oscbp_{\a\dot{\g}}\,
\oscbm_{\dot\g\b}+\sfrac{1}{2}\delta_{\a\b}\big),
\qquad  {\rm B}_{\a\dot\b}=\oscbp_{\a\dot\b},
\qquad  {\rm C}_{\dot\a\b}=-\oscbm_{\dot\a\b}\,,
\eeq
where the hat in the notation $\oE_{\a\b}$ denotes the
transposition of the indices $\a$ and $\b$, 
\beq
\oE_{{\a}{\b}}\equiv \E_{{\b\a}}.
\eeq
The corresponding  L-operator can be written as a block matrix
\beq 
\Lbf_{\{1,2,\ldots,\p\}}(z)=\left(
\begin{BMAT}(r){c.c}{c.c}
z\,\delta_{\a\b}+\oE_{\a\b}-\sum_{\dot\g=\p+1}^n\big( \oscbp_{\a\dot{\g}}\,
\oscbm_{\dot\g\b}+\sfrac{1}{2}\delta_{\a\b}\big)\ &\ \ \oscbp_{\a\dot\b}\\
-\oscbm_{\dot\a\b}&\delta_{\dot \a \dot\b} 
\end{BMAT}\, \right)\ ,\label{Lcanon}
\eeq
where the rows are labeled by the indices $\a$ or $\dot\a$ and the 
columns by $\b$ or $\dot\b$, in similarity to \eqref{L1}. Note that the
$p\times p$ matrix of the generators of $\gl{p}$, which
enters the upper left block, is transposed, i.e.~the $\a$-th row and $\b$-th column in this block contains the element
$\oE_{ab}=\E_{\b\a}$. 

The matrix \eqref{Lcanon} contains the parameter $z$ only in its first $p$
diagonal elements. By simultaneous permutations of rows
and columns in \eqref{Lcanon} one can move these $z$-containing elements to $p$ arbitrary positions on the
diagonal, labeled by a set of integers $I=\{a_1,a_2,\ldots,a_p\}$. 
We shall denote the L-operator obtained in this way by
${\Lbf}_I(z)$. Within this convention the operator \eqref{Lcanon}
corresponds to the set $I=\{1,2,\ldots,\p\}$, as indicated by the subscript in
the left hand side of this equation. Furthermore, employing the notation of Chapter~\ref{ch:ybe} we note that the first space of the operator ${\Lbf}_I(z)$ is in the fundamental representation ${\Lbf}_I(z)={\Lbf}_{\ffbox,I}(z)$.

The ``partonic'' L-operator
\begin{equation}
\label{boselax}
\Lbf_a(z)=\left( \begin{array}{ccccccc}
1 \, \, & \, \, & \, \, & -\oscb_{1,a} \, \,& \, \,&  \, \, & \, \,\\
\, \, &\ddots  \, \, &\, \, & \vdots  \, \,&  \, \,&\, \,& \, \, \\
\, \, &   \, \, & 1\, \, & -\oscb_{a-1,a} \, \,&  \, \,&  \, \, & \, \,\\
\oscbp_{a,1}  \, \, &   \cdots \, \, & \oscbp_{a,a-1}  \, \,& \specbaz -\osch_a \, \,& \oscbp_{a,a+1} \, \,& \cdots\, \,&\oscbp_{a,n} \, \, \\
\, \, &    \, \,&  \, \,& -\oscb_{a+1,a} \, \,& 1 \, \,&  \, \,& \, \,\\
\, \, &   \, \,&  \, \,& \vdots \, \,&   \, \,& \ddots \, \,& \, \,\\
\, \, &   \, \,&  \, \,& -\oscb_{n,a}\, \,&   \, \,& \, \,& 1\, \,\\
\end{array} \right)\,,
\end{equation}
with 
\begin{equation}
 \osch_a=\sum_{b\neq a}\left(\oscbp_{ab}\,\oscbm_{ba}+\sfrac{1}{2}\right)\,,
\end{equation} 
is a particular case of \eqref{Lcanon} with $p=1$,
while the standard Lax operator \eqref{lax} corresponds to $p=n$. 
For $n=2$, $p=1$ these solutions were studied in e.g. \cite{koornwinder,Kuznetsov:1999tk,kovalsky,Bogoliubov1996}. See also \cite{Doikou2013,Khoroshkin2014} for more recent developments. For $n=3$ these solutions can be obtained in the rational limit of trigonometric solutions obtained in \cite{Boos2010a}.

The L-operator in \eqref{Lcanon} provides an evaluation homomorphism of the
infinite-dimensional Yangian algebra \eqref{yangalg} into the
finite-dimensional algebra \eqref{product},
\beq
Y(\gln)\to\gl{\p}\otimes
\oscalg^{\otimes \p(n-\p)}\label{eval-hom}\,,\qquad 1\le \p \le n\,.
\eeq
This means that for any representation of this finite-dimensional
algebra the equation \eqref{Lcanon}
automatically defines a representation of the Yangian
and a matrix solution of the Yang-Baxter
equation \eqref{YB-main}. 
Conversely, 
any first order matrix L-operator with a rank $\p$ leading
term $\mon^{[0]}$ and a non-degenerate matrix ${\rm D}$ in \eqref{L1} is, up to a transformation \eqref{multa}, equivalent to the 
 canonical L-operator \eqref{Lcanon} with some particular 
representation of the algebra \eqref{product}. 
It is worth noting that the transformation \eqref{multa} 
\beq
\Lbf_{\{1,2,\ldots,\p\}}(z)\mapsto B\,\Lbf_{\{1,2,\ldots,\p\}}(z)\, B^{-1}\, , 
\eeq
where $B$ is a block diagonal matrix containing the $p\times p$ matrix
$B_p$ and the $(n-p)\times (n-p)$ matrix $B_{n-p}$ on
the diagonal,  
leaves the form of \eqref{Lcanon} unchanged.


The analysis of this section extends the previous results of 
\cite{koornwinder} devoted to the $n=2$ case. 
The properties of the finite-dimensional
representation of the Yangian ${\mathcal Y}(\gl2)$ 
associated with the L-operator \eqref{lax} can be
found in \cite{Tarasov1, Tarasov2, molevbook} and were also reviewed in Chapter~\ref{ch:yangian}. 
%


\subsection{Fusion and factorization of L-operators}
\label{sec:fusion}
An essential part of our analysis in the following is based 
on some remarkable decomposition properties of the product of two
L-operators of the form \eqref{Lcanon}. 
The coproduct of the Yangian ${\mathcal Y}(\gln)$
\beq
{\mathcal Y}(\gln)\to {\mathcal Y}(\gln)\otimes 
{\mathcal Y}(\gln)
\label{comul}
\eeq
is generated by the matrix product 
of two L-operators corresponding to two different copies of
${\mathcal Y}(\gln)$ appearing on the right-hand side of \eqref{comul}, cf. Section~\ref{sec:cop}.

Our main observation is related to the coproduct of two
operators ${\Lbf}_{I}(z)$ and ${\Lbf}_{J}(z)$, defined by \eqref{Lcanon} 
for two non-intersecting sets $I\cap J=\varnothing$,
\beq
\Lbf(z)=\Lbf_I^{[1]}(z+z_1)\,\Lbf_J^{[2]}(z+z_2),
\label{coprodl}
\eeq 
where the quantities $z_{1,2}$ denote arbitrary constants. 
Let the sets $I$ and $J$ contain $\p_1$ and $\p_2$ elements, respectively. 
The product \eqref{coprodl} is of the first order in the variable
$z$, and that the matrix rank of the term linear in $z$ in \eqref{coprodl}
is equal to $\p_1+\p_2$.
The meaning of the comultiplication is that the matrix product of two
L-operators, each of which satisfies by itself the Yang-Baxter equation
\eqref{YB-main}, solves this equation as well, see \eqref{rttcop}.
All solutions which possess the above properties were classified in the previous section.
Therefore, by using a transformation of type \eqref{gl2inv},  
the right-hand side of \eqref{coprodl} can be 
brought to a particular case of the canonical form \eqref{Lcanon} with
$\p=\p_1+\p_2$. 
It turns out, however, that the
expressions for the matrix elements of the resulting L-operator 
are rather complicated and their
explicit connection to those of
\eqref{Lcanon} is far from obvious, even though these elements
satisfy the same commutation relations. In order to make this
connection more transparent we apply a suitable operatorial similarity transformation
${\mathcal S}$ to each matrix element such that it
rearranges the basis of the oscillator algebras contained in ${\mathcal
 A}_{n,\p_1}\otimes{\mathcal A}_{n,\p_2}$, cf.\ \eqref{gl2sim}.  
Furthermore, the formula \eqref{coprodl} contains two constants $z_1$ and
$z_2$. Only their difference is an essential
parameter, whereas  the sum may be absorbed into the spectral parameter
$z$. Therefore, without loss of generality, one can set
\beq
z_1=\lambda+\sfrac{p_2}{2},\qquad 
z_2=-\sfrac{p_1}{2},
\eeq
where $\lambda$ is arbitrary. This particular parametrization is chosen to 
simplify the subsequent formulas. 

Proceeding as described above, one obtains,    
\beq
\Lbf(z)=\Lbf^{[1]}_I(z+\lambda+\sfrac{p_2}{2})
\,\Lbf^{[2]}_J(z-\sfrac{p_1}{2})={\mathcal S}\,\Big(
\Lbf_{I\cup   J}
(z)\ \Gbb\Big)\,{\mathcal S}^{-1} , \label{product2}
\eeq
where $\Gbb$ is a $z$-independent matrix, whose elements
commute among themselves and with all elements of $\Lbf_{I\cup
 J}^{\phantom{{[1]}}}(z)$. 
It should be stressed that the resulting L-operator $\Lbf_{I\cup
 J}^{\phantom{{[1]}}}(z)$ is only a special case of \eqref{Lcanon}, 
since it is connected to some specific realization of the algebra ${\mathcal
 A}_{n,\p_1+\p_2}$ in terms of the direct product of the two 
algebras ${\mathcal A}_{n,\p_1}\otimes{\mathcal A}_{n,\p_2}$ as
defined in \eqref{product}. Note that the right-hand side of \eqref{product2}
is a particular case of the transformation \eqref{multa}
with $B_1\equiv\mathbb{I}$ and $B_2=\mathbb{G}$. The explicit expressions for the matrices appearing in
\eqref{product2} are presented below.

By permuting rows and columns any two non-intersecting sets $I$ and
$J$ can be reduced to the case when $I=\{1,\ldots,\p_1\}$ and
$J=\{\p_1+1,\ldots,\p_1+\p_2\}$. So it is suffficient to consider this
case only. We introduce three types of indices 
\beq
\a,\b,\in I,\qquad\dot\a,\dot\b\in J,\qquad 
\ddot\a,\ddot\b\in\{\p_1+\p_2+1,\ldots,n\}\,.\label{three}
\eeq
It is convenient to rewrite \eqref{Lcanon} as a $3\times3$ block
matrix
\begin{equation}
\Lbf_I^{[1]}(z)=\left(
\begin{BMAT}(r)[0.15cm,0cm,0cm]{c.c.c}{c.c.c}
{z\,\delta_{\a\b}+\oE^{[1]}_{\a\b}-\sum_{c\not\in I} (\oscbpo_{\a c}\ 
\oscbmo_{c\b}}+\sfrac{1}{2}\delta_{\a\b}){}^{\phantom{|}}\ &\ \oscbpo_{\a\dot\b}\ &\ \oscbpo_{\a\ddot\b}\ \\
-\oscbmo_{\dot\a\b}&\delta_{\dot \a \dot\b}&0\\
-\oscbmo_{\ddot\a\b} \, \, &  0 \, \,& \, \, \delta_{\ddot\a\ddot\b} 
\end{BMAT}\, \right)\ ,\label{Lop1}
\end{equation}
where $I=\{1,2,\ldots,\p_1\}$ and the size of the diagonal blocks is equal to $\p_1\times\p_1$,\ 
$\p_2\times\p_2$ and $(n-\p_1-\p_2)\times(n-\p_1-\p_2)$, respectively.
As before, the superscript ``$[1]$'' indicates that the corresponding operators
belong to the ``first'' algebra in the comultiplication
\eqref{comul}, which in the considered case is realized by the algebra
${\mathcal A}_{n,p_1}$ defined in
\eqref{product}.  Similarly, one can write $\Lbf_J^{[2]}(z)$ as
\begin{equation}
\Lbf_J^{[2]}(z)=
\left( 
\begin{BMAT}(r)[0.15cm,0cm,0cm]{c.c.c}{c.c.c}
\delta_{\a\b} \, & -\oscbmt_{\a\dot\b}& \, \,0\\
\oscbpt_{\dot\a\b} \, \, & z\,  \delta_{\dot\a\dot\b} 
+ \oE_{\dot\a\dot\b}^{[2]} 
-\sum_{ c\not\in J}
(\oscbpt_{\dot\a c}\oscbmt_{c\dot\b}+\sfrac{1}{2}
\delta_{\dot\a\dot\b} ){}^{\phantom{|}}\, \, 
&\, \, \oscbpt_{\dot\a\ddot\b}\\
0 \, \, &  -\oscbmt_{\ddot\a\dot\b}\, \,& \, \, \delta_{\ddot\a\ddot\b} 
\end{BMAT}\, 
\right)\,,\label{Lop2}
\end{equation}
where $J=\{p_1+1, \dots ,p_1+p_2\}$ and the superscript ``$[2]$'' labels operators
from the ``second'' algebra, which is the algebra ${\mathcal A}_{n,p_2}$.
By construction, all operators labeled by the superscript ``${[1]}$''
commute with 
those labeled by the superscript ``${[2]}$''. Recall also that the algebra
\eqref{product} has a direct product structure, so the generators
$\E_{\a\b}^{[1]}$ and $\E_{\dot\a\dot\b}^{[2]}$ commute with all oscillator
operators. 

With the notation introduced above the similarity transformation
${\mathcal S}$ in \eqref{product2} has the form 

\beq \label{sim0}
{\mathcal S}={\mathcal S}_1\,{\mathcal S}_2\, ,
\eeq
where 
\beq\label{sim1}
{\mathcal S}_1=\exp\left(\sum_{c\in I} \sum_{ \dot c\in J}\oscbpo_{c\dot c} \oscbpt_{\dot c c}\right),
\eeq
and 
\beq\label{sim2}
{\mathcal S}_2=\exp\left(\sum_{c\in I}\sum_{\dot c\in J}\sum_{\ddot c\not \in I\cup J}\oscbpo_{c\dot c}\, \oscbpt_{\dot c\ddot c}\, \oscbmo_{\ddot c c} \right).
\eeq
The matrix $\Gbb$ has the form
\beq
\Gbb=\left( 
\begin{BMAT}(r){c.c.c}{c.c.c}
\delta_{\a\b} \, & -\oscbmt_{\a\dot\b}& \, \,0\\
0 &   \delta_{\dot\a\dot\b} 
&\, \, 0\\
0 \, \, & 0\,& \, \,\delta_{\ddot\a\ddot\b} 
\end{BMAT}\, 
\right)\,.\label{matG}
\end{equation}
The similarity transform $\mathcal{S}_1$ serves to expose the fact that the matrix entries of $\Lbf_{I\cup J}(z)$
commute with the entries of   $\Gbb$. 
The similarity transform $\mathcal{S}_2$ brings $\Lbf_{I\cup J}(z)$ to
the form \eqref{Lcanon}. 

Finally, we want to write the operator $\Lbf_{I\cup
 J}^{\phantom{{[1]}}}(z)$ in \eqref{product2} in the form
 \eqref{Lcanon} with $\p=\p_1+\p_2$. To do this we need to make the
 following identifications for the generators in
 the upper diagonal block of \eqref{Lcanon}
\bea
\oJ_{\a\b}&=&\oE_{\a\b}^{[1]}-\sum_{\dot\g\in J} \oscbpo_{\a\dot{\g}}\,
\oscbmo_{\dot\g\b}+\lambda\,
\delta_{\a\b}\,,\nonumber\\[.3cm]
\oJ_{\dot\a\dot\b}&=&\oE_{\dot\a\dot\b}^{[2]}+\sum_{\g\in I}
 \oscbpo_{{\g}\dot\a}\, \oscbmo_{\dot\b\g}\,,\label{J12}\\[.3cm]
\oJ_{\dot\a\b}&=&-\oscbmo_{\dot\a\b}\,,\nonumber\\[.3cm]
\oJ_{\a\dot\b}&=&\left(\sum_{c\in I}\sum_{\dot c\in J} 
\oscbpo_{a\dot c}\,\oscbpo_{c\dot b}\,\oscbmo_{\dot c c}\right)- 
\lambda\,\oscbpo_{a \dot b}+\sum_{\dot c\in J}\oscbpo_{ a \dot c}\,\oE_{\dot c\dot a}^{[2]}\, 
-\sum_{c\in I}\oE_{a c}^{[1]}\,\oscbpo_{c\dot b}\, ,
\nonumber
\eea
where we have used the conventions \eqref{three} for numerating the indices. 

Furthermore, let the indices ${ \ac},\bc$ run
 over the values $1,2,\ldots,\p_1+\p_2$ and $\dot\ac,\dot\bc$ over
 the values $p_1+p_2+1,\ldots,n$. Introduce operators 
\beq\label{cosc-def}
{\bf a}_{\dot\ac\bc}=\begin{cases}\ \oscbmo_{\dot\ac\bc},\quad
\bc\in I,\\[.3cm]
\ \oscbmt_{\dot\ac\bc},\quad
\bc\in J,\\
\end{cases}
\qquad
\bar{\bf a}_{\ac\dot\bc}=\begin{cases}\ \oscbpo_{\ac\dot\bc},\quad
\ac\in I,\\[.3cm]
\ \oscbpt_{\ac\dot\bc},\quad
\ac\in J.\\
\end{cases}
\eeq
Then the L-operator $\Lbf_{I\cup   J}^{\phantom{{[1]}}}(z)$ from
\eqref{product2} can be written as  
\beq 
\Lbf_{\{1,2,\ldots,\p_1+\p_2\}}(z)=\left(
\begin{BMAT}(r){c.c}{c.c}
z\,\delta_{\ac\bc}+{\oJ}_{\ac\bc}-\sum_{\dot\gc\not\in I\cup J}
(\bar{\bf a}_{\ac\dot\gc}\,
{\bf a}_{\dot\gc\bc}+\sfrac{1}{2}\delta_{\ac\bc}),\ &\ \ \bar{\bf a}_{\ac\dot\bc}\\
-{\bf a}_{\dot\ac\bc}&\delta_{\dot \ac \dot\bc} 
\end{BMAT}\, \right)\ ,\label{Lcanon2}
\eeq
which has the required form as in \eqref{Lcanon}. 

The formulas in \eqref{J12} 
give a homomorphism of the algebra $\gl{\p_1+\p_2}$ 
into the direct product 
\beq
\gl{\p_1+\p_2}\to
\gl{\p_1}\otimes\gl{\p_2}\otimes
\oscalg^{\otimes\,\p_1\p_2}={\mathcal B}_{\p_1,\p_2},
\label{B12}
\eeq
which for $\p_{1,2}\not=0$ has only infinite-dimensional representations (similar representations appeared in \cite{Derkachov2008a}).
An important feature of this map is that if one choses highest weight
representations for both algebras $\gl{p_1}$ and $\gl{p_2}$ 
then \eqref{J12} defines a highest weight representation 
of $\gl{p_1+p_2}$. The highest weight conditions 
\eqref{hwcon} are satisfied on the product of the corresponding highest
weight vectors and the standard Fock
vacuum for all oscillator algebras appearing in \eqref{J12}. The 
$\gl{p_1+p_2}$-weight of the resulting representation is  obtained from \eqref{J12}
\beq
\rep^{p_1+p_2}=\left(\m^1_{[1]}+\lambda,\,
\m^2_{[1]}+\lambda,\,
\ldots, \m^{p_1}_{[1]}+\lambda,\,
\m^1_{[2]},\,\m^2_{[2]},\,
\ldots,\m^{p_2}_{[2]}\right),\label{lam12}
\eeq
where $\lambda$ is an arbitrary parameter, cf. \eqref{product2}.

The L-operators in the first product in \eqref{product2} have the
superscripts $[1]$ and $[2]$, which indicate that they belong to different
algebras \eqref{product} with $p=p_1$ and $p=p_2$, respectively.
By the same reason it is useful to rewrite the RHS of \eqref{product2}
supplying similar superscripts 
\beq
\eqref{product2}={\mathcal S}\,\Big(
\Lbf_{I\cup   J}^{{{[1']}}}
(z)\ \Gbb^{[2']}\Big)\,{\mathcal S}^{-1} , \label{product3}
\eeq
where the superscript $[1']$ indicates the algebra \eqref{B12} and the
superscript $[2']$ indicates the product of oscillator algebras
${\mathcal H}^{\otimes p_1p_2}$.  Note that the matrix $\Gbb$ could
be considered as a $z$-independent L-operator, also satisfying the
Yang-Baxter equation \eqref{YB-main}. In this sense \eqref{product3} 
also describes the comultiplication of two representations of the Yangian.  

Consider now some particular consequences of formula \eqref{product2}.  
Using it iteratively with $p_1=1$ and taking into account
\eqref{lam12} one can obtain an arbitrary product of the elementary 
L-operators \eqref{boselax}. Let $I=(a_1,\ldots,a_p)$ be an ordered
integer set, $1\le a_1<\ldots<a_p\le n$, and ${
 \Lbf}^+_I(z\,|\,\rep^p)$ a specialization of the L-operator
\eqref{Lcanon} to the infinite-dimensional highest weight
representation $\pi^+_{\rep^p}$ of the algebra $\gl{p}$ 
\beq
{\Lbf}^+_I(z\,|\,\rep^p)=\pi^+_{\rep^p}\big[\Lbf_I(z)\big],\qquad
\rep^p=
(\m^1,\m^2,\ldots,\m^p)\,.\label{Lp-plus}
\eeq
Define also the shifted weights, cf. \eqref{rho-def},
\beq
\rep'^p=
(\m'^1,\m'^2,\ldots,\m'^p)\,,\qquad \m'^j=\m^j+\sfrac{p-2j+1}{2},\qquad
j=1,\ldots,p\,.\label{shifted}
\eeq 
Then it follows from \eqref{product2} that
\beq
\Lbf_{a_1}(z+\lambda'^1)
\Lbf_{a_2}(z+\lambda'^2)
\cdots
\Lbf_{a_p}(z+\lambda'^p)
={\mathcal S}_I \,{\Lbf}^+_I(z\,|\,\rep^p) \,{\Gbb}_I {\mathcal
 S}_I^{-1}\label{factor1}
\eeq 
where the matrices ${\mathcal S}_I$ and $\Gbb_I$ are products of the
expressions of the type \eqref{sim0} and \eqref{matG} arising from the
repeated use of the formula \eqref{product2}. In the particular case 
$p=n$ the last formula provides the factorization for the
Lax operator \eqref{lax},
\beq
\Lop^+(z\,|\,\rep^n)=\pi^+_{\rep^n}\big[{\bf L}(z)\big],
\eeq
evaluated for the infinite-dimensional  
highest weight representation $\pi^+_{\rep^n}$ in the auxiliary space,
\beq\label{factor2}
\Lbf_{1}(z+\lambda'^1)
\Lbf_{2}(z+\lambda'^2)
\cdots
\Lbf_{n}(z+\lambda'^n)
={\mathcal S}_{\Lop} \,{\Lop}^+(z\,|\,\rep^n) \,{\Gbb}_\Lop {\mathcal
 S}_\Lop^{-1}\,,
\eeq
cf.\ \eqref{J12}.
\subsection{Construction of the Q-operators}
\label{sec:traces}
The purpose of this section is to define the operators $\bar{\bf T}$ and $\bar{\bf
Q}$. They have to commute with the Hamiltonian \eqref{sln-ham} of
the twisted compact $\gln$-spin chain of length  
$L$.  These operators act on the quantum space 
which is an $L$-fold tensor product of the fundamental representations 
of the algebra $\mathfrak{gl}(n)$,
\begin{equation}\label{quantumspace2}
\underbrace{
{\mathbb C}^n\otimes {\mathbb C}^n\otimes \cdots \otimes  
{\mathbb C}^n}_{L}
\ ,
\end{equation}
cf. Chapter~\ref{ch:ybe}.
As before, solutions of the Yang-Baxter equation 
\eqref{YB-main} are considered to be $n\times n$ matrices, acting in the
quantum space of a single spin. Their matrix elements are operators   
in some representation space $V$ of the Yangian algebra \eqref{yangalg}.
This representation space will be called here the auxiliary space. 
For each solution of \eqref{YB-main} one can define a transfer matrix, 
\begin{equation}
\bar{\mathbb T}_{V}(z)={\tr}_{V} \Big\{{\mathbb D}\, {\mon}(z)\otimes
{\mon}(z)\otimes\cdots \otimes{\mon}(z)\Big\},
\label{T-gen}
\end{equation}
where the tensor product is taken with respect to the quantum spaces
${\mathbb C}^n$, while the operator product and the trace is taken
with respect to the auxiliary space $V$. The quantity 
${\mathbb D}$ is a ``boundary twist'' operator acting only in
auxiliary space, i.e.~it acts trivially in the quantum space.
This boundary operator is completely determined by the requirement of
commutativity of the 
transfer matrix \eqref{T-gen} with the Hamiltonian \eqref{sln-ham} and its generating transfer matrix, cf. \eqref{logder}. Following the same logic as in Section~\ref{twist} we obtain the condition
\begin{equation}\label{D-def}
[M(z),{\mathbb D}\otimes\mathcal{D}]=0\,.
\end{equation}
It is convenient to parametrize the diagonal twist $\twist$ as
\begin{equation}\label{twistfund}
\twist=\operatorname{diag}(e^{i\Phi_1},e^{i\Phi_2},\ldots,e^{i\Phi_n})\,,
\end{equation} 
using the fields $\Phi_a$ which enter in \eqref{D-def} only through their differences with
\begin{equation}
\sum_{a=1}^n \Phi_a=0\, .
\end{equation}
Solving \eqref{D-def} for the general L-operator \eqref{Lcanon} with an
arbitrary set $I=\{a_1,a_2,\ldots,a_p\}$ one obtains  
\beq\label{D-exp}
\Dbf_I=\exp\Big\{i\sum_{a\in I} \Phi_a \E_{aa}-
i\sum_{a\in I}\sum_{\dot\b\not\in I} (\Phi_a-\Phi_{\dot\b}) \,\oscbp_{a\dot\b}
\oscbm_{\dot\b\a}\Big\}\,,
\eeq
which reduces for the fundamental representation and $\vert I\vert=n$ to \eqref{twistfund}.

In the following we will use some important properties of the 
trace over the Fock representations of the oscillator algebra 
\beq
[\,\oscb,\bar\oscb\,]=1,\quad [\osch,\oscb]=-\oscb,
\quad [\osch,\bar\oscb\,]=\bar\oscb,\quad
\osch=\oscbp\,\oscb+\sfrac{1}{2}\ .\label{osc3}
\eeq
This algebra has two Fock representations, 
\beq
{\mathcal F}_+:\qquad \oscb\,|0\rangle=0, \qquad
|k+1\rangle=\bar\oscb\,|k\rangle, 
\eeq
and 
\beq
{\mathcal F}_-:\qquad \bar\oscb\,|0\rangle=0, \qquad
|k+1\rangle=\oscb \,|k\rangle, 
\eeq
spanned on the vectors
$|k\rangle$,\ \ $k=0,1,2,\ldots \infty$.
These representations can be obtained from each other via a
simple automorphism of \eqref{osc3} also refered to as particle-hole transformation
\beq
\oscb \to -\bar\oscb,\qquad \bar\oscb \to \oscb,\qquad 
\osch\to-\osch\,.
\eeq
Let $P(\oscb,\oscbp)$ be an arbitrary polynomial of the
operators $\oscb$ and $\oscbp$. 
Below it will be convenient to use a normalized trace over the
representations ${\mathcal F}_\pm$,
\beq
\ntr_{\mathcal F} \Big\{e^{i\Phi \osch}
P(\oscb,\oscbp)\Big\}\ \ \mathop{=}^{\mbox{\small 
def}}\ \ \frac { {\tr}_{\mathcal F} \Big\{e^{i\Phi\osch}
P(\oscb,\oscbp)\Big\}_{\phantom{|}}}{{\tr}_{\mathcal F} 
\Big\{e^{i\Phi\osch}\Big\}^{\phantom{|}}}\ ,\qquad {\mathcal F}={\mathcal
 F}_\pm\, ,
\label{norm-tr}
\eeq
where ${\mathcal F}$ is either ${\mathcal F}_+$ or ${\mathcal F}_-$,
and ${\tr}_{\mathcal F}$ denotes the standard trace, cf. \eqref{trehce}.  An
important feature of the normalized trace \eqref{norm-tr} is that it is
completely determined by the commutation relations \eqref{osc3}
and the cyclic property of the trace. It is therefore independent of a particular 
choice of representation as long as the traces in the RHS
of \eqref{norm-tr} converge.  Alternatively, one can reproduce the same
result by using explicit expressions for the matrix elements of the oscillator operators in 
\eqref{norm-tr}. Then the trace over ${\mathcal F}_+$ converges when
$\mbox{Im}\,\Phi>0$ and the trace over ${\mathcal F}_-$ when
$\mbox{Im}\,\Phi<0$. Both ways of calculation lead to the same analytic
expression for the normalized trace. Thus it is not necessary to
specify which of the two representations ${\mathcal F}_\pm$ is used.

We are now in the position to define various transfer matrices all commuting
with the Hamiltonian \eqref{sln-ham}. Consider the most general
L-operator \eqref{Lcanon} with an arbitrary set
$I=\{a_1,a_2,\ldots,a_p\}$, where $p=1,2,\ldots,n$. Recall that the
matrix elements of \eqref{Lcanon} belong to the direct product
\eqref{product} of the algebra $\gl{\p}$ and of $\p(n-\p)$
oscillator algebras. Choose a finite-dimensional representation
$\pi_{\Lambda^{p}}^{\phantom{+}}$ with the highest weight
${\Lambda^{p}}$ for the $\gl{p}$-factor of this
product.  Then substituting \eqref{Lcanon} and \eqref{D-exp} into
\eqref{T-gen} one can define rather general transfer matrices
\beq
{\Xbf}_I(z,\Lambda^{p})=
e^{i z\,(\,\sum_{a\in I}\Phi_a)
}\ {\tr}_{\pi_{\Lambda^{p}}}
\ntr_{{\mathcal F}^{\p(n-\p)}} 
\big\{\/{\mathcal M}_I(z)\big\}\,,\label{Z-def}
\eeq
where ${\mathcal M}_I(z)$ is the corresponding monodromy matrix, 
\beq\label{M-def}
{\mathcal M}_I(z)=
{\Dbf}_I\, { L}_I(z)\otimes
{ L}_I(z)\otimes\cdots \otimes{ L}_I(z)\,.
\eeq
Here $\ntr_{{\mathcal F}^{\p(n-\p)}}$ denotes the normalized trace
\eqref{norm-tr} for all involved oscillator algebras\,\footnote{%
Note that all possible expressions under the trace in
\eqref{Z-def} for each oscillator algebra are exactly as in the LHS of
\eqref{norm-tr} for some polynomial $P$ and some 
value of $\Phi$. Thus the definition \eqref{norm-tr} is sufficient to
calculate all oscillator traces in \eqref{Z-def}.}, 
while ${\tr}_{\pi_{\Lambda^p}}$
denotes the standard trace over the representation $\pi_{\Lambda^\p}$
of $\gl{\p}$.  The exponential scalar factor in front of the
trace is introduced for later convenience. 

Similarly, one can define a related quantity where the
$\gl{p}$-trace is taken over an infinite-dimensional highest
weight representation $\pi_{\Lambda^{p}}^{{+}}$,
\beq
{\Xbf}_I^+(z,\Lambda^{p})=
e^{i z\,(\,\sum_{a\in I}\Phi_a)
}\ {\tr}_{\pi_{\Lambda^{p}}^+}
\ntr_{{{\mathcal F}^{\p(n-\p)}}^{\phantom{|}}} 
\big\{\/{\mathcal M}_I(z)\big\}\,,\label{Zp-def}
\eeq
while the rest of the expression remains the same as in
\eqref{Z-def}. Note that in the case of \eqref{Z-def} the weights
$\Lambda^p=(\lambda^1,\lambda^2,\ldots,\lambda^{p})$ satisfy the 
conditions \eqref{domint}. In contradistinction, in
\eqref{Zp-def} these weights are arbitrary.   

In the limiting case $p=n$, the general L-operator \eqref{Lcanon}
simplifies to \eqref{lax}, while, as discussed above, the twist \eqref{D-exp} simplifies to 
\beq
\Dbf_{\Lambda^n}=\Dbf_{\{1,2,\ldots,n\}}
=\exp\Big\{i\sum_{a=1}^n \Phi_a \E_{aa}\Big\}\label{sln-exp}\,.
\eeq
In this case the definition \eqref{Z-def} reduces to that for the standard
T-operator
\beq
\bar{\bf T}_{\Lambda^n}(z)\equiv{\Xbf}_{\{1,2,\ldots,n\}}(z,{\Lambda^n})=
{\tr}_{\pi_{\Lambda^n}}\Big\{\Dbf_{\Lambda^n}\,{\bf L}_{\Lambda^n}(z)\otimes{\bf L}_{\Lambda^n}(z)
\cdots \otimes{\bf L}_{\Lambda^n}(z)\Big\}\, ,
\label{t-def}
\eeq
associated with the finite-dimensional representation
${\pi_{\Lambda^n}}$ of the algebra $\gln$ in the auxiliary
space. Here ${\bf L}_{\Lambda^n}(z)$ denotes
the Lax operator \eqref{lax}. However, we like to stress again that opposite to the definitions in Chaper~\ref{ch:ybe}, the spaces in the fundamental representation yield the quantum space and the auxiliary space is labeled by the representation label $\Lambda^n$, cf.~\eqref{tbart}. Likewise, the formula \eqref{Zp-def} reduces to the T-operator
\beq
\bar{\bf T}^+_{\Lambda^n}(z)={\Xbf}^+_{\{1,2,\ldots,n\}}(z,{\Lambda^n}) 
\label{tplus}
\eeq
associated with the infinite-dimensional representation
$\pi_{\Lambda^n}^+$. The above two T-operators are connected
due the Bernstein Gel'fand Gel'fand (BGG) resolution of the finite
dimensional modules \cite{BGG}. The BGG result allows one to express
finite-dimensional highest weight modules in terms of an alternating sum of 
infinite-dimensional highest weight modules. This implies that 
the T-operator \eqref{t-def} for a finite-dimensional module 
can be written in terms of \eqref{tplus} as
\beq
\bar{\bf T}_{\Lambda^n}(z)=\sum_{\sigma\in S_n}(-1)^{l(\sigma)}\, 
\bar{\bf T}^+_{\sigma(\Lambda^n+\rho^n)-\rho^n}(z),
\label{bgg1}
\eeq
where $\rho_n$ is a constant $n$-component vector
\beq
\rho^n=\Big(\sfrac{n-1}{2},\sfrac{n-3}{2},\ldots,\sfrac{1-n}{2}\Big)\,.
\label{rho-def}
\eeq
The summation in \eqref{bgg1} 
is taken over all permutations of $n$
elements, $\sigma\in S_n$,  and $l(\sigma)$ is the parity of the permutation
$\sigma$. The relation \eqref{bgg1} and its connection to the BGG
resolution were first obtained in \cite{Bazhanov:2001xm} in the context of
$U_q(\widehat{sl}(3))$, while the $n=2$ case was previously considered in
\cite{Bazhanov1997,Bazhanov1999}, see also Section~\ref{sec:qnut}. 

Similarly, for \eqref{Z-def} one has
\beq
{\Xbf}_I(z,\Lambda^p)=\sum_{\sigma\in S_p}(-1)^{l(\sigma)}\, 
{\Xbf}_I^+(z,\sigma(\Lambda^p+\rho^p)-\rho^p),
\label{bgg2}
\eeq
where $\rho^{\,p}$ is a $p$-component vector defined as in \eqref{rho-def}
with $n$ replaced by $p$.

Another limiting case of \eqref{Z-def} corresponds to the representation
$\pi^{\phantom{+}}_{\Lambda^\p}$ turning into the trivial
one-dimensional representation of $\gl{p}$ with weight 
$\Lambda=(0,0,\ldots,0)$. As we shall see below the
resulting operators 
\beq
\bar{\bf Q}_I(z)={\Xbf}_I(z,(0))\,,
\label{QI-def}
\eeq
are actually the Q-operators, whose eigenvalues appear in the
nested Bethe ansatz equations.  Let us enumerate these Q-operators. 
It is convenient to start formally from the exceptional case $p=0$,
corresponding to an empty set $I=\varnothing$. By definition we set 
\beq
\bar{\bf Q}_\varnothing(z)\equiv 1.\label{Q0}
\eeq
For the next level $p=1$ there are obviously
$n$ sets $I$ consisting of just one
element $I=\{a\}$, \ $a=1,2,\ldots,n$. The general L-operator
\eqref{Lcanon} in this case takes the simple form \eqref{boselax}
and the twist operator \eqref{D-exp} simplifies to 
\beq \Dbf_{a}\equiv\Dbf_{\{a\}}=
\exp\Big\{-i\sum_{\dot\g\not\in I} (\Phi_a-\Phi_{\dot\g}) \,\oscbp_{a\dot\g}
\oscbm_{\dot\g\a}\Big\}\,,
\qquad a=1,2\ldots,n.\label{D-parton} \eeq
In this way one obtains from \eqref{Z-def}
\beq
\bar{\bf Q}_a(z)=
e^{i z\,\Phi_a
}\ \ntr_{{\mathcal F}^{(n-1)}} 
\Big\{{\Dbf}_a\, { L}_a(z)\otimes\cdots \otimes{ L}_a(z)\Big\}
\,,\label{Qa-def}
\eeq
where $a=1,2\ldots,n$, ${L}_a(z)$ given by \eqref{boselax} and 
\begin{equation}
\bar{\bf Q}_a(z)= \bar{\bf Q}_{\{a\}}(z)={\Xbf}_{\{a\}}(z,(0))\,.
\end{equation} 

More generally, for the level $p$ there are $\binom{n}{p}$ increasing
integer sets $I=\{a_1,\ldots,a_p\}\subseteq \{1,2,\ldots,n\}$, which 
numerate the Q-operators \eqref{QI-def}. For the highest
level $p=n$ the definitions \eqref{QI-def} and \eqref{Z-def}
immediately lead to the result 
\beq
\bar{\bf Q}_{\{1,2,\ldots,n\}}(z)=z^L,\label{Qn}
\eeq
where $L$ is the length of the chain. Altogether there are $2^n$
different Q-operators including \eqref{Q0} and \eqref{Qn}. As discussed in the next section, they can conveniently be associated with nodes of a hyercubical Hasse diagram. 

The  Q-operators form in conjunction with all T- and X-operators a commuting family and therefore can be simultaneously
diagonalized. One sees that their eigenvalues have to be of the form
\beq\label{Q-eigen}
\bar{Q}_I(z)=e^{i z\,(\,\sum_{a\in I}\Phi_a)} \, \prod_{k=1}^{m_I}
(z-z^I_k),\qquad m_I=\sum_{a\in I} m_a\,,
\eeq
where, for each eigenstate, the numbers 
$m_a$ are the conserved occupation numbers,
\beq
m_1+m_2+\cdots+m_n=L\,,
\eeq
cf.~Section~\ref{sec:naba}.

We would like to stress, 
that in general the operators $\bar{\bf X}_I(z,\rep^p)$, defined in 
\eqref{Z-def}, involve the trace over
a representation of the Lie algebra $\gl{p}$ and the trace over a
number of Fock
representations of the oscillator algebra, whereas the T-operators involve only the Lie algebra trace and the Q-operator only the oscillator traces. This is why we denoted the 
hybrid operators \eqref{Z-def} by a distinct symbol $\bar{\bf X}$.

\subsection{Functional relations}
\label{sec:func}

The results of Section~\ref{sec:fusion} imply various functional relations 
for the Q-operators. To derive them we need to use some
additional properties of the twist operators \eqref{D-exp} which are 
not immediately obvious from their definition \eqref{D-def}. 
Let $\Dbf_I$ and $\Dbf_J$ be the operators \eqref{D-exp},
corresponding to $\Lbf^{[1]}_I(z)$ and $\Lbf^{[2]}_J(z)$ from the LHS
of \eqref{product2}. By explicit calculation one can check that 
the product of these operators commutes 
with the similarity
transformation ${\mathcal S}$ defined in \eqref{sim0},
\beq
\big[\Dbf_I\,\Dbf_J\,,\, {\mathcal S}\big]=0\,.\label{D1}
\eeq
Moreover, this product can be rewritten in the form
\beq
\Dbf_I\,\Dbf_J=\Dbf_{I\cup J}\,\Dbf_{\mathbb G}\, ,
\label{D2}
\eeq
where $\Dbf_{I\cup J}$ and  $\Dbf_{\mathbb G}$ are the twist
 operators obtained from \eqref{D-def} for the operator 
$\Lbf_{I\cup   J}(z)$ and 
the $z$-independent L-operator $\Gbb$ from the RHS of
\eqref{product2}. Again the relation \eqref{D2} 
is verified by direct calculation, where one needs to take into
account the explicit form of \eqref{matG}, \eqref{J12},
\eqref{cosc-def} and \eqref{Lcanon2}.

Next, define a scalar factor, {\it cf.}~\eqref{qqrels},
\beq\label{vandermond}
\Delta_I(\Phi)=\Delta_{\{a_1,a_2,\ldots,a_p\}}(\Phi)=
\prod_{1\le i<j\le p} 2i\,\sin\bigg(\frac{\Phi_{a_i}-\Phi_{a_j}}{2}\bigg),
\eeq
which depends on the set $I$ and the fields
$\Phi_1,\Phi_2,\ldots,\Phi_n$. Combining \eqref{product2} with the
definition \eqref{Zp-def} and taking into account \eqref{D1} and
\eqref{D2} one obtains
\begin{equation}
\Delta_I\,
\Xbf^+_{I}(z+\sfrac{p_2}{2},\Lambda^{p_1})\ 
\Delta_{ J}\,\Xbf^+_{J}(z+\lambda-\sfrac{p_1}{2},\Lambda^{p_2})=
{\Delta_{{I}\cup{J}}}\,
\Xbf^+_{{I}\cup{J}}(z,\Lambda^{p_1+p_2}).
\label{func-main}
\end{equation}   
There are two non-trivial steps in the derivation of the last formula which
are explained in the following. First, the simple transfer matrix 
\beq
\bar{\bf T}_{\Gbb}=\ntr_{{\mathcal F}^{p_1p_2}}\Big\{\Dbb_{\Gbb}\,
\Gbb\otimes\Gbb\otimes\cdots\otimes\Gbb\Big\}=1
\eeq
that arises in the calculations is equal to the identity
operator. Second, the scalar factors in \eqref{func-main} arise due to
the difference in the definition of the trace over the oscillator
algebras (normalized trace \eqref{norm-tr}) and over the
representation of $\gl{p}$ (standard trace). From \eqref{B12} it
is clear that $p_1p_2$ oscillator pairs have to be ``redistributed'' to
support the Holstein-Primakoff realization of the
infinite-dimensional representation $\pi^+_{\rep^{p_1+p_2}}$ of the
algebra $\gl{p_1+p_2}$.

A particular simple case of \eqref{func-main} arises when $p_2=1$ and 
$J=\{a_{p+1}\}$,
\begin{equation}
\Delta_I\,
\Xbf^+_{I}(z+\sfrac{1}{2},\Lambda^{p})\ 
\bar{\bf Q}_{a_{p+1}}(z+\lambda^{p+1}-\sfrac{p}{2})=
{\Delta_{{I}\cup{a_{p+1}}}}\,
\Xbf^+_{{I}\cup{a_{p+1}}}(z,\Lambda^{p+1}),
\end{equation}   
where by the definition \eqref{Qa-def} one has
$\bar{\bf Q}_a(z)\equiv{\Xbf}_{\{a\}}(z,(0))$. Iterating the last formula
one obtains 
\beq
\Delta_I\ \Xbf_I^+(z,\rep^p) = \bar{\bf Q}_{a_1}(z+\m'^1)\,{\bf
 Q}_{a_2}(z+\m'^2)\cdots \bar{\bf Q}_{a_p}(z+\m'^p),
\eeq
where the notation here is the same as in \eqref{shifted} and
\eqref{factor1}. Next, applying \eqref{bgg2} one gets 
\beq\label{Xasdet}
\Delta_I\ \Xbf_I(z,\rep^p) = {\det} \Vert\,
\bar{\bf Q}_{a_i}(z+\m'^j)\,\Vert_{1\le i,j \le p}, 
\eeq
and setting $\rep^p=(0)$ one finally arrives at 
\beq\label{Qasdet}
\Delta_I\ \bar{\bf Q}_I(z) = {\det} \Vert\,
\bar{\bf Q}_{a_i}(z-j+\sfrac{p+1}{2})\,\Vert_{1\le i,j \le p}\, . 
\eeq
Note also that in the particular case $I=\{1,2,\ldots,n\}$ the formula 
\eqref{Xasdet}  
leads to the determinant expression \cite{Bazhanov1997,Krichever:1996qd,
Bazhanov1999,
 Bazhanov:2001xm,Kojima:2008zza} 
for transfer matrix \eqref{t-def},
\beq
\Delta_{\{1,2,\ldots,n\}}\ \bar{\bf T}_{\Lambda^n}(z)=
{\det} \Vert\,
\bar{\bf Q}_{i}(z+\m'^j)\,\Vert_{1\le i,j \le n}\ ,
\eeq
where $\Delta_{\{\ldots\}}$ is defined in \eqref{vandermond} and
$\lambda'^j$ in \eqref{shifted}.


As previously mentioned the $2^n$ different operators
$\Qop_I$ can be assigned to the nodes of a hypercubic Hasse diagram.  We will
now show that four Q-operators belonging to the same
quadrilateral as shown in Figure~\ref{Hasse0} satisfy a remarkably simple
functional equation, which can be identified with the famous Hirota
equation from the theory of classical discrete evolution equation, cf.~\eqref{hiroota}. 
Define a matrix
\beq
\label{matyrixofQ}
M_{ij}\equiv \Qop_{a_i}(z-j+\sfrac{p+1}{2})\,,
\quad i,j\in \{0,\ldots,p+1\}\, .
\eeq
where $\{a_0,a_1,\ldots,a_p,a_{p+1}\}$ is an increasing sequence of
 $p+2$ integers which contains the subsequence
 $I=\{a_1,\ldots,a_p\}$ with $a\equiv a_0$ and $b\equiv a_{p+1}$.
\begin{figure}[t]
\begin{center}
 \includegraphics[scale=1.2]{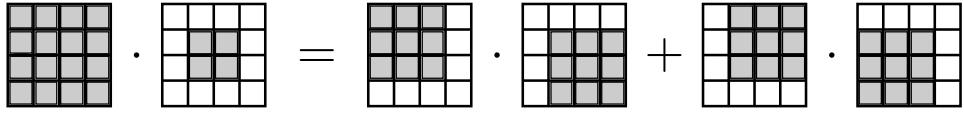}
\end{center}
\caption{Schematic representation of the Desnanot-Jacobi determinant formula for a $4\times 4$ matrix. The shaded area indicates the matrix of which the determinant is taken.}
\label{Fig:Jacobi}
\end{figure}
Let us now use
the Desnanot-Jacobi determinant identity, see Figure~\ref{Fig:Jacobi}, for the 
matrix \eqref{matyrixofQ}. Applying \eqref{Qasdet} for the
subdeterminants one obtains the following
operatorial functional relation\,\footnote{%
More general relations involving operators \eqref{Xasdet} can be
obtained in the same way by replacing the arguments of the Q-operators in \eqref{matyrixofQ} 
with arbitrary constants $z_j$, $j=0,\ldots,p+1$.}
\beq
\label{FormJacobi}
\Delta_{\{a,b\}}\,\Qop_{I\cup a \cup
b}(\specbaz)\,\Qop_{I }(\specbaz)= \Qop_{I \cup
a}(\specbaz-\half)\,\Qop_{I \cup b}(\specbaz+\half)- \Qop_{I
\cup b}(\specbaz-\half)\,\Qop_{I \cup a}(\specbaz+\half)\, ,
\eeq
where $\Delta_{\{a,b\}}$ is given by \eqref{vandermond}. 
Since all Q-operators commute with each other the same relation
\eqref{FormJacobi} holds also for the corresponding eigenvalues. In the next section, we use this to derive the Bethe equations and the energy formula. 
Furthermore, the quadratic functional relations \eqref{FormJacobi} possess an
interesting graphical interpretation, cf.~Figure~\ref{Hasse0}. The full
set of functional equations is nicely depicted in so-called Hasse
diagrams, cf.~\cite{Tsuboi2009} and references therein.
 The Hasse diagram corresponding to the $\gln$ algebra forms an $n$-dimensional ordered hypercube.
Hasse
diagrams for $n=2,3,4$ are presented in the Figure~\ref{Hasse2} and Figure~\ref{Hasse3}, respectively. To read off the
functional relations it is enough to take any 4-cycle in these
diagrams, using the equivalence depicted in Figure~\ref{Hasse0}.

Every path in the Hasse diagram which leads from ${\Qop}_{\varnothing}$ to ${\Qop}_{\{1,\ldots, n\}}$ defines a system of equivalent but distinct nested Bethe equations. To find each such system, it is enough to take all Q-operators on a given path and write one relation for any three subsequent functions on the path. Such relation can be written for every three subsequent Q-operators because there always exists a unique 4-cycle containing them.

\begin{figure}[ht]
\begin{center}
 \includegraphics[scale=0.8]{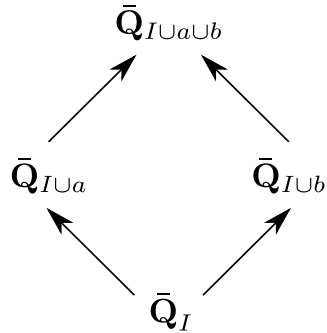}
\end{center}
\caption{Graphical depiction of the functional relations \eqref{FormJacobi}}
\label{Hasse0}
\end{figure}

\begin{figure}[ht]
        \centering
        \begin{subfigure}[b]{0.4\textwidth}
                \includegraphics[width=\textwidth]{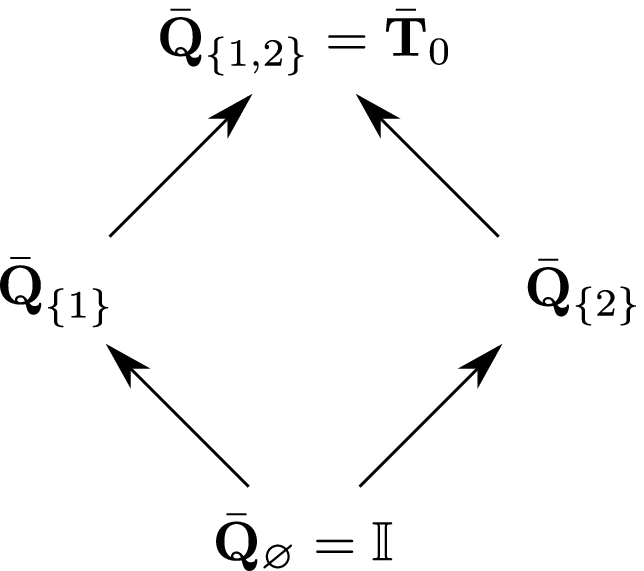}
                \vspace{10pt}
                \caption{}
                   \label{Hasse2}
                   \end{subfigure}
                   \quad\quad\quad\quad
         \begin{subfigure}[b]{0.4\textwidth}
                \includegraphics[width=\textwidth]{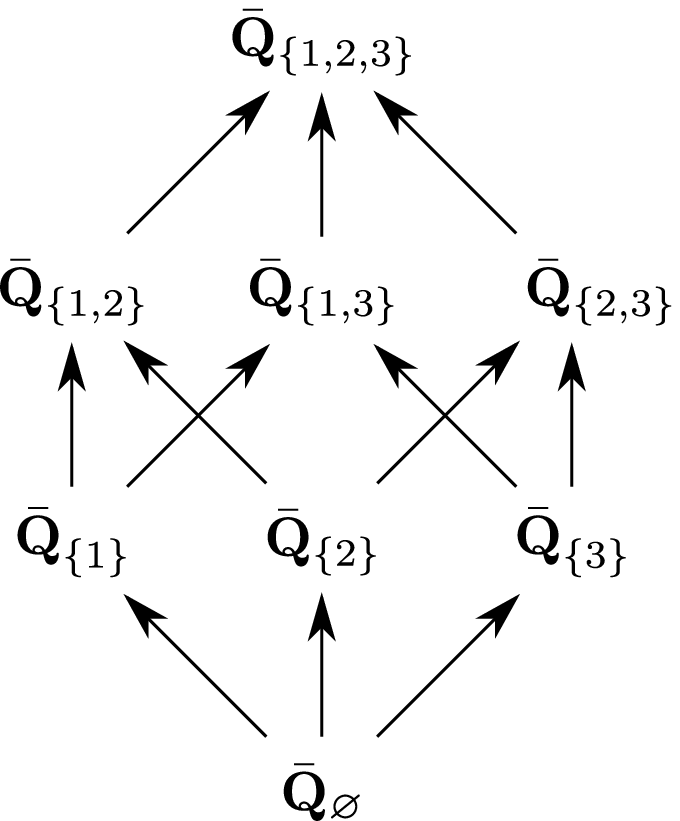}
                \caption{}
                \label{Hasse3}
                \end{subfigure}
                \caption{(a) Hasse diagram for  $\gl2$. Compare also to the discussion in Section~\protect\ref{sec:qnut} where the notation $\bar{\bf Q}_{\{1\},\{2\}}=\bar{\bf Q}_\pm$ has been employed. (b) Cubic Hasse diagram for $\gl3$. }
                \label{Hasse23}
\end{figure}


\subsection{Bethe equations and energy formula}\label{sec:bees}

The connection between the Hirota equations \eqref{FormJacobi} and the
Bethe ansatz equations is well understood 
\cite{Bazhanov1997,Pronko:1999gh,Bazhanov:2001xm,Dorey:2000ma,
  Kazakov:2007fy,Tsuboi2009}.    
Let us consider the QQ-relations \eqref{FormJacobi} at the eigenvalue level. The
reader might find it useful to look at the examples of Hasse diagrams in Figure~\ref{Hasse23} when following the upcoming derivation.  Let us denote the zeros, i.e. the Bethe roots, of $\bar{Q}_{I \cup a}(\specbaz)$ by
$\specbaz_k^{I \cup a}$.  Taking $\specbaz+\half=\specbaz_k^{I \cup
a}$ and $\specbaz-\half=\specbaz_k^{I \cup a}$, equation
\eqref{FormJacobi} yields
\beq
\bar{Q}_{I\cup a \cup b}(\specbaz_k^{I \cup a}-\half)\,\bar{Q}_{I }(\specbaz_k^{I \cup a}-\half)\sim 
\bar{Q}_{I \cup a}(\specbaz_k^{I \cup a}-1)\,\bar{Q}_{I \cup b}(\specbaz_k^{I \cup b})\,,
\eeq
\beq
\bar{Q}_{I\cup a \cup b}(\specbaz_k^{I \cup a}+\half)\,\bar{Q}_{I }(\specbaz_k^{I \cup a}+\half)\sim 
- \bar{Q}_{I \cup a}(\specbaz_k^{I \cup a}+1)\,\bar{Q}_{I \cup b}(\specbaz_k^{I \cup b})\,,
\eeq
respectively, where $a,b\not \in I$.
Taking the ratio of these two equations above one obtains
\beq
\label{formalBETHE}
-1=\frac{\bar{Q}_{I }(\specbaz_k^{I \cup a}-\half)}{\bar{Q}_{I }(\specbaz_k^{I \cup a}+\half)}\,
\frac{ \bar{Q}_{I \cup a}(\specbaz_k^{I \cup a}+1)}{ \bar{Q}_{I \cup a}(\specbaz_k^{I \cup a}-1)}\,
\frac{\bar{Q}_{I\cup a \cup b}(\specbaz_k^{I \cup a}-\half)}{\bar{Q}_{I\cup a \cup b}(\specbaz_k^{I \cup a}+\half)}\,.
\eeq
Here $I$ can also be the empty set. In this case we can remove $\bar{Q}_{\varnothing}(z)$ from
the equation using \eqref{Q0} . The number of elements in $I\cup a\cup b$ cannot
exceed $n$, therefore $I$ contains at most $n-2$ elements. 
Thus one obtains $n-1$ different relations of the type
\eqref{formalBETHE}, with various cardinalities of the set $I$. This
exactly matches the number of levels of nested Bethe equations for the
$\mathfrak{gl}(n)$ spin chain, cf.~Section~\ref{sec:naba}. 

Let us consider a sequence $(a_1,\ldots,a_n)$ of elements of the set
$\{1,\ldots,n\}$, i.e. a path in the Hasse diagram. We construct a sequence of ascending sets
$\varnothing=I_0\subset I_1\subset\ldots\subset I_n=\{1,\ldots,n\}$ such
that
$I_i=I_{i-1}\cup a_i$. Then for each $I_i$, $i=1,\ldots, n-1$, we can rewrite
\eqref{formalBETHE} as
\begin{equation}\label{formalBETHE2}
-1=\frac{\bar{Q}_{I _{i-1}}(\specbaz_k^{I_i}-\half)}{\bar{Q}_{I_{i-1} }(\specbaz_k^{I _i}+\half)}\,
\frac{\bar{Q}_{I_i}(\specbaz_k^{I_i}+1)}{\bar{Q}_{I_i}(\specbaz_k^{I_i}-1)}\,
\frac{\bar{Q}_{I_{i+1}}(\specbaz_k^{I_i}-\half)}{\bar{Q}_{I_{i+1}}(\specbaz_k^{I_i}+\half)}\,.
\end{equation}
Substituting the explicit form of the Q-functions as given in \eqref{Q0}, \eqref{Qn} and \eqref{Q-eigen} yields the Bethe equations
\begin{equation*}\label{BE3}
\left(\frac{z_l^{I_{n-1}}+\sfrac{1}{2}}{z_l^{I_{n-1}}-\sfrac{1}{2}}\right)^L=e^{i(\Phi_{a_{n-1}}-\Phi_{a_{n}})}\prod_k\frac{z_l^{I_{n-1}}-z_k^{I_{n-2}}-\sfrac{1}{2}}{z_l^{I_{a-{n-1}}}-z_k^{I_{n-2}}+\sfrac{1}{2}}\prod_{k\neq l}\frac{z_l^{I_{n-1}}-z_k^{I_{n-1}}+1}{z_l^{I_{n-1}}-z_k^{I_{n-1}}-1}\,.
\end{equation*}
\begin{equation*}
1=e^{i(\Phi_{a_{i}}-\Phi_{a_{i+1}})}\prod_k\frac{z_l^{I_i}-z_k^{I_{i-1}}-\sfrac{1}{2}}{z_l^{I_i}-z_k^{I_{i-1}}+\sfrac{1}{2}}\prod_{k\neq l}\frac{z_l^{I_i}-z_k^{I_i}+1}{z_l^{I_i}-z_k^{I_i}-1}\prod_k\frac{z_l^{I_i}-z^{I_{i+1}}-\sfrac{1}{2}}{z_k^{I_i}-z^{I_{i+1}}+\sfrac{1}{2}}\,,
\end{equation*}
\begin{equation}\label{BE1}
1=e^{i(\Phi_{a_{1}}-\Phi_{a_2})}\prod_{k\neq l}\frac{z_l^{I_1}-z_k^{I_1}+1}{z_l^{I_1}-z_k^{I_1}-1}\prod_k\frac{z_l^{I_1}-z^{I_{2}}-\sfrac{1}{2}}{z_k^{I_1}-z^{I_{2}}+\sfrac{1}{2}}\,.
\end{equation}
Here, as expected for the representation $\Lambda=(1,0,\ldots,0)$ at each site of the spin chain, only one driving term appears on the LHS of the equations, cf. \eqref{nbaes}. 
There are $n!$ alternative forms of the above Bethe ansatz equations,
corresponding to $n!$ permutations of the elements of the set $I$, which in turn are associated with the $n!$ different bottom-to-top paths on the Hasse diagram. We like to stress that in contrast to the analytic Bethe ansatz the analytic structure of \eqref{Q-eigen} 
is rigorously derived without any assumptions. 
It follows from the
explicit construction of the Q-operators in Section~\ref{sec:traces}. 

To conclude we give the
expression for the eigenvalues of the Hamiltonian \eqref{sln-ham}, cf.~\eqref{abaenergy}.
It only involves the roots $z^{I_{n-1}}$ on the last-level 
\begin{equation}\label{energy.formula}
E=2\sum_{k=1}^{\magm_{I_{n-1}}}
\frac{1}{\frac{1}{4}-\left(z_{k}^{I_{n-1}}\right)^2}\,,
\end{equation}
where $\magn_{I_{n-1}}$ is the number of roots of the eigenvalue
$\bar Q_{I_{n-1}}(z)$, 
which according to \eqref{Q-eigen} is equal to 
\beq
\magm_{I_{n-1}}=\magm_{a_1}+\magm_{a_2}+\cdots+\magm_{a_{n-1}}=L-\magm_{a_n}.
\eeq
It can be obtained using the Baxter equation following the logic in Section~\ref{sec:obs}, see \cite{Bazhanov2010} as well as \cite{Tarasov1983}. Note however, that here the numeration of the levels has been reversed and the Bethe roots are shifted as already pointed out in Section~\ref{sec:qnut}.  We present an independent derivation that does not rely on the knowledge of the functional form of the transfer matrix with fundamental representation in the auxiliary space at the end of this chapter.

\subsection{Further solutions of the Yang-Baxter equation}
\label{sec:oscpart}
In the previous sections we constructed Q-operators for integrable spin chains with  fundamental representations of  $\gln$ at each site of the quantum space. In the following we extend our analysis to the general case where the representation at each site of the quantum space is given by any $\gln$ irreducible representation. According to the idea presented above  we construct Q-operators as special transfer matrices, where the auxiliary space is taken to be an appropriately chosen representation of the Yangian algebra. The functional relations do not differ significantly from the case discussed previously. In particular as we only change the representation in the quantum space, the construction of the finite-dimensional module following the BGG resolution \eqref{bgg1} remains unchanged. 

While the L-operators $\Rfi$ are sufficient to construct Q-operators for the fundamental representation, the first step in the generalization of \cite{Bazhanov2010} is to find ``R-operators for Q-operators''. In a series of papers \cite{Bazhanov2010a,Bazhanov2010,Frassek2011}
new solutions to the Yang-Baxter equation were derived. They allow
to construct Baxter Q-operators for $\gln$ invariant
spin-chains. These R-operators are of remarkably
compact form and can be written as\,\footnote{For reasons that will become clear in the next section we explicitly
denoted the product by ``$\cdot$''.%
}
\begin{equation}
\mathcal{R}_{I}(z)\,=e^{\bar{\mathbf{a}}_{c}^{\dot{c}}\, J_{\dot{c}}^{c}}\cdot\mathcal{R}_{0,I}(z)\cdot e^{-\mathbf{a}_{\dot{c}}^{c}\, J_{c}^{\dot{c}}}\,,\label{eq:RI}
\end{equation}
with 
\begin{equation}
\mathcal{R}_{0,I}(z)=\,\rho_{I}(z)\,\prod_{k=1}^{\vert\bar{I}\vert}\,\Gamma(z-\sfrac{\vert\bar{I}\vert}{2}-\hat{\ell}_{k}^{\bar{I}}+1)\,.\label{eq:R0}
\end{equation}
These equations require some explanations. As before, the letter $I$ denotes
a subset of the set $\{1,\ldots,n\}$ of cardinality $\vert I\vert$.
The undotted indices take values from the set $I$ and the dotted
ones from its complement $\bar{I}$
\begin{equation}
a,b,c\in I,\quad\quad\dot{a},\dot{b},\dot{c}\in\bar{I},\quad\quad A,B,C\in I\cup\bar{I}\,.\label{eq:sets}
\end{equation}

The R-operators are composed
out of $\vert I\vert\cdot\vert\bar{I}\vert$ families of oscillators
\begin{equation}
[\mathbf{a}_{\dot{b}}^{a},\bar{\mathbf{a}}_{d}^{\dot{c}}]=\delta_{d}^{a}\delta_{\dot{b}}^{\dot{c}}\label{eq:osc_alg}
\end{equation}
and $\gln$generators $J_{B}^{A}$ with 
\begin{equation}
[J_{B}^{A},J_{D}^{C}]=\delta_{B}^{C}J_{D}^{A}-\delta_{D}^{A}J_{B}^{C}\,.\label{eq:gln_alg}
\end{equation}
For convenience we raised the first index of the $\gln$-generators and the second index of the oscillators
\begin{equation}
 J_{AB}\rightarrow J^A_B\,,\quad\quad (\oscb_{AB},\bar\oscb_{AB})\rightarrow(\osca_{A}^B,\bar\osca_{A}^B)\,.
\end{equation}
Furthermore, we renamed the oscillators and introduced the convention that repeated indices imply the corresponding index summation.
The choice of the set $I$ naturally identifies a subalgebra $\gl{I}$
of $\gln$, i.e.~the subalgebra spanned by $J^a_b$, see (\ref{eq:sets}).

The quantities $\hat{\ell}_{k}^{K}$ are important building blocks
for the R-operators above. They are operatorial shifted weights of the subalgebra $\gl{\bar{I}}$. Spelling out
their characteristics is an essential step in the study of the properties
of Q-operators. The labels of $\hat{\ell}_{k}^{K}$ correspond to
a subset $K$ of $\{1,2,\dots,n\}$ and an index $k=1,2,\dots,|K|$.
For $\gln$ there are $n\cdot2^{n-1}$ such $\hat{\ell}_{k}^{K}$.
The set $K$ identifies a natural embedding of $\gl{K}$
in $\gln$. The Casimirs of $\gl{K}$ defined
as 
\begin{equation}
C_{i}^{(K)}\,=\, J_{a_{i}}^{a_{1}}\, J_{a_{1}}^{a_{2}}\,\dots J_{a_{i-1}}^{a_{i}}\quad\quad\text{with }a_{j}\in K\,,\label{eq:Casimir}
\end{equation}
 are symmetric polynomials of $\hat{\ell}_{k}^{K}$ via the following
formula %
\footnote{We refer the reader to \cite{Frassek2011,Zhelobenko,molevbook,Green:1971rp,Okubo1975} for further details.%
} 
\begin{equation}
C_{i}^{(K)}\,=\,\sum_{k\in K}\,\prod_{j\neq k}\,\left(1+\frac{1}{\hat{\ell}_{k}^{K}-\hat{\ell}_{j}^{K}}\right)\,(\hat{\ell}_{k}^{K})^{i}\,.\label{eq:casimirofl}
\end{equation}
In general not all $\hat{\ell}_{k}^{K}$ do commute among themselves.
For a chosen path in the Hasse diagram, cf. Figure~\ref{Hasse23} i.e. a sequence of sets $\mathcal{P}\equiv\emptyset\subset\{a\}\subset\{a,b\}\subset\dots\subset\{1,2,\dots,n\}$
ordered by inclusion, all the $\frac{n(n+1)}{2}$ corresponding $\hat{\ell}_{k}^{K}$
commute among themselves. In particular, for a given irreducible representation
of $\gln$ there exists a basis such that all $\hat{\ell}_{k}^{K}$
corresponding to the chosen path $\mathcal{P}$ act diagonally. This
basis coincides with the Gelfand-Tsetlin basis, see e.g. \cite{MolevM.HazewinkelEd.2006pp.109-170}
for a nice review and collection of references.

The operators $\mathcal{R}_{I}$ are elements of a suitable extension
of the product space of the universal enveloping algebra $U(\gln)$ and the oscillator algebra denoted by $\mathcal{H}^{(I,\bar{I})}$ which we denote in the following by $\mathfrak{A}_{I}$. The normalization
$\rho_{I}$ in \eqref{eq:R0} is not determined by the Yang-Baxter relation and is discussed
in the next sections.

As shown in \cite{Frassek2011,carlophd}, the R-operators above satisfy the Yang Baxter equation\,\footnote{Here we use the notation ${\bf L}_{\Lambda}(z)=R_{\,\ffbox,\Lambda}(z)$, $\Rfi(z)=R_{\,\ffbox,I}(z)$ and $\Rli(z)=R_{\Lambda,I}(z)$, where $I$ denotes the oscillator space and we defined $x=z_1-z_2$ and $y=z_1-z_3$, cf. \eqref{ybe1}.}
\begin{equation}
{\bf L}_\Lambda(z_1-z_2)\,{L}_{I}(z_1-z_3)\cdot\mathcal{R}_{I}(z_2-z_3)=\mathcal{R}_{I}(z_2-z_3)\cdot{L}_{I}(z_1-z_3)\,\mathbf{L}_\Lambda(z_1-z_2)\,.\label{eq:YBE}
\end{equation}
Here ${L}_{I}$ denotes the operator $\mathcal{R}_{I}$ with
fundamental representation of $\gln$ as introduced in \eqref{Lcanon}
\begin{equation}
{L}_{I}(z)=\begin{pmatrix}z\delta_{b}^{a}+H_{b}^{a} & \bar{\mathbf{a}}_{b}^{\dot{a}}\\
-\mathbf{a}_{\dot{b}}^{a} & \delta_{\dot{b}}^{\dot{a}}
\end{pmatrix}\quad\text{for}\quad I=\{1,\ldots,\vert I\vert\}\,,\label{eq:R-fundamental}
\end{equation}
with $H_{b}^{a}=-\bar{\mathbf{a}}_{b}^{\dot{c}}\mathbf{a}_{\dot{c}}^{a}-\half\delta_{\dot{c}}^{\dot{c}}\delta_{b}^{a}$
where the summation over the dotted indices is understood. The operator
$\mathbf{L}_\Lambda$ denotes the well-known Lax matrix as introduced in \eqref{lax}
\begin{equation}
\mathbf{L}_\Lambda(z)=\begin{pmatrix}z\delta_{b}^{a}+J_{b}^{a} & J_{b}^{\dot{a}}\\
J_{\dot{b}}^{a} & z\delta_{\dot{b}}^{\dot{a}}+J_{\dot{b}}^{\dot{a}}
\end{pmatrix}\,.\label{eq:Lax}
\end{equation}
As before, Baxter Q-operators are constructed as regularized traces over the
oscillator space of monodromies built from the operators $\mathcal{R}_{I}$ with representation $\Lambda$ at each site of the quantum space.
Following \cite{Frassek2011}, they are given by 

\begin{equation}
\bar{\mathbf{Q}}_{I}(z)=e^{iz\,\sum_{a\in I}\phi_{a}}\,\widehat{\tr}_{{\mathcal{F}}^{p(n-p)}}\big\{\mathcal{D}_{I}\,\mathcal{R}_{I}(z)\otimes\ldots\otimes\mathcal{R}_{I}(z)\big\}\,.\label{eq:Qop}
\end{equation}
Here the quantum space consists out of $L$ sites and will be denoted
by $\mathcal{V}=\mathcal{V}_{1}\otimes\ldots\otimes\mathcal{V}_{L}$.
In the following each $\mathcal{V}_{i}$ corresponds to the same representation
$\Lambda$ of $\gln$. The regulator $\mathcal{D}_{I}$ in (\ref{eq:Qop})
is defined in \eqref{D-parton} and the normalized trace in \eqref{norm-tr}.

Applying the reasoning as in the previous section we express all Q-operators using only partonic Q-operators
\begin{equation}
\Delta_I(\Phi) \Qop_I(z)=\det || \Qop_{a_i}(z-j+\sfrac{p+1}{2})||_{1\leq i,j\leq |I|}\,,
\end{equation}
where $\Delta_I$ depends only on the twist angles and was defined in \eqref{vandermond}.
As a direct consequence of these relations one can write down the QQ-relations \eqref{qqrels}.
Just as in the earlier case of the fundamental representation in the quantum space we can associate all Q-operators to the vertices of a Hasse diagram, which for $\gln$ takes the form of an $n$-dimensional hypercube. The functional form of Bethe equations is given by \eqref{formalBETHE2}.

So far we did not use any information about the analytic structure of the Q-operators, apart from the fact that there exists at least one root of $\bar{Q}_I(z)$. For compact representations in the quantum space the eigenvalues of $\Qop_I(z)$ can be written as
\beq
\label{anal}
\bar{Q}_I(z)\,=\,e^{\,iz\,\left(\sum_{i\in\,I}\,\Phi_i\right)}\,\,\left( \prod_{a=1}^q\,\Gamma(z-\sfrac{q}{2}-\ell_a+1)\right)^L\,\prod_{k=1}^{m_I}\,\left(z\,-\,z^{I}_k \right)\,,
\eeq 
where we set the normalization of the R-operators to $\rho_I(z)=1$.
Substituting \eqref{anal} into \eqref{formalBETHE2} we obtain the Bethe equations
\begin{eqnarray}\label{betheequations}
\left( \frac{z^{I_i}_{l}-\sfrac{n-i-1}{2}-\ell_{a_{n-i+1}}-\sfrac{1}{2}}{z^{I_i}_{l}-\sfrac{n-i-1}{2}-\ell_{a_{n-i}}+\sfrac{1}{2}}\right)^L=&&\\\nonumber
&& \hspace{-5cm}e^{i(\Phi_{a_{i}}-\Phi_{a_{i+1}})}\prod_k\frac{z_l^{I_i}-z_k^{I_{i-1}}-\sfrac{1}{2}}{z_l^{I_i}-z_k^{I_{i-1}}+\sfrac{1}{2}}\prod_{k\neq l}\frac{z_l^{I_i}-z_k^{I_i}+1}{z_l^{I_i}-z_k^{I_i}-1}\prod_k\frac{z_l^{I_i}-z^{I_{i+1}}-\sfrac{1}{2}}{z_k^{I_i}-z^{I_{i+1}}+\sfrac{1}{2}}\,,
\end{eqnarray}
whith eigenvalues the shifted weights 
\begin{equation}\label{shiftwei}
 \ell_a=\lambda^a-a+1\,.
\end{equation} 
 For compact representations, due to Weyl permutation symmetry, we have that for fixed $|I|$ all $\Qop_I(z)$ are isospectral. This means that all $n!$ systems of Bethe equations have exactly the same form. This is not true for non-compact representations -- Weyl symmetry is broken, a fact which leads to different analytic properties of the Q-operators. For representations with a lowest/highest weigh state, however, there still exist paths on the Hasse diagram for which all $\Qop_I(z)$ connected by the path can be normalized to be polynomial. In that case the formula \eqref{anal} holds true and we are again able to rewrite \eqref{formalBETHE2} into the form \eqref{betheequations}, see also \eqref{nbaes}. 
  \section{From Q-operators to local charges}\label{sec:qtoh}

In this section we show how local charges can be extracted
directly from the Q-operators built in Section~\ref{sec:Q-opgln}. Here no reference
to the transfer matrix as discussed in Section~\ref{spinchainham} is made. The procedure discussed here avoids the
notoriously complicated construction of the transfer matrix with equal
representation in quantum and auxiliary space. In Section~\ref{sec:Alternative-presentation-of} we introduce an
opposite product on the auxiliary space and obtain alternative presentations
of the degenerate solutions used for the construction of Q-operators,
hereafter referred to as R-operators $\mathcal{R}$. Section~\ref{sec:Projection-property-of}
is dedicated to the extremely important projection properties of the
degenerate Lax operators. Their alternative presentation obtained
in Section~\ref{sec:Alternative-presentation-of} is essential in
order to fully exploit these properties. After developing these techniques
we introduce a convenient diagrammatic language for R-operators $\mathcal{R}$
which extends to Q-operators in Section~\ref{sec:Diagrammatics-and-local}.
It further concerns the derivation of the shift operator and the Hamiltonian
from Q-operators. Equation (\ref{eq:haction}) is one of the main
results of this section. It defines the Hamiltonian density in terms
of the R-operators $\mathcal{R}$ for Q-operators introduced in
Section~\ref{sec:Q-opgln}. Furthermore,
we provide specific examples in Appendix~\ref{sec:The-spin-half} and \ref{fundhamq}.

\subsection{Alternative presentation of R-operators\label{sec:Alternative-presentation-of}}

The solution (\ref{eq:RI}) to the Yang-Baxter equation (\ref{eq:YBE})
is presented as a normal ordered expression in the oscillators $(\bar{\mathbf{a}}_{c}^{\dot{c}},\mathbf{a}_{\dot{c}}^{c})$.
For reasons that will become clear in the next section we are also
interested in its expression which is anti-normal ordered in the oscillators
of the auxiliary space. The anti-normal ordered form of the R-operators $\mathcal{R}$
can be obtained either from the Yang-Baxter equation or directly by
reordering the oscillators in (\ref{eq:RI}). As we will see the approach
from the Yang-Baxter equation will be very powerful to obtain the
desired expressions. However, it is not possible to fix the relative
normalization by this method.

\subsubsection{Yang-Baxter approach\label{sub:Yang-Baxter-approach}}

To derive the expression for the anti-normal ordered R-operators $\mathcal{R}$
directly from the Yang-Baxter equation it is convenient
to introduce an opposite product on $\mathfrak{A}_{I}$. Let $\mathcal{O}\in\mathfrak{A}_{I}$
be written as 
\begin{equation}
\mathcal{O}=\sum_{k}a(k)\otimes b(k)\,,\label{eq:opdecomp}
\end{equation}
with $a(k)\in U(\gln)$ and $b(k)\in\mathcal{H}^{(I,\bar{I})}$.
Given two elements of $\mathfrak{A}_{I}$ the product used in (\ref{eq:YBE})
is defined as 
\begin{equation}
\mathcal{O}_{1}\cdot\mathcal{O}_{2}=\sum_{k,l}a_{1}(k)\, a_{2}(l)\otimes b_{1}(k)\, b_{2}(l)\,.\label{eq:prod}
\end{equation}
We now define the opposite product $\circ$ as 
\begin{equation}
\mathcal{O}_{1}\circ\mathcal{O}_{2}=\sum_{k,l}a_{1}(k)\, a_{2}(l)\otimes b_{2}(l)\, b_{1}(k)\,.\label{eq:prod_op}
\end{equation}
This product is associative. The Yang-Baxter
equation (\ref{eq:YBE}) can then be written as
\begin{equation}
\mathcal{L}(z_{1})\,\mathcal{R}_{I}(z_{2}-z_{1})\circ\mathbf{L}_{I}(z_{2})=\mathbf{L}_{I}(z_{2})\circ\mathcal{R}_{I}(z_{2}-z_{1})\,\mathcal{L}(z_{1})\,.\label{eq:YBE_opposite}
\end{equation}
We like to stress that this is exactly the same equation as (\ref{eq:YBE})
only rewritten in terms of the opposite product. Now, in analogy to
\cite{Frassek2011} we substitute the ansatz 
\begin{equation}
\mathcal{R}_{I}(z)\,=\, e^{\bar{\mathbf{a}}_{c}^{\dot{c}}\, J_{\dot{c}}^{c}}\,\circ\,\tilde{\mathcal{R}}_{0,I}(z)\,\circ\, e^{-\mathbf{a}_{\dot{c}}^{c}\, J_{c}^{\dot{c}}}\label{eq:RI_anti}
\end{equation}
into (\ref{eq:YBE_opposite}) and obtain four defining
relations for $\tilde{\mathcal{R}}_{0,I}$. As there is some redundancy
in these equations we only present one of them 
\begin{equation}
\tilde{\mathcal{R}}_{0,I}(z)\left((z+\frac{\bar{I}}{2})J_{b}^{\dot{a}}-J_{b}^{c}J_{c}^{\dot{a}}\right)=J_{b}^{\dot{a}}\,\tilde{\mathcal{R}}_{0,I}(z)\,.\label{eq:antinormalybe}
\end{equation}
Comparing this equation to \cite{Frassek2011} it is easy to recognize
that $\tilde{\mathcal{R}}_{0,I}(z)$ satisfies the same defining relation
as $\mathcal{R}_{0,\bar{I}}^{-1}\,(z+\sfrac{n}{2})$. We conclude
that 

\begin{equation}
\tilde{\mathcal{R}}_{0,I}(z)=\,\tilde{\rho_{I}}(z)\,\prod_{k=1}^{\text{\ensuremath{\vert}}I\vert}\,\frac{1}{\Gamma(z+\sfrac{\vert\bar{I}\vert}{2}-\hat{\ell}_{k}^{I}+1)}\,.\label{Rt0-2}
\end{equation}
The ratio of $\rho_{I}$ and $\tilde{\rho}_{I}$ entering (\ref{eq:R0})
and (\ref{Rt0-2}) can be determined by requiring that (\ref{eq:RI})
and (\ref{eq:RI_anti}) are the same operators. It is investigated
in the next subsection.

\subsubsection{Direct approach\label{sub:Direct-approach}}

The R-operators $\mathcal{R}$ (\ref{eq:RI}) and (\ref{eq:RI_anti})
can be written as 
\begin{equation}\label{Rliexpand}
 \mathcal{R}_{I}(z)=\sum_{n,m=0}^{\infty}\frac{(-1)^{m}}{n!\, m!}\,\bar{\mathbf{a}}_{a_{1}}^{\dot{a}_{1}}\cdots\bar{\mathbf{a}}_{a_{n}}^{\dot{a}_{n}}\,\mathbf{a}_{\dot{b}_{1}}^{b_{1}}\cdots\mathbf{a}_{\dot{b}_{m}}^{b_{m}}\, J_{\dot{a}_{1}}^{a_{1}}\cdots J_{\dot{a}_{n}}^{a_{n}}\,\mathcal{R}_{0,I}(z)\, J_{b_{1}}^{\dot{b}_{1}}\cdots J_{b_{m}}^{\dot{b}_{m}}\,,
\end{equation} 
\begin{equation}
 \mathcal{R}_{I}(z) = \sum_{n,m=0}^{\infty}\frac{(-1)^{m}}{n!\, m!}\,\mathbf{a}_{\dot{b}_{1}}^{b_{1}}\cdots\mathbf{a}_{\dot{b}_{m}}^{b_{m}}\,\bar{\mathbf{a}}_{a_{1}}^{\dot{a}_{1}}\cdots\bar{\mathbf{a}}_{a_{n}}^{\dot{a}_{n}}\, J_{\dot{a}_{1}}^{a_{1}}\cdots J_{\dot{a}_{n}}^{a_{n}}\,\tilde{\mathcal{R}}_{0,I}(z)\, J_{b_{1}}^{\dot{b}_{1}}\cdots J_{b_{m}}^{\dot{b}_{m}}\,.\label{Rtliexpand}
\end{equation} 
Here we expanded the exponentials using the definition of the products
in (\ref{eq:prod}) and (\ref{eq:prod_op}), respectively. To obtain
the relation between $\mathcal{R}_{0,I}$ and $\tilde{\mathcal{R}}_{0,I}$
we have to reorder the oscillators in one of the two expressions.
We find that for each pair $\bar{\mathbf{a}}_{c}^{\dot{c}}$, $\mathbf{a}_{\dot{c}}^{c}$
that is reordered in (\ref{Rliexpand}) $\mathcal{R}_{0,I}$ is {}``conjugated''
by the corresponding $\mathfrak{gl}(\{c,\dot{c}\})$ generators as
\begin{equation}
\mathcal{R}_{0,I}\longrightarrow\sum_{k=0}^{\infty}\frac{1}{k!}\,(J_{\dot{c}}^{c})^{k}\,\mathcal{R}_{0,I}(z)\,(J_{c}^{\dot{c}})^{k}.\label{eq:R_conjugation}
\end{equation}
This relation is obtained using (\ref{eq:reorder2}) and does not
rely on the precise form of $\mathcal{R}_{0,I}$. After subsequent
conjugation of $\mathcal{R}_{0,I}$ with all $\vert I\vert\cdot\vert\bar{I}\vert$
$\mathfrak{gl}(\{c,\dot{c}\})$ subalgebra generators of $\gln$
one obtains an expression for $\tilde{\mathcal{R}}_{0,I}$:
\begin{equation}
\tilde{\mathcal{R}}_{0,I}(z)=\sum_{\{k_{c\dot{c}}\}=0}^{\infty}\prod_{c\in I,\dot{c}\in\bar{I}}\frac{1}{\sqrt{k_{c\dot{c}}!}}\,(J_{\dot{c}}^{c})^{k_{c\dot{c}}}\,\mathcal{R}_{0,I}(z)\,\prod_{c\in I,\dot{c}\in\bar{I}}\frac{1}{\sqrt{k_{c\dot{c}}!}}(J_{c}^{\dot{c}})^{k_{c\dot{c}}}.\label{eq:R_conjugation-1}
\end{equation}
This fixes the ratio of the prefactors $\rho$ and $\tilde{\rho}$
appearing in (\ref{eq:R0}) and (\ref{Rt0-2}). Naively, (\ref{eq:R_conjugation-1})
appears rather different from (\ref{Rt0-2}). However, they must coincide
since they satisfy the same defining relations. This is explicitly
demonstrated for the case of $\gl2$ and further discussed in Section~\ref{sec:Projection-property-of}.

\subsection{Projection properties of R-operators}\label{sec:Projection-property-of}

The construction of local charges in the conventional QISM relies
on the R-matrix for which the auxiliary
space is the same as the quantum space at each site. It requires the
existence of a special point $z_{*}$ where the R-matrix reduces to
the permutation operator 
\begin{equation}
\mathbf{R}(z_{*})=\mathbf{P}\,,\label{eq:regularity}
\end{equation}
see Chapter~\ref{ch:ybe}. This property is often referred to as regularity condition. The construction
of local charges presented here bypasses the use of the 
R-matrix and is based on remarkable properties of the R-operators $\mathcal{R}$
in (\ref{eq:RI}) for special values of the spectral parameter. For
the example of the fundamental representation in the quantum space
we find that the operator \textbf{$\mathbf{L}_{I}$ }given in (\ref{eq:R-fundamental})
degenerates at \textit{two} special points. Using \textit{both} products
introduced in the previous section we find that 
\begin{equation}
\mathbf{L}_{I}\left(+\sfrac{\vert\bar{I}\vert}{2}\right)=\begin{pmatrix}\bar{\mathbf{a}}_{b}^{\dot{c}}\\
\delta_{\dot{b}}^{\dot{c}}
\end{pmatrix}\cdot\begin{pmatrix}-\mathbf{a}_{\dot{c}}^{a} & \delta_{\dot{c}}^{\dot{a}}\end{pmatrix},\quad\quad\mathbf{L}_{I}\left(-\sfrac{\vert\bar{I}\vert}{2}\right)=\begin{pmatrix}\bar{\mathbf{a}}_{b}^{\dot{c}}\\
\delta_{\dot{b}}^{\dot{c}}
\end{pmatrix}\circ\begin{pmatrix}-\mathbf{a}_{\dot{c}}^{a} & \delta_{\dot{c}}^{\dot{a}}\end{pmatrix}\,.\label{eq:degenerate_fund}
\end{equation}
Interestingly, properties similar to (\ref{eq:degenerate_fund}) appear quite naturally in the derivation of the Baxter equation, see e.g.~\cite{Pronko:2000,Pronko2000}. This in turn is strictly connected to Baxter's original idea \cite{Baxter:1972hz}. The degeneration can be understood from the spectral parameter
dependent part of the R-operators $\mathcal{R}$, compare to (\ref{eq:RI}),
(\ref{eq:RI_anti}). For the fundamental representation, it originates
from a reduction of the rank of $\mathcal{R}_{0,I}$ and $\tilde{\mathcal{R}}_{0,I}$
at the special points $\hat{z}=+\sfrac{\vert\bar{I}\vert}{2}$ and
$\check{z}=-\sfrac{\vert\bar{I}\vert}{2}$, respectively. Of distinguished
importance are $\mathbf{L}_{I}$-operators with $\vert\bar{I}\vert=1$.
In this case the rank of the oscillator independent part reduces to
$1$. 

Relations of the type (\ref{eq:degenerate_fund}) hold for any highest/lowest
weight representation of $\gln$. Their precise form can
be obtained via a careful analysis of the spectrum of the shifted
weight operators $\hat{\ell}_{i}^{K}$ entering $\mathcal{R}_{I}$.
However, the analysis is technically involved. In the following we restrict
to representations corresponding to rectangular Young diagrams and
their infinite dimensional generalization. 

A rectangular Young diagram is labeled by two parameters $(s,a)$
with $s,a\in\mathbb{N}$ according to 

\begin{equation}
\psset{xunit=.5pt,yunit=.5pt,runit=.5pt}
\begin{pspicture}[shift=-47](94.35027313,76.98535919)
{
\newrgbcolor{curcolor}{0 0 0}
\pscustom[linewidth=1,linecolor=curcolor]
{
\newpath
\moveto(33.85028,76.48535657)
\lineto(53.85028,76.48535657)
\lineto(53.85028,56.48535657)
\lineto(33.85028,56.48535657)
\closepath
}
}
{
\newrgbcolor{curcolor}{0 0 0}
\pscustom[linewidth=1,linecolor=curcolor]
{
\newpath
\moveto(53.85028,76.48535657)
\lineto(73.85028,76.48535657)
\lineto(73.85028,56.48535657)
\lineto(53.85028,56.48535657)
\closepath
}
}
{
\newrgbcolor{curcolor}{0 0 0}
\pscustom[linewidth=1,linecolor=curcolor]
{
\newpath
\moveto(73.85028,76.48535657)
\lineto(93.85028,76.48535657)
\lineto(93.85028,56.48535657)
\lineto(73.85028,56.48535657)
\closepath
}
}
{
\newrgbcolor{curcolor}{0 0 0}
\pscustom[linewidth=1,linecolor=curcolor]
{
\newpath
\moveto(33.85028,56.48535657)
\lineto(53.85028,56.48535657)
\lineto(53.85028,36.48535657)
\lineto(33.85028,36.48535657)
\closepath
}
}
{
\newrgbcolor{curcolor}{0 0 0}
\pscustom[linewidth=1,linecolor=curcolor]
{
\newpath
\moveto(53.85028,56.48535657)
\lineto(73.85028,56.48535657)
\lineto(73.85028,36.48535657)
\lineto(53.85028,36.48535657)
\closepath
}
}
{
\newrgbcolor{curcolor}{0 0 0}
\pscustom[linewidth=1,linecolor=curcolor]
{
\newpath
\moveto(73.85028,56.48535657)
\lineto(93.85028,56.48535657)
\lineto(93.85028,36.48535657)
\lineto(73.85028,36.48535657)
\closepath
}
}
{
\newrgbcolor{curcolor}{0 0 0}
\pscustom[linestyle=none,fillstyle=solid,fillcolor=curcolor]
{
\newpath
\moveto(21.80596548,61.53359502)
\curveto(21.80596548,59.22899059)(20.65366326,57.97193362)(17.24913399,56.61012191)
\curveto(17.09200187,56.55774454)(16.98724712,56.40061242)(16.98724712,56.19110293)
\curveto(16.98724712,55.92921606)(17.0396245,55.82446131)(17.30151136,55.71970657)
\curveto(21.80596548,53.99125324)(21.80596548,52.21042255)(21.80596548,50.3248371)
\lineto(21.80596548,41.10641938)
\curveto(21.85834285,40.21600403)(22.32973921,38.95894707)(23.79630567,37.96377697)
\curveto(25.47238162,36.86385213)(28.14362767,36.18294627)(28.45789191,36.18294627)
\curveto(28.77215615,36.18294627)(28.87691089,36.28770102)(28.87691089,36.60196526)
\curveto(28.87691089,36.9162295)(28.82453352,36.9162295)(28.30075979,37.07336162)
\curveto(27.35796706,37.44000324)(24.73909839,38.48755071)(24.21532466,40.53026827)
\curveto(24.11056991,40.94928726)(24.11056991,41.00166463)(24.11056991,42.25872159)
\lineto(24.11056991,50.42959185)
\curveto(24.11056991,52.15804517)(24.11056991,54.61978172)(18.66332307,56.19110293)
\curveto(24.11056991,57.86717888)(24.11056991,60.11940594)(24.11056991,62.00499138)
\lineto(24.11056991,70.17586164)
\curveto(24.11056991,72.00906971)(24.11056991,73.73752303)(28.6674014,75.41359898)
\curveto(28.87691089,75.46597635)(28.87691089,75.72786322)(28.87691089,75.78024059)
\curveto(28.87691089,76.09450483)(28.77215615,76.25163695)(28.45789191,76.25163695)
\curveto(28.19600504,76.25163695)(23.5344188,75.15171211)(22.32973921,72.95186243)
\curveto(21.80596548,71.95669233)(21.80596548,71.64242809)(21.80596548,70.22823901)
\closepath
\moveto(21.80596548,61.53359502)
}
}
{
\newrgbcolor{curcolor}{0 0 0}
\pscustom[linestyle=none,fillstyle=solid,fillcolor=curcolor]
{
\newpath
\moveto(56.05446011,20.85471568)
\curveto(57.69511365,20.85471568)(60.74204167,20.85471568)(63.32021153,14.44835421)
\curveto(63.55459061,13.82334334)(63.55459061,13.74521698)(64.02334876,13.74521698)
\curveto(64.17960148,13.74521698)(64.49210692,13.74521698)(64.57023328,14.05772241)
\curveto(67.2265295,20.85471568)(69.96095208,20.85471568)(72.77350101,20.85471568)
\lineto(86.52374026,20.85471568)
\curveto(89.25816284,20.93284204)(90.74256367,23.1203801)(91.28944819,23.97977005)
\curveto(92.77384901,26.08918176)(93.9457444,30.30800516)(93.9457444,30.85488968)
\curveto(93.9457444,31.32364784)(93.63323897,31.47990055)(93.24260717,31.47990055)
\curveto(92.77384901,31.47990055)(92.69572265,31.24552148)(92.6175963,30.85488968)
\curveto(90.1175528,24.29227549)(87.6175093,24.29227549)(84.80496036,24.29227549)
\lineto(72.6172483,24.29227549)
\curveto(69.88282572,24.29227549)(66.52339226,24.29227549)(64.02334876,16.24526048)
\curveto(61.36705255,24.29227549)(58.47637725,24.29227549)(55.35132287,24.29227549)
\lineto(43.16361081,24.29227549)
\curveto(40.58544095,24.29227549)(37.92914473,24.29227549)(35.50722759,30.69863696)
\curveto(35.27287351,31.24552148)(35.27287351,31.47990055)(34.80409036,31.47990055)
\curveto(34.41343356,31.47990055)(34.17907948,31.32364784)(34.17907948,30.85488968)
\curveto(34.17907948,30.77676332)(35.42910123,24.21414913)(38.71040833,21.87035835)
\curveto(40.1166828,20.85471568)(41.21045183,20.85471568)(43.08548445,20.85471568)
\closepath
\moveto(56.05446011,20.85471568)
}
}
{
\newrgbcolor{curcolor}{0 0 0}
\pscustom[linestyle=none,fillstyle=solid,fillcolor=curcolor]
{
\newpath
\moveto(6.62500367,59.01786041)
\curveto(6.25000367,59.76786041)(5.68750367,60.29911041)(4.78125367,60.29911041)
\curveto(2.46875367,60.29911041)(0.00000367,57.36161041)(0.00000367,54.45536041)
\curveto(0.00000367,52.58036041)(1.09375367,51.26786041)(2.62500367,51.26786041)
\curveto(3.03125367,51.26786041)(4.03125367,51.36161041)(5.21875367,52.76786041)
\curveto(5.37500367,51.92411041)(6.09375367,51.26786041)(7.03125367,51.26786041)
\curveto(7.75000367,51.26786041)(8.18750367,51.73661041)(8.53125367,52.36161041)
\curveto(8.84375367,53.08036041)(9.12500367,54.29911041)(9.12500367,54.33036041)
\curveto(9.12500367,54.54911041)(8.93750367,54.54911041)(8.87500367,54.54911041)
\curveto(8.68750367,54.54911041)(8.65625367,54.45536041)(8.59375367,54.17411041)
\curveto(8.25000367,52.89286041)(7.90625367,51.70536041)(7.09375367,51.70536041)
\curveto(6.53125367,51.70536041)(6.50000367,52.23661041)(6.50000367,52.61161041)
\curveto(6.50000367,53.04911041)(6.53125367,53.23661041)(6.75000367,54.11161041)
\curveto(6.96875367,54.92411041)(7.00000367,55.14286041)(7.18750367,55.89286041)
\lineto(7.90625367,58.67411041)
\curveto(8.03125367,59.23661041)(8.03125367,59.26786041)(8.03125367,59.36161041)
\curveto(8.03125367,59.70536041)(7.81250367,59.89286041)(7.46875367,59.89286041)
\curveto(6.96875367,59.89286041)(6.68750367,59.45536041)(6.62500367,59.01786041)
\closepath
\moveto(5.34375367,53.86161041)
\curveto(5.21875367,53.48661041)(5.21875367,53.45536041)(4.93750367,53.11161041)
\curveto(4.06250367,52.01786041)(3.25000367,51.70536041)(2.68750367,51.70536041)
\curveto(1.68750367,51.70536041)(1.40625367,52.79911041)(1.40625367,53.58036041)
\curveto(1.40625367,54.58036041)(2.03125367,57.01786041)(2.50000367,57.95536041)
\curveto(3.12500367,59.11161041)(4.00000367,59.86161041)(4.81250367,59.86161041)
\curveto(6.09375367,59.86161041)(6.37500367,58.23661041)(6.37500367,58.11161041)
\curveto(6.37500367,57.98661041)(6.34375367,57.86161041)(6.31250367,57.76786041)
\closepath
\moveto(5.34375367,53.86161041)
}
}
{
\newrgbcolor{curcolor}{0 0 0}
\pscustom[linestyle=none,fillstyle=solid,fillcolor=curcolor]
{
\newpath
\moveto(66.90625767,7.68750041)
\curveto(66.37500767,7.65625041)(65.96875767,7.21875041)(65.96875767,6.78125041)
\curveto(65.96875767,6.50000041)(66.15625767,6.18750041)(66.59375767,6.18750041)
\curveto(67.03125767,6.18750041)(67.50000767,6.53125041)(67.50000767,7.31250041)
\curveto(67.50000767,8.21875041)(66.65625767,9.03125041)(65.12500767,9.03125041)
\curveto(62.50000767,9.03125041)(61.75000767,7.00000041)(61.75000767,6.12500041)
\curveto(61.75000767,4.56250041)(63.21875767,4.28125041)(63.81250767,4.15625041)
\curveto(64.84375767,3.93750041)(65.87500767,3.71875041)(65.87500767,2.62500041)
\curveto(65.87500767,2.12500041)(65.43750767,0.43750041)(63.03125767,0.43750041)
\curveto(62.75000767,0.43750041)(61.21875767,0.43750041)(60.75000767,1.50000041)
\curveto(61.53125767,1.40625041)(62.03125767,2.00000041)(62.03125767,2.56250041)
\curveto(62.03125767,3.00000041)(61.68750767,3.25000041)(61.28125767,3.25000041)
\curveto(60.75000767,3.25000041)(60.15625767,2.84375041)(60.15625767,1.93750041)
\curveto(60.15625767,0.81250041)(61.31250767,0.00000041)(63.00000767,0.00000041)
\curveto(66.25000767,0.00000041)(67.03125767,2.40625041)(67.03125767,3.31250041)
\curveto(67.03125767,4.03125041)(66.65625767,4.53125041)(66.40625767,4.75000041)
\curveto(65.87500767,5.31250041)(65.28125767,5.43750041)(64.40625767,5.59375041)
\curveto(63.68750767,5.75000041)(62.90625767,5.90625041)(62.90625767,6.81250041)
\curveto(62.90625767,7.37500041)(63.37500767,8.59375041)(65.12500767,8.59375041)
\curveto(65.62500767,8.59375041)(66.62500767,8.43750041)(66.90625767,7.68750041)
\closepath
\moveto(66.90625767,7.68750041)
}
}
\end{pspicture}\,.\label{eq:tableaux}
\end{equation}
For representations of this type, also known as Kirillov-Reshetikhin modules \cite{Kirillov1990}, there exist two values of the spectral
parameter such that $\mathcal{R}_{0,I}$ and $\tilde{\mathcal{R}}_{0,I}$
respectively are projectors on a highest weight state\,%
\footnote{The notion of highest weight state depends on the choice of the raising
generators. For rectangular representations $a!(n-a)!$ such choices
correspond to the same highest weight state.%
} for $\vert I\vert=n-a$. The number of highest/lowest weight states
for such representations is $\binom{n}{a}$ and exactly coincides
with the number of operators $\mathcal{R}_{I}$ with $\vert I\vert=n-a$.
Each $\mathcal{R}_{0,I}$ and $\tilde{\mathcal{R}}_{0,I}$ of cardinality
$n-a$ projects on a different highest weight state depending on the
elements in $I$. With an appropriate normalization discussed in Section~\ref{sub:On-the-normalization}
we find for $\vert I\vert=n-a$ that 
\begin{equation}
\mathcal{R}_{0,I}(\hat{z})=\vert hws\rangle\langle hws\vert\quad\text{and\quad}\tilde{\mathcal{R}}_{0,I}(\check{z})=\vert hws\rangle\langle hws\vert\,.\label{eq:hwsProjector-1}
\end{equation}
 As a direct consequence of (\ref{eq:hwsProjector-1}) we obtain that
the $\mathcal{R}$-operator at the special points $\hat{z}$ and $\check{z}$
can be written as 
\begin{equation}
\mathcal{R}_{I}(\hat{z})=e^{\bar{\mathbf{a}}_{c}^{\dot{c}}J_{\dot{c}}^{c}}\cdot\vert hws\rangle\langle hws\vert\cdot e^{-\mathbf{a}_{\dot{c}}^{c}J_{c}^{\dot{c}}}\quad\text{and\quad}\mathcal{R}_{I}(\check{z})=e^{\bar{\mathbf{a}}_{c}^{\dot{c}}J_{\dot{c}}^{c}}\circ\vert hws\rangle\langle hws\vert\circ e^{-\mathbf{a}_{\dot{c}}^{c}J_{c}^{\dot{c}}}.\label{eq:RProjector}
\end{equation}
As we will see, these properties carry over to non-compact representations
with highest weight that fulfil a generalized rectangularity
condition which we encountered previously in Chapter~\ref{ch:2dlattice} when studying the inverse of the Lax operators ${\bf L}_\Lambda$. 
However, as discussed in \cite{Frassek2013}, not all R-operators $\mathcal{R}$ of a certain cardinality
$\vert I\vert$ share the projection property. It will become clear in Section~\ref{sec:Diagrammatics-and-local}
that as a consequence of (\ref{eq:RProjector}) the Q-operators at
the special points $\hat{z}$ and $\check{z}$ are related by the
shift operator \eqref{shiftop}, see (\ref{eq:UQgQ-1}).

\subsubsection{Reduction\label{sub:Reduction}}

It emerged in the discussion of the fundamental representation that
at special values of the spectral parameter the operators $\mathcal{R}_{0,I}$
and $\tilde{\mathcal{R}}_{0,I}$ become projectors on a certain subspace.
In the following we show when and how this happens. The analysis is connected to the pattern of the decomposition of the $\gln$
representation at a site 
\begin{equation}
\Lambda\rightarrow\oplus_{\alpha}\, m_{\alpha}\,(\Lambda_{\alpha}^{I},\Lambda_{\alpha}^{\bar{I}})
\end{equation}
under the restriction $\gln\downarrow\gl{I}\oplus\gl{\bar I}$. Here $m_\alpha$ denotes the multiplicity of each decomposition $(\Lambda_{\alpha}^{I},\Lambda_{\alpha}^{\bar{I}})$.
We are specifically interested in representations $\Lambda$ and a
set $I$ such that $\hat{\ell}_{k}^{\bar{I}}$ are bounded from above
and $\hat{\ell}_{k}^{I}$ are bounded from below\,%
\footnote{For any finite-dimensional representation this is true for any set
$I$.%
}. The bound is saturated for all $k$ by the subspace $(\Lambda_{\alpha_{0}}^{I},\Lambda_{\alpha_{0}}^{\bar{I}})$
of $\Lambda$ annihilated by the action of generators $J_{a}^{\dot{a}}$

\begin{equation}
J_{a}^{\dot{a}}\vert\Lambda_{\alpha_{0}}^{I},\Lambda_{\alpha_{0}}^{\bar{I}}\rangle=0\,,
\end{equation}
where the indices take values according to (\ref{eq:sets}). The subspace
$(\Lambda_{\alpha_{0}}^{I},\Lambda_{\alpha_{0}}^{\bar{I}})$ is nothing
but the $\gl{I}\oplus\gl{\bar I}$ irreducible
representation generated by the action of \textbf{$J_{b}^{a}$} and
$J_{\dot{b}}^{\dot{a}}$ on the $\gln$ highest weight
state\,%
\footnote{We chose the raising generators entering the $\gln$ highest
weight condition to be $J_{a}^{\dot{a}}$. %
} $\vert hws\rangle$ labelled by $\Lambda$. Moreover, the eigenvalues of any fixed $\hat{\ell}_{k}^{K}$
are integer spaced. The fact that the operators $\mathcal{R}_{0,I}$
and $\tilde{\mathcal{R}}_{0,I}$ become projectors on this subspace
for special values of the spectral parameter is an immediate consequence
of the properties of the operators $\hat{\ell}_{k}^{K}$ together
with the pole structure of the gamma function.

The class of representations considered at the end of the previous
section, here referred to as generalized rectangular representations,
have a number of remarkable features. In particular, for generalized rectangular representations there is at least one set $I$ such that the subspace
on which $\mathcal{R}_{0,I}$ and $\tilde{\mathcal{R}}_{0,I}$ project
is one-dimensional. This fact is equivalent to the existence of a
state such that 
\begin{equation}
J_{a}^{\dot{a}}\vert\Lambda_{0}^{I},\Lambda_{0}^{\bar{I}}\rangle=0,\quad J_{b}^{a}\vert\Lambda_{0}^{I},\Lambda_{0}^{\bar{I}}\rangle=\lambda_{I}\,\delta_{b}^{a}\vert\Lambda_{0}^{I},\Lambda_{0}^{\bar{I}}\rangle,\quad J_{\dot{b}}^{\dot{a}}\vert\Lambda_{0}^{I},\Lambda_{0}^{\bar{I}}\rangle=\bar{\lambda}_{I}\,\delta_{\dot{b}}^{\dot{a}}\vert\Lambda_{0}^{I},\Lambda_{0}^{\bar{I}}\rangle\,,\label{eq:sate}
\end{equation}
for a properly chosen set $I$ and some $\bar{\lambda}_{I},\lambda_{I}$. For convenience the state defined in (\ref{eq:sate})
is denoted as $\vert hws\rangle$.

In \cite{Frassek2011} the generalized rectangularity condition was defined as
\begin{equation}\label{genrec2}
 J_{C}^{A}J_{B}^{C}=\mu J_{B}^{A}+\nu\delta_{B}^{A}\mathbb{I}\,.
\end{equation}
In this case, using (\ref{eq:Casimir}) and (\ref{eq:casimirofl})
for the full set $K=\{1,\ldots,n\}$, one can show that the shifted weights are given by
\begin{equation}
\ell_{i}=\begin{cases}
(\bar{\lambda}_{I}-i+1)\, & i\leq a\\
(\lambda_{I}-i+1)\, & i>a
\end{cases}\,,\label{eq:ellls}
\end{equation}
where $a$ is an integer with $0\leq a\leq n$, compare to (\ref{eq:tableaux}) and \eqref{shiftwei},
and $\lambda_{I},\bar{\lambda}_{I}$ are in general complex numbers
related to $\mu$ and $\nu$ via
\begin{equation}
 \mu=\lambda_{I}+\bar{\lambda}_{I}+\vert I\vert\,,\quad\quad \nu=-\lambda_{I}(\bar{\lambda}_{I}+\vert I\vert)\,.
\end{equation} 
The label
$I$ is introduced for consistency with Section~\ref{sub:Reduction},
where $\vert I\vert=n-a$. The careful reader might have noticed that a condition similar to \eqref{genrec2} appeared in Chapter~\ref{ch:2dlattice} when studying the inverse of the Lax operators ${\bf L}_\Lambda$. The relation between \eqref{genrec} and \eqref{genrec2} was spelled out in \eqref{genrel}.

\subsubsection{On the normalization of R-operators \label{sub:On-the-normalization}}

Besides the ratio of $\rho_{I}$ and $\tilde{\rho}_{I}$ which is
fixed by (\ref{eq:R_conjugation-1}) an overall normalization of the
R-operators $\mathcal{R}$ was not yet chosen. In our previous analysis
we determined the one-dimensional subspace which saturates the bound
of $\hat{\ell}_{k}^{\bar{I}}$ and $\hat{\ell}_{k}^{I}$. The action
of the shifted weights on this subspace is given in the previous section.
As already mentioned, for our purposes it is convenient to choose
a normalization such that (\ref{eq:RProjector}) holds, i.e.
\begin{equation}
 \mathcal{R}_{0,I}(z)=\kappa_{I}(z)\,\prod_{k=1}^{\vert\bar{I}\vert}\,\frac{\Gamma(z-\sfrac{\vert\bar{I}\vert}{2}-\hat{\ell}_{k}^{\bar{I}}+1)}{\Gamma(z-\sfrac{\vert\bar{I}\vert}{2}+k-\bar{\lambda}_{I})}\,,
\end{equation} 
\begin{equation}
\tilde{\mathcal{R}}_{0,I}(z)=\tilde{\kappa}_{I}(z)\,\prod_{k=1}^{\vert I\vert}\,\frac{\Gamma(z+\sfrac{\vert\bar{I}\vert}{2}+k-\lambda_{I})}{\Gamma(z+\sfrac{\vert\bar{I}\vert}{2}-\hat{\ell}_{k}^{I}+1)}\,,
\end{equation}
compare to (\ref{eq:R0}) and (\ref{Rt0-2}). Above, $\kappa_{I}$
and $\tilde{\kappa}_{I}$ are periodic functions of $\hat{\ell}_{k}^{\bar{I}}$,
$\hat{\ell}_{k}^{I}$ of period one, respectively. Furthermore, they
coincide on the highest weight $\kappa_{I}(z)\vert hws\rangle=\tilde{\kappa_{I}}(z)\vert hws\rangle=\vert hws\rangle$
and in analogy to $\rho_{I}$ and $\tilde{\rho}_{I}$ are dependent
by (\ref{eq:R_conjugation-1}), see also Section~\ref{sec:Example:gl2}
for the example of $\gl2$. As discussed in Section~\ref{sub:Yang-Baxter-approach},
from the study of the Yang-Baxter equation it is natural
to fix the overall normalization such that

\begin{equation}
\tilde{\mathcal{R}}_{0,\bar{I}}(z-\frac{\vert\bar{I}\vert}{2})=\mathcal{R}_{0,I}^{-1}(z+\frac{\vert I\vert}{2})\,.\label{eq:crossing-R0}
\end{equation}
Interestingly, this relation implies the crossing equation

\begin{equation}
\left(\mathcal{R}_{\bar{I}}(z-\frac{\vert\bar{I}\vert}{2})\right)^{*}=\mathcal{R}_{I}^{-1}(z+\frac{\vert I\vert}{2})\label{eq:crossing}
\end{equation}
with $\left(\bar{\mathbf{a}}_{\dot{a}_{1}}^{a_{1}}\cdots\bar{\mathbf{a}}_{\dot{a}_{m}}^{a_{m}}\,\mathbf{a}_{b_{1}}^{\dot{b}_{1}}\cdots\mathbf{a}_{b_{n}}^{\dot{b}_{n}}\right)^{*}=\bar{\mathbf{a}}_{b_{1}}^{\dot{b}_{1}}\cdots\bar{\mathbf{a}}_{b_{m}}^{\dot{b}_{m}}\,\mathbf{a}_{\dot{a}_{1}}^{a_{1}}\cdots\mathbf{a}_{\dot{a}_{n}}^{a_{n}}$.
However, an explicit study of these relations is left to the future.

\subsubsection{The $\gl2$ case: Reordering and projection in full detail\label{sec:Example:gl2}}

In this section we exploit the properties mentioned in the previous
sections for the example of $\gl2$ with $\vert I\vert=\{\dot{c}\}$,
$I=\{c\}$ and $c,\dot{c}=1,2$. In this case $\mathcal{R}_{0,\{c\}}$
and $\tilde{\mathcal{R}}_{0,\{c\}}$ are given by

\begin{equation}
\mathcal{R}_{0,\{c\}}(z)=\kappa_{\{c\}}(z)\,\frac{\Gamma(z+\frac{1}{2}-\ell_{1}^{\{\dot{c}\}})}{\Gamma(z+\frac{1}{2}-\bar{\lambda}_{c})},\quad\quad\tilde{\mathcal{R}}_{0,\{c\}}(z)=\tilde{\kappa}_{\{c\}}(z)\,\frac{\Gamma(z+\frac{3}{2}-\lambda_{c})}{\Gamma(z+\frac{3}{2}-\ell_{1}^{\{c\}})}\,;\label{eq:gl2_rtilde}
\end{equation}
see (\ref{eq:RI}) and (\ref{eq:RI_anti}), respectively. We will
now determine the explicit relation between $\rho$ and $\tilde{\rho}$
as discussed in Section~\ref{sub:Direct-approach}. For the $\gl2$
case $\mathcal{R}_{I}$ contains only one pair of oscillators. From
this follows that the conjugation in (\ref{eq:R_conjugation-1}) has
to be performed only once 
\begin{equation}
\tilde{\mathcal{R}}_{0,\{c\}}(z)=\sum_{n=0}^{\infty}\frac{1}{n!}(J_{\dot{c}}^{c})^{n}\mathcal{R}_{0,\{c\}}(z)(J_{c}^{\dot{c}})^{n}\,.\label{eq:gl2_conjugation}
\end{equation}
We can sum up this expression using the relation 
\begin{equation}
(J_{\dot{a}}^{a})^{k}(J_{a}^{\dot{a}})^{k}=(-1)^{k}\frac{\Gamma(J_{\dot{a}}^{\dot{a}}-\ell_{1}+k)\Gamma(J_{\dot{a}}^{\dot{a}}-\ell_{2}+k)}{\Gamma(J_{\dot{a}}^{\dot{a}}-\ell_{1})\Gamma(J_{\dot{a}}^{\dot{a}}-\ell_{2})},\label{eq:zenter-1}
\end{equation}
where $\ell_{i}=\ell_{i}^{\{a,\dot{a}\}}=\ell_{i}^{\{1,2\}}$. Applying
the reflection formula for Gamma functions 
\begin{equation}
\Gamma(1-z)\Gamma(z)=\frac{\pi}{\sin\pi z}\label{eq:Gamma_reflection}
\end{equation}
one finds that
\begin{equation}
\tilde{\mathcal{R}}_{0,\{c\}}(z)=-\kappa_{\{c\}}(z)\,\frac{\sin\pi(z+\frac{1}{2}-\ell_{1})\sin\pi(z+\frac{1}{2}-\ell_{2})}{\sin\pi(z+\frac{1}{2}-\ell_{1}^{\{\dot{c}\}})\sin\pi(z+\frac{1}{2}-\ell_{1}^{\{c\}})}\,\frac{\Gamma(z+\frac{3}{2}-\lambda_{c})}{\Gamma(z+\frac{3}{2}-\ell_{1}^{\{c\}})}\,,\label{eq:rtgl2exp}
\end{equation}
using $C_{1}=J_{c}^{c}+J_{\dot{c}}^{\dot{c}}=1+\ell_{1}+\ell_{2}$
and that up to permutation of $\ell_{1}$ and $\ell_{2}$ it holds
that $\ell_{1}=\bar{\lambda}_{c}$, $\ell_{2}=\lambda_{c}-1$, see
(\ref{eq:ellls}). This is exactly what we expected from the analysis
of the Yang-Baxter equation, compare (\ref{eq:gl2_rtilde}). Furthermore,
it fixes the relative normalization 
\begin{equation}
\tilde{\kappa}_{\{c\}}(z)=-\,\frac{\sin\pi(z+\frac{1}{2}-\ell_{1})\sin\pi(z+\frac{1}{2}-\ell_{2})}{\sin\pi(z+\frac{1}{2}-\ell_{1}^{\{\dot{c}\}})\sin\pi(z+\frac{1}{2}-\ell_{1}^{\{c\}})}\kappa_{\{c\}}(z)\,.\label{eq:rho_relation}
\end{equation}
Let us now look for the projection point as discussed Section~\ref{sec:Projection-property-of}.
For $\hat{z}=\lambda_{\dot{c}}-\frac{1}{2}$ one obtains%
\footnote{Here we take $\kappa_{\{c\}}(z)\vert hws\rangle=\vert hws\rangle$. %
} 
\begin{equation}
\mathcal{R}_{0,\{c\}}(\hat{z})=\vert hws\rangle\langle hws\vert\,.
\end{equation}
On the other hand at $\check{z}=\lambda_{c}-\frac{3}{2}$ we find
\begin{equation}
\tilde{\mathcal{R}}_{0,\{c\}}(\check{z})=\vert hws\rangle\langle hws\vert.
\end{equation}
The total trigonometric prefactor reduces to $1$. Note that this
is the case for arbitrary $z$ on any state if the spectrum of $\ell^{\{\dot{c}\}}$
and $\ell^{\{c\}}$ is integer spaced. 

\subsection{Diagrammatics and local charges\label{sec:Diagrammatics-and-local}}

As it is customary we denote R-matrices by two crossing lines. In
the construction of generalized transfer matrices each vertical line
corresponds to the quantum space associated to a spin-chain site.
Likewise, horizontal lines represent the auxiliary space. This notation can be related to the one introduced in Chapter~\ref{ch:ybe} by assigning ingoing arrows to the bottom lines. In the following
R-operators $\mathcal{R}$ generating Q-operators are depicted as 
\begin{equation}
\mathcal{R}_{I}(z)\;
\psset{xunit=.5pt,yunit=.5pt,runit=.5pt}
\begin{pspicture}[shift=-50](125.5,111)
{
\newrgbcolor{curcolor}{0 0 1}
\pscustom[linewidth=1,linecolor=curcolor]
{
\newpath
\moveto(75,35.5)
\lineto(75,0.5)
}
}
{
\newrgbcolor{curcolor}{0 0 1}
\pscustom[linewidth=1,linecolor=curcolor]
{
\newpath
\moveto(75,110.5)
\lineto(75,75.5)
}
}
{
\newrgbcolor{curcolor}{1 0 0}
\pscustom[linewidth=1,linecolor=curcolor]
{
\newpath
\moveto(95,55.5)
\curveto(121.97592,55.82787)(125,70.5)(125,110.5)
}
}
{
\newrgbcolor{curcolor}{1 0 0}
\pscustom[linewidth=1,linecolor=curcolor]
{
\newpath
\moveto(55,55.5)
\curveto(28.02408,55.17213)(25,40.5)(25,0.5)
}
}
{
\newrgbcolor{curcolor}{0 0 0}
\pscustom[linestyle=none,fillstyle=solid,fillcolor=curcolor]
{
\newpath
\moveto(12.56250367,57.48466502)
\curveto(12.87500367,57.48466502)(13.25000367,57.48466502)(13.25000367,57.85966502)
\curveto(13.25000367,58.26591502)(12.87500367,58.26591502)(12.59375367,58.26591502)
\lineto(0.65625367,58.26591502)
\curveto(0.37500367,58.26591502)(0.00000367,58.26591502)(0.00000367,57.85966502)
\curveto(0.00000367,57.48466502)(0.37500367,57.48466502)(0.65625367,57.48466502)
\closepath
\moveto(12.59375367,53.60966502)
\curveto(12.87500367,53.60966502)(13.25000367,53.60966502)(13.25000367,54.01591502)
\curveto(13.25000367,54.39091502)(12.87500367,54.39091502)(12.56250367,54.39091502)
\lineto(0.65625367,54.39091502)
\curveto(0.37500367,54.39091502)(0.00000367,54.39091502)(0.00000367,54.01591502)
\curveto(0.00000367,53.60966502)(0.37500367,53.60966502)(0.65625367,53.60966502)
\closepath
\moveto(12.59375367,53.60966502)
}
}
{
\newrgbcolor{curcolor}{1 1 1}
\pscustom[linestyle=none,fillstyle=solid,fillcolor=curcolor]
{
\newpath
\moveto(95,55.5)
\curveto(95,44.454305)(86.045695,35.5)(75,35.5)
\curveto(63.954305,35.5)(55,44.454305)(55,55.5)
\curveto(55,66.545695)(63.954305,75.5)(75,75.5)
\curveto(86.045695,75.5)(95,66.545695)(95,55.5)
\closepath
}
}
{
\newrgbcolor{curcolor}{0 0 0}
\pscustom[linewidth=1,linecolor=curcolor]
{
\newpath
\moveto(95,55.5)
\curveto(95,44.454305)(86.045695,35.5)(75,35.5)
\curveto(63.954305,35.5)(55,44.454305)(55,55.5)
\curveto(55,66.545695)(63.954305,75.5)(75,75.5)
\curveto(86.045695,75.5)(95,66.545695)(95,55.5)
\closepath
}
}
{
\newrgbcolor{curcolor}{0 0 0}
\pscustom[linestyle=none,fillstyle=solid,fillcolor=curcolor]
{
\newpath
\moveto(70.67383367,61.60367732)
\curveto(74.51758367,61.60367732)(75.51758367,60.66617732)(75.51758367,59.35367732)
\curveto(75.51758367,58.13492732)(74.54883367,55.72867732)(71.26758367,55.60367732)
\curveto(70.29883367,55.57242732)(69.61133367,54.88492732)(69.61133367,54.66617732)
\curveto(69.61133367,54.54117732)(69.67383367,54.54117732)(69.70508367,54.54117732)
\curveto(70.54883367,54.38492732)(70.95508367,54.00992732)(71.98633367,51.63492732)
\curveto(72.89258367,49.54117732)(73.42383367,48.63492732)(74.58008367,48.63492732)
\curveto(77.01758367,48.63492732)(79.29883367,51.07242732)(79.29883367,51.50992732)
\curveto(79.29883367,51.66617732)(79.14258367,51.66617732)(79.08008367,51.66617732)
\curveto(78.83008367,51.66617732)(78.04883367,51.38492732)(77.64258367,50.82242732)
\curveto(77.33008367,50.35367732)(76.89258367,49.72867732)(75.92383367,49.72867732)
\curveto(74.89258367,49.72867732)(74.26758367,51.13492732)(73.61133367,52.69742732)
\curveto(73.17383367,53.69742732)(72.83008367,54.44742732)(72.36133367,54.97867732)
\curveto(75.23633367,56.04117732)(77.20508367,58.13492732)(77.20508367,60.19742732)
\curveto(77.20508367,62.69742732)(73.86133367,62.69742732)(70.83008367,62.69742732)
\curveto(68.86133367,62.69742732)(67.73633367,62.69742732)(66.04883367,61.97867732)
\curveto(63.39258367,60.79117732)(63.01758367,59.13492732)(63.01758367,58.97867732)
\curveto(63.01758367,58.85367732)(63.11133367,58.82242732)(63.23633367,58.82242732)
\curveto(63.54883367,58.82242732)(64.01758367,59.10367732)(64.17383367,59.19742732)
\curveto(64.58008367,59.47867732)(64.64258367,59.60367732)(64.76758367,59.97867732)
\curveto(65.04883367,60.79117732)(65.61133367,61.47867732)(68.11133367,61.60367732)
\curveto(68.01758367,60.38492732)(67.83008367,58.50992732)(67.14258367,55.60367732)
\curveto(66.61133367,53.38492732)(65.89258367,51.19742732)(65.01758367,49.04117732)
\curveto(64.89258367,48.82242732)(64.89258367,48.79117732)(64.89258367,48.75992732)
\curveto(64.89258367,48.63492732)(65.04883367,48.63492732)(65.11133367,48.63492732)
\curveto(65.51758367,48.63492732)(66.33008367,49.10367732)(66.54883367,49.47867732)
\curveto(66.61133367,49.60367732)(69.14258367,55.22867732)(69.73633367,61.60367732)
\closepath
\moveto(70.67383367,61.60367732)
}
}
{
\newrgbcolor{curcolor}{0 0 0}
\pscustom[linestyle=none,fillstyle=solid,fillcolor=curcolor]
{
\newpath
\moveto(85.24732,54.49067073)
\curveto(85.37232,54.99067073)(85.40357,55.11567073)(86.46607,55.11567073)
\curveto(86.80982,55.11567073)(86.96607,55.11567073)(86.96607,55.39692073)
\curveto(86.96607,55.52192073)(86.87232,55.61567073)(86.74732,55.61567073)
\curveto(86.46607,55.61567073)(86.09107,55.55317073)(85.80982,55.55317073)
\lineto(84.87232,55.55317073)
\lineto(83.90357,55.55317073)
\curveto(83.59107,55.55317073)(83.24732,55.61567073)(82.93482,55.61567073)
\curveto(82.84107,55.61567073)(82.65357,55.61567073)(82.65357,55.30317073)
\curveto(82.65357,55.11567073)(82.80982,55.11567073)(83.12232,55.11567073)
\curveto(83.12232,55.11567073)(83.40357,55.11567073)(83.62232,55.08442073)
\curveto(83.90357,55.05317073)(84.02857,55.02192073)(84.02857,54.86567073)
\curveto(84.02857,54.80317073)(83.99732,54.70942073)(83.96607,54.58442073)
\lineto(82.09107,47.17817073)
\curveto(81.99732,46.74067073)(81.93482,46.58442073)(80.90357,46.58442073)
\curveto(80.52857,46.58442073)(80.40357,46.58442073)(80.40357,46.27192073)
\curveto(80.40357,46.27192073)(80.40357,46.08442073)(80.62232,46.08442073)
\curveto(81.02857,46.08442073)(82.09107,46.14692073)(82.49732,46.14692073)
\lineto(83.46607,46.11567073)
\curveto(83.74732,46.11567073)(84.12232,46.08442073)(84.40357,46.08442073)
\curveto(84.49732,46.08442073)(84.71607,46.08442073)(84.71607,46.39692073)
\curveto(84.71607,46.58442073)(84.52857,46.58442073)(84.27857,46.58442073)
\curveto(84.24732,46.58442073)(83.93482,46.58442073)(83.65357,46.61567073)
\curveto(83.34107,46.64692073)(83.34107,46.70942073)(83.34107,46.83442073)
\curveto(83.34107,46.83442073)(83.34107,46.92817073)(83.40357,47.11567073)
\closepath
\moveto(85.24732,54.49067073)
}
}
\end{pspicture}\,,
\end{equation}
compare to (\ref{eq:Qop}). Now, we will develop a pictorial language
for the R-operators $\mathcal{R}$, which incorporates all
properties discussed in Section~\ref{sec:Alternative-presentation-of} and
\ref{sec:Projection-property-of}. One of its main advantages is that
the opposite product (\ref{eq:prod_op}), which might look unfamiliar
in the equations, is translated to a rather natural composition rule.
It is a key ingredient to reveal the emergence of local charges from
Q-operators.

\subsubsection{Two multiplication rules}

As discussed in Section~\ref{sec:Alternative-presentation-of}, it
is natural to introduce two different multiplication rules. Diagrammatically
the product $\cdot$ can then be understood as

\begin{equation}
\mathcal{O}_{1}\cdot\mathcal{O}_{2}\;
\psset{xunit=.5pt,yunit=.5pt,runit=.5pt}
\begin{pspicture}[shift=-45](227.5,111)
{
\newrgbcolor{curcolor}{0 0 1}
\pscustom[linewidth=1,linecolor=curcolor]
{
\newpath
\moveto(72,35.5)
\lineto(72,0.5)
}
}
{
\newrgbcolor{curcolor}{0 0 1}
\pscustom[linewidth=1,linecolor=curcolor]
{
\newpath
\moveto(72,75.5)
\curveto(72,85.5)(77,95.5)(92,95.5)
\curveto(107,95.5)(111.72773,85.49629)(112,75.5)
\curveto(112.46098,58.57523)(112.0806,46.03468)(120.34307,34.92015)
\curveto(132.72797,18.2602)(145.93358,3.0848)(155.39765,2.68558)
\curveto(173.76061,1.91098)(177.29455,12.8264)(177.55724,24.67513)
\curveto(177.79807,35.53778)(177.42811,24.72053)(176.75,110.25)
}
}
{
\newrgbcolor{curcolor}{1 0 0}
\pscustom[linewidth=1,linecolor=curcolor]
{
\newpath
\moveto(92,55.5)
\lineto(102.5,55.5)
\curveto(112.5,55.5)(107.7,59.08579)(112.7,60.5)
\curveto(117.87677,58.73224)(112.83876,55.60523)(122.5,55.5)
\lineto(157,55.5)
}
}
{
\newrgbcolor{curcolor}{1 0 0}
\pscustom[linewidth=1,linecolor=curcolor]
{
\newpath
\moveto(52,55.5)
\curveto(25.02408,55.17213)(22,40.5)(22,0.5)
}
}
{
\newrgbcolor{curcolor}{0 0 0}
\pscustom[linestyle=none,fillstyle=solid,fillcolor=curcolor]
{
\newpath
\moveto(12.56250367,56.98466502)
\curveto(12.87500367,56.98466502)(13.25000367,56.98466502)(13.25000367,57.35966502)
\curveto(13.25000367,57.76591502)(12.87500367,57.76591502)(12.59375367,57.76591502)
\lineto(0.65625367,57.76591502)
\curveto(0.37500367,57.76591502)(0.00000367,57.76591502)(0.00000367,57.35966502)
\curveto(0.00000367,56.98466502)(0.37500367,56.98466502)(0.65625367,56.98466502)
\closepath
\moveto(12.59375367,53.10966502)
\curveto(12.87500367,53.10966502)(13.25000367,53.10966502)(13.25000367,53.51591502)
\curveto(13.25000367,53.89091502)(12.87500367,53.89091502)(12.56250367,53.89091502)
\lineto(0.65625367,53.89091502)
\curveto(0.37500367,53.89091502)(0.00000367,53.89091502)(0.00000367,53.51591502)
\curveto(0.00000367,53.10966502)(0.37500367,53.10966502)(0.65625367,53.10966502)
\closepath
\moveto(12.59375367,53.10966502)
}
}
{
\newrgbcolor{curcolor}{1 1 1}
\pscustom[linestyle=none,fillstyle=solid,fillcolor=curcolor]
{
\newpath
\moveto(92.000004,55.5)
\curveto(92.000004,44.454305)(83.045699,35.5)(72.000004,35.5)
\curveto(60.954309,35.5)(52.000004,44.454305)(52.000004,55.5)
\curveto(52.000004,66.545695)(60.954309,75.5)(72.000004,75.5)
\curveto(83.045699,75.5)(92.000004,66.545695)(92.000004,55.5)
\closepath
}
}
{
\newrgbcolor{curcolor}{0 0 0}
\pscustom[linewidth=1,linecolor=curcolor]
{
\newpath
\moveto(92.000004,55.5)
\curveto(92.000004,44.454305)(83.045699,35.5)(72.000004,35.5)
\curveto(60.954309,35.5)(52.000004,44.454305)(52.000004,55.5)
\curveto(52.000004,66.545695)(60.954309,75.5)(72.000004,75.5)
\curveto(83.045699,75.5)(92.000004,66.545695)(92.000004,55.5)
\closepath
}
}
{
\newrgbcolor{curcolor}{0 0 1}
\pscustom[linewidth=1,linecolor=curcolor]
{
\newpath
\moveto(177,110.5)
\lineto(177,75.5)
}
}
{
\newrgbcolor{curcolor}{1 0 0}
\pscustom[linewidth=1,linecolor=curcolor]
{
\newpath
\moveto(197,55.5)
\curveto(223.97592,55.82787)(227,70.5)(227,110.5)
}
}
{
\newrgbcolor{curcolor}{1 1 1}
\pscustom[linestyle=none,fillstyle=solid,fillcolor=curcolor]
{
\newpath
\moveto(197,55.5)
\curveto(197,44.454305)(188.045695,35.5)(177,35.5)
\curveto(165.954305,35.5)(157,44.454305)(157,55.5)
\curveto(157,66.545695)(165.954305,75.5)(177,75.5)
\curveto(188.045695,75.5)(197,66.545695)(197,55.5)
\closepath
}
}
{
\newrgbcolor{curcolor}{0 0 0}
\pscustom[linewidth=1,linecolor=curcolor]
{
\newpath
\moveto(197,55.5)
\curveto(197,44.454305)(188.045695,35.5)(177,35.5)
\curveto(165.954305,35.5)(157,44.454305)(157,55.5)
\curveto(157,66.545695)(165.954305,75.5)(177,75.5)
\curveto(188.045695,75.5)(197,66.545695)(197,55.5)
\closepath
}
}
{
\newrgbcolor{curcolor}{0 0 0}
\pscustom[linestyle=none,fillstyle=solid,fillcolor=curcolor]
{
\newpath
\moveto(75.99001367,58.90453412)
\curveto(75.99001367,61.43578412)(74.86501367,63.34203412)(72.42751367,63.34203412)
\curveto(70.36501367,63.34203412)(68.61501367,61.65453412)(68.45876367,61.52953412)
\curveto(66.86501367,59.96703412)(66.27126367,58.18578412)(66.27126367,58.12328412)
\curveto(66.27126367,57.99828412)(66.36501367,57.99828412)(66.45876367,57.99828412)
\curveto(66.86501367,57.99828412)(67.17751367,58.18578412)(67.45876367,58.40453412)
\curveto(67.83376367,58.65453412)(67.83376367,58.71703412)(68.05251367,59.18578412)
\curveto(68.24001367,59.59203412)(68.70876367,60.56078412)(69.42751367,61.40453412)
\curveto(69.92751367,61.93578412)(70.33376367,62.24828412)(71.11501367,62.24828412)
\curveto(73.05251367,62.24828412)(74.27126367,60.65453412)(74.27126367,58.02953412)
\curveto(74.27126367,53.96703412)(71.39626367,49.93578412)(67.70876367,49.93578412)
\curveto(64.86501367,49.93578412)(63.33376367,52.18578412)(63.33376367,55.02953412)
\curveto(63.33376367,57.74828412)(64.70876367,60.74828412)(67.61501367,62.46703412)
\curveto(67.83376367,62.59203412)(68.42751367,62.93578412)(68.42751367,63.18578412)
\curveto(68.42751367,63.34203412)(68.27126367,63.34203412)(68.24001367,63.34203412)
\curveto(67.52126367,63.34203412)(61.61501367,60.15453412)(61.61501367,54.18578412)
\curveto(61.61501367,51.40453412)(63.08376367,48.84203412)(66.39626367,48.84203412)
\curveto(70.27126367,48.84203412)(75.99001367,52.74828412)(75.99001367,58.90453412)
\closepath
\moveto(75.99001367,58.90453412)
}
}
{
\newrgbcolor{curcolor}{0 0 0}
\pscustom[linestyle=none,fillstyle=solid,fillcolor=curcolor]
{
\newpath
\moveto(81.00826318,55.16652753)
\curveto(81.00826318,55.54152753)(81.00826318,55.54152753)(80.60201318,55.54152753)
\curveto(79.69576318,54.66652753)(78.44576318,54.66652753)(77.88326318,54.66652753)
\lineto(77.88326318,54.16652753)
\curveto(78.19576318,54.16652753)(79.13326318,54.16652753)(79.88326318,54.54152753)
\lineto(79.88326318,47.44777753)
\curveto(79.88326318,46.97902753)(79.88326318,46.79152753)(78.50826318,46.79152753)
\lineto(77.97701318,46.79152753)
\lineto(77.97701318,46.29152753)
\curveto(78.22701318,46.29152753)(79.94576318,46.35402753)(80.44576318,46.35402753)
\curveto(80.88326318,46.35402753)(82.63326318,46.29152753)(82.94576318,46.29152753)
\lineto(82.94576318,46.79152753)
\lineto(82.41451318,46.79152753)
\curveto(81.00826318,46.79152753)(81.00826318,46.97902753)(81.00826318,47.44777753)
\closepath
\moveto(81.00826318,55.16652753)
}
}
{
\newrgbcolor{curcolor}{0 0 0}
\pscustom[linestyle=none,fillstyle=solid,fillcolor=curcolor]
{
\newpath
\moveto(180.64181367,59.61164092)
\curveto(180.64181367,62.14289092)(179.51681367,64.04914092)(177.07931367,64.04914092)
\curveto(175.01681367,64.04914092)(173.26681367,62.36164092)(173.11056367,62.23664092)
\curveto(171.51681367,60.67414092)(170.92306367,58.89289092)(170.92306367,58.83039092)
\curveto(170.92306367,58.70539092)(171.01681367,58.70539092)(171.11056367,58.70539092)
\curveto(171.51681367,58.70539092)(171.82931367,58.89289092)(172.11056367,59.11164092)
\curveto(172.48556367,59.36164092)(172.48556367,59.42414092)(172.70431367,59.89289092)
\curveto(172.89181367,60.29914092)(173.36056367,61.26789092)(174.07931367,62.11164092)
\curveto(174.57931367,62.64289092)(174.98556367,62.95539092)(175.76681367,62.95539092)
\curveto(177.70431367,62.95539092)(178.92306367,61.36164092)(178.92306367,58.73664092)
\curveto(178.92306367,54.67414092)(176.04806367,50.64289092)(172.36056367,50.64289092)
\curveto(169.51681367,50.64289092)(167.98556367,52.89289092)(167.98556367,55.73664092)
\curveto(167.98556367,58.45539092)(169.36056367,61.45539092)(172.26681367,63.17414092)
\curveto(172.48556367,63.29914092)(173.07931367,63.64289092)(173.07931367,63.89289092)
\curveto(173.07931367,64.04914092)(172.92306367,64.04914092)(172.89181367,64.04914092)
\curveto(172.17306367,64.04914092)(166.26681367,60.86164092)(166.26681367,54.89289092)
\curveto(166.26681367,52.11164092)(167.73556367,49.54914092)(171.04806367,49.54914092)
\curveto(174.92306367,49.54914092)(180.64181367,53.45539092)(180.64181367,59.61164092)
\closepath
\moveto(180.64181367,59.61164092)
}
}
{
\newrgbcolor{curcolor}{0 0 0}
\pscustom[linestyle=none,fillstyle=solid,fillcolor=curcolor]
{
\newpath
\moveto(188.03506318,49.52988433)
\lineto(187.56631318,49.52988433)
\curveto(187.53506318,49.21738433)(187.37881318,48.40488433)(187.19131318,48.27988433)
\curveto(187.09756318,48.18613433)(186.03506318,48.18613433)(185.81631318,48.18613433)
\lineto(183.25381318,48.18613433)
\curveto(184.72256318,49.46738433)(185.22256318,49.87363433)(186.03506318,50.52988433)
\curveto(187.06631318,51.34238433)(188.03506318,52.21738433)(188.03506318,53.52988433)
\curveto(188.03506318,55.21738433)(186.56631318,56.24863433)(184.78506318,56.24863433)
\curveto(183.06631318,56.24863433)(181.87881318,55.02988433)(181.87881318,53.74863433)
\curveto(181.87881318,53.06113433)(182.47256318,52.96738433)(182.62881318,52.96738433)
\curveto(182.94131318,52.96738433)(183.34756318,53.21738433)(183.34756318,53.71738433)
\curveto(183.34756318,53.96738433)(183.25381318,54.46738433)(182.53506318,54.46738433)
\curveto(182.97256318,55.43613433)(183.91006318,55.74863433)(184.56631318,55.74863433)
\curveto(185.97256318,55.74863433)(186.69131318,54.65488433)(186.69131318,53.52988433)
\curveto(186.69131318,52.31113433)(185.81631318,51.37363433)(185.37881318,50.87363433)
\lineto(182.03506318,47.52988433)
\curveto(181.87881318,47.40488433)(181.87881318,47.37363433)(181.87881318,46.99863433)
\lineto(187.62881318,46.99863433)
\closepath
\moveto(188.03506318,49.52988433)
}
}
\end{pspicture}\,.\label{eq:O1O2}
\end{equation}
Here the oscillator (\textcolor{red}{red}) and the ${U}(\gln)$
(\textcolor{blue}{blue}) components of $\mathcal{O}_{1}$ are both
multiplied from the left to $\mathcal{O}_{2}$. On the other hand,
the product $\circ$ is denoted by
\begin{equation}
\mathcal{O}_{1}\circ\mathcal{O}_{2}\;
\psset{xunit=.5pt,yunit=.5pt,runit=.5pt}
\begin{pspicture}[shift=-45](227.5,111)
{
\newrgbcolor{curcolor}{0 0 1}
\pscustom[linewidth=1,linecolor=curcolor]
{
\newpath
\moveto(177,35.5)
\lineto(177,0.5)
}
}
{
\newrgbcolor{curcolor}{0 0 1}
\pscustom[linewidth=1,linecolor=curcolor]
{
\newpath
\moveto(177,75.5)
\curveto(177,85.5)(172,95.5)(157,95.5)
\curveto(142,95.5)(137.27227,85.49629)(137,75.5)
\curveto(136.53902,58.57523)(136.9194,46.03468)(128.65693,34.92015)
\curveto(116.27203,18.2602)(103.06642,3.0848)(93.60235,2.68558)
\curveto(75.23939,1.91098)(71.70545,12.8264)(71.44276,24.67513)
\curveto(71.20193,35.53778)(71.57189,24.72053)(72.25,110.25)
}
}
{
\newrgbcolor{curcolor}{1 0 0}
\pscustom[linewidth=1,linecolor=curcolor]
{
\newpath
\moveto(92,55.5)
\lineto(125.3,55.5)
\curveto(135.3,55.5)(131.3,59.08579)(136.3,60.5)
\curveto(141.47677,58.73224)(137.33876,55.60523)(147,55.5)
\lineto(157,55.5)
}
}
{
\newrgbcolor{curcolor}{1 0 0}
\pscustom[linewidth=1,linecolor=curcolor]
{
\newpath
\moveto(52,55.5)
\curveto(25.02408,55.17213)(22,40.5)(22,0.5)
}
}
{
\newrgbcolor{curcolor}{0 0 0}
\pscustom[linestyle=none,fillstyle=solid,fillcolor=curcolor]
{
\newpath
\moveto(12.56250367,56.98466502)
\curveto(12.87500367,56.98466502)(13.25000367,56.98466502)(13.25000367,57.35966502)
\curveto(13.25000367,57.76591502)(12.87500367,57.76591502)(12.59375367,57.76591502)
\lineto(0.65625367,57.76591502)
\curveto(0.37500367,57.76591502)(0.00000367,57.76591502)(0.00000367,57.35966502)
\curveto(0.00000367,56.98466502)(0.37500367,56.98466502)(0.65625367,56.98466502)
\closepath
\moveto(12.59375367,53.10966502)
\curveto(12.87500367,53.10966502)(13.25000367,53.10966502)(13.25000367,53.51591502)
\curveto(13.25000367,53.89091502)(12.87500367,53.89091502)(12.56250367,53.89091502)
\lineto(0.65625367,53.89091502)
\curveto(0.37500367,53.89091502)(0.00000367,53.89091502)(0.00000367,53.51591502)
\curveto(0.00000367,53.10966502)(0.37500367,53.10966502)(0.65625367,53.10966502)
\closepath
\moveto(12.59375367,53.10966502)
}
}
{
\newrgbcolor{curcolor}{1 0 0}
\pscustom[linewidth=1,linecolor=curcolor]
{
\newpath
\moveto(197,55.5)
\curveto(223.97592,55.82787)(227,70.5)(227,110.5)
}
}
{
\newrgbcolor{curcolor}{1 1 1}
\pscustom[linestyle=none,fillstyle=solid,fillcolor=curcolor]
{
\newpath
\moveto(52,55.5)
\curveto(52,44.454305)(60.954305,35.5)(72,35.5)
\curveto(83.045695,35.5)(92,44.454305)(92,55.5)
\curveto(92,66.545695)(83.045695,75.5)(72,75.5)
\curveto(60.954305,75.5)(52,66.545695)(52,55.5)
\closepath
}
}
{
\newrgbcolor{curcolor}{0 0 0}
\pscustom[linewidth=1,linecolor=curcolor]
{
\newpath
\moveto(52,55.5)
\curveto(52,44.454305)(60.954305,35.5)(72,35.5)
\curveto(83.045695,35.5)(92,44.454305)(92,55.5)
\curveto(92,66.545695)(83.045695,75.5)(72,75.5)
\curveto(60.954305,75.5)(52,66.545695)(52,55.5)
\closepath
}
}
{
\newrgbcolor{curcolor}{1 1 1}
\pscustom[linestyle=none,fillstyle=solid,fillcolor=curcolor]
{
\newpath
\moveto(157,55.5)
\curveto(157,44.454305)(165.954305,35.5)(177,35.5)
\curveto(188.045695,35.5)(197,44.454305)(197,55.5)
\curveto(197,66.545695)(188.045695,75.5)(177,75.5)
\curveto(165.954305,75.5)(157,66.545695)(157,55.5)
\closepath
}
}
{
\newrgbcolor{curcolor}{0 0 0}
\pscustom[linewidth=1,linecolor=curcolor]
{
\newpath
\moveto(157,55.5)
\curveto(157,44.454305)(165.954305,35.5)(177,35.5)
\curveto(188.045695,35.5)(197,44.454305)(197,55.5)
\curveto(197,66.545695)(188.045695,75.5)(177,75.5)
\curveto(165.954305,75.5)(157,66.545695)(157,55.5)
\closepath
}
}
{
\newrgbcolor{curcolor}{0 0 0}
\pscustom[linestyle=none,fillstyle=solid,fillcolor=curcolor]
{
\newpath
\moveto(181.66977367,59.73296122)
\curveto(181.66977367,62.26421122)(180.54477367,64.17046122)(178.10727367,64.17046122)
\curveto(176.04477367,64.17046122)(174.29477367,62.48296122)(174.13852367,62.35796122)
\curveto(172.54477367,60.79546122)(171.95102367,59.01421122)(171.95102367,58.95171122)
\curveto(171.95102367,58.82671122)(172.04477367,58.82671122)(172.13852367,58.82671122)
\curveto(172.54477367,58.82671122)(172.85727367,59.01421122)(173.13852367,59.23296122)
\curveto(173.51352367,59.48296122)(173.51352367,59.54546122)(173.73227367,60.01421122)
\curveto(173.91977367,60.42046122)(174.38852367,61.38921122)(175.10727367,62.23296122)
\curveto(175.60727367,62.76421122)(176.01352367,63.07671122)(176.79477367,63.07671122)
\curveto(178.73227367,63.07671122)(179.95102367,61.48296122)(179.95102367,58.85796122)
\curveto(179.95102367,54.79546122)(177.07602367,50.76421122)(173.38852367,50.76421122)
\curveto(170.54477367,50.76421122)(169.01352367,53.01421122)(169.01352367,55.85796122)
\curveto(169.01352367,58.57671122)(170.38852367,61.57671122)(173.29477367,63.29546122)
\curveto(173.51352367,63.42046122)(174.10727367,63.76421122)(174.10727367,64.01421122)
\curveto(174.10727367,64.17046122)(173.95102367,64.17046122)(173.91977367,64.17046122)
\curveto(173.20102367,64.17046122)(167.29477367,60.98296122)(167.29477367,55.01421122)
\curveto(167.29477367,52.23296122)(168.76352367,49.67046122)(172.07602367,49.67046122)
\curveto(175.95102367,49.67046122)(181.66977367,53.57671122)(181.66977367,59.73296122)
\closepath
\moveto(181.66977367,59.73296122)
}
}
{
\newrgbcolor{curcolor}{0 0 0}
\pscustom[linestyle=none,fillstyle=solid,fillcolor=curcolor]
{
\newpath
\moveto(186.68802318,55.99495463)
\curveto(186.68802318,56.36995463)(186.68802318,56.36995463)(186.28177318,56.36995463)
\curveto(185.37552318,55.49495463)(184.12552318,55.49495463)(183.56302318,55.49495463)
\lineto(183.56302318,54.99495463)
\curveto(183.87552318,54.99495463)(184.81302318,54.99495463)(185.56302318,55.36995463)
\lineto(185.56302318,48.27620463)
\curveto(185.56302318,47.80745463)(185.56302318,47.61995463)(184.18802318,47.61995463)
\lineto(183.65677318,47.61995463)
\lineto(183.65677318,47.11995463)
\curveto(183.90677318,47.11995463)(185.62552318,47.18245463)(186.12552318,47.18245463)
\curveto(186.56302318,47.18245463)(188.31302318,47.11995463)(188.62552318,47.11995463)
\lineto(188.62552318,47.61995463)
\lineto(188.09427318,47.61995463)
\curveto(186.68802318,47.61995463)(186.68802318,47.80745463)(186.68802318,48.27620463)
\closepath
\moveto(186.68802318,55.99495463)
}
}
{
\newrgbcolor{curcolor}{0 0 0}
\pscustom[linestyle=none,fillstyle=solid,fillcolor=curcolor]
{
\newpath
\moveto(76.16977367,58.73296122)
\curveto(76.16977367,61.26421122)(75.04477367,63.17046122)(72.60727367,63.17046122)
\curveto(70.54477367,63.17046122)(68.79477367,61.48296122)(68.63852367,61.35796122)
\curveto(67.04477367,59.79546122)(66.45102367,58.01421122)(66.45102367,57.95171122)
\curveto(66.45102367,57.82671122)(66.54477367,57.82671122)(66.63852367,57.82671122)
\curveto(67.04477367,57.82671122)(67.35727367,58.01421122)(67.63852367,58.23296122)
\curveto(68.01352367,58.48296122)(68.01352367,58.54546122)(68.23227367,59.01421122)
\curveto(68.41977367,59.42046122)(68.88852367,60.38921122)(69.60727367,61.23296122)
\curveto(70.10727367,61.76421122)(70.51352367,62.07671122)(71.29477367,62.07671122)
\curveto(73.23227367,62.07671122)(74.45102367,60.48296122)(74.45102367,57.85796122)
\curveto(74.45102367,53.79546122)(71.57602367,49.76421122)(67.88852367,49.76421122)
\curveto(65.04477367,49.76421122)(63.51352367,52.01421122)(63.51352367,54.85796122)
\curveto(63.51352367,57.57671122)(64.88852367,60.57671122)(67.79477367,62.29546122)
\curveto(68.01352367,62.42046122)(68.60727367,62.76421122)(68.60727367,63.01421122)
\curveto(68.60727367,63.17046122)(68.45102367,63.17046122)(68.41977367,63.17046122)
\curveto(67.70102367,63.17046122)(61.79477367,59.98296122)(61.79477367,54.01421122)
\curveto(61.79477367,51.23296122)(63.26352367,48.67046122)(66.57602367,48.67046122)
\curveto(70.45102367,48.67046122)(76.16977367,52.57671122)(76.16977367,58.73296122)
\closepath
\moveto(76.16977367,58.73296122)
}
}
{
\newrgbcolor{curcolor}{0 0 0}
\pscustom[linestyle=none,fillstyle=solid,fillcolor=curcolor]
{
\newpath
\moveto(83.56302318,48.65120463)
\lineto(83.09427318,48.65120463)
\curveto(83.06302318,48.33870463)(82.90677318,47.52620463)(82.71927318,47.40120463)
\curveto(82.62552318,47.30745463)(81.56302318,47.30745463)(81.34427318,47.30745463)
\lineto(78.78177318,47.30745463)
\curveto(80.25052318,48.58870463)(80.75052318,48.99495463)(81.56302318,49.65120463)
\curveto(82.59427318,50.46370463)(83.56302318,51.33870463)(83.56302318,52.65120463)
\curveto(83.56302318,54.33870463)(82.09427318,55.36995463)(80.31302318,55.36995463)
\curveto(78.59427318,55.36995463)(77.40677318,54.15120463)(77.40677318,52.86995463)
\curveto(77.40677318,52.18245463)(78.00052318,52.08870463)(78.15677318,52.08870463)
\curveto(78.46927318,52.08870463)(78.87552318,52.33870463)(78.87552318,52.83870463)
\curveto(78.87552318,53.08870463)(78.78177318,53.58870463)(78.06302318,53.58870463)
\curveto(78.50052318,54.55745463)(79.43802318,54.86995463)(80.09427318,54.86995463)
\curveto(81.50052318,54.86995463)(82.21927318,53.77620463)(82.21927318,52.65120463)
\curveto(82.21927318,51.43245463)(81.34427318,50.49495463)(80.90677318,49.99495463)
\lineto(77.56302318,46.65120463)
\curveto(77.40677318,46.52620463)(77.40677318,46.49495463)(77.40677318,46.11995463)
\lineto(83.15677318,46.11995463)
\closepath
\moveto(83.56302318,48.65120463)
}
}
\end{pspicture}\,.\label{eq:O1circO2}
\end{equation}
Here the ${U}(\gln)$ part of $\mathcal{O}_{1}$
is also multiplied from the left to $\mathcal{O}_{2}$, but the oscillator
part is multiplied to the right. In summary, once the operators $\mathcal{O}_{i}$
are written as (\ref{eq:opdecomp}) the order of the factors (from
left to right) in (\ref{eq:O1O2}) and (\ref{eq:O1circO2}) is obtained
by following the lines from bottom to top.

\subsubsection{R-operators $\mathcal{R}$}

We will now develop a diagrammatic expression for $\mathcal{R}_{I}$
for both (\ref{eq:RI}) and (\ref{eq:RI_anti}). To be pedagogical
we proceed slowly. It is clear that $\mathcal{R}_{I}$ can be regarded
as a composite object of four parts, namely 
\begin{equation}
e^{\bar{\mathbf{a}}_{c}^{\dot{c}}J_{\dot{c}}^{c}}\;
\psset{xunit=.5pt,yunit=.5pt,runit=.5pt}
\begin{pspicture}[shift=-50](125.5,111)
{
\newrgbcolor{curcolor}{0 0 1}
\pscustom[linewidth=1,linecolor=curcolor]
{
\newpath
\moveto(75,35.5)
\lineto(75,0.5)
}
}
{
\newrgbcolor{curcolor}{0 0 1}
\pscustom[linewidth=1,linecolor=curcolor]
{
\newpath
\moveto(75,110.5)
\lineto(75,75.5)
}
}
{
\newrgbcolor{curcolor}{1 0 0}
\pscustom[linewidth=1,linecolor=curcolor]
{
\newpath
\moveto(95,55.5)
\curveto(121.97592,55.82787)(125,70.5)(125,110.5)
}
}
{
\newrgbcolor{curcolor}{1 0 0}
\pscustom[linewidth=1,linecolor=curcolor]
{
\newpath
\moveto(55,55.5)
\curveto(28.02408,55.17213)(25,40.5)(25,0.5)
}
}
{
\newrgbcolor{curcolor}{0 0 0}
\pscustom[linestyle=none,fillstyle=solid,fillcolor=curcolor]
{
\newpath
\moveto(12.56250367,57.48466502)
\curveto(12.87500367,57.48466502)(13.25000367,57.48466502)(13.25000367,57.85966502)
\curveto(13.25000367,58.26591502)(12.87500367,58.26591502)(12.59375367,58.26591502)
\lineto(0.65625367,58.26591502)
\curveto(0.37500367,58.26591502)(0.00000367,58.26591502)(0.00000367,57.85966502)
\curveto(0.00000367,57.48466502)(0.37500367,57.48466502)(0.65625367,57.48466502)
\closepath
\moveto(12.59375367,53.60966502)
\curveto(12.87500367,53.60966502)(13.25000367,53.60966502)(13.25000367,54.01591502)
\curveto(13.25000367,54.39091502)(12.87500367,54.39091502)(12.56250367,54.39091502)
\lineto(0.65625367,54.39091502)
\curveto(0.37500367,54.39091502)(0.00000367,54.39091502)(0.00000367,54.01591502)
\curveto(0.00000367,53.60966502)(0.37500367,53.60966502)(0.65625367,53.60966502)
\closepath
\moveto(12.59375367,53.60966502)
}
}
{
\newrgbcolor{curcolor}{1 1 1}
\pscustom[linestyle=none,fillstyle=solid,fillcolor=curcolor]
{
\newpath
\moveto(95,55.5)
\curveto(95,44.454305)(86.045695,35.5)(75,35.5)
\curveto(63.954305,35.5)(55,44.454305)(55,55.5)
\curveto(55,66.545695)(63.954305,75.5)(75,75.5)
\curveto(86.045695,75.5)(95,66.545695)(95,55.5)
\closepath
}
}
{
\newrgbcolor{curcolor}{0 0 0}
\pscustom[linewidth=1,linecolor=curcolor]
{
\newpath
\moveto(95,55.5)
\curveto(95,44.454305)(86.045695,35.5)(75,35.5)
\curveto(63.954305,35.5)(55,44.454305)(55,55.5)
\curveto(55,66.545695)(63.954305,75.5)(75,75.5)
\curveto(86.045695,75.5)(95,66.545695)(95,55.5)
\closepath
}
}
{
\newrgbcolor{curcolor}{0 0 0}
\pscustom[linestyle=none,fillstyle=solid,fillcolor=curcolor]
{
\newpath
\moveto(69.12862967,59.97931122)
\curveto(72.97237967,59.97931122)(73.97237967,59.04181122)(73.97237967,57.72931122)
\curveto(73.97237967,56.51056122)(73.00362967,54.10431122)(69.72237967,53.97931122)
\curveto(68.75362967,53.94806122)(68.06612967,53.26056122)(68.06612967,53.04181122)
\curveto(68.06612967,52.91681122)(68.12862967,52.91681122)(68.15987967,52.91681122)
\curveto(69.00362967,52.76056122)(69.40987967,52.38556122)(70.44112967,50.01056122)
\curveto(71.34737967,47.91681122)(71.87862967,47.01056122)(73.03487967,47.01056122)
\curveto(75.47237967,47.01056122)(77.75362967,49.44806122)(77.75362967,49.88556122)
\curveto(77.75362967,50.04181122)(77.59737967,50.04181122)(77.53487967,50.04181122)
\curveto(77.28487967,50.04181122)(76.50362967,49.76056122)(76.09737967,49.19806122)
\curveto(75.78487967,48.72931122)(75.34737967,48.10431122)(74.37862967,48.10431122)
\curveto(73.34737967,48.10431122)(72.72237967,49.51056122)(72.06612967,51.07306122)
\curveto(71.62862967,52.07306122)(71.28487967,52.82306122)(70.81612967,53.35431122)
\curveto(73.69112967,54.41681122)(75.65987967,56.51056122)(75.65987967,58.57306122)
\curveto(75.65987967,61.07306122)(72.31612967,61.07306122)(69.28487967,61.07306122)
\curveto(67.31612967,61.07306122)(66.19112967,61.07306122)(64.50362967,60.35431122)
\curveto(61.84737967,59.16681122)(61.47237967,57.51056122)(61.47237967,57.35431122)
\curveto(61.47237967,57.22931122)(61.56612967,57.19806122)(61.69112967,57.19806122)
\curveto(62.00362967,57.19806122)(62.47237967,57.47931122)(62.62862967,57.57306122)
\curveto(63.03487967,57.85431122)(63.09737967,57.97931122)(63.22237967,58.35431122)
\curveto(63.50362967,59.16681122)(64.06612967,59.85431122)(66.56612967,59.97931122)
\curveto(66.47237967,58.76056122)(66.28487967,56.88556122)(65.59737967,53.97931122)
\curveto(65.06612967,51.76056122)(64.34737967,49.57306122)(63.47237967,47.41681122)
\curveto(63.34737967,47.19806122)(63.34737967,47.16681122)(63.34737967,47.13556122)
\curveto(63.34737967,47.01056122)(63.50362967,47.01056122)(63.56612967,47.01056122)
\curveto(63.97237967,47.01056122)(64.78487967,47.47931122)(65.00362967,47.85431122)
\curveto(65.06612967,47.97931122)(67.59737967,53.60431122)(68.19112967,59.97931122)
\closepath
\moveto(69.12862967,59.97931122)
}
}
{
\newrgbcolor{curcolor}{0 0 0}
\pscustom[linestyle=none,fillstyle=solid,fillcolor=curcolor]
{
\newpath
\moveto(81.889616,60.83432221)
\lineto(86.170866,60.83432221)
\curveto(86.358366,60.83432221)(86.670866,60.83432221)(86.670866,61.14682221)
\curveto(86.670866,61.52182221)(86.358366,61.52182221)(86.170866,61.52182221)
\lineto(81.889616,61.52182221)
\lineto(81.889616,65.80307221)
\curveto(81.889616,65.95932221)(81.889616,66.30307221)(81.577116,66.30307221)
\curveto(81.233366,66.30307221)(81.233366,65.99057221)(81.233366,65.80307221)
\lineto(81.233366,61.52182221)
\lineto(76.952116,61.52182221)
\curveto(76.764616,61.52182221)(76.420866,61.52182221)(76.420866,61.17807221)
\curveto(76.420866,60.83432221)(76.733366,60.83432221)(76.952116,60.83432221)
\lineto(81.233366,60.83432221)
\lineto(81.233366,56.55307221)
\curveto(81.233366,56.36557221)(81.233366,56.02182221)(81.545866,56.02182221)
\curveto(81.889616,56.02182221)(81.889616,56.36557221)(81.889616,56.55307221)
\closepath
\moveto(81.889616,60.83432221)
}
}
\end{pspicture}\,,\quad\quad e^{-\mathbf{a}_{\dot{c}}^{c}J_{c}^{\dot{c}}}\;
\psset{xunit=.5pt,yunit=.5pt,runit=.5pt}
\begin{pspicture}[shift=-50](125.5,111)
{
\newrgbcolor{curcolor}{0 0 1}
\pscustom[linewidth=1,linecolor=curcolor]
{
\newpath
\moveto(75,35.5)
\lineto(75,0.5)
}
}
{
\newrgbcolor{curcolor}{0 0 1}
\pscustom[linewidth=1,linecolor=curcolor]
{
\newpath
\moveto(75,110.5)
\lineto(75,75.5)
}
}
{
\newrgbcolor{curcolor}{1 0 0}
\pscustom[linewidth=1,linecolor=curcolor]
{
\newpath
\moveto(95,55.5)
\curveto(121.97592,55.82787)(125,70.5)(125,110.5)
}
}
{
\newrgbcolor{curcolor}{1 0 0}
\pscustom[linewidth=1,linecolor=curcolor]
{
\newpath
\moveto(55,55.5)
\curveto(28.02408,55.17213)(25,40.5)(25,0.5)
}
}
{
\newrgbcolor{curcolor}{0 0 0}
\pscustom[linestyle=none,fillstyle=solid,fillcolor=curcolor]
{
\newpath
\moveto(12.56250367,57.48466502)
\curveto(12.87500367,57.48466502)(13.25000367,57.48466502)(13.25000367,57.85966502)
\curveto(13.25000367,58.26591502)(12.87500367,58.26591502)(12.59375367,58.26591502)
\lineto(0.65625367,58.26591502)
\curveto(0.37500367,58.26591502)(0.00000367,58.26591502)(0.00000367,57.85966502)
\curveto(0.00000367,57.48466502)(0.37500367,57.48466502)(0.65625367,57.48466502)
\closepath
\moveto(12.59375367,53.60966502)
\curveto(12.87500367,53.60966502)(13.25000367,53.60966502)(13.25000367,54.01591502)
\curveto(13.25000367,54.39091502)(12.87500367,54.39091502)(12.56250367,54.39091502)
\lineto(0.65625367,54.39091502)
\curveto(0.37500367,54.39091502)(0.00000367,54.39091502)(0.00000367,54.01591502)
\curveto(0.00000367,53.60966502)(0.37500367,53.60966502)(0.65625367,53.60966502)
\closepath
\moveto(12.59375367,53.60966502)
}
}
{
\newrgbcolor{curcolor}{1 1 1}
\pscustom[linestyle=none,fillstyle=solid,fillcolor=curcolor]
{
\newpath
\moveto(95,55.5)
\curveto(95,44.454305)(86.045695,35.5)(75,35.5)
\curveto(63.954305,35.5)(55,44.454305)(55,55.5)
\curveto(55,66.545695)(63.954305,75.5)(75,75.5)
\curveto(86.045695,75.5)(95,66.545695)(95,55.5)
\closepath
}
}
{
\newrgbcolor{curcolor}{0 0 0}
\pscustom[linewidth=1,linecolor=curcolor]
{
\newpath
\moveto(95,55.5)
\curveto(95,44.454305)(86.045695,35.5)(75,35.5)
\curveto(63.954305,35.5)(55,44.454305)(55,55.5)
\curveto(55,66.545695)(63.954305,75.5)(75,75.5)
\curveto(86.045695,75.5)(95,66.545695)(95,55.5)
\closepath
}
}
{
\newrgbcolor{curcolor}{0 0 0}
\pscustom[linestyle=none,fillstyle=solid,fillcolor=curcolor]
{
\newpath
\moveto(69.61317367,61.25012392)
\curveto(73.45692367,61.25012392)(74.45692367,60.31262392)(74.45692367,59.00012392)
\curveto(74.45692367,57.78137392)(73.48817367,55.37512392)(70.20692367,55.25012392)
\curveto(69.23817367,55.21887392)(68.55067367,54.53137392)(68.55067367,54.31262392)
\curveto(68.55067367,54.18762392)(68.61317367,54.18762392)(68.64442367,54.18762392)
\curveto(69.48817367,54.03137392)(69.89442367,53.65637392)(70.92567367,51.28137392)
\curveto(71.83192367,49.18762392)(72.36317367,48.28137392)(73.51942367,48.28137392)
\curveto(75.95692367,48.28137392)(78.23817367,50.71887392)(78.23817367,51.15637392)
\curveto(78.23817367,51.31262392)(78.08192367,51.31262392)(78.01942367,51.31262392)
\curveto(77.76942367,51.31262392)(76.98817367,51.03137392)(76.58192367,50.46887392)
\curveto(76.26942367,50.00012392)(75.83192367,49.37512392)(74.86317367,49.37512392)
\curveto(73.83192367,49.37512392)(73.20692367,50.78137392)(72.55067367,52.34387392)
\curveto(72.11317367,53.34387392)(71.76942367,54.09387392)(71.30067367,54.62512392)
\curveto(74.17567367,55.68762392)(76.14442367,57.78137392)(76.14442367,59.84387392)
\curveto(76.14442367,62.34387392)(72.80067367,62.34387392)(69.76942367,62.34387392)
\curveto(67.80067367,62.34387392)(66.67567367,62.34387392)(64.98817367,61.62512392)
\curveto(62.33192367,60.43762392)(61.95692367,58.78137392)(61.95692367,58.62512392)
\curveto(61.95692367,58.50012392)(62.05067367,58.46887392)(62.17567367,58.46887392)
\curveto(62.48817367,58.46887392)(62.95692367,58.75012392)(63.11317367,58.84387392)
\curveto(63.51942367,59.12512392)(63.58192367,59.25012392)(63.70692367,59.62512392)
\curveto(63.98817367,60.43762392)(64.55067367,61.12512392)(67.05067367,61.25012392)
\curveto(66.95692367,60.03137392)(66.76942367,58.15637392)(66.08192367,55.25012392)
\curveto(65.55067367,53.03137392)(64.83192367,50.84387392)(63.95692367,48.68762392)
\curveto(63.83192367,48.46887392)(63.83192367,48.43762392)(63.83192367,48.40637392)
\curveto(63.83192367,48.28137392)(63.98817367,48.28137392)(64.05067367,48.28137392)
\curveto(64.45692367,48.28137392)(65.26942367,48.75012392)(65.48817367,49.12512392)
\curveto(65.55067367,49.25012392)(68.08192367,54.87512392)(68.67567367,61.25012392)
\closepath
\moveto(69.61317367,61.25012392)
}
}
{
\newrgbcolor{curcolor}{0 0 0}
\pscustom[linestyle=none,fillstyle=solid,fillcolor=curcolor]
{
\newpath
\moveto(88.81166,59.10513491)
\curveto(89.03041,59.10513491)(89.37416,59.10513491)(89.37416,59.41763491)
\curveto(89.37416,59.79263491)(89.03041,59.79263491)(88.81166,59.79263491)
\lineto(80.49916,59.79263491)
\curveto(80.28041,59.79263491)(79.93666,59.79263491)(79.93666,59.44888491)
\curveto(79.93666,59.10513491)(80.24916,59.10513491)(80.49916,59.10513491)
\closepath
\moveto(88.81166,59.10513491)
}
}
\end{pspicture},
\end{equation}

\begin{equation}
\mathcal{R}_{0,I}\;
\psset{xunit=.5pt,yunit=.5pt,runit=.5pt}
\begin{pspicture}[shift=-50](125.5,111)
{
\newrgbcolor{curcolor}{0 0 1}
\pscustom[linewidth=1,linecolor=curcolor]
{
\newpath
\moveto(85,35)
\lineto(85,0)
}
}
{
\newrgbcolor{curcolor}{0 0 1}
\pscustom[linewidth=1,linecolor=curcolor]
{
\newpath
\moveto(85,110)
\lineto(85,75)
}
}
{
\newrgbcolor{curcolor}{0 0 0}
\pscustom[linestyle=none,fillstyle=solid,fillcolor=curcolor]
{
\newpath
\moveto(12.56250367,57.48466502)
\curveto(12.87500367,57.48466502)(13.25000367,57.48466502)(13.25000367,57.85966502)
\curveto(13.25000367,58.26591502)(12.87500367,58.26591502)(12.59375367,58.26591502)
\lineto(0.65625367,58.26591502)
\curveto(0.37500367,58.26591502)(0.00000367,58.26591502)(0.00000367,57.85966502)
\curveto(0.00000367,57.48466502)(0.37500367,57.48466502)(0.65625367,57.48466502)
\closepath
\moveto(12.59375367,53.60966502)
\curveto(12.87500367,53.60966502)(13.25000367,53.60966502)(13.25000367,54.01591502)
\curveto(13.25000367,54.39091502)(12.87500367,54.39091502)(12.56250367,54.39091502)
\lineto(0.65625367,54.39091502)
\curveto(0.37500367,54.39091502)(0.00000367,54.39091502)(0.00000367,54.01591502)
\curveto(0.00000367,53.60966502)(0.37500367,53.60966502)(0.65625367,53.60966502)
\closepath
\moveto(12.59375367,53.60966502)
}
}
{
\newrgbcolor{curcolor}{1 1 1}
\pscustom[linestyle=none,fillstyle=solid,fillcolor=curcolor]
{
\newpath
\moveto(105,55)
\curveto(105,43.954305)(96.045695,35)(85,35)
\curveto(73.954305,35)(65,43.954305)(65,55)
\curveto(65,66.045695)(73.954305,75)(85,75)
\curveto(96.045695,75)(105,66.045695)(105,55)
\closepath
}
}
{
\newrgbcolor{curcolor}{0 0 0}
\pscustom[linewidth=1,linecolor=curcolor]
{
\newpath
\moveto(105,55)
\curveto(105,43.954305)(96.045695,35)(85,35)
\curveto(73.954305,35)(65,43.954305)(65,55)
\curveto(65,66.045695)(73.954305,75)(85,75)
\curveto(96.045695,75)(105,66.045695)(105,55)
\closepath
}
}
{
\newrgbcolor{curcolor}{1 0 0}
\pscustom[linewidth=1,linecolor=curcolor]
{
\newpath
\moveto(40,109.5)
\lineto(40,-0.5)
}
}
{
\newrgbcolor{curcolor}{0 0 0}
\pscustom[linestyle=none,fillstyle=solid,fillcolor=curcolor]
{
\newpath
\moveto(80.92688367,61.60367732)
\curveto(84.77063367,61.60367732)(85.77063367,60.66617732)(85.77063367,59.35367732)
\curveto(85.77063367,58.13492732)(84.80188367,55.72867732)(81.52063367,55.60367732)
\curveto(80.55188367,55.57242732)(79.86438367,54.88492732)(79.86438367,54.66617732)
\curveto(79.86438367,54.54117732)(79.92688367,54.54117732)(79.95813367,54.54117732)
\curveto(80.80188367,54.38492732)(81.20813367,54.00992732)(82.23938367,51.63492732)
\curveto(83.14563367,49.54117732)(83.67688367,48.63492732)(84.83313367,48.63492732)
\curveto(87.27063367,48.63492732)(89.55188367,51.07242732)(89.55188367,51.50992732)
\curveto(89.55188367,51.66617732)(89.39563367,51.66617732)(89.33313367,51.66617732)
\curveto(89.08313367,51.66617732)(88.30188367,51.38492732)(87.89563367,50.82242732)
\curveto(87.58313367,50.35367732)(87.14563367,49.72867732)(86.17688367,49.72867732)
\curveto(85.14563367,49.72867732)(84.52063367,51.13492732)(83.86438367,52.69742732)
\curveto(83.42688367,53.69742732)(83.08313367,54.44742732)(82.61438367,54.97867732)
\curveto(85.48938367,56.04117732)(87.45813367,58.13492732)(87.45813367,60.19742732)
\curveto(87.45813367,62.69742732)(84.11438367,62.69742732)(81.08313367,62.69742732)
\curveto(79.11438367,62.69742732)(77.98938367,62.69742732)(76.30188367,61.97867732)
\curveto(73.64563367,60.79117732)(73.27063367,59.13492732)(73.27063367,58.97867732)
\curveto(73.27063367,58.85367732)(73.36438367,58.82242732)(73.48938367,58.82242732)
\curveto(73.80188367,58.82242732)(74.27063367,59.10367732)(74.42688367,59.19742732)
\curveto(74.83313367,59.47867732)(74.89563367,59.60367732)(75.02063367,59.97867732)
\curveto(75.30188367,60.79117732)(75.86438367,61.47867732)(78.36438367,61.60367732)
\curveto(78.27063367,60.38492732)(78.08313367,58.50992732)(77.39563367,55.60367732)
\curveto(76.86438367,53.38492732)(76.14563367,51.19742732)(75.27063367,49.04117732)
\curveto(75.14563367,48.82242732)(75.14563367,48.79117732)(75.14563367,48.75992732)
\curveto(75.14563367,48.63492732)(75.30188367,48.63492732)(75.36438367,48.63492732)
\curveto(75.77063367,48.63492732)(76.58313367,49.10367732)(76.80188367,49.47867732)
\curveto(76.86438367,49.60367732)(79.39563367,55.22867732)(79.98938367,61.60367732)
\closepath
\moveto(80.92688367,61.60367732)
}
}
{
\newrgbcolor{curcolor}{0 0 0}
\pscustom[linestyle=none,fillstyle=solid,fillcolor=curcolor]
{
\newpath
\moveto(96.93787,50.52192073)
\curveto(96.93787,52.05317073)(96.75037,53.17817073)(96.12537,54.14692073)
\curveto(95.68787,54.77192073)(94.81287,55.33442073)(93.71912,55.33442073)
\curveto(90.46912,55.33442073)(90.46912,51.52192073)(90.46912,50.52192073)
\curveto(90.46912,49.52192073)(90.46912,45.80317073)(93.71912,45.80317073)
\curveto(96.93787,45.80317073)(96.93787,49.52192073)(96.93787,50.52192073)
\closepath
\moveto(93.71912,46.20942073)
\curveto(93.06287,46.20942073)(92.21912,46.58442073)(91.93787,47.70942073)
\curveto(91.75037,48.52192073)(91.75037,49.67817073)(91.75037,50.70942073)
\curveto(91.75037,51.74067073)(91.75037,52.80317073)(91.93787,53.55317073)
\curveto(92.25037,54.64692073)(93.12537,54.95942073)(93.71912,54.95942073)
\curveto(94.46912,54.95942073)(95.18787,54.49067073)(95.43787,53.67817073)
\curveto(95.65662,52.92817073)(95.68787,51.92817073)(95.68787,50.70942073)
\curveto(95.68787,49.67817073)(95.68787,48.64692073)(95.50037,47.77192073)
\curveto(95.21912,46.49067073)(94.28162,46.20942073)(93.71912,46.20942073)
\closepath
\moveto(93.71912,46.20942073)
}
}
\end{pspicture}\,,\quad\quad\tilde{\mathcal{R}}_{0,I}\;
\psset{xunit=.5pt,yunit=.5pt,runit=.5pt}
\begin{pspicture}[shift=-50](125.5,111)
{
\newrgbcolor{curcolor}{0 0 1}
\pscustom[linewidth=1,linecolor=curcolor]
{
\newpath
\moveto(85,35)
\lineto(85,0)
}
}
{
\newrgbcolor{curcolor}{0 0 1}
\pscustom[linewidth=1,linecolor=curcolor]
{
\newpath
\moveto(85,110)
\lineto(85,75)
}
}
{
\newrgbcolor{curcolor}{0 0 0}
\pscustom[linestyle=none,fillstyle=solid,fillcolor=curcolor]
{
\newpath
\moveto(12.56250367,57.48466502)
\curveto(12.87500367,57.48466502)(13.25000367,57.48466502)(13.25000367,57.85966502)
\curveto(13.25000367,58.26591502)(12.87500367,58.26591502)(12.59375367,58.26591502)
\lineto(0.65625367,58.26591502)
\curveto(0.37500367,58.26591502)(0.00000367,58.26591502)(0.00000367,57.85966502)
\curveto(0.00000367,57.48466502)(0.37500367,57.48466502)(0.65625367,57.48466502)
\closepath
\moveto(12.59375367,53.60966502)
\curveto(12.87500367,53.60966502)(13.25000367,53.60966502)(13.25000367,54.01591502)
\curveto(13.25000367,54.39091502)(12.87500367,54.39091502)(12.56250367,54.39091502)
\lineto(0.65625367,54.39091502)
\curveto(0.37500367,54.39091502)(0.00000367,54.39091502)(0.00000367,54.01591502)
\curveto(0.00000367,53.60966502)(0.37500367,53.60966502)(0.65625367,53.60966502)
\closepath
\moveto(12.59375367,53.60966502)
}
}
{
\newrgbcolor{curcolor}{1 1 1}
\pscustom[linestyle=none,fillstyle=solid,fillcolor=curcolor]
{
\newpath
\moveto(105,55)
\curveto(105,43.954305)(96.045695,35)(85,35)
\curveto(73.954305,35)(65,43.954305)(65,55)
\curveto(65,66.045695)(73.954305,75)(85,75)
\curveto(96.045695,75)(105,66.045695)(105,55)
\closepath
}
}
{
\newrgbcolor{curcolor}{0 0 0}
\pscustom[linewidth=1,linecolor=curcolor]
{
\newpath
\moveto(105,55)
\curveto(105,43.954305)(96.045695,35)(85,35)
\curveto(73.954305,35)(65,43.954305)(65,55)
\curveto(65,66.045695)(73.954305,75)(85,75)
\curveto(96.045695,75)(105,66.045695)(105,55)
\closepath
}
}
{
\newrgbcolor{curcolor}{1 0 0}
\pscustom[linewidth=1,linecolor=curcolor]
{
\newpath
\moveto(40,109.5)
\lineto(40,-0.5)
}
}
{
\newrgbcolor{curcolor}{0 0 0}
\pscustom[linestyle=none,fillstyle=solid,fillcolor=curcolor]
{
\newpath
\moveto(87.32827243,66.78613466)
\lineto(87.01577243,67.06738466)
\curveto(87.01577243,67.06738466)(86.26577243,66.12988466)(85.39077243,66.12988466)
\curveto(84.92202243,66.12988466)(84.42202243,66.41113466)(84.07827243,66.62988466)
\curveto(83.54702243,66.94238466)(83.20327243,67.06738466)(82.85952243,67.06738466)
\curveto(82.10952243,67.06738466)(81.70327243,66.62988466)(80.70327243,65.50488466)
\lineto(81.01577243,65.22363466)
\curveto(81.01577243,65.22363466)(81.76577243,66.16113466)(82.64077243,66.16113466)
\curveto(83.10952243,66.16113466)(83.60952243,65.87988466)(83.95327243,65.69238466)
\curveto(84.48452243,65.34863466)(84.85952243,65.22363466)(85.20327243,65.22363466)
\curveto(85.95327243,65.22363466)(86.32827243,65.66113466)(87.32827243,66.78613466)
\closepath
\moveto(87.32827243,66.78613466)
}
}
{
\newrgbcolor{curcolor}{0 0 0}
\pscustom[linestyle=none,fillstyle=solid,fillcolor=curcolor]
{
\newpath
\moveto(81.98754367,61.25012392)
\curveto(85.83129367,61.25012392)(86.83129367,60.31262392)(86.83129367,59.00012392)
\curveto(86.83129367,57.78137392)(85.86254367,55.37512392)(82.58129367,55.25012392)
\curveto(81.61254367,55.21887392)(80.92504367,54.53137392)(80.92504367,54.31262392)
\curveto(80.92504367,54.18762392)(80.98754367,54.18762392)(81.01879367,54.18762392)
\curveto(81.86254367,54.03137392)(82.26879367,53.65637392)(83.30004367,51.28137392)
\curveto(84.20629367,49.18762392)(84.73754367,48.28137392)(85.89379367,48.28137392)
\curveto(88.33129367,48.28137392)(90.61254367,50.71887392)(90.61254367,51.15637392)
\curveto(90.61254367,51.31262392)(90.45629367,51.31262392)(90.39379367,51.31262392)
\curveto(90.14379367,51.31262392)(89.36254367,51.03137392)(88.95629367,50.46887392)
\curveto(88.64379367,50.00012392)(88.20629367,49.37512392)(87.23754367,49.37512392)
\curveto(86.20629367,49.37512392)(85.58129367,50.78137392)(84.92504367,52.34387392)
\curveto(84.48754367,53.34387392)(84.14379367,54.09387392)(83.67504367,54.62512392)
\curveto(86.55004367,55.68762392)(88.51879367,57.78137392)(88.51879367,59.84387392)
\curveto(88.51879367,62.34387392)(85.17504367,62.34387392)(82.14379367,62.34387392)
\curveto(80.17504367,62.34387392)(79.05004367,62.34387392)(77.36254367,61.62512392)
\curveto(74.70629367,60.43762392)(74.33129367,58.78137392)(74.33129367,58.62512392)
\curveto(74.33129367,58.50012392)(74.42504367,58.46887392)(74.55004367,58.46887392)
\curveto(74.86254367,58.46887392)(75.33129367,58.75012392)(75.48754367,58.84387392)
\curveto(75.89379367,59.12512392)(75.95629367,59.25012392)(76.08129367,59.62512392)
\curveto(76.36254367,60.43762392)(76.92504367,61.12512392)(79.42504367,61.25012392)
\curveto(79.33129367,60.03137392)(79.14379367,58.15637392)(78.45629367,55.25012392)
\curveto(77.92504367,53.03137392)(77.20629367,50.84387392)(76.33129367,48.68762392)
\curveto(76.20629367,48.46887392)(76.20629367,48.43762392)(76.20629367,48.40637392)
\curveto(76.20629367,48.28137392)(76.36254367,48.28137392)(76.42504367,48.28137392)
\curveto(76.83129367,48.28137392)(77.64379367,48.75012392)(77.86254367,49.12512392)
\curveto(77.92504367,49.25012392)(80.45629367,54.87512392)(81.05004367,61.25012392)
\closepath
\moveto(81.98754367,61.25012392)
}
}
{
\newrgbcolor{curcolor}{0 0 0}
\pscustom[linestyle=none,fillstyle=solid,fillcolor=curcolor]
{
\newpath
\moveto(97.99853,50.16836733)
\curveto(97.99853,51.69961733)(97.81103,52.82461733)(97.18603,53.79336733)
\curveto(96.74853,54.41836733)(95.87353,54.98086733)(94.77978,54.98086733)
\curveto(91.52978,54.98086733)(91.52978,51.16836733)(91.52978,50.16836733)
\curveto(91.52978,49.16836733)(91.52978,45.44961733)(94.77978,45.44961733)
\curveto(97.99853,45.44961733)(97.99853,49.16836733)(97.99853,50.16836733)
\closepath
\moveto(94.77978,45.85586733)
\curveto(94.12353,45.85586733)(93.27978,46.23086733)(92.99853,47.35586733)
\curveto(92.81103,48.16836733)(92.81103,49.32461733)(92.81103,50.35586733)
\curveto(92.81103,51.38711733)(92.81103,52.44961733)(92.99853,53.19961733)
\curveto(93.31103,54.29336733)(94.18603,54.60586733)(94.77978,54.60586733)
\curveto(95.52978,54.60586733)(96.24853,54.13711733)(96.49853,53.32461733)
\curveto(96.71728,52.57461733)(96.74853,51.57461733)(96.74853,50.35586733)
\curveto(96.74853,49.32461733)(96.74853,48.29336733)(96.56103,47.41836733)
\curveto(96.27978,46.13711733)(95.34228,45.85586733)(94.77978,45.85586733)
\closepath
\moveto(94.77978,45.85586733)
}
}
\end{pspicture}.\label{eq:diag:RRt0}
\end{equation}
$\mathcal{R}_{I,0}$ and $\tilde{\mathcal{R}}_{I,0}$ act trivially
in the auxiliary space, this is depicted by the straight line in (\ref{eq:diag:RRt0}).
The label $I$ is suppressed in the pictures. Let us now construct
the two expressions of $\mathcal{R}_{I}$ given in (\ref{eq:RI})
and (\ref{eq:RI_anti}) . Using the ingredients above and the multiplication
rules (\ref{eq:O1O2}) and (\ref{eq:O1circO2}) one finds

\begin{equation}
\mathcal{R}_{I}\;\input{content/pictures/NewDiagrams/Rdec},\quad\quad\mathcal{R}_{I}\;\input{content/pictures/NewDiagrams/Rtdec}\,.\label{eq:rdec}
\end{equation}
When reading the diagrams from bottom to top it becomes clear that
the expression on the left hand side is normal ordered, while the
expression on the right hand side is anti-normal ordered in the oscillator
space.

\subsubsection{Projection properties{\normalsize \label{sub:Projection-property}}}

In Section~\ref{sec:Projection-property-of} we argued that at the
special points $\hat{z}$ and $\check{z}$ some R-operators $\mathcal{R}$
of certain cardinality $\vert I\vert$ decompose into an outer product,
see (\ref{eq:RProjector}). This fact is a consequence of the degeneration
(to rank $1$) of $\mathcal{R}_{0,I}$ and $\tilde{\mathcal{R}}_{0,I}$
for generalized rectangular representations. In the diagrammatics
introduced, the building blocks of (\ref{eq:RProjector}) are denoted
by 
\begin{equation}
e^{\bar{\mathbf{a}}_{c}^{\dot{c}}J_{\dot{c}}^{c}}\vert hws\rangle\;
\psset{xunit=.5pt,yunit=.5pt,runit=.5pt}
\begin{pspicture}[shift=-50](123,110.5)
{
\newrgbcolor{curcolor}{1 0 0}
\pscustom[linewidth=1,linecolor=curcolor]
{
\newpath
\moveto(92.14645,55)
\curveto(119.12237,55.32787)(122.14645,70)(122.14645,110)
}
}
{
\newrgbcolor{curcolor}{0 0 1}
\pscustom[linewidth=1,linecolor=curcolor]
{
\newpath
\moveto(72,35.5)
\lineto(72,0.5)
}
}
{
\newrgbcolor{curcolor}{1 0 0}
\pscustom[linewidth=1,linecolor=curcolor]
{
\newpath
\moveto(52,55.5)
\curveto(25.02408,55.17213)(22,40.5)(22,0.5)
}
}
{
\newrgbcolor{curcolor}{0 0 0}
\pscustom[linestyle=none,fillstyle=solid,fillcolor=curcolor]
{
\newpath
\moveto(12.56250367,56.98466502)
\curveto(12.87500367,56.98466502)(13.25000367,56.98466502)(13.25000367,57.35966502)
\curveto(13.25000367,57.76591502)(12.87500367,57.76591502)(12.59375367,57.76591502)
\lineto(0.65625367,57.76591502)
\curveto(0.37500367,57.76591502)(0.00000367,57.76591502)(0.00000367,57.35966502)
\curveto(0.00000367,56.98466502)(0.37500367,56.98466502)(0.65625367,56.98466502)
\closepath
\moveto(12.59375367,53.10966502)
\curveto(12.87500367,53.10966502)(13.25000367,53.10966502)(13.25000367,53.51591502)
\curveto(13.25000367,53.89091502)(12.87500367,53.89091502)(12.56250367,53.89091502)
\lineto(0.65625367,53.89091502)
\curveto(0.37500367,53.89091502)(0.00000367,53.89091502)(0.00000367,53.51591502)
\curveto(0.00000367,53.10966502)(0.37500367,53.10966502)(0.65625367,53.10966502)
\closepath
\moveto(12.59375367,53.10966502)
}
}
{
\newrgbcolor{curcolor}{1 1 1}
\pscustom[linestyle=none,fillstyle=solid,fillcolor=curcolor]
{
\newpath
\moveto(92.000004,55.5)
\curveto(92.000004,44.454305)(83.045699,35.5)(72.000004,35.5)
\curveto(60.954309,35.5)(52.000004,44.454305)(52.000004,55.5)
\curveto(52.000004,66.545695)(60.954309,75.5)(72.000004,75.5)
\curveto(83.045699,75.5)(92.000004,66.545695)(92.000004,55.5)
\closepath
}
}
{
\newrgbcolor{curcolor}{0 0 0}
\pscustom[linewidth=1,linecolor=curcolor]
{
\newpath
\moveto(92.000004,55.5)
\curveto(92.000004,44.454305)(83.045699,35.5)(72.000004,35.5)
\curveto(60.954309,35.5)(52.000004,44.454305)(52.000004,55.5)
\curveto(52.000004,66.545695)(60.954309,75.5)(72.000004,75.5)
\curveto(83.045699,75.5)(92.000004,66.545695)(92.000004,55.5)
\closepath
}
}
{
\newrgbcolor{curcolor}{0 0 0}
\pscustom[linestyle=none,fillstyle=solid,fillcolor=curcolor]
{
\newpath
\moveto(72.57602367,54.95171122)
\lineto(78.13852367,54.95171122)
\curveto(78.41977367,54.95171122)(78.79477367,54.95171122)(78.79477367,55.35796122)
\curveto(78.79477367,55.73296122)(78.41977367,55.73296122)(78.13852367,55.73296122)
\lineto(72.57602367,55.73296122)
\lineto(72.57602367,61.32671122)
\curveto(72.57602367,61.60796122)(72.57602367,61.98296122)(72.16977367,61.98296122)
\curveto(71.76352367,61.98296122)(71.76352367,61.60796122)(71.76352367,61.32671122)
\lineto(71.76352367,55.73296122)
\lineto(66.20102367,55.73296122)
\curveto(65.91977367,55.73296122)(65.54477367,55.73296122)(65.54477367,55.35796122)
\curveto(65.54477367,54.95171122)(65.91977367,54.95171122)(66.20102367,54.95171122)
\lineto(71.76352367,54.95171122)
\lineto(71.76352367,49.35796122)
\curveto(71.76352367,49.07671122)(71.76352367,48.70171122)(72.16977367,48.70171122)
\curveto(72.57602367,48.70171122)(72.57602367,49.07671122)(72.57602367,49.35796122)
\closepath
\moveto(72.57602367,54.95171122)
}
}
\end{pspicture},\quad\quad\langle hws\vert e^{-\mathbf{a}_{\dot{c}}^{c}J_{c}^{\dot{c}}}\;
\psset{xunit=.5pt,yunit=.5pt,runit=.5pt}
\begin{pspicture}[shift=-50](122.64644623,111.19999695)
{
\newrgbcolor{curcolor}{1 0 0}
\pscustom[linewidth=1,linecolor=curcolor]
{
\newpath
\moveto(92.14645,54.99999695)
\curveto(119.12237,55.32786695)(122.14645,69.99999695)(122.14645,109.99999695)
}
}
{
\newrgbcolor{curcolor}{0 0 1}
\pscustom[linewidth=1,linecolor=curcolor]
{
\newpath
\moveto(71.8,110.69999695)
\lineto(71.8,75.69999695)
}
}
{
\newrgbcolor{curcolor}{1 0 0}
\pscustom[linewidth=1,linecolor=curcolor]
{
\newpath
\moveto(52,55.49999695)
\curveto(25.02408,55.17212695)(22,40.49999695)(22,0.49999695)
}
}
{
\newrgbcolor{curcolor}{0 0 0}
\pscustom[linestyle=none,fillstyle=solid,fillcolor=curcolor]
{
\newpath
\moveto(12.56250367,56.98466197)
\curveto(12.87500367,56.98466197)(13.25000367,56.98466197)(13.25000367,57.35966197)
\curveto(13.25000367,57.76591197)(12.87500367,57.76591197)(12.59375367,57.76591197)
\lineto(0.65625367,57.76591197)
\curveto(0.37500367,57.76591197)(0.00000367,57.76591197)(0.00000367,57.35966197)
\curveto(0.00000367,56.98466197)(0.37500367,56.98466197)(0.65625367,56.98466197)
\closepath
\moveto(12.59375367,53.10966197)
\curveto(12.87500367,53.10966197)(13.25000367,53.10966197)(13.25000367,53.51591197)
\curveto(13.25000367,53.89091197)(12.87500367,53.89091197)(12.56250367,53.89091197)
\lineto(0.65625367,53.89091197)
\curveto(0.37500367,53.89091197)(0.00000367,53.89091197)(0.00000367,53.51591197)
\curveto(0.00000367,53.10966197)(0.37500367,53.10966197)(0.65625367,53.10966197)
\closepath
\moveto(12.59375367,53.10966197)
}
}
{
\newrgbcolor{curcolor}{1 1 1}
\pscustom[linestyle=none,fillstyle=solid,fillcolor=curcolor]
{
\newpath
\moveto(92.000004,55.49999695)
\curveto(92.000004,44.45430195)(83.045699,35.49999695)(72.000004,35.49999695)
\curveto(60.954309,35.49999695)(52.000004,44.45430195)(52.000004,55.49999695)
\curveto(52.000004,66.54569194)(60.954309,75.49999695)(72.000004,75.49999695)
\curveto(83.045699,75.49999695)(92.000004,66.54569194)(92.000004,55.49999695)
\closepath
}
}
{
\newrgbcolor{curcolor}{0 0 0}
\pscustom[linewidth=1,linecolor=curcolor]
{
\newpath
\moveto(92.000004,55.49999695)
\curveto(92.000004,44.45430195)(83.045699,35.49999695)(72.000004,35.49999695)
\curveto(60.954309,35.49999695)(52.000004,44.45430195)(52.000004,55.49999695)
\curveto(52.000004,66.54569194)(60.954309,75.49999695)(72.000004,75.49999695)
\curveto(83.045699,75.49999695)(92.000004,66.54569194)(92.000004,55.49999695)
\closepath
}
}
{
\newrgbcolor{curcolor}{0 0 0}
\pscustom[linestyle=none,fillstyle=solid,fillcolor=curcolor]
{
\newpath
\moveto(77.50410367,54.95538787)
\curveto(77.84785367,54.95538787)(78.22285367,54.95538787)(78.22285367,55.36163787)
\curveto(78.22285367,55.73663787)(77.84785367,55.73663787)(77.50410367,55.73663787)
\lineto(66.72285367,55.73663787)
\curveto(66.37910367,55.73663787)(66.03535367,55.73663787)(66.03535367,55.36163787)
\curveto(66.03535367,54.95538787)(66.37910367,54.95538787)(66.72285367,54.95538787)
\closepath
\moveto(77.50410367,54.95538787)
}
}
\end{pspicture}.\label{eq:ApAm}
\end{equation}
As the above expressions are {}``vector'' and {}``covector'' in
the quantum space, a quantum space operator acts on them by left and
right multiplication, respectively. This is indicated by one missing
ingoing/outgoing vertical line. Using the notions of the two defined
products we see that according to (\ref{eq:RProjector}) at the special
points (\ref{eq:rdec}) simplifies to 

\begin{equation}
\mathcal{R}_{I}(\hat{z})\;
\psset{xunit=.5pt,yunit=.5pt,runit=.5pt}
\begin{pspicture}[shift=-50](193,111)
{
\newrgbcolor{curcolor}{0 0 1}
\pscustom[linewidth=1,linecolor=curcolor]
{
\newpath
\moveto(71.99999633,35.5)
\lineto(71.99999633,0.5)
}
}
{
\newrgbcolor{curcolor}{1 0 0}
\pscustom[linewidth=1,linecolor=curcolor]
{
\newpath
\moveto(91.99999633,55.5)
\curveto(114.41354633,55.63359)(138.29021633,55.508)(156.99999633,55.5)
}
}
{
\newrgbcolor{curcolor}{1 0 0}
\pscustom[linewidth=1,linecolor=curcolor]
{
\newpath
\moveto(51.99999633,55.5)
\curveto(25.02407633,55.17213)(21.99999633,40.5)(21.99999633,0.5)
}
}
{
\newrgbcolor{curcolor}{0 0 0}
\pscustom[linestyle=none,fillstyle=solid,fillcolor=curcolor]
{
\newpath
\moveto(12.5625,56.98466502)
\curveto(12.875,56.98466502)(13.25,56.98466502)(13.25,57.35966502)
\curveto(13.25,57.76591502)(12.875,57.76591502)(12.59375,57.76591502)
\lineto(0.65625,57.76591502)
\curveto(0.375,57.76591502)(0,57.76591502)(0,57.35966502)
\curveto(0,56.98466502)(0.375,56.98466502)(0.65625,56.98466502)
\closepath
\moveto(12.59375,53.10966502)
\curveto(12.875,53.10966502)(13.25,53.10966502)(13.25,53.51591502)
\curveto(13.25,53.89091502)(12.875,53.89091502)(12.5625,53.89091502)
\lineto(0.65625,53.89091502)
\curveto(0.375,53.89091502)(0,53.89091502)(0,53.51591502)
\curveto(0,53.10966502)(0.375,53.10966502)(0.65625,53.10966502)
\closepath
\moveto(12.59375,53.10966502)
}
}
{
\newrgbcolor{curcolor}{1 1 1}
\pscustom[linestyle=none,fillstyle=solid,fillcolor=curcolor]
{
\newpath
\moveto(92.00000033,55.5)
\curveto(92.00000033,44.454305)(83.04569532,35.5)(72.00000033,35.5)
\curveto(60.95430533,35.5)(52.00000033,44.454305)(52.00000033,55.5)
\curveto(52.00000033,66.545695)(60.95430533,75.5)(72.00000033,75.5)
\curveto(83.04569532,75.5)(92.00000033,66.545695)(92.00000033,55.5)
\closepath
}
}
{
\newrgbcolor{curcolor}{0 0 0}
\pscustom[linewidth=1,linecolor=curcolor]
{
\newpath
\moveto(92.00000033,55.5)
\curveto(92.00000033,44.454305)(83.04569532,35.5)(72.00000033,35.5)
\curveto(60.95430533,35.5)(52.00000033,44.454305)(52.00000033,55.5)
\curveto(52.00000033,66.545695)(60.95430533,75.5)(72.00000033,75.5)
\curveto(83.04569532,75.5)(92.00000033,66.545695)(92.00000033,55.5)
\closepath
}
}
{
\newrgbcolor{curcolor}{0 0 0}
\pscustom[linestyle=none,fillstyle=solid,fillcolor=curcolor]
{
\newpath
\moveto(72.57602,54.95171122)
\lineto(78.13852,54.95171122)
\curveto(78.41977,54.95171122)(78.79477,54.95171122)(78.79477,55.35796122)
\curveto(78.79477,55.73296122)(78.41977,55.73296122)(78.13852,55.73296122)
\lineto(72.57602,55.73296122)
\lineto(72.57602,61.32671122)
\curveto(72.57602,61.60796122)(72.57602,61.98296122)(72.16977,61.98296122)
\curveto(71.76352,61.98296122)(71.76352,61.60796122)(71.76352,61.32671122)
\lineto(71.76352,55.73296122)
\lineto(66.20102,55.73296122)
\curveto(65.91977,55.73296122)(65.54477,55.73296122)(65.54477,55.35796122)
\curveto(65.54477,54.95171122)(65.91977,54.95171122)(66.20102,54.95171122)
\lineto(71.76352,54.95171122)
\lineto(71.76352,49.35796122)
\curveto(71.76352,49.07671122)(71.76352,48.70171122)(72.16977,48.70171122)
\curveto(72.57602,48.70171122)(72.57602,49.07671122)(72.57602,49.35796122)
\closepath
\moveto(72.57602,54.95171122)
}
}
{
\newrgbcolor{curcolor}{0 0 1}
\pscustom[linewidth=1,linecolor=curcolor]
{
\newpath
\moveto(142.49999633,110.5)
\lineto(142.49999633,75.5)
}
}
{
\newrgbcolor{curcolor}{1 0 0}
\pscustom[linewidth=1,linecolor=curcolor]
{
\newpath
\moveto(162.49999633,55.5)
\curveto(189.47591633,55.82787)(192.49999633,70.5)(192.49999633,110.5)
}
}
{
\newrgbcolor{curcolor}{1 1 1}
\pscustom[linestyle=none,fillstyle=solid,fillcolor=curcolor]
{
\newpath
\moveto(162.49999633,55.5)
\curveto(162.49999633,44.454305)(153.54569132,35.5)(142.49999633,35.5)
\curveto(131.45430133,35.5)(122.49999633,44.454305)(122.49999633,55.5)
\curveto(122.49999633,66.545695)(131.45430133,75.5)(142.49999633,75.5)
\curveto(153.54569132,75.5)(162.49999633,66.545695)(162.49999633,55.5)
\closepath
}
}
{
\newrgbcolor{curcolor}{0 0 0}
\pscustom[linewidth=1,linecolor=curcolor]
{
\newpath
\moveto(162.49999633,55.5)
\curveto(162.49999633,44.454305)(153.54569132,35.5)(142.49999633,35.5)
\curveto(131.45430133,35.5)(122.49999633,44.454305)(122.49999633,55.5)
\curveto(122.49999633,66.545695)(131.45430133,75.5)(142.49999633,75.5)
\curveto(153.54569132,75.5)(162.49999633,66.545695)(162.49999633,55.5)
\closepath
}
}
{
\newrgbcolor{curcolor}{0 0 0}
\pscustom[linestyle=none,fillstyle=solid,fillcolor=curcolor]
{
\newpath
\moveto(149.04275584,54.70171122)
\curveto(149.38650584,54.70171122)(149.76150584,54.70171122)(149.76150584,55.10796122)
\curveto(149.76150584,55.48296122)(149.38650584,55.48296122)(149.04275584,55.48296122)
\lineto(138.26150584,55.48296122)
\curveto(137.91775584,55.48296122)(137.57400584,55.48296122)(137.57400584,55.10796122)
\curveto(137.57400584,54.70171122)(137.91775584,54.70171122)(138.26150584,54.70171122)
\closepath
\moveto(149.04275584,54.70171122)
}
}
\end{pspicture},\quad\quad\mathcal{R}_{I}(\check{z})\;
\psset{xunit=.5pt,yunit=.5pt,runit=.5pt}
\begin{pspicture}[shift=-50](193,111)
{
\newrgbcolor{curcolor}{0 0 1}
\pscustom[linewidth=1,linecolor=curcolor]
{
\newpath
\moveto(71.99999633,110.5)
\lineto(71.99999633,75.5)
}
}
{
\newrgbcolor{curcolor}{1 0 0}
\pscustom[linewidth=1,linecolor=curcolor]
{
\newpath
\moveto(91.99999633,55.5)
\curveto(114.41354633,55.63359)(138.29021633,55.508)(156.99999633,55.5)
}
}
{
\newrgbcolor{curcolor}{1 0 0}
\pscustom[linewidth=1,linecolor=curcolor]
{
\newpath
\moveto(51.99999633,55.5)
\curveto(25.02407633,55.17213)(21.99999633,40.5)(21.99999633,0.5)
}
}
{
\newrgbcolor{curcolor}{0 0 0}
\pscustom[linestyle=none,fillstyle=solid,fillcolor=curcolor]
{
\newpath
\moveto(12.5625,56.98466502)
\curveto(12.875,56.98466502)(13.25,56.98466502)(13.25,57.35966502)
\curveto(13.25,57.76591502)(12.875,57.76591502)(12.59375,57.76591502)
\lineto(0.65625,57.76591502)
\curveto(0.375,57.76591502)(0,57.76591502)(0,57.35966502)
\curveto(0,56.98466502)(0.375,56.98466502)(0.65625,56.98466502)
\closepath
\moveto(12.59375,53.10966502)
\curveto(12.875,53.10966502)(13.25,53.10966502)(13.25,53.51591502)
\curveto(13.25,53.89091502)(12.875,53.89091502)(12.5625,53.89091502)
\lineto(0.65625,53.89091502)
\curveto(0.375,53.89091502)(0,53.89091502)(0,53.51591502)
\curveto(0,53.10966502)(0.375,53.10966502)(0.65625,53.10966502)
\closepath
\moveto(12.59375,53.10966502)
}
}
{
\newrgbcolor{curcolor}{1 1 1}
\pscustom[linestyle=none,fillstyle=solid,fillcolor=curcolor]
{
\newpath
\moveto(92.00000033,55.5)
\curveto(92.00000033,44.454305)(83.04569532,35.5)(72.00000033,35.5)
\curveto(60.95430533,35.5)(52.00000033,44.454305)(52.00000033,55.5)
\curveto(52.00000033,66.545695)(60.95430533,75.5)(72.00000033,75.5)
\curveto(83.04569532,75.5)(92.00000033,66.545695)(92.00000033,55.5)
\closepath
}
}
{
\newrgbcolor{curcolor}{0 0 0}
\pscustom[linewidth=1,linecolor=curcolor]
{
\newpath
\moveto(92.00000033,55.5)
\curveto(92.00000033,44.454305)(83.04569532,35.5)(72.00000033,35.5)
\curveto(60.95430533,35.5)(52.00000033,44.454305)(52.00000033,55.5)
\curveto(52.00000033,66.545695)(60.95430533,75.5)(72.00000033,75.5)
\curveto(83.04569532,75.5)(92.00000033,66.545695)(92.00000033,55.5)
\closepath
}
}
{
\newrgbcolor{curcolor}{0 0 1}
\pscustom[linewidth=1,linecolor=curcolor]
{
\newpath
\moveto(142.49999633,35.5)
\lineto(142.49999633,0.5)
}
}
{
\newrgbcolor{curcolor}{1 0 0}
\pscustom[linewidth=1,linecolor=curcolor]
{
\newpath
\moveto(162.49999633,55.5)
\curveto(189.47591633,55.82787)(192.49999633,70.5)(192.49999633,110.5)
}
}
{
\newrgbcolor{curcolor}{1 1 1}
\pscustom[linestyle=none,fillstyle=solid,fillcolor=curcolor]
{
\newpath
\moveto(162.49999633,55.5)
\curveto(162.49999633,44.454305)(153.54569132,35.5)(142.49999633,35.5)
\curveto(131.45430133,35.5)(122.49999633,44.454305)(122.49999633,55.5)
\curveto(122.49999633,66.545695)(131.45430133,75.5)(142.49999633,75.5)
\curveto(153.54569132,75.5)(162.49999633,66.545695)(162.49999633,55.5)
\closepath
}
}
{
\newrgbcolor{curcolor}{0 0 0}
\pscustom[linewidth=1,linecolor=curcolor]
{
\newpath
\moveto(162.49999633,55.5)
\curveto(162.49999633,44.454305)(153.54569132,35.5)(142.49999633,35.5)
\curveto(131.45430133,35.5)(122.49999633,44.454305)(122.49999633,55.5)
\curveto(122.49999633,66.545695)(131.45430133,75.5)(142.49999633,75.5)
\curveto(153.54569132,75.5)(162.49999633,66.545695)(162.49999633,55.5)
\closepath
}
}
{
\newrgbcolor{curcolor}{0 0 0}
\pscustom[linestyle=none,fillstyle=solid,fillcolor=curcolor]
{
\newpath
\moveto(77.79275584,55.20171122)
\curveto(78.13650584,55.20171122)(78.51150584,55.20171122)(78.51150584,55.60796122)
\curveto(78.51150584,55.98296122)(78.13650584,55.98296122)(77.79275584,55.98296122)
\lineto(67.01150584,55.98296122)
\curveto(66.66775584,55.98296122)(66.32400584,55.98296122)(66.32400584,55.60796122)
\curveto(66.32400584,55.20171122)(66.66775584,55.20171122)(67.01150584,55.20171122)
\closepath
\moveto(77.79275584,55.20171122)
}
}
{
\newrgbcolor{curcolor}{0 0 0}
\pscustom[linestyle=none,fillstyle=solid,fillcolor=curcolor]
{
\newpath
\moveto(143.07602,54.45171122)
\lineto(148.63852,54.45171122)
\curveto(148.91977,54.45171122)(149.29477,54.45171122)(149.29477,54.85796122)
\curveto(149.29477,55.23296122)(148.91977,55.23296122)(148.63852,55.23296122)
\lineto(143.07602,55.23296122)
\lineto(143.07602,60.82671122)
\curveto(143.07602,61.10796122)(143.07602,61.48296122)(142.66977,61.48296122)
\curveto(142.26352,61.48296122)(142.26352,61.10796122)(142.26352,60.82671122)
\lineto(142.26352,55.23296122)
\lineto(136.70102,55.23296122)
\curveto(136.41977,55.23296122)(136.04477,55.23296122)(136.04477,54.85796122)
\curveto(136.04477,54.45171122)(136.41977,54.45171122)(136.70102,54.45171122)
\lineto(142.26352,54.45171122)
\lineto(142.26352,48.85796122)
\curveto(142.26352,48.57671122)(142.26352,48.20171122)(142.66977,48.20171122)
\curveto(143.07602,48.20171122)(143.07602,48.57671122)(143.07602,48.85796122)
\closepath
\moveto(143.07602,54.45171122)
}
}
\end{pspicture}\,.\label{eq:rdecproj}
\end{equation}

\subsubsection{Baxter Q-operators\label{sub:Baxter-Q-operators}}

Baxter Q-operators can be built from the monodromy of the R-operators $\mathcal{R}$
as introduced in Section~\ref{sec:Q-opgln}. Here, we will concentrate
on the Q-operators constructed out of the R-operators $\mathcal{R}$
that satisfy condition (\ref{eq:RProjector}), see Section~\ref{sec:Projection-property-of}
for more details. Hereafter, the index $I$ of the chosen Q-operator
will be suppressed. The diagrammatics for these R-operators $\mathcal{R}$
was developed above. From (\ref{eq:rdecproj}) it is clear that the
corresponding Q-operators at the special points are given by 
\begin{equation}
\bar{\bf Q}(\hat{z})\;\input{content/pictures/NewDiagrams/Qpresize}\label{eq:qp}
\end{equation}
and
\begin{equation}
\bar{\bf Q}(\check{z})\;\input{content/pictures/NewDiagrams/Qmresize}.\label{eq:qm}
\end{equation}
Here $\mathcal{D}$ denotes the regulator in (\ref{D-parton}).
For convenience we recall that in (\ref{eq:qp}) and (\ref{eq:qm})
there is one ingoing and one outgoing vertical line for each spin-chain
site. As indicated in the picture, the auxiliary space is closed by
the trace, see (\ref{eq:Qop}).

\subsubsection{Shift mechanism\label{sub:Shift-mechanism-and}}

The homogeneous spin-chain has the property of being translationally
invariant; the shift operator defined as 
\begin{equation}
\mathbf{U}\, X_{n}\,\mathbf{U}^{-1}=X_{n-1}\,,\quad\quad\mathbf{U}\, X_{1}\,\mathbf{U}^{-1}=f_{L}(\Phi)\, X_{L}\, f_{L}^{-1}(\Phi)\,,\label{eq:shiftX}
\end{equation}
commutes with the Hamiltonian and all generalized transfer matrices.
The operator $f_{L}(\Phi)$ arises from the twisted boundary conditions
and is explicitly given below.

The shift operator can be written as 
\begin{equation}
\mathbf{U}=f_{L}(\Phi)\,\mathbf{P}_{L,L-1}\mathbf{P}_{L-1,L-2}\cdots\mathbf{P}_{2,1}\,,
\end{equation}
cf. \eqref{shiftop}, where $\mathbf{P}_{i+1,i}$ acts as a permutation on site $i$ and
$i+1$ in the quantum space. The main result of this subsection is
to show the relation 
\begin{equation}
\bar{\bf Q}(\check{z})=\mathbf{U}\,\bar{\bf Q}(\hat{z})\,.\label{eq:UQgQ-1}
\end{equation}
The label $I$ has been omitted following the logic as in Section~\ref{sub:Projection-property}
and \ref{sub:Baxter-Q-operators}. This equation is immediately proven
once it is rewritten in the diagrammatic language developed previously, cf. \eqref{shiftpic} and \eqref{eq:qp}:
\begin{equation}
\input{content/pictures/NewDiagrams/QeqUQ}\,.\label{eq:uqpic}
\end{equation}
The only non-trivial step in the proof is to move the last term $
\psset{xunit=.2pt,yunit=.2pt,runit=.2pt}
\begin{pspicture}(41,41)
{
\newrgbcolor{curcolor}{1 1 1}
\pscustom[linestyle=none,fillstyle=solid,fillcolor=curcolor]
{
\newpath
\moveto(40.5,20.5)
\curveto(40.5,9.454305)(31.545695,0.5)(20.5,0.5)
\curveto(9.454305,0.5)(0.5,9.454305)(0.5,20.5)
\curveto(0.5,31.545695)(9.454305,40.5)(20.5,40.5)
\curveto(31.545695,40.5)(40.5,31.545695)(40.5,20.5)
\closepath
}
}
{
\newrgbcolor{curcolor}{0 0 0}
\pscustom[linewidth=1,linecolor=curcolor]
{
\newpath
\moveto(40.5,20.5)
\curveto(40.5,9.454305)(31.545695,0.5)(20.5,0.5)
\curveto(9.454305,0.5)(0.5,9.454305)(0.5,20.5)
\curveto(0.5,31.545695)(9.454305,40.5)(20.5,40.5)
\curveto(31.545695,40.5)(40.5,31.545695)(40.5,20.5)
\closepath
}
}
{
\newrgbcolor{curcolor}{0 0 0}
\pscustom[linestyle=none,fillstyle=solid,fillcolor=curcolor]
{
\newpath
\moveto(20.86829663,19.34981438)
\lineto(31.3141774,19.34981438)
\curveto(31.84233991,19.34981438)(32.54655659,19.34981438)(32.54655659,20.11271578)
\curveto(32.54655659,20.81693246)(31.84233991,20.81693246)(31.3141774,20.81693246)
\lineto(20.86829663,20.81693246)
\lineto(20.86829663,31.32149796)
\curveto(20.86829663,31.84966047)(20.86829663,32.55387715)(20.10539522,32.55387715)
\curveto(19.34249382,32.55387715)(19.34249382,31.84966047)(19.34249382,31.32149796)
\lineto(19.34249382,20.81693246)
\lineto(8.89661304,20.81693246)
\curveto(8.36845053,20.81693246)(7.66423385,20.81693246)(7.66423385,20.11271578)
\curveto(7.66423385,19.34981438)(8.36845053,19.34981438)(8.89661304,19.34981438)
\lineto(19.34249382,19.34981438)
\lineto(19.34249382,8.84524888)
\curveto(19.34249382,8.31708637)(19.34249382,7.61286969)(20.10539522,7.61286969)
\curveto(20.86829663,7.61286969)(20.86829663,8.31708637)(20.86829663,8.84524888)
\closepath
\moveto(20.86829663,19.34981438)
}
}
\end{pspicture}$
in the left hand side of (\ref{eq:uqpic}) through the regulator $
\psset{xunit=.2pt,yunit=.2pt,runit=.2pt}
\begin{pspicture}(41,41)
{
\newrgbcolor{curcolor}{1 1 1}
\pscustom[linestyle=none,fillstyle=solid,fillcolor=curcolor]
{
\newpath
\moveto(40.5,20.5)
\curveto(40.5,9.454305)(31.545695,0.5)(20.5,0.5)
\curveto(9.454305,0.5)(0.5,9.454305)(0.5,20.5)
\curveto(0.5,31.545695)(9.454305,40.5)(20.5,40.5)
\curveto(31.545695,40.5)(40.5,31.545695)(40.5,20.5)
\closepath
}
}
{
\newrgbcolor{curcolor}{0 0 0}
\pscustom[linewidth=1,linecolor=curcolor]
{
\newpath
\moveto(40.5,20.5)
\curveto(40.5,9.454305)(31.545695,0.5)(20.5,0.5)
\curveto(9.454305,0.5)(0.5,9.454305)(0.5,20.5)
\curveto(0.5,31.545695)(9.454305,40.5)(20.5,40.5)
\curveto(31.545695,40.5)(40.5,31.545695)(40.5,20.5)
\closepath
}
}
{
\newrgbcolor{curcolor}{0 0 0}
\pscustom[linestyle=none,fillstyle=solid,fillcolor=curcolor]
{
\newpath
\moveto(14.64935033,8.91556116)
\curveto(22.65924018,8.91556116)(33.94834668,15.04393326)(33.94834668,24.39776436)
\curveto(33.94834668,27.46195041)(32.49689013,29.34346816)(30.83040298,30.41862116)
\curveto(27.87373223,32.35389656)(24.70203088,32.35389656)(21.42281423,32.35389656)
\curveto(18.51990113,32.35389656)(16.47711043,32.35389656)(13.57419733,31.11747061)
\curveto(9.05855473,29.07467991)(8.35970528,26.22552446)(8.35970528,25.95673621)
\curveto(8.35970528,25.74170561)(8.52097823,25.68794796)(8.73600883,25.68794796)
\curveto(9.27358533,25.68794796)(10.07995008,26.17176681)(10.34873833,26.33303976)
\curveto(11.04758778,26.81685861)(11.15510308,27.03188921)(11.37013368,27.67698101)
\curveto(11.85395253,29.07467991)(12.82159023,30.25734821)(17.12220223,30.47237881)
\curveto(16.58462573,23.26885371)(14.81062328,16.71042041)(12.44528668,10.74332126)
\curveto(11.15510308,10.31326006)(10.34873833,9.50689531)(10.34873833,9.13059176)
\curveto(10.34873833,8.96931881)(10.34873833,8.91556116)(11.10134543,8.91556116)
\closepath
\moveto(14.81062328,10.79707891)
\curveto(18.73493173,20.36594061)(19.43378118,26.33303976)(19.91760003,30.47237881)
\curveto(22.22917898,30.47237881)(30.99167593,30.47237881)(30.99167593,22.89255016)
\curveto(30.99167593,16.17284391)(24.97081913,10.79707891)(16.69214103,10.79707891)
\closepath
\moveto(14.81062328,10.79707891)
}
}
\end{pspicture}$.
This is done using the relation

\begin{equation}
\input{content/pictures/NewDiagrams/eqn}\,.
\end{equation}
A direct computation shows that 
\begin{equation}
f(\Phi)=\exp\left\{ i\sum_{c\in I}\Phi_{c}(J_{c}^{c}-\lambda_{I})+i\sum_{\dot{c}\in\bar{I}}\Phi_{\dot{c}}(J_{\dot{c}}^{\dot{c}}-\bar{\lambda}_{I})\right\} \,.
\end{equation}
This proves relation (\ref{eq:UQgQ-1}) for the large class of generalized
rectangular representations. Using (\ref{eq:UQgQ-1}) and the form
of the Q-operator eigenvalues in terms of Bethe roots $\left\{ z_{i}\right\} _{i=1}^{\magm}$
\begin{equation}
\bar Q(z)=e^{iz\Phi_{I}}\prod_{i=1}^{\magm}(z-z_{i})\,,\label{eq:qbethe}
\end{equation}
the eigenvalues of the shift operators are written as
\begin{equation}
U=e^{i(\check{z}-\hat{z})\Phi_{I}}\prod_{i=1}^{\magm}\frac{\check{z}-z_{i}}{\hat{z}-z_{i}}\,,\label{eq:shiftfunction-1}
\end{equation}
compare \eqref{abashift}.
The identification of the special points $\hat{z}$ and $\check{z}$
is particularly important as it reveals how higher local charges may
be extracted from Q-operators. In the next subsection this is elucidated
for the case of the nearest-neighbor Hamiltonian.

\subsubsection{The nearest-neighbor Hamiltonian and its action\label{sub:The-Hamiltonian-density}}

We identified two special points $\hat{z}$ and $\check{z}$ at which
the Q-operators are related by the shift operator, see (\ref{eq:UQgQ-1}).
This enables us to give a direct operatorial derivation of (\ref{abaenergy}) for the representations under study.
The main result of this section is that 
\begin{equation}
\mathbf{H}=\frac{\bar{\bf Q}'(\check{z})}{\bar{\bf Q}(\check{z})}-\frac{\bar{\bf Q}'(\hat{z})}{\bar{\bf Q}(\hat{z})}\,,\label{eq:H=00003DQ/Q-Q/Q}
\end{equation}
is a nearest-neighbor Hamiltonian, i.e. 
\begin{equation}
\mathbf{H}=\sum_{i=1}^{L}\mathcal{H}_{i,i+1}\,.\label{eq:density}
\end{equation}
It is of prime importance as it yields the total energy of the spin-chain
and determines the time-evolution of the system. An important step
in the derivation of the Hamiltonian (\ref{eq:density}) is to rewrite
(\ref{eq:H=00003DQ/Q-Q/Q}) as 
\begin{equation}
\mathbf{H}\,\bar{\bf Q}(\check{z})=\bar{\bf Q}'(\check{z})-\mathbf{U}\,\bar{\bf Q}'(\hat{z})\,.\label{eq:HQ-1}
\end{equation}
Here we used (\ref{eq:UQgQ-1}). The derivation of (\ref{eq:H=00003DQ/Q-Q/Q})
is a direct consequence of the truly remarkable identity 
\begin{equation}
\input{content/pictures/NewDiagrams/Haction}\,\,.\label{eq:haction}
\end{equation}
The significance of this equation is twofold. Firstly, it contains
the non-trivial statement that the right-hand side of (\ref{eq:haction})
can be written as the left-hand side for {}``some'' $\mathcal{H}$
that, as encoded in the picture, acts non-trivially only in the quantum
space. This is proven in \cite{Frassek2013} using
the so called Sutherland equation, originally introduced to provide
a criterion for a local Hamiltonian to commute with a given tranfer
matrix \cite{Sutherland1970}, and the special properties of the $\mathcal{R}$-operator.
Secondly, (\ref{eq:haction}) defines $\mathcal{H}$ uniquely in terms
of the R-operators $\mathcal{R}$ for Q-operators. The fact that this
way of defining $\mathcal{H}$ can be particularly convenient for
practical purposes is supported by the non-trivial example of $\mathfrak{sl}(2)$
spin~$-\frac{1}{2}$ in Appendix~\ref{sec:The-spin-half}. 

Using (\ref{eq:haction}), (\ref{eq:HQ-1}) can be shown quickly.
The derivative of the Q-operators follows immediately from the definition
in (\ref{eq:Qop}) for any set $I$ 
\begin{equation}
\bar{\bf Q}_{I}^{'}(z)=i\,\Phi_{I}\,\bar{\bf Q}_{I}(z)+e^{iz\,\Phi_{I}}\,\sum_{k=1}^{L}\,\widehat{\tr}_{{\mathcal{H}}^{(I,\bar{I})}}\big\{\mathcal{D}_{I}\,\mathcal{R}_{I}(z)\otimes\ldots\otimes\underbrace{\mathcal{R}'_{I}(z)}_{\text{k-th side}}\otimes\ldots\otimes\mathcal{R}_{I}(z)\big\}\,.\label{eq:Qopprime-1}
\end{equation}
Taking a closer look at (\ref{eq:HQ-1}), one finds that the right
hand side can be rearranged as a sum of local contributions corresponding
to the Hamiltonian density $\mathcal{H}$. This is done by pairing
terms according to (\ref{eq:haction}). Furthermore, from (\ref{eq:HQ-1})
it follows that 
\begin{equation}
\mathcal{H}_{L,L+1}=f_{1}(\Phi)\mathcal{H}_{L,1}f_{1}^{-1}(\Phi)\,.\label{eq:htwist}
\end{equation}
Thus, we have shown (\ref{eq:HQ-1}). It is worth to mention that
an analogous and equivalent relation as (\ref{eq:haction}) holds
for the action of $\mathcal{H}$ from the right.
On the level of eigenvalues, upon using (\ref{eq:qbethe}), (\ref{eq:H=00003DQ/Q-Q/Q})
gives the famous energy formula 
\begin{equation}
E=\sum_{i=1}^{\magm}\left(\frac{1}{\check{z}-z_{i}}-\frac{1}{\hat{z}-z_{i}}\right)\,.\label{eq:energyformula}
\end{equation}
We would like to
stress again that the auxiliary and quantum space of the R-operators $\mathcal{R}$
are of different nature. The mechanism by which the Hamiltonian density
can be extracted from the R-operators $\mathcal{R}$ is encoded in (\ref{eq:haction}).
The explicit expression for $\mathcal{H}$ for generalized rectangular
representations in the quantum space is provided in the Appendix~\ref{sub:Hamiltonian-Density}.
If we further restrict to certain representations one obtains rather
convenient expressions for the Hamiltonian density. This is done for
the fundamental representation and the $\mathfrak{sl}(2)$ spin~$-\frac{1}{2}$
case in Appendix~\ref{sec:The-spin-half} and \ref{fundhamq}.

\chapter{Bethe ansatz for Yangian invariants}\label{ch:BAforYI}
  In this chapter we study the nature of Yangian symmetry as
it appears in tree-level scattering amplitudes of $\mathcal{N}=4$ super Yang-Mills theory from the view point of
integrability and the quantum inverse scattering method. It was shown in \cite{Drummond2010c,Drummond2009} that the superconformal symmetry and the dual superconformal symmetry of those amplitudes combines into a Yangian symmetry. More precisely, using Drinfeld's first realization reviewed in Appendix~\ref{app:1rel} and discussed at the end of Section~\ref{sec:cop}, it can be shown that all tree-level amplitudes are annihilated by a certain realization of the first and second level generators.
In order to apply the machinery of the quantum inverse scattering method we reformulate the Yangian invariance  condition as
\begin{equation}
  \label{mon_ev1}
  \mon_{ab}(\spec)|\Psi\rangle = \delta_{ab}|\Psi\rangle\,, 
\end{equation}
where  $|\Psi\rangle$  is a Yangian invariant. This is the key formula to connect the scattering amplitudes and the Bethe
ansatz. Here $\mon(\spec)$ is a monodromy matrix, given by a product
of suitable R-matrices, $\spec$ is a spectral parameter, and $a,b$ are
indices in an auxiliary space, taking values in the fundamental
representation of the underlying symmetry algebra. The generators of
the Yangian algebra are obtained as the coefficients $\mon_{ab}^{[r]}$
of an expansion of the monodromy matrix $\mon(\spec)$ in powers $r$ of
the inverse spectral parameter $\spec^{-1}$, cf. Chapter~\ref{ch:yangian}. From \eqref{mon_ev1} with
$M_{ab}^{[0]}=\delta_{ab}$ one then sees that $|\Psi\rangle$ is
annihilated by all Yangian generators. By definition, $|\Psi\rangle$
is thus a Yangian invariant. Furthermore, finding all solutions of
\eqref{mon_ev1} for all suitable $\mon(\spec)$ should then lead
to the complete set of such invariants.

In the simplest case of $\gl2$ equation
\eqref{mon_ev1} can be derived from the rational limit of a
two-dimensional integrable model, the so-called Z-invariant six-vertex
model introduced by Baxter. It has a description as an inhomogeneous
spin chain \cite{Baxter:1987}. Introducing a certain oscillator
formalism, and thereby considering more general representations, cf. Chapter~\ref{ch:ybe} and \ref{ch:2dlattice}, one
can then obtain $\mathfrak{gl}(2)$ Yangian generators. Excitingly,
they take forms analogous to the ones acting on the scattering
amplitudes in $\mathcal{N}=4$ super Yang-Mills theory
\cite{Drummond2010a,Drummond2010b}. Furthermore, the procedure generalizes
to higher rank cases. 

A further interesting aspect of \eqref{mon_ev1} is that it allows one
to consider the Bethe ansatz for the spin chain. Equation \eqref{mon_ev1} represents a system
of eigenvalue problems for the matrix elements $\mon_{ab}(\spec)$ of
the monodromy matrix $\mon(\spec)$, with rather trivial eigenvalues
$0$ or $1$ for a common eigenvector $|\Psi\rangle$. In addition, by
taking a trace on both sides, \eqref{mon_ev1} becomes an eigenvalue
problem for the transfer matrix $\tm(\spec)$ of the spin chain,
\begin{equation}
  \label{trans_ev1}
  \tm(\spec)=\tr\mon(\spec)\,,\quad 
  \tm(\spec)|\Psi\rangle= n|\Psi\rangle\,,
\end{equation}
where we already generalized from $\mathfrak{gl}(2)$ to
$\gln$, hence $\mon(\spec)$ is a $n\times n$ matrix in the
auxiliary space. We conclude that any such Yangian invariant
$|\Psi\rangle$ must then be a special eigenvector of the transfer
matrix $\tm(\spec)$ with prescribed eigenvalue $n$. It is important to
stress that the Yangian invariant $|\Psi\rangle$ in \eqref{trans_ev1}
and thus also in \eqref{mon_ev1} does not depend on the
spectral parameter $\spec$.

In this study we concentrate on compact representations
of $\gln$. This should play the role of a toy model of the
$\mathcal{N}=4$ scattering amplitudes, where suitable non-compact
representations of $\gl{(4|4)}$ are needed instead. The latter
are built from continuous generalizations of the oscillators mentioned
above, which are essentially the spinor-helicity variables and their
derivatives. The basic philosophy based on \eqref{mon_ev1} should
nevertheless remain applicable, at least in the case of the tree-level
amplitudes, where Yangian invariance is unequivocal. Each
$\sites$-particle tree-level amplitude should then be identical to an
invariant $|\Psi\rangle$ solving \eqref{mon_ev1} with a monodromy of
``length'' $\sites$, and thus amenable to analysis by the QISM. The
monodromy is built from $L$ suitable R-matrices, just as in the case
of integrable spin chains. Thus amplitudes should turn into
``special'' spin chain states, similar, as we shall see, to
$\gln$ symmetric antiferromagnetic ground states of the
chain. The spin chain monodromy is again inhomogeneous, and the
external scattering data is encoded in the representing
``oscillators'' $\hat =$ spinor-helicity variables. Alternatively, we can
think of the tree-level amplitudes as appropriately generalized Baxter
lattices, i.e.\ special vertex models.

Just like in the toy model, it is imperative that the Yangian
invariants and therefore the tree-level amplitudes do not depend on
the spectral parameter $\spec$. The latter merely serves as a suitable
device for applying the QISM and for employing the Bethe ansatz to the problem. On the other hand,
spectral parameters were recently introduced as certain natural
``helicity'' deformations of $\mathcal{N}=4$ scattering amplitudes in
\cite{Ferro2012,Ferro:2013dga}. This is not a contradiction. In the
present framework, these parameters simply correspond to a freedom in
the choice of the inhomogeneities of the monodromy in
\eqref{mon_ev1}. In the $\gln$ toy model, the
representation labels in general do not fix the inhomogeneities
completely. The same holds true in the $\mathcal{N}=4$ case. In fact,
R-matrices of rational models in arbitrary representations are also
Yangian invariants. They may therefore be found from special solutions
of \eqref{mon_ev1}. For instance, a standard four-legged
$\gln$ R-matrix acting on the tensor product of two
arbitrary compact representations may be deduced from the
eigenvector $|\Psi\rangle$ of a length-four monodromy
$\mon(\spec)$. Here a difference of inhomogeneities, denoted by
$\inhdiff$, in the monodromy $\mon(\spec)$ is to be interpreted as a
spectral parameter of the R-matrix $R(\inhdiff)$. But
$\inhdiff$ is not the spectral parameter $\spec$ used to solve the
spectral problem \eqref{mon_ev1} for Yangian invariants.

In section~\ref{sec:pba} we
review Baxter's Z-invariant six-vertex model in the $\mathfrak{gl}(2)$
limit, as well as its remarkable solution through the little-known
perimeter Bethe ansatz \cite{Baxter:1987}. Its key feature is, rather
unusually, that the Bethe equations may be explicitly solved with
comparative ease. In section~\ref{sec:yi} we show that Baxter's
approach may be generalized to an important class of compact
representations of $\gln$, and reinterpreted as a
systematic way to define and derive Yangian invariants. This opens the
way to derive a perimeter Bethe ansatz for the latter.  In
section~\ref{sec:osc} we illustrate the method for the case of compact
oscillator representations of $\gln$ by presenting
explicit Yangian invariants for three specific examples. Pictorially
they correspond to a line, a three-vertex and a four-vertex. The
invariants are expressed in oscillator notation. They look somewhat
different from the Yangian-invariant tree-level scattering amplitudes,
which is surely due to the different nature of the representations
under investigation. However, in section~\ref{sec:amp} we demonstrate
that our examples may be rewritten as Graßmannian contour
integrals. Interestingly, this manifestly turns them into close
analogues of the scattering amplitudes, see
\cite{ArkaniHamed:2009si}. An added benefit of our approach is that
the (multi)-contours are precisely defined by the construction.  In
section~\ref{sec:bethe} we then discuss the perimeter Bethe ansatz for
the Yangian invariants of our toy model. We illustrate it for
$\mathfrak{gl}(2)$ for the sample invariants of section~\ref{sec:osc}
and section~\ref{sec:amp}. Remarkably, the Bethe roots assemble into
exact strings in the complex spectral parameter plane, and are thus
explicitly determined. We also sketch the generalization to
$\gln$, where a nested perimeter Bethe ansatz is
required. In addition to \cite{Frassek2013b}, we provide a diagrammatic study of the invariants in Appendix~\ref{app:comp} where a remarkable relation between special points of the R-matrix and the BCFW recursion relation, see e.g. \cite{Elvang2013a}, is pointed out.

\section{Perimeter Bethe ansatz}
\label{sec:pba}

In Section~\ref{sec:6vmodel} we introduced the six-vertex model. We discussed its partition function on several lattices with certain boundary conditions, cf. Figure~\ref{fig:rect}, Figure~\ref{Figguu} and Figure~\ref{subfig:uncrazya}. Furthermore, we introduced the concept of Z-invariance in Section~\ref{sec:zinv}. In the current section we will study the partition function of so-called Baxter lattices introduced in \cite{Baxter1978}. As we will see, they belong to a slightly more general class of lattices than the ones introduced previously. Instead of following the usual prescription to calculate the partition function as given in \eqref{pfunn} we focus on the method proposed by Baxter in 1987 and is known as the perimeter Bethe ansatz \cite{Baxter:1987}. It relies on the knowledge of the off-shell wave-function of the $\gl2$-invariant spin chain with inhomogeneities which we presented in \eqref{su2inh}. For certain choices of the inhomogeneities and Bethe roots the components of the Bethe vectors, cf. \eqref{barecbavec1}, yield the partition function. 
\subsection{Baxter Lattice}
\label{sec:baxter-lattice}

\begin{figure}[!t]
  \begin{center}  
    \includegraphics[width=0.5\textwidth]{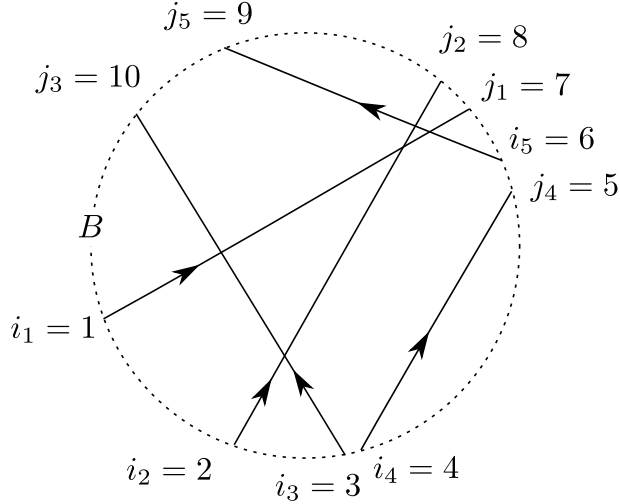}
       \caption{An example of a Baxter lattice with $N=5$ lines and the line configuration
         $\graph=((1,7),(2,8),(3,10),(4,5),(6,9))$ with respect to the reference point $B$.}
    \label{fig:pba-lattice}
  \end{center}  
\end{figure}
The Baxter lattice is constructed from $\lines$ straight lines intersecting in an arbitrary way in the interior of a
circle. Each line starts and ends at points on the perimeter of the circle. Furthermore, only two lines are
allowed to intersect in one point. An example can be found in Figure~\ref{fig:pba-lattice}, where we represented the perimeter by a dotted circle. The $\lines$ lines and their $2\lines$ endpoints are labeled
counterclockwise starting from the reference point $B$ on the
perimeter. To each line we assign an orientation, i.e. for the $k^\text{th}$ line
with endpoints $(\epe_k, \epb_k)$ and $1\leq \epe_k <
\epb_k\leq2\lines$ we introduce an arrow pointing from $\epe_k$ towards
$\epb_k$. Additionally, each line with $(\epe_k, \epb_k)$ carries a complex rapidity variable $\rap_k$. 
A Baxter lattice is then specified by the ordered sets
\begin{align}
  \label{eq:pba-g-theta}
  \graph=((\epe_1,\epb_1),\ldots,(\epe_\lines, \epb_\lines))\,,
  \quad
  \rapset=(\rap_1,\ldots,\rap_\lines)\,.
\end{align}
Following Chapter~\ref{ch:2dlattice}, we assign the
Boltzmann weights of the six-vertex model in the rational limit to each vertex in the Baxter lattice, see \eqref{6vbw}. They depend
on the rapidities, orientations of the lines and state labels $1$ or $2$ as shown in \eqref{pic:6vstate1} and \eqref{pic:6vstate2}. 
\begin{figure}[!t]
  \begin{center}  
   \includegraphics[width=0.5\textwidth]{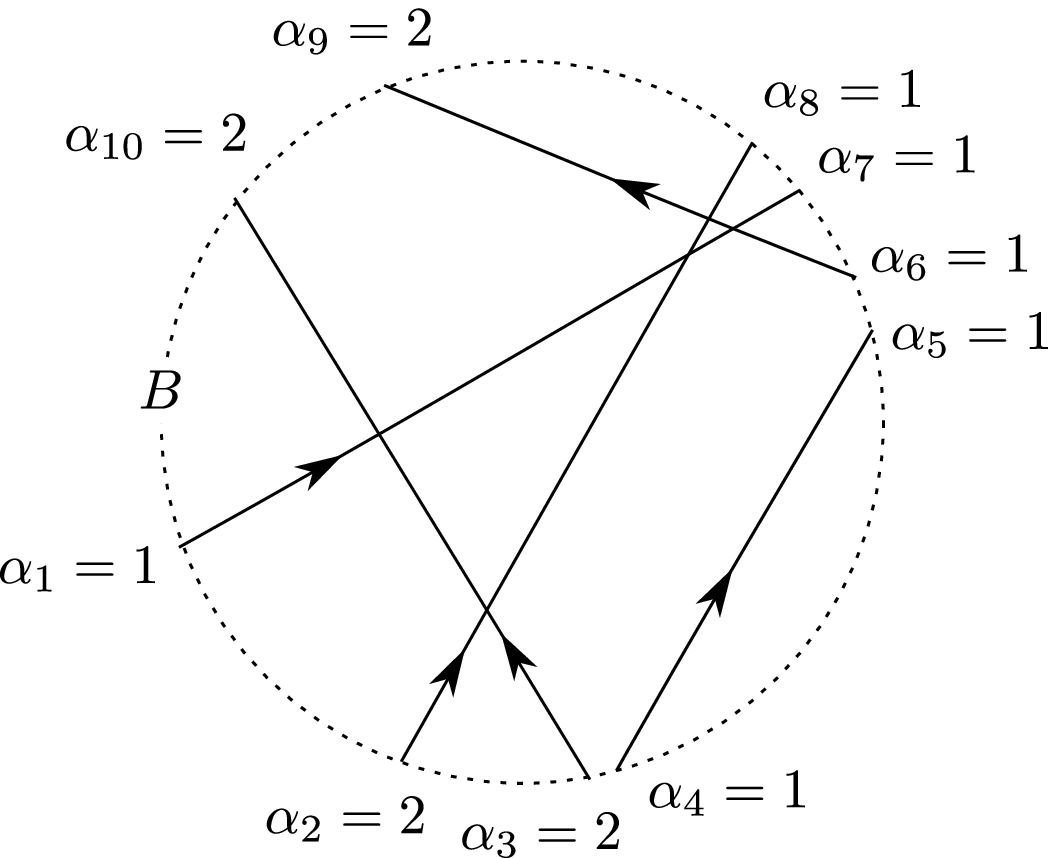}
       \caption{The Baxter lattice shown in 
         Figure~\ref{fig:pba-lattice} with fixed boundary configuration labeled by the set $\stateset$. In this case we have $\tilde\magnset=(1,4,6,9,10)$. These variables will be used to express the
         partition function $Z(\graph,\rapset,\stateset)$ in
         terms of the wave-function in
         \eqref{su2inh}.}
    \label{fig:pba-lattice-states}
  \end{center}  
\end{figure}
The boundary conditions of the lattice are given by the labels
$\alpha_{\epe_k}$ and $\alpha_{\epb_k}$ that take the values $1$
or $2$. They are assigned to the endpoints $(\epe_k, \epb_k)$ of the
lines, compare Figure~\ref{fig:pba-lattice-states}. We denote them as
\begin{align}
  \label{eq:pba-alpha}
  \stateset=(\alpha_1, \ldots, \alpha_{2 \lines})\,.
\end{align}
The partition function of a Baxter lattice is a function of the line configuration $\graph$, the rapidities $\rapset$ and the external state configuration $\stateset$ 
\begin{align}
  \label{eq:pba-partition}
  Z=
  Z(\graph,\rapset,\stateset)\,.
\end{align} 
Furthermore, we add an additional rule for lines that do not intersect with any others, see e.g.\ the line with endpoints $(4,5)$ in
Figure~\ref{fig:pba-lattice-states}.  If such a line $k$ has equal
state labels $\alpha_{\epe_k}=\alpha_{\epb_k}$ at the endpoints, it
contributes a factor of unity to the partition function. In case of
differing labels $\alpha_{\epe_k} \neq \alpha_{\epb_k}$, the partition
function is set to zero.

 As a consequence of the ice-rule the number of ingoing states $1$ ($2$) agrees with the number of outgoing states $1$ ($2$), cf. Section~\ref{sec:relr}.
This allows us to label the states at the boundary by $N$ parameters
\begin{equation}
\tilde\magnset=(\tilde\magn_1,\ldots,\tilde\magn_N)\quad\text{with}\quad 1\leq
  \tilde\magn_1<\ldots<\tilde\magn_\lines\leq 2\lines\,,
\end{equation} 
not taking into account the vanishing configurations that do not satisfy the ice-rule.
Starting from the reference point $B$ they label the
  endpoint positions $\epe_k$ at outward pointing edges with
  $\alpha_{\epe_k}=1$ and $\epb_k$ at edges directed inwards with
  $\alpha_{\epb_k}=2$, i.e.
  
  \begin{align}
    \label{eq:pba-alpha-position}
    \{\tilde\magn_k\}
    =
    \{\epe_k|\alpha_{\epe_k}\!=\!1\}
    \cup
    \{\epb_k|\alpha_{\epb_k}\!=\!2\}\,,
  \end{align}
  compare Figure~\ref{fig:pba-lattice-states}.

The R-matrix \eqref{rweight} and thus the Boltzmann weights at
the vertices \eqref{6vbw}  satisfy the Yang-Baxter equation, see Chapter~\ref{ch:2dlattice}. Thus, as discussed in Section~\ref{sec:zinv}, the partition function is not changed when moving lines in the interior without changing the order of the endpoints due to Z-invariance.

\subsection{Baxter's solution}
\label{sec:pba-solution}

An expression for the partition function of the six-vertex model on the Baxter lattice
\eqref{eq:pba-partition} was obtained in terms of the Bethe wave
function \eqref{su2inh} in \cite{Baxter:1987}. We derived the wave-function of the homogenous spin~$\frac{1}{2}$ Heisenberg
spin chain using the coordinate Bethe ansatz in Section~\ref{sec:cba}, see also \cite{Karbach1997,Sutherland:2004}.
The generalization to a spin chain with inhomogeneities can be found in Section~\ref{sec:morecba}. As discussed there, the wave-function of a
spin chain of length $\sites$ with $\brts$ excitations or magnons
 is parametrized by
\begin{align}
  \label{eq:pba-psi-param}
  \inhtset=(\inht_1,\ldots,\inht_{\sites})\,,
  \quad
  \brtset=(\brt_1,\ldots,\brt_\brts)\,,
  \quad
  \magnset=(\magn_1,\ldots,\magn_{\brts})\,,
\end{align}
denoting the inhomogeneities, Bethe roots and positions
of the magnons with $1\leq \magn_1<\ldots<\magn_\brts\leq \sites$, respectively. In Section~\ref{sec:morecba} we suppressed the dependence on the Bethe roots and indicated the inhomogeneities with the additional label ``$\text{inh}$'', cf. \eqref{su2inh}. For convenience we spell out the dependence on these variables explicitly and write the inhomogenous wave-function as
\begin{align}
  \label{eq:pba-psi}
  \Phi(\inhtset,\brtset,\magnset)
  =
  \sum_P
  {\rm A}(z_{P(1)},\ldots,z_{P(\brts)})
  \prod_{k=1}^{\brts}\phi_{\magn_k}(\brt_{P(k)},\inhtset)\,,
\end{align} 
where the sum is over all permutations $P$ of $\brts$ elements, the amplitude part $ {\rm A}(z_{P(1)},\ldots,z_{P(\brts)})={\rm A}(P)$ is given in \eqref{cbaampli} and the
single particle wave-function takes the form
\begin{align}
  \label{eq:pba-phi}
  \phi_{\magn}(\brt,\inhtset)
  =
  \prod_{j=1}^{\magn-1}(\brt-\inht_j+1)
  \prod_{j=\magn+1}^{\sites}(\brt-\inht_j)\,.
\end{align} 

Now, we are in the position to express the partition function
\eqref{eq:pba-partition} in terms of the Bethe wave-function
\eqref{eq:pba-psi}. 
After identifying the positions of the magnons of the wave-function $x_i$ with the variables $\tilde x_i$ introduced above, the partition function is calculated in two simple steps
\begin{enumerate}
\item For a Baxter lattice with $\lines$ lines, we employ a wave
  function with length $\sites=2\lines$ and $\brts=\lines$
  excitations.
\item The inhomogeneities $\inhtset$ and the Bethe
  roots $\brtset$ are given in terms of the rapidities $\rapset$ and
  $\graph$. For each line $k$ with endpoints $(\epe_k,\epb_k)$ we set
  \begin{align}
    \label{eq:pba-inhomo-rap}
    \inht_{\epe_k}=\rap_k+1\,,
    \quad
    \inht_{\epb_k}=\rap_k+2\,,
    \quad
    \brt_k=\rap_k+1\,.
  \end{align}
  Note that the wave
  function \eqref{eq:pba-psi} is invariant under permutations of the
  Bethe roots.
\end{enumerate}
Under these identifications, we obtain the desired expression
for the partition function \eqref{eq:pba-partition} in terms of the
Bethe wave-function \eqref{eq:pba-psi}:
\begin{align}
  \label{eq:pba-partition-wave}
   Z (\graph,\rapset,\stateset)
  =
  \mathcal{C}(\graph,\rapset)^{-1}
  (-1)^{\mathcal{K}(\graph,\stateset)}
  \Phi(\inhtset,\brtset,\magnset)\,.
\end{align}
The exponent $\mathcal{K}(\graph,\stateset)$ counts the
number of endpoints $\epe_k$ with state label $\alpha_{\epe_k}=2$,
\begin{align}
  \label{eq:pba-partition-wave-exp}
  \mathcal{K}(\graph,\stateset)
  =
  \big|\{\epe_k\,|\,\alpha_{\epe_k}\!=\!2\}\big|\,.
\end{align}
It will become clear in the framework of the quantum inverse scattering method that this factor is an artefact of the crossing relation in \eqref{selfcros}, see Section~\ref{sec:osc}.
The $\stateset$-independent normalization is given by
\begin{align}
  \label{eq:-partition-wave-norm}
  \mathcal{C}(\graph,\rapset)
  =
  \Phi(\inhtset,\brtset,\magnset_0)\,,
\end{align}
where $\magnset_0=(\epe_1,\ldots,\epe_\lines)$ is obtained from
\eqref{eq:pba-alpha-position} with $\stateset_0=(1,\ldots,1)$, which
means the state labels are $1$ at all $2\lines$ endpoints.

Remarkably, the choice of inhomogeneities and Bethe roots \eqref{eq:pba-inhomo-rap} yields a solution of the corresponding Bethe equations~\eqref{bae}. This can be seen after writing the
  Bethe equations in polynomial form in order to avoid divergencies,
  cf. \eqref{prebae}. 

\section{From vertex models to Yangian invariance}
\label{sec:yi}


In this section we study the partition function of Baxter
lattices for a wider class of Boltzmann weights assigned to the vertices. We extend
the algebra from $\mathfrak{gl}(2)$ to
$\gln$  and allow for more general representations $\Lambda$ assigned to each line. Note that the representations may differ from line to line. We derive a characteristic identity satisfied by these
partition functions and translate it into a set of
eigenvalue equations within the context of the QISM. 
In particular, the
partition function for a fixed line configuration $ Z $ is identified with a component of a
vector $|\Psi\rangle$ which is an eigenvector of all elements
of a certain spin chain monodromy $\mathcal{M}_{ab}(\spec)$  as introduced in \eqref{moninh}. 
As discussed in Chapter~\ref{ch:yangian}, such monodromies provide realizations of the Yangian algebra
$\mathcal{Y}(\gln)$. We will see, that the aforemetioned set of eigenvalue
equations characterizes vectors $|\Psi\rangle$ that are Yangian
invariant. In Section~\ref{sec:osc-3vertices}, we will encounter examples of Yangian invariants outside the framework
of Baxter lattices. Their existence can be seen as a consequence of the Bootstrap equation \eqref{bootsym1}.

\subsection{Vertex models on Baxter lattices}
\label{sec:yi-baxter-lattice}

As discussed in Section~\ref{sec:baxter-lattice}, a Baxter lattice is determined by the lattice configuration $\graph$, the set of rapidities $\rapset$ and the state configuration at the boundary $\stateset$. When generalizing to representations of $\gln$ we employ the Boltzmann weights given by the R-matrix $R_{\Lambda_1,\Lambda_2}$ as introduced in \eqref{rcomps}
\begin{align}\label{rcomps2}
 \langle \alpha_1,\alpha_2\vert R_{\Lambda_1,\Lambda_2}(u_1-u_2)\vert \beta_1, \beta_2\rangle=
 \begin{aligned}
  \includegraphics[scale=0.80]{content/pictures/vertexrep.eps}
 \end{aligned}
 \,.
\end{align}
In this case, the variables in the set $\stateset$ take the values $1,\ldots,\dim\Lambda_i$ depending on the representation $\Lambda_i$ carried by the corresponding line $i$. To keep track of the different representations carried by the lines in the Baxter lattice we introduce the ordered set
\begin{equation}
 \repset=(\Lambda_{1},\ldots,\Lambda_{\lines})\,.
\end{equation} 
Following the ordinary prescription introduced in \eqref{pfunn}, we can calculate the partition function from the explicit Boltzmann weights. It depends on the variables
\begin{equation}
 Z=Z(\graph,\repset,
        \rapset,\stateset)\,.
\end{equation} 
An example is shown in Figure~\ref{fig:yi-baxter-lattice}.
Again, we impose that if there is a line without any intersection but different states
at the boundary, the partition function vanishes.

\begin{figure}[h]
  \begin{center}  
    \includegraphics[width=0.4\textwidth]{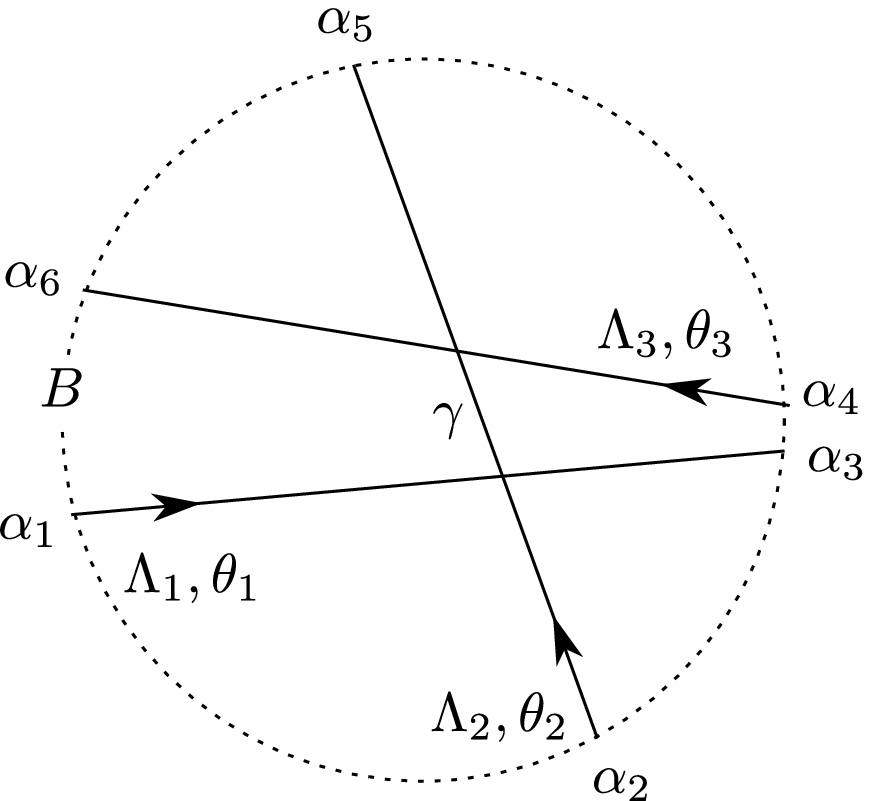}
    \caption{Baxter lattice with $\lines=3$
      lines and $\graph=((1,3),(2,5),(4,6))$. To each line
      we assign a $\gln$ representation $\Lambda_k$ and a
      spectral parameter $\rap_k$. The
     corresponding partition function is calculated via
      $
        Z 
        (\graph,\repset,
        \rapset,\stateset)
        =
        \sum_\gamma\,
        \langle\alpha_1,\alpha_2|
        R_{\Lambda_1,\Lambda_2}(\rap_1-\rap_2)
        |\alpha_3,\gamma\rangle
        \langle\gamma,\alpha_4|
        R_{\Lambda_2,\Lambda_3}(\rap_2-\rap_3)
        |\alpha_5,\alpha_6\rangle$
      }     
    \label{fig:yi-baxter-lattice}
  \end{center}  
\end{figure}

\subsection{Partition function as an eigenvalue problem}
\label{sec:yi-monodromy}

To understand the partition function of a Baxter
lattice as an eigenvalue problem within the framework of the quantum inverse scattering method we introduce a dashed arc which
is opened at the reference point $B$ at the boundary that was previously indicated by the dotted line. In contrast to Baxter's original construction it represents an actual
space that we term auxiliary space.\,\footnote{Note that the lines of the
Baxter lattice are slightly extended such that they intersect the
arc.} Such a lattice is depicted Figure~\ref{fig:yi-baxter-lattice-aux}.
\begin{figure}[!t]
  \begin{center}
    \includegraphics[width=0.4\textwidth]{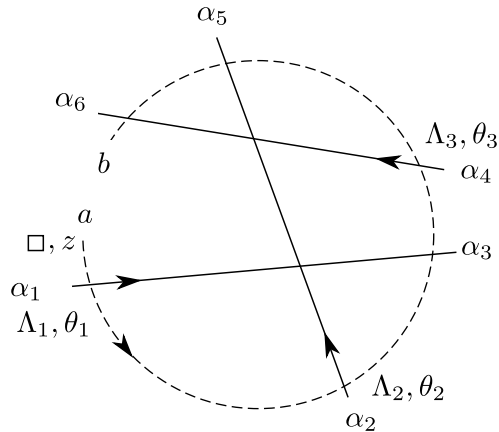}
    \caption{Baxter lattice  where an additional auxiliary space indicated by the dashed line has been introduced at the boundary. The auxiliary space is in the
      fundamental representation $\square$ with spectral parameter
      $\spec$ and has states labels $a$, $b$ assigned to its endpoints.}
    \label{fig:yi-baxter-lattice-aux}
  \end{center}
\end{figure}
The auxiliary space $V_\square=\mathbb{C}^n$ carries the fundamental
representation $\square$ of $\gln$ as well as a spectral
parameter $\spec$. The orientation of the line is chosen counterclockwise. The states at the endpoints of the auxiliary line are labeled by the
indices $a$ and $b$ which may take the values
$1,\ldots,n$.  As the auxiliary space intersects all other lines twice, it
introduces an additional layer of $2N$ vertices at the boundary of the
Baxter lattice. The Boltzmann weights at these vertices correspond to
elements of R-matrices of the type $R_{\square,\Lambda}$
or $R_{\Lambda,\square}$ introduced in Section~\ref{sec:croslax}. As discussed in Section~\ref{sec:rmatrices}, the Lax operators $R_{\square,\Lambda}$ satisfies a Yang-Baxter equation of
the form
\begin{align}
  \label{eq:yi-ybe-boundary}
  \begin{aligned}
    \includegraphics[width=0.6\textwidth]{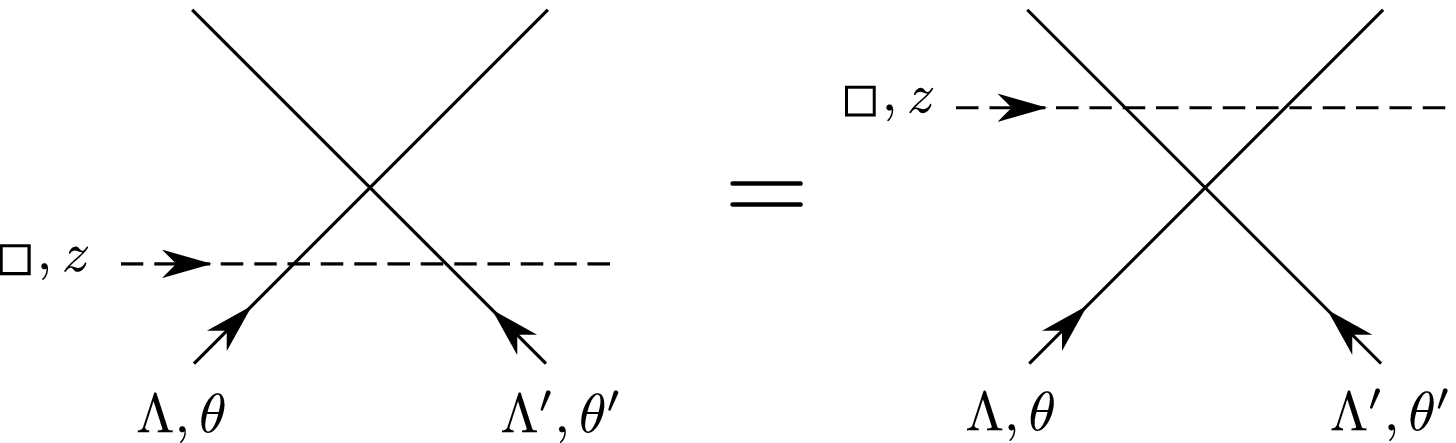}
  \end{aligned}
  \begin{aligned}
    .\\\phantom{}
  \end{aligned}
\end{align}
In addition,
we demand the unitarity condition introduced in Section~\ref{sec:unitarity}
\begin{align}
  \label{eq:yi-unitarity}
  \begin{aligned}
    R_{\square,\Lambda}(\spec-\rap)
    R_{\Lambda,\square}(\rap-\spec)
    =
    1\,,
  \end{aligned}
  \quad\text{i.e.}\quad
  \begin{aligned}
        \includegraphics[width=0.3\textwidth]{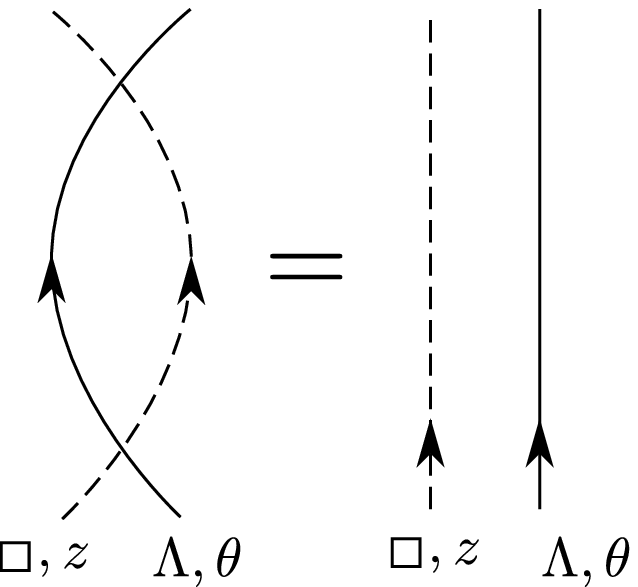}
  \end{aligned}
\end{align}
using the graphical notation. As outlined in Section~\ref{sec:croslax}, the unitarity condition \eqref{eq:yi-unitarity} implies certain restrictions on the representations in the quantum space \eqref{genrec} and constrains the normalization of the R-matrices $R_{\square,\Lambda}$ and $R_{\Lambda,\square}$.

Making use of the Yang-Baxter equation \eqref{eq:yi-ybe-boundary} and
the unitarity condition \eqref{eq:yi-unitarity} we can
disentangle the auxiliary space (dashed line) from the $\lines$ spaces
defining the Baxter lattice (solid lines). Graphically one easily sees
that this yields a non-trivial identity for the partition function
$ Z (\graph,\repset,\rapset,\stateset)$ of a Baxter lattice,
see Figure~\ref{fig:yi-disentangle}.
\begin{figure}[!t]
  \begin{center}
   \includegraphics[width=1\textwidth]{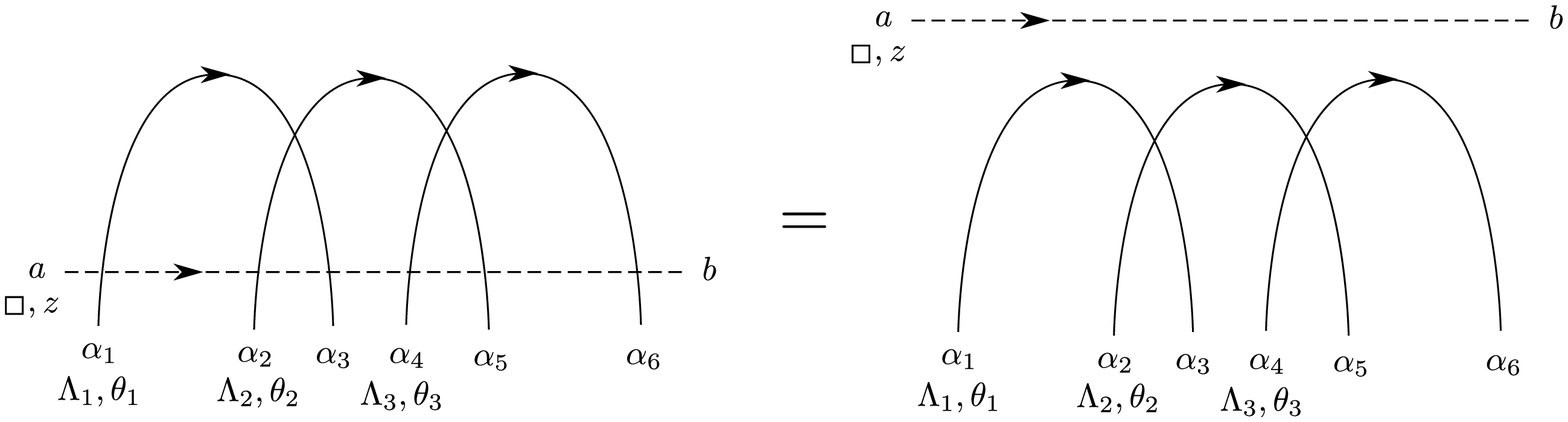}
    \caption{An identity for $ Z (\graph,\repset,
      \rapset,\stateset)$ of the Baxter lattice in
      Figure~\ref{fig:yi-baxter-lattice} is obtained by disentangling
      the dashed auxiliary line from the solid lines using
      \eqref{eq:yi-ybe-boundary} and \eqref{eq:yi-unitarity}. Here, the
      lattice has been deformed to emphasize that the row of vertices
      involving the auxiliary line will be written as a monodromy
      shortly, cf. Figure~\ref{fig:yi-mono-element}.}
    \label{fig:yi-disentangle}
  \end{center}
\end{figure}
To obtain this identity for a general Baxter lattice,
we study the expression for the partition function of the vertices sharing the auxiliary space
\begin{align}
  \label{eq:yi-add-boltzmann}
    \begin{aligned}
      Z^\text{arc}_{ab}(\spec,\graph,
      \repset,
      \rapset,
      \stateset,
      \boldsymbol{\beta})\\
      =
      \smash[b]{\sum_{c_1,\ldots,c_{2\lines-1}=1}^n}\;
      \Bigg(\;
      &
      \smash[b]{\prod_{k=1}^\lines}\;
      \langle c_{\epe_k-1},\alpha_{\epe_k}|
      R_{\square,\Lambda_k}(\spec-\rap_k)
      |c_{\epe_k},\beta_{\epe_k}\rangle\\
      &\quad\cdot
      \langle \beta_{\epb_k},c_{\epb_k-1}|
      R_{\Lambda_k,\square}(\rap_k-\spec)
      |\alpha_{\epb_k},c_{\epb_k}\rangle
      \smash[t]{
        \Bigg)_{
          \begin{subarray}{l}
            c_0:=a\\c_{2\lines}:=b
          \end{subarray}
        }
      }\,,
    \end{aligned}
\end{align}
where each of the $\lines$ lines of the lattice contributes two
weights. While the variables  $c_i$ with $i=1,\ldots,2\lines-1$ are assigned to the internal edges of
the  auxiliary space, the states $c_0:=a$ and $c_{2\lines}:=b$ are associated to the external ones. The variables 
$\boldsymbol{\beta}=(\beta_1,\ldots,\beta_{2\lines})$ label the states at
the edges that connect the layer of $2N$ vertices to the Baxter lattice, cf. Figure~\ref{fig:yi-baxter-lattice-aux}.  Equating the Baxter lattice entangled with the auxiliary
space to the disentangled situation, we find
\begin{align}
  \label{eq:yi-condition-z-general}
  \sum_{\boldsymbol{\beta}}
  Z^\text{arc}_{ab}(\spec,\graph,
  \repset,\rapset,
  \stateset,\boldsymbol{\beta})\,
   Z 
  (\graph,
  \repset,\rapset,\boldsymbol{\beta})
  =
  \delta_{ab}\,
   Z 
  (\graph,
  \repset,\rapset,\stateset)\,.
\end{align}
Here, the unraveled auxiliary line simply translates into a Kronecker delta
$\delta_{ab}$ on the right-hand-side of \eqref{eq:yi-condition-z-general}, compare Figure~\ref{fig:yi-disentangle}, which is compatible with the way we treated the non-intersecting lines of the Baxter lattice.

The summed-over Boltzmann weights in
$Z^\text{arc}_{ab}$
can be written as matrix elements of an inhomogeneous spin chain
monodromy $\mathcal{M}(\spec)$ with $\sites=2N$ sites. This allows to establish a link
with the QISM. We encountered a rather similar situation at the end of Section~\ref{pfaqism}. The monodromy is introduced as
\begin{align}
  \label{eq:yi-mono}
  \begin{aligned}
    \mathcal{M}(\spec)
    =
    R_{\square\,\Xi_1}(\spec-\inh_1)
    \cdots R_{\square,\Xi_\sites}(\spec-\inh_\sites)
    =
    \,\,\,\\\phantom{}
  \end{aligned}
  \begin{aligned}
   \includegraphics[width=0.3\textwidth]{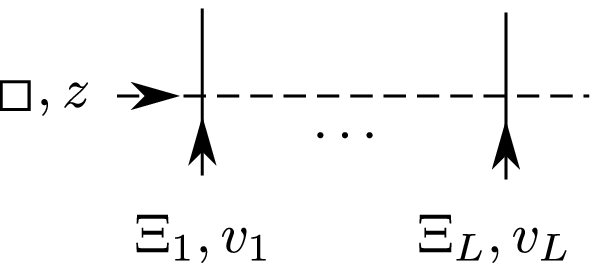}
  \end{aligned}
  \begin{aligned}
    ,\\\phantom{}
  \end{aligned}
\end{align}
cf. \eqref{moninh}.
To avoid possible confusion with the representation labels of the lines in the Baxter lattice we call the representation labels at the $j^\text{th}$ site $\Xi_j$. The
auxiliary space (dashed line) carries the fundamental representation $\square$ and
the matrix elements of the monodromy with respect to this space are
denoted as
\begin{align}
  \label{eq:yi-mono-elements}
  \mathcal{M}_{ab}(\spec):=\langle a|\mathcal{M}(\spec)|b\rangle\,.
\end{align}
They are still operators in the quantum space. In what follows,
we require the Lax operators associated to the Boltzmann weights in
\eqref{eq:yi-add-boltzmann} to satisfy the crossing relation \eqref{croslaxx}. Furthermore, using the symmetry relation \eqref{symree}, see also Figure~\ref{fig:symrel}, we obtain 
\begin{equation}\label{crossslaxx}
   \langle a,\beta\vert R_{\ffbox,\bar\Lambda}(\spec-\theta+\gamma)\vert b,\alpha\rangle=\langle \alpha,a\vert R_{\Lambda,\ffbox}(\theta-\spec)\vert \beta,b\rangle\,,
\end{equation} 
with $\gamma$ defined in \eqref{genrec}. Graphically we denote this relation as
\begin{align}\label{crossslaxxpic}
 \begin{aligned}
   \includegraphics[width=0.5\textwidth]{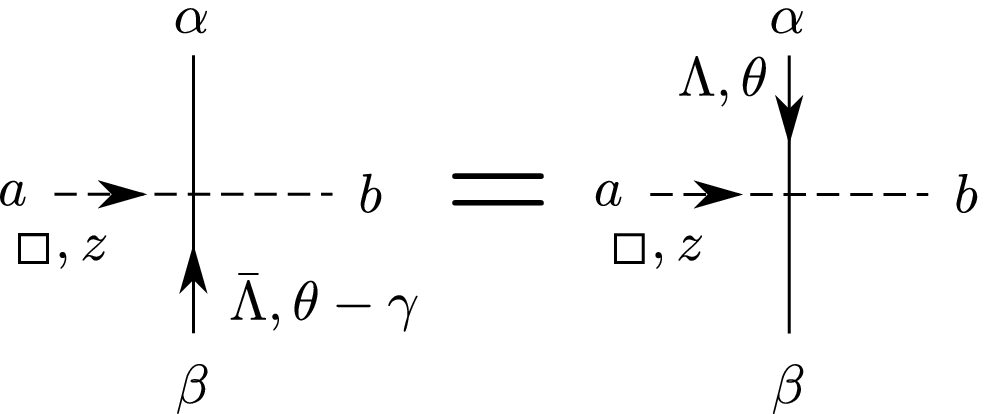}\,.
 \end{aligned}
\end{align}
Applying the crossing relation \eqref{crossslaxx} to the weights
in the second line of \eqref{eq:yi-add-boltzmann} yields
\begin{align}
  \label{eq:yi-add-boltzmann-cross}
  \begin{aligned}
    Z^\text{arc}_{ab}(\spec,\graph,
    \repset,\rapset,
    \stateset,\boldsymbol{\beta})\\
    =
    \smash[b]{\sum_{c_1,\ldots,c_{2\lines-1}=1}^n}\;
    \Bigg(\;
    &
    \smash[b]{\prod_{k=1}^\lines}\;
    \langle c_{\epe_k-1},\alpha_{\epe_k}|
    R_{\square,\Lambda_k}(\spec-\rap_k)
    |c_{\epe_k},\beta_{\epe_k}\rangle\\
    &\quad\cdot
    \langle c_{\epb_k-1},\alpha_{\epb_k}|
    R_{\square,\bar\Lambda_k}(\spec-\rap_k+\gamma_k)
    |c_{\epb_k},\beta_{\epb_k}\rangle
    \smash[t]{
      \Bigg)_{
        \begin{subarray}{l}
          c_0:=a\\c_{2\lines}:=b
        \end{subarray}
      }
    }\,.
  \end{aligned}
\end{align}
In this form the index structure is such that all weights combine into
matrix elements of the monodromy \eqref{eq:yi-mono} with
$\sites=2\lines$ sites
\begin{align}
  \label{eq:yi-mono-element}
  \begin{aligned}
    Z^\text{arc}_{ab}(\spec,\graph,
    \repset,\rapset,
    \stateset,\boldsymbol{\beta})
    =
    \langle\stateset|
    \mathcal{M}_{ab}(\spec)
    |\boldsymbol{\beta}\rangle\,,
  \end{aligned}
\end{align}
satisfying the RTT-relation.

We used the notation
$|\boldsymbol{\beta}\rangle:=|\beta_1\rangle\otimes\cdots\otimes|\beta_{2\lines}\rangle\in
V_1\otimes\cdots\otimes V_{2\lines}$ for the basis vectors of the quantum space.
The labels of the total quantum space of the monodromy $\mathcal{M}$ are
hidden. Thus there is no analogue of the labels $\graph$, $\repset$,
$\rapset$ on the right-hand-side of \eqref{eq:yi-mono-element}. However, for each line $k$ of the Baxter
lattice with endpoints $\epe_k<\epb_k$ specified in $\graph$ we obtain two spin chain sites with
representations and inhomogeneities given by
\begin{align}
  \label{eq:yi-ident-rep-inhomo}
  \Xi_{\epe_k}=\Lambda_k\,,
  \quad
  \inh_{\epe_k}=\rap_k
  \quad
  \text{and}
  \quad
  \Xi_{\epb_k}=\bar\Lambda_k\,,
  \quad
  \inh_{\epb_k}=\rap_k-\gamma_k\,.
\end{align}
As an example, the spin chain corresponding to the Baxter lattice in Figure~\ref{fig:yi-disentangle} can be found in Figure~\ref{fig:yi-mono-element}.
\begin{figure}[!t]
  \begin{center}
    \begin{align*}
      \begin{aligned}
         \includegraphics[width=0.8\textwidth]{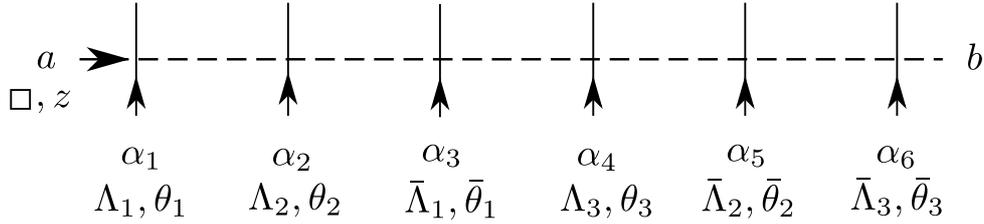}
      \end{aligned}
    \end{align*}
    \caption{Applying
      \eqref{crossslaxxpic} in Figure~\ref{fig:yi-disentangle}, we can transform all vertical lines
      such that they have the same orientation. The representations are given by are given by $\Lambda_i$ and $\bar\Lambda_i$ and the inhomogeneities take the values $\theta_i$ and $\bar\theta_i=\theta_i-\gamma_i$, cf. \eqref{eq:yi-ident-rep-inhomo}. This determines the corresponding spin chain.}
    \label{fig:yi-mono-element}
  \end{center}
\end{figure}

The partition function of the vertex model defines a vector
$|\Psi\rangle$ in the quantum space of the spin chain via
\begin{align}
  \label{eq:yi-partition-psi}
  \langle\stateset|\Psi\rangle
  :=
   Z (\graph,
  \repset,\rapset,\stateset)\,.
\end{align}
This allows us to translate the identity
\eqref{eq:yi-condition-z-general} partition function into
\begin{align}
  \label{eq:yi-eigenvalue-bra}
  \langle\stateset|\mathcal{M}_{ab}(\spec)|\Psi\rangle
  =
  \delta_{ab}\langle\stateset|\Psi\rangle\,,
\end{align}
using \eqref{eq:yi-mono-elements}, \eqref{eq:yi-partition-psi} and the
orthonormality of the states $|\boldsymbol{\beta}\rangle$.
Dropping the bra $\langle\stateset|$ in
\eqref{eq:yi-eigenvalue-bra}, we obtain the previously mentioned set of eigenvalue
equations. These equations characterize the vector $|\Psi\rangle$,
which, according to \eqref{eq:yi-partition-psi}, is built out of the
partition functions of a Baxter lattice for all possible boundary
configurations $\stateset$. Equation
\eqref{eq:yi-eigenvalue-bra} tells us that $|\Psi\rangle$ is a simultaneous eigenvector of all matrix elements of the monodromy \eqref{eq:yi-mono} with representation labels and inhomogeneities given
\eqref{eq:yi-ident-rep-inhomo}. The eigenvector $|\Psi\rangle$ is
special because its eigenvalues are fixed to be $1$ for the diagonal
monodromy elements and $0$ for the off-diagonal ones. Remarkably,
\eqref{eq:yi-eigenvalue-bra} is an eigenvalue problem that can be solved within the framework
of the quantum inverse scattering method.

\subsection{Yangian algebra and invariants}
\label{sec:yi-yangian}

In this section we will analyze the characteristic eigenvalue equation \eqref{eq:yi-eigenvalue-bra} in the
context of the Yangian discussed in Chapter~\ref{ch:yangian}. We continue to employ the spin chain monodromy
\eqref{eq:yi-mono}, but allow for general
representations $\Xi_i$ subject to the condition \eqref{genrec} and inhomogeneities $\inh_i$, which in general do
not have to obey the restrictions
\eqref{eq:yi-ident-rep-inhomo}. Furthermore, an odd number of sites
$\sites$ is now also permitted which was not meaningful in the context of
Section~\ref{sec:yi-monodromy} as each line of the Baxter lattice
gave rise to exactly two sites.

Let us
recall the eigenvalue equation \eqref{eq:yi-eigenvalue-bra} in the more general context of the
current section.  Omitting the bra $\langle\stateset|$, the set of
eigenvalue equations\footnote{In \cite{deVega:1984wk,Destri:1993qh}
  such a set of equations is shown to be satisfied by the physical
  vacuum state of integrable two-dimensional quantum field theories.} reads
\begin{align}
  \label{eq:yi-inv-rmm}
  \mathcal{M}_{ab}(\spec)|\Psi\rangle
  =
  \delta_{ab}|\Psi\rangle\,.
\end{align}
It may be depicted as
\begin{align}
  \label{eq:yi-inv-rmm-pic}
  \begin{aligned}
             \includegraphics[width=0.8\textwidth]{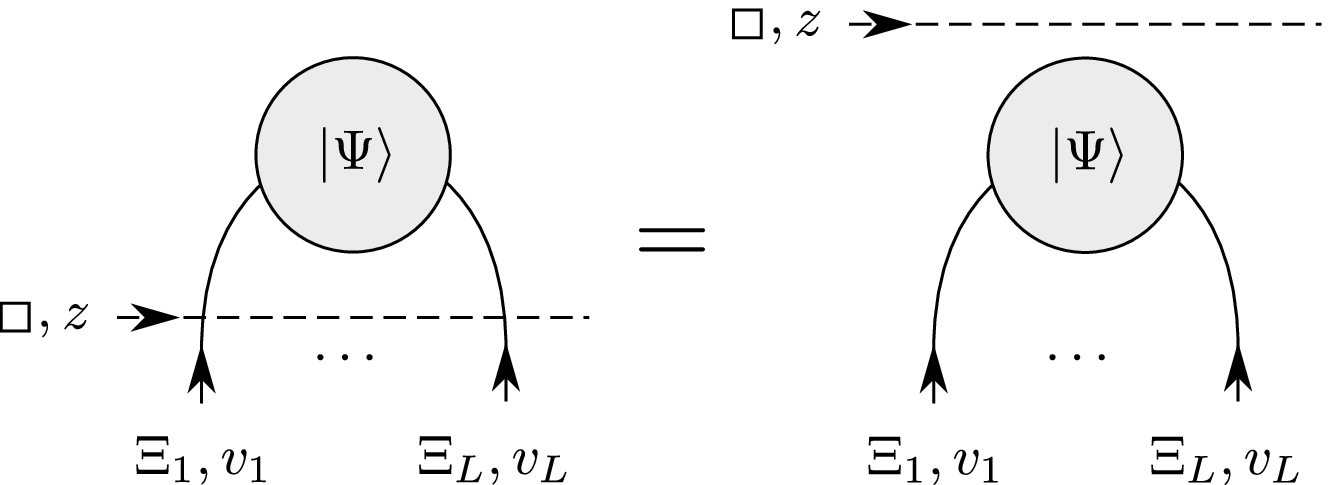}
  \end{aligned}\,,
\end{align}
following the graphical notation for the monodromy \eqref{eq:yi-mono} and Figure~\ref{fig:yi-disentangle}. It follows from \eqref{eq:yi-inv-rmm} that all Yangian generators in the inverse expansion of the spectral parameter $\spec$  annihilate the vector $|\Psi\rangle$, i.e.
\begin{align}
  \label{eq:yi-inv-expanded}
  \mathcal{M}^{(r)}_{ab}|\Psi\rangle=0
\end{align}
for $r\geq 1$, cf. \eqref{monexpansion}. This means that $|\Psi\rangle$ forms a one-dimensional
representation of the Yangian. Thus, we call $|\Psi\rangle$ Yangian
  invariant. The observation that \eqref{eq:yi-inv-rmm} characterizes
Yangian invariants is the main result of this section. Compared to the
expanded version \eqref{eq:yi-inv-expanded}, which is on the level of
Drinfeld's first realization presented in Appendix~\ref{app:1rel}, equation \eqref{eq:yi-inv-rmm} has the
advantage that it may be understood within the QISM. As a result,
powerful mathematical tools become applicable. The
formulation \eqref{eq:yi-inv-rmm} will be exploited in
Section~\ref{sec:bethe}, where this eigenvalue problem is solved using the
algebraic Bethe ansatz.

Note that we have seen that the partition
function $ Z $ of a vertex model on a Baxter lattice can be interpreted as a component of a Yangian invariant vector $|\Psi\rangle$, cf.\
\eqref{eq:yi-partition-psi}. However, a generic solution
$|\Psi\rangle$ of \eqref{eq:yi-inv-rmm} with a more general monodromy
$\mathcal{M}(\spec)$ built from representations $\Xi_i$ and inhomogeneities
$\inh_i$ not obeying the conditions \eqref{eq:yi-ident-rep-inhomo} and possibly
containing an odd number of sites $\sites$ will in general not correspond to a
partition function in the sense of
Section~\ref{sec:yi-monodromy}. Hence, we symbolically denote $|\Psi\rangle$ in
\eqref{eq:yi-inv-rmm-pic} by a ``black box'' without specifying
the interior. In Section~\ref{sec:osc-3vertices} we will indeed find
solutions of the Yangian invariance condition \eqref{eq:yi-inv-rmm}
that go beyond the Baxter lattices of
Section~\ref{sec:yi-monodromy}. The graphical representation of these
solutions not only contains lines and four-valent vertices, i.e\
R-matrices, but also trivalent-vertices discussed previously in Section~\ref{sec:three} and \ref{sec:bstrap}.

We end this section with a remark on a reformulation of Yangian
invariance. The condition in the form \eqref{eq:yi-inv-rmm} can
naturally be understood as an intertwining relation of the tensor
product of the first $\dsites$ with the remaining $\sites-\dsites$
spaces of the total quantum space. For this purpose we split the
monodromy \eqref{eq:yi-mono} as
\begin{align}
  \label{eq:yi-mono-split}
  \begin{aligned}
    \mathcal{M}(\spec)
    =
    &R_{\square,\Xi_1}(\spec-\inh_1)\cdots R_{\square,\Xi_\dsites}(\spec-\inh_\dsites)\\
    &\cdot R_{\square,\Xi_{\dsites+1}}(\spec-\inh_{\dsites+1})\cdots R_{\square,\Xi_\sites}(\spec-\inh_\sites)\,.
  \end{aligned}
\end{align}
Conjugating \eqref{eq:yi-inv-rmm} in the first $\dsites$ spaces and using
\eqref{eq:yi-unitarity} and \eqref{crossslaxx}
yields the intertwining relation
\begin{align}
  \label{eq:yi-intertwiner}
  \begin{aligned}
    &R_{\square,\Xi_{\dsites+1}}(\spec-\inh_{\dsites+1})\cdots R_{\square,\Xi_\sites}(\spec-\inh_\sites)
    \mathcal{O}_\Psi\\
    &=
    \mathcal{O}_\Psi
    R_{\square,\bar\Xi_\dsites}(\spec-\inh_\dsites+\gamma_{\dsites})\cdots
    R_{\square,\bar\Xi_1}(\spec-\inh_1+\gamma_{1})\,,
  \end{aligned}
\end{align}
where $\mathcal{O}_\Psi:=|\Psi\rangle^{\dagger_1\cdots\dagger_\dsites}$.
Using the notation $\bar v_i=v_i-\gamma_i$, this relation is graphically depicted as
\begin{align}
  \label{eq:yi-intertwiner-pic}
  \begin{aligned}
               \includegraphics[width=0.8\textwidth]{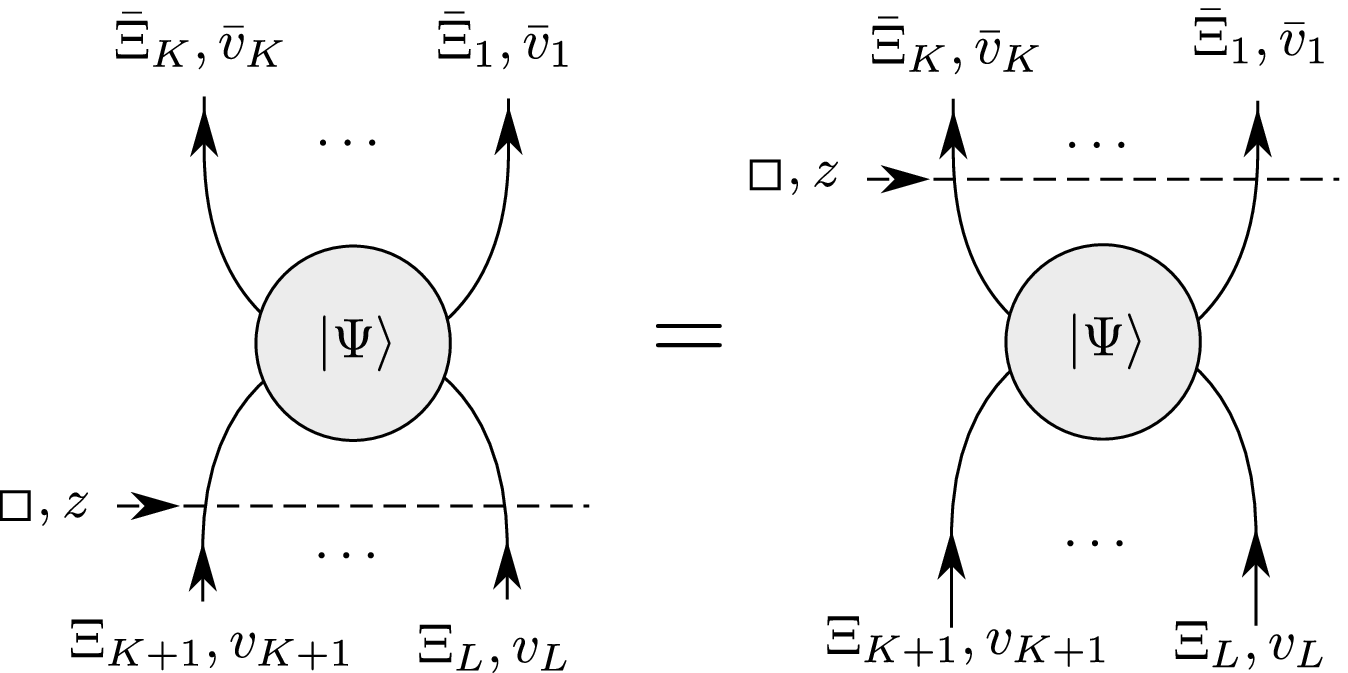}
  \end{aligned}.
\end{align}
In the case where $\mathcal{O}_\Psi$ corresponds to the partition function
$ Z $ of a Baxter lattice, this equation is nothing but a
consequence of Z-invariance, cf. \cite{Baxter1978} and also
Section~\ref{sec:zinv}. An equation of the type
\eqref{eq:yi-intertwiner} also appeared in the context of a spectral parameter
deformation of planar $\mathcal{N}=4$ super Yang-Mills scattering
amplitudes \cite{Ferro:2013dga}. There it was referred to as ``generalized Yang-Baxter
equation''. In the scattering problem, Yangian invariance of
undeformed tree-level amplitudes is usually formulated in the sense of
\eqref{eq:yi-inv-expanded}, see \cite{Drummond2009} and
e.g. \cite{Beisert2010}. At this point, we like to refer the reader to Appendix~\ref{app:comp} where a curious relation between the invariants under study and the {\small BCFW} recursion relation is pointed out.

\section{Yangian invariants in oscillator formalism}
\label{sec:osc}

In Section~\ref{sec:yi-yangian} we characterized Yangian invariants by
the eigenvalue equations \eqref{eq:yi-inv-rmm} for matrix
elements of a monodromy and equivalently by the associated
intertwining relation \eqref{eq:yi-intertwiner}. Here we will begin
our study of \eqref{eq:yi-inv-rmm} by working out explicit solutions
$|\Psi\rangle$ in a number of concrete examples. We restrict our
analysis to monodromies $\mathcal{M}(\spec)$, where the total quantum space
is built by tensoring finite-dimensional totally symmetric
representations $\s$ and their conjugates $\bs$, i.e.\ $\Xi_i=\s_i$ or
$\Xi_i=\bs_i$ for all $i=1, \ldots \sites$ in \eqref{eq:yi-mono}. We
need these conjugate representations to make sure that the total
quantum space contains a $\gln$ singlet, which is a
necessary criterion for Yangian invariants, see
\eqref{eq:yi-inv-expanded} for the case $r=1$. Note that as discussed in Section~\ref{sec:crosfun} , the conjugate representation $\bs$ for $\gl2$ can be related to the representations $\s$ via a similarity transform, cf. \eqref{selfcros}. This allowed us to write the partition function in terms of the wave-function with fundamental representation at each site. A detailed study of this relation can be found in Section~6.4 of \cite{Frassek2013b}.

The representations $\s$ and $\bs$ are realized in terms of oscillator
algebras, see Section~\ref{sec:osc-rep-lax}. Since the non-zero
eigenvalues appearing in \eqref{eq:yi-inv-rmm} are identical to $1$,
the normalization of the Lax operators used in the construction is
clearly important and will be discussed in some detail. After that we
are in place to construct the sought solutions in
Section~\ref{sec:osc-examples}. Our first and simplest examples are
the two-site monodromies of Section~\ref{sec:osc-line}, where the
representations of the two sites are necessarily conjugate to each
other. The inhomogeneities are then fixed by demanding Yangian
invariance, i.e.\ \eqref{eq:yi-inv-rmm}. This solution $|\Psi\rangle$
is graphically represented by a Baxter lattice consisting of a single
line. In Section~\ref{sec:osc-3vertices} we construct three-site
invariants. The corresponding intertwiner $\mathcal{O}_\Psi$
satisfying \eqref{eq:yi-intertwiner} is interpreted as a
solution of a bootstrap equation discussed in Section~\ref{sec:bstrap}. Although these invariants leave the
framework of Section~\ref{sec:yi-monodromy}, they are naturally
included in our definition of Yangian invariants. Finally, in
Section~\ref{sec:osc-4vertex} we study the Yangian invariant related
to the first non-trivial Baxter lattice consisting of two intersecting
lines. The associated intertwiner $\mathcal{O}_\Psi$ contains a free
parameter $\inhdiff$ and turns out to be the
$\gln$-invariant R-matrix $R_{\s,\s'}(\inhdiff)$ for
arbitrary totally symmetric representations $\s$, $\s'$. We obtain a
compact expression for this R-matrix in a certain oscillator
basis. The spectral parameter $\inhdiff$ of the R-matrix should not be
confused with that of the auxiliary space in Section~\ref{sec:yi}
denoted by $\spec$.

\subsection{Oscillators, Lax operators and monodromies}
\label{sec:osc-rep-lax}

We start by specifying the two types of oscillator realizations of the $\gln$ algebra \eqref{glnalg},
which will be used for the local quantum spaces of the monodromy
\eqref{eq:yi-mono}. These representations are labeled by their highest
weight, cf. Appendix~\ref{app:gln}. Consider the totally symmetric representation of
$\gln$ with highest weight $\s=(s,0,\ldots,0)$, where $s$
is a non-negative integer. We build these representations from a
single family of oscillators $\osca_a$ with $a=1,\ldots,n$.
Furthermore, the highest weight $\bs=(0,\ldots,0,-s)$ is constructed using a second
family of $n$ oscillators $\oscb_a$. The $n^2$ generators $J_{ab}$ of
the representation $\s$ and the second set of $n^2$ generators $\bar
J_{ab}$ of $\bs$ are given by
\begin{align}
  \label{eq:osc-gen-s-bs}
  \begin{aligned}
    J_{ab}&=+\bar\osca_a\osca_b&
    \quad\text{with}\quad&&
    [\osca_a,\bar\osca_b]=\delta_{ab}\,,&&
    \osca_a|0\rangle=0\,,&&
    \bar\osca_a=\osca_a^\dagger\,,&\\
    \bar J_{ab}&=-\bar\oscb_b\oscb_a&
    \quad\text{with}\quad&&
    [\oscb_a,\bar\oscb_b]=\delta_{ab}\,,&&
    \oscb_a|0\rangle=0\,,&&
    \bar\oscb_a=\oscb_a^\dagger\,.&
  \end{aligned}
\end{align}
It can be checked that this realization of the $\gln$ generators satisfies the commutation relations \eqref{glnalg}.
Commutators of oscillators that are not specified by these relations
vanish. See e.g.\ \cite{Biedenharn:1981} for a review of such
Jordan-Schwinger-type realizations of the $\gln$
algebra. The generators \eqref{eq:osc-gen-s-bs} act on the
representation spaces $V_{\s}$ and $V_{\bs}$. These spaces consist of
homogeneous polynomials of degree $s$ in, respectively, the creation
operators $\bar\osca_a$ and $\bar\oscb_a$ acting on the Fock vacuum
$|0\rangle$. Therefore the number operators
$\sum_{a=1}^n\bar\osca_a\osca_a$ and $\sum_{a=1}^n\bar\oscb_a \oscb_a$
both take the value $s$. The highest weight states in $V_\s$ and
$V_{\bs}$ are
\begin{align}
  \label{eq:osc-hws}
  \begin{aligned}
    |\sigma\rangle&=(\bar\osca_1)^s|0\rangle
    &\quad\text{with}\quad&&
    J_{aa}|\sigma\rangle&=+s\,\delta_{1\,a}|\sigma\rangle\,,\quad&
    J_{ab}|\sigma\rangle&=0\quad\text{for}\quad a<b\,,\\
    |\bar\sigma\rangle&=(\bar\oscb_n)^s|0\rangle
    &\quad\text{with}\quad&&
    \bar J_{aa}|\bar\sigma\rangle&=-s\,\delta_{n\,a}|\bar\sigma\rangle\,,\quad&
    \bar J_{ab}|\bar\sigma\rangle&=0\quad\text{for}\quad a<b\,,
  \end{aligned}
\end{align}
respectively.
We find that for $J_{ab}=\bar\osca_a\osca_b$ the condition \eqref{genrec} is satisfied
\begin{equation}\label{genrecosc}
 \sum_{c=1}^n J_{ca}J_{bc}=(s-1) J_{ba}+s\,\delta_{ab}\mathbb{I}\,,\quad\leftrightarrow\quad  \gamma=s-1\,,\, \sigma =s\,.
\end{equation} 
on the space of homogeneous polynomials.
The representation $\bs$ is conjugate to
$\s$ in the sense of \eqref{outaut},
\begin{align}
  \label{eq:osc-gen-conj}
  \bar J_{ab}\big|_{\oscb_a\mapsto\osca_a}
  =
  -J_{ab}^\dagger\,.
\end{align}
The Lax operators for the two
realizations defined in \eqref{eq:osc-gen-s-bs} read
\begin{align}
  \label{eq:osc-lax-fund-s}
  \begin{aligned}
    R_{\square,\s}(\spec-\inh)
    &=
    f_{\s}(\spec-\inh)
    \Bigg(1+(\spec-\inh)^{-1}\sum_{a,b=1}^ne_{ab}\bar\osca_b\osca_a\Bigg)
    =
    \,\,\,\\\phantom{}
  \end{aligned}
  \begin{aligned}
                  \includegraphics[width=0.18\textwidth]{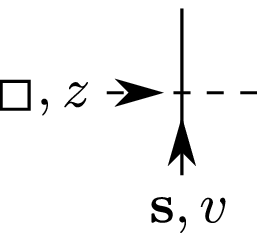}
  \end{aligned}
  \begin{aligned}
    ,\\\phantom{}
  \end{aligned}
  \\
  \label{eq:osc-lax-fund-bs}
  \begin{aligned}
    R_{\square,\bs}(\spec-\inh)
    &=
    f_{\bs}(\spec-\inh)
    \Bigg(1-(\spec-\inh)^{-1}\sum_{a,b=1}^ne_{ab}\bar\oscb_a\oscb_b\Bigg)
    =
    \,\,\,\\\phantom{}
  \end{aligned}
  \begin{aligned}
                  \includegraphics[width=0.18\textwidth]{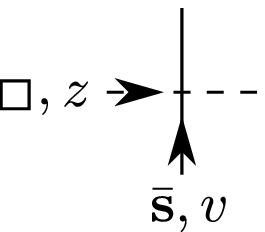}
  \end{aligned}
  \begin{aligned}
    ,\\\phantom{}
  \end{aligned}
\end{align}
compare \eqref{lax}.
As discussed in Section~\ref{sec:yi-monodromy}, we require these Lax
operators to possess the properties of unitarity
\eqref{eq:yi-unitarity} and crossing \eqref{crossslaxx}. These conditions impose constraints on the normalizations $f_{\s}(\spec)$. Following Section~\ref{sec:croslax}, we obtain
$f_{\bs}(\spec)$
\begin{equation}\label{eq:osc-norm-unitarity}
 f_\s(-u)f_\s(u-s+1)=\frac{u(u-s+1)}{u(u-s+1)-s}\,,\quad\quad 
 f_{\bs}(-u)f_{\bs}(u-n-s+1)=1\,,
\end{equation} 
using the characteristic relation \eqref{genrecosc} in \eqref{unit_fund} and \eqref{laxnb}, respectively.
The two normalizations in \eqref{eq:osc-lax-fund-s} are related via
\begin{equation}\label{eq:osc-crossing-norm}
 f_\s(\spec)=f_{\bs}(-\spec)\,,
\end{equation} 
cf. \eqref{crossss}.
We have derived the appropriate normalization $f_{\s}(\spec)$ in \eqref{symcnorm} using the fusion procedure at the end of Section~\ref{sec:bstrap}. It is specified up to a periodic function also called CDD-factor \cite{Castillejo1956}. 

We concentrate on monodromies $\mathcal{M}(\spec)$ of the form
\eqref{eq:yi-mono}, which are built entirely out of the two types of
Lax operators \eqref{eq:osc-lax-fund-s} and \eqref{eq:osc-lax-fund-bs}
with the crossing normalization as discussed above. Consequently, at the $i^\text{th}$ site of the
monodromy the representation of the local quantum space is
$\Xi_i=\s_i$ or $\Xi_i=\bs_i$ and the oscillator families building
these representations are labeled $\osca_a^i$ or $\oscb_a^i$,
respectively. Further restricting to monodromies that allow for
solutions $|\Psi\rangle$ of the Yangian invariance condition
\eqref{eq:yi-inv-rmm}, one finds severe constraints on the
representation labels $s_i$ and inhomogeneities $\inh_i$.

A class of such monodromies is obtained by considering Baxter
lattices in the sense of Section~\ref{sec:yi-monodromy}, where each
line carries either a symmetric representation or a conjugate one.  If
the $k^\text{th}$ line of the Baxter lattice with endpoints $\epe_k<\epb_k$
and spectral parameter $\rap_k$ carries a symmetric representation
labeled by $\Lambda_k=\s_{\epe_k}$, then according to
\eqref{eq:yi-ident-rep-inhomo}, the monodromy $\mathcal{M}(\spec)$ contains the two sites
\begin{align}
  \label{eq:osc-line-s}
  \begin{gathered}
  \Xi_{\epe_k}=\s_{\epe_k}\,,
  \quad
  \inh_{\epe_k}=\rap_k
  \quad
  \text{and}
  \quad
  \Xi_{\epb_k}=\bs_{\epb_k}\,,
  \quad
  \inh_{\epb_k}=\rap_k-s_{\epe_k}+1\\
  \text{with}
  \quad
  s_{\epe_k}=s_{\epb_k}\,.
  \end{gathered}
\end{align}
As a consequence, in $\mathcal{M}(\spec)$ the Lax operator
$R_{\square,\s_{\epe_k}}(\spec-\inh_{\epe_k})$ with the symmetric
representation is placed left of
$R_{\square,\bs_{\epb_k}}(\spec-\inh_{\epb_k})$ with the conjugate
representation. If instead the $k^\text{th}$ line carries the conjugate
representation $\Lambda_k=\bs_{\epe_k}$, we obtain 
\begin{align}
  \label{eq:osc-line-bs}
  \begin{gathered}
    \Xi_{\epe_k}=\bs_{\epe_k}\,,
    \quad
    \inh_{\epe_k}=\rap_k
    \quad
    \text{and}
    \quad
    \Xi_{\epb_k}=\s_{\epb_k}\,,
    \quad
    \inh_{\epb_k}=\rap_k+s_{\epe_k}-1+n\\
    \text{with}
    \quad
    s_{\epe_k}=s_{\epb_k}\,,
  \end{gathered}
\end{align}
from
\eqref{eq:yi-ident-rep-inhomo}. In this case the Lax operator with the conjugate representation is to
the left of the one with the symmetric representation. 

Let us comment on the normalization of the monodromies considered in
the remainder of Section~\ref{sec:osc}. The constraints on their
representation labels and inhomogeneities guarantee that the gamma
functions in the normalizations of the different Lax operators cancel
and only a rational function in $\spec$ remains.

\subsection{Sample invariants}
\label{sec:osc-examples}

After these preparations, we are in the position to actually solve
\eqref{eq:yi-inv-rmm} for a number of simple examples. From now on, we
label the monodromies $\mathcal{M}_{\sites,\dsites}(\spec)$ and the Yangian
invariants $|\Psi_{\sites,\dsites}\rangle$ by the total number of
sites $\sites$ and the number $\dsites$ of sites carrying a conjugate
representation of type $\bs$. This is motivated by
Section~\ref{sec:amp}, where the invariant
$|\Psi_{\sites,\dsites}\rangle$ is compared with the $\sites$-particle
$\text{N}^{\dsites-2}\text{MHV}$ tree-level scattering amplitude of
planar $\mathcal{N}=4$ super Yang-Mills theory. In addition, we focus
on monodromies $\mathcal{M}_{\sites,\dsites}(\spec)$ whose sites with
conjugate representations of type $\bs$ are all to the left of the
sites with $\s$. This order corresponds to the gauge fixing used in
the Graßmannian integral formulation in Section~\ref{sec:amp}.

\subsubsection{Line and identity operator}
\label{sec:osc-line}

The simplest Yangian invariant $|\Psi_{2,1}\rangle$ solving
\eqref{eq:yi-inv-rmm} corresponds to a Baxter lattice consisting of a
single line. In order to obtain the associated monodromy
$\mathcal{M}_{2,1}(\spec)$ where the site with the conjugate representation
is situated to the left of the symmetric one, we choose the line in
the Baxter lattice to carry a conjugate representation, cf.\
\eqref{eq:osc-line-bs} and see Figure~\ref{fig:osc-psi21}.
\begin{figure}[!t]
  \begin{center}  
      \includegraphics[width=0.8\textwidth]{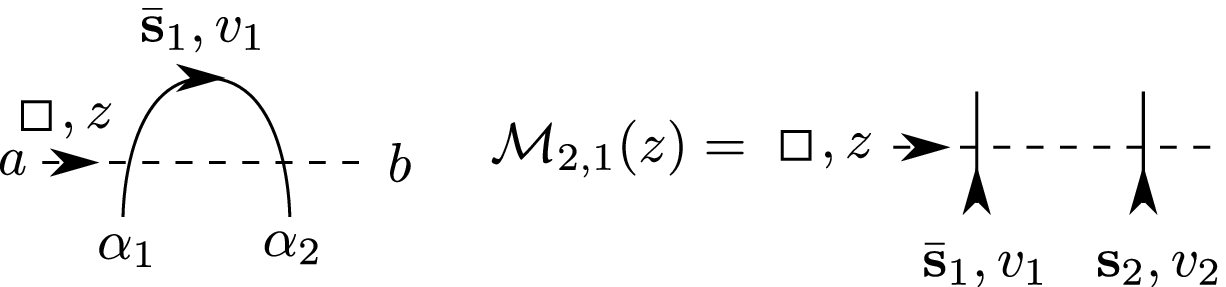}
    \caption{A Baxter lattice with one line specified by
      $\graph=((1,2))$, $\repset=(\bs_1)$, $\rapset=(\inh_1)$,
      $\stateset=(\alpha_1,\alpha_2)$ and intersected by a dashed auxiliary space, left part. This
      arrangement of Boltzmann weights corresponds to the l.h.s.\ of
      the invariance condition \eqref{eq:yi-inv-rmm} for
      $|\Psi_{2,1}\rangle$, i.e.\ to
      $\mathcal{M}_{2,1}(\spec)|\Psi_{2,1}\rangle$. The elements of the
      monodromy $\mathcal{M}_{2,1}(\spec)$ in the right part of the figure
      are obtained from the Boltzmann weights on the left side using
      the crossing relation \eqref{crossslaxx}. The
      representation labels and inhomogeneities of this monodromy obey
      \eqref{eq:osc-m21-vs}.}
    \label{fig:osc-psi21}
  \end{center} 
\end{figure} 

This leads to the length-two monodromy
\begin{align}
  \label{eq:osc-m21}
  \mathcal{M}_{2,1}(\spec)
  =
  R_{\square,\bs_1}(\spec-\inh_1)R_{\square,\s_2}(\spec-\inh_2)
\end{align}
with the constraints on the representation labels and
inhomogeneities
\begin{align}
  \label{eq:osc-m21-vs}
    \inh_1=\inh_2-n-s_2+1\,,
    \quad
    s_1=s_2\,.
\end{align}
Recalling the Baxter lattice associated to this particular monodromy,
we happily notice that \eqref{eq:osc-m21-vs} agrees with
\eqref{eq:osc-line-bs}. The overall normalization of the monodromy
\eqref{eq:osc-m21} originating from those of the Lax operators
\eqref{eq:osc-lax-fund-s} and \eqref{eq:osc-lax-fund-bs} trivializes,
\begin{align}
  \label{eq:osc-m21-norm}
  f_{\bs_1}(\spec-\inh_1)f_{\s_2}(\spec-\inh_2)=1\,,
\end{align}
where we used \eqref{eq:osc-m21-vs} and subsequently the unitarity
condition for $f_{\bs}(\spec)$ in \eqref{eq:osc-norm-unitarity} and
the relation between the two normalizations $f_{\s}(\spec)$ and
$f_{\bs}(\spec)$ in \eqref{eq:osc-crossing-norm}. We can now easily
solve \eqref{eq:yi-inv-rmm} to obtain the explicit form of the
invariant,
\begin{align}
  \label{eq:osc-psi21}
  |\Psi_{2,1}\rangle
  =
  (\bar\oscb^1\cdot \bar\osca^2)^{s_2}|0\rangle
  \quad
  \text{with}
  \quad
  \bar\oscb^i\cdot\bar\osca^j:=\sum_{a=1}^n\bar\oscb^i_a\bar\osca^j_a\,,
\end{align} 
where we recall that the upper indices on the oscillators refer to the
sites of the monodromy. This solution is unique up to a scalar factor,
which clearly drops out of \eqref{eq:yi-inv-rmm}. To obtain the
intertwiner associated to the invariant $|\Psi_{2,1}\rangle$ we employ
\eqref{eq:yi-intertwiner} with $\dsites=1$ and use the value of the
crossing parameter according to \eqref{genrecosc}. This leads to
\begin{align}
  \label{eq:osc-int-psi21}  
  R_{\square,\s_2}(\spec-\inh_2)\mathcal{O}_{\Psi_{2,1}}
  =
  \mathcal{O}_{\Psi_{2,1}}R_{\square,\s_1}(\spec-\inh_2)
\end{align}
with
\begin{align}
  \label{eq:osc-opsi21}
  \mathcal{O}_{\Psi_{2,1}}:=|\Psi_{2,1}\rangle^{\dagger_1}
  =
  \sum_{a_1,\ldots,a_{s_2}=1}^n
  \!\!\!
  \bar\osca^2_{a_1}\cdots\bar\osca^2_{a_{s_2}}
  |0\rangle\langle 0|
  \oscb^1_{a_1}\cdots\oscb^1_{a_{s_2}}\,.
\end{align} 
After identifying the representation spaces $V_{\s_1}$ and $V_{\s_2}$,
which is possible because of $s_1=s_2$ in \eqref{eq:osc-m21-vs}, we
see that $\mathcal{O}_{\Psi_{2,1}}$ reduces to $s_2!$ times the
identity operator.

\subsubsection{Three-vertices and bootstrap equations}
\label{sec:osc-3vertices}

The next simplest Yangian invariants are characterized by monodromies
with three sites and are of the type $|\Psi_{3,1}\rangle$ or
$|\Psi_{3,2}\rangle$. We restrict once more to the case where the
sites with conjugate representations are to the left of those with
symmetric ones. These three-site invariants clearly leave the
framework of Section~\ref{sec:yi-monodromy}. We represent them
graphically by an extension of the Baxter lattice, which in this case
consists of a trivalent vertex, compare Section~\ref{sec:bstrap}. See Figure~\ref{fig:osc-psi31} and
\ref{fig:osc-psi32} for the invariants $|\Psi_{3,1}\rangle$ and
$|\Psi_{3,2}\rangle$, respectively.

\begin{figure}[!t]
  \begin{center}
    \includegraphics[width=0.9\textwidth]{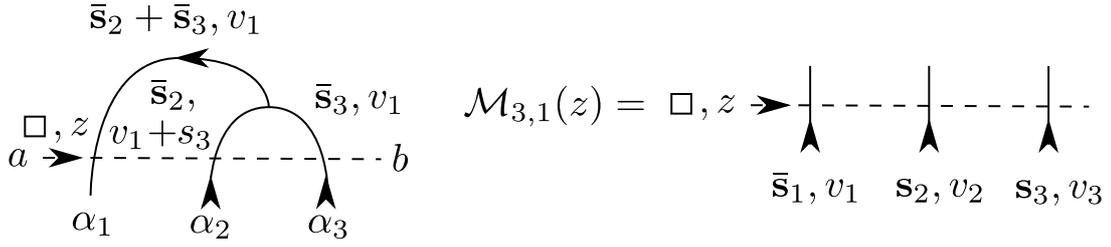}
    \caption{The left part corresponds to the l.h.s.\ of
      \eqref{eq:yi-inv-rmm} for $|\Psi_{3,1}\rangle$, i.e.\
      $\mathcal{M}_{3,1}(\spec)|\Psi_{3,1}\rangle$. It contains a (solid)
      trivalent vertex, which is an extension of the usual Baxter
      lattice, and a dashed auxiliary line. Using the crossing
      relation \eqref{crossslaxx} and the crossing
      parameters in \eqref{eq:osc-crossing-norm}, the Boltzmann
      weights involving the auxiliary line can be reformulated as
      elements of the monodromy $\mathcal{M}_{3,1}(\spec)$, as is shown on the
      right side. The necessary constraints on the representation
      labels and inhomogeneities of the monodromy may be found in
      \eqref{eq:osc-m31-vs}.}
    \label{fig:osc-psi31}
  \end{center} 
\end{figure} 
We start with a monodromy containing one conjugate site,
\begin{align}
  \label{eq:osc-m31}
  \mathcal{M}_{3,1}(\spec)
  =
  R_{\square,\bs_1}(\spec-\inh_1)
  R_{\square,\s_2}(\spec-\inh_2)
  R_{\square,\s_3}(\spec-\inh_3)\,,
\end{align}
see also the right part of Figure~\ref{fig:osc-psi31}. Now the Yangian
invariance condition \eqref{eq:yi-inv-rmm} can be easily solved if the
parameters obey
\begin{equation}
  \label{eq:osc-m31-vs}
  \inh_2=\inh_1+n+s_2+s_3-1\,,
  \quad
  \inh_3=\inh_1+n+s_3-1\,,
  \quad
  s_1=s_2+s_3\,.
\end{equation}
In this case the normalizations of the Lax operators of type
\eqref{eq:osc-lax-fund-s} and \eqref{eq:osc-lax-fund-bs} appearing in
\eqref{eq:osc-m31} trivializes using the relation
\begin{equation}\label{eq:osc-norm-add-eq} 
 f_{\s}(u)f_{\s'}(u + s) = f_{\s+\s'}(u)\,,
\end{equation} 
which follows from \eqref{symcnorm}, the unitarity
condition for $f_{\bs}(\spec)$ and finally expressing $f_{\bs}(\spec)$
in terms of $f_{s}(\spec)$ with the help of
\eqref{eq:osc-crossing-norm}:
\begin{align}
  \label{eq:osc-m31-norm}
  f_{\bs_1}(\spec-\inh_1)f_{\s_2}(\spec-\inh_2)f_{\s_3}(\spec-\inh_3)=1\,.
\end{align}
Then one immediately checks that the solution of \eqref{eq:yi-inv-rmm}
is given by
\begin{equation}
  \label{eq:osc-psi31}
  |\Psi_{3,1}\rangle
  =
  (\bar\oscb^1\cdot\bar\osca^2)^{s_2}
  (\bar\oscb^1\cdot\bar\osca^3)^{s_3}
  |0\rangle\,,
\end{equation}
where we fixed a possible scalar prefactor. We once again proceed to
the corresponding intertwining relation. From its general form in
\eqref{eq:yi-intertwiner} we obtain for $\dsites=1$ the relation
\begin{equation}
  \label{eq:osc-int-psi31}
  R_{\square,\s_2}(\spec-\inh_2)
  R_{\square,\s_3}(\spec-\inh_2+s_2)
  \mathcal{O}_{\Psi_{3,1}}
  =
  \mathcal{O}_{\Psi_{3,1}}
  R_{\square,\s_1}(\spec-\inh_2)
\end{equation} 
with
\begin{equation}
  \label{eq:osc-opsi31}
  \mathcal{O}_{\Psi_{3,1}}
  :=
  |\Psi_{3,1}\rangle^{\dagger_1}
  =
  \sum_{\substack{a_1,\ldots,a_{s_2}\\b_1,\ldots,b_{s_3}}}
  \!\!\!
  \bar\osca_{a_1}^2\cdots\bar\osca_{a_{s_2}}^2
  \bar\osca_{b_1}^3\cdots\bar\osca_{b_{s_3}}^3
  |0\rangle
  \langle 0|
  \oscb_{a_1}^1\cdots\oscb_{a_{s_2}}^1
  \oscb_{b_1}^1\cdots\oscb_{b_{s_3}}^1\,.
\end{equation}
The intertwining relation \eqref{eq:osc-int-psi31} is usually referred to as the
bootstrap equation, cf. \eqref{bootsym1}.

\begin{figure}[!t]
  \begin{center}
      \includegraphics[width=0.9\textwidth]{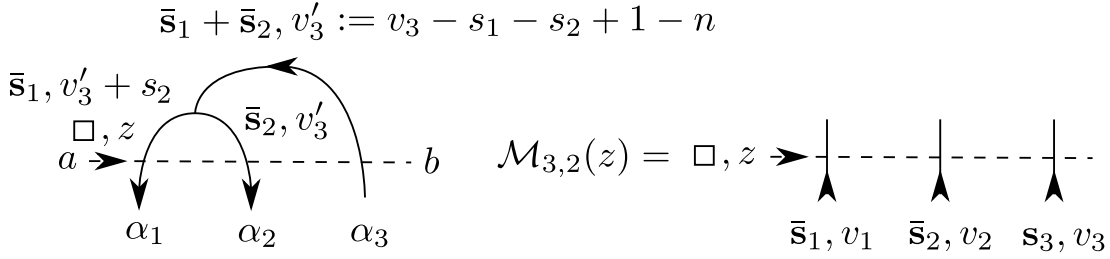}
    \caption{The l.h.s.\ $\mathcal{M}_{3,2}(\spec)|\Psi_{3,2}\rangle$ of
      \eqref{eq:yi-inv-rmm} for $|\Psi_{3,2}\rangle$ corresponds to
      the lattice in the left part. It consists of an extended Baxter
      lattice in form of a trivalent vertex and a dashed auxiliary
      space. The Boltzmann weights containing the auxiliary space can
      be formulated as elements of a monodromy $\mathcal{M}_{3,2}(\spec)$ using
      the crossing relation \eqref{crossslaxx} with
      \eqref{eq:osc-crossing-norm}. This monodromy is shown in the
      right part and the parameters of the monodromy obey the
      constraints \eqref{eq:osc-m32-vs}.}
    \label{fig:osc-psi32}
  \end{center} 
\end{figure} 
We move on to a monodromy with two conjugate sites on the left,
\begin{equation}
  \label{eq:osc-m32}  
  \mathcal{M}_{3,2}(\spec)
  =
  R_{\square,\bs_1}(\spec-\inh_1)
  R_{\square,\bs_2}(\spec-\inh_2)
  R_{\square,\s_3}(\spec-\inh_3)\,,
\end{equation}
see also the right part of Figure~\ref{fig:osc-psi32}. Looking for
solutions $|\Psi_{3,2}\rangle$ of \eqref{eq:yi-inv-rmm} with this
monodromy again leads to constraints on the representation labels and
inhomogeneities,
\begin{equation}
  \label{eq:osc-m32-vs}
  \inh_1=\inh_3-n-s_1+1\,,
  \quad
  \inh_2=\inh_3-n-s_3+1\,,
  \quad
  s_3=s_1+s_2\,.
\end{equation}
Analogously to the discussion of the other three-site invariant, the
normalization of the monodromy \eqref{eq:osc-m32} trivializes using
\eqref{eq:osc-m32-vs}:
\begin{align}
  \label{eq:osc-m32-norm}
  f_{\bs_1}(\spec-\inh_1)f_{\bs_2}(\spec-\inh_2)f_{\s_3}(\spec-\inh_3)=1\,.
\end{align}
The explicit expression for the solution of \eqref{eq:yi-inv-rmm}
turns out to be
\begin{align}
  \label{eq:osc-psi32}
  |\Psi_{3,2}\rangle
  =
  (\bar\oscb^1\cdot\bar\osca^3)^{s_1}
  (\bar\oscb^2\cdot\bar\osca^3)^{s_2}
  |0\rangle\,.
\end{align}
Again we fixed a scalar prefactor. We employ the intertwining relation
\eqref{eq:yi-intertwiner} in this case with $\dsites=2$ to derive the bootstrap equation
\begin{equation}
  \label{eq:osc-int-psi32}
  R_{\square,\s_3}(\spec-\inh_3)
  \mathcal{O}_{\Psi_{3,2}}
  =
  \mathcal{O}_{\Psi_{3,2}}
  R_{\square,\s_2}(\spec-\inh_3+s_1)
  R_{\square,\s_1}(\spec-\inh_3)
\end{equation}
with the solution
\begin{equation}
  \label{eq:osc-opsi32}
  \mathcal{O}_{\Psi_{3,2}}
  :=
  |\Psi_{3,2}\rangle^{\dagger_1\dagger_2}
  =
  \sum_{\substack{a_1,\ldots,a_{s_1}\\b_1,\ldots,b_{s_2}}}
  \!\!\!
  \bar\osca_{a_1}^3\cdots\bar\osca_{a_{s_1}}^3
  \bar\osca_{b_1}^3\cdots\bar\osca_{b_{s_2}}^3
  |0\rangle \langle 0|
  \oscb_{a_1}^1\cdots\oscb_{a_{s_1}}^1
  \oscb_{b_1}^2\cdots\oscb_{b_{s_2}}^2\,.
\end{equation}
The invariants $\mathcal{O}_{\Psi_{3,1}}$ and $\mathcal{O}_{\Psi_{3,2}}$ can be understood as the generalizations of $Y_L$ and $Y_R$ introduced in Section~\ref{sec:three}, respectively.
\subsubsection{Four-vertex and Yang-Baxter equation}
\label{sec:osc-4vertex}

Let us proceed to Yangian invariants associated to four-site
monodromies. As an important check of our formalism we will rederive
the well-known $\gln$ invariant
R-matrix~\cite{Kulish1981} in the form as presented in \cite{Beisert:2004ry}, see also Section~\ref{sec:rmatrices}. We leave aside the cases where the Baxter lattice consists of two
non-intersecting lines and focus on the invariants of type
$|\Psi_{4,2}\rangle$, where the Baxter lattice is a four-vertex\,\footnote{In addition to these invariants there may also exist solutions where all representations are chosen intependently and are not the conjugates of one and another but still form a singlet. We expect that these invariants are combinations of the three-vertices discussed above.}. Once
again, we may a priori vary the positions of the conjugate sites
within the monodromy. We picked a particular assignment, where all
sites with conjugate representations are left of those with symmetric
representations, see figure \ref{fig:osc-psi42}.

We use the four-site monodromy
\begin{equation}
  \label{eq:osc-m42}
  \mathcal{M}_{4,2}(\spec)
  =
  R_{\square,\bs_1}(\spec-\inh_1)R_{\square,\bs_2}(\spec-\inh_2)
  R_{\square,\s_3}(\spec-\inh_3)R_{\square,\s_4}(\spec-\inh_4)
\end{equation}
with
\begin{equation}
  \label{eq:osc-m42-vs}
  \inh_1=\inh_3-n-s_1+1\,,
  \quad
  \inh_2=\inh_4-n-s_2+1\,,
  \quad
  s_1=s_3\,,
  \quad
  s_2=s_4\,.
\end{equation}
This identification of the inhomogeneities and representation labels
corresponds to a Baxter lattice with two lines of type
\eqref{eq:osc-line-bs}. In order to simplify the normalizations of the
Lax operators in \eqref{eq:osc-m42}, we note that the relations in
\eqref{eq:osc-m42-vs} are two sets of conditions of the form appearing
for the two site-invariant in \eqref{eq:osc-m21-vs}. Hence the
normalization factors are simplified analogously to the discussion in
Section~\ref{sec:bethe-gl2-sol-line}, which leads to
\begin{align}
  \label{eq:osc-m42-norm}
  f_{\bs_1}(\spec-\inh_1)f_{\bs_2}(\spec-\inh_2)
  f_{\s_3}(\spec-\inh_3)f_{\s_4}(\spec-\inh_4)=1\,.
\end{align}
For the solution of the eigenvalue equation \eqref{eq:yi-inv-rmm} with
this monodromy we make the $\gln$ invariant ansatz
\begin{equation}
  \label{eq:osc-psi42}
  |\Psi_{4,2}(\inh_3-\inh_4)\rangle
  :=
  |\Psi_{4,2}\rangle
  =
  \sum_{k=0}^{\min(s_3,s_4)}
  \!\!\!
  d_k(\inh_3-\inh_4)
  |\Upsilon_k\rangle
\end{equation}
with
\begin{align}
  \label{eq:osc-phi}
  |\Upsilon_k\rangle
  =
  (\bar\oscb^1\cdot\bar\osca^3)^{s_3-k}
  (\bar\oscb^2\cdot \bar\osca^4)^{s_4-k}
  (\bar\oscb^2\cdot \bar\osca^3)^{k}
  (\bar\oscb^1\cdot \bar\osca^4)^{k}
  |0\rangle\,.
\end{align}
In our formalism, the spectral parameter dependence of this four-site
invariant emerges in a natural fashion as the difference of two
inhomogeneities which from now on is denoted by
\begin{align}
  \label{eq:osc-diffinh}
  \inhdiff:=\inh_3-\inh_4.
\end{align}
We have made this manifest by using the notation
$|\Psi_{4,2}(\inhdiff)\rangle$. Substituting \eqref{eq:osc-psi42} into
\eqref{eq:yi-inv-rmm} yields a recursion relation for the coefficients
$d_k$,
\begin{equation}
  \label{eq:osc-recrel}
  \frac{d_k(\inhdiff)}{d_{k+1}(\inhdiff)}
  =
  \frac{(k+1)(\inhdiff-s_3+k+1)}{(s_3-k)(s_4-k)}\,.
\end{equation}
It is solved, up to a function periodic in the index $k$ with period
$1$, by
\begin{equation}
  \label{eq:osc-coeff}
 d_k(\inhdiff)
 =
 \frac{1}{(s_3-k)!(s_4-k)!k!^2}\,
 \frac{k!}{\Gamma(\inhdiff-s_3+k+1)}\,.
\end{equation}

\begin{figure}[!t]
  \begin{center}
   \includegraphics[width=1\textwidth]{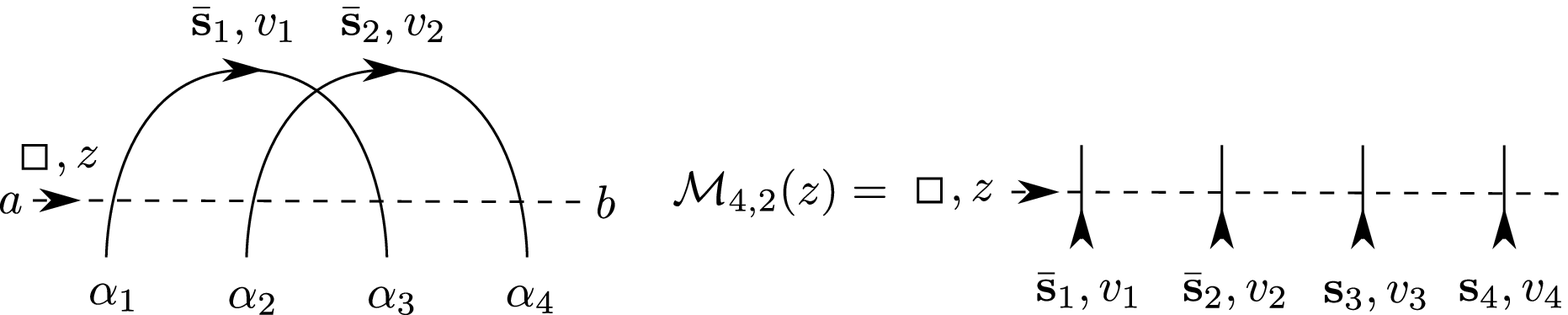}
  \end{center}
  \caption{The left part shows a Baxter lattice with two lines
    specified by $\graph=((1,3),(2,4))$,
    $\repset=(\bs_1,\bs_2)$,
    $\rapset=(\inh_1,\inh_2)$,
    $\stateset=(\alpha_1,\alpha_2,\alpha_3,\alpha_4)$ and a
    dashed auxiliary space. It corresponds to
    $\mathcal{M}_{4,2}(\spec)|\Psi_{4,2}\rangle$ as the l.h.s.\ of
    \eqref{eq:yi-inv-rmm}. The right part contains the monodromy
    $\mathcal{M}_{4,2}(\spec)$, which is associated to this Baxter lattice. The
    necessary identifications of the representation labels and the
    inhomogeneities are written in \eqref{eq:osc-m42-vs}.}
  \label{fig:osc-psi42}
\end{figure}

Following the same logic as before we obtain the equation, which
determines the intertwiner corresponding to
$|\Psi_{4,2}(\inhdiff)\rangle$, from \eqref{eq:yi-intertwiner} with
$\dsites=2$. This yields the Yang-Baxter like equation in
the form
\begin{equation}
  \label{eq:osc-int-psi42}
 R_{\square,\s_3}(\spec-\inh_3)
 R_{\square,\s_4}(\spec-\inh_4)
 \mathcal{O}_{\Psi_{4,2}(\inhdiff)}
 =
 \mathcal{O}_{\Psi_{4,2}(\inhdiff)}
 R_{\square,\s_2}(\spec-\inh_4)
 R_{\square,\s_1}(\spec-\inh_3)\,,
\end{equation} 
where
\begin{equation}
  \label{eq:osc-opsi42}
  \mathcal{O}_{\Psi_{4,2}(\inhdiff)}
  :=
  |\Psi_{4,2}(\inhdiff)\rangle^{\dagger_1\dagger_2}
  =
  \sum_{k=0}^{\min(s_3,s_4)}d_k(\inhdiff)
  \mathcal{O}_{\Upsilon_k}\,,
\end{equation}
with
\begin{align}
  \label{eq:osc-ophi}
  \begin{aligned}
    \mathcal{O}_{\Upsilon_k}
    :=
    |\Upsilon_k\rangle^{\dagger_1\dagger_2}
    =
    \smash{\sum_{\substack{a_1,\ldots,a_{s_3}\\b_1,\ldots,b_{s_4}}}}
    &\bar\osca^3_{a_1}\cdots\bar\osca^3_{a_{s_3}}
    \bar\osca^4_{b_1}\cdots\bar\osca^4_{b_{s_4}}
    |0\rangle\\
    &\quad\cdot
    \langle 0| \oscb^1_{a_1}\cdots\oscb^1_{a_{s_3-k}}
    \oscb^1_{b_{s_4-k+1}}\cdots\oscb^1_{b_{s_4}}\\
    &\quad\quad\cdot
    \oscb^2_{b_1}\cdots\oscb^2_{b_{s_4-k}}
    \oscb^2_{a_{s_3-k+1}}\cdots\oscb^2_{a_{s_3}}\,.
  \end{aligned}
\end{align}
In order to rewrite this form of the Yang-Baxter equation in the
standard way, we identify space $V_{\s_1}$ with $V_{\s_3}$ and
$V_{\s_2}$ with $V_{\s_4}$, and simultaneously
$\mathcal{O}_{\Psi_{4,2}(\inhdiff)}$ with
$R_{\s_3,\s_4}(\inhdiff)$. This then yields
\begin{equation}
  \label{eq:osc-psi42-ybe}
  \begin{aligned}
    R_{\square,\s_3}(\spec-\inh_3)
    R_{\square,\s_4}(\spec-\inh_4)
    R_{\s_3,\s_4}(\inhdiff)
    =
    R_{\s_3,\s_4}(\inhdiff)
    R_{\square,\s_4}(\spec-\inh_4)
    R_{\square,\s_3}(\spec-\inh_3)\,,
  \end{aligned}
\end{equation} 
cf. \eqref{rinter}.
Indeed, \eqref{eq:osc-psi42-ybe} establishes that
$R_{\s_3,\s_4}(\inhdiff)$ must be the $\gln$ invariant
R-matrix \cite{Kulish1981} for symmetric representations.

In our approach $R_{\s_3,\s_4}(\inhdiff)$ is expressed in an oscillator
basis. To be as explicit as possible, it is convenient to introduce
the hopping operators
\begin{equation}
  \label{eq:osc-hop}
  \text{Hop}_k
  =
  \frac{1}{k!^2}
  \sum_{\substack{a_1,\ldots,a_k\\b_1,\ldots,b_k}}
  \!
  \bar\osca^3_{a_1}\cdots\bar\osca^3_{a_k}
  \bar\osca^4_{b_1}\cdots\bar\osca^4_{b_k}
  \osca^3_{b_1}\cdots\osca^3_{b_k}
  \osca^4_{a_1}\cdots\osca^4_{a_k}\,.
\end{equation}
On $V_{\s_3}\otimes V_{\s_4}$ the operator $\text{Hop}_k$ agrees with
$\mathcal{O}_{\Upsilon_k}$, after the said identification of spaces,
up to a trivial combinatorial factor.  This hopping basis allows us to
express the R-matrix\,\footnote{This expression for the R-matrix has been obtained in collaboration with Tomasz {\L}ukowski.} in the form
\begin{equation}
  \label{eq:osc-hoppingr}
  R_{\s_3,\s_4}(\inhdiff)
  =
  \sum_{k=0}^{\min(s_3,s_4)}
  \!\!\!
  \frac{k!}{\Gamma(\inhdiff-s_3+k+1)}\,\text{Hop}_k\,.
\end{equation} 
The operator $\text{Hop}_k$ produces a sum of states containing all
possibilities to exchange $k$ of the oscillators in space $V_{\s_3}$
with $k$ of the oscillators in space $V_{\s_4}$, i.e.\ it hops $k$
oscillators between the two spaces. See also \cite{Ferro:2013dga} for
its supersymmetric and non-compact version and Appendix~\ref{app:hop} for an expression of the Hamiltonian using the operators $\text{Hop}_k$. Note that we can extend
the summation range in \eqref{eq:osc-hoppingr} to infinity as
$\text{Hop}_k$ with $k>\min(s_3,s_4)$ will annihilate any state.
Note also that in \eqref{eq:osc-hoppingr} the dependence on the
representation labels of the coefficients can be absorbed by a shift
of the spectral parameter. Taken in conjunction, these two
observations allow to interpret the expression \eqref{eq:osc-hoppingr}
in a way that does not depend on a specific symmetric representation
$\s$. 

As mentioned above, apart from the invariant \eqref{eq:osc-psi42} which
corresponds to the R-matrix there exists another class of invariants
based on the monodromy \eqref{eq:osc-m42}. Relaxing the conditions in
\eqref{eq:osc-m42-vs} one finds further solutions with
$s_1+s_2=s_3+s_4$. However, in the general case with $s_1\neq s_3$
these invariants do not depend on a free complex spectral parameter.

\section{Toy model for super Yang-Mills scattering amplitudes}
\label{sec:amp}

The main result of Section~\ref{sec:osc} is summarized by the explicit
formulas for the sample invariants \eqref{eq:osc-psi21},
\eqref{eq:osc-psi31}, \eqref{eq:osc-psi32} and \eqref{eq:osc-psi42} of
the Yangian $\mathcal{Y}(\gln)$. The aim of this section
is to establish a relation between these expressions and tree-level
scattering amplitudes of planar $\mathcal{N}=4$ super Yang-Mills
theory, which will often simply be referred to as ``scattering
amplitudes''. See e.g.\ \cite{Drummond:2011ic} for a recent review of
the latter.

The essential connection between the expressions of
Section~\ref{sec:osc}, which are formulated using oscillator algebras,
and these amplitudes is Yangian invariance. For the amplitudes this
was shown in \cite{Drummond2009} employing spinor-helicity
variables.\footnote{Special diligence is required if the particle
  momenta are not in generic position, but there are collinear
  particles \cite{Bargheer2009a}, see also \cite{Bargheer2011}.}
A formal relation between these variables and certain oscillators was
indicated in \cite{Beisert2010}.  Nevertheless, the Yangian is
different in both cases. Here we are focusing on
finite-dimensional representations of $\mathcal{Y}(\gln)$
and not on the infinite-dimensional representation of the Yangian of
$\mathfrak{psu}(2,2|4)\subset\mathfrak{gl}(4|4)$, which is the one
relevant for amplitudes. However, the results obtained are strikingly similar.

In order to compare and relate these two different types of Yangian
invariants, it turns out to be most appropriate to formulate the
scattering amplitudes as Graßmannian integrals in terms of super
twistors \cite{Arkani-Hamed2012}. In these variables the generators
of the superconformal algebra, i.e.\ the lowest level Yangian
generators, are realized as first order differential operators
\cite{Witten:2003nn}. The Yangian invariance of these integrals was
proven in \cite{Drummond2010a}, see also
\cite{Drummond2010b,Ferro:2011ph,Korchemsky:2010ut}. Furthermore,
the super twistor variables together with the associated differential
operators obey the commutation relations of the oscillator algebra. In
the way the invariants of Section~\ref{sec:osc} are formulated within
the framework of the QISM, they naturally contain spectral parameters
in the form of inhomogeneities. Hence, we are led to compare these
invariants to a recent spectral parameter deformation
\cite{Ferro2012,Ferro:2013dga} of these amplitudes.

Those aspects of the Graßmannian integral for undeformed and deformed
scattering amplitudes which are important for our discussion are
briefly summarized in Section~\ref{sec:amp-grass}. In
Section~\ref{sec:amp-osc} we reformulate the invariants obtained in
Section~\ref{sec:osc} with the aim of comparing them to the deformed
amplitudes. As a first step, the invariants are expressed as
multi-dimensional contour integrals over exponential functions of
creation operators. Next, the oscillator algebras are realized in
terms of multiplication and differentiation operators. This turns the exponential
functions into certain delta functions, which are characteristic of
Graßmannian integrals.

Rewritten in this way, the Yangian invariants of Section~\ref{sec:osc}
are essentially $\gln$ analogues of the deformed
tree-level scattering amplitudes of planar $\mathcal{N}=4$ super
Yang-Mills theory. Hence, we may think of them as a ``toy model'' for
scattering amplitudes. Note that our construction allows to explicitly specify
the multi-dimensional integration contour for the sample invariants at
hand which are in general not known for amplitudes.

\subsection{Graßmannian integral for scattering
  amplitudes}
\label{sec:amp-grass}

All tree-level scattering amplitudes of planar $\mathcal{N}=4$ super
Yang-Mills theory can be packaged into a single compact Graßmannian
integral formula using super twistor variables, see
\cite{Arkani-Hamed2012} for a recent formulation, and
\cite{Arkani-Hamed2010a} for the original proposal. In this formalism
the $\sites$-point $\text{N}^{\dsites-2}\text{MHV}$
amplitude is formally given by
\begin{align}
  \label{eq:amp-grassint}
  \mathcal{A}_{\sites,\dsites}
  =
  \int
  \frac{\prod_{k=1}^\dsites\prod_{i=\dsites+1}^\sites\dint c_{ki}}
  {(1\hdots \dsites)
    \cdots 
    (\sites\hdots \sites+\dsites-1)}
  \prod_{k=1}^\dsites
  \delta^{4|4}
  \Bigg(
  \mathcal{W}^k+\sum_{i=\dsites+1}^\sites\!\!\! c_{ki}\mathcal{W}^i
  \Bigg)\,.
\end{align}
These amplitudes are organized by the deviation $\dsites-2$ from the
maximally helicity violating (MHV) configuration. The minor $(i\ldots
i+\dsites)$ , i.e.\ the $\dsites\times \dsites$ subdeterminant, is
built from the columns $i,\ldots,i+\dsites$ of the $\dsites\times
\sites$ matrix
\begin{align}
  \label{eq:amp-gaugefix}
  \begin{pmatrix}
    1&&0&c_{1\,\dsites+1}&\hdots&c_{1\sites}\\
    &\ddots&&\vdots&&\vdots\\
    0&&1&c_{\dsites\,\dsites+1}&\hdots&c_{\dsites\sites}\\
  \end{pmatrix}.
\end{align}
A $\mathfrak{gl}(\dsites)$ gauge symmetry of the Graßmannian integral
\eqref{eq:amp-grassint} has already been fixed by the choice of the
first $\dsites$ columns in \eqref{eq:amp-gaugefix}. The delta functions
$\delta^{4|4}$ in \eqref{eq:amp-grassint} are given by the product of
four bosonic and four fermionic delta functions depending on the super
twistor variables $\mathcal{W}_a^i$ with a point index $i$ and a
fundamental $\mathfrak{gl}(4|4)$ index $a$,
\begin{align}
  \label{eq:amp-delta44}
  \delta^{4|4}
  \Bigg(
  \mathcal{W}^k+\sum_{i=\dsites+1}^\sites\!\!\! c_{ki}\mathcal{W}^i
  \Bigg)
  :=
  \prod_{a=1}^{4+4}
  \delta\Bigg(
  \mathcal{W}_a^k+\sum_{i=\dsites+1}^\sites\!\!\! c_{ki}\mathcal{W}_a^i
  \Bigg)\,.
\end{align}
The Graßmannian integral \eqref{eq:amp-grassint} is often treated in a
formal sense, neither explicitly specifying the domain of integration
nor the meaning of the delta functions of \emph{complex}
variables. See, however, e.g.\ \cite{Mason:2009qx} for a
mathematically more rigorous approach.

Recently, a spectral parameter deformation of the Graßmannian integral
for scattering amplitudes has been introduced
\cite{Ferro2012,Ferro:2013dga} in order to establish connections
with the common language of quantum integrable systems. Here we
consider the deformations of the $3$-point $\overline{\text{MHV}}$
amplitude $\mathcal{A}_{3,1}$, the $3$-point $\text{MHV}$ amplitude
$\mathcal{A}_{3,2}$, and the $4$-point $\text{MHV}$ amplitude
$\mathcal{A}_{4,2}$. These will shortly be compared with the Yangian
invariants constructed in Section~\ref{sec:osc}. The two $3$-point
amplitudes are of special importance as they provide the building
blocks for general $\sites$-point amplitudes. The $4$-point $\text{MHV}$
amplitude is the first non-trivial example that can be constructed
using these building block. The deformations of these amplitudes read
\cite{Ferro:2013dga}
\begin{align}
  \label{eq:amp-def-a31}
  \tilde{\mathcal{A}}_{3,1}
  &=
  \int
  \frac{\dint c_{12}\dint c_{13}}{c_{12}^{s_2+1}c_{13}^{s_3+1}}
  \delta^{n|m}(\mathcal{W}^1+c_{12}\mathcal{W}^2+c_{13}\mathcal{W}^3)\,,\\
  \label{eq:amp-def-a32}
  \tilde{\mathcal{A}}_{3,2}
  &=
  \int
  \frac{\dint c_{13}\dint c_{23}}{c_{13}^{s_1+1}c_{23}^{s_2+1}}
  \delta^{n|m}(\mathcal{W}^1+c_{13}\mathcal{W}^3)
  \delta^{n|m}(\mathcal{W}^2+c_{23}\mathcal{W}^3)\,,\\
  \label{eq:amp-def-a42}
  \begin{split}
    \tilde{\mathcal{A}}_{4,2}(\inhdiff)
    &=
    \int
    \frac{\dint c_{13}\dint c_{14}\dint c_{23}\dint c_{24}}
    {c_{13}c_{24}(c_{13}c_{24}-c_{23}c_{14})}
    \left(-\frac{c_{13}c_{24}}{c_{13}c_{24}-c_{23}c_{14}}\right)^\inhdiff
    \frac{c_{24}^{s_3-s_4}}{(-c_{13}c_{24}+c_{23}c_{14})^{s_3}}\\
    &\quad\quad\quad\quad
    \cdot\,
    \delta^{n|m}(\mathcal{W}^1+c_{13}\mathcal{W}^3+c_{14}\mathcal{W}^4)
    \delta^{n|m}(\mathcal{W}^2+c_{23}\mathcal{W}^3+c_{24}\mathcal{W}^4)\,,
  \end{split}
\end{align}
where the deformation parameters $s_i\in\mathbb{C}$ can be understood
as representation labels. For these low values of $\sites$ and
$\dsites$ a spectral parameter $\inhdiff$ appears only in the last
expression \eqref{eq:amp-def-a42}. In addition, in these deformations
the super twistors are generalized to variables $\mathcal{W}_a^i$ with
a fundamental $\mathfrak{gl}(n|m)$ index $a$ and the delta functions
are to be understood as the corresponding extension of
\eqref{eq:amp-delta44}. In case of $n|m=4|4$, $s_i=0$ and $\inhdiff=0$
the deformations $\tilde{\mathcal{A}}_{\sites,\dsites}$ reduce to the
undeformed scattering amplitudes $\mathcal{A}_{\sites,\dsites}$
obtained from the Graßmannian integral \eqref{eq:amp-grassint}. For
our comparison in the next section we will need the case $n|0$ with
positive integer values of $s_i$, because we will be dealing with
finite-dimensional, purely bosonic representations, and generic
complex $\inhdiff$.

\subsection{Sample invariants as Graßmannian-like
  integrals}
\label{sec:amp-osc}

Let us collect the invariants \eqref{eq:osc-psi21},
\eqref{eq:osc-psi31}, \eqref{eq:osc-psi32}, \eqref{eq:osc-psi42} of
the Yangian $\mathcal{Y}(\gln)$ constructed in
Section~\ref{sec:osc} in terms of oscillators:
\begin{align}
  \label{eq:amp-osc21}
  |\Psi_{2,1}\rangle
  &=
  (\bar\oscb^1\cdot \bar\osca^2)^{s_2}
  |0\rangle\,,\\
  \label{eq:amp-osc31}
  |\Psi_{3,1}\rangle
  &=
  (\bar\oscb^1\cdot\bar\osca^2)^{s_2}
  (\bar\oscb^1\cdot\bar\osca^3)^{s_3}
  |0\rangle\,,\\
  \label{eq:amp-osc32}
  |\Psi_{3,2}\rangle
  &=
  (\bar\oscb^1\cdot\bar\osca^3)^{s_1}
  (\bar\oscb^2\cdot\bar\osca^3)^{s_2}
  |0\rangle\,,\\
  \label{eq:amp-osc42}
  \begin{split}
    |\Psi_{4,2}(\inhdiff)\rangle
    &=
    \sum_{k=0}^{\min(s_3,s_4)}
    \!\!\!
    \frac{1}{(s_3-k)!(s_4-k)!k!^2}\,
    \frac{k!}{\Gamma(\inhdiff-s_3+k+1)}\\
    &\quad\quad\quad\quad
    \cdot\,
    (\bar\oscb^1\cdot\bar\osca^3)^{s_3-k}
    (\bar\oscb^2\cdot \bar\osca^4)^{s_4-k}
    (\bar\oscb^2\cdot \bar\osca^3)^{k}
    (\bar\oscb^1\cdot \bar\osca^4)^{k}
    |0\rangle\,.
  \end{split}
\end{align}
At first sight there seems to be little resemblance between these
formulas and the deformed amplitudes \eqref{eq:amp-def-a31},
\eqref{eq:amp-def-a32} and \eqref{eq:amp-def-a42}. Nevertheless, in
this section we will reformulate these sample
$\mathcal{Y}(\gln)$ invariants
$|\Psi_{\sites,\dsites}\rangle$ and compare to the $\gln$
version of the deformed amplitudes
$\tilde{\mathcal{A}}_{\sites,\dsites}$.

We start by introducing complex contour integrals in some auxiliary
variables $c_{ki}$. In case of the simplest two-site invariant
\eqref{eq:amp-osc21} this is particularly easy and we write
\begin{align}
  \label{eq:amp-int21}
  |\Psi_{2,1}\rangle
  &=
  (\bar\oscb^1\cdot \bar\osca^2)^{s_2}|0\rangle
  =
  \frac{s_2!(-1)^{s_2}}{2\pi i}
  \oint
  \frac{\dint c_{12}}{c_{12}^{s_2+1}}
  e^{-c_{12}\,\bar\oscb^1\cdot \bar\osca^2}
  |0\rangle\,,
\end{align}
where the closed contour encircles the pole at the origin of the
complex $c_{12}$-plane counterclockwise. In the same way each product
$\bar\oscb^k\cdot\bar\osca^i$ of oscillators appearing in the further
invariants \eqref{eq:amp-osc31}, \eqref{eq:amp-osc32} and
\eqref{eq:amp-osc42} is translated into one complex contour integral
in the variable $c_{ki}$,
\begin{align}
  \label{eq:amp-int31}
  |\Psi_{3,1}\rangle
  &=
  \frac{s_2!s_3!(-1)^{s_2+s_3}}{(2\pi i)^2}
  \oint
  \frac{\dint c_{12}\dint c_{13}}{c_{12}^{s_2+1}c_{13}^{s_3+1}}
  e^{-c_{12}\bar\oscb^1\cdot\bar\osca^2-c_{13}\bar\oscb^1\cdot\bar\osca^3}
  |0\rangle\,,\\
  \label{eq:amp-int32}
  |\Psi_{3,2}\rangle
  &=
  \frac{s_1!s_2!(-1)^{s_1+s_2}}{(2\pi i)^2}
  \oint
  \frac{\dint c_{13}\dint c_{23}}{c_{13}^{s_1+1}c_{23}^{s_2+1}}
  e^{-c_{13}\bar\oscb^1\cdot\bar\osca^3-c_{23}\bar\oscb^2\cdot\bar\osca^3}
  |0\rangle\,,\\
  \label{eq:amp-int42} 
  \small 
  \begin{split}
    |\Psi_{4,2}(\inhdiff)\rangle
    &=  
    \frac{(-1)^{s_3+s_4}}{(2\pi i)^4}
    \oint
    \frac{\dint c_{13}\dint c_{14}\dint c_{23}\dint c_{24}}
    {c_{13}^{s_3+1}c_{24}^{s_4+1}c_{14}c_{23}}
    \sum_{k=0}^{\min(s_3,s_4)}
    \!\!\!
    \frac{k!}{\Gamma(\inhdiff-s_3+k+1)}
    \left(\frac{c_{13}c_{24}}{c_{14}c_{23}}\right)^k\\
    &\quad\quad\quad\quad\quad\quad\quad\quad\quad\quad\quad\quad\quad\quad
    \cdot\,
    e^{-c_{13}\bar\oscb^1\cdot\bar\osca^3-c_{14}\bar\oscb^1\cdot\bar\osca^4
      -c_{23}\bar\oscb^2\cdot\bar\osca^3-c_{24}\bar\oscb^2\cdot\bar\osca^4}
    |0\rangle\, ,
  \end{split}
\end{align}
where the contour in each of the variables $c_{ki}$ is again a closed
counterclockwise circle around the origin. The four-site invariant
\eqref{eq:amp-int42} can also be expressed in a slightly more compact
form. We notice that the range of the summation in
\eqref{eq:amp-int42} can be extended to infinity without changing the
value of the integral because the additional terms have a vanishing
residue. Furthermore, choosing a contour that satisfies
$|c_{13}c_{24}|<|c_{14}c_{23}|$, the infinite sum is a series
expansion of a hypergeometric function leading to\footnote{Naively
  this expression does not seem to be valid at the special points
  $s_3-\inhdiff=1,2,3,\ldots$ because in this case the series
  expansion of the hypergeometric function is not defined. However,
  the divergence of the expansion is regularized by the gamma
  function, see e.g. \cite{AbramowitzStegun:1964}, and
  \eqref{eq:amp-int42-2f1} is also valid at these points.}
\begin{align}
  \label{eq:amp-int42-2f1}
  \begin{split}
    |\Psi_{4,2}(\inhdiff)\rangle
    &=
    \frac{(-1)^{s_3+s_4}}{(2\pi i)^4}
    \oint
    \frac{\dint c_{13}\dint c_{14}\dint c_{23}\dint c_{24}}
    {c_{13}^{s_3+1}c_{24}^{s_4+1}c_{14}c_{23}}\,
    \frac{{}_2F_1\Big(1,1;\inhdiff-s_3+1;\frac{c_{13}c_{24}}{c_{14}c_{23}}\Big)}
    {\Gamma(\inhdiff-s_3+1)}\\
    &\quad\quad\quad\quad\quad\quad\quad\quad\quad\quad
    \cdot\,
    e^{-c_{13}\bar\oscb^1\cdot\bar\osca^3-c_{14}\bar\oscb^1\cdot\bar\osca^4
      -c_{23}\bar\oscb^2\cdot\bar\osca^3-c_{24}\bar\oscb^2\cdot\bar\osca^4}
    |0\rangle\,.
  \end{split}
\end{align}
After these reformulations the integral structure of the invariants
$|\Psi_{\sites,\dsites}\rangle$ already matches the one of the
deformed amplitudes $\tilde{\mathcal{A}}_{\sites,\dsites}$, in the
sense that in both cases there are $\dsites(\sites- \dsites)$ integration
variables. The exponential functions of creation operators in the
integrands of the sample invariants $|\Psi_{\sites,\dsites}\rangle$
are reminiscent of the link representation of scattering amplitudes
\cite{Arkani-Hamed2010a,ArkaniHamed:2009si}.

Next, we turn to the form of the integrand with the aim to express the
exponential functions of creation operators as appropriate delta
functions like those in \eqref{eq:amp-def-a31}, \eqref{eq:amp-def-a32}
and \eqref{eq:amp-def-a42}. For this purpose we employ different
representations of the oscillator algebras at sites carrying symmetric
representations of type $\s$ and at sites with conjugate
representations of type $\bs$, respectively:
\begin{align}
  \label{eq:amp-osc-delta}
  \begin{aligned}
    \bar\osca&
    \mathrel{\widehat{=}}\mathcal{W}\,,&\quad
    \osca&
    \mathrel{\widehat{=}}\partial_{\mathcal{W}}\,,&\quad 
    |0\rangle&
    \mathrel{\widehat{=}}1&\text{for sites with}&\quad
    \s\,,\\
    \bar\oscb&
    \mathrel{\widehat{=}}-\partial_{\mathcal{W}}\,,&\quad
    \oscb&
    \mathrel{\widehat{=}}\mathcal{W}\,,&\quad 
    |0\rangle&
    \mathrel{\widehat{=}}\delta(\mathcal{W})&
    \text{for sites with}&\quad
    \bs\,.
  \end{aligned}
\end{align}
The oscillators are realized as multiplication and differentiation
operators in a complex variable $\mathcal{W}$. Consequently, as we
already stressed above, $\delta(\mathcal{W})$ is a delta function of a
complex variable. These representations of the oscillator algebra are
discussed in detail in Appendix~A of \cite{Frassek2013b}.

Before we apply \eqref{eq:amp-osc-delta} to the integral expressions
of the invariants $|\Psi_{\sites,\dsites}\rangle$ given in
\eqref{eq:amp-int21}, \eqref{eq:amp-int31}, \eqref{eq:amp-int32} and
\eqref{eq:amp-int42-2f1}, it is instructive to first look at the form
of the Yangian generators annihilating these invariants, recall
\eqref{eq:yi-inv-expanded}. The corresponding monodromies all have a
trivial overall normalization factor, cf.\ \eqref{eq:osc-m21-norm},
\eqref{eq:osc-m31-norm}, \eqref{eq:osc-m32-norm} and
\eqref{eq:osc-m42-norm}. Hence, their expansion
\eqref{monexpansion} leads to the common Yangian generators
\begin{align}
  \label{eq:amp-gen-yangian}
  \mathcal{M}_{ab}^{[1]}
  =
  \sum_{i=1}^\sites J_{ba}^i\,,
  \quad
  \mathcal{M}_{ab}^{[2]}
  =
  \sum_{1\leq i<j\leq\sites}\sum_{c=1}^nJ_{ca}^i J_{bc}^j
  +\sum_{k=1}^\sites\inh_k\,J_{ba}^k\,,
\end{align}
where the $\gln$ generators at the sites are
\begin{align}
  \label{eq:amp-gen-gln}
  J^i_{ab}=
  \left\{
  \begin{aligned}
    \bar\osca_a^i\osca_b^i&
    \mathrel{\widehat{=}}\mathcal{W}_a^i\partial_{\mathcal{W}_b^i}&
    \text{for sites with}&
    \quad\s_i\,,\\
    -\bar\oscb_b^i\oscb_a^i&
    \mathrel{\widehat{=}}\mathcal{W}_a^i\partial_{\mathcal{W}_b^i}+\delta_{ab}&
    \text{for sites with}&
    \quad\bs_i\,.
  \end{aligned}
  \right.\quad
\end{align}
The inhomogeneities $\inh_i$ depend on the chosen invariant and are
specified in Section~\ref{sec:osc}. In this formulation, the variables
$\mathcal{W}_a^i$ can be thought of as analogous to the super twistors
used in scattering amplitudes, where in case of the latter $a$ is a
fundamental $\mathfrak{gl}(4|4)$ index. While the oscillator form of
the $\gln$ generators in \eqref{eq:amp-gen-gln} has a
different structure at the two distinct types of sites, the generators
are, up to the shift $\delta_{ab}$, identical when written in terms of
$\mathcal{W}_a^i$. The two distinct types of representations, $\s_i$
and $\bs_i$, nevertheless manifest themselves in the structure of the
states: The invariants are polynomials in $\mathcal{W}_a^i$ if the
$i^\text{th}$ site carries a representation $\s_i$, and they contain delta
functions with argument $\mathcal{W}_a^i$ and derivatives thereof for
a site with $\bs_i$. In discussions of the Yangian invariance of
scattering amplitudes the $\mathfrak{gl}(4|4)$ generators also take an
identical form for all points of the amplitude, see e.g.\
\cite{Drummond2010b}.

Let us return to our main goal of applying \eqref{eq:amp-osc-delta} to
the sample invariants $|\Psi_{\sites,\dsites}\rangle$ in the form
\eqref{eq:amp-int21}, \eqref{eq:amp-int31}, \eqref{eq:amp-int32} and
\eqref{eq:amp-int42-2f1}. Note that with \eqref{eq:amp-osc-delta} an
exponential of creation operators becomes
\begin{align}
  \label{eq:amp-shift}
  e^{-c_{ki}\bar\osca_a^i\bar\oscb_b^k}|0\rangle
  \mathrel{\widehat{=}}
  e^{c_{ki}\mathcal{W}_a^i\partial_{\mathcal{W}_b^k}}\delta(\mathcal{W}_b^k)
  =
  \delta(\mathcal{W}_b^k+c_{ki}\mathcal{W}_a^i)\,.
\end{align}
Here $|0\rangle$ denotes the tensor product of the Fock vacua of the
two oscillator algebras. The vacuum of the oscillators $\osca_a^i$ is
realized as $1$ and that of $\oscb_b^k$ as a delta function. For the
invariants $|\Psi_{\sites,\dsites}\rangle$, the symbol $|0\rangle$
stands more generally for the tensor product of the Fock vacua of all
involved oscillator algebras. This means, using
\eqref{eq:amp-osc-delta}, that
\begin{align}
  \label{eq:amp-fock-vac}
  |0\rangle
  \mathrel{\widehat{=}}
  \prod_{k\in\{\text{sites with $\bs$}\}}
  \!\!\!\!\!\!
  \delta^n(\mathcal{W}^k)
  \quad
  \text{with}
  \quad
  \delta^n(\mathcal{W}^k)
  :=
  \prod_{a=1}^n\delta(\mathcal{W}_a^k)\,,
\end{align}
where the range of the first product extends over all sites carrying a
conjugate representation of type $\bs$. Using
\eqref{eq:amp-osc-delta}, \eqref{eq:amp-shift} and
\eqref{eq:amp-fock-vac} the two-site invariant \eqref{eq:amp-int21} is
expressed as
\begin{align}
  \label{eq:amp-delta21}
  |\Psi_{2,1}\rangle
  \mathrel{\widehat{=}}
  \bigg(-\sum_{a=1}^n \mathcal{W}^2_a\partial_{\mathcal{W}^1_a}\bigg)^{s_2}
  \delta^n(\mathcal{W}^1)
  =
  \frac{s_2!(-1)^{s_2}}{2\pi i}
  \oint
  \frac{\dint c_{12}}{c_{12}^{s_2+1}}\,
  \delta^n(\mathcal{W}^1+c_{12}\mathcal{W}^2)\,.
\end{align}
To show the equality of the middle and the right expression in this
formula explicitly, we have to evaluate a contour integral where the
integrand contains a delta function. Proceeding
analogously in the cases of the invariants \eqref{eq:amp-int31},
\eqref{eq:amp-int32}, \eqref{eq:amp-int42-2f1} we
obtain\footnote{Similar formulas for invariants of the Yangian of
  $\gln$ were also obtained recently in
  \cite{Chicherin:2013sqa}. This was extended in
  \cite{Chicherin:2013ora} to $\mathfrak{gl}(n|m)$, which includes the
  $\mathfrak{gl}(4|4)$ case relevant to scattering amplitudes.}
\begin{align}
  \label{eq:amp-delta31}
  |\Psi_{3,1}\rangle
  &\mathrel{\widehat{=}}
  \frac{s_2!s_3!(-1)^{s_2+s_3}}{(2\pi i)^2}
  \oint
  \frac{\dint c_{12}\dint c_{13}}{c_{12}^{s_2+1}c_{13}^{s_3+1}}\,
  \delta^n(\mathcal{W}^1+c_{12}\mathcal{W}^2+c_{13}\mathcal{W}^3)\,,
  \end{align}
  \begin{align}
  \label{eq:amp-delta32}
  |\Psi_{3,2}\rangle
  &\mathrel{\widehat{=}}
  \frac{s_1!s_2!(-1)^{s_1+s_2}}{(2\pi i)^2}
  \oint
  \frac{\dint c_{13}\dint c_{23}}{c_{13}^{s_1+1}c_{23}^{s_2+1}}\,
  \delta^n(\mathcal{W}^1+c_{13}\mathcal{W}^3)
  \delta^n(\mathcal{W}^2+c_{23}\mathcal{W}^3)\,,
    \end{align}
  \begin{align}
  \label{eq:amp-delta42}
  \begin{split}
    |\Psi_{4,2}(\inhdiff)\rangle
    &\mathrel{\widehat{=}}
    \frac{(-1)^{s_3+s_4}}{(2\pi i)^4}
    \oint
    \frac{\dint c_{13}\dint c_{14}\dint c_{23}\dint c_{24}}
    {c_{13}^{s_3+1}c_{24}^{s_4+1}c_{14}c_{23}}\,
    \frac{{}_2F_1\Big(1,1;\inhdiff-s_3+1;\frac{c_{13}c_{24}}{c_{14}c_{23}}\Big)}
    {\Gamma(\inhdiff-s_3+1)}\\
    &\quad\quad\quad\quad\quad
    \cdot\,
    \delta^n(\mathcal{W}^1+c_{13}\mathcal{W}^3+c_{14}\mathcal{W}^4)
    \delta^n(\mathcal{W}^2+c_{23}\mathcal{W}^3+c_{24}\mathcal{W}^4)\,.
  \end{split}
\end{align}
Recall that for these invariants the integrations in all variables
$c_{ki}$ are closed counterclockwise contours encircling the origin
and for \eqref{eq:amp-delta42} we have to assume in addition
$|c_{13}c_{24}|<|c_{14}c_{23}|$.

Finally, we want to compare this version of the invariants to the
deformed amplitudes summarized in Section~\ref{sec:amp-grass}. The
integrations and the delta functions appearing in these deformed
amplitudes are normally only understood in a formal sense, cf.\
\cite{Ferro:2013dga}. To be able to make the comparison, we chose
closed counterclockwise circles around the coordinate origins for the
integration contours in \eqref{eq:amp-def-a31}, \eqref{eq:amp-def-a32}
and \eqref{eq:amp-def-a42}. 

First of all, no deformed amplitude $\tilde{\mathcal{A}}_{2,1}$ is
presented in \cite{Ferro:2013dga}. However, at least for $s_2=0$ the
two-site invariant \eqref{eq:amp-delta21} is contained up to a
normalization factor in the general formula \eqref{eq:amp-grassint}
for $\mathcal{A}_{\sites,\dsites}$ after replacing the delta function
$\delta^{4|4}$ by $\delta^n$. Both three-site invariants
\eqref{eq:amp-delta31} and \eqref{eq:amp-delta32} agree (again up to a
constant normalization) with the $\mathfrak{gl}(n|0)$ version of the
deformed amplitudes provided in \eqref{eq:amp-def-a31} and
\eqref{eq:amp-def-a32},
\begin{align}
  \label{eq:amp-compare-3sites}
  |\Psi_{3,1}\rangle
  \propto
  \tilde{\mathcal{A}}_{3,1}\Big|_{n|0}\,,
  \quad
  |\Psi_{3,2}\rangle
  \propto
  \tilde{\mathcal{A}}_{3,2}\Big|_{n|0}\,.
\end{align}
As already mentioned, the $3$-point amplitudes can be understood as
the basic building blocks for more general amplitudes. Hence,
\eqref{eq:amp-compare-3sites} is an important check of our formalism.
Interestingly, however, the integrand of the deformed amplitude
$\tilde{\mathcal{A}}_{4,2}(\inhdiff)$ given in \eqref{eq:amp-def-a42}
does not fully agree with that of $|\Psi_{4,2}(\inhdiff)\rangle$ found
in \eqref{eq:amp-delta42}. To relate these two expressions we note
that at the special points $s_3-\inhdiff=1,2,3,\ldots$ of the spectral
parameter the series expansion of the hypergeometric function in
\eqref{eq:amp-delta42} simplifies to
\begin{align}
  \label{eq:amp-delta42-special}
  \begin{split}
    |\Psi_{4,2}(\inhdiff)\rangle
    &\mathrel{\widehat{=}}
    \frac{(-1)^{s_3+s_4}}{(2\pi i)^4}
    \oint
    \frac{\dint c_{13}\dint c_{14}\dint c_{23}\dint c_{24}}
    {c_{13}c_{24}(c_{13}c_{24}-c_{23}c_{14})}\,
    \frac{1}{c_{13}^{s_3}c_{24}^{s_4}}
    \left(\frac{-c_{13}c_{24}}{c_{13}c_{24}-c_{23}c_{14}}\right)^{\inhdiff-s_3}\\
    &
    \quad\quad\quad\quad\quad
    \cdot\,
    \delta^n(\mathcal{W}^1+c_{13}\mathcal{W}^3+c_{14}\mathcal{W}^4)
    \delta^n(\mathcal{W}^2+c_{23}\mathcal{W}^3+c_{24}\mathcal{W}^4)\,.
    \end{split}
\end{align}
This agrees up to a shift of the spectral parameter (and again a
normalization factor) with the deformed amplitude:
\begin{align}
  \label{eq:amp-compare-4site}
  |\Psi_{4,2}(\inhdiff)\rangle
  \propto
  \tilde{\mathcal{A}}_{4,2}(\inhdiff-2s_3)\Big|_{n|0}\,
  \quad
  \text{for}
  \quad
  s_3-\inhdiff=1,2,3,\ldots
\end{align}
In \cite{Ferro:2013dga} $\tilde{\mathcal{A}}_{4,2}(\inhdiff)$ is also used
for generic values of $\inhdiff$. However, at least for our choice of the
integration contours around zero this is problematic due to the branch
cut of the complex power function in \eqref{eq:amp-def-a42}. We want
to stress that in the present formulation
\eqref{eq:amp-delta42-special} is only valid at the special points of
the spectral parameter and the full four-site invariant, i.e.\ the
invariant corresponding to the R-matrix, is given by
\eqref{eq:amp-delta42} involving a hypergeometric function. The
interesting question whether a $\mathfrak{gl}(4|4)$ version of
\eqref{eq:amp-delta42} might be a more appropriate deformation of the
four-point $\text{MHV}$ amplitude $\mathcal{A}_{4,2}$ than
\eqref{eq:amp-def-a42} should definitely be clarified.

\section{Bethe ansatz for Yangian invariants}
\label{sec:bethe}

In Section~\ref{sec:osc} we discussed some sample Yangian
invariants. Their relation to (deformed) super Yang-Mills scattering
amplitudes was then established in Section~\ref{sec:amp}. We will
proceed to a systematic construction of Yangian invariants based on
their characterization as solutions of the set of eigenvalue equations
\eqref{eq:yi-inv-rmm}, which involves the monodromy matrix elements
$\mathcal{M}_{ab}(\spec)$. This characterization shows that the invariant
$|\Psi\rangle$ is a special eigenstate of the transfer matrix
\begin{align}
  \label{eq:bethe-trans}
  \tm(\spec)=\tr\mathcal{M}(\spec)\,,
\end{align}
where the trace is taken over the auxiliary space
$V_\square=\mathbb{C}^n$, see \eqref{transf}. Indeed, \eqref{eq:yi-inv-rmm} implies
\begin{align}
  \label{eq:bethe-inv-eigen}
  \tm(\spec)|\Psi\rangle=n|\Psi\rangle
\end{align} 
with the fixed eigenvalue $n$. The transfer matrix
\eqref{eq:bethe-trans} can be diagonalized by means of a Bethe ansatz, cf. Chapter~\ref{ch:ybe}. Therefore a
Yangian invariant $|\Psi\rangle$ is a special Bethe vector. This is
the key observation leading to the construction of $|\Psi\rangle$ by a
Bethe ansatz for Yangian invariants in this section.

For simplicity, we first focus on $\mathfrak{gl}(2)$ monodromies with
finite-dimensional highest weight representations in the quantum
space, as discussed in Section~\ref{sec:gl2} and specialize to the case of Yangian invariant Bethe
vectors  in
Section~\ref{sec:bethe-gl2}. This leads to a set of functional relations characterizing
Yangian invariants, which are equivalent to a degenerate case of the
Baxter equation \cite{Baxter:1972hz}. These equations determine the
Bethe roots and, in addition, constrain the allowed representation
labels and inhomogeneities of the monodromy. Remarkably, a large class
of explicit solutions of these functional relations can be
obtained. They show an interesting structure which is discussed in
Section~\ref{sec:bethe-gl2-sol}. The Bethe roots form exact
strings in the complex plane. The positions of these strings depend on
the inhomogeneities of the monodromy. The length of the strings, i.e.\
the number of Bethe roots per string, is determined by the
representation labels. We illustrate this structure using the sample
invariants already known from Section~\ref{sec:osc}. We also present
solutions to the functional relations corresponding to Baxter lattices
with $\lines$ lines. In particular, this includes lattices where all lines
carry the spin~$\frac{1}{2}$ representation of $\mathfrak{su}(2)$.
Finally, in Section~\ref{sec:bethe-gln} we
sketch the generalization of the set of functional relations
characterizing Yangian invariants from the $\mathfrak{gl}(2)$ to the
$\gln$ case.

\subsection{Bethe ansatz for invariants of Yangian
  $\mathcal{Y}(\mathfrak{gl}(2))$}
\label{sec:bethe-gl2}

Let us explicitly spell out the definition \eqref{eq:yi-inv-rmm} of
Yangian invariants for the $\mathfrak{gl}(2)$ case using the notation
\eqref{gl2mon} for the monodromy elements,
\begin{align}
  \label{eq:bethe-gl2-diag}
  A(\spec)|\Psi\rangle&=|\Psi\rangle\,,&
  D(\spec)|\Psi\rangle&=|\Psi\rangle\,,\\
  \label{eq:bethe-gl2-offdiag}
  B(\spec)|\Psi\rangle&=0\,,&
  C(\spec)|\Psi\rangle&=0\,.
\end{align}
Here we separated the equations into \eqref{eq:bethe-gl2-diag}
involving the diagonal monodromy elements and
\eqref{eq:bethe-gl2-offdiag} with the off-diagonal elements. To
construct Yangian invariants $|\Psi\rangle$ we first solve
\eqref{eq:bethe-gl2-diag} by specializing the Bethe ansatz of
Section~\ref{sec:gl2}. In a second step, we show that for
finite-dimensional representations the Bethe vectors $|\Psi\rangle$
obtained in this way automatically obey also
\eqref{eq:bethe-gl2-offdiag}. The result of this procedure yields a
characterization of Yangian invariants in terms of functional
relations that will be summarized at the end of this section.

Let us first concentrate on the diagonal part
\eqref{eq:bethe-gl2-diag}. Usually, cf.\
Section~\ref{sec:gl2}, one wants to construct
eigenvectors of the transfer matrix, i.e.\ eigenvectors of the sum
$A(\spec)+D(\spec)$. However, here we additionally require that
$|\Psi\rangle$ is a common eigenvector of $A(\spec)$ and
$D(\spec)$. As in Section~\ref{sec:gl2} we make the
ansatz \eqref{su2BV} for the eigenvector and use
the commutation relations \eqref{abcom} and \eqref{dbcom} to derive
\eqref{acta} and \eqref{actd} with
\begin{equation}\label{eq:bethe-gl2-alphadelta}
 \alpha(\spec)=\prod_{i=1}^L f_{\Xi_i}(\spec-\inh_i)\frac{\spec-\inh_i+\xi_i^{(1)}}{\spec-\inh_i}\,,\quad\quad  \delta(\spec)=\prod_{i=1}^L f_{\Xi_i}(\spec-\inh_i)\frac{\spec-\inh_i+\xi_i^{(2)}}{\spec-\inh_i}\,,
\end{equation} 
with the representation labels  $\Xi=(\xi^{(1)},\xi^{(2)})$.
However, we now need to demand that
the ``unwanted terms'' in \eqref{acta} and \eqref{actd} are identical to zero separately. This is
guaranteed by
\begin{align}
  \label{eq:bethe-gl2-specbethe}
  \alpha(\brt_k)Q(\brt_k-1)=0\,,
  \quad
  \delta(\brt_k)Q(\brt_k+1)=0\,,
\end{align}
which is the degenerate case of the Bethe equations
\eqref{prebae} where each term vanishes individually.
In order to fix the eigenvalues of $A(\spec)$ and $D(\spec)$ to be $1$, the 
equations in \eqref{acta} and \eqref{actd} imply that we have to
require
\begin{align}  
  \label{eq:bethe-gl2-specbaxter}
  1=\alpha(\spec)\frac{Q(\spec-1)}{Q(\spec)}\,,
  \quad
  1=\delta(\spec)\frac{Q(\spec+1)}{Q(\spec)}\,.
\end{align}
This is the degenerate case of the Baxter equation
\eqref{baxeq} where each term on the r.h.s.\ is equal
to $1$. It leads to the required transfer matrix eigenvalue
$\tau(\spec)=1+1=2$, which is the rank of $\mathfrak{gl}(2)$. Assuming the
regularity of $\alpha(\spec)$ and $\delta(\spec)$ at the Bethe roots
$\spec=\brt_k$, one shows by taking the residue as in
Section~\ref{sec:gl2} that
\eqref{eq:bethe-gl2-specbaxter} implies
\eqref{eq:bethe-gl2-specbethe}. Consequently, the problem of
constructing common solutions of the eigenvalue equations in
\eqref{eq:bethe-gl2-diag} has been reduced to solving
\eqref{eq:bethe-gl2-specbaxter}.

To address \eqref{eq:bethe-gl2-offdiag} involving the off-diagonal
monodromy elements, we use \eqref{eq:bethe-gl2-diag}, which we already
solved. We expand \eqref{eq:bethe-gl2-offdiag} using
\eqref{monexpansion} to obtain
\begin{align}
  \label{eq:bethe-gl2-gl2-weight}
  \mathcal{M}_{11}^{[1]}|\Psi\rangle=0\,,
  \quad
  \mathcal{M}_{22}^{[1]}|\Psi\rangle=0\,.
\end{align}
As indicated in Section~\ref{sec:lax}, the
generators $-\mathcal{M}^{[1]}_{ab}(\spec)$ form a $\mathfrak{gl}(2)$ algebra
and thus \eqref{eq:bethe-gl2-gl2-weight} means that $|\Psi\rangle$ has
$\mathfrak{gl}(2)$ weight $(0,0)$. The expansion of $C(\spec)\bvac=0$
from \eqref{actmon} implies
$\mathcal{M}_{21}^{[1]}\bvac=0$. Using the commutation relations of the Yangian algebra
\eqref{yangalg} and the relations in \eqref{acta} and \eqref{actd} one
shows
\begin{align}
  \label{eq:bethe-gl2-m21psi}
  \mathcal{M}^{[1]}_{21}|\Psi\rangle
  =
  -\sum_{k=1}^\brts(\alpha(\brt_k)Q(\brt_k-1)+\delta(\brt_k)Q(\brt_k+1))
  \prod_{\substack{i=1\\i\neq k}}^\brts\frac{B(\brt_i)}{\brt_k-\brt_i}\bvac
  =
  0\,,
\end{align}
where we needed \eqref{eq:bethe-gl2-specbethe} for the last
equality. As we are dealing with a finite-dimensional
$\mathfrak{gl}(2)$ representation, \eqref{eq:bethe-gl2-gl2-weight} and
\eqref{eq:bethe-gl2-m21psi} imply that $|\Psi\rangle$ is a
$\mathfrak{gl}(2)$ singlet. Hence, also
\begin{align}
  \label{eq:bethe-gl2-m12psi}
  \mathcal{M}^{[1]}_{12}|\Psi\rangle=0\,.
\end{align}
Finally, we obtain the relations
\begin{align}
  \label{eq:bethe-gl2-comm}
  [\mathcal{M}_{12}^{[1]},A(\spec)-D(\spec)]=2B(\spec)\,,
  \quad
  [\mathcal{M}_{21}^{[1]},D(\spec)-A(\spec)]=2C(\spec)\,,
\end{align}
using \eqref{yangalg}.
Acting with these on $|\Psi\rangle$ and using
\eqref{eq:bethe-gl2-diag}, \eqref{eq:bethe-gl2-m21psi} and
\eqref{eq:bethe-gl2-m12psi}, we see that also the off-diagonal
part \eqref{eq:bethe-gl2-offdiag} of the Yangian invariance condition
is satisfied.

In conclusion, we have reduced the problem of constructing invariants
$|\Psi\rangle$ of the Yangian $\mathcal{Y}(\mathfrak{gl}(2))$ to the
problem of solving the functional relations
\eqref{eq:bethe-gl2-specbaxter}. Given a solution
$(\alpha(\spec),\delta(\spec),Q(\spec))$ of
\eqref{eq:bethe-gl2-specbaxter}, where the Q-function is of the form
\eqref{qfct}, and both $\alpha(\spec)$ and
$\delta(\spec)$ are regular at the Bethe roots $\brt_k$, the Bethe vector
$|\Psi\rangle$ given in \eqref{su2BV} is Yangian
invariant. It is convenient to represent the functional relations
\eqref{eq:bethe-gl2-specbaxter} in a slightly different
form. Remarkably, the system of two equations in
\eqref{eq:bethe-gl2-specbaxter} can be decoupled into an equation that
depends only on the eigenvalues \eqref{actmon} of the
monodromy acting on the reference state and not on the Bethe roots,
\begin{align}
  \label{eq:bethe-gl2-ad}
  1=\alpha(\spec)\delta(\spec-1)\,,
\end{align} 
and a further equation also involving the Bethe roots contained in
the Q-function,
\begin{align}
  \label{eq:bethe-gl2-qaq}
  \frac{Q(\spec)}{Q(\spec+1)}=\delta(\spec)\,.
\end{align} 
As $\alpha(\spec)$ and $\delta(\spec)$
contain the representation labels and inhomogeneities, cf.\
\eqref{eq:bethe-gl2-alphadelta} where a different normalization of the Lax operators has been used, this equation determines those
monodromies that correspond to a Yangian invariant, i.e.\ for which
\eqref{eq:yi-inv-rmm} admits a solution $|\Psi\rangle$. Once a
suitable solution of \eqref{eq:bethe-gl2-ad} is found\,\footnote{Note that \eqref{eq:bethe-gl2-qaq} is equivalent to the $\mathcal{Y}(\mathfrak{sl}(2))$ constraint when acting on the vacuum state $\vert\Omega\rangle$, cf. \eqref{ysl}.}, the difference
equation \eqref{eq:bethe-gl2-qaq} can typically be solved with ease
for the Bethe roots $\brt_k$. This is in stunning contradistinction to
the usual situation in most spin chain spectral problems, where the
Bethe equations are very hard to solve. Substituting the Bethe roots
into \eqref{su2BV} yields the Bethe state, and
hence the invariant $|\Psi\rangle$. It would be interesting to study the conditions \eqref{eq:bethe-gl2-ad} and \eqref{eq:bethe-gl2-qaq} more carefully from a representation theoretical point of view and understand how they are related to Drinfeld polynomials which provide a systematic way to classify representations of the Yangian, see e.g. \cite{Kuramoto2009}.

\subsection{Sample solutions of
$\mathfrak{gl}(2)$ functional relations}
\label{sec:bethe-gl2-sol}

At present, we lack a complete understanding of the set of solutions
to the functional relations \eqref{eq:bethe-gl2-ad} and
\eqref{eq:bethe-gl2-qaq}. Gaining it should lead to a classification
of invariants of the Yangian $\mathcal{Y}(\mathfrak{gl}(2))$, clearly
an interesting problem for future research. It was shown in \cite{Kanning2014} that deformation parameters as introduced in \cite{Beisert2014} of the tree-level scattering amplitudes satisfy the relation \eqref{eq:bethe-gl2-ad}. Here, we show how the
$\mathfrak{gl}(2)$ versions of the invariants in oscillator
form with representations of type $\s=(s,0)$ and $\bs=(0,-s)$ studied in
Section~\ref{sec:osc} fit into the framework of the Bethe ansatz for
Yangian invariants. In particular, we again discuss the invariant
$|\Psi_{2,1}\rangle$, which was represented by a Baxter lattice with a
single line, the three-vertices $|\Psi_{3,1}\rangle$ and
$|\Psi_{3,2}\rangle$, as well as the four-vertex (R-matrix)
$|\Psi_{4,2}(\inhdiff)\rangle$. We also consider the invariant
associated to a Baxter lattice of $\lines$ lines. For all these examples
the Bethe roots are given explicitly. They arrange themselves into
strings in the complex plane.

\subsubsection{Line}
\label{sec:bethe-gl2-sol-line}

\begin{figure}[!t]
  \begin{center}
    \begin{tikzpicture}
      \draw[thick] (0,0) circle (3pt) node[above=0.1cm]{$\inh_2$};
      \filldraw[thick] (-1,0) circle (1pt) node[below]{$\brt_1$};
      \filldraw[thick] (-2,0) circle (1pt) node[below]{$\brt_2$};
      \node at (-3,0) {\ldots};
      \filldraw[thick] (-4,0) circle (1pt) node[below]{$\brt_{s_2-1}$};
      \filldraw[thick] (-5,0) circle (1pt) node[below]{$\brt_{s_2}$};
      \draw[thick] (-6,0) circle (3pt) node[above=0.1cm]{$\inh_1$};
      \draw[thick,|-|] (-6,1) -- node[midway,above] {$s_2+1$} (0,1);
    \end{tikzpicture}
    \caption{The Bethe roots $\brt_k$ associated to the Yangian invariant
      $|\Psi_{2,1}\rangle$ of Section~\ref{sec:osc-line} arrange into 
      a string between the two inhomogeneities $\inh_1$ and $\inh_2$,
      cf.\ \eqref{eq:bethe-gl2-sol-line-roots} in the complex
      plane. This string consists of $s_2$ roots with a uniform real
      spacing of $1$.}
    \label{fig:bethe-gl2-sol-line}
  \end{center} 
\end{figure}
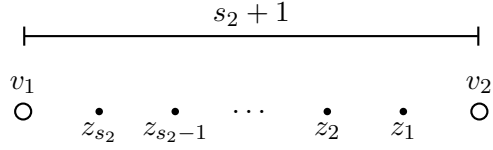 

Let us recall the representation labels and inhomogeneities for the
$\mathfrak{gl}(2)$ case of the invariant $|\Psi_{2,1}\rangle$
discussed in Section~\ref{sec:osc-line} and associated to the spin
chain monodromy $\mathcal{M}_{2,1}(\spec)$ with $\sites=2$ sites, cf.\
\eqref{eq:osc-m21} and \eqref{eq:osc-m21-vs}:
\begin{align}
  \label{eq:bethe-gl2-sol-line-constr}
  \begin{gathered}
    \Xi_1=\bs_1\,,
    \quad
    \Xi_2=\s_2\,,\\
    \inh_1=\inh_2-1-s_2\,,
    \quad
    s_1=s_2\,.
  \end{gathered}
\end{align}
With these relations and the trivial normalization
\eqref{eq:osc-m21-norm} of the monodromy,
\eqref{eq:bethe-gl2-alphadelta} simplifies to
\begin{align}
  \label{eq:bethe-gl2-sol-line-ad-eval}  
  \alpha(\spec)=\frac{\spec-\inh_2+s_2}{\spec-\inh_2}\,,
  \quad
  \delta(\spec)=\frac{\spec-\inh_2+1}{\spec-\inh_2+1+s_2}\,.
\end{align}
In this form one directly sees that the first functional relation
\eqref{eq:bethe-gl2-ad} holds. The remaining relation
\eqref{eq:bethe-gl2-qaq} is solved by
\begin{align}
  \label{eq:bethe-gl2-sol-line-q}  
  Q(\spec)
  =
  \frac{\Gamma(\spec-\inh_2+s_2+1)}{\Gamma(\spec-\inh_2+1)}
  =
  \prod_{k=1}^{s_2}(\spec-\inh_2+k)\,,
\end{align} 
where the freedom of multiplying this solution by a function of period
$1$ in $\spec$ has been fixed by imposing the polynomial form of the Q-function \eqref{qfct}. Because $s_2$ is a
positive integer, the gamma functions in
\eqref{eq:bethe-gl2-sol-line-q} indeed reduce to a polynomial and we
can read off the Bethe roots as zeros of the Q-function,
\begin{align}
  \label{eq:bethe-gl2-sol-line-roots}  
  \brt_k=\inh_2-k\quad\text{for}\quad k=1,\ldots,s_2\,.
\end{align}
They form a string in the complex plane, see
Figure~\ref{fig:bethe-gl2-sol-line}. Note that, as is usual for a
$\mathfrak{gl}(2)$ Bethe ansatz, the labels of the Bethe roots can be
permuted because the operators $B(\spec)$ appearing in the Bethe
vector \eqref{su2BV} commute for different values
of the spectral parameter $\spec$, cf.\
\eqref{Bcom}. Finally, we want to construct the
Yangian invariant Bethe vector \eqref{su2BV}
corresponding to this solution of the functional relations. For this
purpose we need the reference state \eqref{bethevac} for
the representations specified in
\eqref{eq:bethe-gl2-sol-line-constr}. It is given by a tensor product
of the highest weight states \eqref{eq:osc-hws}:
\begin{align}
  \label{eq:bethe-gl2-sol-line-vac}  
  \bvac
  =
  (\bar\oscb_2^1)^{s_2}
  (\bar\osca_1^2)^{s_2}
  |0\rangle\,.
\end{align}
Then we can evaluate \eqref{su2BV} using
\eqref{eq:bethe-gl2-sol-line-constr},
\eqref{eq:bethe-gl2-sol-line-roots} and
\eqref{eq:bethe-gl2-sol-line-vac}, where we note that because of
\eqref{eq:osc-m21-norm} also the normalization of the operators
$B(\brt_k)$ is trivial. Some details of this straightforward
computation for general $s_2\in\mathbb{N}$ are given in \cite{Frassek2013b}. One finds
\begin{align}
  \label{eq:bethe-gl2-sol-line-inv}  
  |\Psi \rangle
  =
  B(\brt_1)\cdots B(\brt_{s_2})\bvac
  =
  (-1)^{s_2}
  (\bar\oscb^1\cdot \bar\osca^2)^{s_2}|0\rangle
  \propto |\Psi_{2,1}\rangle\,.
\end{align}
Thus, our Bethe ansatz for Yangian invariants nicely matches $|\Psi_{2,1}\rangle$ as given in \eqref{eq:osc-psi21}. 

\subsubsection{Three-vertices}
\label{sec:bethe-gl2-sol-three}

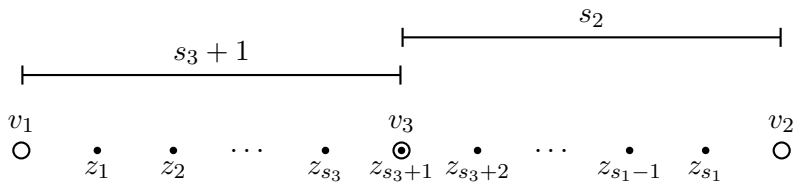
\begin{figure}[!t]
  \begin{center}
    \begin{tikzpicture}
      \draw[thick] (0,0) circle (3pt) node[above=0.1cm]{$\inh_1$};
      \filldraw[thick] (1,0) circle (1pt) node[below]{$\brt_1$};
      \filldraw[thick] (2,0) circle (1pt) node[below]{$\brt_2$};
      \node at (3,0) {\ldots};
      \filldraw[thick] (4,0) circle (1pt) node[below]{$\brt_{s_3}$};
      \filldraw[thick] (5,0) circle (1pt) node[below]{$\brt_{s_3+1}$};
      \draw[thick] (5,0) circle (3pt) node[above=0.1cm]{$\inh_3$};
      \filldraw[thick] (6,0) circle (1pt) node[below]{$\brt_{s_3+2}$};
      \node at (7,0) {\ldots};
      \filldraw[thick] (8,0) circle (1pt) node[below]{$\brt_{s_1-1}$};
      \filldraw[thick] (9,0) circle (1pt) node[below]{$\brt_{s_1}$};
      \draw[thick] (10,0) circle (3pt) node[above=0.1cm]{$\inh_2$};
      \draw[thick,|-|] (0,1) -- node[midway,above] {$s_3+1$} (5,1);
      \draw[thick,|-|] (5,1.5) -- node[midway,above] {$s_2$} (10,1.5);
    \end{tikzpicture}
    \caption{The invariant $|\Psi_{3,1}\rangle$ gives rise to a real
      string of $s_1$ uniformly spaced Bethe roots $\brt_k$ in the
      complex plane, see \eqref{eq:bethe-gl2-sol-three1-roots}. They
      lie in between the inhomogeneities $\inh_1$, $\inh_2$ and one root
      coincides with $\inh_3$.}
    \label{fig:bethe-gl2-three1-line}
  \end{center} 
\end{figure} 

In Section~\ref{sec:osc-3vertices} we discussed two different
three-site invariants. For the $\mathfrak{gl}(2)$ case the monodromy
$\mathcal{M}_{3,1}(\spec)$ associated to the first invariant
$|\Psi_{3,1}\rangle$ is defined by, cf.\ \eqref{eq:osc-m31} and
\eqref{eq:osc-m31-vs},
\begin{align}
  \label{eq:bethe-gl2-sol-three1-constr}
  \begin{gathered}
    \Xi_1=\bs_1\,,
    \quad
    \Xi_2=\s_2\,,
    \quad
    \Xi_3=\s_3\,,\\
    \inh_2=\inh_1+1+s_2+s_3\,,
    \quad
    \inh_3=\inh_1+1+s_3\,,
    \quad
    s_1=s_2+s_3\,.
  \end{gathered}
\end{align}
With \eqref{eq:bethe-gl2-sol-three1-constr} and the trivial
normalization of the monodromy \eqref{eq:osc-m31-norm}, the
eigenvalues of the monodromy on the reference state of the Bethe
ansatz in \eqref{eq:bethe-gl2-alphadelta} turn into
\begin{align}
  \label{eq:bethe-gl2-sol-three1-ad-eval}  
  \alpha(\spec)=\frac{\spec-\inh_1-1}{\spec-\inh_1-s_1-1}\,,
  \quad
  \delta(\spec)=\frac{\spec-\inh_1-s_1}{\spec-\inh_1}\,.
\end{align} 
Obviously, they obey \eqref{eq:bethe-gl2-ad}. The other functional
relation \eqref{eq:bethe-gl2-qaq} is uniquely solved by
\begin{align}
  \label{eq:bethe-gl2-sol-three1-q}  
  Q(\spec)
  =
  \frac{\Gamma(\spec-\inh_1)}{\Gamma(\spec-\inh_1-s_1)}
  =
  \prod_{k=1}^{s_1}(\spec-\inh_1-k)\,,
\end{align}
because the Q-function is of the form \eqref{qfct}. The
zeros of \eqref{eq:bethe-gl2-sol-three1-q} yield the Bethe roots
\begin{align}
  \label{eq:bethe-gl2-sol-three1-roots}  
  \brt_k=\inh_1+k\quad\text{for}\quad k=1,\ldots,s_1\,.
\end{align}
For this invariant the Bethe roots again form a string in the complex
plane, see Figure~\ref{fig:bethe-gl2-three1-line}. We now turn to the
construction of the associated Bethe vector. With \eqref{eq:osc-hws}
the reference state \eqref{bethevac} for the Bethe ansatz
with the representation labels found in
\eqref{eq:bethe-gl2-sol-three1-constr} becomes
\begin{align}
  \label{eq:bethe-gl2-sol-three1-vac}  
  \bvac
  =
  (\bar\oscb_2^1)^{s_2+s_3}
  (\bar\osca_1^2)^{s_2}
  (\bar\osca_1^3)^{s_3}|0\rangle\,.
\end{align}
Notice that one Bethe root is identical to an inhomogeneity,
$\brt_{s_3+1}=\inh_3$. Consequently, the Lax operator
$R_{\square,{\s}_3}(\brt_{s_3+1}-\inh_3)$, cf.\
\eqref{eq:osc-lax-fund-s}, contributing to $B(\brt_{s_3+1})$ in the
Bethe vector \eqref{su2BV} diverges. Nevertheless,
we obtain a finite Bethe vector using an ad hoc prescription, which we
verified for small values of $s_2$ and $s_3$: First, all
non-problematic Bethe roots are inserted into
\eqref{su2BV}, while $\brt_{s_3+1}$ is kept
generic. In the resulting expression the divergence at
$\brt_{s_3+1}=\inh_3$ disappears. Hence, in a second step, we can
safely insert the last root, leading to
\begin{align}
  \label{eq:bethe-gl2-sol-three1-inv}  
  |\Psi\rangle=B(\brt_1)\cdots B(\brt_{s_1})\bvac
  =
  (-1)^{s_2+s_3}
  (\bar\oscb^1\cdot\bar\osca^2)^{s_2}
  (\bar\oscb^1\cdot\bar\osca^3)^{s_3}
  |0\rangle
  \propto
  |\Psi_{3,1}\rangle\,.
\end{align}
Therefore, we have obtained also the three-site Yangian invariant
$|\Psi_{3,1}\rangle$ presented in \eqref{eq:osc-psi31} from a Bethe
ansatz. A derivation of \eqref{eq:bethe-gl2-sol-three1-inv} for
general $s_2,s_3\in\mathbb{N}$.

So-called ``singular solutions'' of the Bethe equations leading
naively to divergent Bethe vectors are well known for the homogeneous
$\mathfrak{su}(2)$ spin~$\frac{1}{2}$ chain, see e.g.\ the recent
discussion \cite{Nepomechie:2013mua}, \cite{Baxter:2001sx} and the
references therein. Such solutions were already known to Bethe himself
\cite{Bethe1931} and appeared also early on in the planar
$\mathcal{N}=4$ super Yang-Mills spectral problem
\cite{Beisert:2003xu}. There are several ways to treat them properly,
cf.\ \cite{Nepomechie:2013mua}, which might also be applicable for the
inhomogeneous spin chain with mixed representations needed for the
three-site invariant $|\Psi_{3,1}\rangle$.

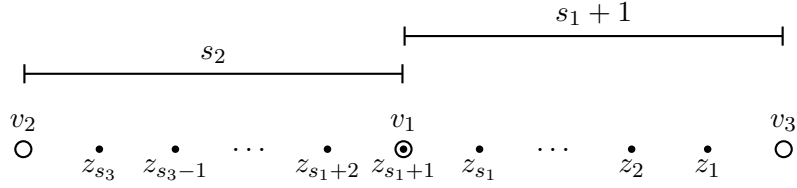
\begin{figure}[!t]
  \begin{center}
    \begin{tikzpicture}
      \draw[thick] (0,0) circle (3pt) node[above=0.1cm]{$\inh_2$};
      \filldraw[thick] (1,0) circle (1pt) node[below]{$\brt_{s_3}$};
      \filldraw[thick] (2,0) circle (1pt) node[below]{$\brt_{s_3-1}$};
      \node at (3,0) {\ldots};
      \filldraw[thick] (4,0) circle (1pt) node[below]{$\brt_{s_1+2}$};
      \filldraw[thick] (5,0) circle (1pt) node[below]{$\brt_{s_1+1}$};
      \draw[thick] (5,0) circle (3pt) node[above=0.1cm]{$\inh_1$};
      \filldraw[thick] (6,0) circle (1pt) node[below]{$\brt_{s_1}$};
      \node at (7,0) {\ldots};
      \filldraw[thick] (8,0) circle (1pt) node[below]{$\brt_2$};
      \filldraw[thick] (9,0) circle (1pt) node[below]{$\brt_1$};
      \draw[thick] (10,0) circle (3pt) node[above=0.1cm]{$\inh_3$};
      \draw[thick,|-|] (0,1) -- node[midway,above] {$s_2$} (5,1);
      \draw[thick,|-|] (5,1.5) -- node[midway,above] {$s_1+1$} (10,1.5);
    \end{tikzpicture}
    \caption{The string of Bethe roots $\brt_k$ belonging to the
      invariant $|\Psi_{3,2}\rangle$. The roots lie between the
      inhomogeneities $\inh_2$ and $\inh_3$. One of them coincides with
      $\inh_1$.}
    \label{fig:bethe-gl2-three2-line}
  \end{center} 
\end{figure} 

The $\mathfrak{gl}(2)$ version of the second three-site invariant
discussed in Section~\ref{sec:osc-3vertices}, $|\Psi_{3,2}\rangle$, is
characterized by the monodromy $\mathcal{M}_{3,2}(\spec)$ defined in
\eqref{eq:osc-m32} and \eqref{eq:osc-m32-vs},
\begin{align}
  \label{eq:bethe-gl2-sol-three2-constr}
  \begin{gathered}
    \Xi_1=\bs_1\,,
    \quad
    \Xi_2=\bs_2\,,
    \quad
    \Xi_3=\s_3\,,\\
    \inh_1=\inh_3-1-s_1\,,
    \quad
    \inh_2=\inh_3-1-s_1-s_2\,,
    \quad
    s_3=s_1+s_2\,.
  \end{gathered}
\end{align}
The trivial normalization of this monodromy, cf.\
\eqref{eq:osc-m32-norm}, together with
\eqref{eq:bethe-gl2-sol-three2-constr} implies that
\eqref{eq:bethe-gl2-alphadelta} simplifies to
\begin{align}
  \label{eq:bethe-gl2-sol-three2-ad-eval}  
  \alpha(\spec)=\frac{\spec-\inh_3+s_3}{\spec-\inh_3}\,,
  \quad
  \delta(\spec)=\frac{\spec-\inh_3+1}{\spec-\inh_3+1+s_3}\,,
\end{align} 
which is a solution of the functional relation
\eqref{eq:bethe-gl2-ad}. The second relation \eqref{eq:bethe-gl2-qaq}
is then solved by 
\begin{align}
  \label{eq:bethe-gl2-sol-three2-q}  
  Q(\spec)
  =
  \frac{\Gamma(\spec-\inh_3+s_3+1)}{\Gamma(\spec-\inh_3+1)}
  =
  \prod_{k=1}^{s_3}(\spec-\inh_3+k)\,.
\end{align}
Demanding this solution to be of the form \eqref{qfct}
guarantees its uniqueness and allows us to read off the Bethe roots
\begin{align}
  \label{eq:bethe-gl2-sol-three2-roots}  
  \brt_k=\inh_3-k\quad\text{for}\quad k=1,\ldots,s_3\,.
\end{align}
Once again, they form a string, see
Figure~\ref{fig:bethe-gl2-three2-line}. To obtain the corresponding
Bethe vector, we first evaluate the reference state
\eqref{bethevac} with \eqref{eq:osc-hws} and the
representations labels given in
\eqref{eq:bethe-gl2-sol-three2-constr}. This leads to
\begin{align}
  \label{eq:bethe-gl2-sol-three2-vac}  
  \bvac
  =
  (\bar\oscb_2^1)^{s_1}
  (\bar\oscb_2^2)^{s_2}
  (\bar\osca_1^3)^{s_1+s_2}|0\rangle\,.
\end{align}
Just like the other three-site invariant, the operators $B(\brt_{s_1+1})$
diverges because $\brt_{s_1+1}=\inh_1$. With the same ad hoc prescription as
above, we obtain again a finite Bethe vector that, for small values of
$s_1$ and $s_2$, has the explicit form
\begin{align}
  \label{eq:bethe-gl2-sol-three2-inv}  
  |\Psi\rangle=B(\brt_1)\cdots B(\brt_{s_1})\bvac
  =(-1)^{s_1+s_2}
  (\bar\oscb^1\cdot\bar\osca^3)^{s_1}
  (\bar\oscb^2\cdot\bar\osca^3)^{s_2}
  |0\rangle
  \propto
  |\Psi_{3,2}\rangle\,.
\end{align}
This matches the form of the three-site invariant $|\Psi_{3,2}\rangle$
given in \eqref{eq:osc-psi32}.

\subsubsection{Four-vertex}
\label{sec:bethe-gl2-sol-four}

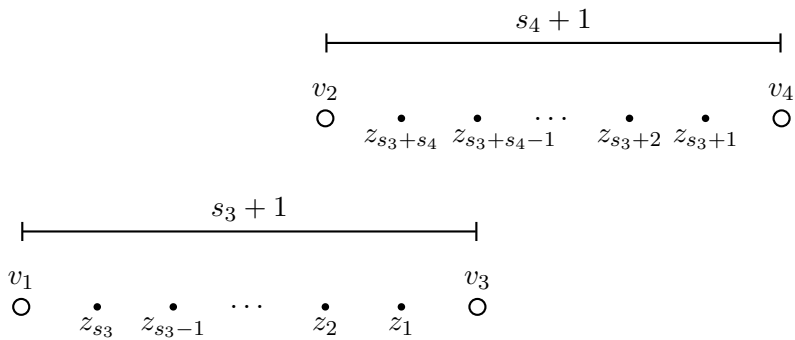
\begin{figure}[!t]
  \begin{center}
    \begin{tikzpicture}
      \begin{scope}
        \draw[thick] (0,0) circle (3pt) node[above=0.1cm]{$\inh_3$};
        \filldraw[thick] (-1,0) circle (1pt) node[below]{$\brt_1$};
        \filldraw[thick] (-2,0) circle (1pt) node[below]{$\brt_2$};
        \node at (-3,0) {\ldots};
        \filldraw[thick] (-4,0) circle (1pt) node[below]{$\brt_{s_3-1}$};
        \filldraw[thick] (-5,0) circle (1pt) node[below]{$\brt_{s_3}$};
        \draw[thick] (-6,0) circle (3pt) node[above=0.1cm]{$\inh_1$};
        \draw[thick,|-|] (-6,1) -- node[midway,above] {$s_3+1$} (0,1);
      \end{scope}
      \begin{scope}[shift={(4,2.5)}]
        \draw[thick] (0,0) circle (3pt) node[above=0.1cm]{$\inh_4$};
        \filldraw[thick] (-1,0) circle (1pt) node[below]{$\brt_{s_3+1}$};
        \filldraw[thick] (-2,0) circle (1pt) node[below]{$\brt_{s_3+2}$};
        \node at (-3,0) {\ldots};
        \filldraw[thick] (-4,0) circle (1pt) 
        node[below]{\hspace{20pt}$\brt_{s_3+s_4-1}$};
        \filldraw[thick] (-5,0) circle (1pt) node[below]{$\brt_{s_3+s_4}$};
        \draw[thick] (-6,0) circle (3pt) node[above=0.1cm]{$\inh_2$};
        \draw[thick,|-|] (-6,1) -- node[midway,above] {$s_4+1$} (0,1);
      \end{scope}
    \end{tikzpicture}
    \caption{The Bethe roots $\brt_k$ corresponding to the four site
      invariant $|\Psi_{4,2}(\inhdiff)\rangle$, i.e.\ to the R-matrix
      $R_{\s_3,\s_4}(\inhdiff)$, arrange into two real strings in the
      complex plane. The strings consist of $s_3$ and $s_4$ roots,
      respectively. The difference of their endpoints
      $\inhdiff:=\inh_3-\inh_4$, cf.\ \eqref{eq:osc-diffinh}, is the
      spectral parameter of the R-matrix.}
    \label{fig:bethe-gl2-four-line}
  \end{center} 
\end{figure} 

The $\mathfrak{gl}(2)$ version of the four site invariant
$|\Psi_{4,2}(\inh_3-\inh_4)\rangle$ of section \ref{sec:osc-4vertex} is
characterized by a monodromy matrix $\mathcal{M}_{4,2}(\spec)$ that is specified by,
cf.\ \eqref{eq:osc-m42} and \eqref{eq:osc-m42-vs},
\begin{align}
  \label{eq:bethe-gl2-sol-four-constr}
  \begin{gathered}
    \Xi_1=\bs_1\,,
    \quad
    \Xi_2=\bs_2\,,
    \quad
    \Xi_3=\s_3\,,
    \quad
    \Xi_4=\s_4\,,\\
    \inh_1=\inh_3-1-s_3\,,
    \quad
    \inh_2=\inh_4-1-s_4\,,
    \quad
    s_1=s_3\,,
    \quad
    s_2=s_4\,.
  \end{gathered}
\end{align}
For this monodromy the overall normalization
\eqref{eq:osc-m42-norm} is once again trivial and with
\eqref{eq:bethe-gl2-sol-four-constr} the eigenvalues
\eqref{eq:bethe-gl2-alphadelta} become
\begin{align}
  \label{eq:bethe-gl2-sol-four-ad-eval}
 \alpha(\spec)
 =
 \frac{\spec-\inh_3+s_3}{\spec-\inh_3}\,
 \frac{\spec-\inh_4+s_4}{\spec-\inh_4}\,,
 \quad
 \delta(\spec)
 =
 \frac{\spec-\inh_3+1}{\spec-\inh_3+1+s_3}\,
 \frac{\spec-\inh_4+1}{\spec-\inh_4+1+s_4}\,.
\end{align} 
They obey the functional relation \eqref{eq:bethe-gl2-ad}. A
solution of \eqref{eq:bethe-gl2-qaq} is given by
\begin{align}
  \label{eq:bethe-gl2-sol-four-q}  
  \begin{aligned}
    Q(\spec)
    &=
    \frac{\Gamma(\spec-\inh_3+s_3+1)}{\Gamma(\spec-\inh_3+1)}\,
    \frac{\Gamma(\spec-\inh_4+s_4+1)}{\Gamma(\spec-\inh_4+1)}\\
    &=
    \prod_{k=1}^{s_3}(\spec-\inh_3+k)\prod_{k=1}^{s_4}(\spec-\inh_4+k)\,.
  \end{aligned}
\end{align}
Because of \eqref{qfct} this solution is unique. The
Bethe roots
\begin{align}
  \label{eq:bethe-gl2-sol-four-roots}  
  \begin{aligned}
    \brt_k&=\inh_3-k\quad\text{for}\quad k=1,\ldots,s_3\,,\\
    \brt_{k+s_3}&=\inh_4-k\quad\text{for}\quad k=1,\ldots,s_4\,,
  \end{aligned}
\end{align}
which we read off as the zeros of \eqref{eq:bethe-gl2-sol-four-q},
form two strings, see Figure~\ref{fig:bethe-gl2-four-line}. To
construct the Bethe vector \eqref{su2BV} we need
the reference state \eqref{bethevac} for the
representation labels found in \eqref{eq:bethe-gl2-sol-four-constr}:
\begin{align}
  \label{eq:bethe-gl2-sol-four-vac}  
  \bvac
  =
  (\bar\oscb_2^1)^{s_3}(\bar\osca_1^3)^{s_3}
  (\bar\oscb_2^2)^{s_4}(\bar\osca_1^4)^{s_4}
  |0\rangle\,.
\end{align}
Then the explicit evaluation of \eqref{su2BV} for
small values of $s_3$ and $s_4$ yields
\begin{align}
  \label{eq:bethe-gl2-sol-four-inv}  
  \begin{aligned}
    |\Psi\rangle
    &=
    B(\brt_1)\cdots B(\brt_{s_3})B(\brt_{s_3+1})\cdots 
    B(\brt_{s_3+s_4})\bvac\\
    &=
    (-1)^{s_3+s_4}s_3!s_4!
    \!\!
    \prod_{l=1}^{\min(s_3,s_4)}
    \!\!\!
    (\inh_3-\inh_4+s_4-l+1)^{-1}
    \!\!
    \sum_{k=0}^{\min(s_3,s_4)}
    \!\!\!
    \frac{1}
    {(s_3-k)!(s_4-k)!k!}
    \\
    &\quad
    \cdot\,
    \!\!
    \prod_{l=k+1}^{\min(s_3,s_4)}
    \!\!\!
    (\inh_3-\inh_4-s_3+l)\;
    (\bar\oscb^1\cdot\bar\osca^3)^{s_3-k}
    (\bar\oscb^2\cdot \bar\osca^4)^{s_4-k}
    (\bar\oscb^2\cdot \bar\osca^3)^{k}
    (\bar\oscb^1\cdot \bar\osca^4)^{k}
    |0\rangle\\
    &\propto|\Psi_{4,2}(\inh_3-\inh_4)\rangle\,,
  \end{aligned}
\end{align}
which coincides with the expression for $|\Psi_{4,2}(\inhdiff)\rangle$ from
\eqref{eq:osc-psi42} with \eqref{eq:osc-phi}, \eqref{eq:osc-diffinh}
and \eqref{eq:osc-coeff}. As the invariant
$|\Psi_{4,2}(\inhdiff)\rangle$ can be understood as the R-matrix
$R_{\s_3,\s_4}(\inhdiff)$, we might say that this R-matrix is a
special Bethe vector.

\subsubsection{Baxter lattice with $\lines$ lines}
\label{sec:bethe-gl2-sol-lattice}

We know from section \ref{sec:osc} that the invariants
$|\Psi_{2,1}\rangle$ and $|\Psi_{4,2}(\inhdiff)\rangle$ can be
understood as a Baxter lattice with, respectively, one and two lines
carrying conjugate symmetric representations. Here we work out the
solution to the functional relations \eqref{eq:bethe-gl2-ad} and
\eqref{eq:bethe-gl2-qaq} for a Baxter lattice consisting of $\lines$
lines of this type. In this case the monodromy has $\sites=2\lines$
sites. According to the $\mathfrak{gl}(2)$ version of
\eqref{eq:osc-line-bs}, the $k^\text{th}$ line of the Baxter lattice with
endpoints $\epe_k<\epb_k$, the representation $\Lambda_k=\bs_{\epe_k}$
and a spectral parameter $\rap_k$ gives rise to the two spin chain
sites
\begin{align}
  \label{eq:bethe-gl2-sol-lattice-constr}
  \begin{gathered}
  \Xi_{\epe_k}=\bs_{\epe_k}\,,
  \quad
  \Xi_{\epb_k}=\s_{\epb_k}\,,\\
  \inh_{\epe_k}=\rap_k\,,
  \quad
  \inh_{\epb_k}=\rap_k+s_{\epe_k}+1\,,
  \quad
  s_{\epe_k}=s_{\epb_k}\,.
  \end{gathered}
\end{align}
This turns the monodromy eigenvalues \eqref{eq:bethe-gl2-alphadelta}
into
\begin{align}
  \label{eq:bethe-gl2-sol-lattice-ad-eval}
  \begin{aligned}
    \alpha(\spec)
    &=
    \prod_{k=1}^\lines
    f_{\bs_{\epe_k}}(\spec-\inh_{\epe_k})f_{\s_{\epb_k}}(\spec-\inh_{\epb_k})
    \frac{\spec-\inh_{\epb_k}+s_{\epb_k}}{\spec-\inh_{\epb_k}}
    =\prod_{k=1}^\lines\frac{\spec-\inh_{\epb_k}+s_{\epb_k}}{\spec-\inh_{\epb_k}}\,,\\
    \delta(\spec)
    &=
    \prod_{k=1}^\lines
    f_{\bs_{\epe_k}}(\spec-\inh_{\epe_k})f_{\s_{\epb_k}}(\spec-\inh_{\epb_k})
    \frac{\spec-\inh_{\epe_k}-s_{\epe_k}}{\spec-\inh_{\epe_k}}
    =\prod_{k=1}^\lines\frac{\spec-\inh_{\epb_k}+1}{\spec-\inh_{\epb_k}+1+s_{\epb_k}}\,.
  \end{aligned}
\end{align}
For the last equality in both equations we used that each factor of
the products corresponds to one line of the Baxter lattice. Using
\eqref{eq:bethe-gl2-sol-lattice-constr} the normalization factors
belonging to each of these lines reduce to $1$ analogously to the case
of a single line explained before~\eqref{eq:osc-m21-norm}.  Obviously,
the eigenvalues in \eqref{eq:bethe-gl2-sol-lattice-ad-eval} satisfy
\eqref{eq:bethe-gl2-ad}. The relation \eqref{eq:bethe-gl2-qaq} is
solved by
\begin{align}
  \label{eq:bethe-gl2-sol-lattice-q}  
  \begin{aligned}
    Q(\spec)
    =
    \prod_{k=1}^\lines
    \frac{\Gamma(\spec-\inh_{\epb_k}+s_{\epb_k}+1)}{\Gamma(\spec-\inh_{\epb_k}+1)}
    =
    \prod_{k=1}^\lines\prod_{l=1}^{s_{\epb_k}}(\spec-\inh_{\epb_k}+l)\,,
  \end{aligned}
\end{align} 
which is the unique solution because we also demand the Q-function to
be of the form \eqref{qfct}. We read off the Bethe
roots as zeros of \eqref{eq:bethe-gl2-sol-lattice-q},
\begin{align}
  \label{eq:bethe-gl2-sol-lattice-roots}  
  \begin{aligned}
    \brt_k&
    =
    \inh_{\epb_1}-k\quad 
    \text{for}\quad 
    k=1,\ldots,s_{\epb_1}\,,\\
    \brt_{k+s_{\epb_1}}&
    =
    \inh_{\epb_2}-k\quad 
    \text{for}\quad 
    k=1,\ldots,s_{\epb_2}\,,\\
    &\hspace{6pt}\vdots\\
    \brt_{k+s_{\epb_{\lines-1}}}&
    =
    \inh_{\epb_\lines}-k\quad 
    \text{for}\quad 
    k=1,\ldots,s_{\epb_\lines}\,.
  \end{aligned}
\end{align}
They arrange into $\lines$ strings. The $k^\text{th}$ line of the Baxter
lattice with representation $\Lambda_k=\bs_{\epe_k}$ leads to one string
of $s_{\epe_k}=s_{\epb_k}$ Bethe roots with a uniform real spacing of $1$
lying between the inhomogeneities $\inh_{\epe_k}$ and $\inh_{\epb_k}$. The
arrangement of these strings in the complex plane is determined by the
spectral parameters $\rap_k=\inh_{\epe_k}$ of the lines. Next, we
concentrate on the associated Bethe vector. With the form of the
highest weight states \eqref{eq:osc-hws} and
\eqref{eq:bethe-gl2-sol-lattice-constr} the reference state
\eqref{bethevac} turns into
\begin{align}
  \label{eq:bethe-gl2-sol-lattice-vac}  
  \bvac
  =
  \prod_{k=1}^\lines
  (\bar\oscb_2^{\epe_k})^{s_{\epb_k}}(\bar\osca_1^{\epb_k})^{s_{\epb_k}}|0\rangle\,.
\end{align}
The Yangian invariant is then given by the Bethe vector
\eqref{su2BV}. Note that as in the special cases of
one- and two-line Baxter lattices discussed, respectively, in
Section~\ref{sec:bethe-gl2-sol-line} and
Section~\ref{sec:bethe-gl2-sol-four}, for generic values of
$\rap_k=\inh_{\epe_k}$ no Bethe root coincides with an
inhomogeneity. Consequently, these Bethe vectors with an even number
of spin chain sites are manifestly non-divergent.

We finish with a remark on the general structure of the set of
solutions to the functional relations \eqref{eq:bethe-gl2-ad} and
\eqref{eq:bethe-gl2-qaq}. Notice that the solution of these relations
defined by \eqref{eq:bethe-gl2-sol-lattice-ad-eval} and
\eqref{eq:bethe-gl2-sol-lattice-q} is actually the product of $\lines$ line
solutions of the type discussed in
Section~\ref{sec:bethe-gl2-sol-line}. More generally, given two
solutions $(\alpha_1(\spec),\delta_1(\spec),Q_1(\spec))$ and
$(\alpha_2(\spec),\delta_2(\spec),Q_2(\spec))$ of the functional
relations, a new one is obtained as the product
\begin{align}
  \label{eq:bethe-gl2-superpos}
  (\alpha_1(\spec)\alpha_2(\spec),
  \delta_1(\spec)\delta_2(\spec),
  Q_1(\spec)Q_2(\spec))\,.
\end{align}
Using this method one can construct new Yangian invariants by
superposing known ones. For example, it should be possible to
combine line solutions with the three-vertices discussed in
Section~\ref{sec:bethe-gl2-sol-three}.

\subsection{Outline of $\gln$
  functional relations}
\label{sec:bethe-gln}

In Section~\ref{sec:bethe-gl2} we discussed in detail how the Bethe ansatz for $\mathfrak{gl}(2)$ spin chains can be specialized in such a
way that the resulting Bethe vector $|\Psi\rangle$ is Yangian
invariant. This leads to functional relations
\eqref{eq:bethe-gl2-specbaxter} which restrict the allowed
representations and inhomogeneities of the monodromy and determine the
Bethe roots. The derivation was based on the observation
\eqref{eq:bethe-inv-eigen} that a Yangian invariant $|\Psi\rangle$ is
a special eigenvector of a transfer matrix. Of course, this
observation is also valid more generally for invariants of the Yangian
of $\gln$. We briefly discussed the nested algebraic Bethe ansatz in Section~\ref{sec:naba}. In generalization of the discussion of the
$\mathfrak{gl}(2)$ situation in section \ref{sec:bethe-gl2}, it can be
specialized to the case where the Bethe vectors are Yangian
invariant.

Here we only state one of the main results, the set of functional
relations determining the representation labels, inhomogeneities and
Bethe roots of Yangian invariants in the $\gln$ case:
\begin{align}
  \label{eq:bethe-gln-specbaxter}
  \begin{aligned}
    1&=\mu_1(\spec)
    \frac{Q_1(\spec-1)}{Q_1(\spec)}\,,\\
    1&=\mu_2(\spec)
    \frac{Q_1(\spec+1)}{Q_1(\spec)}\,
    \frac{Q_2(\spec-1)}{Q_2(\spec)}\,,\\
    1&=\mu_3(\spec)
    \frac{Q_2(\spec+1)}{Q_2(\spec)}\,
    \frac{Q_3(\spec-1)}{Q_3(\spec)}\,,\\
    &\;\;\vdots\\
    1&=\mu_{n-1}(\spec)
    \frac{Q_{n-2}(\spec+1)}{Q_{n-2}(\spec)}\,
    \frac{Q_{n-1}(\spec-1)}{Q_{n-1}(\spec)}\,,\\
    1&=\mu_{n}(\spec)
    \frac{Q_{n-1}(\spec+1)}{Q_{n-1}(\spec)}\,.
  \end{aligned}
\end{align}
Here $\mu_1(\spec),\ldots,\mu_n(\spec)$ are the eigenvalues of the diagonal
monodromy elements on the
pseudo vacuum of the Bethe ansatz, cf. \eqref{baxtergln}. For a monodromy \eqref{eq:yi-mono},
which is composed out of the Lax operators \eqref{lax}
with a finite-dimensional $\gln$ representation of highest
weight $\Xi_i=(\xi_{i}^{(1)},\ldots,\xi_{i}^{(n)})$ at the local
quantum space of the $i^\text{th}$ site, these eigenvalues are given by
\begin{align}
  \label{eq:bethe-gln-mu}
  \mu_{a}(\spec)
  =
  \prod_{i=1}^{\sites}f_{\Xi_i}(\spec-\inh_i)
  \frac{\spec-\inh_{i}+\xi_{i}^{(a)}}{\spec-\inh_{i}}\,.
\end{align} 
The Bethe roots are encoded into the Q-functions
\begin{align}
  \label{eq:bethe-gln-q}
  Q_k(\spec)=\prod_{i=1}^{\brts_k}(\spec-\brt_{i}^{(k)})\,,
\end{align}
where $k=1,\ldots,n-1$ is the nesting level with $\brts_k$ Bethe roots
$\brt_i^{(k)}$. Obviously, for $n=2$
equation~\eqref{eq:bethe-gln-specbaxter} reduces to the functional
relations \eqref{eq:bethe-gl2-specbaxter}. As one can see from the
Baxter equation for $\gln$ \eqref{baxtergln}, \eqref{eq:bethe-gln-specbaxter} is compatible
with the fixed eigenvalue in \eqref{eq:bethe-inv-eigen}. More
precisely, each term in the Baxter equation is equal to one.

Interestingly, the functional relations
\eqref{eq:bethe-gln-specbaxter} can also be written in the form
\begin{align}
  \label{eq:bethe-gln-separated-mu}
  1&=\;\prod_{\mathclap{a=1}}^n\;\mu_a(\spec-a+1)\,,\\
  \label{eq:bethe-gln-separated-qk}
  \frac{Q_k(\spec)}{Q_k(\spec+1)}
  &=
  \;\prod_{\mathclap{a=k+1}}^n\;\mu_a(\spec-a+k+1)
\end{align}
for $k=1,\ldots,n-1$. The first equation
\eqref{eq:bethe-gln-separated-mu} does not involve the Bethe roots and
only constrains the representation labels and inhomogeneities of the
monodromy. Each of the remaining equations
\eqref{eq:bethe-gln-separated-qk} only involves the Bethe roots of one
nesting level $k$. The equations \eqref{eq:bethe-gln-separated-mu} and
\eqref{eq:bethe-gln-separated-qk} generalize \eqref{eq:bethe-gl2-ad}
and \eqref{eq:bethe-gl2-qaq}, respectively, to the $\gln$
case.
\chapter{Conclusion and outlook}
We reviewed the basic framework of the quantum inverse scattering method focusing on rational spin chains. Apart from being the fundamental building blocks of the monodomy matrix for spin chains, the solutions of the Yang-Baxter equation, R-matrices, play an important role in the study of two-dimensional integrable lattice models. Here, their entries are identified with Boltzmann weights of the lattice. This identification leads to an intimate relation between the two subjects. We discussed how unitarity and crossing relations relate certain R-matrices to each other and constrain their normalization which is not fixed by the Yang-Baxter equation. In addition, we introduced certain three-vertices which naturally arise in the study of R-matrices. In $1+1$-dimensional integrable quantum field theory these projectors are usually interpreted as the process of forming a bound states. The basic mathematical structure underlying the Yang-Baxter equation is known as the Yangian which we recalled focusing on the {\small RTT}-realization. 
In Chapter~\ref{ch:qop}, which is based on the authors' original results obtained in \cite{Bazhanov2010,Frassek2010,Frassek2011,Frassek2013}, we constructed Q-operators for rational homogeneous spin chain within the framework of the quantum inverse scattering method by employing certain degenerate solutions of the Yang-Baxter equation. This construction allowed us to derive functional relations among the Q-operators and the Bethe equations without making an ansatz for the wave-function. To underline the strength of the Q-operator construction we studied how the nearest-neighbor Hamiltonian enters the hierarchy of Q-operators. In this way, one can circumvent the standard procedure where the Hamiltonian is extracted from the transfer matrix with equal representation for quantum and auxiliary space.
In Chapter~\ref{ch:BAforYI}, we studied Yangian invariance in the context of the {\small RTT}-realization. This chapter is based on the authors' publication \cite{Frassek2013b}. We argued that Yangian invariants can be understood as special eigenstates of certain inhomogeneous spin chains. As a consequence, we were able to apply Bethe ansatz methods to construct Yangian invariants. Furthermore, we investigated the relation between the Yangian invariant spin chain eigenstates whose components can be interpreted as the partition functions of certain lattice models and tree-level scattering amplitudes in $\mathcal{N}=4$ super Yang-Mills theory. Additional material can be found in the Appendices. In the following, we discuss some open problems, further directions and provide an outlook.
\newline

An important question that immediately arises in the study of Chapter~\ref{ch:BAforYI} is whether there is an efficient way to obtain Yangian invariants explicitly.
The self-evident idea is to evaluate the notoriously intricate expressions for the off-shell Bethe vectors using the knowledge of the Bethe roots.
There are different approaches to tackle the complexity of the off-shell Bethe vectors as e.g. discussed in \cite{Tarasov2013,Ragoucy2014,Gier2011}. However, in general, the off-shell expressions are much more cumbersome than the final Yangian invariant, see also Appendix~\ref{app:su3}.
Nevertheless, the arrangements of the Bethe roots into strings suggest that there is a systematic way to deduce the invariants analytically from the off-shell Bethe vectors.  
Possibly, further insight may be gained from other topics studied in integrable models. As already mentioned in Baxters original work, all Yangian invariants satisfy certain bootstrap like equations. Excitingly, these relations bear certain similarities with the qKZ-equation or Watson equation, see e.g. \cite{Jimbo1995,Mussardo2010}. However, to our knowledge, the precise connection between them has not been worked out yet.
\newline

Besides the application of the Q-operator to planar $\mathcal{N}=4$ super Yang-Mills theory, see Chapter~\ref{ch:intro} and the discussion below, it is interesting to study whether the way of constructing Q-operators presented here has implications on topics studied in the field of integrable models. Recently, there has been renewed interest in open spin-chains, see \cite{Cao2013} and references thereof. The construction of the transfer matrices for these types of spin chains is well understood in the framework of the quantum inverse scattering method, see e.g. \cite{Sklyanin2008a}. However, it is unclear if the Q-operator construction in Chapter~\ref{ch:qop} can be generalized for this case. The construction may lead to a thorough understanding of the new type of functional equations appearing in this context. Furthermore, Q-operators play an important role in the recently established fermionic basis for the XXZ spin chain \cite{BoosCommun.Math.Phys.272:263-2812007,Boos2008,Jimbo2008,Boos2009,Jimbo2010}, see also \cite{Boos2012}. It would be very interesting to study whether this method can be generalized to spin chains with representations of higher rank and supersymmetry using the Q-operator construction presented in this thesis. Also, the connection between B\"acklund transformations in the theory of classical integrable models and the Q-operators introduced here, see e.g. \cite{Sklyanin2007}, and its implications on {\small AdS/CFT}, remain to be understood.
\newline

Our findings and developments bring us closer towards the understanding of weakly coupled planar $\mathcal{N}=4$ super Yang-Mills theory as an integrable model. However, the final step of adapting our results to the realization of symmetry algebra $\mathfrak{psu}(2,2|4)$ remains outstanding. 
The R-operators for the Q-operators can be deduced from their general expression for $\gl{n|m}$ given in Appendix~\ref{susylax}, while the functional relations should remain the same as for the supersymmetric case studied in \cite{Frassek2010}. We have seen, that the form of the Yangian invariants for the non-supersymmetric case already reflect the structure of their supersymmetric counterparts, i.e. the tree-level scattering amplitudes in the Gra\ss mannian formulation with an appropriate contour. Furthermore, it was shown in \cite{Kanning2014} that the deformation parameters introduced in \cite{Beisert2014} satisfy the same relation as in the case of $\gl2$ studied here. Thus, we do not expect any major changes in the form of the invariants nor in the Bethe equations. However, as in principle non-compact representations will appear it is not clear whether the algebraic Bethe ansatz can still be applied.
\newline

The deformation of scattering amplitudes in $\mathcal{N}=4$ super Yang-Mills theory \cite{Ferro2012,Ferro:2013dga} sparked a lot of interest in the last years \cite{Ferro2014,Bargheer2014,Chicherin:2013ora,Broedel2014a,Broedel2014,Kanning2014,Beisert2014}. Interestingly, without any reference to integrability, such deformations were already considered several decades ago in order to regulate scattering processes in quantum field theories \cite{Penrose1973,Hodges1985}. Following \cite{Drummond2010a,Drummond2010b}, these results led to the fully deformed Gra\ss mannian integral formula \cite{Ferro2014,Bargheer2014} employing the appropriate form of the Yangian generators \cite{Beisert2014,Frassek2013b}. However, the integration contours relevant to evaluate a given amplitude from the (deformed) Gra\ss mannian integral formula are in general still unknown. For the representations studied in this thesis we obtained well defined expressions for the corresponding Yangian invariants. This gives hope that the contours relevant for amplitudes can be obtained using our approach at least at tree-level by adjusting the latter to an appropriate representation. It remains to see whether the deformation combined with integrability can be used at loop-level to obtain the naively {\small IR}-divergent scattering amplitudes.
\newline

In the long run, we expect that the results obtained in this thesis may be helpful to understand planar $\mathcal{N}=4$ super Yang-Mills theory from first principles. However, even at weak coupling a coherent reformulation of the theory as an integrable model is unknown and it is not clear how correlation function, Wilson loops, amplitudes and form factors fit in a general picture\,\footnote{Yangian invariance of smooth Maldacena-Wilson loops was discussed in \cite{Mueller2013} and it is known that the structure constants of three-point functions can be extracted from certain contractions of Bethe eigenstates \cite{OkuyamaJHEP0408:0552004,RoibanJHEP0409:0322004,Escobedo2010}.}.  
Apart from studying whether other observables can be understood within the framework discussed here it is certainly tempting to speculate that our approach to construct tree-level scattering amplitudes can be uplifted to all-loop planar $\mathcal{N}=4$ super Yang-Mills theory as it was done in the spectral problem. 
We have seen, however, that the spectrum of the commuting family of operators is not sufficient to construct Yangian invariants and information about the eigenspace is required. This suggests that the {\small TBA}-methods currently used to study the spectral problem, see \cite{Gromov2014a} in references therein, have to be extended to describe other observables like amplitudes, higher-point correlation functions and Wilson loops.
Such methods are available in integrable models. The construction of Q-operators to all orders of the coupling would yield the full information about the eigensystem. 
In addition to the Q-operator approach, there exist powerful bootstrap methods for form factors in integrable models, see e.g. \cite{Mussardo2010}. The bootstrap is based on a set of axioms satisfied by the form factors. 
As mentioned above, there is hope that the Yangian invariants studied in this thesis can be understood as solutions of a qKZ- or Watson equation, see also \cite{Babujian2008}. 
The fact that Yangian invariants can be obtained by combining minimal solutions as discussed in \cite{Chicherin:2013ora,Broedel2014a,Kanning2014} supports this idea. Certainly, it would be very interesting to study this connection also at loop level and try to understand the relation to the bootstrap equations that appeared in \cite{Basso:2013vsa,Basso:2013aha,Basso2014b,Basso2014} and \cite{Klose2013,Klose2013a}.
Finally, we hope that the fruitful interplay between integrable models and planar $\mathcal{N}=4$ super Yang-Mills theory will ultimately lead to a new understanding of gauge theories.

\chapter{Acknowledgments}
 First of all, I like to thank Matthias Staudacher for supervising this thesis. I am indebted to Zolt\'{a}n Bajnok,  J\'{a}nos Balog, Patrick Dorey, Nadav Drukker, Jaume Gomis,  Paul Heslop, Jan Plefka, Matthias Staudacher and Konstantin Zarembo for their support. I like to thank Vladimir Bazhanov, Nils Kanning, Yumi Ko, Tomasz \L{}ukowski, Carlo Meneghelli and Matthias Staudacher for collaboration. I also thank the referees of the dissertation, Patrick Dorey, Frank G\"{o}hmann and  Matthias Staudacher for their time and useful comments, as well as Jan Plefka and Ulrich Wolff for being part of the committee. Furthermore, I like to thank 
Burkhard Eden, 
Jan Fokken, 
Sergey Frolov,
Raquel G\'{o}mez-Ambrosio,
Philipp Hähnel,
Stephan Kirsten,
Laura Koster,
Pedro Liendo,
David Meidinger, 
Vladimir Mitev, 
Dhritiman Nandan,
Elli Pomoni, 
Gregor Richter, 
Jan Schlenker,
Alessandro Sfondrini,
Christoph Sieg, 
Stijn van Tongeren,
Zengo Tsuboi, 
Vitaly Velizhanin,
Matthias Wilhelm, 
as well as 
 Harald Dorn, 
  Valentina Forini, 
 Geroge Jorjadze, 
 Thomas Klose
and the whole Quantum Field and String Theory Group at the Humboldt-University as well as 
Chamgrim Ahn, 
Till Bargheer,
Niklas Beisert,
Diego Bombardelli, 
Johannes Br\"odel,
Reza Doobary, 
Claude Duhr, 
Parikshit Dutta, 
Philipp Fleig, 
Stefan Fredenhagen, 
Frank G\"{o}hmann,
Chrysostomos Kalousios,
Vladimir Kazakov,
Pan Kessel,
Minkyoo Kim, 
Michael Koehn, 
Bum-Hoon Lee,  
Marius de Leeuw,
Paulo Liebgott, 
Florian Loebbert,
Paul Mansfield, 
Roberto Mozara,
Stefano Negro, 
Georgios Papathanasiou,
Stefan Pfenninger, 
Christopher Prior,
Cosimo Restuccia, 
Burkhard Schwab,
Vladimir Smirnov,
Douglas Smith,
István Szécsényi,
Kolja Them,
and
Stefan Zieme for discussions, advice, support and or company.
Also I like to thank Nils Kanning and David Meidinger for carefully reading and commenting the manuscript. 
I like to thank my parents for their endless support as well as my whole family. 
Last but not least, I like to thank Konstantin, Jennifer and Sarah.

The research leading to these results has received funding from the People Programme (Marie Curie Actions) of the European Union's Seventh Framework Programme FP7/2007-2013/ under REA Grant Agreement No 317089 (GATIS).


%
\appendix
\chapter{Representations of $\protect\gln$}\label{app:gln}
In this appendix we introduce the $\gln$ algebra and Dynkin labels to fix the conventions used throughout the thesis.
The commutation relations of the $\gln$ algebra are defined by the relation
\begin{equation}\label{glnalgapp}
 [J_{ab},J_{cd}]=\delta_{cb}\,J_{ad}-\delta_{ad}\,J_{cb}\,,
\end{equation} 
where commutator is given as usual by
\begin{equation}
 [X,Y]=XY-YX\,,
\end{equation} 
for any $X,Y$. Furthermore, $J_{ab}$ are the $\gln$-generators with $a,b=1,\ldots,n$. For irreducible highest weight representations there exists a state such that
\begin{equation}\label{hwcon}
 J_{ab}\vert hws\rangle=0\quad\quad\text{for}\quad\quad 1\leq a<b\leq n\,.
\end{equation} 
Such representations are labeled by the Dynkin labels $\Lambda=\Lambda^n=(\lambda^1,\ldots,\lambda^n)$ with
\begin{equation}
 J_{aa}\vert hws\rangle=\lambda^a\vert hws\rangle\,.
\end{equation}
In particular, for finite-dimensional representations, i.e. representations with highest and lowest weight, we have
\beq
\m^1\ge\m^2\ge\ldots\m^n,\qquad \m^a-\m^b\in{\mathbb Z}\,.\label{domint}
\eeq
In Chapter~\ref{ch:qop}, we denote these representations by $\pi_{\rep^n}$ and the infinite-dimen\-sional representations without lowest weight state by $\pi^+_{\rep^n}$.
\chapter{Drinfeld's first realization}\label{app:1rel}
As outlined in Chapter~\ref{ch:yangian}, there are three different realizations of the Yangian. For completeness we present Drinfeld's first realization as this realization was used in \cite{Drummond2009} to check the Yangian invariance of tree-level scattering amplitudes.
This realization of the Yangian does not make any reference to the Yang-Baxter equation, see e.g. \cite{Bernard1993,MacKay2005, Torrielli2011}. For convenience we restrict ourselves to the case of $\mathfrak{sl}(n)$ and only spell out the main properties of this realization. It is defined by the ordinary $\mathfrak{sl}(n)$ commutation relations
\begin{equation}
 [I_\mu,I_\nu]=f_{\mu\nu\lambda}I_\lambda\,,
\end{equation} 
where $I_\mu$ denote the $\mathfrak{sl}(n)$-generators (level-one generators) with $\mu=1,\ldots,n-1$ and $f_{\mu\nu\lambda}$ the structure constants. The coproduct satisfying the compatibility condition
\begin{equation}
 \Delta([x,y])=[\Delta(x),\Delta(y)]\,,
\end{equation} 
is defined for these generators as
\begin{equation}
 \Delta(I_\mu)=I_\mu\otimes 1+1\otimes I_\mu\,.
\end{equation} 
Additionally we have a second set of generators (level-two generators)
\begin{equation}
 [I_\mu,J_\mu]=f_{\mu\nu\lambda}J_\lambda\,,
\end{equation} 
with the coproduct
\begin{equation}
 \Delta(J_\mu)=J_\mu\otimes 1+1\otimes J_\mu+\frac{h}{2}f_{\mu\nu\lambda}\,I_\lambda\otimes I_\nu\,.
\end{equation} 
Additionally, the Serre relations\,\footnote{note that we can bring it in a slightly more symmetric form using the Jacobi identity and $f_{abc}=-f_{bac}$: $ [J_\mu[J_\nu,I_\lambda]]+[J_\lambda[J_\mu,I_\nu]]+[J_\nu[J_\lambda,I_\mu]]=h^2 a_{\mu\nu\lambda\delta\gamma\sigma}\left\{I_\delta,I_\gamma,I_\sigma\right\}$}
\begin{equation}
 [J_\mu[J_\nu,I_\lambda]]-[I_\mu[J_\nu,J_\lambda]]=h^2 a_{\mu\nu\lambda\delta\gamma\sigma}\left\{I_\delta,I_\gamma,I_\sigma\right\}
\end{equation} 
and 
\begin{equation}
\begin{split}
 [[J_\mu,J_\nu],[I_\lambda,J_\gamma]]+[[J_\lambda,J_\gamma],[I_\mu,J_\nu]]=\\h^2 \left(a_{\mu\nu\delta\omega\sigma\kappa}f_{\lambda\gamma\delta}
+a_{\lambda\gamma\delta\omega\sigma\kappa}f_{\mu\nu\delta}\right)\left\{I_\omega,I_\sigma,J_\kappa\right\}
\end{split}
\end{equation} 
with
\begin{equation}
 a_{\mu\nu\delta\omega\sigma\kappa}=\frac{1}{24}f_{\mu\omega\gamma}f_{\nu\sigma\lambda}f_{\delta\kappa\tau}f_{\gamma\lambda\tau}\,,\quad\quad\{x_1,x_2,x_3\}=\sum_{\mu\neq\nu\neq\lambda}x_\mu x_\nu x_\lambda
\end{equation} 
have to be satisfied.The antipode of the first and second level generators is defined as
\begin{equation}
 S(I_\mu)=-I_\mu\,,\quad\quad S(J_\mu)=-J_\mu+f_{\mu\nu\rho}I_\nu I_\rho\,.
\end{equation} 

\chapter{CBA for higher rank}\label{cba_high}
To generalize Baxter's perimeter Bethe ansatz to higher rank the knowledge of the coordinate wave function for the appropriate spin chain, cf. Chapter~\ref{ch:BAforYI}, is needed. In particular, this spin chain will have different representations associated to its sites. 

The wave-function for the fundamental $\gl3$ spin chain with representation $\Lambda=(1,0,0)$ at each site can be obtained by combining two $\gl2$ spin chains as introduced in \eqref{su2inh}. The excitations on the chain carry different ``colors'' (charges). The two colors are taken into account by the nesting procedure. Let us first consider only one type of excitations. Then the wave function is the same as in spin $1/2$ case
\begin{equation}
\psi_1({\bf x}_1,P_1)=\prod_{k=1}^{m_1}\prod_{j=1}^{x_{1,k}-1}\left(u_{1,P_1(k)}-v_j+1\right)\prod^{L}_{j=x_{1,k}+1}\left(u_{1,P_1(k)}-v_j\right)\,,
\end{equation} 
with
\begin{equation}
 {\bf x}_i=\{x_{i,1},\ldots,x_{i,m_i}\}\,.
\end{equation} 
Now, we the other type of excitations on top of these excitations, but as these excitations can only exist on top of the others we have to consider an inhomogeneous chain with the inhomogeneities given by the rapidities of the first excitations
\begin{equation}
\begin{split}
\psi_2({\bf x}_2,P_1,P_2)=\prod_{k=1}^{m_2}&\prod^{L}_{j=1}\left(u_{2,P_2(k)}-v_j\right) \prod_{j=1}^{x_{2,k}-1} \left(u_{2,P_2(k)}-u_{1,P_1(j)}+1\right)\cdot\\&\cdot\prod_{j=x_{2,k}+1}^{m_1} \left(u_{2,P_2(k)}-u_{1,P_1(j)}\right)\,.
\end{split}
\end{equation} 
The wave function reads
\begin{equation}
\psi({\bf x}_1,{\bf x}_2)=\sum_{P_1}\left(A(P_1) \, \psi_1({\bf x}_1,P_1)\,\sum_{P_2}\bigg(A(P_2)\,\psi_2({\bf x}_2,P_1,P_2)\bigg)\right)\,.
\end{equation} 
The wave function of a $\gl3$ spin chain involving the fundamental and the antifundamental representations\,\footnote{We like to thank Yumi Ko for collaboration.} is given in Figure~\ref{app:su3}. This wave function can be used to generalize Baxter's perimeter Bethe ansatz to the case where the bulk R-matrices are given by the $\gl3$-invariant R-matrix of the type \eqref{fundr}. A generalization to totally symmetric/antisymmetric representations should be possible in the same manner as discussed in Section~\ref{sec:morecba}. Furthermore, we like to refer the reader to  \cite{Ragoucy2014} and references therein as well as to \cite{Tarasov2013} where expressions for Bethe vectors of $\gln$ spin chains are given. It would be interesting to evaluate those to proof the rather explicit form of the wave function in Figure~\ref{app:su3}.
\newpage
\thispagestyle{empty}
\begin{sidewaysfigure}
\begin{center}
\scriptsize
 \begin{equation*}
 \psi_{1,R}(L_f,L_d,{\bf x}_1,\sigma_1)=\prod_{j=m_1^-+1}^{m_1^-+m_1^+}\prod_{k=1}^{L_d}\left(u_{1,\sigma_1(j)}+1-w_k\right) \prod_{k=1}^{{\bf x}_{1,j}-1}\left(u_{1,\sigma_1(j)}+1-v_k\right)\prod_{k={\bf x}_{1,j}+1}^{L_f}\left(u_{1,\sigma_1(j)}-v_k\right)
\end{equation*} 
\begin{equation*}
 \psi_{2,L}(L_f,L_d,{\bf x}_2,\sigma_2)=\prod_{j=1}^{m_2^-} \prod_{k=1}^{L_f}\left(u_{2,\sigma_2(j)}-v_k\right)\prod_{k=1}^{\vert {\bf x}_{2,j}\vert-1}\left(u_{2,\sigma_2(j)}-w_{L_d-k+1}\right)\prod_{k=\vert {\bf x}_{2,j}\vert+1}^{L_d}\left(u_{2,\sigma_2(j)}+1-w_{L_d-k+1}\right)
\end{equation*} 
\begin{equation*}
 \psi_{2,R}({\bf x}_1,{\bf x}_2,\sigma_1,\sigma_2)=\prod_{j=m_2^-+1}^{m_2^-+m_2^+} \prod_{k=1}^{L_d}\left(u_{2,\sigma_2(j)}+1-w_k\right)\prod_{k=1}^{L_f}\left(u_{2,\sigma_2(j)}-v_k\right)\prod_{k=1}^{{\bf x}_{2,j}-1}\left(u_{2,\sigma_2(j)}-u_{1,\sigma_1(m_1^-+k)}+1\right)\prod_{k={\bf x}_{2,j}+1}^{m_1^+}\left(u_{2,\sigma_2(j)}-u_{1,\sigma_1(m_1^-+k)}\right)
\end{equation*} 
\begin{equation*}
 \psi_{1,L}({\bf x}_1,{\bf x}_2,\sigma_1,\sigma_2)=\prod_{j=1}^{m_1^-} \prod_{k=1}^{L_d}\left(u_{1,\sigma_1(j)}+1-w_k\right)\prod_{k=1}^{L_f}\left(u_{1,\sigma_1(j)}-v_k\right)\prod_{k=1}^{\vert {\bf x}_{1,j}\vert-1}\left(u_{1,\sigma_1(j)}-u_{2,\sigma_2(m_2^--k+1)}-1\right)\prod_{k=\vert {\bf x}_{1,j}\vert+1}^{m_2^-}\left(u_{1,\sigma_1(j)}-u_{2,\sigma_1(m_2^--k+1)}\right)
\end{equation*}
\begin{equation*}
 \Xi({\bf x}_1,{\bf x}_2,\sigma_1,\sigma_2)=\prod_{k=m_2^-+1}^{m_2^-+m_2^+}\prod_{j=1}^{m_1^-}\left(u_{2,\sigma_2(k)}-u_{1,\sigma_1(j)}+1\right)\prod_{k=1}^{m_2^-}\prod_{j=m_1^-+1}^{m_1^-+m_1^+}\left(u_{2,\sigma_2(k)}-u_{1,\sigma_1(j)}\right)
\end{equation*} 
\begin{equation*}
 A_i(\sigma_i)=\prod_{j<k}\frac{u_{i,\sigma_i(j)}-u_{i,\sigma_i(k)}+1}{u_{i,\sigma_i(j)}-u_{i,\sigma_i(k)}}
\end{equation*} 
\begin{equation*}
 \Psi(L_f,L_d,{\bf x}_1,{\bf x}_2)=\sum_{\sigma_1,\sigma_2}\,A_1(\sigma_1)\,A_2(\sigma_2)\, \Xi({\bf x}_1,{\bf x}_2,\sigma_1,\sigma_2)\, \psi_{1,R}(L_f,L_d,{\bf x}_1,\sigma_1)\, \psi_{2,L}(L_f,L_d,{\bf x}_2,\sigma_2)\, \psi_{2,R}({\bf x}_1,{\bf x}_2,\sigma_1,\sigma_2)\, \psi_{1,L}({\bf x}_1,{\bf x}_2,\sigma_1,\sigma_2)
\end{equation*}
\caption{Wave function of the $\gl{3}$-invariant spin chain with inhomogeneities. Here the first $L_d$ sites carry the representation $\Lambda_d=(1,1,0)$ and the last $L_f$ sites $\Lambda_f=(1,0,0)$. These representations are dual to each other. As before, the variable ${\bf x}_1$ denotes the position of the level-one excitations and ${\bf x}_2$ the level-two excitations on top of them. This picture is only valid on the part of the chain with representation $\Lambda_f$. On the other part the variable ${\bf x}_1$ denotes the position of the level-two excitations and ${\bf x}_2$ the level-one excitations. The magnon numbers $m_{1,2}^-$ denote the number of level-one or level-two excitations on the left side of the chain, while $m_{1,2}^+$  denote the number of level-one or level-two excitations on the right side of the chain. This result has been checked using \texttt{Mathematica} for certain cases. A similar spin chain appeared in the context of a three dimensional Chern Simons theory \cite{Aharony2008} in \cite{Minahan2008}, see also \cite{Ahn2009a}.}
\label{app:su3}
\end{center}
 \end{sidewaysfigure}

\chapter{Bethe vectors as partition functions}\label{app:bvaspf}
It is rather interesting that the elements of Bethe vectors can be represented as partition functions of certain lattices. In this sense, any Bethe vector can be computed through Baxter's perimeter Bethe ansatz, cf. Chapter~\ref{ch:BAforYI}, which itself yields a Bethe vector. Here, we only state the lattices of interest for the $\gl2$ and $\gl3$ case and leave further studies of this matryoshka principle as mentioned above for the future.

Using the diagrammatic language introduced in \eqref{pic:smon}, see also Section~\ref{sec:relr}, we can represent the elements of the monodromy \eqref{gl2mon} as
\begin{align}
 \begin{aligned}
A(z)
 \end{aligned}
 =
 \begin{aligned}
  \includegraphics[scale=0.70]{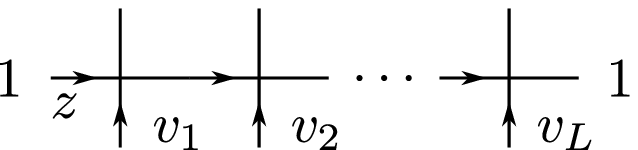}
 \end{aligned}
 \,,\quad
  \begin{aligned}
B(z)
 \end{aligned}
 =
 \begin{aligned}
  \includegraphics[scale=0.70]{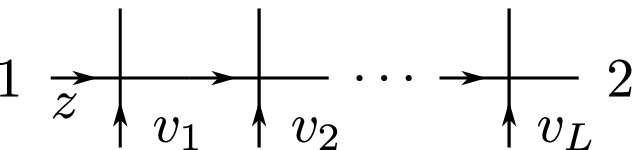}
 \end{aligned}
 \,,
\end{align}
\begin{align}
 \begin{aligned}
C(z)
 \end{aligned}
 =
 \begin{aligned}
  \includegraphics[scale=0.70]{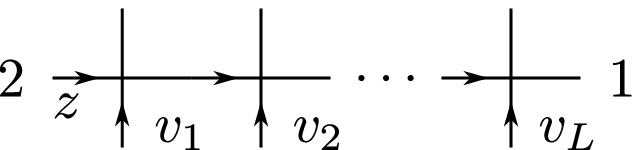}
 \end{aligned}
 \,,\quad
  \begin{aligned}
D(z)
 \end{aligned}
 =
 \begin{aligned}
  \includegraphics[scale=0.80]{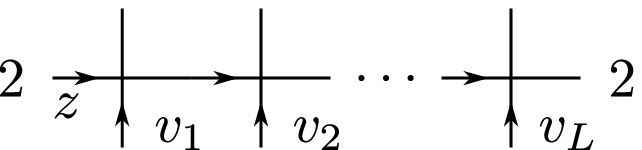}
 \end{aligned}
 \,,
\end{align}
where we did not specify the representation at the different sites of the quantum space. The inhomogeneities are assigned to the vertical and the spectral parameter of the auxiliary space to the horizontal line. The components of the $\gl2$ off-shell Bethe vector \eqref{su2BV} are given by the lattice
\begin{align}\label{su2pic}
 \begin{aligned}
  \includegraphics[scale=0.80]{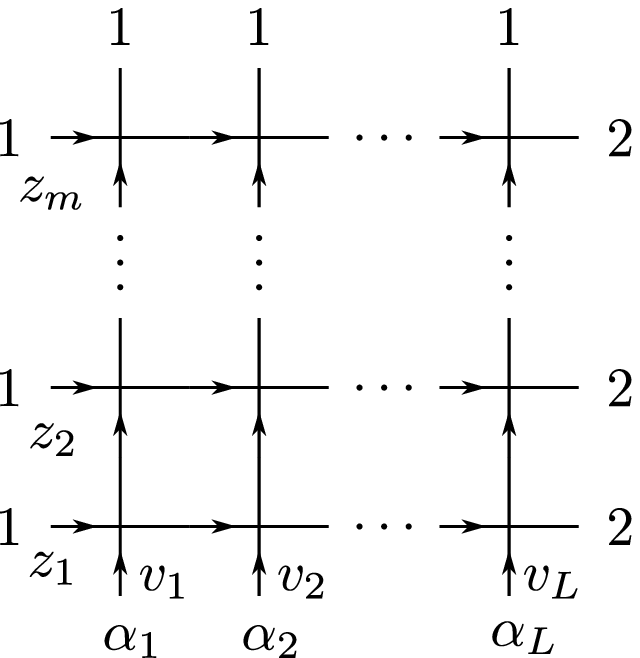}
 \end{aligned}
 \,,
\end{align}
where for simplicity we restrict to fundamental representations at each site of the spin chain. The upper part represents the vacuum of our choice $\vert \Omega\rangle$. In particular, we note that we assigned a Bethe root $z_i$ to each of the  horizontal B-lines as given on the right-hand side, cf. \eqref{actmon}.

For $\gl3$ the Bethe vectors are constructed from three different $B$-type operators corresponding to the upper/lower triangular entries of the monodromy in the auxiliary space.  Diagrammatically it can be depicted as
\begin{align}\label{su3bv}
 \begin{aligned}
  \includegraphics[scale=0.80]{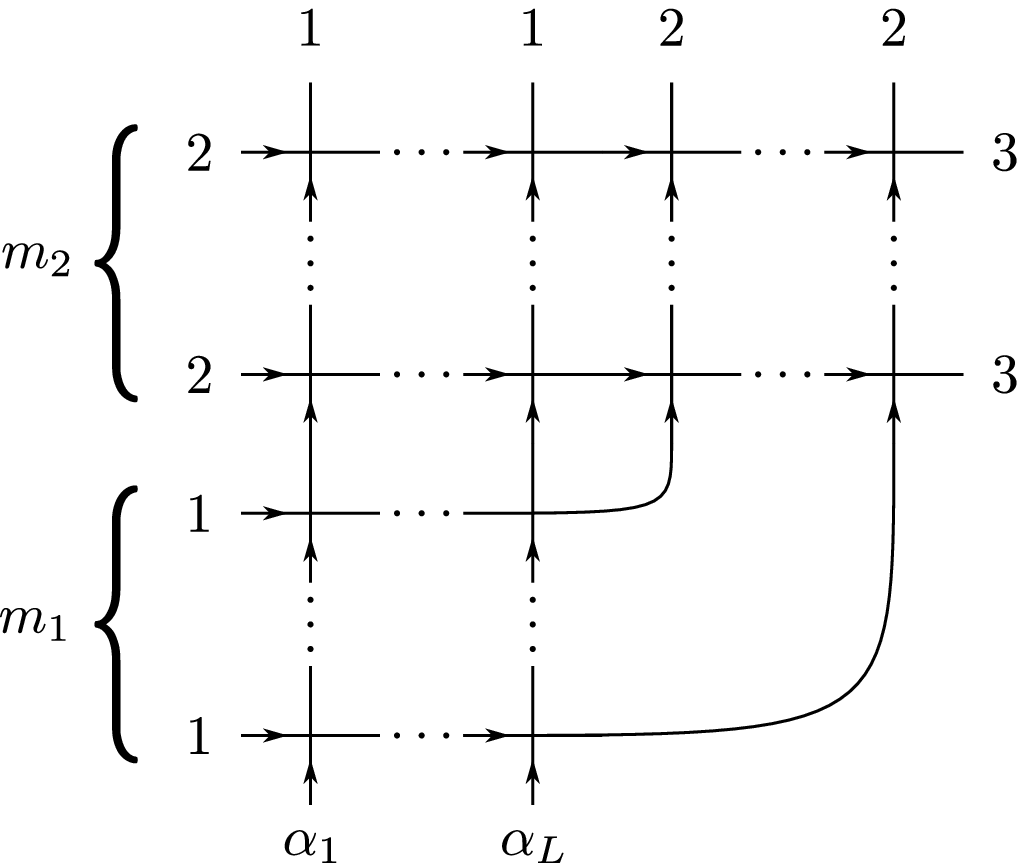}
 \end{aligned}
 \,,
\end{align}
where we restricted to the fundamental representations at each site of the spin chain by choosing the vacuum state. We find that for this case the partition function simplifies as the $L\times m_2$ vertices in the upper left corner freeze, cf. Section~\ref{sec:6vmodel}. However, in general we see that the case where we do not have any second level excitations $m_2=0$ coincides with \eqref{su2pic}. For each level we have a set of Bethe roots assigned to the lines on the left-hand side of \eqref{su3bv}. The inhomogeneities are associated to the $L$ vertical lines. As discussed in Appendix~\ref{cba_high}, the upper part of \eqref{su3bv} can be understood as an inhomogeneous $\gl2$ spin chain where the first $L$ inhomogeneities are given by the inhomogeneities of the $\gl3$ chain and the last $m_1$ inhomogeneities by the Bethe roots of the previous level. The diagram in \eqref{su3bv} generalizes to the case of $\gl{n}$ where the total length of the horizontal line at the top of the lattice is $L+m_1+\ldots+m_{n-2}$. The vacuum state has to be chosen appropriately. Further details can be found in \cite{Kulish1983a,Reshetikhin1989,Foda2011,Foda2013b}.

\chapter{Partonic Lax operators}\label{otherlax}
In Chapter~\ref{ch:qop} we only considered Q-operators constructed from the Lax operators $L_{\ffbox,+}(z)$ and $L_{\ffbox,-}(z)$. Another family of Q-operators can be constructed from the Lax operators $L_{-,\ffbox}(z)$ and $L_{-,\ffbox}(z)$. Both types of Lax operators are presented here for the case of $\gl2$
\begin{equation}\label{plaxpp}
L_{\ffbox,+}(z)=\left(\begin{array}{cc}
              z-\osch&\bar\osca\\
	      -\osca&1
             \end{array}\right)\quad  \text{and} \quad L_{+,\ffbox}(z)=\frac{1}{z+\half}\left(\begin{array}{cc}
              -1&\bar\osca\\
	      -\osca&z+1+\osch
             \end{array}\right) \,,
\end{equation}
\begin{equation}\label{plaxmm}
L_{\ffbox,-}(z)=\left(\begin{array}{cc}
              1&-\osca\\
	      \bar\osca&z-\osch
             \end{array}\right)\quad  \text{and} \quad
L_{-,\ffbox}(z)=\frac{1}{z+\half}\left(\begin{array}{cc}
              z+1+\osch&-\osca\\
	      \bar\osca&-1
             \end{array}\right) \,,
\end{equation}
where 
\begin{equation}
 L_{-,\ffbox}(z)L_{\ffbox,-}(-z)=1\,,\quad\quad L_{+,\ffbox}(-z)L_{\ffbox,+}(z)=1\,.
\end{equation} 
 It might be interesting to understand the relation between the Q-operators constructed from the Lax operators on the left- and right-hand side of \eqref{plaxpp} and \eqref{plaxmm}.

For the Lax operator in Holstein-Primakoff realization
\begin{equation}\label{laxhpp}
\mathcal{L}_{\ffbox,j}(z)=\left(\begin{array}{cc}
              z+j-\bar\osca\osca&\bar\osca(\bar\osca\osca-2j)\\
	      -\osca&z-j+\bar\osca\osca
             \end{array}\right) \,,
\end{equation}
 it holds that
\begin{equation}\label{laxhpp2}
\mathcal{L}_{j,\ffbox}(z)=-\frac{1}{(z-j)(z+j+1)}\mathcal{L}_{\ffbox,j}(z+1) \,.
\end{equation}
We can determine the coefficients in \eqref{genrec} and find
\begin{equation}
 \gamma=-1\,,\quad\quad\sigma=j(j+1)\,.
\end{equation} 
\chapter{Supersymmetric Lax operators}\label{susylax}
The generalization of the Q-operator construction for the fundamental representation of $\gln$ at each site of the quantum space, see Section~\ref{sec:Q-opgln}, to the superalgebra $\gl{n\vert m}$ was carried out in \cite{Frassek2010}, see also \cite{Bazhanov:2008yc}. As the construction is rather similar to the $\gln$-case we only give the supersymmetric counterparts of the Lax operators \eqref{Lcanon}, see also \eqref{eq:R-fundamental}. It can be written as
 \begin{equation}
\mathfrak{L}_I(z)=\left(\begin{BMAT}(@,30pt,30pt)
{c.c}{c.c}
z\,\delta_{B}^A-(-1)^{p(B)}\,\left(E_{B}^A+\big(H^I\big)_{B}^{A}\right)&\bar\xi_{\dot{B}}^A\\
-(-1)^{p(B)}\,\xi_{B}^{\dot{A}}&\delta_{\dot B}^{\dot A}
\end{BMAT}
\right)\,,
\end{equation}
with
 \begin{equation}
 \big(H^I\big)_{B}^{A}=\bar\xi_{\dot{D}}^A\,\xi^{\dot{D}}_{B}+\half(-1)^{p(A)+p(\dot D)}\delta_{AB}\,.
\end{equation} 
Here $E_{B}^A$ denote the $\gl{n\vert m}$-generators satisfying
 \begin{equation}
 [E^{A}_{B},E^{C}_{D}\}=\delta^{C}_{B}E^{A}_{D}-(-1)^{(p(A)+p(B))(p(C)+p(D))}\delta^{A}_{D}E^{C}_{B}
 \end{equation}
 where the commutator is defined as
 \begin{equation}
   [X,Y\}=XY-(-1)^{p(X)p(Y)}YX
 \end{equation} 
 for any $X,Y$ and the parity function 
 \begin{equation}
  p:\{1,\cdots,m+n\}\rightarrow \{0,1\},
 \end{equation} 
 depending on the grading of the algebra. Furthermore, the graded oscillators obey the commutation relations
  \begin{equation}
[\xi^{A}_{B},\bar\xi^{C}_{D}\}=\delta^{A}_{D}\delta_{C}^{B}\,.
 \end{equation}
As before we can use these Lax operators to derive the R-operators for general representations. Solving a supersymmetric counterpart of the Yang-Baxter equation \eqref{eq:YBE}, see also \cite{Kulish:1985bj}, we find the supersymmetric version of the R-operators $\mathcal{R}$ presented in \eqref{eq:RI}. It is of the form
 \begin{equation}\label{superlax}
 \mathfrak{R}_I(z)=\rho(z)e^{E^{d}_{\dot c}\xi^{\dot c}_{d}}\prod_{a=1}^r\prod_{\alpha=1}^q\frac{\Gamma( z-\tilde\mu_a)}{\Gamma(z +\tilde\nu_{\alpha})}e^{(-1)^{p(\dot c)}\bar\xi^{d}_{\dot c}E^{\dot c}_{d}}\,,
 \end{equation}
 where $p$ counts the bosonic elements in $\bar I$ and $r$ the fermionic. The shifted weights are given by
 \begin{eqnarray}
 \tilde\mu_a&=&\mu_a+r-q-a\,,\\
 \tilde\nu_\alpha&=&\nu_\alpha+q-\alpha\,,
 \end{eqnarray} 
 see \cite{Jarvis1979,Bincer1983,Scheunert1983a} for a definition of the shifted weights and further details on the Cayley-Hamilton theorem used in the derivation\,\footnote{This result was obtained in collaboration with Tomasz \L{}ukowski, Carlo Meneghelli and Matthias Staudacher. The details of the derivation are left to a future publication.}.

\chapter{Q-operators and the Hamiltonian}\label{app:hfq}
This appendix contains additional material relevant for Section~\ref{sec:qtoh}. In particular, we provide a direct relation between the R-operators $\mathcal{R}$ and the Hamiltonian density in analogy to \eqref{logderR} and exemplify it for fundamental representations of $\gln$ as well as for the non-compact spin $-\frac{1}{2}$ chain.

\section{A plug-in formula for the Hamiltonian density\label{sub:Hamiltonian-Density}}

For practical purposes we give a plug-in formula for the Hamiltonian
density in this appendix. By multiplying (\ref{eq:haction}) in the
auxiliary space with $
\psset{xunit=.3pt,yunit=.3pt,runit=.3pt}
\begin{pspicture}[shift=-30](71.0246048,71.0246048)
{
\newrgbcolor{curcolor}{0 0 1}
\pscustom[linestyle=none,fillstyle=solid,fillcolor=curcolor]
{
\newpath
\moveto(35.5123,70.5123048)
\curveto(35.39332,44.0928048)(35.5053,22.5660648)(35.5123,0.5123048)
}
}
{
\newrgbcolor{curcolor}{0 0 1}
\pscustom[linewidth=1.02460253,linecolor=curcolor]
{
\newpath
\moveto(35.5123,70.5123048)
\curveto(35.39332,44.0928048)(35.5053,22.5660648)(35.5123,0.5123048)
}
}
{
\newrgbcolor{curcolor}{1 1 1}
\pscustom[linestyle=none,fillstyle=solid,fillcolor=curcolor]
{
\newpath
\moveto(30.5123,25.51230204)
\lineto(40.5123,25.51230204)
\lineto(40.5123,0.51230204)
\lineto(30.5123,0.51230204)
\closepath
}
}
{
\newrgbcolor{curcolor}{1 1 1}
\pscustom[linewidth=1,linecolor=curcolor]
{
\newpath
\moveto(30.5123,25.51230204)
\lineto(40.5123,25.51230204)
\lineto(40.5123,0.51230204)
\lineto(30.5123,0.51230204)
\closepath
}
}
{
\newrgbcolor{curcolor}{1 0 0}
\pscustom[linewidth=1.02460253,linecolor=curcolor]
{
\newpath
\moveto(0.5123,35.5123048)
\curveto(26.9318,35.3933248)(48.45854,35.5053048)(70.5123,35.5123048)
}
}
{
\newrgbcolor{curcolor}{1 1 1}
\pscustom[linestyle=none,fillstyle=solid,fillcolor=curcolor]
{
\newpath
\moveto(50.14644305,35.14644896)
\curveto(50.14644305,43.22866473)(43.59451262,49.78059516)(35.51229685,49.78059516)
\curveto(27.43008108,49.78059516)(20.87815065,43.22866473)(20.87815065,35.14644896)
\curveto(20.87815065,27.06423318)(27.43008108,20.51230276)(35.51229685,20.51230276)
\curveto(43.59451262,20.51230276)(50.14644305,27.06423318)(50.14644305,35.14644896)
\closepath
}
}
{
\newrgbcolor{curcolor}{0 0 0}
\pscustom[linewidth=0.73170731,linecolor=curcolor]
{
\newpath
\moveto(50.14644305,35.14644896)
\curveto(50.14644305,43.22866473)(43.59451262,49.78059516)(35.51229685,49.78059516)
\curveto(27.43008108,49.78059516)(20.87815065,43.22866473)(20.87815065,35.14644896)
\curveto(20.87815065,27.06423318)(27.43008108,20.51230276)(35.51229685,20.51230276)
\curveto(43.59451262,20.51230276)(50.14644305,27.06423318)(50.14644305,35.14644896)
\closepath
}
}
{
\newrgbcolor{curcolor}{0 0 0}
\pscustom[linestyle=none,fillstyle=solid,fillcolor=curcolor]
{
\newpath
\moveto(39.3409106,35.67968206)
\curveto(39.59243498,35.67968206)(39.86682522,35.67968206)(39.86682522,35.38242597)
\curveto(39.86682522,35.10803574)(39.59243498,35.10803574)(39.3409106,35.10803574)
\lineto(31.45219127,35.10803574)
\curveto(31.20066689,35.10803574)(30.9491425,35.10803574)(30.9491425,35.38242597)
\curveto(30.9491425,35.67968206)(31.20066689,35.67968206)(31.45219127,35.67968206)
\closepath
\moveto(39.3409106,35.67968206)
}
}
\end{pspicture}$ from the right
one finds that 

\begin{equation}
\begin{split}
 \mathcal{H}_{i,i+1}\mathcal{R}_{I,i}(\check{z})\,\mathcal{R}_{I,i+1}(\check{z})=&\mathcal{R}_{I,i}(\check{z})\,\mathcal{R}'_{I,i+1}(\check{z})\\&-\mathbf{P}_{i,i+1}e^{\bar{\mathbf{a}}_{c}^{\dot{c}}J(i)_{\dot{c}}^{c}}\circ\vert hws\rangle_{i}\mathcal{R}'_{I,i+1}(\hat{z})\langle hws\vert_{i}\circ e^{-\mathbf{a}_{\dot{c}}^{c}J(i)_{c}^{\dot{c}}}.\label{eq:HLL}
\end{split}
\end{equation}
Interestingly, $\mathcal{R}_{I,i}(\check{z})\,\mathcal{R}_{I,i+1}(\check{z})$
can be inverted under $\cdot$ to obtain $\mathcal{H}_{i,i+1}$. As
$\mathcal{H}_{i,i+1}$ does not depend on the auxiliary space all
oscillators can be removed in a consistent way. In this way one can
write

\begin{multline}
\mathcal{H}_{i,i+1}\mathcal{R}_{0,i}(\check{z})e^{-J(i)_{a}^{\dot{a}}J(i+1)_{\dot{a}}^{a}}\mathcal{R}_{0,i+1}(\check{z})=\mathcal{R}_{0,i}(\check{z})e^{-J(i)_{a}^{\dot{a}}J(i+1)_{\dot{a}}^{a}}\mathcal{R}'_{0,i+1}(\check{z})\\
-\mathbf{P}_{i,i+1}\sum_{\{k_{c\dot{c}}\},\{m_{c\dot{c}}\}=0}^{\infty}\bigg(\prod_{c\in I,\dot{c}\in\bar{I}}\frac{1}{k_{c\dot{c}}!m_{c\dot{c}}!}\,(J_{\dot{c}}^{c}(i)+J_{\dot{c}}^{c}(i+1))^{k_{\dot{c}c}+m_{c\dot{c}}}\,\mathcal{R}_{0,I}(\hat{z})\,\\
\cdot\mathcal{\tilde{R}}'_{0,i+1}(\hat{z})\,\prod_{c\in I,\dot{c}\in\bar{I}}(J_{c}^{\dot{c}}(i))^{k_{c\dot{c}}}(J_{c}^{\dot{c}}(i+1))^{m_{c\dot{c}}}\bigg)\,.
\end{multline}
In analogy to (\ref{eq:HLL}) this yields the Hamiltonian density.

\section{Hamiltonian action for non-compact spin
chains\label{sec:The-spin-half} }

In this appendix we study how the Hamiltonian density for the non-compact
spin $-\frac{1}{2}$ spin-chain emerges in the presented formalism.
This spin-chain received special interest in the context of the $AdS_{5}/CFT_{4}$
correspondence \cite{BeisertNucl.Phys.B676:3-422004,Bellucci2005,Stefanski2004,Romagnoni2011a}.
The $\mathcal{R}$-operators for $\gl2$ were discussed
in Section~\ref{sec:Example:gl2} in great detail. Restricting to
$\mathfrak{sl}(2)$ one finds that one of the two $\mathcal{R}$-operators
can be written as
\begin{equation}
\mathcal{R}_{+}(z)\,=e^{\bar{\mathbf{a}}\, J_{+}}\,\,\mathcal{R}_{0,+}(z)\,\, e^{\mathbf{a}\, J_{-}}\quad\quad\text{with}\quad\quad\mathcal{R}_{0,+}(z)=\,\frac{\Gamma(z+J_{0})}{\Gamma(z+\sfrac{1}{2})}\,,\label{eq:RI-half}
\end{equation}
compare Section~\ref{sec:Example:gl2}. The usual $\mathfrak{sl}(2)$ commutation
relations are 
\begin{equation}
[J_{0},J_{\pm}]=\pm J_{\pm}\quad\quad[J_{+},J_{-}]=-2J_{0}\,.
\end{equation} 
Furthermore, we define the action on module via the common relations
\begin{equation}
J_{+}\vert m\rangle=(m+1)\vert m+1\rangle\quad\quad J_{-}\vert m\rangle=m\vert m-1\rangle\quad\quad J_{0}\vert m\rangle=(m+\sfrac{1}{2})\vert m\rangle\,.
\end{equation}
It follows that 
\begin{align}
\mathcal{R}_{0,+}(z)&=\sum_{m=0}^{\infty}\frac{\Gamma(z+\sfrac{1}{2}+m)}{\Gamma(z+\sfrac{1}{2})}\,\vert m\rangle\langle m\vert\,,\\
\mathcal{\tilde{R}}_{0,+}(z)&=\sum_{m=0}^{\infty}(-1)^{m}\frac{\Gamma(-z+\sfrac{1}{2}+m)}{\Gamma(-z+\sfrac{1}{2})}\,\vert m\rangle\langle m\vert\,,\nonumber
\end{align}
and
\begin{align}
\mathcal{R}'_{0,+}(\hat{z})&=\sum_{m=1}^{\infty}\Gamma(m)\,\vert m\rangle\langle m\vert\,,\\
\mathcal{\tilde{R}}'_{0,+}(\check{z})&=\sum_{m=1}^{\infty}(-1)^{m+1}\Gamma(m)\,\vert m\rangle\langle m\vert\,,\nonumber
\end{align}
with $\hat{z}=-\sfrac{1}{2}$ and $\check{z}=\sfrac{1}{2}$. The relevant
terms in (\ref{eq:haction}) are then given by

\begin{equation}
\psset{xunit=.5pt,yunit=.5pt,runit=.5pt}
\begin{pspicture}[shift=-80](170.890625,170.890625)
{
\newrgbcolor{curcolor}{0 0 1}
\pscustom[linewidth=0.89062893,linecolor=curcolor]
{
\newpath
\moveto(85.44531447,170.44531053)
\lineto(85.00000447,99.99999053)
}
}
{
\newrgbcolor{curcolor}{1 0 0}
\pscustom[linewidth=0.89062893,linecolor=curcolor]
{
\newpath
\moveto(112.55443447,86.33595053)
\curveto(132.51659447,86.45493053)(153.78184447,86.34295053)(170.44531447,86.33595053)
}
}
{
\newrgbcolor{curcolor}{1 0 0}
\pscustom[linewidth=0.89062893,linecolor=curcolor]
{
\newpath
\moveto(40.44531447,86.33595053)
\curveto(60.40746447,86.45493053)(81.67272447,86.34295053)(98.33619447,86.33595053)
}
}
{
\newrgbcolor{curcolor}{1 0 0}
\pscustom[linewidth=0.89062893,linecolor=curcolor]
{
\newpath
\moveto(0.44531447,86.33595053)
\curveto(20.40746447,86.45493053)(36.67272447,86.34295053)(53.33619447,86.33595053)
}
}
{
\newrgbcolor{curcolor}{1 1 1}
\pscustom[linestyle=none,fillstyle=solid,fillcolor=curcolor]
{
\newpath
\moveto(49.71361145,86.33595329)
\curveto(49.71361145,78.25373747)(43.16168099,71.70180701)(35.07946517,71.70180701)
\curveto(26.99724936,71.70180701)(20.4453189,78.25373747)(20.4453189,86.33595329)
\curveto(20.4453189,94.4181691)(26.99724936,100.97009956)(35.07946517,100.97009956)
\curveto(43.16168099,100.97009956)(49.71361145,94.4181691)(49.71361145,86.33595329)
\closepath
}
}
{
\newrgbcolor{curcolor}{0 0 0}
\pscustom[linewidth=0.73170731,linecolor=curcolor]
{
\newpath
\moveto(49.71361145,86.33595329)
\curveto(49.71361145,78.25373747)(43.16168099,71.70180701)(35.07946517,71.70180701)
\curveto(26.99724936,71.70180701)(20.4453189,78.25373747)(20.4453189,86.33595329)
\curveto(20.4453189,94.4181691)(26.99724936,100.97009956)(35.07946517,100.97009956)
\curveto(43.16168099,100.97009956)(49.71361145,94.4181691)(49.71361145,86.33595329)
\closepath
}
}
{
\newrgbcolor{curcolor}{0 0 0}
\pscustom[linestyle=none,fillstyle=solid,fillcolor=curcolor]
{
\newpath
\moveto(35.26063126,85.93476537)
\lineto(39.33075315,85.93476537)
\curveto(39.53654583,85.93476537)(39.81093606,85.93476537)(39.81093606,86.23202147)
\curveto(39.81093606,86.50641171)(39.53654583,86.50641171)(39.33075315,86.50641171)
\lineto(35.26063126,86.50641171)
\lineto(35.26063126,90.59939945)
\curveto(35.26063126,90.80519213)(35.26063126,91.07958237)(34.96337516,91.07958237)
\curveto(34.66611907,91.07958237)(34.66611907,90.80519213)(34.66611907,90.59939945)
\lineto(34.66611907,86.50641171)
\lineto(30.59599718,86.50641171)
\curveto(30.3902045,86.50641171)(30.11581427,86.50641171)(30.11581427,86.23202147)
\curveto(30.11581427,85.93476537)(30.3902045,85.93476537)(30.59599718,85.93476537)
\lineto(34.66611907,85.93476537)
\lineto(34.66611907,81.84177763)
\curveto(34.66611907,81.63598495)(34.66611907,81.36159471)(34.96337516,81.36159471)
\curveto(35.26063126,81.36159471)(35.26063126,81.63598495)(35.26063126,81.84177763)
\closepath
\moveto(35.26063126,85.93476537)
}
}
{
\newrgbcolor{curcolor}{1 0 0}
\pscustom[linewidth=0.89062893,linecolor=curcolor]
{
\newpath
\moveto(95.44531447,86.33595053)
\curveto(115.40746447,86.45493053)(136.67271447,86.34295053)(153.33619447,86.33595053)
}
}
{
\newrgbcolor{curcolor}{1 1 1}
\pscustom[linestyle=none,fillstyle=solid,fillcolor=curcolor]
{
\newpath
\moveto(99.71361138,86.33595329)
\curveto(99.71361138,78.25373747)(93.16168092,71.70180701)(85.07946511,71.70180701)
\curveto(76.99724929,71.70180701)(70.44531883,78.25373747)(70.44531883,86.33595329)
\curveto(70.44531883,94.4181691)(76.99724929,100.97009956)(85.07946511,100.97009956)
\curveto(93.16168092,100.97009956)(99.71361138,94.4181691)(99.71361138,86.33595329)
\closepath
}
}
{
\newrgbcolor{curcolor}{0 0 0}
\pscustom[linewidth=0.73170731,linecolor=curcolor]
{
\newpath
\moveto(99.71361138,86.33595329)
\curveto(99.71361138,78.25373747)(93.16168092,71.70180701)(85.07946511,71.70180701)
\curveto(76.99724929,71.70180701)(70.44531883,78.25373747)(70.44531883,86.33595329)
\curveto(70.44531883,94.4181691)(76.99724929,100.97009956)(85.07946511,100.97009956)
\curveto(93.16168092,100.97009956)(99.71361138,94.4181691)(99.71361138,86.33595329)
\closepath
}
}
{
\newrgbcolor{curcolor}{0 0 0}
\pscustom[linestyle=none,fillstyle=solid,fillcolor=curcolor]
{
\newpath
\moveto(89.35863951,86.09549997)
\curveto(89.6101639,86.09549997)(89.88455414,86.09549997)(89.88455414,86.39275606)
\curveto(89.88455414,86.6671463)(89.6101639,86.6671463)(89.35863951,86.6671463)
\lineto(81.46992012,86.6671463)
\curveto(81.21839574,86.6671463)(80.96687135,86.6671463)(80.96687135,86.39275606)
\curveto(80.96687135,86.09549997)(81.21839574,86.09549997)(81.46992012,86.09549997)
\closepath
\moveto(89.35863951,86.09549997)
}
}
{
\newrgbcolor{curcolor}{1 1 1}
\pscustom[linestyle=none,fillstyle=solid,fillcolor=curcolor]
{
\newpath
\moveto(150.07946431,85.97009929)
\curveto(150.07946431,77.88788347)(143.52753385,71.33595301)(135.44531804,71.33595301)
\curveto(127.36310222,71.33595301)(120.81117176,77.88788347)(120.81117176,85.97009929)
\curveto(120.81117176,94.0523151)(127.36310222,100.60424556)(135.44531804,100.60424556)
\curveto(143.52753385,100.60424556)(150.07946431,94.0523151)(150.07946431,85.97009929)
\closepath
}
}
{
\newrgbcolor{curcolor}{0 0 0}
\pscustom[linewidth=0.73170731,linecolor=curcolor]
{
\newpath
\moveto(150.07946431,85.97009929)
\curveto(150.07946431,77.88788347)(143.52753385,71.33595301)(135.44531804,71.33595301)
\curveto(127.36310222,71.33595301)(120.81117176,77.88788347)(120.81117176,85.97009929)
\curveto(120.81117176,94.0523151)(127.36310222,100.60424556)(135.44531804,100.60424556)
\curveto(143.52753385,100.60424556)(150.07946431,94.0523151)(150.07946431,85.97009929)
\closepath
}
}
{
\newrgbcolor{curcolor}{0 0 0}
\pscustom[linestyle=none,fillstyle=solid,fillcolor=curcolor]
{
\newpath
\moveto(135.62648576,85.56891137)
\lineto(139.69660765,85.56891137)
\curveto(139.90240033,85.56891137)(140.17679057,85.56891137)(140.17679057,85.86616747)
\curveto(140.17679057,86.14055771)(139.90240033,86.14055771)(139.69660765,86.14055771)
\lineto(135.62648576,86.14055771)
\lineto(135.62648576,90.23354545)
\curveto(135.62648576,90.43933813)(135.62648576,90.71372837)(135.32922967,90.71372837)
\curveto(135.03197358,90.71372837)(135.03197358,90.43933813)(135.03197358,90.23354545)
\lineto(135.03197358,86.14055771)
\lineto(130.96185169,86.14055771)
\curveto(130.75605901,86.14055771)(130.48166877,86.14055771)(130.48166877,85.86616747)
\curveto(130.48166877,85.56891137)(130.75605901,85.56891137)(130.96185169,85.56891137)
\lineto(135.03197358,85.56891137)
\lineto(135.03197358,81.47592363)
\curveto(135.03197358,81.27013095)(135.03197358,80.99574071)(135.32922967,80.99574071)
\curveto(135.62648576,80.99574071)(135.62648576,81.27013095)(135.62648576,81.47592363)
\closepath
\moveto(135.62648576,85.56891137)
}
}
{
\newrgbcolor{curcolor}{0 0 1}
\pscustom[linewidth=0.89062893,linecolor=curcolor]
{
\newpath
\moveto(35.44531447,71.34563053)
\lineto(35.00000447,0.90031053)
}
}
{
\newrgbcolor{curcolor}{0 0 1}
\pscustom[linewidth=0.89062893,linecolor=curcolor]
{
\newpath
\moveto(135.44531447,71.34563053)
\lineto(135.00000447,0.90031053)
}
}
\end{pspicture}=\sum_{m_{1},m_{2}=0}^{\infty}\bar{\mathbf{a}}^{m_{1}}\left(\vert m_{1},m_{2}\rangle\right)\langle0\vert e^{\mathbf{a}J_{2}^{-}}\bar{\mathbf{a}}^{m_{2}}\,,
\end{equation}
\begin{equation}
\begin{split}
 \input{content/pictures/NewDiagrams/Rpp}=-\sum_{m_{1},m_{2}=0}^{\infty}\bar{\mathbf{a}}^{m_{1}}\bigg(&h(m_{1})\,\vert m_{1},m_{2}\rangle\\
 &-\sum_{\ell=1}^{m_{2}}\frac{1}{\ell}\vert m_{1}+\ell,m_{2}-\ell\rangle\bigg)\langle0\vert e^{\mathbf{a}J_{2}^{-}}\bar{\mathbf{a}}^{m_{2}}\,,
\end{split}
\end{equation}

\begin{equation}
\begin{split}
 \input{content/pictures/NewDiagrams/pR}=\sum_{m_{1},m_{2}=0}^{\infty}\bar{\mathbf{a}}^{m_{1}}\bigg(&h(m_{2})\,\vert m_{1},m_{2}\rangle\\&-\sum_{\ell=1}^{m_{1}}\frac{1}{\ell}\vert m_{1}-\ell,m_{2}+\ell\rangle\bigg)\langle0\vert e^{\mathbf{a}J_{2}^{-}}\bar{\mathbf{a}}^{m_{2}}\,.
\end{split}
\end{equation}
From this we find that
\begin{equation}
\begin{split}
 \mathcal{H}\vert m_{1}m_{2}\rangle=&\big(h(m_{1})+h(m_{2})\big)\vert m_{1},m_{2}\rangle\\&-\sum_{\ell=1}^{m_{1}}\frac{1}{\ell}\vert m_{1}-\ell,m_{2}+\ell\rangle-\sum_{\ell=1}^{m_{2}}\frac{1}{\ell}\vert m_{1}+\ell,m_{2}-\ell\rangle\,.
\end{split}
\end{equation}

\section{The Hamiltonian for the fundamental representation}\label{fundhamq}

For the fundamental representation one has $a=1$, see (\ref{eq:tableaux}).
Therefore the R-operators $\mathcal{R}$ of cardinality $\vert I\vert=n-1$
carry the information about the Hamiltonian. In this case the special
points are located at $\hat{z}=+\frac{1}{2}$ and $\check{z}=-\frac{1}{2}$,
compare (\ref{eq:degenerate_fund}). The derivative ${L}_{I}'$ does not depend on the spectral parameter $z$ and does not contain
oscillators. It follows that equation (\ref{eq:HLL}) simplifies to

\begin{equation}
\mathcal{H}_{i,i+1}{L}_{I,i}(\check{z})\,{L}_{I,i+1}(\check{z})=(\mathbf{1}-\mathbf{P})_{i,i+1}{L}_{I,i}(\check{z}){L}'_{I,i+1}\,.\label{eq:HLL_restricted}
\end{equation}
Furthermore, ${L}'_{I}$ can be written as 
\begin{equation}
{L}'_{I}={L}_{I}(\hat{z})-{L}_{I}(\check{z})\,.
\end{equation}
The expression for the Hamiltonian density
\begin{equation}
\mathcal{H}_{i,i+1}=(\mathbf{P}-\mathbf{1})_{i,i+1}
\end{equation}
follows noting that
\begin{equation}\label{eq:ident}
(\mathbf{1}-\mathbf{P})_{i,i+1}{L}_{I,i}(\check{z}){L}_{I,i+1}(\hat{z})=0\,.
\end{equation}
This coincides with the Hamiltonian density in \eqref{hamden} up to a term proportional to the identity.
Interestingly, as a consequence of (\ref{eq:rdecproj}), identity (\ref{eq:ident})
holds true for any generalized rectangular representation.

\section{Reordering formula\label{sub:Reorder}}

The reordering of the oscillators in the auxiliary space we are interested
in is of the form
\begin{equation}
e^{\bar{\mathbf{a}}A}\cdot B\cdot e^{\mathbf{a}C}=e^{\bar{\mathbf{a}}A}\circ\tilde{B}\circ e^{\mathbf{a}C}\,.\label{eq:reorder2}
\end{equation}
Using $e^{-\bar{\mathbf{a}}A}\mathbf{a}{}^{n}e^{\bar{\mathbf{a}}A}=(\mathbf{a}+A)^{n}$
we find that 
\begin{equation}
\tilde{B}=\sum_{n=0}^{\infty}\frac{(-1)^{n}}{n!}\, A^{n}\, B\, C^{n}\,,\quad\quad B=\sum_{n=0}^{\infty}\frac{1}{n!}\, A^{n}\,\tilde{B}\, C^{n}\,.
\end{equation}
Here we did not specifying any commutation relations among $A,B,C$.

\chapter{Glued invariants and special points of the R-matrix}\label{app:comp}
It is well known that there exist special points of the R-matrix where 
it becomes a projector on a lower-dimensional subspace. In 
particular, the formation of bound states is associated to such a 
special point, see Section~\ref{sec:three}. However, there are other values of 
the spectral parameter that can be interpreted as on-shell rescattering 
processes along the lines of \cite{Coleman1978}. Typically, these special points are associated to 
poles in the S-matrix. To make the pole-structure of the R-matrix 
manifest we rewrite \eqref{eq:osc-psi42} as
\begin{align}\label{polexp}
  \Gamma(z-s_3) \vert\Psi_{4,2}(z)\rangle
   =\sum_{k=0}^{\min(s_3,s_4)}
   \frac{1}{z+k-s_3}
   \vert\tilde\Psi_{4,2}(s_3-k)\rangle\,.
\end{align}
Naturally, \eqref{polexp} contains only poles of first order ignoring 
the overall normalization. For $z=s_3$ the residue of \eqref{polexp} is then given by 
\begin{align}
  \begin{aligned}
             \includegraphics[width=0.2\textwidth]{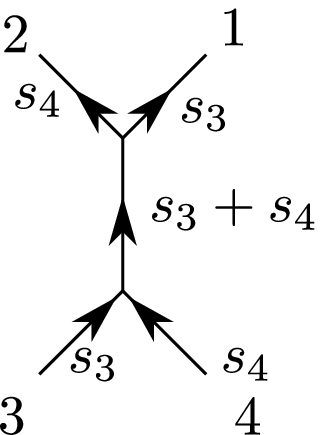}
  \end{aligned}\,,
\end{align}
Furthermore, for $z=s_3-\min(s_3,s_4)$ we obtain the decomposition 
\begin{align}
  \begin{aligned}
             \includegraphics[width=0.3\textwidth]{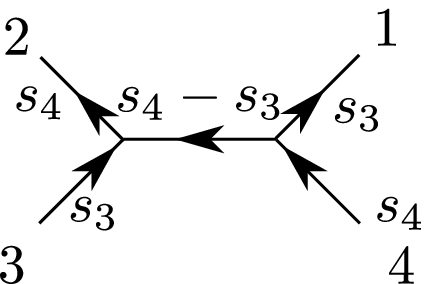}
  \end{aligned}\,,
\end{align}
or
\begin{align}
  \begin{aligned}
             \includegraphics[width=0.3\textwidth]{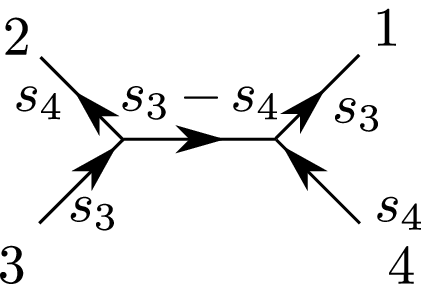}
  \end{aligned}\,,
\end{align}
depending on wether $s_3>s_4$ or $s_4<s_3$. If $s_3=s_4$ the middle line vanishes and we obtain a permutation.
The points between the ones discussed above 
obviously also develop poles. However, there does not seem to be a 
fundamental principle to determine the corresponding on-shell diagrams. 
 From \eqref{polexp} we find that the general decomposition of the R-matrix at the 
special points $z=s_3,\ldots,s_3-\min(s_3,s_4)$ is given by
\begin{align}
  \begin{aligned}
             \includegraphics[width=0.4\textwidth]{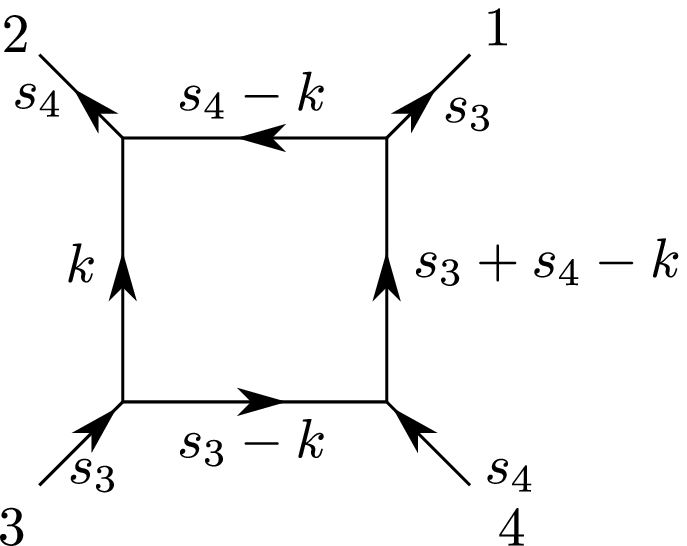}
  \end{aligned}\,,
\end{align}
The details of this calculation can be found in the following section.

We would like to point out that the diagram above exactly coincides 
with the BCFW formula for the four point tree-level amplitude in terms of three point 
amplitudes. Furthermore, we note that a three vertex can be interpreted as two lines, cf. \eqref{eq:osc-psi21}, \eqref{eq:osc-psi31} and \eqref{eq:osc-psi32}. A construction of Yangian invariants only involving line solutions was proposed in \cite{Chicherin:2013ora} and further explored in \cite{Kanning2014,Broedel2014a}. 

\section{Gluing prescription}
\label{sec:gluing-procedure}
We start by gluing two of the lines introduced in
\eqref{eq:osc-psi21}. This is done most conveniently in the
operator form of the invariants \eqref{eq:osc-opsi21}. The representation
labels of both lines have to be identical. Then, after the appropriate
identification of the spaces, gluing is just multiplication of the
operators. Using
\begin{align}
  \label{eq:gluing_sumperm}
  \langle 0\vert\osca_{c_1}\cdots\osca_{c_{s_2}}
  \bar\osca_{d_1}\cdots\bar\osca_{d_{s_2}}\vert0\rangle
  =\sum_{\sigma}\delta_{c_1 d_{\sigma(1)}}\cdots\delta_{c_{s_2} d_{\sigma(s_2)}}\,,
\end{align}
where we sum over all permutations $\sigma$ of $s_2$ elements, we obtain
\begin{align}
  \begin{aligned}
   \includegraphics[width=0.2\textwidth]{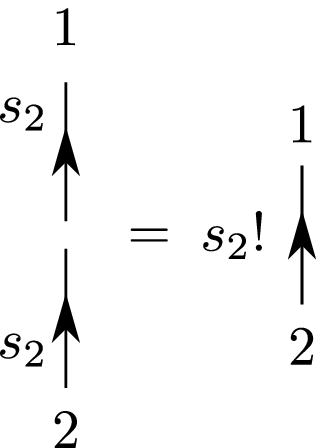}
  \end{aligned}.
\end{align}
The result is not a different invariant but again a line with some
combinatorial prefactor that could be absorbed into the normalization.

In the following we have to
evaluate expressions of the type
\begin{align}
  \label{eq:glue_details_start}
  \begin{aligned}
    &\sum_{a_i,b_i,c_i,d_i=1}^n
    \bar\osca^3_{a_1}\cdots\bar\osca^3_{a_{s_3}}
    \bar\osca^4_{b_1}\cdots\bar\osca^4_{b_{s_4}}
    \vert 0\rangle
    \langle 0\vert \osca_{a_1}\cdots\osca_{a_{s_3}}
    \osca_{b_1}\cdots\osca_{b_{s_4}}\\
    &\hspace{4cm}
    \cdot\,
    \bar\osca_{c_1}\cdots\bar\osca_{c_{s_1}}
    \bar\osca_{d_1}\cdots\bar\osca_{d_{s_2}}
    \vert 0\rangle
    \langle 0\vert \osca^1_{c_1}\cdots\osca^1_{c_{s_1}}
    \osca^2_{d_1}\cdots\osca^2_{d_{s_2}}
  \end{aligned}
\end{align}
with $s_1+s_2=s_3+s_4$. The middle part becomes a sum like in
\eqref{eq:gluing_sumperm} ranging over all permutations of $s_1+s_2$
objects. Each term in this sum represents one way to contract the
indices of the oscillators acting in spaces $1$ and $2$ with those
acting in $3$ and $4$. The expression \eqref{eq:glue_details_start},
for notational convenience conjugated in spaces $1$ and $2$, evaluates
to
{
\small
\begin{align}\label{combii}
  \begin{aligned}
  &\sum_{q=\max(0,s_1-s_3)}^{\min(s_1,s_4)}\hspace{-0.5cm}
    p!q!v!w!\binom{s_1}{q}\binom{s_2}{v}\binom{s_3}{v}\binom{s_4}{q}
    (\bar\osca^1\cdot\bar\osca^3)^{p}
    (\bar\osca^1\cdot\bar\osca^4)^{q}
    (\bar\osca^2\cdot\bar\osca^3)^{v}
    (\bar\osca^2\cdot\bar\osca^4)^{w} \vert 0\rangle\,,
  \end{aligned}
\end{align}
}
where
\begin{align}
  p=s_1-q\,,\quad w=s_4-q\,,\quad v=s_3-s_1+q\,.
\end{align}
The combinatorial coefficient corresponds to the number of possible
contractions of $q$ of the oscillators in space $1$ with oscillators
in space $4$ where the remaining $p=s_1-q$ oscillators are contracted
with those in space $3$ and in addition the oscillators in space $2$
are contracted with those that are still left in spaces $3$ and
$4$. This is illustrated in
Figure~\ref{fig:glue_details_combinatorics}.

\begin{figure}
  \begin{center}
    \includegraphics{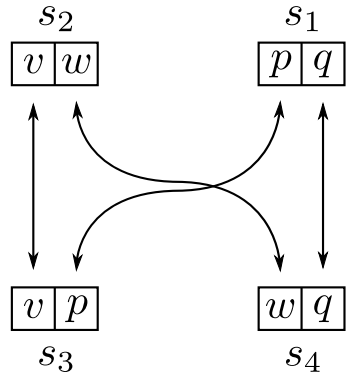}    
  \end{center}
  \caption{Contractions of oscillators for gluing of three-site
    invariants.}
  \label{fig:glue_details_combinatorics}
\end{figure}

Instead of immediately composing four three-site invariants
analogously to the composition of the four-leg amplitude, it is
helpful to study parts of this composition first. We glue the two
three-site invariants \eqref{eq:osc-psi31} and \eqref{eq:osc-psi32} in the
following way:
\begin{align}
  \label{eq:gluing_3vertex1}
  \mathcal{O}_{\Delta(s_2,s_3,k)}=
  \begin{aligned}
   \includegraphics[width=0.3\textwidth]{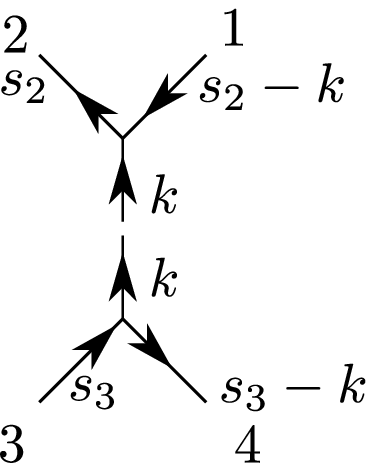}
  \end{aligned}.
\end{align}
In order for the representation labels of all lines to be positive we
have to require $k=0,1,\ldots,\min(s_2,s_3)$. Again the gluing is done
by a direct computation in the operatorial form of the invariants. To
get a compact expression for the result we conjugate in spaces $1$ and
$2$ and obtain
\begin{align}
  \vert \Delta(s_2,s_3,k)\rangle
  =k!
  (\bar\oscb^2\cdot\bar\osca^1)^{s_3-k}
  (\bar\oscb^2\cdot\bar\osca^3)^k
  (\bar\oscb^4\cdot\bar\osca^3)^{s_3-k}
  \vert 0\rangle\,.
\end{align}
The two three-site invariants can also be combined in a different way:
\begin{align}
  \label{eq:gluing_3vertex2}
  \mathcal{O}_{\Theta(s_1,s_4,k)}=
  \begin{aligned}
     \includegraphics[width=0.4\textwidth]{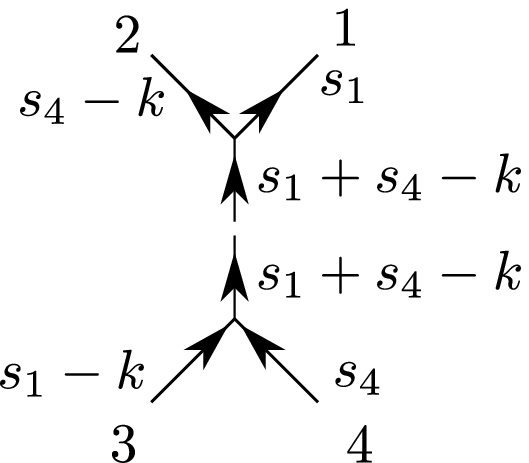}
  \end{aligned}.
\end{align}
In this case the combinatorics is more involved due to the double
line. Using \eqref{combii} we find
\begin{align}
  \begin{aligned}
    \vert \Theta(s_1,s_4,k)\rangle&=\sum_{q=k}^{\min(s_1,s_4)}
    \frac{(s_1-k)!(s_4-k)!s_1!s_4!}{(s_1-q)!(s_4-q)!q!(q-k)!}\\
    &\hspace{2.5cm}\cdot
    (\bar\oscb^1\cdot\bar\osca^3)^{s_1-q}
    (\bar\oscb^1\cdot\bar\osca^4)^{q}
    (\bar\oscb^2\cdot\bar\osca^3)^{q-k}
    (\bar\oscb^2\cdot\bar\osca^4)^{s_4-q} \vert 0\rangle\,,
  \end{aligned}
\end{align}
where we have to restrict to $k=0,1,\ldots,\min(s_1,s_4)$.  

Finally, we combine the two parts \eqref{eq:gluing_3vertex1} and
\eqref{eq:gluing_3vertex2} into
\begin{align}
  \label{eq:gluing_3vertex_bothparts}
  \mathcal{O}_{\Omega(s_3,s_4,k)}=
  \begin{aligned}
    \includegraphics[width=0.4\textwidth]{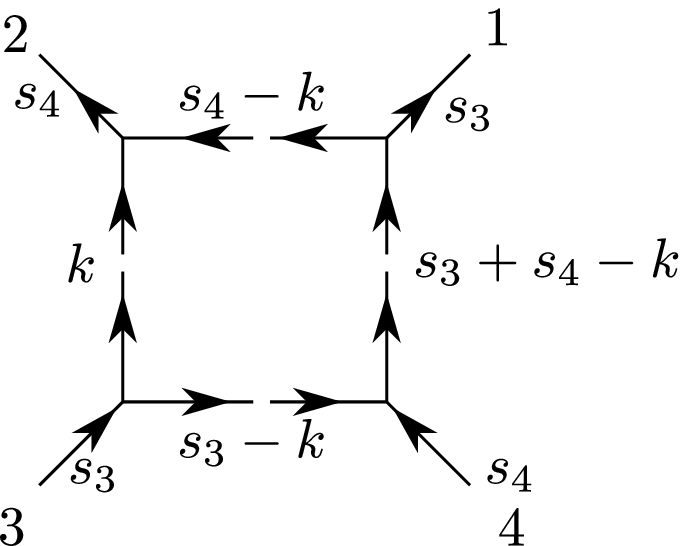}
  \end{aligned}.
\end{align}
The two remaining contractions connecting the parts can again be considered
as special cases of \eqref{combii} leading to
\begin{align}
  \label{eq:gluing_3vertex_four}
  \begin{aligned}
    \vert \Omega(s_3,s_4,k)\rangle&= \sum_{q=k}^{\min(s_3,s_4)}
    \frac{(s_3-k)!^2(s_4-k)!^2s_3!s_4!k!}{(s_3-q)!(s_4-q)!q!(q-k)!}\\
    &\hspace{2.5cm}\cdot (\bar\oscb^1\cdot\bar\osca^3)^{s_3-q}
    (\bar\oscb^1\cdot\bar\osca^4)^{q}
    (\bar\oscb^2\cdot\bar\osca^3)^{q}
    (\bar\oscb^2\cdot\bar\osca^4)^{s_4-q} \vert 0\rangle\,.
  \end{aligned}
\end{align}
These invariants correspond to the R-matrix in vector form
\eqref{eq:osc-psi42} at special points of the spectral parameter
\begin{align}
  \label{eq:gluing_3vertex_identification}
  \vert \Omega(s_3,s_4,k)\rangle=
  s_3!s_4!k!(s_3-k)!^2(s_4-k)!^2\vert\Psi_{4,2}(s_3-k)\rangle\,,
\end{align}
for $k=0,1,\ldots,\min(s_3,s_4)$. If we allow also complex values for
$k$ in \eqref{eq:gluing_3vertex_four} and thus replace the factorials
by gamma functions, the expression literally agrees with
\eqref{eq:osc-psi42}. This suggests that, analogously to the four-leg
amplitude, the R-matrix with a generic spectral parameter might be
understood as a composition of four trivalent vertices. In such an
interpretation the internal lines in
\eqref{eq:gluing_3vertex_bothparts} would carry complex labels and
thus leave the realm of finite-dimensional representations. 

Alternatively, the complete R-matrix can be reconstructed from the
values at the special points,
\begin{align}
  \vert\Psi_{4,2}(z)\rangle
  =\frac{1}{\Gamma{(z-s_3)}}\sum_{k=0}^{\min(s_3,s_4)}
  \frac{(-1)^k}{k!(z+k-s_3)}
  \vert\Psi_{4,2}(s_3-k)\rangle\,.
\end{align}
With \eqref{eq:gluing_3vertex_identification} this is understood as an
expansion of the R-matrix in terms of the diagrams in
\eqref{eq:gluing_3vertex_bothparts}. So if we stick to
finite-dimensional representations also for the internal lines in
\eqref{eq:gluing_3vertex_bothparts}, in contrast to the four-leg
amplitude, the R-matrix is not a simple combination of four
three-site invariants but a weighted sum over all diagrams.\footnote{We like to thank Nils Kanning for checking and discussing about the results presented in this appendix.} 


\chapter{Hopping Hamiltonians}\label{app:hop}
Fixing the normalization of the R-matrix in \eqref{eq:osc-hoppingr} to be given by
\begin{equation}
  R_{\s,\s}(\inhdiff)
  =\rho(\inhdiff)\,\frac{\Gamma(\inhdiff +1)^2}{\Gamma(\inhdiff +1+s)}
  \sum_{k=0}^{s}
  \frac{k!}{\Gamma(\inhdiff-s+k+1)}\,\text{Hop}_k\,,
\end{equation} 
we obtain the unitarity relation
\begin{equation}
   R_{\s,\s}(\inhdiff)  R_{\s,\s}(-\inhdiff)=\rho(u)\rho(-u)\,.
\end{equation} 
To extract the Hamiltonian we take the logarithmic derivative, see \eqref{logderR}, and find
\begin{equation}\label{hophami}
 \mathcal{H}=\sum_{k=0}^s c_{k,s}\,\text{Hop}_k\,,
\end{equation} 
with 
\begin{equation}
 c_{0,s}=-h(s)\,\quad\quad c_{k,s}=(-1)^{k+1}\frac{\Gamma(k)\Gamma(1+s-k)}{\Gamma(1+s)}\,.
\end{equation} 
This type of Hamiltonian was discussed in \cite{Beisert:2004ry} and referred to as Harmonic Action. Apart from the underlying symmetry, the major difference is the explicit realization\,\footnote{This formalism was developed with Tomasz \L{}ukowski.} of the operators $\text{Hop}_k$. In \cite{Zwiebel2007} it was shown that the action of the Hamiltionian on a general state
\begin{equation}\label{genstate}
 \vert {\bf i},{\bf j}\rangle=\prod_{k=1}^{s}\prod_{l=1}^{s}{\bar{\mathbf{a}}}_{i_{k}}{\bar{\mathbf{b}}}_{j_{l}}\vert0\rangle\,,
\end{equation} 
can be written as $\mathcal{H}=\mathcal{H}_0+\mathcal{H}'$ with
\begin{equation}\label{zwiebelh}
  \mathcal{H} \vert {\bf i},{\bf j}\rangle= \mathcal{H}_0 \vert {\bf i},{\bf j}\rangle-\frac{1}{\pi}\int_0^{\pi/2}d\theta\int_0^{2\pi}d\varphi\cot\theta\prod_{k=1}^{s}\prod_{l=1}^{s}\hat{\bar{\mathbf{a}}}_{i_{k}}(\theta,\varphi)\hat{\bar{\mathbf{b}}}_{j_{l}}(\theta,\varphi)\vert0\rangle\,.
\end{equation} 
Here the hatted oscillators denote the rotation 
\begin{equation}\label{rota}
\hat{\bar{\mathbf{a}}}(\theta,\varphi)=\cos\theta\,\bar\osca-e^{+i\varphi}\,\sin\theta\,\bar\oscb\,,
\end{equation} 
\begin{equation}\label{rotb}
\hat{\bar{\mathbf{b}}}(\theta,\varphi)=\cos\theta\,\bar\oscb+e^{-i\varphi}\,\sin\theta\,\bar\osca\,,
\end{equation} 
and $\mathcal{H}_0 $ is introduced to regulate the diagonal contribution
\begin{equation}
 \mathcal{H}_0 =\frac{1}{\pi}\int_0^{\pi/2}d\theta\int_0^{2\pi}d\varphi\left(\cot\theta\cos^{2s}\theta-\frac{h_s}{\pi}\right) \,,
\end{equation} 
with the harmonic numbers $h_s$.
\subsection*{$
\psset{xunit=.1pt,yunit=.1pt,runit=.1pt}
\begin{pspicture}(226.71427917,323.85714722)
{
\newrgbcolor{curcolor}{0.80000001 0.80000001 0.80000001}
\pscustom[linestyle=none,fillstyle=solid,fillcolor=curcolor]
{
\newpath
\moveto(226.21428,111.92857722)
\curveto(226.21428,26.71892722)(178.54355,0.49999722)(116.21428,0.49999722)
\curveto(53.885,0.49999722)(0.5,26.71892722)(0.5,111.92857722)
\curveto(0.5,197.13822722)(56.74214,323.35714722)(119.07142,323.35714722)
\curveto(181.40069,323.35714722)(226.21428,197.13822722)(226.21428,111.92857722)
\closepath
}
}
{
\newrgbcolor{curcolor}{0 0 0}
\pscustom[linewidth=1,linecolor=curcolor]
{
\newpath
\moveto(226.21428,111.92857722)
\curveto(226.21428,26.71892722)(178.54355,0.49999722)(116.21428,0.49999722)
\curveto(53.885,0.49999722)(0.5,26.71892722)(0.5,111.92857722)
\curveto(0.5,197.13822722)(56.74214,323.35714722)(119.07142,323.35714722)
\curveto(181.40069,323.35714722)(226.21428,197.13822722)(226.21428,111.92857722)
\closepath
}
}
{
\newrgbcolor{curcolor}{1 1 1}
\pscustom[linestyle=none,fillstyle=solid,fillcolor=curcolor]
{
\newpath
\moveto(102.28571,1.77277722)
\curveto(101.10714,1.93669722)(97.08928,2.41727722)(93.35714,2.84073722)
\curveto(45.48852,8.27212722)(14.7099,34.17233722)(5.14481,77.07142722)
\curveto(2.45943,89.11524722)(1.68868,97.19357722)(1.73733,112.78570722)
\curveto(1.84548,147.45080722)(10.29532,186.43260722)(25.9314,224.40059722)
\curveto(47.92273,277.80060722)(78.41727,313.69683722)(108,321.00652722)
\curveto(114.19596,322.53750722)(124.32145,322.55081722)(130.14286,321.03562722)
\curveto(135.97036,319.51885722)(146.29063,314.30249722)(151.95119,310.01265722)
\curveto(186.72983,283.65574722)(216.79892,211.48179722)(224.15073,136.71426722)
\curveto(225.26138,125.41894722)(225.24741,99.27456722)(224.12583,90.28569722)
\curveto(221.5919,69.97613722)(216.26444,54.02120722)(207.54974,40.64284722)
\curveto(203.6527,34.66029722)(191.46517,22.47674722)(185.49993,18.60023722)
\curveto(173.68232,10.92056722)(159.38605,5.84736722)(141.9285,3.13840722)
\curveto(134.81678,2.03484722)(107.23347,1.08464722)(102.28564,1.77276722)
\lineto(102.28567,1.77276722)
\closepath
}
}
{
\newrgbcolor{curcolor}{0 0 0}
\pscustom[linewidth=0.71428573,linecolor=curcolor]
{
\newpath
\moveto(102.28571,1.77277722)
\curveto(101.10714,1.93669722)(97.08928,2.41727722)(93.35714,2.84073722)
\curveto(45.48852,8.27212722)(14.7099,34.17233722)(5.14481,77.07142722)
\curveto(2.45943,89.11524722)(1.68868,97.19357722)(1.73733,112.78570722)
\curveto(1.84548,147.45080722)(10.29532,186.43260722)(25.9314,224.40059722)
\curveto(47.92273,277.80060722)(78.41727,313.69683722)(108,321.00652722)
\curveto(114.19596,322.53750722)(124.32145,322.55081722)(130.14286,321.03562722)
\curveto(135.97036,319.51885722)(146.29063,314.30249722)(151.95119,310.01265722)
\curveto(186.72983,283.65574722)(216.79892,211.48179722)(224.15073,136.71426722)
\curveto(225.26138,125.41894722)(225.24741,99.27456722)(224.12583,90.28569722)
\curveto(221.5919,69.97613722)(216.26444,54.02120722)(207.54974,40.64284722)
\curveto(203.6527,34.66029722)(191.46517,22.47674722)(185.49993,18.60023722)
\curveto(173.68232,10.92056722)(159.38605,5.84736722)(141.9285,3.13840722)
\curveto(134.81678,2.03484722)(107.23347,1.08464722)(102.28564,1.77276722)
\lineto(102.28567,1.77276722)
\closepath
}
}
{
\newrgbcolor{curcolor}{0.80000001 0.80000001 0.80000001}
\pscustom[linestyle=none,fillstyle=solid,fillcolor=curcolor]
{
\newpath
\moveto(95.50001165,88.00001347)
\curveto(95.50001165,75.17911794)(84.3071304,64.78572857)(70.50001165,64.78572857)
\curveto(56.6928929,64.78572857)(45.50001165,75.17911794)(45.50001165,88.00001347)
\curveto(45.50001165,100.82090899)(56.6928929,111.21429836)(70.50001165,111.21429836)
\curveto(84.3071304,111.21429836)(95.50001165,100.82090899)(95.50001165,88.00001347)
\closepath
}
}
{
\newrgbcolor{curcolor}{0 0 0}
\pscustom[linewidth=1,linecolor=curcolor]
{
\newpath
\moveto(95.50001165,88.00001347)
\curveto(95.50001165,75.17911794)(84.3071304,64.78572857)(70.50001165,64.78572857)
\curveto(56.6928929,64.78572857)(45.50001165,75.17911794)(45.50001165,88.00001347)
\curveto(45.50001165,100.82090899)(56.6928929,111.21429836)(70.50001165,111.21429836)
\curveto(84.3071304,111.21429836)(95.50001165,100.82090899)(95.50001165,88.00001347)
\closepath
}
}
{
\newrgbcolor{curcolor}{0.80000001 0.80000001 0.80000001}
\pscustom[linestyle=none,fillstyle=solid,fillcolor=curcolor]
{
\newpath
\moveto(204.7857134,144.78571903)
\curveto(204.7857134,130.97860029)(194.55222173,119.78571903)(181.92857,119.78571903)
\curveto(169.30491827,119.78571903)(159.0714266,130.97860029)(159.0714266,144.78571903)
\curveto(159.0714266,158.59283778)(169.30491827,169.78571903)(181.92857,169.78571903)
\curveto(194.55222173,169.78571903)(204.7857134,158.59283778)(204.7857134,144.78571903)
\closepath
}
}
{
\newrgbcolor{curcolor}{0 0 0}
\pscustom[linewidth=1,linecolor=curcolor]
{
\newpath
\moveto(204.7857134,144.78571903)
\curveto(204.7857134,130.97860029)(194.55222173,119.78571903)(181.92857,119.78571903)
\curveto(169.30491827,119.78571903)(159.0714266,130.97860029)(159.0714266,144.78571903)
\curveto(159.0714266,158.59283778)(169.30491827,169.78571903)(181.92857,169.78571903)
\curveto(194.55222173,169.78571903)(204.7857134,158.59283778)(204.7857134,144.78571903)
\closepath
}
}
{
\newrgbcolor{curcolor}{0.80000001 0.80000001 0.80000001}
\pscustom[linestyle=none,fillstyle=solid,fillcolor=curcolor]
{
\newpath
\moveto(121.92857954,196.92857182)
\curveto(121.92857954,183.51594173)(111.05549452,172.64285671)(97.64286443,172.64285671)
\curveto(84.23023434,172.64285671)(73.35714933,183.51594173)(73.35714933,196.92857182)
\curveto(73.35714933,210.34120191)(84.23023434,221.21428692)(97.64286443,221.21428692)
\curveto(111.05549452,221.21428692)(121.92857954,210.34120191)(121.92857954,196.92857182)
\closepath
}
}
{
\newrgbcolor{curcolor}{0 0 0}
\pscustom[linewidth=1,linecolor=curcolor]
{
\newpath
\moveto(121.92857954,196.92857182)
\curveto(121.92857954,183.51594173)(111.05549452,172.64285671)(97.64286443,172.64285671)
\curveto(84.23023434,172.64285671)(73.35714933,183.51594173)(73.35714933,196.92857182)
\curveto(73.35714933,210.34120191)(84.23023434,221.21428692)(97.64286443,221.21428692)
\curveto(111.05549452,221.21428692)(121.92857954,210.34120191)(121.92857954,196.92857182)
\closepath
}
}
\end{pspicture}\quad$Beam splitter Hamiltonian}

We will now derive an expression for the Hamiltonian from its action given in \eqref{zwiebelh} as we did in \eqref{hophami} for the Harmonic Action in \cite{Beisert:2004ry}, i.e. find an explicit form of $\mathcal{H}'$ in \eqref{zwiebelh}. Furthermore, as we will see, there is an amusing way to interpret the action of the Hamiltonian density and in particular the coefficients appearing in the expansion \eqref{hophami}. 

Let us first  define the operator
\begin{equation}
S\left(z,\bar{z}\right)=e^{z\,\bar{\mathbf{a}}\mathbf{b}-\bar{z}\,\mathbf{a}\bar{\mathbf{b}}}\,,
\end{equation}
where $\bar z$ denotes the complex conjugate of the varianble $z$.
It can be used to write the rotation of the oscillators in \eqref{rota} and \eqref{rotb} as a similarity transformation 
\begin{equation}\label{transa}
\hat{\bar{\mathbf{a}}}(z,\bar{z})=S\left(z,\bar{z}\right)\bar{\mathbf{a}}\,S^{-1}\left(z,\bar{z}\right)=\cos\vert z\vert\bar\osca-\bar{z}\,\frac{\sin\vert z\vert}{\vert z\vert}\bar\oscb
\end{equation} 
\begin{equation}\label{transb}
\hat{\bar{\mathbf{b}}}(z,\bar{z})=S\left(z,\bar{z}\right)\bar\oscb\, S^{-1}\left(z,\bar{z}\right)=\cos\vert z\vert\bar\oscb+z\,\frac{\sin\vert z\vert}{\vert z\vert}\bar\osca
\end{equation} 
with $z=\theta\, e^{-i\,\varphi}$ where $0\leq\theta\leq\pi/2$ and $0\leq\varphi\leq2\pi$. Its action on the Fock vacuum $\vert0\rangle$ is given by
\begin{equation}\label{transvac}
S\left(\theta\right)\vert00\rangle=S^{-1}\left(\theta\right)\vert00\rangle=\vert00\rangle
\end{equation} 
for all $\theta.$
Using
\begin{equation}
 S_2(z,\bar z)=e^{\sum_{i=1}^{2}\left(z\,\bar{\mathbf{a}}_{i}\mathbf{b}_{i}-\bar{z}\,\mathbf{a}_{i}\bar{\mathbf{b}}_{i}\right)}\,,
\end{equation} 
the transformation rules \eqref{transa} and \eqref{transb} and the action of $S$ on the Fock vacuum \eqref{transvac} we obtain 
\begin{equation}
e^{\sum_{i=1}^{2}\left(z\,\bar{\mathbf{a}}_{i}\mathbf{b}_{i}-\bar{z}\,\mathbf{a}_{i}\bar{\mathbf{b}}_{i}\right)} \vert {\bf i},{\bf j}\rangle=\prod_{k=1}^{s}\prod_{l=1}^{s}\hat{\bar{\mathbf{a}}}_{i_{k}}(z,\bar{z})\hat{\bar{\mathbf{b}}}_{j_{l}}(z,\bar{z})\vert00\rangle.
\end{equation} 
and the explicit form of the non-diagonal part of the Hamiltonian 
\begin{equation}
 \mathcal{H}'=-\frac{1}{\pi}\int_0^{\pi/2}d\theta\int_0^{2\pi}d\varphi\cot\theta\, S_2(\theta,\varphi)\,.
\end{equation} 

The operator $S_2(z,\bar{z})=S_2(\theta,\varphi)$ can be interpreted
as a quantum gate, see e.g. \cite{Gerry2005}, as shown in Figure~\ref{fig:oneparticle}. To clarify
this statement we act with it on the tensor state with one particle
at site ``1'' and no particles at site ``2'', i.e. $\bar{\mathbf{a}}\vert00\rangle=\vert10\rangle$.
\begin{figure}
\begin{centering}
\includegraphics[clip,scale=0.5]{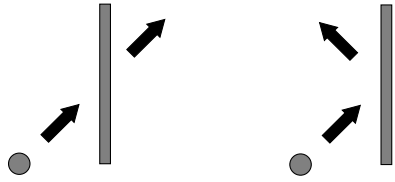}
\par\end{centering}

\protect\caption{Possible scattering with a particle at one site and none on the other.\label{fig:oneparticle} }
\end{figure}
We find that 
\[
S_2(\theta,\varphi)\vert10\rangle=\cos\theta\vert10\rangle-e^{i\varphi}\,\sin\theta\vert01\rangle.
\]
This is a normalized entangled state between the two sites of the
chain parametrized through the angles of the Bloch sphere. The $\theta$
angle can be seen as the angle under which the quanta hit the splitter.
The angle $\varphi$ denotes a phase difference that the quanta pick
up when they pass through the gate. We can define the transmission
coefficient as $T=\sin^{2}\theta$ and the reflection coefficient
as $R=\cos^{2}\theta$. In the $\gl2$ case we have two different kinds of oscillators
and the total number of them at each site is fixed by the representation
label. Each kind of oscillator we assign one color white or gray and
interpret them as particles. Take the state where two red particles
are on the first site and two gray on the second. As
we will integrate over $\varphi$ all configurations that do not preserve
particle numbers at each site will drop out, compare Figure~\ref{fig:4particles}.

\begin{figure}
\begin{centering}
\includegraphics[scale=0.5]{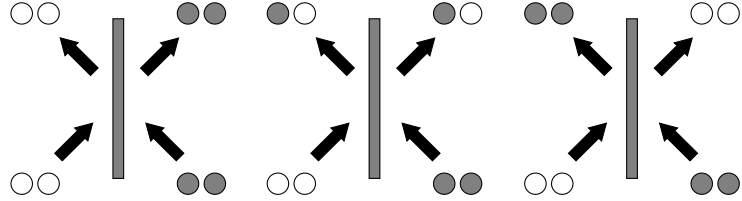}
\par\end{centering}

\protect\caption{Possible scattering with two particles at each site. \label{fig:4particles}}
\end{figure}

\addcontentsline{toc}{chapter}{References}
\bibliographystyle{hieeetr}
\bibliography{diss}
\end{document}